\documentclass[graybox,envcountchap]{svmono-mod}
\usepackage{helvet}
\usepackage{courier}
\usepackage{type1cm}         

\usepackage{makeidx}         % allows index generation
\usepackage{graphicx}        % standard LaTeX graphics tool
\usepackage{multicol}        % used for the two-column index
\usepackage[bottom]{footmisc}% places footnotes at page bottom

\makeindex             % used for the subject index
\usepackage{bm,color,amsmath,amssymb,mathrsfs,latexsym,graphicx,psfrag}
\usepackage{pst-all,epsfig,wrapfig,subfigure,rotating,wasysym} 

\newcommand{\bs}[1]{\boldsymbol{#1}}
\newcommand{\braket}[2]{\left\langle #1 | #2 \right\rangle}

\newcommand{\bra}[1]{\left\langle#1\right|}
\newcommand{\ket}[1]{\left|#1\right\rangle}
\newcommand{\bigket}[1]{\bigl|#1\bigr\rangle}
\newcommand{\textket}[1]{|#1\rangle}

\newcommand{\comm}[2]{\left[#1,#2\right]}
\newcommand{\bigcomm}[2]{\bigl[#1,#2\bigr]}
\newcommand{\Bigcomm}[2]{\Bigl[#1,#2\Bigr]}

\newcommand{\anticomm}[2]{\left\{#1,#2\right\}}
\newcommand{\biganticomm}[2]{\bigl\{#1,#2\bigr\}}

\newcommand{\half}{$\frac{1}{2}$ }
\renewcommand{\i}{\text{i}}
\renewcommand{\r}{\text{r}}
\newcommand{\x}{\text{x}}
\newcommand{\y}{\text{y}}
\newcommand{\z}{\text{z}}

\newcommand{\up}{\uparrow}
\newcommand{\dw}{\downarrow}
\newcommand{\vac}{\left|0\right\rangle}
\newcommand{\bac}{\left\langle 0\right|}
\newcommand{\abs}[1]{\left|#1\right|}

\newcommand{\parder}[2]{\frac{\partial #1}{\partial #2}}

\DeclareMathOperator{\sign}{sign}

\newcommand {\calpha}[1]{\alpha_{#1}}
\newcommand {\etba}{\bar\eta}
\newcommand {\et}[1]{\eta_{\calpha{#1}}}
\newcommand {\etb}[1]{\bar\eta_{\calpha{#1}}}

\newcommand{\Om}[1]{\Omega_{\calpha{#1}}}

\renewcommand{\a}{\alpha}
\renewcommand{\b}{\beta}
\renewcommand{\c}{\gamma}
\newcommand {\ea}{\eta_{\alpha}}
\newcommand {\eb}{\eta_{\beta}}
\newcommand {\ec}{\eta_{\gamma}}

\newcommand {\eab}{\bar\eta_{\alpha}}
\newcommand {\ebb}{\bar\eta_{\beta}}
\newcommand {\ecb}{\bar\eta_{\gamma}}

\newcommand {\Sa}{S_{\alpha}}
\newcommand {\Sb}{S_{\beta}}
\newcommand {\Sc}{S_{\gamma}}

\newcommand {\bsS}{\bs{S}}
\newcommand {\bSa}{\bs{S}_{\alpha}}
\newcommand {\bSb}{\bs{S}_{\beta}}
\newcommand {\bSc}{\bs{S}_{\gamma}}
\newcommand {\bStot}{\bs{S}_{\text{tot}}}
\newcommand {\OaHS}{\Omega_{\alpha}^{\s\text{HS}}}
\newcommand {\OabHS}{\bar\Omega_{\alpha}^{\s\text{HS}}}
\newcommand {\OaSone}{\Omega_{\alpha}^{S=1}}
\newcommand {\OabSone}{\bar\Omega_{\alpha}^{S=1}}
\newcommand {\OaS}{\Omega_{\alpha}^{S}}
\newcommand {\OabS}{\bar\Omega_{\alpha}^{S}}
\renewcommand{\strut}{\rule[-.9\baselineskip]{0pt}{2.5\baselineskip}}
\def\s{\scriptscriptstyle}
\def\chap{chapter}
\def\Chap{Chapter}
\def\Chaps{Chapters}
\def\defining{defining }
\def\paper{monograph}
\def\book{book}
\newlength{\ytlength}
\def\ie{{i.e.},\ }
\def\eg{{e.g.}\ }
\def\etal{{et al.}}
\def\etc{{etc.}\ }

\def\cf{{cf.}\ }

\renewcommand{\theequation}{\thesubsection.\arabic{equation}}
\numberwithin{equation}{section}

\allowdisplaybreaks[3]

\begin{document}

\author{Martin Greiter}
\title{Mapping of parent Hamiltonians:\\ from Abelian and non-Abelian
  quantum Hall states to\\ exact models of critical spin chains}
\subtitle{-- Monograph --}
%\subtitle{-- Habilitationsschrift --}
\date{April 2011}
\date{}
\maketitle

\frontmatter%%%%%%%%%%%%%%%%%%%%%%%%%%%%%%%%%%%%%%%%%%%%%%%%%%%%%%%%%
% Creates the german title
% \dttitlehead{Deutscher Titel} 
% \dttitlehead{\large\rm -- Deutscher Titel --} 
% \dttitle{Abbildungen von Hamiltonoperatoren: von abelschen und
%   nichtabelschen Quantenhallzust\"anden zu exakten Modellen
%   kritischer Spinketten}
% \dtsubtitle{-- Habilitationsschrift --} 
% \date{}
% \declaration{Gem\"a{\ss} \S 4(1) der Habilitationsordnung des KIT vom
%   23. November 1999 versichere ich, dass diese Habilitationsleistung
%   selbst\"anding verfasst und keine anderen als die angegebenen
%   Quellen und Hilfsmittel benutzt wurden.}  
% \dtmaketitle

% \include{dedic}
% \include{foreword}
% \include{zfassung}

%%%%%%%%%%%%%%%%%%%%%%preface.tex%%%%%%%%%%%%%%%%%%%%%%%%%%%%%%%%%%%%%%%%%
% sample preface
%
% Use this file as a template for your own input.
%
%%%%%%%%%%%%%%%%%%%%%%%% Springer %%%%%%%%%%%%%%%%%%%%%%%%%%

\preface

%% Please write your preface here
% Use the template \emph{preface.tex} together with the Springer document class SVMono (monograph-type books) or SVMult (edited books) to style your preface in the Springer layout.
% 
% A preface\index{preface} is a book's preliminary statement, usually written by the \textit{author or editor} of a work, which states its origin, scope, purpose, plan, and intended audience, and which sometimes includes afterthoughts and acknowledgments of assistance. 
% 
% When written by a person other than the author, it is called a foreword. The preface or foreword is distinct from the introduction, which deals with the subject of the work.
% 
% Customarily \textit{acknowledgments} are included as last part of the preface.

The immediate advance we communicate with this \paper\ is the
discovery of an exact model for a critical spin chain with arbitrary
spin $S$, which includes the Haldane--Shastry model as the special
case $S=\frac{1}{2}$.  For $S\ge 1$, we propose that the spinon
excitations obey a one-dimensional version of non-Abelian statistics,
where the topological degeneracies are encoded in the fractional
momentum spacings for the spinons.  The model and its properties,
however, are not the only, and possibly not even the most important
thing one can learn from the analysis we present.

The benefit of science may be that it honors the human spirit, gives
pleasure to those who immerse themselves in it, and pragmatically,
contributes to the improvement of the human condition in the long
term.  The purpose of the individual scientific work can hence be
either a direct contribution to this improvement, or more often an
indirect contribution by making an advance which inspires further
advances in a field.  When we teach Physics, be it in lectures, books,
monographs, or research papers, we usually teach what we understand,
but rarely spend much effort on teaching how this understanding was
obtained.  The first volume of the famed course of theoretical physics
by L.D.~Landau and E.M.~Lifshitz~\cite{Landau1}, for example, begins
by stating the principle of least action, but does nothing to motivate
how it was discovered historically or how one could be led to discover
it from the study of mechanical systems.
%\footnote{The newer editions actually begin with a 19 page biography 
%of the senior author.}. %, but this does not completely erase the point.}.  
This reflects that we teach our students how to apply certain
principles, but not how to discover or extract such principles from a
given body of observations.  The reason for this is not that we are
truely content to teach students of physics as if they were students
of engineering, but that the creative process in physics is usually
erratic and messy, if not plainly embarrassing to those actively
involved, and hence extremely difficult to recapture.  As with most of
what happens in reality, the actual paths of discovery are usually
highly unlikely.  Since we enjoy the comfort of perceiving actions
and events as more likely and sensible, our minds subconsciously
filter our memory to this effect.

One of the first topics I immersed myself in after completing my
graduate coursework was Laughlin's theory of the fractionally
quantized Hall effect~\cite{laughlin83prl1395}.  I have never
completely moved away from it, as this work testifies, and take
enormous delight whenever I recognize quantum Hall physics in other
domains of physics.  More important than the theory itself, however,
was to me to understand and learn from the way R.B. Laughlin actually
discovered the wave function.  He numerically diagonalized a system of
three electrons in a magnetic field in an open plane, and observed
that the total canonical angular momentum around the origin jumped by
a factor of three (from $3\hbar$ to $9\hbar$) when he implemented a
Coulomb interaction between the electrons.  At the same time, no
lesser scientists than D.~Yoshioka, B.I.~Halperin, and
P.A.~Lee~\cite{yoshioka-83prl1219} had, in an heroic effort,
diagonalized up to six electrons with periodic boundary conditions, and
concluded that their data were ``supportive of the idea that the
ground state is not crystalline, but a translationally invariant
``liquid.''\hspace{1pt}'' Their analysis was much more distinguished
and scholarly, but unfortunately, did not 
%anoint the authors with a Nobel price.
yield the wave function.

The message I learned from this episode is that it is often beneficial
to leave the path of scholarly analysis, and play with the simplest
system of which one may hope that it might give away natures thoughts.
For the Laughlin series of quantized Hall states, this system
consisted of three electrons.  I spend most of my scientific life
adapting this approach to itinerant antiferromagnets in two
dimensions, where I needed to go to twelve lattice sites until I could
grasp what nature had in mind.  But I am digressing.  To complete the
story about the discovery of the quantum Hall effect, Laughlin gave a
public lecture in Amsterdam within a year of having received the Nobel
price.  He did not mention how he discovered the state, and at first
couldn't recall it when I asked him in public after the lecture.  As
he was answering other questions, he recalled the answer to mine and
weaved it into the answer of another question.  During the %same
evening in a cafe, a very famous Russian colleague whom I regard with
the utmost respect commented the story of the discovery with the words
%``But that's stupid!''.
``But this is stupid!''.

Maybe it is.  If it is so, however, the independent discoveries of the
spin \half\ model by F.D.M.~Haldane~\cite{haldane88prl635} and
B.S.~Shastry~\cite{shastry88prl639} may fall into the same category.
Unfortunately, I do not know much about these discoveries.  Haldane told
me that he first observed striking degeneracies when he looked at the
model for $N=6$ sites numerically, motivated by the fact that the
$1/r^2$ exchange is the discrete Fourier transform of $\epsilon(k)=k\,(k-2\pi)$
in one dimension.  Shastry told me that he discovered it ``by doing
calculations'', which is not overly instructive to future generations.
If my discovery of the general model I document in this \paper\ will
be perceived in the spirit of my friends comment, I will at least have
made no attempt to evade the charge.

In short, what I document on these pages is not just an exact model,
but a precise and reproducible account of how I discovered this model.
This reflects my belief that the path of discovery can be as
instructive to future generations as the model itself.  Of course, the
analysis I document does not fully reflect the actual path of
discovery, but what would have been the path if my thinking had
followed a straight line.  It took me about four weeks to obtain all the
results and about four months to write this \paper .  The reason for
this discrepancy is not that my writing proceeds slowly, but that I
had left out many intermediate steps when I did the
calculation.  The actual path of discovery must have been highly
unlikely.  In any event, it is comforting to me that, now that I have
written a scholarly and coherent account of it, there is little
need to recall what actually might have happened.

I am deeply grateful to Ronny Thomale for countless discussions and
his critical reading of the manuscript, to %Max F\"uhringer,
Burkhard Scharfenberger, Dirk Schuricht, and Stephan Rachel for
collaborations on various aspects of quantum spin chains,
to Rose Schrempp and
the members of the Institute for Theory of Condensed Matter at KIT for
providing me with a pleasant and highly stimulating atmosphere, and
especially to Peter W\"olfle for his continued encouragement and
support.

\vspace{\baselineskip}
\begin{flushright}\noindent
% Place(s),\hfill {\it Firstname  Surname}\\
% month year\hfill {\it Firstname  Surname}\\
Karlsruhe, April 2011\hfill {\it Martin Greiter}\\
\end{flushright}

\tableofcontents

\mainmatter%%%%%%%%%%%%%%%%%%%%%%%%%%%%%%%%%%%%%%%%%%%%%%%%%%%%%%%%%%

\numberwithin{equation}{chapter}
\chapter{Introduction and summary}

Fractional quantization, and in particular {fractional
statistics}~\cite{wilczek90,stern08ap204}, in two-di\-men\-sio\-nal
quantum liquids is witnessing a renaissance of interest in present
times.  The field started more than a quarter of a century ago with the
discovery of the fractional quantum Hall effect, which was explained
by Laughlin~\cite{laughlin83prl1395} in terms of an incompressible
quantum liquid supporting fractionally charged (vortex or)
quasiparticle excitations.  When formulating a hierarchy of quantized
Hall states~\cite{haldane83prl605,halperin84prl1583,greiter94plb48} to
explain the observation of quantized Hall states at other filling
fractions fractions, Halperin~\cite{halperin84prl1583} noted that
these excitations obey fractional statistics, and are hence
conceptually similar to the charge-flux tube composites introduced by
Wilczek two years earlier~\cite{wilczek82prl957}.  Physically, the
fractional statistics manifests itself through fractional quantization
of the kinematical relative angular momenta of the anyons.

The interest was renewed a few years later, when
Anderson~\cite{anderson87s1196} proposed that hole-doped Mott
insulators, and in particular the $t$--$J$
model~\cite{zhang-88prb3759,eskes-88prl1415} universally believed to
describe the CuO planes in high $T_\text{c}$
superconductors~\cite{zaanen-06np138,orenstein-00s468}, can be
described in terms of a spin liquid (\ie a state with strong, local
antiferromagnetic correlations but without long range order), which
would likewise support fractionally quantized excitations.  In this
proposal, the excitations are spinons and holons, which carry spin
\half and no charge or no spin and charge $+e$, respectively.  The
fractional quantum number of the spinon is the spin, which is half
integer while the Hilbert space (for the undoped system) is built up
of spin flips, which carry spin one.  One of the earliest proposals
for a spin liquid supporting deconfined spinon and holon excitations
is the (Abelian) chiral spin
liquid~\cite{kalmeyer-87prl2095,kalmeyer-89prb11879,schroeter-07prl097202,thomale-09prb104406}.
Following up on an idea by D.H.~Lee, Kalmeyer and
Laughlin~\cite{kalmeyer-87prl2095,kalmeyer-89prb11879} proposed that a
quantized Hall wave function for bosons could be used to describe the
amplitudes for spin-flips on a lattice.  The chiral spin liquid state
did not turn out to be relevant to CuO superconductivity, but remains
one of very few examples of two-dimensional spin liquids with
fractional statistics.  Other established examples of two-dimensional
spin liquids
%with fractional quantization 
include the resonating valence bond (RVB) phases of the
Rokhsar-Kivelson model~\cite{kivelson-87prb8865} on the triangular
lattice identified by Moessner and Sondhi~\cite{moessner-01prl1881},
of the Kitaev model~\cite{Kitaev06ap2}, and of the Hubbard model on
the honeycomb lattice~\cite{meng-10n847}.

While usually associated with two-dimensional systems, fractional
statistics is also possible in one dimension.  The paradigm for
one-dimensional anyons are the spinon excitations in the
Haldane--Shastry model~\cite{haldane88prl635,shastry88prl639}, a spin
chain model with $S=\frac{1}{2}$ and long-ranged Heisenberg
interactions.
%, which fall off as $1/r^2$ with the distance $r$ between spins.  
The ground state can be generated by Gutzwiller projection of
half-filled bands of free fermions, and is equivalent to a chiral spin
liquid in one dimension.  The unique feature of the model is that the
spinons are free in the sense that they only interact through their
fractional statistics~\cite{haldane94proc,greiter-05prb224424}.  The
half-fermi statistics was originally discovered and formulated through
a fractional exclusion or generalized Pauli
principle~\cite{haldane91prl937}, according to which the creation of
two spinons reduces the number of single particle states available for
further spinons by one.  It manifests itself physically through
fractional shifts in the spacings between the kinematical momenta of
the individual spinons~\cite{%greiter-05prb224424,
  greiter-06prl059701,greiter-07prl237202,greiter09prb064409}.
%spacings between the kinematical spinon momenta.

%\vspace{50pt}\newpage

The present renaissance of interest in fractional statistics is due to
possible applications of states supporting excitations with
{non-Abelian statistics}~\cite{stern10n187} to the rapidly evolving
field of quantum computation and
cryptography~\cite{kitaev02ap2,nayak-08rmp1083}.  The paradigm for
this universality class, is the Pfaffian state introduced by Moore and
Read~\cite{moore-91npb362} in 1991.  The state was proposed to be
realized at the experimentally observed fraction
$\nu=\frac{5}{2}$~\cite{willet-87prl1776} (\ie at $\nu=\frac{1}{2}$ in
the second Landau level) by Wen, Wilczek, and
ourselves~\cite{greiter-91prl3205,greiter-92npb567}, a proposal which
recently received experimental support through the direct measurement
of the quasiparticle charge~\cite{dolev-08n829,radu-08s899}.  The
Moore--Read state possesses $p+\i p$-wave pairing correlations.  The
flux quantum of the vortices is one half of the Dirac quantum, which
implies a quasiparticle charge of $e/4$.  Like the vortices in a
$p$-wave superfluid, these quasiparticles possess Majorana-fermion
states~\cite{read-00prb10267} at zero energy (\ie one fermion state
per pair of vortices, which can be occupied or unoccupied).  A
Pfaffian state with $2L$ spatially separated quasiparticle excitations
is hence $2^L$ fold degenerate~\cite{nayak-96npb529}, in accordance
with the dimension of the internal space spanned by the zero energy
states.  While adiabatic interchanges of quasiparticles yield only
overall phases in Abelian quantized Hall states, braiding of
half-vortices of the Pfaffian state will in general yield non-trivial
changes in the occupations of the zero energy
states~\cite{ivanov01prl268,stern-04prb205338}, which render the
interchanges non-commutative or non-Abelian.  In particular, the
internal state vector is insensitive to local perturbations---it can
only be manipulated through non-local operations like braiding of the
vortices or measurements involving two or more vortices
simultaneously.  For a sufficiently large number of vortices, on the
other hand, any unitary transformation in this space can be
approximated to arbitrary accuracy through successive braiding
operations~\cite{freedman-02cmp587}.  These properties together render
non-Abelions preeminently suited for applications as protected qubits
in quantum
computation~\cite{dasSarma-05prl166802,nayak-08rmp1083,bishara-09prb155303,moore09p82,stern10n187}.
Non-Abelian anyons are further established in certain other quantum
Hall states described by Jack
polynomials~\cite{greiter93baps137,simon-07prb075317,bernevig-08prl246802}
including Read-Rezayi states~\cite{Read-99prb8084}, in the non-Abelian
phase of the Kitaev model~\cite{Kitaev06ap2}, in the Yao--Kivelson
model~\cite{yao-07prl247203}, and in the non-Abelian chiral spin
liquid proposed by Thomale and ourselves~\cite{greiter-09prl207203}.
In this liquid, the amplitudes for renormalized spin-flips on a
lattice with spins $S=1$ are described by a bosonic Pfaffian state.

The connection between the Haldane-Shastry ground state, the chiral
spin liquid, and a bosonic Laughlin state at Landau level filling
fraction $\nu=\frac{1}{2}$ suggests that one may consider the
non-Abelian chiral spin liquid in one dimension as a ground state for
a spin chain with $S=1$.  This state is related to a bosonic
Moore--Read state at filling fraction $\nu=1$.  In this \paper, we
will introduce and elaborate on this one-dimensional spin liquid
state, construct a parent Hamiltonian, and generalize the model to
arbitrary spin $S$.  We further propose that the spinon excitations of
the states for $S\ge 1$ will obey a novel form of ``non-Abelian''
statistics, where the internal, protected Hilbert space associated
with the statistics is spanned by topological shifts in the spacings
of the single spinon momenta when spinons are present.

Most of the \book\ will be devoted to the construction of the model
Hamiltonian for spin $S$.  In \Chap\ \ref{sec:3mod}, we introduce
three exact models, and the ground state for the $S=1$ spin chain for which 
we wish to construct a parent Hamiltonian.  The exact models consist
of Hamiltonians, their ground states, and the elementary excitations,
which are in some cases exact and in others approximate eigenstates of the
Hamiltonian.  In Section \ref{sec:qh}, we review the Laughlin $\nu
=\frac{1}{m}$ state for quantized Hall liquids, 
\begin{equation}
  \label{eq:i:int:qhpsiLaugh}
  \psi_{0}(z_1,z_2,\ldots ,z_M)
  =\prod^M_{i<j}(z_i-z_j)^m\prod_{i=1}^M e^{-\frac{1}{4}|z_i|^2},
\end{equation}
where the $z_i$'s are the coordinates of $M$ electrons in the
complex plane, and $m$ is odd for fermions and even for bosons.
For $m=2$, its parent Hamiltonian is given by the kinetic term giving
rise to Landau level quantization supplemented by a $\delta$-function
potential, which excludes the component with relative angular momentum
zero between pairs of bosons.  The ground state wave function for a
bosonic $m=2$ Laughlin state is similar to the ground state of the
Haldane--Shastry model we review in Section \ref{sec:hs},
\begin{equation}
  \label{eq:i:int:hspsi0}
  \psi^{\s\text{HS}}_{0}(z_1,z_2,\ldots ,z_M) = 
  \prod_{i<i}^M\,(z_i-z_j)^2\,\prod_{i=1}^M\,z_i\,, 
\end{equation}
where the $z_i$'s are now coordinates of spin flips for a spin chain
with $N$ sites on a unit circle embedded in the complex plane, and
$M=\frac{N}{2}$.  The Haldane--Shastry Hamiltonian,
\begin{equation}
  \label{eq:i:int:hsham}
  {H}^{\s\text{HS}} = \left(\frac{2\pi}{N}\right)^2
  \sum^N_{\alpha <\beta}\,
  \frac{{\boldsymbol{S}}_\alpha {\boldsymbol{S}}_\beta 
  }{\left|\eta_\alpha-\eta_\beta \right|^2}\,,
\end{equation}
where $\eta_\alpha=e^{\text{i}\frac{2\pi}{N}\alpha }$ are the
coordinates of the $N$ sites on the unit circle, however, bears no
resemblance to the $\delta$-function Hamiltonian for the Laughlin
states.  We will elaborate in Section \ref{sec:a:gen} that these
models are both physically and mathematically sufficiently different
to consider them unrelated.  Even the ground state wave functions,
when adapted as far as any possible by formulating the bosonic
Laughlin state on the sphere and by inserting a quasihole at the south
pole, differ due to different Hilbert space normalizations.  From a
scholarly point of view, there just appears to be no connection.

From a pragmatic point of view, however, we may view both Hamiltonians
as devices to obtain the coefficients of the polynomial
\begin{equation*}
  \prod_{i<i}^N\,(z_i-z_j)^2
\end{equation*}
for particle numbers such that the Hamiltonians can be diagonalized
numerically. %($N=18$ at the time of this writing).
In fact, Haldane~\cite{haldane83prl605} introduced the parent
Hamiltonian for the Laughlin state in order to obtain the coefficients
of all the configurations of the state vector for $N=6$, which he
could then compare numerically to the exact ground state for Coulomb
interactions.  This raises the question whether the recipes used by
both Hamiltonians for obtaining these coefficients are really
different.  If one wishes to attribute the results we presented to a
discovery, this discovery is that they are not.

When we ``derive'' the Haldane--Shastry model from the bosonic $m=2$
Laughlin state and its $\delta$-function parent Hamiltonian in
\Chap\ \ref{sec:l2hs}, we really first extract this recipe from the
quantum Hall Hamiltonian, and then use it to construct a parent
Hamiltonian for the quantum spin chain, which has to be Hermitian,
local, and invariant under translations, parity, time reversal, and
SU(2) spin rotations.  Written in the language of the spin system, the
recipe is the condition that the Haldane--Shastry ground state is
annihilated by the operator
\begin{equation}
  \label{eq:i:a:Omegadef}
  \OaHS =\sum_{\substack{\beta=1\\[2pt]\beta\ne\alpha}}^N 
  \frac{1}{\ea-\eb} \Sa^- \Sb^-,\qquad 
  \OaHS \ket{\psi^{\s\text{HS}}_{0}} = 0 \quad\forall\, \alpha.
\end{equation}
The Haldane--Shastry model has been known for more than two decades,
but while Haldane and Shastry independently
% \footnote{The attentive
%   reader may wonder why, if these discoveries were independent, both
%   manuscripts were received by Physical Review Letters within 10 days.
%   The story is that B.S.~Shastry was spending his sebatical at
%   Princeton University in the fall term of 1987, where he discovered
%   the model and enthusiastically went through the hallways announcing
%   his breakthrough.  When he arrived at the office of F.D.M.~Haldane,
%   the latter went through the drawers of his desk, pulled out a
%   preprint he had written in the early 1980s, and handed it to
%   Shastry: ``Do you mean this model?''} 
discovered it, we derive it.  Unlike the discoveries, this derivation
lends itself to a generalization to higher spins.  The construction of
exact models of critical spin chains following the line of reasoning
we use in our derivation of the Haldane--Shastry model is the subject of
this \paper .

In Section \ref{sec:3mod-pf}, we review the properties of the
Moore--Read state~\cite{moore-91npb362,greiter-91prl3205,greiter-92npb567},
\begin{equation}
  \label{eq:i:pfpsi0}
  \psi_0(z_1,z_2,\ldots ,z_N) %[z_i]
  =\text{Pf}\left(\frac{1}{z_{i}-z_{j}}\right)
  \prod_{i<j}^{N}(z_i-z_j)^m
  \prod_{i=1}^N e^{-\frac{1}{4}|z_i|^2},          
\end{equation}
at Landau level filling fraction $\nu =\frac{1}{m}$, where $m$ is even
for fermions and odd for bosons, with emphasis on the non-Abelian
statistics of the half-vortex quasiparticle excitations.  For $m=1$,
the Pfaffian state is the exact ground state of the kinetic
Hamiltonian supplemented by the three-body interaction
term~\cite{greiter-92npb567}
\begin{equation}
  \label{eq:i:pfGWWham}
  V=\sum_{i,j<k}^N \delta^{(2)}(z_i-z_j)\delta^{(2)}(z_i-z_k).
\end{equation}
The bosonic $m=1$ ground state is similar to the ground state
wave function of the critical $S=1$ spin liquid state we introduce in
Section \ref{sec:3mod-na},
\begin{equation}
  \label{eq:i:napsi0}
  \psi^{S=1}_0(z_1,z_2,\ldots ,z_N) %[z_i]
  =\text{Pf}\left(\frac{1}{z_{i}-z_{j}}\right)
  \prod_{i<j}^{N}(z_i-z_j)\prod_{i=1}^{N}\,z_i,
\end{equation}
which describes the amplitudes of renormalized spin flips 
\begin{equation}
  \label{eq:i:b:tildeS+}
  \tilde{S}_{\alpha}^{+}=\frac{{S}^{\rm{z}}_{\alpha}+1}{2} S_\alpha^{+},
\end{equation}
on sites $\eta_\alpha=e^{\text{i}\frac{2\pi}{N}\alpha}$ on a unit
circle embedded in the complex plane.  These spin flips act on a
vacuum where all the $N$ spins are in the $S^\z=-1$ state.  In Section
\ref{sec:nana}, we propose that the momentum spacings between the
individual spinon excitations of this liquid alternate between being
odd multiples of $\frac{\pi}{N}$ and being either even or odd
multiples of $\frac{\pi}{N}$.  (Since the spacings for bosons or
fermions are multiples of $\frac{2\pi}{N}$, an odd multiply of
$\frac{\pi}{N}$ corresponds to half-fermion, and an even multiple to
boson or fermion statistics.)  When we have a choice between even and
odd, this choice represents a topological quantum number.
%, which is insensitive to local perturbations.  
The momentum spacings hence span an internal or topological Hilbert
space of dimension $2^L$ when $2L$ spinons are present, as appropriate
for Ising anyons.  These spacings constitute the analog of the
Majorana fermion states in the cores of the half-vortex excitations of
the Moore--Read state.

In \Chap\ \ref{sec:pf2s1}, we derive a parent Hamiltonian for the
$S=1$ spin liquid state \eqref{eq:i:pfpsi0} from the three-body parent
Hamiltonian \eqref{eq:i:pfGWWham} of the Moore--Read state.  The
steps are similar to those taken for the Haldane--Shastry model, but
technically more involved.  The \defining condition for the state, \ie
the recipe used by the quantum Hall Hamiltonian to specify the
coefficients of the polynomial
\begin{equation*}
  \text{Pf}\left(\frac{1}{z_{i}-z_{j}}\right)\prod_{i<j}^{N}(z_i-z_j),
\end{equation*}
is in the language of the $S=1$ spin model given by
\begin{equation}
  \label{eq:i:b:Omegadef}
    \OaSone =\sum_{\substack{\beta=1\\[2pt]\beta\ne\alpha}}^N 
    \frac{1}{\ea-\eb} (\Sa^-)^2 \Sb^-,\qquad 
    \OaSone \ket{\psi^{S=1}_{0}} = 0 \quad\forall\, \alpha.
\end{equation}
As an aside, we also find that the state is annihilated by the operator
\begin{align}
  \label{eq:i:b:Thetadef}
    \Xi_{\a} 
     &=\sum_{\substack{\beta,\gamma=1\\\beta,\gamma\ne\alpha}}^N 
    \frac{\Sa^- \Sb^- \Sc^-}{(\ea-\eb)(\ea-\ec)}
    -\sum_{\substack{\beta=1\\[2pt]\beta\ne\alpha}}^N
    \frac{(\Sa^-)^2 \Sb^-}{(\ea-\eb)^2} ,\quad 
    \Xi_{\a} \ket{\psi^{S=1}_{0}} = 0 \ \forall\, \alpha,
\end{align}
which we do not consider further.  A Hermitian and translationally
invariant annihilation operator for the $S=1$ spin liquid state
\eqref{eq:i:pfpsi0} is given by
\begin{align}
  \label{eq:i:b:h0}
  H_0=\frac{1}{2}\sum_{\a=1}^N {\OaSone}^\dagger\OaSone .
\end{align}
Since the state is a spin singlet, \ie invariant under SU(2) spin
rotations, all the different tensor components of \eqref{eq:i:b:h0}
must annihilate it individually.  In Section \ref{sec:b:ham}, we
obtain the desired parent Hamiltonian for the $S=1$ spin liquid state
\eqref{eq:i:napsi0},
\begin{align}
  \label{eq:i:b:h}
%   H^{S=1}&=\sum_{\substack{\a\ne \b}}\frac{\bSa\bSb}{\vert\ea-\eb\vert^2} 
%   -\frac{1}{20}\sum_{\substack{\a,\b,\c\\ \a\ne\b,\c}}
%   \frac{(\bSa\bSb)(\bSa\bSc) + (\bSa\bSc)(\bSa\bSb)}{(\eab-\ebb)(\ea-\ec)}
  H^{S=1}&=\frac{2\pi^2}{N^2}
  \Bigg[
  \sum_{\substack{\a\ne \b}}^N\frac{\bSa\bSb}{\vert\ea-\eb\vert^2} 
  -\frac{1}{20}\sum_{\substack{\a,\b,\c\\ \a\ne\b,\c}}^N
  \frac{(\bSa\bSb)(\bSa\bSc) + (\bSa\bSc)(\bSa\bSb)}{(\eab-\ebb)(\ea-\ec)}
  \Bigg],
\end{align}
by projecting out the component of $H_0$ which is invariant under
% parity and time reversal, and transforms as a scalar spin rotations.
parity, time reversal, and SU(2) spin rotations.  The energy of the
ground state \eqref{eq:i:napsi0} is given by 
\begin{align}
  \label{eq:i:b:e0}
  E_0^{S=1}=-\frac{2\pi^2}{N^2}\frac{N (N^2+5)}{15}.
%  =-\frac{2\pi^2}{15}\left(N+\frac{5}{N}\right).
\end{align}
Finally, we use the same methods to obtain vector annihilation
operators for the $S=1$ spin liquid state in Section \ref{sec:b:vec}.

In \Chap\ \ref{sec:gen2s}, we generalize the model to arbitrary spin
$S$.  We do, however, no longer start with a quantum Hall state and
its parent Hamiltonian, but generalize the spin liquid states and the
defining conditions for $S=\frac{1}{2}$ and $S=1$, \ie the conditions
\eqref{eq:i:a:Omegadef} and \eqref{eq:i:b:Omegadef}, directly to
higher spins.  To generalize the state vector, we first recall from
Section \ref{sec:napro} that the $S=1$ spin liquid can be obtained by
taking two (identical) Gutzwiller or Haldane--Shastry ground
states and projecting onto the triplet or $S=1$ configuration at each
site~\cite{greiter02jltp1029}.  This projection can be accomplished
conveniently if we write the Haldane--Shastry ground state
\eqref{eq:hspsi0} in terms of Schwinger bosons,
\begin{align}
  \label{eq:i:c:hsket}
  \ket{\psi^{\s\text{HS}}_{0}}
%   &=\sum_{\{z_1,z_2,\ldots ,z_M\}}
%   \psi^{\s\text{HS}}_{0}(z_1,\ldots ,z_M)\;
%   {S}^+_{z_1}\cdot\ldots\cdot {S}^+_{z_M} 
% %  \big|\underbrace{\dw\dw\ldots\ldots\dw}_{\text{all\ } N \text{\ spins\ } \dw}
% %  \big\rangle
%   \ket{\dw\dw\ldots\dw} 
%   \nonumber\\[0.4\baselineskip] 
  &=\sum_{\{z_1,\ldots ,z_M;w_1,\ldots,w_M\}}
  \psi^{\s\text{HS}}_{0}(z_1,\ldots ,z_M)\;
    {a}^+_{z_1}\ldots a^\dagger_{z_M}
    {b}^+_{w_1}\ldots b^\dagger_{w_M}%
  \vac\!
  \nonumber\\[0.4\baselineskip] 
  &\equiv \Psi^{\s\text{HS}}_{0}[a^\dagger,b^\dagger] \vac\!,
\end{align}
where $M=\frac{N}{2}$ and the $w_k$'s are those coordinates on
the unit circle which are not occupied by any of the $z_i$'s.
The $S=1$ spin liquid state \eqref{eq:i:napsi0} can then be written
\begin{equation}
  \label{eq:i:naop}
  \ket{\psi^{S=1}_0}=
  \Big(\Psi^{\s\text{HS}}_0\big[a^\dagger ,b^\dagger\big]\Big)^2\vac.
\end{equation}
To generalize the ground state to arbitrary spin $S$, we just take
$2S$ (identical) copies Haldane--Shastry ground state, and project at
each site onto the completely symmetric representation with total spin
$S$.  In terms of Schwinger bosons,
\begin{equation}
  \label{eq:i:c:psi0schwinger}
  \ket{\psi^{S}_0}
  =\Big(\Psi^{\s\text{HS}}_0\big[a^\dagger ,b^\dagger\big]\Big)^{2S}\vac.
\end{equation} 
This state is related to bosonic Read--Rezayi
states~\cite{Read-99prb8084} in the quantum Hall system.  In Section
\ref{sec:c:defining}, we verify that the state is annihilated by the
operator
\begin{align}
  \label{eq:i:c:Omegadef}
  \OaS 
  &=\sum_{\substack{\beta=1\\[2pt]\beta\ne\alpha}}^N 
  \frac{1}{\ea-\eb} (\Sa^-)^{2S} \Sb^-,\qquad 
  \OaS \ket{\psi^S_{0}} = 0 \quad\forall\, \alpha.
\end{align}
In Section \ref{sec:c:ham}, we follow the same steps as for the $S=1$ state
%\eqref{eq:i:napsi0} 
to construct a parent Hamiltonian for the spin $S$ state
\eqref{eq:i:c:psi0schwinger}, and obtain
\begin{align}
  \label{eq:i:c:h}
  H^{S} =\frac{2\pi^2}{N^2}
  \Bigg[
  &\sum_{\substack{\a\ne\b}}^N \frac{\bSa\bSb}{\vert\ea-\eb\vert^2}
  \nonumber\\[0.2\baselineskip] 
  &  -\frac{1}{2(S+1)(2S+3)}\sum_{\substack{\a,\b,\c\\ \a\ne\b,\c}}^N
  \frac{(\bSa\bSb)(\bSa\bSc) + (\bSa\bSc)(\bSa\bSb)}{(\eab-\ebb)(\ea-\ec)}
  \Bigg].
\end{align}
The energy eigenvalue is given by
\begin{align}
  \label{eq:i:c:E_0}
  E_0^{S} &=-\frac{2\pi^2}{N^2}\frac{S(S+1)^2}{2S+3}\,\frac{N (N^2+5)}{12}.
%  \nonumber\\[0.2\baselineskip] &
%  = -\frac{\pi^2}{6}\frac{s(s+1)^2}{2s+3}\left(N+\frac{5}{N}\right).
\end{align}
This is the main result we present.  In Section \ref{sec:c:vec},
we construct the vector annihilation operators
\begin{align}
  \label{eq:i:c:doperator}
  &\bs{D}_\a^{S}
%   =\frac{\i}{2}\sum_{\substack{\b\\ \b\ne\a}}\frac{\ea+\eb}{\ea-\eb}
%   \bigg[(\bSa\times\bSb) -\i(s+1)\bSb 
%   + \frac{\i}{s+1}\bSa\big(\bSa\bSb\big)\bigg],
  =\frac{1}{2}\sum_{\substack{\b\\ \b\ne\a}}\frac{\ea+\eb}{\ea-\eb}
  \bigg[\i(\bSa\times\bSb) + (S+1)\,\bSb 
  - \frac{1}{S+1}\bSa\big(\bSa\bSb\big)\bigg],
  \nonumber\\[0.2\baselineskip] 
  &\bs{D}_\a^{S}\ket{\psi^{S}_{0}} = 0 \quad\forall\, \alpha,
\end{align}
and
\begin{align}
  \label{eq:i:c:aoperator}
  \bs{A}_\a^{S} 
%   &\equiv \frac{2(2s+3)(s+2)}{3}
%   \Big(\big\{H_\a^{\rm P\bar T=}\big\}_{\bs{1}} +
%   \{H_\a^{\rm P\bar T\ne}\big\}_{\bs{1}}\big)
%   \nonumber\\[0.6\baselineskip] 
  &=\sum_{\substack{\b\\\b\ne\a}} %\omega_{\a\b\b} \Big[
%  \bSa(\bSa\bSb) + (\bSa\bSb)\bSa + 2(s+1)\,\bSb\Big]
  \frac{\bSa(\bSa\bSb) + (\bSa\bSb)\bSa + 2(S+1)\,\bSb}{\vert\ea-\eb\vert^2}
  \nonumber\\[0.2\baselineskip] 
%  &\quad
  & +\sum_{\substack{\b,\c\\ \b,\c\ne\a}}%\omega_{\a\b\c}\bigg[
  \frac{1}{(\eab-\ebb)(\ea-\ec)}\bigg[
  -\frac{(\bSa\bSb)\bSa(\bSa\bSc) + (\bSa\bSc)\bSa(\bSa\bSb)}{S+1}\,
%  -\frac{1}{s+1}\,
%  \bigl[(\bSa\bSb)\bSa(\bSa\bSc) + (\bSa\bSc)\bSa(\bSa\bSb)\bigr]
  \nonumber\\*[-0.4\baselineskip] 
  &\hspace{102pt} +  2(S+2)\,\bSa(\bSb\bSc) 
  - \bSb(\bSa\bSc) - (\bSa\bSb)\bSc\bigg], 
  \nonumber\\[0.2\baselineskip] 
  \bs{A}_\a^{S} &\ket{\psi^{S}_{0}} = 0 \quad \forall\, \alpha.
\end{align}
In Section \ref{sec:c:scalarfromvec}, we evaluate the parity and time
reversal invariant scalar operators
\begin{align}
  \sum_{\a} {\bs{D}_\a^{S}}^\dagger\bs{D}_\a^{S}
  \qquad\text{and}\qquad \sum_{\a} \bSa \bs{A}_\a^{S},
\end{align}
and find that both of them reproduce the model \eqref{eq:i:c:h}.  The
factorization of $H^{S}$ is terms of ${\bs{D}_\a^{S}}^\dagger$ and
$\bs{D}_\a^{S}$ shows that $\ket{\psi^{S}_{0}}$ is not just an
eigenstate of \eqref{eq:i:c:h}, but also a ground state.  Numerical
work~\cite{manuscriptinpreparationTRSG11} indicates that
$\ket{\psi^{S}_{0}}$ is the only ground state of $\ket{\psi^{S}_{0}}$.
In Section \ref{sec:c:hs}, we show  that the model \eqref{eq:i:c:h}
reduces to the Haldane--Shastry model if we take $S=\frac{1}{2}$.
%\nopagebreak

We conclude with a brief discussion of several unresolved issues as
well as possible generalizations of the model in
\Chap~\ref{sec:concl}.  These include the quest for integrability, the
correctness and universality of our assignments for the SU(2) level
$k=2S$ anyon-type momentum spacings of the spinon excitations and the
feasibility of applications as protected cubits in quantum
computation.  We outline how to generalize the model to symmetric
representations of SU($n$), where the non-abelian statistics of the
spinons appears to have no correspondence in a quantum Hall system.

\numberwithin{equation}{section}

% lp -dtkmsek -P 7-9 map.ps 

\chapter{Three models and a ground state}
\label{sec:3mod}
\section{The Laughlin state and its parent Hamiltonian}
\label{sec:qh}

Laughlin's
theory~\cite{laughlin83prl1395,halperin83hpa75,haldane83prl605,laughlin84ss163,PrangeGirvin90,ChakrabortyPietilainen95}
for a series of fractionally quantized Hall states is first and
foremost the key to an explanation for the experimentally observed,
fractionally quantized plateaus in the Hall resistivity of a
spin-polarized, two-dimensional electron gas realized in semiconductor
inversion
layers~\cite{tsui-82prl1559,chang-84prl997,clark-86ss141,willet-87prl1776,PrangeGirvin90}.
For our purposes here, however, we will view it primarily as an exact
model, that is, a ground state which supports fractionally quantized
excitations, and a model Hamiltonian for which this ground state is
exact.  %To explain the theory starting from first principles, we
%We begin with a review of Landau level quantization in the planar
%geometry.

We will first review the theory in a planar geometry with open
boundary conditions, and then turn to the spherical geometry,
which will turn out to be the relevant geometry for the mapping of 
quantized Hall system onto a spin chain. We begin with a review 
of Landau level quantization in the plane.

\subsection{Landau level quantization in the planar geometry}
\label{sec:qhp}

To describe the dynamics of charged particles (\eg spin-polarized
electrons) in a two-dimensional plane subject to a perpendicular
magnetic field $\bs{B}=-B\bs{e}_\z $, it is convenient to introduce
complex particles coordinates $z=x+\text{i}y$ and $\bar
z=x-\text{i}y$~\cite{landau30zp629,Arovas86}.  The associated derivative
operators are
%conjugate derivative operators are given by
\begin{equation}
  \label{eq:qhpartialz}
  \parder{}{z}
%  =\parder{x}{z}\parder{}{x}+\parder{y}{z}\parder{}{y}
  =\frac{1}{2}\left(\parder{}{x}-\text{i}\parder{}{y}\right),
  \quad
  \parder{}{\bar z}
%  =\parder{x}{\bar z}\parder{}{x}+\parder{y}{\bar z}\parder{}{y}
  =\frac{1}{2}\left(\parder{}{x}+\text{i}\parder{}{y}\right).
\end{equation}
%
% \begin{align}
%   \label{eq:qhpartialz}
%   \parder{}{z}&
% %  =\parder{x}{z}\parder{}{x}+\parder{y}{z}\parder{}{y}
%   =\frac{1}{2}\left(\parder{}{x}-i\parder{}{y}\right),
%   \\[0.5\baselineskip]
%   \parder{}{\bar z}&
%  % =\parder{x}{\bar z}\parder{}{x}+\parder{y}{\bar z}\parder{}{y}
%   =\frac{1}{2}\left(\parder{}{x}+i\parder{}{y}\right).
% \end{align}
% \begin{align}
%   \label{eq:qhpartialz}
%   &z=x+iy
%   &\parder{}{z}
% %  =\parder{x}{z}\parder{}{x}+\parder{y}{z}\parder{}{y}
%   =\frac{1}{2}\left(\parder{}{x}-i\parder{}{y}\right),
%   \\[0.5\baselineskip]
%   &z=x-iy
%   &\parder{}{\bar z}
%  % =\parder{x}{\bar z}\parder{}{x}+\parder{y}{\bar z}\parder{}{y}
%   =\frac{1}{2}\left(\parder{}{x}+i\parder{}{y}\right).
% \end{align}
Note that hermitian conjugation yields a $-$ sign,
\begin{equation}
  \label{eq:qhhconjparderz}
  \left(\parder{}{z}\right)^\dagger
%  =\frac{1}{2}\left(\parder{}{x}-i\parder{}{y}\right)^\dagger
%  =\frac{1}{2}\left[\left(\parder{}{x}\right)^\dagger
%                  +i\left(\parder{}{y}\right)^\dagger\right]
%  =\frac{1}{2}\left(-\parder{}{x}-i\parder{}{y}\right)
  =-\parder{}{\bar z}.
\end{equation}
We further define the complex momentum 
% \begin{align}
%   \label{eq:qhpdef}
%   p&\equiv p_\x+ip_\y
% %  =\hbar \left(-i\parder{}{x}\right)+i\left(-i\parder{}{y}\right)
%   =-2i\hbar \parder{}{\bar z}
%   \\[0.5\baselineskip]
%   \bar p&= p_\x-ip_\y
% %  =\left(-i\parder{}{x}\right)+i\left(-i\parder{}{y}\right)
%   =-2i\hbar \parder{}{z}
% \end{align}
\begin{equation}
  \label{eq:qhpdef}
  p\equiv p_\x+\text{i}p_\y=-2\text{i}\hbar \parder{}{\bar z},
  \quad
  \bar p= p_\x-\text{i}p_\y  =-2\text{i}\hbar \parder{}{z}.
\end{equation}
The single particle Hamilton operator is obtained by minimally coupling the
gauge field to the canonical momentum,
\begin{equation}
  \label{eq:qhSinglePartHam1}
  H=\frac{1}{2M}\left(\bs{p}+\frac{e}{c}\bs{A}\right)^2,
\end{equation}
where $M$ is the mass of the particle and $e>0$.  In the symmetric
gauge
%\begin{equation}
%  \label{eq:qhgauge}
%  \bs{A}=\frac{B}{2}\,\bs{r}\times\bs{e}_\z =\frac{1}{2}\left(
%    \begin{array}{c}
%      By \\ -Bx\\ 0
%    \end{array}\right),
%\end{equation}
$\bs{A}=\frac{1}{2}B\,\bs{r}\times\bs{e}_\z $,
% $A^i=\frac{1}{2}B\epsilon^{ij}r^j$, 
and with the definition of the magnetic length %$l$, $l^2=\frac{\hbar c}{eB}$ 
\begin{equation}
  \label{eq:qhl^2}
  l=\sqrt{\frac{\hbar c}{eB}},
\end{equation}
we write
\begin{align}
  \label{eq:qhSinglePartHam2}
  H&=\frac{1}{2M}\left[\left(p_\x+\frac{\hbar}{2l^2}y\right)^2
    +\left(p_\y-\frac{\hbar}{2l^2}x\right)^2\right]
  \nonumber\\[0.5\baselineskip]
  &=\frac{1}{2M}\left[\Re^2\!\left(p-\frac{\text{i}\hbar}{2l^2}z\right)
    +\Im^2\!\left(p-\frac{\text{i}\hbar}{2l^2}z\right)\right]
% \end{align}
% \begin{align}
%   \label{eq:qhSinglePartHam2}
%  \pagebreak 
  \nonumber\\[0.5\baselineskip]
  &=\frac{1}{4M}
  \anticomm{p-\frac{\text{i}\hbar}{2l^2}z}%
           {\,\bar p+\frac{\text{i}\hbar}{2l^2}\bar z},
%  \biggl\{ \underbrace{p-\frac{i\hbar}{2l^2}z}_{=\hbar\sqrt{2}/(il)\,a},\,
%  \bar p+\frac{i\hbar}{2l^2}\bar z \biggr\}
  \nonumber\\[0.5\baselineskip]
   &=\frac{\hbar^2}{2Ml^2}\anticomm{a}{a^\dagger}
%   \,=\,\frac{\hbar}{2}\frac{eB}{Mc}\anticomm{a}{\, a^\dagger}
%   \,=\,\frac{\hbar \omega_c}{2}\anticomm{a}{\, a^\dagger}
\end{align}
where $\Re$ and $\Im$ denote the real and imaginary part,
respectively.  In the last line, we have introduced the ladder
operators~\cite{macdonald84prb3550,girvin-83prb4506,Arovas86}\footnote{We
  have not been able to find out who introduced the ladder operators
  for Landau levels in the plane.  The energy eigenfunctions were
  known since Landau~\cite{landau30zp629}.
  MacDonald~\cite{macdonald84prb3550} used the ladder operators in
  1984, but neither gave nor took credit.  Girvin and
  Jach~\cite{girvin-83prb4506} were aware of two independent ladders a
  year earlier, but neither spelled out the formalism, nor pointed to
  references.  It appears that the community had been aware of them,
  but not aware of who introduced them.  The clearest and most
  complete presentation we know of is due to Arovas~\cite{Arovas86}.}
% \begin{align}
%   \label{eq:qhaladder2}
%   a &=
%   \frac{il}{\sqrt{2}}\left(-2i\parder{}{\bar z}-\frac{i}{2l^2}z\right),
%   \nonumber\\[0.5\baselineskip]
%   a^\dagger &=
%   -\frac{il}{\sqrt{2}}\left(-2i\parder{}{z}+\frac{i}{2l^2}\bar z\right),
% \end{align}
% \begin{equation}
%   \label{eq:qhaladder2}
%   a=\frac{\text{i}l}{\sqrt{2}}
%   \left(-2\text{i}\parder{}{\bar z}-\frac{\text{i}}{2l^2}z\right),
%   \quad
%   a^\dagger=-\frac{\text{i}l}{\sqrt{2}}
%   \left(-2\text{i}\parder{}{z}+\frac{\text{i}}{2l^2}\bar z\right),  
% \end{equation}
\begin{equation}
  \label{eq:qhaladder2}
  a=\frac{l}{\sqrt{2}}
  \left(2\parder{}{\bar z}+\frac{1}{2l^2}z\right),
  \quad
  a^\dagger=\frac{l}{\sqrt{2}}
  \left(-2\parder{}{z}+\frac{1}{2l^2}\bar z\right),  
\end{equation}
which obey 
\begin{equation}
  \label{eq:qhCommaa^dagger}
  \comm{a}{a^\dagger}=1.
\end{equation}
With the cyclotron frequency $\omega_\text{c}=eB/Mc$ and 
\eqref{eq:qhCommaa^dagger} we finally obtain
\begin{equation}
  \label{eq:qhSinglePartHam3}
  H=\hbar \omega_\text{c} \left(a^\dagger a+\frac{1}{2}\right).
\end{equation}
The kinetic energy of charged particles in a perpendicular magnetic field
is hence quantized like a harmonic oscillator.  The energy levels are called 
Landau levels.

It is convenient to write the ladder operators describing the
cyclotron variables as
\begin{align}
  \label{eq:qhaladder3}
   a&=       +\sqrt{2}l\exp\!\left(-\frac{1}{4l^2}\bar zz\right)
          \parder{}{\bar z} \exp\!\left(+\frac{1}{4l^2}\bar zz\right),
   \\[0.5\baselineskip] \label{eq:qhadagladder3}
   a^\dagger&=-\sqrt{2}l\exp\!\left(+\frac{1}{4l^2}\bar zz\right)
                \parder{}{z}\exp\!\left(-\frac{1}{4l^2}\bar zz\right),
\end{align}
and introduce a second set of ladder operators for the guiding center
variables,
\begin{align}
  \label{eq:qhbladder3}
   b&=       +\sqrt{2}l\exp\!\left(-\frac{1}{4l^2}\bar zz\right)
               \parder{}{z} \exp\!\left(+\frac{1}{4l^2}\bar zz\right),
   \\[0.5\baselineskip] \label{eq:qhbdagladder3}
   b^\dagger&=-\sqrt{2}l\exp\!\left(+\frac{1}{4l^2}\bar zz\right)
           \parder{}{\bar z}\exp\!\left(-\frac{1}{4l^2}\bar zz\right).
\end{align}
They likewise obey 
\begin{equation}
  \label{eq:qhCommbb^dagger}
  \comm{b}{b^\dagger}=1,
\end{equation}
and commute with the cyclotron ladder operators:  
\begin{equation}
  \label{eq:qhCommab}
    \comm{a^{\phantom{\dagger}}\!\!}{b}=\comm{a}{b^\dagger}=0
%    =\comm{a^\dagger}{b}=\comm{a^\dagger}{b^\dagger}=0. %\rule{0pt}{16pt}.
\end{equation}
A calculation similar to the one presented above for $H$ yields
\begin{equation}
  \label{eq:qhSinglePartL1}
%  L_\z =\epsilon^{ij}p^ir^j
%  =\hbar \left(b^\dagger b - a^\dagger a\right).
  \bs{L}=\bs{r}\times\bs{p}
  =\hbar \left(b^\dagger b - a^\dagger a\right)\bs{e}_\z 
\end{equation}
for the \emph{canonical} angular momentum around the origin.  (The 
 \emph{kinematical} angular momentum is given by the $a^\dagger a$ term in 
\eqref{eq:qhSinglePartL1}).

Since the angular momentum \eqref{eq:qhSinglePartL1} commutes with the
Hamiltonian \eqref{eq:qhSinglePartHam3}, we can use it to classify
the vastly degenerate states within each Landau level.  Specifically,
we introduce the basis states 
\begin{equation}
  \label{eq:qhketnm}
  \ket{n,m}
  =\frac{1}{\sqrt{n!}}\frac{1}{\sqrt{m!}}(a^\dagger)^n (b^\dagger)^m\ket{0,0},
%  \qquad a\vac=b\vac=0,
\end{equation}
where the vacuum state is by definition annihilated by both
destruction operators, %$a\vac=b\vac=0.$
\begin{equation}
  \label{eq:qhvacdef}
  a\ket{0,0}=b\ket{0,0}=0.
\end{equation}
Solving \eqref{eq:qhvacdef} yields the real space representation 
\begin{equation}
  \label{eq:qhvac}
  \phi_0(z)\equiv
  \phi_0(z,\bar z)= %\equiv
  \braket{\bs{r}}{0,0}=
  \frac{1}{\sqrt{2\pi l^2}}\exp\left(-\frac{1}{4l^2}|z|^2\right).
\end{equation}
(In the following, we omit $\bar z$ from the argument of wave functions
as a choice of notation.)
The basis states \eqref{eq:qhketnm} are trivially eigenstates of both
$H$ and $L_\z $,
\begin{align}
  \label{eq:qhHLketnm}
  H\ket{n,m}&= \hbar\omega_c\!\left(n+\frac{1}{2}\right)
  \nonumber\\[0.3\baselineskip]
  L_\z \ket{n,m}&= \hbar(m-n)\ket{n,m}
\end{align}
The particle coordinate and momentum are given in terms of the ladder
operators by  
\begin{equation}
  \label{eq:qhzp}
  z=\sqrt{2}l\left(a+b^{\dagger}\right),
  \quad 
  p=-\frac{\i\hbar}{\sqrt{2}l}\left(a-b^{\dagger}\right).
\end{equation}
This implies that we can write a complete, orthonormal set of basis states in 
the lowest Landau level
($n=0$) as
\begin{align}
  \label{eq:qhLLLm}
%   \phi_m(z,\bar z)
   \phi_m(z)
   &=\braket{\bs{r}}{0,m}
   \nonumber\\
   &=\frac{1}{\sqrt{m!}}
   (b^\dagger)^m\phi_0(z,\bar z)
   \nonumber\\
   &=\frac{1}{\sqrt{2\pi l^2 m!}}
   (a+b^\dagger)^m \exp\left(-\frac{1}{4l^2}|z|^2\right)
   \nonumber\\
   &=%\frac{1}{\sqrt{2^{m+1}\pi l^{2(m+1)} m!}}
   \frac{1}{\sqrt{2^{m+1}\pi m!}\; l^{m+1}}\;
%   \frac{1}{\sqrt{\pi (2l^2)^{m+1} m!}}
   z^m \exp\left(-\frac{1}{4l^2}|z|^2\right).
\end{align}
These states is describe narrow rings centered around the origin, 
with the radius determined by
\begin{equation*}
  \label{eq:qhrm}
  \parder{}{r}\abs{\phi_m(r)}^2\biggl|_{r=r_m}\overset{!}{=}0\biggr.,
%  \quad\Rightarrow\quad r_m=\sqrt{2m}\,l. 
\end{equation*}
which yields $r_m=\sqrt{2m}\,l$.  Since there are also $m$ states
inside the ring, the areal degeneracy is
%number of states per area is
\begin{equation}
  \label{eq:qhArealDeg}
  \frac{\text{number of states}}{\text{area}}
  =\frac{m}{\pi r_m^2}=\frac{1}{2\pi l^2},
%  =\frac{B}{\Phi_0}, 
\end{equation}
% where
% \begin{equation*}
%   \Phi_0=\frac{2\pi \hbar c}{e}
% \end{equation*}
% is the the Dirac flux quantum.
The magnetic flux required for each state,
\begin{equation*}
%  \Phi=
  2\pi l^2 B =\frac{2\pi \hbar c}{e}=\Phi_0,
\end{equation*}
is hence given by the Dirac flux quantum.  This implies that in each
Landau level, there are as many single particle states in a given area
as there are Dirac quanta of magnetic flux going through it.
In the following, we set %$\hbar=c=1$ and 
$l=1$, %(which implies $eB=1$),
and no longer keep track of wave function normalizations.

The $N$ particle wave function for a filled lowest Landau level (LLL)
on a circular disk  %(with radius $R\approx\sqrt{2N}\,l$)
is obtained by antisymmetrizing the basis states \eqref{eq:qhLLLm},
\begin{align}
  \label{eq:qhfilledLLL}
  \psi(z_1,\hdots,z_N)
   &=\mathcal{A}\left\{z_1^0z_2^1\ldots z_N^{N-1}\right\}\cdot
   \prod_{i=1}^N e^{-\frac{1}{4}|z_i|^2}
   \nonumber\\
   &=\prod^N_{i<j}(z_i-z_j)\,
   \prod_{i=1}^N e^{-\frac{1}{4}|z_i|^2}.
\end{align}
The most general form for the single particle wave function in the 
lowest Landau level is
\begin{equation}
  \label{eq:qhSinPartLLL}
  \psi(z)=f(z)\,e^{-\frac{1}{4}|z|^2},
\end{equation}
where $f(z)$ is an analytic function of $z$.  Since $\psi(z)$ is 
annihilated by the destruction operator $a$, 
the energy is trivially $\frac{1}{2}\hbar\omega_\text{c}$ 
%  \eqref{eq:qhSinglePartHam3} implies
%  \begin{equation*}
%    H\psi(z)=\frac{1}{2}\omega_\text{c}\psi(z).
%  \end{equation*}
%
The most general $N$ particle state in the LLL is
given by
\begin{equation}
  \label{eq:qhNPartLLL}
  \psi(z_1,\hdots,z_N)=f(z_1,\hdots,z_N) \prod_{i=1}^N e^{-\frac{1}{4}|z_i|^2},
\end{equation}
where $f(z_1,\hdots,z_N)$ is analytic in all the $z$'s, and symmetric
or antisymmetric for bosons or fermions, respectively.  If we impose
periodic boundary conditions~\cite{haldane-85prb2529}, we find that
$\psi(z_1,z_2,\hdots,z_N)$, when viewed as a function of $z_1$ while
$z_2,\hdots,z_N$ are parameters, has exactly as many zeros as there
states in the LLL, \ie as there are Dirac flux quanta going through
the unit cell or principal region.
If $\psi(z_1,\hdots,z_N)$ describes fermions and is hence
antisymmetric, there will be at least one zero seen by $z_1$ at each of
the other particle positions.  The most general wave function is hence 
% \begin{equation}
%   \label{eq:qhfLLL}
%   f(z_1,\hdots,z_N)=P(z_1,\hdots,z_N) \prod^N_{i<j}(z_i-z_j),
% \end{equation}
\begin{equation}
  \label{eq:qhNPartLLLf}
  \psi(z_1,\hdots,z_N)
  =P(z_1,\hdots,z_N) \prod^N_{i<j}(z_i-z_j) \prod_{i=1}^N e^{-\frac{1}{4}|z_i|^2},
\end{equation}
where $P$ is a symmetric polynomial in the $z_i$'s.  In the case of 
a completely filled Landau level, there are only as many zeros as there
are particles, which implies that all except one of the zeros in $z_1$ 
will be located at the other particle positions  $z_2,\hdots,z_N$.
This yields \eqref{eq:qhfilledLLL} as the unique state for open
boundary conditions.  For periodic boundary conditions,  there is 
one additional zero as there cannot be a zero seen by $z_1$ at $z_1$.  The
location of this zero, which Haldane and Rezayi~\cite{haldane-85prb2529} 
refer to as the center-of-mass zero, encodes the information about the 
boundary phases a test particle acquires as it is taken around one of 
the meridians of the torus. 

To elevate the most general LLL state \eqref{eq:qhNPartLLL} into the
$(n+1)$-th Landau level, we only have to apply
$\left(a^\dagger\right)^n$ to all the particles in the LLL,
\begin{align}
  \label{eq:qhHigherLL}
  \psi_n(z_1,\hdots,z_N)
  &=\prod_{i=1}^N \bigl(a_i^\dagger\bigr)^n \psi(z_1,\hdots,z_N)
  \nonumber\\*[0.3\baselineskip]
  &=\prod_{i=1}^N e^{-\frac{1}{4}|z_i|^2}
  \prod_{i=1}^N \left(2\parder{}{z_i}- \bar z_i\right)^n 
  f(z_1,\hdots,z_N).
\end{align}
The energy per particle in this state is 
% $E_n=\hbar\omega_\text{c}\left(n+\frac{1}{2}\right)$.
$\hbar\omega_\text{c}\!\left(n+\frac{1}{2}\right)$.

\subsection{The Laughlin state}
\label{sec:qhlaughlin}

The experimental observation which Laughlin's
theory~\cite{laughlin83prl1395} explains is a plateau in the Hall
resistivity of a two-dimensional electron gas at a Landau level
filling fraction $\nu=1/3$.
% The latter is defined through
% \begin{equation}
%   \label{eq:qhnu}
%   \frac{1}{\nu}=\parder{N_\Phi}{N},
% \end{equation}
% where $N_\Phi$ is the number of Dirac flux quanta through the sample
% and hence the number of states in each Landau level, and $N$ is the
% number of particles.  The filling fraction hence denotes the number
% of particles divided by the number of  number of states in each Landau level
% in the thermodynamic limit.
The filling fraction denotes the number of
particles divided by the number of number of states in each Landau
level in the thermodynamic limit, and is defined through
\begin{equation}
  \label{eq:qhnu}
  \frac{1}{\nu}=\parder{N_\Phi}{N},
\end{equation}
where $N_\Phi$ is the number of Dirac flux quanta through the sample
and $N$ is the number of particles.  For a wave function at $\nu=1/3$,
we consequently have three times as many zeros seen by $z_1$ as there
are particles, and the polynomial $P(z_1,z_2,\hdots,z_N)$ in
\eqref{eq:qhNPartLLLf} has two zeros per particle.  The experimental
findings, as well as early numerical work by Yoshioka, Halperin, and
Lee~\cite{yoshioka-83prl1219}, are consistent with, if not indicative
of, a quantum liquid state at a preferred filling fraction $\nu=1/3$.
Since the kinetic energy is degenerate in each Landau level, such a
liquid has to be stabilized by the repulsive Coulomb interactions
between the electrons.  This implies that the wave function should be
highly effective in suppressing configurations in which particles
approach each other, as there is a significant potential energy cost
associated with it.  We may hence ask ourselves whether there is any
particular way of efficiently distributing the zeros of
$P(z_1,z_2,\hdots,z_N)$ in this regard.

Laughlin's wave function amounts to attaching the additional zeros onto the
particles, such that each particle coordinate $z_2,\hdots,z_N$ becomes
a triple zero of $z_1$ when $\psi(z_1,z_2,\hdots,z_N)$ is viewed as a
function of $z_1$ with parameter $z_2,\hdots,z_N$.  For filling
fraction $\nu=1/m$, where $m$ is an odd integer if the particles are 
fermions and an even integer if they are bosons, he proposed the
ground state wave function
\begin{equation}
  \label{eq:qhpsiLaugh}
  \psi_m(z_1,\ldots,z_N)
  =\prod^N_{i<j}(z_i-z_j)^m\prod_{i=1}^N e^{-\frac{1}{4}|z_i|^2}.
\end{equation}
There are hence no zeros wasted---all of them contribute in keeping
the particles away from each other effectively, as $\psi_m$ vanishes
as the $m$-th power of the distance when two particles approach each
other.  This is the uniquely defining property of Laughlin's state,
and also the property which enabled Haldane~\cite{haldane83prl605} to
identify a parent Hamiltonian, which singles out the state as its
unique and exact ground state. We discuss the Hamiltonian in Section
\ref{sec:laughsphere} below.  The wave function \eqref{eq:qhpsiLaugh}
describes an incompressible quantum liquid, as the construction is
only possible at filling fractions $\nu=1/m$.

One of the assumptions of the theory is that we can neglect transitions
into higher Landau levels, as the Landau level splitting
$\hbar\omega_\text{c}$ is much larger then the potential energy per
particle, a condition met by the systems amenable to experiment.
Formally, the LLL limit requires $\omega_\text{c}\rightarrow\infty$
while keeping the magnetic length $l^2$ constant, which is achieved by
taking $M\rightarrow 0$.  The LLL limit is hence a zero mass limit.

Even within this limit, which we assume to hold in the following, the
Laughlin state \eqref{eq:qhpsiLaugh} is not the exact ground state for
electrons with (screened) Coulomb interactions at filling fraction
$\nu=1/3$.  It is, however, reasonably close in energy and has a
significant overlap with the exact ground state for finite systems.
The difference between the exact ground state and Laughlin's state is
that in the exact ground state, the zeros of $P(z_1,z_2,\hdots,z_N)$
are attached to the particle coordinates, but do not coincide with
them~\cite{halperin83hpa75,greiter97pe1}.  At long distances, the
physics described by both states is identical.  In particular, the
topological quantum numbers of both states, such as the charge and the
statistics of the (fractionally) charged excitations, or the
degeneracies on closed surfaces of genus one and higher, are
identical.

%One way to characterize the Laughlin state is to say that the
%The characteristic property of the Laughlin state is that 
%The Laughlin state can be characterized by observing that 
%the particles have become 
% One way to characterize the Laughlin state is through the notion of
The Laughlin state can be characterized through the notion of
``superfermions''~\cite{greiter-90mplb1063}.
For fermions (bosons), the relative angular momentum is quantized as
$\hbar l$, where $l$ is an odd (even) integer, due to the antisymmetry
(symmetry) of the wave function under interchange of particles.  In
the LLL, the relative angular momentum between pairs of fermions can
only have components with $l=1,3,5,\ldots$, but no negative values.
If we interchange the particles through winding them counterclockwise
around each other, these components acquire a phase factor $e^{i\pi
  l}$.  The smallest component hence acquires a phase $\pi$, as
required by Fermi statistics.  For the Laughlin state
\eqref{eq:qhpsiLaugh}, the smallest component of relative angular
momentum is $l=m$, and the phase this component acquires upon
interchange is $m\pi$, while only a phase $\pi$ is required by Fermi
statistics.  In this sense, the particles are ``superfermions'' for
$m$ odd, $m>1$.  In the exact ground state for Coulomb interaction,
the electrons are ``approximate superfermions''.

For completeness, we wish to mention that there is a variant of
Haldane's parent Hamiltonian~\cite{haldane83prl605} for the planar
geometry, due to Trugman and Kivelson~\cite{trugman-85prb5280}.  They
noted that since the Laughlin state \eqref{eq:qhpsiLaugh} contains a
term $(z_i-z_j)^m$ for each pair, it is annihilated by the short range
potential interaction
\begin{equation}
  \label{eq:qhTrugmanHfermion}
  V^{(m)}=\sum_{i<j}^N \left(\nabla_i^2\right)^{(m-1)/2}\delta^{(2)}(z_i-z_j)
\end{equation}
for $m$ odd, and
\begin{equation}
  \label{eq:qhTrugmanHboson}
  V^{(m)}=\sum_{i<j}^N \left(\nabla_i^2\right)^{(m-2)/2}\delta^{(2)}(z_i-z_j)
\end{equation}
for $m$ even, as well as by the same terms with any smaller power of
the Laplacian.  If we combine these terms with the kinetic terms
\eqref{eq:qhSinglePartHam3}, the resulting Hamiltonian will single out
\eqref{eq:qhpsiLaugh} as the exact and unique ground state.

\subsection{Fractionally charged quasiparticle excitations}

Laughlin~\cite{laughlin83prl1395} created the elementary, charged
excitations of the fractionally quantized Hall state
\eqref{eq:qhpsiLaugh} through a \emph{Gedankenexperiment}.  If one
adiabatically inserts one Dirac quantum of magnetic flux through an
infinitely thin solenoid at a position $\xi$, and then removes this
flux quanta via a singular gauge transformation, the final Hamiltonian
will be identical to the initial one.  The final state will hence be
an eigenstate of the initial Hamiltonian as well.  The adiabatic
insertion of the flux will induce an electric field
\begin{equation}
  \oint \bs{E}\,\mathrm{d}\bs{s}
  =E_{\varphi}\cdot 2\pi r
  =\frac{1}{c}\parder{\phi}{t},
\end{equation}
which in turn will change the canonical angular momentum $L_\z $ around
$\xi$ by
\begin{equation}
  \label{eq:qhDeltaLz}
  \Delta L_\z 
  =\int F_{\varphi}\cdot r\,\mathrm{d}t
  =\frac{e}{2\pi c}\int \parder{\phi}{t}\,\mathrm{d}t
  =\frac{e}{2\pi c}\cdot \phi_0=\hbar .
\end{equation}
If we choose a basis of eigenstates of angular momentum around $\xi$, the 
basis states evolve according to 
\begin{equation}
  \label{eq:qhLzBasisEvo}
  (z-\xi)^m \,e^{-\frac{1}{4}|z|^2}
  \rightarrow (z-\xi)^{m+1} \,e^{-\frac{1}{4}|z|^2}.
\end{equation}
Note that the kinematical angular momentum, which is given by the
second term in \eqref{eq:qhSinglePartL1}, has eigenvalue $-\hbar n$,
where $n$ labels the Landau level.  In this process, it remains zero
as the states remain in the lowest Landau level---as there are no
states with positive kinematical angular momentum, the insertion of
the flux just shifts the states within the LLL.

The Laughlin ground state \eqref{eq:qhpsiLaugh} evolves in the process
into
\begin{equation}
  \label{eq:qhpsiQH}
  \psi_\xi^{\s\text{QH}}(z_1,\hdots,z_N)
  =\prod^N_{i=1}(z_i-\xi)
  \prod^N_{i<j}(z_i-z_j)^m\prod_{i=1}^N e^{-\frac{1}{4}|z_i|^2},
\end{equation}
which describes a quasihole excitation at $\xi$.  It is easy to see
that if the electron charge is $-e$, the charge of the quasihole is
$+e/m$.  If we were to create $m$ quasiholes at $\xi$ by inserting
$m$ Dirac quanta, the final wave function would be
\begin{equation}
  \label{eq:qhpsiQHThree}
  \psi_\xi^{\s m\,\text{QH's}}(z_1,\hdots,z_N)
  =\prod^N_{i=1}(z_i-\xi)^m
  \prod^N_{i<j}(z_i-z_j)^m\prod_{i=1}^N e^{-\frac{1}{4}|z_i|^2},
\end{equation}
\ie we would have created a true hole in the liquid, which is screened
as all the other particles.  Since the hole has charge $+e$,
the quasihole has charge $+e/m$.  One may view the quasihole as a zero
in the wave function which is not attached to any of the electrons.

The quasielectron, \ie the antiparticle of the quasihole, has charge
$-e/m$ and is created by inserting the flux adiabatically in the
opposite direction, thus lowering the angular momentum around some
position $\xi$ by $\hbar$, or alternatively, by removing one of the
zeros from the wave function.  To accomplish this formally, we first
rewrite \eqref{eq:qhpsiQH} in terms of ladder operators:
\begin{equation}
  \label{eq:qhpsiQHladder}
  \psi_\xi^{\s\text{QH}}(z_1,\hdots,z_N)
  =\prod^N_{i=1}\left(\sqrt{2}b^\dagger_i-\xi\right)
  \prod^N_{i<j}(z_i-z_j)^m\prod_{i=1}^N e^{-\frac{1}{4}|z_i|^2}.
\end{equation}
The insertion of a flux quanta in the opposite direction, or the
lowering of angular momentum around $\xi$, will then correspond to the
Hermitian conjugate operation.  Laughlin~\cite{laughlin84ss163} hence
proposed for the quasielectron wave function
\begin{align}
  \label{eq:qhpsiQE}
  \psi_{\bar\xi}^{\s\text{QE}}(z_1,\hdots,z_N)
%  \psi_m^{\s\text{QE}}(\bar\xi;z_1,\hdots,z_N)
  &=\prod^N_{i=1}\left(\sqrt{2}b_i-\bar\xi\right)
  \prod^N_{i<j}(z_i-z_j)^m\prod_{i=1}^N e^{-\frac{1}{4}|z_i|^2}
  \nonumber\\[0.3\baselineskip]
  &=\prod_{i=1}^N e^{-\frac{1}{4}|z_i|^2}
  \prod^N_{i=1}\left(2\parder{}{z_i}-\bar\xi\right)
  \prod^N_{i<j}(z_i-z_j)^m.
\end{align}
While the quasihole excitation \eqref{eq:qhpsiQH} is still an exact
eigenstate of Haldane's parent Hamiltonian, this is not true for the
quasielectron \eqref{eq:qhpsiQE}.  The problem here is that while
there is a clean and unique way of introducing an additional zero (we
just put it somewhere), there is no such clean way of removing one.
One can view the quasielectron as a region, in which $n$ electrons
nearby share $2n-1$ zeros attached to the particles.  In other words,
one zero is missing, but not from any specific electron---rather, the
dearth is distributed among all the electrons nearby.  The charge of
the quasielectron is accordingly not as localized as it is for the
quasihole.

The plateau in the observed Hall resistivity occurs because the
current in the experiments is carried by edge states, which are
sensitive only to the topological quantum numbers of the state.  In
the vicinity of one of the prefered filling fractions $\nu=1/m$, the
excess density of electrons yields to a finite density of
quasielectrons or holes, which get pinned by disorder and hence do not
contribute to the transport properties.

\subsection{Fractional statistics}

Possibly the most interesting property of fractionally quantized Hall
states is that the quasiparticle excitations obey fractional
statistics~\cite{halperin84prl1583,arovas-84prl722}.  The possibility
of fractional
statistics~\cite{leinaas-77ncb1,wilczek82prl1144,wilczek82prl957,wu84prl2103,arovas-85npb117,froehlich-88lmp347,goldhaber-89mpla21,wilczek90,khare05}
arises in two space dimensions because the space of trajectories for
two identical particles consists of an infinite number of
topologically distinct sectors, corresponding to the number of times
the particles wind around each other.  The laws of quantum mechanics
allow us to assign distinct phases to paths belonging these sectors,
which only need to satisfy the composition principle.

In three or more dimensions, by contrast, there are only two
topological distinct sectors, corresponding to interchanging the
particles or not interchanging them.  The group which classifies all
the topologically distinct trajectories is hence the permutation
group, and since amplitudes are complex numbers, the
possibilities for the quantum statistics are limited to the
one-dimensional representations of the permutation group.  There are
only two such representations, the symmetric and the antisymmetric
representation.  These correspond to the familiar choices of Bose and
Fermi statistics.

In two dimensions, the group is the braid group.  The one-dimensional
representations are obtained by assigning an arbitrary phase
$\tau(T_i)=e^{i\theta}$ for each counterclockwise interchange $T_i$ of
the two particles, with statistical parameter $\theta\in ]-\pi,\pi]$.
Particles interpolating between the familiar choices of bosons
($\theta=0$) and fermions ($\theta=\pi$) are generically called
anyons.  We will see in Section \ref{sec:3mod-pf-nastat} that
non-Abelian generalizations exist, where successive interchanges of
anyons do not commute.

\begin{figure}[tb]
  \begin{center}
  \begin{picture}(320,32)(-40,-3)
    \put(0,0){\oval(40,40)[t]}
    \put(0,0){\circle*{3}}
    \put(20,0){\circle*{3}}
    \put(-20,0){{\vector(0,-1){2}}}
    \put(52,20){\makebox(0,0)[l]{\small counterclockwise interchange yields:
        \quad $|\psi\!\rangle \rightarrow e^{\text{i}\theta} |\psi\!\rangle$}}
    \put(52,0){\makebox(0,0)[l]{\small relative angular momentum:\quad 
        $l_z \rightarrow l_z  - \frac{\hbar}{\pi}\theta$}}
%    \put(60,6){\makebox(0,0)[l]
%      {$\displaystyle \ket{\psi}\,\rightarrow \,e^{i\theta}\,\ket{\psi}$}}
  \end{picture}
  \end{center}
  \caption{Fractional statistics in two dimensions.  The many particle
    wave function acquires a statistical phase $\theta$ whenever we
    interchange two anyons conterclockwise.}
  \label{fig:statistics2D}
\end{figure}
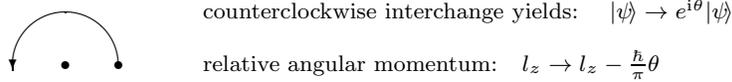

The most direct physical manifestation of the fractional statistics is
the quantization of the relative angular momentum of the anyons 
(see Figure \ref{fig:statistics2D}).
In three dimensions, there are three generators of rotations, and the
relative angular momentum is quantized as $\hbar l$, with $l$ an even
integer for bosons and an odd integer $l$ odd for fermions.  In two
dimensions, the wave function may acquire a phase
$\exp\!\left({\frac{i}{\pi}\theta\,\varphi}\right)$
%$\exp\!\left({{i}\theta\,\varphi/{\pi}}\right)$ 
%$e^{{{i}\theta\,\varphi/{\pi}}}$
%
% \begin{equation}
%   \label{eq:qhfsphase}
%   \exp\!\left({\frac{i}{\pi}\theta\,\varphi}\right) 
% \end{equation}
as two anyons wind counterclockwise around each other with winding
angle $\varphi$, which implies that the relative angular momentum is
quantized as
\begin{equation}
  \label{eq:qhLzrel}
  L_{\text{rel}} = \hbar\left(-\frac{\theta}{\pi}+2n\right),
%  L_\z  = \hbar\biggl(\frac{\theta}{\pi}+2n\biggr),
\end{equation}
where $n$ is an integer.  Note that the possibility of fractional
statistics exists only for particles which are strictly
two-dimensional, like vortices in an (approximately) two-dimensional
quantum fluid.
%
%  The ``two-dimensional'' electron gas in semiconductor inversion layer,
%  for example, would not qualify, as the electrons are only effectively
%  two-dimensional by being locked in the ground state of a potential
%  well in perpendicular direction.  Since there are always virtual
%  transitions into exited states, the particles are topologically
%  three-dimensional and have to be quantized accordingly.
% The excitations of such an electronic state,
% however, are strictly two-dimensional.

\vspace{.5\baselineskip} The only established realization of
fractional statistics is provided by the quasiparticles in the
fractionally quantized Hall
effect~\cite{halperin84prl1583,arovas-84prl722}.  When Laughlin
introduced the quasiparticles, he introduced them as localized defects
or more precisely, vortices in an otherwise uniform quantum liquid.
To address the question of their statistics, however, it is propitious
to view them as particles, with a Hilbert space spanned by the parent
wave function for the electrons.  We consider here a Laughlin
state with two quasiholes in an eigenstate of relative angular momentum
in an ``orbit'' centered at the origin.  Since the quasiholes have
charge $e^*=+e/m$, the effective flux quantum seen by them is
\begin{equation}
  \label{eq:qhPhiQH}
  \Phi_0^*=\frac{2\pi\hbar c}{e^*}=m\Phi_0,
\end{equation}
and the effective magnetic length is
% is $l^*=l\sqrt{m}$, 
\begin{equation}
  \label{eq:qhlQH}
  l^*=\sqrt{\frac{\hbar c}{e^*B}}=\sqrt{m}\,l.
%  (l^*)^2={\frac{\hbar c}{e^*B}}=m l
\end{equation}
We expect the single quasihole wave function to describe a particle of
charge $e^*$ in the LLL, and hence be of the general form
\begin{equation}
  \label{eq:qhQHphi}
  \phi(\bar\xi)=f(\bar\xi)\,e^{-\frac{1}{4m}|\xi|^2}.
\end{equation}
The complex conjugation %in comparison to \eqref{eq:qhSinPartLLL}
reflects that the sign of the quasihole charge is reversed relative to
the electron charge $-e$.  

The electron wave function for the state with two quasiholes 
in an eigenstate of relative angular momentum is given by
\begin{align}
  \label{eq:qh2QH}
  \psi(z_1,\hdots,z_N)
%  =\int\int \text{d}\xi_1\text{d}\bar\xi_1\text{d}\xi_2\text{d}\bar\xi_2
  =\int \text{D}[\xi_1,\xi_2] 
  \,\phi_{p,m}(\bar\xi_1,\bar\xi_2)
%  \,\phi_{\bar\xi_1,\bar\xi_2}
%  \,\psi_{\xi_1,\xi_2}^{m}(z_1,\hdots,z_N)
  \,\psi_{\xi_1,\xi_2}^{\s\text{QHs}}(z_1,\hdots,z_N)
%  \,\psi_m^{\s\text{QHs}}({\xi_1,\xi_2}; z_1,\hdots,z_N)
\end{align}
with
\renewcommand{\strut}{\rule[-.6\baselineskip]{0pt}{2\baselineskip}}
\begin{equation}
  \label{eq:qhphipm}\strut
  \phi_{p,m}(\bar\xi_1,\bar\xi_2)
  =(\bar\xi_1-\bar\xi_2)^{p+\frac{1}{m}} \prod_{k=1,2}e^{-\frac{1}{4m}|\xi_k|^2},
%  \quad p\ \text{even},
\end{equation}
where $p$ is an even integer, and
\begin{align}
  \label{eq:qhpsiXi1Xi2}
  \psi_{\xi_1,\xi_2}^{\s\text{QHs}}(z_1,\hdots,z_N)
  &=(\xi_1-\xi_2)^{\frac{1}{m}}\prod_{k=1,2}e^{-\frac{1}{4m}|\xi_k|^2}
  \nonumber\\[0.2\baselineskip]
  &\quad \hspace{-30pt}\cdot\prod^N_{i=1}(z_i-\xi_1)(z_i-\xi_2)
%  &\quad \hspace{-30pt}\cdot\prod_{k=1,2}\prod^N_{i=1}(z_i-\xi_i)
  \prod^N_{i<j}(z_i-z_j)^m\prod_{i=1}^N e^{-\frac{1}{4}|z_i|^2}.
\end{align}
The quasihole coordinate integration extends over the complex plane,
\begin{equation*}
  \label{eq:qhD}
  \int \text{D}[\xi_1,\xi_2]
  \equiv \int\!\!\ldots\!\!\int 
%  \text{d}\xi_1\text{d}\bar\xi_1\text{d}\xi_2\text{d}\bar\xi_2,
  \text{d}x_1 \text{d}y_1 \text{d}x_2 \text{d}y_2, 
\end{equation*}
where $\xi_1=x_1+\text{i} y_1$ and $\xi_2=x_2+\text{i} y_2$.

This needs explanation.  We see that both
$\phi_{p,m}(\bar\xi_1,\bar\xi_2)$ and 
% We will now motivate this form.  We see that both 
% $\phi_{p,m}(\bar\xi_1,\bar\xi_2)$ and \linebreak
$\psi_{\xi_1,\xi_2}^{\s\text{QHs}}(z_1,\hdots,z_N)$ contain multiple
valued functions of $\bar\xi_1-\bar\xi_2$ and $\xi_1-\xi_2$,
respectively, while the product of them is understood to be single
valued.  The reason for this is that the Hilbert space for the
quasiholes at $\xi_1$ and $\xi_2$ spanned by
$\psi_{\xi_1,\xi_2}^{\s\text{QHs}}(z_1,\hdots,z_N)$ has to be
normalized and is, apart from the exponential, supposed to be analytic
in $\xi_1$ and $\xi_2$.  At the same time, we expect
$\phi_{p,m}(\bar\xi_1,\bar\xi_2)$ to be of the general form
\eqref{eq:qhPhiQH}, \ie to be an analytic function of
$\bar\xi_1$, $\bar\xi_2$ times the exponential. 

The form \eqref{eq:qhphipm} of the quasihole wave function including
its branch cut, is indicative of fractional statistics with
statistical parameter $\theta =\pi/m$.  This indication, however, is
by itself not conclusive, as it is possible to change the
representation of the wave function through singular gauge
transformations~\cite{wilczek82prl957,halperin84prl1583}, where one
removes or adds flux tubes with a fraction of a Dirac flux quanta to
the particles, and hence turn an anyonic representation into a bosonic
or fermionic one and vice versa.  The physically unambivalent quantity
is the relative angular momentum of the quasiholes, which for
\eqref{eq:qhphipm} is given by
\begin{equation}
  \label{eq:qhLrel}
  L_{\text{rel}} %= \hbar\left(\frac{\theta}{\pi}+2n\right),
  = -\hbar\left(p + \frac{1}{m}\right).
\end{equation}
Comparing this with \eqref{eq:qhLzrel} yields $\theta =\pi/m$.  This
result agrees with the results of Halperin~\cite{halperin84prl1583}
and of Arovas, Schrieffer, and Wilczek~\cite{arovas-84prl722}, who
calculated the statistical parameter directly using the adiabatic
theorem~\cite{berry84prsla45,simon83prl2167,wilczek-84prl52,%Schiff55,
  WilczekShapere89}.

%%%%%%%%%%%%%%%%%%%%%%%%%%%%%%%%%%%%%%%%%%%%%%%%%%%%%%%%%%%%%%%%%%%%%%%%%%%
%%%%%%%%%%%%%%%%%%%%%%%%%%%%%%%%%%%%%%%%%%%%%%%%%%%%%%%%%%%%%%%%%%%%%%%%%%%
%%%%%%%%%%%%%%%%%%%%%%%%%%%%%%%%%%%%%%%%%%%%%%%%%%%%%%%%%%%%%%%%%%%%%%%%%%%
%%%%%%%%%%%%%%%%%%%%%%%%%%%%%%%%%%%%%%%%%%%%%%%%%%%%%%%%%%%%%%%%%%%%%%%%%%%
%%%%%%%%%%%%%%%%%%%%%%%%%%%%%%%%%%%%%%%%%%%%%%%%%%%%%%%%%%%%%%%%%%%%%%%%%%%

\subsection{Landau level quantization in the spherical geometry}
\label{sec:qhs}

The formalism for Landau level quantization in a spherical geometry,
\ie for the dynamics of a charged particle on the surface of a sphere
with radius $R$, in a magnetic (monopole) field, was pioneered by
Haldane for the lowest Landau
level~\cite{haldane83prl605,fano-86prb2670}, and only very recently
generalized to higher Landau levels~\cite{greiter11prb115129}.  We will
content ourselves here with a review of the formalism for the lowest
Landau level.

% We first review his formalism and then generalize it to the full
% Hilbert space, which includes higher Landau levels as
% well~\cite{greiter11prb115129}.
% 
% The key insight permitting this generalization is that there is not
% one but that there are two mutually commuting SU(2) algebras with spin $s$,
% one for the cyclotron variables and one for the guiding center
% variables.
%
% \vspace{\baselineskip} 
% \emph{Haldane's formalism.}---%
Following Haldane~\cite{haldane83prl605}, we assume a radial magnetic
field of strength
\begin{equation}
  \label{eq:qhsb}
  B=\frac{\hbar c s_0}{eR^2}\qquad(e>0).
\end{equation}
The number of magnetic Dirac flux quanta through the surface of the sphere is
\begin{equation}
  \label{eq:qhsphitot}
  \frac{\Phi_\text{tot}}{\Phi_0}=\frac{4\pi R^2B}{{2\pi\hbar c}/{e}} 
  =2s_0,
\end{equation}
which must be integer due to Dirac's monopole quantization 
condition~\cite{dirac31prsla60}.  In the following, we take $\hbar=c=1$.

The Hamiltonian is given by 
\begin{equation}
  \label{eq:qhsham1}
  H\;=\;\frac{\bs{\Lambda}^2}{2MR^2}
  \;=\;\frac{\omega_\text{c}}{2 s_0}\bs{\Lambda}^2,
\end{equation}
where $\omega_\text{c}=eB/M$ is the cyclotron frequency,
%where $\omega_\text{c}=\frac{eB}{M}$ and
\begin{equation}
  \label{eq:qhslambda1}
  \bs{\Lambda}=\bs{r}\times\big(-\i\nabla + e\bs{A}(\bs{r})\big)
\end{equation}
is the dynamical angular momentum, $\bs{r}=R\bs{e}_\r$, and
$\nabla\times\bs{A}=B\bs{e}_\r$.  With
\eqref{eq:app-sprhs}--\eqref{eq:app-spnabla} from
Appendix~\ref{sec:app-sphere} we obtain
\begin{equation}
  \label{eq:qhslambda2}
  \bs{\Lambda}=-\i\left(
    \bs{e}_\varphi\parder{}{\theta}
    -\bs{e}_\theta\frac{1}{\sin\theta}\parder{}{\varphi}
  \right) + eR\,\big(\bs{e}_\r\times\bs{A}(\bs{r})\big).
\end{equation}
Note that 
\begin{equation}
  \label{eq:qhsedotlambda}
%  \bs{e}_\r\cdot \bs{\Lambda}=\bs{\Lambda}\cdot\bs{e}_\r=0,
  \bs{e}_\r\bs{\Lambda}=\bs{\Lambda}\bs{e}_\r=0,
\end{equation}
as one can easily verify with \eqref{eq:app-spderivatives}.
%
% The commutators of the Cartesian components of $\bs{\Lambda}$ with themselves
% and with $\bs{e}_\r$ can easily be evaluated using \eqref{eq:qhslambda2} and 
% \eqref{eq:app-spe}, which yields
% \begin{align}
%   \label{eq:qhslambdacom}
%   \comm{\Lambda^i}{\Lambda^j}&
%   =i\epsilon^{ijk}(\Lambda^k- s_0\,e_\r^k),
%   \qquad i,j,k=x,y,\ \text{or}\ z,
%   \\[0.5\baselineskip]
%   \comm{\Lambda^i}{e_\r^j}&=i\epsilon^{ijk}e_\r^k,
% \end{align}
% where $e_\r^k$ is the $k$-th cartesian coordinate of $\bs{e}_\r$.
%
The commutators of the Cartesian components of $\bs{\Lambda}$ with themselves
and with $\bs{e}_\r$ can easily be evaluated using \eqref{eq:qhslambda2} and 
\eqref{eq:app-spe}--\eqref{eq:app-spderivatives}.  This yields
\begin{gather}
  \label{eq:qhslambdacomm}
%   \comm{\Lambda^\alpha}{\Lambda^\beta}
%   =i\epsilon^{\alpha\beta\gamma}(\Lambda^\gamma- s_0\,e_\r^\gamma), 
%   \\[0.5\baselineskip] \label{eq:qhslambdaecomm}
%   \comm{\Lambda^\alpha}{e_\r^\beta}=i\epsilon^{\alpha\beta\gamma}e_\r^\gamma,
  \comm{\Lambda^i}{\Lambda^j}
  =\i\varepsilon^{ijk}(\Lambda^k- s_0\,e_\r^k), 
  \\[0.5\baselineskip] \label{eq:qhslambdaecomm}
  \comm{\Lambda^i}{e_\r^j}=\i\epsilon^{ijk}e_\r^k,
\end{gather}
where $i,j,k=\x,\y,\ \text{or}\ \z$, and $e_\r^k$ is
the $k$-th Cartesian coordinate of $\bs{e}_\r$.  From
%\eqref{eq:qhslambdacom}, \eqref{eq:qhslambdaecom} and 
% \eqref{eq:qhsedotlambda}, 
\eqref{eq:qhsedotlambda}--\eqref{eq:qhslambdaecomm},
we see that that the operator
\begin{equation}
  \label{eq:qhL1}
  \bs{L}=\bs{\Lambda}+ s_0\bs{e}_\r
\end{equation}
%satisfies an SU(2) angular momentum algebra and 
is the generator of rotations around the origin,
\begin{equation}
  \label{eq:qhslambdaplusecomm}
%  \comm{L^\alpha}{X^\beta}=\i\epsilon^{\alpha\beta\gamma}X^\gamma
  \comm{L^i}{X^j}=\i\epsilon^{ijk}X^k
\quad\text{with}\quad \bs{X}=\bs{\Lambda},\;\bs{e}_\r,\;\text{or}\;\bs{L},
\end{equation}
and hence the angular momentum.  As it satisfies the angular momentum
algebra, it can be quantized accordingly.  Note that $\bs{L}$ has a
component in the $\bs{e}_\r$ direction:
\begin{equation}
  \label{eq:qhLdoter}
  \bs{L}\bs{e}_\r=\bs{e}_\r\bs{L}= s_0.
\end{equation}
%Since $\bs{L}$ satisfies the SU(2) angular momentum algebra, it can be
%quantized accordingly.  
If we take the eigenvalue of $\bs{L}^2$ to be $s(s+1)$, this implies
$s=s_0+n$, where $n=0,1,2,\ldots$ is a non-negative integer (while $s$
and $s_0$ can be integer or half integer, according to number of Dirac
flux quanta through the sphere).

With \eqref{eq:qhL1} and \eqref{eq:qhsedotlambda}, we obtain
\begin{equation}
  \label{eq:qhslambda^2L^2}
  \bs{\Lambda}^2=\bs{L}^2-s_0^2.
\end{equation}
The energy eigenvalues of \eqref{eq:qhsham1} are hence
\begin{align}
  \label{eq:qhse}
  E_n&=\frac{ \omega_\text{c}}{2s_0}
   \big[s(s+1) - s_0^2\big]\nonumber\\[0.5\baselineskip]
   &=\frac{ \omega_\text{c}}{2s_0}
   \big[(2n+1)s_0 + n(n+1) \big]\nonumber\\[0.5\baselineskip]
   &= \omega_\text{c}
   \left[\left(n+\frac{1}{2}\right) + \frac{n(n+1)}{2s_0} \right].
\end{align}
The index $n$ hence labels the Landau levels.

\vspace{\baselineskip}
To obtain the eigenstates of \eqref{eq:qhsham1}, we have to choose a
gauge and then explicitly solve the eigenvalue equation.  We choose the
latitudinal gauge
\begin{equation}
  \label{eq:qhsgauge}
  \bs{A}=-\bs{e}_\varphi\frac{s_0}{eR}\cot\theta.
\end{equation}
The singularities of $\bs{B}=\nabla\times\bs{A}$ at the poles are
without physical significance.  They describe infinitly thin solenoids
admitting flux $s_0\Phi_0$ each and reflect our inability to formulate
a true magnetic monopole.

The dynamical angular momentum \eqref{eq:qhslambda2} becomes
\begin{equation}
  \label{eq:qhslambda3}
  \bs{\Lambda}=-\i\left[
    \bs{e}_\varphi\parder{}{\theta}
    -\bs{e}_\theta\frac{1}{\sin\theta}
    \left(\parder{}{\varphi}-\i s_0\cos\theta\right)
  \right].
\end{equation}
With \eqref{eq:app-spderivatives} we obtain
\begin{equation}
  \label{eq:qhslambda^2}
  \bs{\Lambda}^2\;=\;
     -\frac{1}{\sin\theta}\parder{}{\theta}
     \left(\sin\theta\parder{}{\theta}\right)
     -\frac{1}{\sin^2\theta}
     \left(\parder{}{\varphi}-\i s_0\cos\theta\right)^2.
\end{equation}

%\vspace{0.5\baselineskip}% 
To formulate the eigenstates,
Haldane~\cite{haldane83prl605} introduced spinor coordinates for the
particle position,
\begin{equation}
  \label{eq:qhsuv}%\strut 
%  u=\cos\frac{\theta}{2} e^{i\frac{\varphi}{2}},\
%  u=\cos\left(\frac{\theta}{2}\right) \exp\left({i\frac{\varphi}{2}}\right),\
%  u=\cos{\left(\textstyle\frac{\theta}{2}\right)} \exp\left({\textstyle\frac{i\varphi}{2}}\right),\
%  v=\sin{\left(\textstyle\frac{\theta}{2}\right)} \exp\left({\textstyle -\frac{i\varphi}{2}}\right),\
  u=\cos\frac{\theta}{2} \exp\left({\frac{i\varphi}{2}}\right),\
  v=\sin\frac{\theta}{2} \exp\left({-\frac{i\varphi}{2}}\right),
\end{equation}
such that
\begin{equation}
  \label{eq:qhsOmegauv}
  \bs{e}_\r%(\theta,\varphi)
  \;=\;\bs{\Omega}(u,v)\;\equiv\;(u,v)\,\bs{\sigma}\!\left(\!\!
    \begin{array}{c}
      \bar u\\[2pt]\bar v
    \end{array}\!\!\right)\!,
\end{equation}
where $\bs{\sigma}=(\sigma_\x,\sigma_\y,\sigma_\z )$ is the vector consisting of
the three Pauli matrices
\begin{equation}
  \label{eq:qhsPauli}
  \sigma_\x=\!\left(\!\!
    \begin{array}{cc}
      \,0\,&\,1\,\\1&0 
    \end{array}\!\!\right),\quad
  \sigma_\y=\!\left(\!\!
    \begin{array}{cc}
      \,0\,&\,-\i\,\\\i&\,0\, 
    \end{array}\!\!\right),\quad
  \sigma_\z =\!\left(\!\!
    \begin{array}{cc}
      \,1\,&\,0\,\\0&\,-1\, 
    \end{array}\!\!\right).
\end{equation}

In terms of these, a complete, orthogonal
%orthonormal 
basis of the states spanning the lowest Landau
level ($n=0$, $s=s_0$) is given by
\begin{equation}
  \label{eq:qhsLLLbasis}
  \psi_{m,0}^{s}(u,v) = 
%  \sqrt{\frac{(2s+1)!}{4\pi\,(s+m)!\, (s-m)!}}\;
  u^{s+m}v^{s-m}
\end{equation}
with $$m=-s,s+1,\ldots,s.$$  For these states,
\begin{align}
  \label{eq:qhspsi0eq}
  L^\z \psi_{m,0}^{s}&= m\,\psi_{m,0}^{s},
  \nonumber\\[0.2\baselineskip]
  H\psi_{m,0}^{s}&= \frac{1}{2}\omega_{\text{c}}\,\psi_{m,0}^{s}.
\end{align}
% \begin{equation}
%   \label{eq:qhspsi0eq}
%   H\psi_{m,0}^{s}=\frac{1}{2}\omega_{\text{c}}\,\psi_{m,0}^{s}.
% \end{equation}
%
To verify \eqref{eq:qhspsi0eq}, we consider the action of 
\eqref{eq:qhslambda^2} on the more general basis states
\begin{align}
  \label{eq:qhsmpbasis}
  \phi_{m,p}^{s}(u,v)
%  &=\left(\textstyle\cos\frac{\theta}{2}\right)^{s+m} 
%         \left(\textstyle\sin\frac{\theta}{2}\right)^{s-m} e^{i(m-p)\varphi} 
  &=\left(\cos\frac{\theta}{2}\right)^{s+m} 
         \left(\sin\frac{\theta}{2}\right)^{s-m} 
         e^{i(m-p)\varphi}
%         \exp\left({i(m-p)\varphi}\right)
  \nonumber\\[0.5\baselineskip]
  &= \left\{\begin{array}{ll}
    {\bar v}^{-p}\,u^{s+m}\,v^{s-m+p},\quad & \text{for}\ p<0,\\[5pt] 
    {\bar u}^{p}\,u^{s+m-p}\,v^{s-m}, \quad & \text{for}\ p\ge 0.
    \end{array}\right.
\end{align}
This yields
\begin{align}
  \label{eq:qhsLambda^2mpbasis}
  \bs{\Lambda}^2\,\phi_{m,p}^{s}
  &=\left[s
    -\left(\frac{s\cos\theta-m}{\sin\theta}\right)^2
    +\left(\frac{s_0\cos\theta-m+p}{\sin\theta}\right)^2
  \right]\phi_{m,p}^{s}
  \nonumber\\[0.5\baselineskip]
%   &=\left[s
%     +\frac{2(s\cos\theta\!-\!m)(p\!-\!n\cos\theta)
%       +(p\!-\!n\cos\theta)^2}{\sin^2\theta}
%   \right]\psi_{m,p}^{s},\quad
%   \nonumber\\[0.5\baselineskip]
  &=\left[s
    +\frac{2(s\cos\theta\!-\!m\!+\!p)(p\!-\!n\cos\theta)
      -(p^2\!-\!n^2\cos^2\theta)}{\sin^2\theta}
  \right]\phi_{m,p}^{s},\quad
%   \nonumber\\[0.5\baselineskip]
%   \!&=\left[(2n+1)s-n^2\right]\psi_{m,p}^{s}
%   \nonumber\\[0.5\baselineskip]
%   &\quad 
%   +\;\frac{2(sp\!+\!mn\!-\!pn)\cos\theta -2sn\!-\!2mp\!+\!p^2\!+\!n^2}%
%   {\sin^2\theta}\,\phi_{m,p}^{s},\quad
\end{align}
%where we have used $s_0=s-n$.  
For $p=n=0$, this clearly reduces to
$\bs{\Lambda}^2\,\psi_{m,0}^{s}=s\,\psi_{m,0}^{s}$, and hence
\eqref{eq:qhspsi0eq}.
The normalization of \eqref{eq:qhsLLLbasis} can easily be obtained
with the integral
\begin{equation}
  \label{eq:qhspsinorm}
  \frac{1}{4\pi}\int \text{d}\Omega\, \bar u^{S+m'}\bar v^{S-m'} u^{s+m}v^{s-m}
  =\frac{(s+m)!\, (s-m)!}{(2s+1)!}\,\delta_{mm'},
\end{equation}
where $\text{d}\Omega=\sin\theta\, \text{d}\theta\, \text{d}\phi$.

\vspace{0.5\baselineskip} 
To describe particles in the lowest Landau level which are localized at a
point $\bs{\Omega}(\alpha,\beta)$ with spinor coordinates
$(\alpha,\beta)$,
\begin{equation}
  \label{eq:qhsOmegaalphabeta}
  \bs{\Omega}(\alpha,\beta)=(\alpha,\beta)\,\bs{\sigma}\!\left(\!\!
    \begin{array}{c}
      \bar\alpha\\[2pt]\bar\beta
    \end{array}\!\!\right)\!,
\end{equation}
Haldane~\cite{haldane83prl605} introduced ``coherent states'' defined by
\renewcommand{\strut}{\rule[-.6\baselineskip]{0pt}{2\baselineskip}}
\begin{equation}
  \label{eq:qhsCoherentDef}%\strut
  \{\bs{\Omega}(\alpha,\beta)\,\bs{L}\}\,\psi_{(\alpha,\beta),0}^{s}(u,v)
  =s\,\psi_{(\alpha,\beta),0}^{s}(u,v).
\end{equation}
In the lowest Landau level, the angular momentum $\bs{L}$ can be written
\begin{equation}
  \label{eq:qhslLLL}
  \bs{L}=\frac{1}{2}(u,v)\,\bs{\sigma}\!\left(\!\!
    \begin{array}{c}
      \parder{}{u}\\[4pt] \parder{}{v}
    \end{array}\!\!\right)\!.
\end{equation}
Note that $u,v$ may be viewed as Schwinger boson creation, and
$\parder{}{u},\parder{}{v}$ the corresponding annihilation operators
(see Section \ref{sec:naschwinger}).
The solutions of \eqref{eq:qhsCoherentDef} are given by
\begin{equation}
  \label{eq:qhsCoherent}\strut
  \psi_{(\alpha,\beta),0}^{s}(u,v) = (\bar\alpha u + \bar\beta v)^{2s},
\end{equation}
as one can verify easily with the identity
\begin{equation}
  \label{eq:qhsBiedenharn}
%  (\bs{a}\,\bs{\sigma}\,\bs{b})(\bs{c}\,\bs{\sigma}\,\bs{d})
%      =2 (\bs{a}\bs{d})(\bs{c}\bs{b})-(\bs{a}\bs{b})(\bs{c}\bs{d}).
   (\underline{a}\,\bs{\sigma}\,\underline{b})
   (\underline{c}\,\bs{\sigma}\,\underline{d})
      =2 (\underline{a}\,\underline{d})(\underline{c}\,\underline{b})
      -(\underline{a}\,\underline{b})(\underline{c}\,\underline{d}).
\end{equation}
where $\underline{a}$, $\underline{b}$, $\underline{c}$,
$\underline{d}$ are two-component spinors.

\vspace{0.5\baselineskip} 
Haldane~\cite{haldane83prl605} further introduced two-particle 
coherent lowest Landau level states defined by
\renewcommand{\strut}{\rule[-.6\baselineskip]{0pt}{2\baselineskip}}
\begin{equation}
  \label{eq:qhstwoCoherentDef}%\strut
  \{\bs{\Omega}(\alpha,\beta)\,(\bs{L}_1+\bs{L}_2)\}\,
  \psi_{(\alpha,\beta),0}^{s,j}[u,v]
  =j\,\psi_{(\alpha,\beta),0}^{s,j}[u,v],
\end{equation}
where $[u,v]:=(u_1,u_2,v_1,v_2)$ and $j$ is the total
angular momentum,
\begin{equation}
  \label{eq:qhstwoCoherentJtot}%\strut
  (\bs{L}_1+\bs{L}_2)^2\,
  \psi_{(\alpha,\beta),0}^{s,j}[u,v]
  =j(j+1)\,\psi_{(\alpha,\beta),0}^{s,j}[u,v].
\end{equation}
The solution of \eqref{eq:qhstwoCoherentDef} is given by
%The wave functions for these states are given by
\begin{equation}
  \label{eq:qhstwoCoherent}%\strut
  \psi_{(\alpha,\beta),0}^{s,j}[u,v] = 
  (u_1 v_2 - u_2 v_1)^{2s-j}\,
  \prod_{i=1,2}(\bar\alpha  u_i + \bar\beta  v_i)^{j}.
\end{equation}
It describes two particles with relative momentum ${2s-j}$ precessing
about their common center of mass at $\bs{\Omega}(\alpha,\beta)$.

Since $0\le j\le 2s$, the relative momentum quantum number $l=2s-j$
has to be a non-negative integer.  The restriction to non-negative
integers is a consequence of Landau level quantization, and exists in
the plane as well, as we discussed in Section \ref{sec:qhlaughlin}.
For bosons or fermions, $l$ has to be even or odd, respectively.  This
implies that the projection $\Pi_0$ into the lowest Landau level of
any rotationally invariant operator $V(\bs{r}_1\cdot\bs{r}_2)$, such
as two particle interactions, can be expanded as
\begin{equation}
  \label{eq:qhsLLLpro}
  \Pi_0 V(\bs{r}_1\cdot\bs{r}_2) \Pi_0
  =\sum_{l}^{2s} V_l\, P_{2s-l}(\bs{L}_1+\bs{L}_2),
\end{equation}
where the sum over $l$ is restricted to even (odd) integer for bosons
(fermions), $V_l$ denotes the so-called pseudopotential coefficients,
and $P_j(\bs{L})$ % $P_j(\bs{L}_1+\bs{L}_2)$ 
is the projection operator on states with
total momentum $\bs{L}^2=j(j+1)$.

% As mentioned, this formalism was very recently generalized to include
% higher Landau levels as well~\cite{greiter11prb115129}.  The key insight
% permitting this generalization was that there are, in analogy to the
% two mutually commuting ladder algebras $a,a^\dagger$ and $b,b^\dagger$
% in the plane, two mutually commuting SU(2) algebras with spin $s$, one
% for the cyclotron variables and one for the guiding center variables.

As mentioned, this formalism was very recently generalized to include
higher Landau levels as well~\cite{greiter11prb115129}.  The key insight
permitting this generalization was that there are two mutually
commuting SU(2) algebras with spin $s$, one for the cyclotron
variables and one for the guiding center variables.  These algebras
are analogous to the the two mutually commuting ladder algebras
$a,a^\dagger$ and $b,b^\dagger$ in the plane, which we introduced in
Section \ref{sec:qhp}.

\subsection{The Laughlin state and its parent Hamiltonian on the sphere}
\label{sec:laughsphere}

In analogy to \eqref{eq:qhpsiLaugh}, Haldane~\cite{haldane83prl605}
writes the Laughlin $\nu=1/m$ state for $N$ particles on a sphere with
$2s_0=m(N-1)$ as
\begin{equation}
  \label{eq:qhspsiLaugh}
  \psi_m[u,v]=\prod_{i<j}^N(u_iv_j-u_jv_i)^m.
\end{equation}
Since the factors $(u_iv_j-u_jv_i)$ commute with the total angular 
momentum
\begin{equation}
  \label{eq:qhsLtot}
  \bs{L}_{\text{tot}}=\sum_{i=1}^N \bs{L}_i,
\end{equation}
\eqref{eq:qhspsiLaugh} is obviously invariant under spacial rotations
around the sphere:
\begin{equation}
  \label{eq:qhsLtotpsi}
  \bs{L}_{\text{tot}}\psi_m=0.
\end{equation}
The Laughlin droplet wave function centered at $\bs{\Omega}(\alpha,\beta)$
can be recovered by multiplying $\psi_m[u,v]$ by a factor
\begin{equation*}
  \prod_{i=1}^N(\bar\alpha u_i+\bar\beta v_i)^n,
\end{equation*}
and then taking the limit $n\to\infty$, $R\to\infty$, while $4\pi
R^2/n=2\pi l^2=\text{const.}$, where $l^2$ is the magnetic length
\eqref{eq:qhl^2}.

As in the plane, the uniquely specifying property of the Laughlin
state \eqref{eq:qhspsiLaugh} is that the smallest component of
relative angular momentum is $m$, which is even for bosons and odd for
fermions.  Haldane~\cite{haldane83prl605} constructed a model
Hamiltonian, which, together with the kinetic Hamiltonian
\eqref{eq:qhsham1}, singles out \eqref{eq:qhspsiLaugh} as exact and
unique zero energy ground state, by assigning a finite energy cost to
the components of angular momentum smaller than $m$.  With the most
general two-particle interaction Hamiltonian given by
\begin{equation}
  \label{eq:qhsVham2}
  H_{\s\text{int}}=\sum_{i<j}^N 
  \left\{\sum_{l}^{2s} V_l \text{P}_{2s-l}(\bs{L}_i+\bs{L}_j)\right\},
\end{equation}
where the values of $l$ are restricted to even (odd) integers for
bosons (fermions) and $\text{P}_{2s-l}$ is as defined in
\eqref{eq:qhsLLLpro}, Haldane's Hamiltonian amounts to taking
\begin{equation}
  \label{eq:qhsPP}
  V_l=
  \left\{\begin{array}{ll}
    1 &\text{for}\ l<m,\\[2pt]
    0 &\text{for}\ l\ge m .
   \end{array}\right.
\end{equation}
For all practical purposes, we need to rewrite \eqref{eq:qhsVham2} in terms
of boson or fermion creation or annihilation operators,
\begin{align}
  \label{eq:qhsVham2s}
  H_{\s\text{int}}
  &=\sum_{m_1=-s}^s\, \sum_{m_2=-s}^s\, \sum_{m_3=-s}^s\, \sum_{m_4=-s}^s
  \ a_{m_1}^\dagger a_{m_2}^\dagger a_{m_3} a_{m_4}\,
  \delta_{m_1+m_2,m_3+m_4}
  \nonumber\\[0.2\baselineskip]
  &\quad\cdot\sum_{l=0}^{2s}
  \braket{s,m_1;s,m_2}{2s-l,m_1+m_2} V_l \braket{2s-l,m_3+m_4}{s,m_3;s,m_4}\!,
  \hspace{8pt}
  \nonumber\\[-0.2\baselineskip]
\end{align}
where $a_m$ annihilates a boson or fermion in the properly normalized
single particle state
\begin{equation}
  \label{eq:qhsnormLLLbasis}
  \psi_{m,0}^{s}(u,v) = 
  \sqrt{\frac{(2s+1)!}{4\pi\,(s+m)!\, (s-m)!}}\;
  u^{s+m}v^{s-m},
\end{equation}
and $\braket{s,m_1;s,m_2}{j,m_1+m_2}$ \etc are Clebsch--Gordan
coefficients~\cite{baym69}.  Essentially, we take two particles with
$L_\z $ eigenvalues $m_3$ and $m_4$, change the basis into one where
$m_3+m_4$ and the total two particle momentum $2s-l$ are replacing the
quantum numbers $m_3$ and $m_4$, multiply each amplitude by $V_l$, and
convert the two particles states back into a basis of $L_\z $
eigenvalues $m_1$ and $m_2$.

The fractionally charged quasihole and quasielectron excitations of
the Laughlin state \eqref{eq:qhspsiLaugh} localized at
$\bs{\Omega}(\alpha,\beta)$ on the sphere are given by
\begin{align}
  \label{eq:qhspsiLaughQH}
  \psi_{(\alpha,\beta)}^{\s\text{QH}}[u,v]
    &= \prod_{i=1}^N(\beta u_i-\alpha v_i)\,
    \prod_{i<j}^N(u_iv_j-u_jv_i)^m
\end{align}
and
\begin{align}
  \label{eq:qhspsiLaughQE}
  \hspace{30pt}\psi_{(\alpha,\beta)}^{\s\text{QE}}[u,v]
    &= \prod_{i=1}^N(\bar\beta\parder{}{u_i}-\bar\alpha\parder{}{v_i})\,
    \prod_{i<j}^N(u_iv_j-u_jv_i)^m,
\end{align}
which increase or decrease the number of flux quanta $2s_0$ through
the sphere by one, and decrease or increase
$\bs{\Omega}(\alpha,\beta)\bs{L}_{\text{tot}}$ by $\frac{1}{2}N$.

Due to the formal simplicity, the sphere is particularly well suited
to formulate the hierarchy of quantized Hall states, where all
odd-denominator filling fractions can be obtained through successive
condensation of quasiparticles into Laughlin-type
fluids~\cite{haldane83prl605,halperin84prl1583,greiter94plb48}.

%\newpage
\section{The Haldane--Shastry model}
\label{sec:hs}

\subsection{The $1/r^2$ model of Haldane and Shastry}
The Haldane--Shastry
model~\cite{haldane88prl635,shastry88prl639,inozemtsev90jsp1143,haldane91prl1529,shastry92prl164,haldane-92prl2021,kawakami92prb1005,kawakami92prb3191,talstra95,laughlin-00proc,bernevig-01prl3392,bernevig-01prb024425,greiter-07prl237202}
is one of the most important paradigms for a generic spin \half liquid on
a chain.  Consider a spin \half chain with periodic boundary
conditions and an even number of sites $N$ on a unit circle embedded
in the complex plane:
%\begin{equation}
%\begin{picture}(280,70)(-20,-35)
\begin{center}
\begin{picture}(320,70)(-40,-35)
\put(0,0){\circle{100}}
%\put(20,0){\circle*{3}}
\put( 20.0,   .0){\circle*{3}}
\put( 17.3, 10.0){\circle*{3}}
\put( 10.0, 17.3){\circle*{3}}
\put(   .0, 20.0){\circle*{3}}
\put(-10.0, 17.3){\circle*{3}}
\put(-17.3, 10.0){\circle*{3}}
\put(-20.0,   .0){\circle*{3}}
\put(-17.3,-10.0){\circle*{3}}
\put(-10.0,-17.3){\circle*{3}}
\put(   .0,-20.0){\circle*{3}}
\put( 10.0,-17.3){\circle*{3}}
\put( 17.3,-10.0){\circle*{3}}
\qbezier[20]( 20.0,   .0)(  5.0,8.65)(-10.0, 17.3)
\put(50,12){\makebox(0,0)[l]
{$N$\ sites with spin \half on unit circle: 
}}
%$\displaystyle \eta_\alpha=e^{i\frac{2\pi}{N}\alpha }$}}
\put(50,-12){\makebox(0,0)[l]
{$\displaystyle \eta_\alpha=e^{\text{i}\frac{2\pi}{N}\alpha }$
\ \ with\ $\alpha = 1,\ldots ,N$}}
\end{picture}
\end{center}
%\end{equation}
The ${1}/{r^2}$-Hamiltonian
\begin{equation}
  \label{eq:hsham}
  {H}^{\s\text{HS}} = \left(\frac{2\pi}{N}\right)^2
  \sum^N_{\alpha <\beta}\,
  \frac{{\boldsymbol{S}}_\alpha {\boldsymbol{S}}_\beta 
  }{\left|\eta_\alpha-\eta_\beta \right|^2}\,,
\end{equation}
where $\left|\eta_\alpha-\eta_\beta \right|$ is the chord distance between
the sites $\alpha$ and $\beta$, has the exact ground state 
\begin{equation}
  \label{eq:hsket}
  \ket{\psi^{\s\text{HS}}_{0}}\,=
  \sum_{\{z_1,\ldots ,z_M\}}\psi^{\s\text{HS}}_{0} 
  (z_1,\ldots ,z_M)\,{S}^+_{z_1}\cdot\ldots\cdot {S}^+_{z_M} 
  \big|\underbrace{\dw\dw\ldots\ldots\dw}_{\text{all\ } N \text{\ spins\ } \dw}
  \big\rangle,
\end{equation}
where the sum extends over all possible ways to distribute the 
$M=\frac{N}{2}$ $\up$-spin coordinates $z_i$ on the unit circle and
\begin{equation}
  \label{eq:hspsi0}
  \psi^{\s\text{HS}}_{0}(z_1,z_2,\ldots ,z_M) = 
  \prod_{i<i}^M\,(z_i-z_j)^2\,\prod_{i=1}^M\,z_i\,. 
\end{equation}
The ground state has momentum
\begin{equation}
  \label{eq:hsgp0}
  p_0=-\frac{\pi}{2} N,
\end{equation}
where we have adopted a convention according to which the
``vacuum'' state $\ket{\dw\dw\ldots\dw}$ has momentum $p=0$ (and the
empty state $\ket{0}$ has $p=\pi (N-1)$) and energy
%This state is real, a spin singlet, and has ground state energy
%$E_0=-\frac{\pi^2}{24}\left(N-\frac{5}{N}\right)$.
\begin{equation}
  \label{eq:hse0}
  E_0=-\frac{\pi^2}{24}\left(N+\frac{5}{N}\right).
\end{equation}
We will verify \eqref{eq:hsgp0} and \eqref{eq:hse0} in Sections
\ref{sec:hsgs} and \ref{sec:hsexsol}, respectively.

\subsection{Symmetries and integrability}
\label{sec:hssymm}

The Haldane--Shastry Hamiltonian \eqref{eq:hsham} is clearly invariant
under space translations (rotations of the unit circle), time
reversal, parity, and global SU(2) spin rotations generated by
\begin{equation}
  {\bs{S}}_{\text{tot}} = \sum_{\alpha=1}^N {\bs{S}}_\alpha,\quad
  \comm{{H}^{\s\text{HS}}}{{\bs{S}}_{\text{tot}}} = 0.
  \label{eq:hsspinsymmetry}
\end{equation}
The total spin trivially satisfies the standard commutation relations for
angular momentum,
\begin{equation}
  \bigcomm{{S}^i_{\text{tot}}}{{S}^j_{\text{tot}}}
  =\text{i}\,\varepsilon^{ijk}\,{S}^k_{\text{tot}}.
\end{equation}
The model possesses an additional
symmetry~\cite{haldane-92prl2021,ha-93prb12459} generated by 
% another vector operator, 
the rapidity operator
\begin{equation}
  {\bs{\Lambda}}=\frac{\text{i}}{2}
  \sum^N_{\substack{\alpha,\beta=1\\\alpha\neq\beta}}
  \frac{\eta_\alpha + \eta_\beta}{\eta_\alpha - \eta_\beta}\,
  {\bs{S}}_\alpha\times{\bs{S}}_\beta,\quad
  \comm{{H}^{\s\text{HS}}}{{\bs{\Lambda}}} = 0,
  \label{eq:hsrapidityoperator}
\end{equation}
which measures the spin current. 
%which corresponds physically to a spin current operator. 
It transforms as a vector under spin rotations,
\begin{equation}
\comm{{S}^i_{\text{tot}}}{{\Lambda}^j}=\text{i}\,\varepsilon^{ijk}\,{\Lambda}^k.
\label{eq:commSLambda}
\end{equation}
Note that even though both $\bs{S}_{\text{tot}}$ and ${\bs{\Lambda}}$
commute with the Hamiltonian, they do not commute mutually, but
generate an infinite dimensional associative algebra with certain
defining relations and consistency conditions, %.  This algebra is
the Yangian Y(sl$_2$)~\cite{Drinfeld85smd254,ChariPressley98}.
Since the commutator of the total spin squared with the rapidity
operator does not vanish in general, 
\begin{equation}
  \label{eq:hscommS^2Lambda}
  \comm{\bs{S}_{\text{tot}}^2}{\Lambda^i}
  =-\text{i}\,\varepsilon^{ijk}\,\biganticomm{S^j_{\text{tot}}}{\Lambda^k},
\end{equation}
elements of the Yangian algebra connect degenerate eigenstates with
different total spins.  With these elements, it is possible to generate
all the eigenstates of the model from all the completely spin
polarized eigenstates.

The Yangian symmetry of the
model~\cite{haldane-92prl2021,ha-93prb12459} implies significant
degeneracies in the spectrum and hence indicates integrability.  The
model is not integrable in the usual sense, however, as the method of
quantum inverse scattering~\cite{KorepinBogoliubovIzergin93} is not
applicable to models with longe-range interactions.  Talstra and
Haldane~\cite{talstra-95jpa2369} have nonetheless succeeded in
constructing an infinite set of mutually commuting integrals of motion
for the model by using the determinant rather than the trace of the
monodromy matrix.  These integrals provide the framework for the
model's integrability.  The integrability is hence only indirectly
related to the Yangian symmetry.

The model is further amenable to exact solution via the asymtotic
Bethe
Ansatz~\cite{haldane91prl1529,sutherland71jmp246,sutherland71jmp251,sutherland71pra2019,sutherland72pra1372,kawakami92prb1005,ha-93prb12459,ha-94prl2887},
even though the application of this method to models with long-range
interactions is likewise heuristic.

\subsection{Ground state properties}
\label{sec:hsgs}
%The proof of solution is rather lengthy\cite{hs,mimo}.  

The ground state \eqref{eq:hspsi0} is %further
real (and hence both parity and time-reversal invariant), a spin
singlet, and can equivalently be obtained by Gutzwiller
projection~\cite{gutzwiller63prl159,gaudin73jpp511,metha-75jmp1256,kaplan-82prl889,gros-87prb381,metzner-87prl121,gebhardt-87prl1472},
as we will verify now after evaluating the total momentum.

%\vspace{.5\baselineskip} 
\vspace{\baselineskip} 
\emph{Ground state momentum.}---%
To determine the momentum $p_0$ (in units of inverse lattice spacings
$1/a$) we translate the ground state (\ref{eq:hspsi0})
counterclockwise by one lattice spacing around the unit circle,
\begin{equation}
  \label{eq:hst0}
  \boldsymbol{T} \ket{\psi^{\s\text{HS}}_{0}} 
  = e^{\text{i}p_0} \ket{\psi^{\s\text{HS}}_{0}}.
\end{equation}
%With $\boldsymbol{T} z_i=\eta_1 z_i=e^{i\frac{2\pi}{N}} z_i$, 
With $\boldsymbol{T} z_i=e^{\text{i}\frac{2\pi}{N}} z_i$, 
%With $\boldsymbol{T} z_i=\exp\left(i\frac{2\pi}{N}\right) z_i$, 
%With $\boldsymbol{T} z_i=\eta_1 z_i$, 
we find 
\begin{equation*}
  \label{eq:hsgp00}
  p_0=\frac{2\pi}{N} \left(2\,\frac{M(M-1)}{2}+M\right) = \pi M, 
\end{equation*}
and hence \eqref{eq:hsgp0}.  Note that the sign of $p_0$ is irrelevant
for \eqref{eq:hspsi0}, as $N$ is always even, and $p_0$ is $0$ or
$\pi$.  The sign will become significant only in sections
\ref{sec:hsspinons} and \ref{sec:hsyt} below, when we assign spinons
momenta for states with $N$ odd.

\vspace{\baselineskip} 
\emph{Singlet property.}---%
%We now proof that (\ref{eq:hspsi0}) is a singlet.  
Since
${S}^\z _{\text{tot}}\ket{\psi^{\s\text{HS}}_{0}}=0$, it suffices to show that
$\ket{\psi^{\s\text{HS}}_{0}}$ is annihilated by ${S}^-_{\text{tot}}$:
\begin{align}
  \label{eq:hsStot-psi0}
  {S}^-_{\text{tot}}\ket{\psi^{\s\text{HS}}_{0}}&=
  \sum_{\alpha=1}^{N} {S}^-_\alpha 
  \sum_{\{z_1,\ldots z_M\}} \psi^{\s\text{HS}}_{0}(z_1,z_2,\ldots z_M)\, 
  {S}^+_{z_1}\ldots {S}^+_{z_M}\ket{\dw\dw\ldots\dw} 
  \nonumber\\[0.2\baselineskip] 
  &=\sum_{\{z_2,\ldots,z_M\}}\, 
  \underbrace{\sum_{\alpha=1}^{N}\,\psi^{\s\text{HS}}_{0} 
    (\eta_\alpha,z_2,\ldots,z_M)}_{=0} 
  \,{S}^+_{z_2}\ldots {S}^+_{z_M}\ket{\dw\dw\ldots\dw},%=0
\end{align}
since $\psi^{\s\text{HS}}_{0}(\eta_\alpha,z_2,\ldots,z_M)$ contains
only powers $\eta_\alpha^1, \eta_\alpha^2,\ldots , \eta_\alpha^{N-1}$ and
\begin{equation}
  \sum_{\alpha=1}^{N} \eta_\alpha^m = 
  % \left\{\begin{array}{cl} 
  %     N &\ \ \text{for}\ m=0\ \text{mod}\ N\\
  %     0 &\ \ \text{otherwise.}
  %   \end{array}\right.
  N\delta_{m,0}\quad \text{mod}\ N.
  \label{eq:hssumeta^m}
\end{equation}

\vspace{\baselineskip} 
%\emph{Parity (P) and time reversal (T) invariance.}---%
\emph{Parity and time reversal invariance.}---%
We begin by showing that $\psi^{\s\text{HS}}_{0}$ is real.
With $\bar z_i=1/z_i$ and hence
\begin{equation}
(z_i-z_j)^2\,=- z_i z_j\, |z_i-z_j|^2\,,
\end{equation}
we write
\begin{align}
\psi^{\s\text{HS}}_{0}
(z_1,z_2,\ldots ,z_M) &=
\pm\,\prod_{i<j}^M\,|z_i-z_j|^2\;
\prod_{i<j}^M\,z_i z_j\;\prod_{i=1}^M\,z_i\nonumber\\
&=\pm\,\prod_{i<j}^M\,|z_i-z_j|^2\;\prod_{i=1}^M\,G(z_i)
\label{eq:hspsi0real}
\end{align}
where 
\begin{equation}
G(\eta_\alpha)=(\eta_\alpha)^\frac{N}{2}=
\left\{\begin{array}{ll} 
+1 &\quad \alpha\ \text{even}\\
-1 &\quad \alpha\ \text{odd}.
\end{array}\right.
\label{eq:hspsi0g}
\end{equation}
The gauge factor $G(z_i)$ effects that the Marshall sign 
criteria~\cite{marshall55prsla48} 
is fulfilled. 

Since parity tranforms $\eta_\alpha\to\eta_{-\alpha}=\bar\eta_\alpha$
and hence $z_i\to\bar z_i$, the fact that $\psi^{\s\text{HS}}_{0}$ is
real implies that $\ket{\psi^{\s\text{HS}}_{0}}$ is invariant under
parity.  Time reversal transforms~\cite{gottfried66} 
% $\text{i}\to -\text{i}$ and therefore $z_i\to\bar z_i$, but it also
% tranforms $\bs{S}_\alpha\to -\bs{S}_\alpha$ (and hence $S^+_\alpha\to
% -S^-_\alpha$) as well as\footnote{See Appendix Appendix~\ref{sec:app-am} for
%   the notation.}
% \begin{equation*}
%   \ket{j,m}\to \text{i}^{2m}\ket{j,-m},
% \end{equation*}
% which implies that our basis ket with $N$ $\dw$-spins transforms as
\begin{equation*}
  \text{i}\to -\text{i},\quad 
%  z_i\to\bar z_i,\quad
  \bs{S}_\alpha\to -\bs{S}_\alpha,\quad 
  \ket{s,m}\to \text{i}^{2m}\ket{s,-m},
\end{equation*}
which implies $z_i\to\bar z_i$, $S^+_\alpha\to -S^-_\alpha$, and 
$\ket{\dw\dw\ldots\dw}\to(-\text{i})^N \ket{\up\up\ldots\up}$.
% that the basis ket with $N$ $\dw$-spins transforms as
% \begin{equation*}
%   \ket{\dw\dw\ldots\dw}\to(-\text{i})^N \ket{\up\up\ldots\up}.
% \end{equation*}
The basis states in \eqref{eq:hsket} hence transform according to
\begin{equation}
  \label{eq:hsketT}
  {S}^+_{z_1}\cdot\ldots\cdot {S}^+_{z_M} \ket{\dw\dw\ldots\dw}
  \to {S}^-_{z_1}\cdot\ldots\cdot {S}^-_{z_M} \ket{\up\up\ldots\up}.
\end{equation}
Together with the singlet property, this implies that 
$\ket{\psi^{\s\text{HS}}_{0}}$ is invariant under time reversal.

%\vspace{.5\baselineskip}
\vspace{\baselineskip} 
\emph{Generation by Gutzwiller projection.}---%
% Finally, we show that
% $\psi^{\s\text{HS}}_{0}$ can be
% generated by Gutzwiller Projection from filled bands.
The ground state of the model was first obtained by Gutzwiller
projection from a completely filled one-dimensional band which in
total contains as many spin \half fermions as there are lattice
sites~\cite{gutzwiller63prl159,kaplan-82prl889,gros-87prb381,metzner-87prl121,gebhardt-87prl1472}:
\begin{equation}
  \label{eq:hsgw}
  \ket{\psi^{\s\text{HS}}_{0}}
  =\text{P}_{\s\text{GW}}\ket{\psi^{\s N}_{\s\text{SD}}},\qquad
  \ket{\psi^{\s N}_{\s\text{SD}}}\equiv
  \prod_{q\in\mathcal{I}} c_{q\up}^\dagger c_{q\dw}^\dagger\ket{0},
\end{equation}
where the Gutzwiller projector %$P_{\s\text{GW}}$ 
\begin{equation}
  \label{eq:hsgwp}
  \text{P}_{\s\textrm{GW}} \equiv \prod_{i=1}^N
  \big(1-c^\dagger_{i\up}c^{\phantom{\dagger}}_{i\up}
  c^\dagger_{i\dw}c_{i\dw} \big)
\end{equation}
eliminates configurations with more than one particle on any site and
the interval $\mathcal{I}$ contains $M\frac{N}{2}$ adjacent momenta.
We will now show that \eqref{eq:hsgw} is equivalent to
\eqref{eq:hspsi0}.  With lattice constant $a=\frac{2\pi}{N}$,
the allowed momenta are given by integers, $q=0,1,\ldots ,N-1$.
With
\begin{equation}
  \label{eq:hscq}
  c_{q}^\dagger
  =\sum_{\alpha=1}^N e^{\text{i}\frac{2\pi}{N}\alpha q} c_{\alpha}^\dagger
  =\sum_{\alpha=1}^N \eta_{\alpha}^q c_{\alpha}^\dagger,
\end{equation}
the (unnormalized) single particle momentum eigenstates are given by
\begin{equation}
  \label{eq:hssps}
  \phi_q(z)=\braket{z}{q}=\bra{0}c_{z} c_{q}^\dagger\ket{0}=z^q.
\end{equation}
The many particle wave function for $M$ fermions with adjacent momenta
$q\in\mathcal{I}=[q_1,q_1+M-1]$ is hence given by
\begin{equation}
  \label{eq:hsmps}
  \phi_{\mathcal{I}}(z_1,z_2,\ldots,z_M)
  =\prod_{i=1}^M z_i^{q_1}\cdot 
  \mathcal{A}\left\{z_1^0z_2^1\ldots z_M^{M-1}\right\}
  =\prod_{i=1}^M z_i^{q_1}\, \prod_{i<j}^M (z_i-z_j). 
\end{equation}
The Gutzwiller state \eqref{eq:hsgw} is given by
% \begin{equation}
%   \label{eq:hsgww}
%   \begin{array}{r}\displaystyle
%     \ket{\psi^{\s\text{HS}}_{0}}
%     =\sum_{\s\{z_1,\ldots ,z_M; w_1,\ldots,w_{M} \}}\,
%     \phi_{\mathcal{I}}(z_1,\ldots,z_M)\phi_{\mathcal{I}}(w_1,\ldots,w_M)
%     \\ \rule{0pt}{0pt}\displaystyle
%     \cdot\,c^\dagger_{z_1\up}\ldots c^\dagger_{z_M\up}\,
%     c^\dagger_{w_1\dw}\ldots c^\dagger_{w_{M}\dw}\vac ,
%   \end{array}
% \end{equation}
\begin{align}
  \label{eq:hsgww}
    \ket{\psi^{\s\text{HS}}_{0}}
    &=\sum_{\s\{z_1,\ldots ,z_M; w_1,\ldots,w_{M} \}}\,
    \phi_{\mathcal{I}}(z_1,\ldots,z_M)\,\phi_{\mathcal{I}}(w_1,\ldots,w_M)
    \nonumber\\[0.2\baselineskip]
    &\hspace{100pt}\cdot\,c^\dagger_{z_1\up}\ldots c^\dagger_{z_M\up}\,
    c^\dagger_{w_1\dw}\ldots c^\dagger_{w_{M}\dw}\vac ,
\end{align}
where the sum extends over all possible ways to distribute the
coordinates $z_i$ and $w_k$ on mutually distinct lattice sites.  

Let $\tilde{\mathcal{I}}$ contain all those $M$ momenta not contained
in $\mathcal{I}$, and $w_1,\ldots,w_M$ denote the sites which are not occupied by
any of the $z_i$'s.  Then
\begin{align}
  \label{eq:hsphiw}
  \phi_{\mathcal{I}}(w_1,\ldots,w_M)
%  &\equiv \phi_{\mathcal{I}}[w]\\
  &=\bra{0}
  c_{w_M}\ldots c_{w_1} \prod_{q\in\mathcal{I}}c_{q}^\dagger\ket{0}
  \nonumber\\[0.2\baselineskip]
  &=\text{sign}[z;w]\cdot\bra{0}
  \prod_{q\in\mathcal{I}}c_{q}\prod_{q\in\tilde{\mathcal{I}}}c_{q}\;
  c^\dagger_{z_1}\ldots c^\dagger_{z_M}
  \prod_{q\in\mathcal{I}}c_{q}^\dagger\ket{0}\nonumber\\[0.2\baselineskip]
  &=\text{sign}[z;w]\cdot
  \bra{0}\prod_{q\in\tilde{\mathcal{I}}}c_{q}\;
  c^\dagger_{z_1}\ldots c^\dagger_{z_M}\ket{0}\nonumber\\[0.2\baselineskip]
  &=\text{sign}[z;w]\cdot
  {\phi_{\tilde{\mathcal{I}}}}^*(z_1,\ldots,z_M)\nonumber\\[0.2\baselineskip]
  &=\text{sign}[z;w]\cdot\prod_{i=1}^M \bar{z_i}^{M}\cdot
  {\phi_{\mathcal{I}}}^*(z_1,\ldots,z_M), 
%\nonumber\\
%  &=\text{sign}[z;w]\cdot\prod_{i=1}^M G(z_i)\cdot
%  {\phi_{\mathcal{I}}}^*(z_1,\ldots,z_M),
\end{align}
where
\begin{equation}
  \label{eq:hssign}
  \text{sign}[z;w]\equiv\bra{0}
  c_{w_M}\ldots c_{w_1} c_{z_M}\ldots c_{z_1}
  \prod_{q\in\tilde{\mathcal{I}}}c^\dagger_{q}\,\prod_{q\in\mathcal{I}}c^\dagger_{q}
  \ket{0}
\end{equation}
is an overal sign associated with ordering the $z$'s and $w$'s 
according to the lattice sites indices $\alpha$.  Since
\begin{equation}
  \label{eq:hsvacc}
  \text{sign}[z;w]\cdot c^\dagger_{z_1\up}\ldots c^\dagger_{z_M\up}\,
  c^\dagger_{w_1\dw}\ldots c^\dagger_{w_{M}\dw}\vac
  ={S}^+_{z_1}\cdot\ldots\cdot {S}^+_{z_M}, 
%  \big|\underbrace{\dw\dw\dw\ldots\ldots\dw
%  }_{\text{all\ } N \text{\ spins\ } \dw}\big\rangle,
  \ket{\dw\dw\ldots\dw}
\end{equation}
we may write
\begin{multline}
  \label{eq:hsgwww}
    \ket{\psi^{\s\text{HS}}_{0}}
    =\sum_{\s\{z_1,\ldots ,z_M\}}
    |\phi_{\mathcal{I}}(z_1,\ldots,z_M)|^2\,\,\prod_{i=1}^M  G(z_i)\,
    {S}^+_{z_1}\cdot\ldots\cdot {S}^+_{z_M}\ket{\dw\dw\ldots\dw}.
\end{multline}
This is equivalent to \eqref{eq:hspsi0real}.

As an aside, it is very easy to verify the singlet property in the
Gutzwiller formulation \eqref{eq:hsgw} of the ground state.  To begin
with, filling the same single particle states with $\up$ and $\dw$
spin fermions obviously yields a singlet,
\begin{equation}
  \label{eq:hssdsin}
  {\bs{S}}_{\text{tot}} \ket{\psi^{\s N}_{\s\text{SD}}} = 0.
\end{equation}
The Gutzwiller projector \eqref{eq:hsgwp}, however, commutes with
the local spin operators and hence also with the total spin,
\begin{equation}
  \label{eq:hsgwcomm}
  \comm{\text{P}_{\s\textrm{GW}}}{{\bs{S}}_{\alpha}}
  =  \comm{\text{P}_{\s\textrm{GW}}}{{\bs{S}}_{\text{tot}}}
  = 0.
\end{equation}
Hence 
\begin{equation}
  \label{eq:hssdsin2}
  {\bs{S}}_{\text{tot}} \ket{\psi^{\s\text{HS}}_{0}} = 0.
\end{equation}

%\vspace{.5\baselineskip} 
\vspace{\baselineskip} 
\emph{Norm.}---%
The norm of the ground state is~\cite{wilson62jmp1040}
\begin{align}
  \label{eq:hsnorm}
%  C_M &= 
  \sum_{\s\{z_1,\ldots ,z_M\}} \prod_{i<j}^M | z_i - z_j |^4
%  \nonumber\\ 
  &=
  \left( \frac{N}{2 \pi \text{i} }\right)^M 
  \oint \frac{ \text{d} z_1}{  z_1} \ldots \oint
  \frac{ \text{d} z_{M}}{  z_{M} } \prod_{ i \neq j }^M
  \left( 1 - \frac{ z_i }{ z_j} \right)^2 \nonumber\\
  &= \frac{ N^M  ( 2 M )!}{ 2^M}.
\end{align}

\vspace{\baselineskip} \emph{Relation to the chiral spin liquid.}---%
The Haldane--Shastry ground state may be viewed as the one-dimensional
analog of the abelian or $S=\frac{1}{2}$ chiral spin
liquid~\cite{kalmeyer-87prl2095,kivelson-88prl2630,zou-89prb11424,wen-89prb11413,kalmeyer-89prb11879,laughlin-90prb664,schroeter-07prl097202,thomale-09prb104406},
which is essentially a Laughlin $m=2$ quantized Hall
state~\cite{laughlin83prl1395} for spin flips on a two dimensional
lattice.  The spinons in the chiral spin liquid were understood to
obey half-Fermi statistics long before this was realized for the
Haldane-Shastry model.

%\newpage
%\subsection{Verification of the model}
\subsection{Explict solution}
\label{sec:hsexsol}

For the explict calculation presented here to be applicable to the
one- and two-spinon eigenstates investigated in section
\ref{sec:hsspinons} below, we consider wavefunctions of the
form~\cite{haldane91prl1529,laughlin-00proc,bernevig-01prl3392,bernevig-01prb024425}
\begin{equation}
  \label{eq:hspsi}
  \psi%^{\s\text{HS}}
  (z_1,\dots,z_M)
  = \phi(z_1,\dots,z_M)\cdot\psi^{\s\text{HS}}_0(z_1,\dots,z_M),
\end{equation}
where $\psi^{\s\text{HS}}_0$ is given by \eqref{eq:hsket} and
$\phi[z]\equiv\phi(z_1,\dots,z_M)$ a polynomial of degree strictly less than
$N-2M+2$ in each of the $z_i$'s.  This implies that degree of
$\psi^{\s\text{HS}}$ is strictly less than $N+1$.  $N$ can be even or
odd.  This condition enables us to use a Taylor expansion when we
calculate the action of the %Haldane--Shastry
Hamiltonian \eqref{eq:hsham} on the state.  The result is that
\begin{equation}
  \label{eq:hspsieq}
  H^{\s\text{HS}} \ket{\psi}%^{\s\text{HS}}} 
  =\frac{2\pi^2}{N^2}\!\left(\lambda
%  = \frac{1}{2}\left(\frac{2\pi}{N}\right)^2\!\left(\lambda
    +\frac{N}{48}(N^2\!-\!1)+\frac{M}{6}(4M^2\!-\!1)-\frac{N}{2} M^2\right)
  \ket{\psi},%^{\s\text{HS}}},
\end{equation}
provided that $\phi$ satisfies the eigenvalue equation
\begin{equation}
  \label{eq:hsphieq}
  \sum_{j=1}^M\Biggl(\frac{1}{2}z_j^2\frac{\partial^2}{\partial z_j^2} 
    + \sum_{\substack{k=1\\[1pt]k\neq j}}^M 
      \frac{2z_j^2}{z_{j}-z_{k}}\frac{\partial}{\partial z_j}
    - \frac{N-3}{2}z_{j}\frac{\partial}{\partial z_{j}}\Biggr)\phi[z] 
  = \lambda\phi[z]
\end{equation}
for $\lambda$.  The derivative operators in \eqref{eq:hsphieq} and
below are understood to act on the analytic extension of
$\phi(z_1,\ldots,z_{M})$, in which the $z_i$'s are allowed to take any
value in the complex plane.
For $\phi[z]=1$, \eqref{eq:hspsieq} shows that
$\ket{\psi^{\s\text{HS}}_0}$ is an eigenstate of $H^{\s\text{HS}}$
with energy $E_0$ given by \eqref{eq:hse0}.

%\vspace{.5\baselineskip} 
\vspace{\baselineskip} 
\emph{Derivation of \eqref{eq:hspsieq} and \eqref{eq:hsphieq}.}---%
%To derive \eqref{eq:hspsieq} and \eqref{eq:hsphieq}, 
We first use
$S^{\pm}=S^\x \pm i S^\y$
% \begin{equation}
%   \label{eq:hss+-}
% %  S_{\alpha}^{\pm}=S_{\alpha}^\x \pm i S_{\alpha}^\y
%   S^{\pm}=S^\x \pm i S^\y
% \end{equation}
to rewrite \eqref{eq:hsham} as the sum of a ``kinetic'' %$S^+S^-$ 
and a ``potential'' %$S^\z S^\z $ 
term,
\begin{equation}
  \label{eq:hshamladder}
  H^{\s\text{HS}} 
  =\frac{2\pi^2}{N^2}
%  =\frac{1}{2}\left(\frac{2\pi}{N}\right)^2
%  \sum^N_{\substack{\alpha,\beta=1\\\alpha\neq\beta}}
  \sum^N_{\alpha\neq\beta}
  \frac{1}{\vert \eta_{\alpha}-\eta_{\beta}\vert^2}
  \left(S_{\alpha}^+S_{\beta}^-+S_{\alpha}^\z S_{\beta}^\z \right).
%  \equiv\frac{2\pi^2}{N^2}(T+V).
\end{equation}
%
% \begin{equation}
%   \label{eq:hshamladder}
%   H^{\s\text{HS}} 
% %  =\frac{2\pi^2}{N^2}\Biggl(
%   =\frac{1}{2}\left(\frac{2\pi}{N}\right)^2\Biggl(
%   \sum^N_{\alpha\neq\beta}
%   \frac{S_{\alpha}^+S_{\beta}^-}{\vert \eta_{\alpha}-\eta_{\beta}\vert^2}
%   + \sum^N_{\alpha\neq\beta}
%   \frac{S_{\alpha}^\z S_{\beta}^\z }{\vert \eta_{\alpha}-\eta_{\beta}\vert^2}
%   \Biggl)
%   =
% \end{equation}
We first evaluate the action of the kinetic term on $\ket{\psi}$.
%, and then the action of the potential term.
% 
% \vspace{.5\baselineskip}
% \emph{Evaluation of the kinetic term.}---%
Consider first
\begin{align}
  \label{eq:hss+s-psi}
  S_{\alpha}^+S_{\beta}^-\ket{\psi}
  &= S_{\alpha}^+S_{\beta}^-
  \sum_{\{z_2,\ldots ,z_M\}}\,\psi 
  (\eta_{\beta},z_2,\ldots ,z_M)\,
  {S}^+_{\beta}{S}^+_{z_2}\cdot\ldots\cdot {S}^+_{z_M}
  \ket{\dw\dw\ldots\dw}
  \nonumber\\[0.2\baselineskip]
  &=\sum_{\{z_2,\ldots ,z_M\}}\,\psi 
  (\eta_{\beta},z_2,\ldots ,z_M)\,
  {S}^+_{\alpha}{S}^+_{z_2}\cdot\ldots\cdot {S}^+_{z_M}
  \ket{\dw\dw\ldots\dw},
\end{align}
where we have implicitly assumed that each spin configuration in the sum over
${\{z_1,z_2,\ldots ,z_M\}}$ in \eqref{eq:hsket} appears only once (and not
$M!$ times due to permutations of the $z_i$'s).  We write this %abbreviated 
as
\begin{equation}
  \label{eq:hss+s-abb}
  \left[S_{\alpha}^+S_{\beta}^-\psi \right]\!(\eta_{\alpha},z_2,\ldots ,z_M)
  =\psi (\eta_{\beta},z_2,\ldots ,z_M).
\end{equation}
Note in particlular that 
$\bigr[S_{\alpha}^+S_{\beta}^-\psi \bigl](z_1,z_2,\ldots ,z_M)$
vanishes unless $\eta_{\alpha}$ equals one of the $z_i$'s.

The action of the kinetic term on $\psi$ is given by
\begin{align}
  \label{eq:hs1}
% \Biggl[\Biggr.\!\hspace{-6pt}&\quad \hspace{-16pt} 
% \sum_{\substack{\alpha\neq\beta}}^N
% \frac{ S_{\alpha}^{+}S_{\beta}^{-}}{\vert \eta_\alpha - \eta_\beta \vert^2}
% \psi \Biggl]\biggr. (z_{1},\ldots,z_{M})\\
  T\psi[z] &\equiv\Biggl[\,\sum_{\substack{\alpha\neq\beta}}^N
  \frac{ S_{\alpha}^{+}S_{\beta}^{-}}{\vert \eta_\alpha - \eta_\beta \vert^2}
  \psi \Biggr](z_{1},\ldots,z_{M})\nonumber\\[0.2\baselineskip]
  &=\sum_{j=1}^M\sum_{\substack{\beta=1\\[1pt]\eta_\beta \neq z_j}}^N
  \frac{\eta_\beta}{\vert z_j-\eta_\beta\vert^2}
  \frac{\psi (z_{1}, \ldots , z_{j-1}, \eta_\beta ,z_{j+1}, \ldots , z_{M})}
  {\eta_\beta}.
\end{align}
Since the last fraction %$\frac{\psi[z]}{\eta_\alpha}$ 
%${\psi[z]}/{\eta_\alpha}$ 
is a polynomial of degree strictly less than $N$ in $\beta$, we can
Taylor expand it around $z_j$,
\begin{equation}
  \label{eq:hs2}
  \frac{\psi(z_1,\ldots, \eta_\beta,\ldots,z_M)}{\eta_\beta}
  =\sum_{l = 0}^{N-1}\frac{(\eta_\beta - z_j)^l}{l!}
  \frac{\partial^l}{\partial z_j^l} \;%\left( 
  \frac{ \Psi(z_1,\ldots,z_M)}{z_j}.%\right).
\end{equation}
The sum over $\beta$ yields
\begin{equation}
  \label{eq:hs3}
  \sum_{\substack{\beta=1\\[1pt]\eta_\beta \neq z_j}}^N
  \frac{\eta_\beta(\eta_\beta - z_j)^l}{\vert z_j-\eta_\beta\vert^2}
  =z_j^{l+1} A_l,\quad
%  A_l\equiv -\sum_{\beta=1}^{N-1} \eta_\beta^2 (\eta_\beta -1)^{l-2},
  A_l=-\sum_{\alpha=1}^{N-1} \eta_\alpha^2 (\eta_\alpha -1)^{l-2},
\end{equation}
where $A_0$, $A_1$, and $A_2$ are evaluated with \eqref{eq:app-hs11},
\eqref{eq:app-hsfouriersum1}, and \eqref{eq:app-hs2} from
Appendix~\ref{sec:app-hssums}, respectively:
\begin{align*}
  A_0&=-\sum_{\alpha=1}^{N-1} \frac{\eta_\alpha^2}{(\eta_\alpha -1)^2}
  =\frac{(N-1)(N-5)}{12},%\label{eq:app-hsa0}
  \\[0.2\baselineskip]
  A_1&=-\sum_{\alpha=1}^{N-1} \frac{\eta_\alpha^2}{\eta_\alpha -1}
  =-\frac{N-3}{2},%\label{eq:app-hsa1}
  \\[0.2\baselineskip]
  A_2&=-\sum_{\alpha=1}^{N-1}\eta_\alpha^2=1,%\label{eq:app-hsa2}
  \\[0.2\baselineskip]
  A_l&=-\sum_{\alpha=1}^{N}\eta_\alpha^2 (\eta_\alpha -1)^{l-2}=0 
  \qquad\text{for}\ 2 < l\le N-1. %\label{eq:app-hsal3}
\end{align*}
In the last line, we have used that $\eta_\alpha^2 (\eta_\alpha
-1)^{l-2}$ vanishes for $\eta_\alpha=1$ and contains only powers
$\eta_\alpha^2,\ldots\eta_\alpha^{N-1}$ for $2 < l\le N-1$.
Substituion into \eqref{eq:hs1} and \eqref{eq:hs2} yields
\begin{align}
  \label{eq:hs4}
  T\psi[z] &= \sum_{j=1}^{M} \left( \frac{(N\!-\!1)(N\!-\!5)}{12} z_{j}
    - \frac{N\!-\!3}{2} z_{j}^2 \frac{\partial}{\partial z_{j}}
    + \frac{1}{2} z_{j}^3
    \frac{\partial^2} {\partial z_{j}^2} \right)
  \frac{ \psi[z]}{ z_{j}} 
  \nonumber\\[0.2\baselineskip]
  &=\frac{M(N-1)(N-5)}{12}\,\psi[z]
  \;-\;\frac{N-3}{2}
  \underbrace{\sum_{{%\substack{j,k=1\\ 
      j\neq k}}^M\frac{2z_j}{z_j-z_k}}_{=M(M-1)}\psi[z] 
  \nonumber\\*[0.2\baselineskip]
  &\quad+ \sum_{{%\substack{j,k=1\\ 
      j\neq k}}^M\frac{z_j^2}{(z_j-z_k)^2}\,\psi[z] \;
  +\underbrace{\sum_{\substack{j,k,m=1\\[1pt] j \neq k \neq m\neq j}}^{M}
    \frac{2 z_{j}^2}{(z_{j} - z_{k})(z_{j} - z_{m})}}_{=2M(M-1)(M-2)/3}
  \psi[z] \nonumber\\*[0.2\baselineskip]
  &\quad+\sum_{j=1}^M \psi_0^{\s\text{HS}}[z]
   \Biggl(\frac{1}{2}z_j^2\frac{\partial^2}{\partial z_j^2} +
   \sum_{{%\substack{k=1\\
       k\neq j}}^M \frac{2z_j^2}{z_{j}-z_{k}}\psi_0
   \frac{\partial}{\partial z_j}
   -\frac{N-3}{2}z_{j}\frac{\partial}{\partial z_{j}}\Biggr)\phi[z],
\nonumber
\end{align}
where we have used the algebraic identity \eqref{eq:app-hsidentity}
% \begin{equation}
%   \label{eq:app-hsidentity2}
%   \frac{a^2}{(a-b)(a-c)}+\frac{b^2}{(b-a)(b-c)}+\frac{c^2}{(c-a)(c-b)} =1
% \end{equation}
in the evaluation of the triple sum.

For the action of the potential term we write
\begin{displaymath}
  S_\alpha^\z S_\beta^\z =\left(S_\alpha^\z +\frac{1}{2}\right)
  \left(S_\beta^\z +\frac{1}{2}\right)
  -\frac{1}{2}(S_\alpha^\z +S_\beta^\z )-\frac{1}{4}.
\end{displaymath}
This yields
\begin{align}
  V\psi[z] &\equiv  \Biggl[\,\sum_{\substack{\alpha\neq\beta}}^N
  \frac{S_\alpha^\z S_\beta^\z }
  {\vert\eta_\alpha -\eta_\beta\vert^2}\psi\Biggr](z_1,\ldots,z_M)
  \nonumber\\[0.2\baselineskip] 
  &=\sum_{j \neq k}^M \frac{1}{\vert z_j-z_k\vert^2}\,\psi[z]-
  \sum_{\alpha\neq\beta}^N
  \frac{S_\alpha^\z +\frac{1}{2}}{\vert \eta_\alpha - \eta_\beta \vert^2 }\,
  \psi[z]
%  \nonumber\\[0.2\baselineskip] 
%  &\quad
   +\frac{1}{4}\underbrace{\sum_{\alpha\neq\beta}^N
    \frac{1}{\vert \eta_\alpha - \eta_\beta \vert^2 }}_{=N(N^2-1)/12}\,\psi[z].
\end{align}
With
\begin{equation*}
  \sum_{j \neq k}^M \frac{1}{\vert z_j-z_k\vert^2}\,\psi[z]
  +\sum_{j \neq k}^M\frac{z_j^2}{(z_j-z_k)^2}\,\psi[z]
  =\frac{1}{2}M(M-1)\,\psi[z]
\end{equation*}
and
\begin{equation*}
  \sum_{\alpha\neq\beta}^N 
  \frac{S_\alpha^\z +\frac{1}{2}}{\vert\eta_\alpha-\eta_\beta\vert^2 }\,\psi[z] 
  =\sum_{\alpha=1}^N \sum_{\beta=1}^{N-1}
  \frac{S_\alpha^\z +\frac{1}{2}}{\vert 1 - \eta_\beta \vert^2 }\,\psi[z] 
  = M\frac{N^2-1}{12}\psi[z],
\end{equation*}
where we have substituted $\eta_{\beta}\to{\eta_{\beta}}{\eta_{\alpha}}$
and used \eqref{eq:app-hs12}, we obtain \eqref{eq:hspsieq}
and \eqref{eq:hsphieq}.

\subsection{Factorization of the Hamiltonian} 
\label{sec:hsfactorization}

In \ref{sec:hsexsol} we have shown that $\ket{\psi^{\s\text{HS}}_0}$
is an eigenstate of $H^{\s\text{HS}}$ with energy $E_0$ given by
\eqref{eq:hse0}.  To show that $\ket{\psi^{\s\text{HS}}_0}$ is the
ground state (or at least one of several ground states), we factorize
the Haldane--Shastry
Hamiltonian~\cite{shastry92prl164,laughlin-00proc,bernevig-01prb024425}.
For every site $\eta_{\alpha}$, we define an auxiliary operator
$\bs{D}_{\alpha}$ by
\begin{equation}
  \bs{D}_\alpha
  = \frac{1}{2} \sum_{\substack{\beta=1\\\beta\neq\alpha}}^N \frac{\eta_\alpha +
    \eta_\beta}{\eta_\alpha - \eta_\beta} \Bigl[ \text{i} ( \bs{S}_\alpha \times
  \bs{S}_\beta ) + \bs{S}_\beta \Bigr].
  \label{eq:hsdoperator}
\end{equation}
The rapidity operator \eqref{eq:hsrapidityoperator} is given in terms
of these by
\begin{equation}
  \sum_{\alpha=1}^N \bs{D}_{\alpha} = \bs{\Lambda},
\end{equation}
as one can easily see with \eqref{eq:app-hs13}.  

We will show below that $H^{\s\text{HS}}$ can be written as:
\begin{equation}
  \label{eq:hsfactorization}
  H^{\s\text{HS}}=\frac{2\pi^2}{N} \Biggl[ 
  \frac{2}{9}\sum_{\alpha=1}^N \bs{D}_\alpha^\dagger %\cdot 
  \bs{D}_\alpha
%  - \frac{ N ( N^2 + 5 )}{ 48} 
  + \frac{ N +1}{ 12} \bs{S}_{\text{tot}}^2 \Biggr] +E_0,
\end{equation}
which consists of two positive semi-definite operators (\ie operators
with only non-negative eigenvalues) and a constant.  The lowest energy
eigenvalue of $H^{\s\text{HS}}$ is therefore $E_0$, and 
$\ket{\psi^{\s\text{HS}}_{0}}$ is a ground state.  

Taking the ground state expectation value of
\eqref{eq:hsfactorization} implies with
\begin{equation}
  \label{eq:hseigenvalue}
  H^{\s\text{HS}}\ket{\psi^{\s\text{HS}}_{0}}=E_0\ket{\psi^{\s\text{HS}}_{0}}
\end{equation}
that
\begin{equation}
  \label{eq:hsd2}
  \bs{D}_\alpha\ket{\psi^{\s\text{HS}}_{0}}=0,\quad \forall \ \alpha=1,\ldots,N.
\end{equation}
and $\bs{S}_{\text{tot}}\ket{\psi^{\s\text{HS}}_{0}}=0$.  This trivially 
implies 
\begin{equation}
  \label{eq:hslambdapsi0}
  \bs{\Lambda}\ket{\psi^{\s\text{HS}}_{0}}=0,
\end{equation}
\ie there is no spin current in the ground state.  Note that if other
ground states were to exist, \eqref{eq:hsfactorization} shows that
they would have to be singlets and likewise be annihilated by
$\bs{D}_\alpha$.  It is not very difficult to verify \eqref{eq:hsd2}
directly, but since we have verified \eqref{eq:hseigenvalue} in
Section \ref{sec:hsexsol} and will verify \eqref{eq:hsfactorization}
below, there is no need to do so.

%\vspace{.5\baselineskip} 
\vspace{\baselineskip} 
\emph{Verification of \eqref{eq:hsfactorization}.}---%
For convenience, we define the purely imaginary parameter
\begin{displaymath}
  \theta_{\alpha\beta}
  \equiv\frac{\eta_{\alpha}+\eta_{\beta}}{\eta_{\alpha}-\eta_{\beta}}
\end{displaymath}
and recall
\begin{align*}
  \bs{D}_{\alpha}^{\dagger}
  &=\frac{1}{2} 
  \sum_{\substack{\beta=1\\[1pt]\beta\neq\alpha}}^N\theta_{\alpha\beta}
  \big[\text{i} (\bs{S}_\alpha \times\bs{S}_\beta ) - \bs{S}_\beta \big],  
%  \label{eq:hsdoperator1*}
  \\[0.2\baselineskip]
  \bs{D}_\alpha
  &=\frac{1}{2} 
  \sum_{\substack{\gamma=1\\[1pt]\gamma\neq\alpha}}^N \theta_{\alpha\gamma}
  \big[ \text{i} (\bs{S}_\alpha \times\bs{S}_\gamma ) + \bs{S}_\gamma \big].
%  \label{eq:hsdoperator1}
\end{align*}
For $S=\frac{1}{2}$ and $\a\ne\b,\c$, we obtain
\begin{align}
%  \hspace{20pt}&\hspace{-20pt}
  \i(\bSa\times\bSb)\i(\bSa\times\bSc)
  &=\varepsilon^{ijk}\varepsilon^{ilm}\Sb^j\Sa^k\Sa^l\Sc^m
  \nonumber\\[0.2\baselineskip] 
  &=\big(\delta^{jl}\delta^{km}-\delta^{jm}\delta^{kl}\big)
  \Sb^j\bigg(\frac{1}{4}\delta^{kl} + \frac{\i}{2}\varepsilon^{kln}\Sa^n
  \bigg)\Sc^m
%  \nonumber\\[0.2\baselineskip] 
%  &=-\frac{1}{2}\bSb\bSc + \frac{\i}{2}\varepsilon^{mjn}\Sa^n\Sb^j\Sc^m
  \nonumber\\[0.2\baselineskip] 
  &=-\frac{1}{2}\bSb\bSc - \frac{\i}{2}\bSa(\bSb\times\bSc), 
\end{align}
and therewith 
\begin{equation*}
  \big[\text{i} (\bs{S}_\alpha \times\bs{S}_\beta ) - \bs{S}_\beta \big]\cdot
  \big[\text{i} (\bs{S}_\alpha \times\bs{S}_\gamma ) + \bs{S}_\gamma \big]
  =-\frac{3}{2}\big[\bs{S}_{\beta}\bs{S}_{\gamma}-\text{i}\bs{S}_{\alpha}
    (\bs{S}_{\beta}\times\bs{S}_{\gamma})\big].
\end{equation*}
This implies
%This implies
% \begin{align*}
%   &\quad \hspace{-20pt}
%   \sum_{\alpha=1}^N\bs{D}_{\alpha}^{\dagger}\bs{D}_\alpha \\
%   \!\! &=\!\!
%   -\frac{1}{4}\sum_{\alpha=1}^N\sum^N_{\substack{\beta=1\\(\beta\neq\alpha)}}
%   \sum^N_{\substack{\gamma=1\\(\gamma\neq\alpha)}}\theta_{\alpha\beta}
%   \theta_{\alpha\gamma}
%   \left[i(\bs{S}_{\beta}\times\bs{S}_{\alpha})+\bs{S}_{\beta}\right]\cdot
%   \left[i(\bs{S}_{\alpha}\times\bs{S}_{\gamma})+\bs{S}_{\gamma}\right]\\
%   &=\!\! -\frac{3}{8}\sum_{\alpha=1}^N\sum^N_{\substack{\beta=1\\
%       (\beta\neq\alpha)}}
%   \sum^N_{\substack{\gamma=1\\(\gamma\neq\alpha)}}\theta_{\alpha\beta}
%   \theta_{\alpha\gamma}
%   \left[\bs{S}_{\beta}\bs{S}_{\gamma}+i\bs{S}_{\beta}
%     (\bs{S}_{\alpha}\times\bs{S}_{\gamma})\right].
% \end{align*}
\begin{equation*}
  \sum_{\alpha=1}^N\bs{D}_{\alpha}^{\dagger}\bs{D}_\alpha 
  =-\frac{3}{8}\sum_{\alpha=1}^N\sum^N_{\substack{\beta=1\\[1pt]
      \beta\neq\alpha}}
  \sum^N_{\substack{\gamma=1\\[1pt]\gamma\neq\alpha}}\theta_{\alpha\beta}
  \theta_{\alpha\gamma}
  \left[\bs{S}_{\beta}\bs{S}_{\gamma}-\text{i}\bs{S}_{\alpha}
    (\bs{S}_{\beta}\times\bs{S}_{\gamma})\right].
\end{equation*}
For the terms with $\alpha\neq\beta=\gamma$, we use
%$\bs{S_{\beta}}\times\bs{S_{\beta}} = i\bs{S_{\beta}}$ 
$\bs{S}\times\bs{S} = \text{i}\bs{S}$ 
to write
\begin{equation*}
  \label{eq:hsspinproductterm1}
  \bs{S}_{\beta}\bs{S}_{\beta}
  - \text{i} \bs{S}_{\alpha}(\bs{S}_{\beta}\times\bs{S}_{\beta})
  =\frac{3}{4}+\bs{S}_{\alpha}\bs{S}_{\beta},
\end{equation*}
and observe
\begin{equation*}
  \theta_{\alpha\beta}^2=1-\frac{4}{\vert\eta_{\alpha}-\eta_{\beta}\vert^2}.
\end{equation*}
For the terms with $\alpha$,$\beta$, and $\gamma$ all distinct, the
vector product term vanishes as it changes sign under interchange of
the dummy indices $\beta$ and $\gamma$.  For these terms we rearrange
the sums
\begin{displaymath}
  \sum_{\alpha=1}^N\sum^N_{\substack{\beta=1\\[1pt]\beta\neq\alpha}}
  \sum^N_{\substack{\gamma=1\\[1pt]\gamma\neq\alpha}}
  =\sum_{\beta=1}^N\sum_{\gamma=1}^N\sum^N_{\substack{\alpha=1\\[1pt]
      \alpha\neq\beta,\gamma}}
\end{displaymath}
and carry out the summation over $\alpha$.  With
\begin{displaymath}
  \frac{1}{(\eta_{\alpha}-\eta_{\beta})(\eta_{\alpha}-\eta_{\gamma})}=
  \frac{1}{\eta_{\beta}-\eta_{\gamma}}\left(\frac{1}{\eta_{\alpha}-\eta_{\beta}}-
    \frac{1}{\eta_{\alpha}-\eta_{\gamma}}\right)
\end{displaymath}
and
\begin{equation*}
  \sum^N_{\substack{\alpha=1\\[1pt]\alpha\neq\beta,\gamma}}
  \frac{\eta_{\beta}}{\eta_{\alpha}-\eta_{\beta}}
  =-\frac{N-1}{2}-\frac{\eta_{\beta}}{\eta_{\gamma}-\eta_{\beta}},
\end{equation*}
which follows directly from \eqref{eq:app-hs9}, we obtain
\begin{equation*}
  \sum^N_{\substack{\alpha=1\\[1pt]\alpha\neq\beta,\gamma}}\theta_{\alpha\beta}
  \theta_{\alpha\gamma}=
  \sum^N_{\substack{\alpha=1\\[1pt]\alpha\neq\beta,\gamma}}
  \left(1+\frac{2\eta_{\beta}}{\eta_{\alpha}-\eta_{\beta}}\right)
  \left(1+\frac{2\eta_{\gamma}}{\eta_{\alpha}-\eta_{\gamma}}\right)
  = N-\frac{8}{\vert\eta_{\beta}-\eta_{\gamma}\vert^2}.
%%%% Full calculation:
\end{equation*}
% \begin{align*}
%   \sum^N_{\substack{\alpha=1\\\alpha\neq\beta,\gamma}}\theta_{\alpha\beta}
%   \theta_{\alpha\gamma}
%   \!\!&=\!\!\sum^N_{\substack{\alpha=1\\\alpha\neq\beta,\gamma}}
%   \left(1+\frac{2\eta_{\beta}}{\eta_{\alpha}-\eta_{\beta}}\right)
%   \left(1+\frac{2\eta_{\gamma}}{\eta_{\alpha}-\eta_{\gamma}}\right)\\
%   &=\!\! N-2+\sum^N_{\substack{\alpha=1\\\alpha\neq\beta}}
%   \frac{2\eta_{\beta}}{\eta_{\alpha}-\eta_{\beta}}
%   -\frac{2\eta_{\beta}}{\eta_{\gamma}-\eta_{\beta}}
%   +\sum^N_{\substack{\alpha=1\\\alpha\neq\gamma}}
%   \frac{2\eta_{\gamma}}{\eta_{\alpha}-\eta_{\gamma}}
%   -\frac{2\eta_{\gamma}}{\eta_{\beta}-\eta_{\gamma}}+\\*
%   & &\!\!\frac{4\eta_{\beta}\eta_{\gamma}}{\eta_{\beta}-\eta_{\gamma}}\left(
%     \sum^N_{\substack{\alpha=1\\\alpha\neq\beta}}
%     \frac{1}{\eta_{\alpha}-\eta_{\beta}}
%     -\frac{1}{\eta_{\gamma}-\eta_{\beta}}
%     -\sum^N_{\substack{\alpha=1\\\alpha\neq\gamma}}
%     \frac{1}{\eta_{\alpha}-\eta_{\gamma}}
%     +\frac{1}{\eta_{\beta}-\eta_{\gamma}}\right)\\
%   &=\!\! N-\frac{8}{\vert \eta_{\beta}-\eta_{\gamma}\vert^2} .
% \end{align*}
%
Collecting all the terms yields
\begin{align*}
  \hspace{10pt}&\hspace{-10pt}
  \frac{8}{3}\sum_{\alpha=1}^N\bs{D}_{\alpha}^{\dagger}\bs{D}_{\alpha}
  \\[0.2\baselineskip]
  &= \sum^N_{\alpha\neq\beta}
  \left(\frac{4}{\vert\eta_{\alpha}-\eta_{\beta}\vert^2}-1\right)
  \left(\frac{3}{4}+\bs{S}_{\alpha}
    \bs{S}_{\beta}\right)
%\\* & &
%  +\sum^N_{\substack{\beta,\gamma=1\\[1pt] \beta\neq\gamma}}
  +\sum^N_{\beta\neq\gamma}
  \left(\frac{8}{\vert\eta_{\beta}-\eta_{\gamma}\vert^2}-N\right)
  \bs{S}_{\beta}\bs{S}_{\gamma}
  \\[0.2\baselineskip]
  &= 12\sum^N_{\alpha\neq\beta}
  \frac{\bs{S}_{\alpha}
    \bs{S}_{\beta}}{\vert\eta_{\alpha}-\eta_{\beta}\vert^2}-
%  (1+N)\underbrace{\sum^N_{\substack{\alpha,\beta=1\\\alpha\neq\beta}}
%  \bs{S}_{\alpha}\bs{S}_{\beta}}_{\textstyle =\bs{S}_{\text{tot}}^2 -\frac{3}{4}N}
%  (N+1)\sum^N_{\substack{\alpha,\beta=1\\[1pt]\alpha\neq\beta}}
  (N+1)\sum^N_{\alpha\neq\beta}
  \bs{S}_{\alpha}\bs{S}_{\beta}
%\\* & &
%  +\frac{3}{4}\sum^N_{\substack{\alpha,\beta=1\\\alpha\neq\beta}}
%  \left(\frac{4}{\vert\eta_{\alpha}-\eta_{\beta}\vert^2}-1\right).
%  +\sum^N_{\substack{\alpha,\beta=1\\[1pt]\alpha\neq\beta}}
  +\sum^N_{\alpha\neq\beta}
  \left(\frac{3}{\vert\eta_{\alpha}-\eta_{\beta}\vert^2}-\frac{3}{4}\right).
\end{align*}
With the identities
\begin{equation*}
%  \label{eq:app-hsfactsdots}
%  \sum^N_{\substack{\alpha,\beta=1\\[1pt]\alpha\neq\beta}}
  \sum^N_{\alpha\neq\beta}
  \bs{S}_{\alpha}\bs{S}_{\beta}=\bs{S}_{\text{tot}}^2-\frac{3}{4}N
\end{equation*}
and
\begin{equation*}
%  \sum^N_{\substack{\alpha,\beta=1\\\alpha\neq\beta}}
  \sum^N_{\alpha\neq\beta}
  \left(\frac{3}{\vert\eta_{\alpha}-\eta_{\beta}\vert^2}-\frac{3}{4}\right)
  =\frac{1}{4}N(N^2-1)-\frac{3}{4}N(N-1),
\end{equation*}
where we have used \eqref{eq:app-hs12}, we obtain 
\begin{equation*}
%  \label{eq:app-hsfact}
%  \sum^N_{\substack{\alpha,\beta=1\\\alpha\neq\beta}}
  \sum^N_{\alpha\neq\beta}
  \frac{\bs{S}_{\alpha}\bs{S}_{\beta}}{\vert\eta_{\alpha}-\eta_{\beta}\vert^2}
  =\frac{2}{9}\sum_{\alpha=1}^N\bs{D}_{\alpha}^{\dagger}\bs{D}_{\alpha}+
  \frac{N+1}{12}\bs{S}_{\text{tot}}^2-\frac{N(N^2+5)}{48},
\end{equation*}
and hence \eqref{eq:hsfactorization}.\hfill $\Box$

%\vspace{100pt}

\subsection{Spinon excitations and fractional statistics}
\label{sec:hsspinons}

The elementary excitations for this model are free spinon excitations,
which carry spin \half and no charge.  They constitute an instance of
fractional quantization, which is both conceptually and
mathematically similar to the fractional quantization of charge in the
fractional quantum Hall effect~\cite{laughlin83prl1395}.  Their
fractional quantum number is the spin, which takes the value \half in
a Hilbert space (\ref{eq:hsket}) made out of spin flips ${S}^+$,
which carry spin 1.

\vspace{\baselineskip}
\emph{One-spinon states.}---%
To write the wave function for a $\dw$-spin spinon localized at site 
$\eta_\alpha$, consider a chain with an odd number of sites $N$ and
let $M=\frac{N-1}{2}$ be the number of $\up$ or $\dw$ spins condensed
in the uniform liquid.  The spinon wave function is then given by
\begin{equation}
  \label{eq:hsp1sp}
  \psi_{\alpha\dw}
  (z_1,z_2,\ldots ,z_M) =\prod_{i=1}^M\,(\eta_\alpha -z_i)\,
%  \prod_{i<j}^M\,(z_i-z_j)^2\,\prod_{i=1}^M\,z_i,
  \psi^{\s\text{HS}}_{0}(z_1,z_2,\ldots ,z_M),
\end{equation}
which we understand substituted into (\ref{eq:hsket}).  
%Using Laughlin's plasma analogy\cite{fqhe}, the reader may easily
%convince himself that the spinon is localized in space.
It is easy to verify 
${S}^\z _{\text{tot}} \psi_{\alpha\dw} = 
-\frac{1}{2} \psi_{\alpha\dw}$ and 
${S}^-_{\text{tot}} \psi_{\alpha\dw} = 0$,
which shows that the spinon transforms as a spinor under rotations.  

The localized spinon (\ref{eq:hsp1sp}) is not an eigenstate of the
%Haldane--Shastry 
Hamiltonian (\ref{eq:hsham}).  To obtain exact
eigenstates, we construct momentum eigenstates according to
\begin{equation}
  \label{eq:hspsim}
  \psi_{m\dw}(z_1,z_2,\ldots ,z_M) =
  \sum_{\alpha=1}^{N} (\bar\eta_\alpha)^m\,
  \psi_{\alpha\dw}(z_1,z_2,\ldots ,z_M), 
\end{equation}
where the integer $m$ corresponds to a momentum quantum number.
%(Note that 
%$\overline{\eta}_\alpha^m=\exp\left(-i\frac{2\pi}{N}\alpha m \right)$.)
Since $\psi_{\alpha\dw} (z_1,z_2,\ldots ,z_M)$ contains
only powers $\eta_\alpha^0, \eta_\alpha^1,\ldots , \eta_\alpha^M$ and
\begin{equation}
  \label{eq:hsdelta}
  \sum_{\alpha=1}^{N} \overline{\eta}_\alpha^m \eta_\alpha^n = \delta_{mn}
  \quad \text{mod}\ N,
\end{equation}
$\psi_{m\dw} (z_1,z_2,\ldots ,z_M)$ will vanish unless $m=0,1,\ldots
,M$.  There are only roughly half as many spinon orbitals as there are
sites.  Spinons on neighboring sites hence cannot be orthogonal.  With
\eqref{eq:hspsieq} and \eqref{eq:hsphieq}, we obtain

\begin{equation}
  H^{\s\text{HS}}\ket{\psi_{m\dw}}
  =\left[ -\frac{\pi^2}{24}\left( N-\frac{1}{N}\right)+
    \frac{2\pi^2}{N^2} m(M-m) \right] \ket{\psi_{m\dw}}.
\label{eq:hs1spinonenergy}
\end{equation}

To make a correspondence between $m$ and the spinon momentum $p_m$, we
translate (\ref{eq:hspsim}) counterclockwise by one lattice spacing
(which we set to unity for present purposes) around the unit circle,
\begin{equation}
  \label{eq:hst}
  \boldsymbol{T}\, \ket{\psi_{m\dw}} 
  = e^{\text{i}(p_0+p_m)}\ket{\psi_{m\dw}}.
\end{equation}
With $p_0=-\frac{\pi}{2}N$, we find
\begin{equation}
  \label{eq:hsq}
%  p_m=\pi\!-\!\frac{2\pi}{N}\!\left(m\!+\!\frac{1}{4}\right)\!.
  p_m=\pi-\frac{2\pi}{N}\!\left(m+\frac{1}{4}\right).
\end{equation}

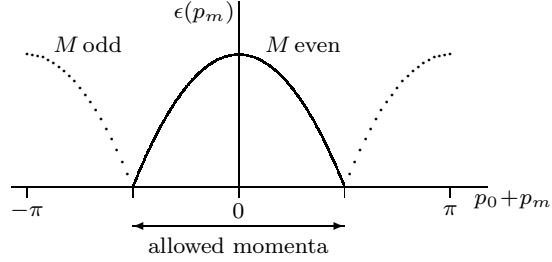
\begin{figure}[tb]
  \begin{center}
    \begin{picture}(200,110)(-100,-30)
      \qbezier[2000](-40,0)(0,100)(40,0)
      \put(4,0){\makebox(0,0){\rule{180.pt}{0.3pt}}}
%      \put(4,0){\makebox(0,0){\rule{200.pt}{0.3pt}}}
      % \put(0,40.57){\makebox(0,0[r]){\rule{4.pt}{0.3pt}}}
      \put(0,0){\makebox(0,0)[b]{\rule{0.3pt}{70.pt}}}
      \put(-80,0){\makebox(0,0)[t]{\rule{0.3pt}{4pt}}}
      \put(-80,-9){\makebox(0,0){\small $-\pi$}}
      \put(-40,0){\makebox(0,0)[t]{\rule{0.3pt}{4pt}}}
      \put(0,0){\makebox(0,0)[t]{\rule{0.3pt}{4pt}}}
      \put(0,-9){\makebox(0,0){\small 0}}
      \put(40,0){\makebox(0,0)[t]{\rule{0.3pt}{4pt}}}
      \put(80,0){\makebox(0,0)[t]{\rule{0.3pt}{4pt}}}
      \put(80,-9){\makebox(0,0){\small $\pi$}}
%      \put(92,-5){\makebox(0,0){\small $p$}}
      \put(104,-5){\makebox(0,0){\small $p_0\!+\!p_m$}}
%      \put(-8,66){\makebox(0,0){\small $E_p$}}
      \put(-13,66){\makebox(0,0){\small $\epsilon(p_m)$}}
      \put(0,-15){{\vector(-1,0){40}}}
      \put(0,-15){{\vector(1,0){40}}}
      \put(0,-22){\makebox(0,0){\small allowed momenta}}
%      \put(0,-22){\makebox(0,0){\small allowed momenta for $M$ even}}
      \put(10,54){\makebox(0,0)[l]{\small $M\!$ even}}
      \put(-70,54){\makebox(0,0)[l]{\small $M\!$ odd}}
      \put(-80.0, 50.0){\circle*{1}}
      \put(-78.0, 49.9){\circle*{1}}
      \put(-76.0, 49.5){\circle*{1}}
      \put(-74.0, 48.9){\circle*{1}}
      \put(-72.0, 48.0){\circle*{1}}
      \put(-70.0, 46.9){\circle*{1}}
      \put(-68.0, 45.5){\circle*{1}}
      \put(-66.0, 43.9){\circle*{1}}
      \put(-64.0, 42.0){\circle*{1}}
      \put(-62.0, 39.9){\circle*{1}}
      \put(-60.0, 37.5){\circle*{1}}
      \put(-58.0, 34.9){\circle*{1}}
      \put(-56.0, 32.0){\circle*{1}}
      \put(-54.0, 28.9){\circle*{1}}
      \put(-52.0, 25.5){\circle*{1}}
      \put(-50.0, 21.9){\circle*{1}}
      \put(-48.0, 18.0){\circle*{1}}
      \put(-46.0, 13.9){\circle*{1}}
      \put(-44.0,  9.5){\circle*{1}}
      \put(-42.0,  4.9){\circle*{1}}
      \put( 80.0, 50.0){\circle*{1}}
      \put( 78.0, 49.9){\circle*{1}}
      \put( 76.0, 49.5){\circle*{1}}
      \put( 74.0, 48.9){\circle*{1}}
      \put( 72.0, 48.0){\circle*{1}}
      \put( 70.0, 46.9){\circle*{1}}
      \put( 68.0, 45.5){\circle*{1}}
      \put( 66.0, 43.9){\circle*{1}}
      \put( 64.0, 42.0){\circle*{1}}
      \put( 62.0, 39.9){\circle*{1}}
      \put( 60.0, 37.5){\circle*{1}}
      \put( 58.0, 34.9){\circle*{1}}
      \put( 56.0, 32.0){\circle*{1}}
      \put( 54.0, 28.9){\circle*{1}}
      \put( 52.0, 25.5){\circle*{1}}
      \put( 50.0, 21.9){\circle*{1}}
      \put( 48.0, 18.0){\circle*{1}}
      \put( 46.0, 13.9){\circle*{1}}
      \put( 44.0,  9.5){\circle*{1}}
      \put( 42.0,  4.9){\circle*{1}}
    \end{picture}
    \caption{Dispersion of a single spinon in a Haldane--Shastry chain.}
    \label{fig:hs1spd}
  \end{center}
\end{figure}

The energy \eqref{eq:hs1spinonenergy} can be written as
$E=E_0+\epsilon(p_m)$, with the spinon dispersion given by
\begin{equation}
  \label{eq:hsep}
  \epsilon(p)=\frac{1}{2}p\left(\pi-p\right)+\frac{\pi^2}{8N^2},
\end{equation}
as depicted in Figure \ref{fig:hs1spd}.  The interval of allowed spinon
momenta spans only half of the Brillouin zone, and alternates with $M$
even vs.\ $M$ odd.
% OLD:
% \begin{equation}
%   \label{eq:hsq}
%   p=\pi M + \pi\frac{M-2m}{2M+1}.
% \end{equation}
% The energy of \eqref{eq:hspsim} is $E=E_0+E_p$, with $E_0$ given by 
% \eqref{eq:hse0} and 
% \begin{equation}
%   \label{eq:hseq}
%   E_p = \frac{1}{2}\left[\left(\frac{\pi}{2}\right)^2-p^2\right]
%   +\frac{\pi^2}{8N^2}\quad\text{mod}\ \pi.
% \end{equation}
% This spinon dispersion is depicted in Figure \ref{fig:hs1spd}.

% The construction (\ref{eq:hspsisp},\ref{eq:hspsim}) can be generalized 
% to many spinon states. 
% The construction can, however, still be used to count the number of 
% available orbitals for spinons, which leads us to the subject of 
% fractional statistics.

%\pagebreak
\vspace{\baselineskip}
\emph{Two-spinon states.}---%
To write the wave function for two $\dw$-spin spinons localized at sites 
$\eta_\alpha$ and $\eta_\beta$, consider a chain with $N$ even and 
$M=\frac{N-2}{2}$.  The two-spinon state is then given by
\begin{equation}
  \label{eq:hsp2sp}
  \psi_{\alpha\beta}(z_1,z_2,\ldots ,z_M) =
  \prod_{i=1}^M\,(\eta_\alpha -z_i)(\eta_\beta -z_i)\,
  \psi^{\s\text{HS}}_{0}(z_1,z_2,\ldots ,z_M).
\end{equation}
A momentum basis for the two-spinon states is given by
\begin{equation}
  \label{eq:hspsimn}
  \psi_{mn}(z_1,z_2,\ldots ,z_M) =
  \sum_{\alpha,\beta=1}^{N} (\bar\eta_\alpha)^m\,(\bar\eta_\beta)^m\,
  \psi_{\alpha\beta}(z_1,z_2,\ldots ,z_M), 
\end{equation}
where $M\ge m\ge n\ge 0$.  For $m$ or $n$ outside this range,
$\psi_{mn}$ vanishes identically, reflecting the overcompleteness of
the position space basis.
%Acting with \eqref{eq:hsham} on \eqref{eq:hspsimn}, we
With \eqref{eq:hspsieq}, \eqref{eq:hsphieq}, and the algebraic identity
  \begin{equation}
    \frac{x + y}{x - y} (x^{m} y^{n} - x^{n} y^{m})
    =  2 \sum_{l = 0}^{m-n} x^{m-l} y^{n+l}- (x^{m} y^{n} + x^{n} y^{m}),
    \label{eq:hsxyformula}
\end{equation}
we obtain~\cite{haldane91prl1529,laughlin-00proc,bernevig-01prl3392,bernevig-01prb024425}
\begin{equation}
  \label{eq:hsscattspinon}
  H^{\s\text{HS}}\ket{\psi_{mn}}=
  E_{mn}\ket{\psi_{mn}}+\sum_{l=1}^{l_{\text{max}}}V_l^{mn}\ket{\psi_{m+l,n-l}}
\end{equation}
with
% \begin{align}
%   \label{eq:hs2spE}
%   E_{mn}&=\frac{2\pi^2}{N^2}\left[ m\left(\frac{N}{2} - 1 - m\right) 
%     +n\left(\frac{N}{2} - 1 - n\right)- \frac{m-n}{2}\right]\nonumber\\
%   &-&\frac{\pi^2}{24}\left(N-\frac{19}{N}+\frac{24}{N^2}\right)\\
%   V_l^{mn}&=-\frac{2\pi^2}{N^2}(m-n+2l),
% \end{align}
%
\begin{align}
  \label{eq:hs2spE}
  E_{mn}\ =
  &-\frac{\pi^2}{24}\left(N-\frac{19}{N}+\frac{24}{N^2}\right)
  \nonumber\\[0.2\baselineskip]
  &+\frac{2\pi^2}{N^2}\left[ m\left(\frac{N}{2} - 1 - m\right) 
    +n\left(\frac{N}{2} - 1 - n\right)- \frac{m-n}{2}\right],
  \\[0.2\baselineskip]
  V_l^{mn}\ =
  &-\frac{2\pi^2}{N^2}(m-n+2l),
\end{align}
and $l_{\text{max}}=\min(M-m,n)$.  Since the ``scattering'' of the
non-ortho\-gonal basis states $\ket{\psi_{mn}}$ in
\eqref{eq:hsscattspinon} only occurs in one direction, increasing
$m-n$ while keeping $m+n$ fixed, the eigenstates of $H^{\s\text{HS}}$
have energy eigenvalues $E_{mn}$, and are of the form
\begin{equation}
  \label{eq:hsphinm}
  \ket{\phi_{mn}}
  =\sum_{l=0}^{l_M} a_l^{mn}\ket{\psi_{m+l,n-l}}.
\end{equation}
A recursion relation for the coefficients $a_l^{mn}$ is readily 
obtained from \eqref{eq:hsscattspinon}.

If we identify the single-spinon momenta for $m\ge n$ according
to
\begin{equation}
  \label{eq:hsqmqnspinon}
  p_m=\pi-\frac{2\pi}{N}\left(m+\frac{1}{2}+s\right),\quad
  p_n=\pi-\frac{2\pi}{N}\left(n+\frac{1}{2}-s\right),
\end{equation}
with a {statistical shift}
$s=\frac{1}{4}$~\cite{greiter-05prb224424,greiter-06prl059701}, we can write
the energy
\begin{equation}
  \label{eq:hsetotspinon}
  E_{mn}=E_0+\epsilon(p_m)+\epsilon(p_n),
\end{equation}
where $E_0$ is the ground state energy \eqref{eq:hse0} and
$\epsilon(p)$ the spinon dispersion \eqref{eq:hsep}.

\vspace{\baselineskip}
\emph{Fractional statistics.}---%
The mutual half-fermi statistics of the spinons manifests itself in
the fractional shift $s$ in the
single-spinon momenta \eqref{eq:hsqmqnspinon}, as we will elaborate 
now~\cite{greiter09prb064409}.  
The Ansatz \eqref{eq:hspsimn} unambiguously implies that the sum of
the two spinon momenta is given by
% \begin{equation}
%   \label{eq:hsqn+qm}
%   e^{i(q_m+q_n)}
%   =\frac{\psi_{mn}[\eta_1z_i]}{\psi_{mn}[z_i]}
%   \frac{\psi_0[z_i]}{\psi[\eta_1z_i]},
% \end{equation}
% %with $[\eta_1z]\equiv(\eta_1z_1,\ldots,\eta_1z_M)$, 
% which implies 
$q_m+q_n=2\pi-\frac{2\pi}{N}(m+n+1)$, and hence
\eqref{eq:hsqmqnspinon}.  The shift $s$ is determined by demanding
that the excitation energy \eqref{eq:hsetotspinon} of the two-spinon
state is a sum of single-spinon energies, which in turn is required
for the explicit solution here to be consistent with the models
solution via the asymptotic Bethe
ansatz~\cite{ha-93prb12459,essler95prb13357,greiter-05prb224424}.
% \begin{figure}[t]
%   \begin{center}
%     \begin{minipage}[c]{0.75\textwidth}
%       \includegraphics[width=\linewidth]{./statistics.eps}
%     \end{minipage}
%     \caption{(Color online) Fractional statistics in two and one space
%       dimensions.}
%   \end{center}
%   \label{fig:statistics}
% \end{figure}

The shift decreases the momentum $p_m$ of spinon 1 and increases
momentum $p_n$ of spinon 2.  This may surprise at first as the basis
states \eqref{eq:hspsimn} are constructed symmetrically with regard to
interchanges of $m$ and $n$.  To understand this asymmetry, note that
$M\ge m\ge n\ge 0$ implies $0<p_m<p_n<\pi$.  The dispersion
 \eqref{eq:hsep} implies that the group velocity of the spinons is
given by
\begin{equation}
%  \label{eq:hsvg}
  v_\text{g}(p)=\partial_p\epsilon(p)=\frac{\pi}{2}-p,
\end{equation}
%$v_\text{g}(q)=\partial_q\epsilon(q)=\frac{\pi}{2}-q$,
%
which in turn implies that $v_\text{g}(p_m)>v_\text{g}(p_n)$.  
%The physical significance of this result can hardly be overstated.  
This means that the \emph{relative motion} of spinon 1 (with $q_m$) with
respect to spinon 2 (with $q_n$) is \emph{always counterclockwise} on
the unit circle %. 
(see Figure \ref{fig:statistics}).  
\begin{figure}[tb]
\begin{center}
  \begin{picture}(320,60)(-32,-15) \put(0,0){\circle{100}} \put(
    17.3,10.0){\circle*{4}} \put(-10.0, 17.3){\circle*{4}}
    \put(26.8,14){\makebox(0,0)[l]{\vector(-2,3){7}}}
    \put(-18,24){\makebox(0,0)[l]{\vector(3,2){10}}}
    \put(26,18){\makebox(0,0)[l]{\small $1$}}
    \put(-19,30){\makebox(0,0)[l]{\small $2$}}
    \put(52,30){\makebox(0,0)[l] {\small relative motion of
        one-dimensional anyons is unidirectional}}
    \put(52,18){\makebox(0,0)[l]{(\small e.g.\ 2 moves clockwise
        relative to 1)}} \put(52,0){\makebox(0,0)[l]{\small when
        anyons cross:\quad $|\psi\!> \rightarrow e^{\text{i}\theta}
        |\psi\!>$}} \put(52,-18){\makebox(0,0)[l]{\small momentum
        spacing:\quad $\displaystyle p_1\!-\!p_2 = \Delta p\rightarrow
        \Delta p - \theta$}}
  \end{picture}
  \end{center}
  \caption{Fractional statistics in one dimension. The crossings of
    the anyons are unidirectional, and the many particle wave function
    acquires a statistical phase $\theta$ whenever they cross.}
  \label{fig:statistics}
\end{figure}
The shifts in the individual spinon momenta can
hence be explained by assuming that the two-spinon state
acquires a statistical phase $\theta=2\pi s$ whenever the spinons pass
through each other.  This phase implies that $q_m$ is shifted by
$-\frac{2\pi}{N}s$ since we have to translate spinon 1
counterclockwise through spinon 2 and hence counterclockwise around
the unit circle when obtaining the allowed values for $q_m$ from the
PBCs.  Similarly, $q_n$ is shifted by $+\frac{2\pi}{N}s$ since we have
to translate spinon 2 clockwise through spinon 1 and hence clockwise
around the unit circle when obtaining the quantization of $q_n$.

That the crossing of the spinons occurs only in one direction is a
necessary requirement for fractional statistics to exist in one
dimension.  If the spinons could cross in both directions, the fact
that paths interchanging them twice (\ie once in each direction) are
topologically equivalent to paths not interchanging them at all would
imply $2\theta=0\ \text{mod}\ 2\pi$ for the statistical phase, \ie
only allow for the familiar choices of bosons or fermions.  With the
scattering occurring in only one direction, arbitrary values for
$\theta$ are possible.  Note that the one-dimensional anyons break
neither time-reversal symmetry (T) nor parity (P).

\vspace{.5\baselineskip} The fractional statistics of the spinons
manifests itself further in the fractional exclusion (or generalized
Pauli) principle introduced by Haldane~\cite{haldane91prl937}.  If we
consider a state with $L$ spinons, we can easily see from
\eqref{eq:hspsim}, \eqref{eq:hsdelta}, and \eqref{eq:hspsimn} that the
number of orbitals available for further spinons we may wish to create
is $M+1$, where $M=\frac{N-L}{2}$ is the number of $\up$ or $\dw$
spins in the remaining uniform liquid.  (In this representation, the
spinon wave functions are symmetric; two or more spinons can have the
same value for $m$.)  In other words, the creation of {\em two}
spinons reduces the number of available single spinon states by {\em
  one}.  They hence obey half-fermi statistics in the sense of
Haldane's exclusion principle.  (For fermions, the creation of two
particles would decrease the number of available single particle by
two, while this number would not change for bosons.)

%\subsection{Many spinon states}
\subsection{Young tableaux and many spinon states}
\label{sec:hsyt} 

The easiest way to obtain the spectrum of the model is through the
one-to-one correspondence between the Young tableaux classifying the
total spin representations of $N$ spins and the exact eigenstates of
the the Haldane-Shastry model for a chain with $N$ sites, which are
classified by the total spins and the fractionally spaced
single-particle momenta of the spinons~\cite{greiter-07prl237202}.  

This correspondence yields the allowed sequences of single-spinon
momenta $p_1,\ldots,p_L$ as well as the allowed representations for
the total spin of the states such that the eigenstates of the Haldane
Shastry model have momenta and energies
\begin{equation}
  p=p_0+\sum_{i=1}^L p_i,\quad E=E_0+\sum_{i=1}^L \epsilon(p_i),
  \label{eq:hsLspinonenergy}
\end{equation} 
where $p_0$ and $E_0$ denote the ground state momentum and energy,
respectively, and $\epsilon(p)$ is the single-spinon dispersion.  The
correspondence hence does not only provide the quantum numbers of all
the states in the spectrum, but also shows that it is sensible to view
the individual spinons as particles, rather than just as solitons or
collective excitations in many body condensates.  We now proceed by
stating these rules without further motivating or even deriving them.

\begin{figure}[tb]
  \begin{center}
    \setlength{\unitlength}{\ytlength}
    \begin{picture}(32,8)(1.5,-1) \linethickness{0.5pt}
      \put(3,6){\line(1,0){1}} \put(3,5){\line(1,0){1}}
      \put(3,5){\line(0,1){1}} \put(4,5){\line(0,1){1}}
      \put(3,5){\makebox(1,1){1}} \put(4.6,5.3){$\otimes$}
      \put(6,6){\line(1,0){1}} \put(6,5){\line(1,0){1}}
      \put(6,5){\line(0,1){1}} \put(7,5){\line(0,1){1}}
      \put(6,5){\makebox(1,1){2}}
      \put(3,4.4){$\underbrace{\phantom{iiiiiiiiii}}$}
      \put(3,2.9){\line(1,0){1}} \put(3,1.9){\line(1,0){1}}
      \put(3,0.9){\line(1,0){1}} \put(3,0.9){\line(0,1){2}}
      \put(4,0.9){\line(0,1){2}} \put(3,1.9){\makebox(1,1){1}}
      \put(3,0.9){\makebox(1,1){2}} \put(2,-0.5){$S\!=\!0$}
      \put(4.6,1.5){$\oplus$} \put(6,2.4){\line(1,0){2}}
      \put(6,1.4){\line(1,0){2}} \put(6,1.4){\line(0,1){1}}
      \put(7,1.4){\line(0,1){1}} \put(8,1.4){\line(0,1){1}}
      \put(6,1.4){\makebox(1,1){1}} \put(7,1.4){\makebox(1,1){2}}
      \put(5.7,-0.2){$S\!=\!1$} \put(7.6,5.3){$\otimes$}
      \put(9,6){\line(1,0){1}} \put(9,5){\line(1,0){1}}
      \put(9,5){\line(0,1){1}} \put(10,5){\line(0,1){1}}
      \put(9,5){\makebox(1,1){3}} \put(11.6,5.3){$=$}
      \put(14,6){\line(1,0){1}} \put(14,5){\line(1,0){1}}
      \put(14,4){\line(1,0){1}} \put(14,3){\line(1,0){1}}
      \put(14,3){\line(0,1){3}} \put(15,3){\line(0,1){3}}
      \put(14,5){\makebox(1,1){1}} \put(14,4){\makebox(1,1){2}}
      \put(14,3){\makebox(1,1){3}} \put(16.1,5.3){$\oplus$}
      \put(18,6){\line(1,0){2}} \put(18,5){\line(1,0){2}}
      \put(18,4){\line(1,0){1}} \put(18,4){\line(0,1){2}}
      \put(19,4){\line(0,1){2}} \put(20,5){\line(0,1){1}}
      \put(18,5){\makebox(1,1){1}} \put(19,5){\makebox(1,1){2}}
      \put(18,4){\makebox(1,1){3}} \put(17.5,2.2){$S\!=\!\frac{1}{2}$}
      \put(21.1,5.3){$\oplus$} \put(23,6){\line(1,0){2}}
      \put(23,5){\line(1,0){2}} \put(23,4){\line(1,0){1}}
      \put(23,4){\line(0,1){2}} \put(24,4){\line(0,1){2}}
      \put(25,5){\line(0,1){1}} \put(23,5){\makebox(1,1){1}}
      \put(23,4){\makebox(1,1){2}} \put(24,5){\makebox(1,1){3}}
      \put(22.5,2.2){$S\!=\!\frac{1}{2}$} \put(26.1,5.3){$\oplus$}
      \put(28,6){\line(1,0){3}} \put(28,5){\line(1,0){3}}
      \put(28,5){\line(0,1){1}} \put(29,5){\line(0,1){1}}
      \put(30,5){\line(0,1){1}} \put(31,5){\line(0,1){1}}
      \put(28,5){\makebox(1,1){1}} \put(29,5){\makebox(1,1){2}}
      \put(30,5){\makebox(1,1){3}} \put(28,2.2){$S\!=\!\frac{3}{2}$}
      \put(14.5,4.5){\makebox(0,0){\line(1,2){1.8}}}
      \put(14.5,4.5){\makebox(0,0){\line(1,-2){1.8}}}
    \end{picture}
    \caption{Total spin representations of three $S=\frac{1}{2}$ spins
      with Young tableaux. For SU($n$) with $n>2$, the tableaux with
      three boxes on top of each other would exist as well.}
    \label{fig:hsyoungdiagram}
  \end{center}
\end{figure}

To begin with, the Hilbert space of a system of $N$ identical SU($n$)
spins can be decomposed into representations of the total spin, which
commutes with \eqref{eq:hsham} and hence can be used to classify the
eigenstates.  
% This decomposition can be obtained using Young tableaux~(see \eg
% \cite{InuiTanabeOnodera96}), as illustrated for three $S=\frac{1}{2}$
% spins in Fig.~\ref{fig:hsyoungdiagram}.
These representations are compatible with the representations of the
symmetric group S$_N$ of $N$ elements, which may be expressed in terms
of Young tableaux~\cite{Hamermesh62,InuiTanabeOnodera96}.  The general
rule for obtaining Young tableaux is illustrated for three
$S=\frac{1}{2}$ spins in Fig.~\ref{fig:hsyoungdiagram}.
%The general rule is as follows.  
For each of the $N$ spins, draw a box
numbered consecutively from left to right.  The representations of
SU($n$) are constructed by putting the boxes together such that the
numbers assigned to them increase in each row from left to right and
in each column from top to bottom.  Each tableau indicates
symmetrization over all boxes in the same row, and antisymmetrization
over all boxes in the same column.  This implies that we cannot have
more than $n$ boxes on top of each other for SU($n$) spins.  For
SU(2), each tableau corresponds to a spin
$S=\frac{1}{2}(\lambda_1-\lambda_2)$ representation, with $\lambda_i$
the number of boxes in the $i\,\text{th}$ row, and stands for a
multiplet $S^\z =-S,\ldots,S$.

\begin{figure}[tb]
  \begin{center}
    \setlength{\unitlength}{\ytlength}
    \begin{picture}(32,6)(0,0) \linethickness{0.5pt}
      \put(5.25,4){\makebox(1,1){$S_{\text{tot}}$}}
      \put(21.5,4){\makebox(1,1){$L$}} \put(24,4.1){$a_1,\dots,a_L$}
      \put(0,3){\line(1,0){2}} \put(0,2){\line(1,0){2}}
      \put(0,1){\line(1,0){2}} \put(0,1){\line(0,1){2}}
      \put(1,1){\line(0,1){2}} \put(2,1){\line(0,1){2}}
      \put(0,2){\makebox(1,1){1}} \put(0,1){\makebox(1,1){2}}
      \put(1,2){\makebox(1,1){3}} \put(1,1){\makebox(1,1){4}}
      \put(5.25,1.5){\makebox(1,1){0}} \put(7,1.8){$\rightarrow$}
      \put(9,3){\line(1,0){2}} \put(9,2){\line(1,0){2}}
      \put(9,1){\line(1,0){2}} \put(9,1){\line(0,1){2}}
      \put(10,1){\line(0,1){2}} \put(11,1){\line(0,1){2}}
      \put(9,2){\makebox(1,1){1}} \put(9,1){\makebox(1,1){2}}
      \put(10,2){\makebox(1,1){3}} \put(10,1){\makebox(1,1){4}}
      \put(14,1.8){$\rightarrow$} \put(16,3){\line(1,0){2}}
      \put(16,2){\line(1,0){2}} \put(16,1){\line(1,0){2}}
      \put(16,1){\line(0,1){2}} \put(17,1){\line(0,1){2}}
      \put(18,1){\line(0,1){2}} \put(16,2){\makebox(1,1){1}}
      \put(16,1){\makebox(1,1){2}} \put(17,2){\makebox(1,1){3}}
      \put(17,1){\makebox(1,1){4}} \put(21.5,1.5){\makebox(1,1){0}}
      \put(24,2){\line(1,0){5}}
      \multiput(25,1.85)(1,0){4}{\rule{0.5pt}{2pt}}
% Martins addition
      \put(31,3.7){\makebox(1,1){$p_{\text{tot}}$}}
      \put(31,1.5){\makebox(1,1){$0$}}
    \end{picture}

    \begin{picture}(32,3.5)(0,0) \linethickness{0.5pt}
      \put(0,3){\line(1,0){2}} \put(0,2){\line(1,0){2}}
      \put(0,1){\line(1,0){2}} \put(0,1){\line(0,1){2}}
      \put(1,1){\line(0,1){2}} \put(2,1){\line(0,1){2}}
      \put(0,2){\makebox(1,1){1}} \put(1,2){\makebox(1,1){2}}
      \put(0,1){\makebox(1,1){3}} \put(1,1){\makebox(1,1){4}}
      \put(5.25,1.5){\makebox(1,1){0}} \put(7,1.8){$\rightarrow$}
      \put(9,3){\line(1,0){2}} \put(9,2){\line(1,0){3}}
      \put(10,1){\line(1,0){2}} \put(9,2){\line(0,1){1}}
      \put(10,1){\line(0,1){2}} \put(11,1){\line(0,1){2}}
      \put(12,1){\line(0,1){1}} \put(9,2){\makebox(1,1){1}}
      \put(10,2){\makebox(1,1){2}} \put(10,1){\makebox(1,1){3}}
      \put(11,1){\makebox(1,1){4}} \put(14,1.8){$\rightarrow$}
      \put(16,3){\line(1,0){2}} \put(16,2){\line(1,0){3}}
      \put(17,1){\line(1,0){2}} \put(16,2){\line(0,1){1}}
      \put(17,1){\line(0,1){2}} \put(18,1){\line(0,1){2}}
      \put(19,1){\line(0,1){1}} \put(16,2){\makebox(1,1){1}}
      \put(17,2){\makebox(1,1){2}} \put(17,1){\makebox(1,1){3}}
      \put(18,1){\makebox(1,1){4}} \put(16.5,1.5){\circle*{0.4}}
      \put(18.5,2.5){\circle*{0.4}} \put(21.5,1.5){\makebox(1,1){2}}
      \put(24,2){\line(1,0){5}}
      \multiput(25,1.85)(1,0){4}{\rule{0.5pt}{2pt}}
      \multiput(25,2)(3,0){2}{\circle*{0.5}}
      \put(24.5,0.5){\makebox(1,1){1}}
      \put(27.5,0.5){\makebox(1,1){4}}
      \put(31,1.5){\makebox(1,1){$\pi$}}
    \end{picture}

    \begin{picture}(32,3.5)(0,0) \linethickness{0.5pt}
      \put(0,3){\line(1,0){3}} \put(0,2){\line(1,0){3}}
      \put(0,1){\line(1,0){1}} \put(0,1){\line(0,1){2}}
      \put(1,1){\line(0,1){2}} \put(2,2){\line(0,1){1}}
      \put(3,2){\line(0,1){1}} \put(0,2){\makebox(1,1){1}}
      \put(0,1){\makebox(1,1){2}} \put(1,2){\makebox(1,1){3}}
      \put(2,2){\makebox(1,1){4}} \put(5.25,1.5){\makebox(1,1){1}}
      \put(7,1.8){$\rightarrow$} \put(9,3){\line(1,0){3}}
      \put(9,2){\line(1,0){3}} \put(9,1){\line(1,0){1}}
      \put(9,1){\line(0,1){2}} \put(10,1){\line(0,1){2}}
      \put(11,2){\line(0,1){1}} \put(12,2){\line(0,1){1}}
      \put(9,2){\makebox(1,1){1}} \put(9,1){\makebox(1,1){2}}
      \put(10,2){\makebox(1,1){3}} \put(11,2){\makebox(1,1){4}}
      \put(14,1.8){$\rightarrow$} \put(16,3){\line(1,0){3}}
      \put(16,2){\line(1,0){3}} \put(16,1){\line(1,0){1}}
      \put(16,1){\line(0,1){2}} \put(17,1){\line(0,1){2}}
      \put(18,2){\line(0,1){1}} \put(19,2){\line(0,1){1}}
      \put(16,2){\makebox(1,1){1}} \put(16,1){\makebox(1,1){2}}
      \put(17,2){\makebox(1,1){3}} \put(18,2){\makebox(1,1){4}}
      \put(17.5,1.5){\circle*{0.4}} \put(18.5,1.5){\circle*{0.4}}
      \put(21.5,1.5){\makebox(1,1){2}} \put(24,2){\line(1,0){5}}
      \multiput(25,103.85)(1,0){4}{\rule{0.5pt}{2pt}}
      \multiput(27,2)(1,0){2}{\circle*{0.5}}
      \put(26.5,0.5){\makebox(1,1){3}}
      \put(27.5,0.5){\makebox(1,1){4}}
      \put(31,1.5){\makebox(1,1){$\frac{3\pi}{2}$}}
    \end{picture}

    \begin{picture}(32,3.5)(0,0) \linethickness{0.5pt}
      \put(0,3){\line(1,0){3}} \put(0,2){\line(1,0){3}}
      \put(0,1){\line(1,0){1}} \put(0,1){\line(0,1){2}}
      \put(1,1){\line(0,1){2}} \put(2,2){\line(0,1){1}}
      \put(3,2){\line(0,1){1}} \put(0,2){\makebox(1,1){1}}
      \put(1,2){\makebox(1,1){2}} \put(0,1){\makebox(1,1){3}}
      \put(2,2){\makebox(1,1){4}} \put(5.25,1.5){\makebox(1,1){1}}
      \put(7,1.8){$\rightarrow$} \put(9,3){\line(1,0){3}}
      \put(9,2){\line(1,0){3}} \put(10,1){\line(1,0){1}}
      \put(9,2){\line(0,1){1}} \put(10,1){\line(0,1){2}}
      \put(11,1){\line(0,1){2}} \put(12,2){\line(0,1){1}}
      \put(9,2){\makebox(1,1){1}} \put(10,2){\makebox(1,1){2}}
      \put(10,1){\makebox(1,1){3}} \put(11,2){\makebox(1,1){4}}
      \put(14,1.8){$\rightarrow$} \put(16,3){\line(1,0){3}}
      \put(16,2){\line(1,0){3}} \put(17,1){\line(1,0){1}}
      \put(16,2){\line(0,1){1}} \put(17,1){\line(0,1){2}}
      \put(18,1){\line(0,1){2}} \put(19,2){\line(0,1){1}}
      \put(16,2){\makebox(1,1){1}} \put(17,2){\makebox(1,1){2}}
      \put(17,1){\makebox(1,1){3}} \put(18,2){\makebox(1,1){4}}
      \put(16.5,1.5){\circle*{0.4}} \put(18.5,1.5){\circle*{0.4}}
      \put(21.5,1.5){\makebox(1,1){2}} \put(24,2){\line(1,0){5}}
      \multiput(25,1.85)(1,0){4}{\rule{0.5pt}{2pt}}
      \multiput(25,2)(3,0){2}{\circle*{0.5}}
      \put(24.5,0.5){\makebox(1,1){1}}
      \put(27.5,0.5){\makebox(1,1){4}}
      \put(31,1.5){\makebox(1,1){$\pi$}}
    \end{picture}

    \begin{picture}(32,3.5)(0,0) \linethickness{0.5pt}
      \put(0,3){\line(1,0){3}} \put(0,2){\line(1,0){3}}
      \put(0,1){\line(1,0){1}} \put(0,1){\line(0,1){2}}
      \put(1,1){\line(0,1){2}} \put(2,2){\line(0,1){1}}
      \put(3,2){\line(0,1){1}} \put(0,2){\makebox(1,1){1}}
      \put(1,2){\makebox(1,1){2}} \put(2,2){\makebox(1,1){3}}
      \put(0,1){\makebox(1,1){4}} \put(5.25,1.5){\makebox(1,1){1}}
      \put(7,1.8){$\rightarrow$} \put(9,3){\line(1,0){3}}
      \put(9,2){\line(1,0){3}} \put(11,1){\line(1,0){1}}
      \put(9,2){\line(0,1){1}} \put(10,2){\line(0,1){1}}
      \put(11,1){\line(0,1){2}} \put(12,1){\line(0,1){2}}
      \put(9,2){\makebox(1,1){1}} \put(10,2){\makebox(1,1){2}}
      \put(11,2){\makebox(1,1){3}} \put(11,1){\makebox(1,1){4}}
      \put(14,1.8){$\rightarrow$} \put(16,3){\line(1,0){3}}
      \put(16,2){\line(1,0){3}} \put(18,1){\line(1,0){1}}
      \put(17,2){\line(0,1){1}} \put(16,2){\line(0,1){1}}
      \put(18,1){\line(0,1){2}} \put(19,1){\line(0,1){2}}
      \put(16,2){\makebox(1,1){1}} \put(17,2){\makebox(1,1){2}}
      \put(18,2){\makebox(1,1){3}} \put(18,1){\makebox(1,1){4}}
      \put(16.5,1.5){\circle*{0.4}} \put(17.5,1.5){\circle*{0.4}}
      \put(21.5,1.5){\makebox(1,1){2}} \put(24,2){\line(1,0){5}}
      \multiput(25,1.85)(1,0){4}{\rule{0.5pt}{2pt}}
      \multiput(25,2)(1,0){2}{\circle*{0.5}}
      \put(24.5,0.5){\makebox(1,1){1}}
      \put(25.5,0.5){\makebox(1,1){2}}
      \put(31,1.5){\makebox(1,1){$\frac{\pi}{2}$}}
    \end{picture}

    \begin{picture}(32,3.5)(0,0) \linethickness{0.5pt}
      \put(0,3){\line(1,0){4}} \put(0,2){\line(1,0){4}}
      \put(0,2){\line(0,1){1}} \put(1,2){\line(0,1){1}}
      \put(2,2){\line(0,1){1}} \put(3,2){\line(0,1){1}}
      \put(4,2){\line(0,1){1}} \put(0,2){\makebox(1,1){1}}
      \put(1,2){\makebox(1,1){2}} \put(2,2){\makebox(1,1){3}}
      \put(3,2){\makebox(1,1){4}} \put(5.25,1.5){\makebox(1,1){2}}
      \put(7,1.8){$\rightarrow$} \put(9,3){\line(1,0){4}}
      \put(9,2){\line(1,0){4}} \put(9,2){\line(0,1){1}}
      \put(10,2){\line(0,1){1}} \put(11,2){\line(0,1){1}}
      \put(12,2){\line(0,1){1}} \put(13,2){\line(0,1){1}}
      \put(9,2){\makebox(1,1){1}} \put(10,2){\makebox(1,1){2}}
      \put(11,2){\makebox(1,1){3}} \put(12,2){\makebox(1,1){4}}
      \put(14,1.8){$\rightarrow$} \put(16,3){\line(1,0){4}}
      \put(16,2){\line(1,0){4}} \put(16,2){\line(0,1){1}}
      \put(17,2){\line(0,1){1}} \put(18,2){\line(0,1){1}}
      \put(19,2){\line(0,1){1}} \put(20,2){\line(0,1){1}}
      \put(16,2){\makebox(1,1){1}} \put(17,2){\makebox(1,1){2}}
      \put(18,2){\makebox(1,1){3}} \put(19,2){\makebox(1,1){4}}
      \put(16.5,1.5){\circle*{0.4}} \put(17.5,1.5){\circle*{0.4}}
      \put(18.5,1.5){\circle*{0.4}} \put(19.5,1.5){\circle*{0.4}}
      \put(21.5,1.5){\makebox(1,1){4}} \put(24,2){\line(1,0){5}}
      \multiput(25,2)(1,0){4}{\circle*{0.5}}
      \put(24.5,0.5){\makebox(1,1){1}}
      \put(25.5,0.5){\makebox(1,1){2}}
      \put(26.5,0.5){\makebox(1,1){3}}
      \put(27.5,0.5){\makebox(1,1){4}}
      \put(31,1.5){\makebox(1,1){$0$}}
    \end{picture}
    \caption{Young tableau decomposition and the corresponding spinon
      states for an $S=\frac{1}{2}$ spin chain with $N=4$ sites.  The
      dots represent the spinons.  The spinon momentum
      numbers $a_i$ are given by the numbers in the boxes of the same
      column.  Note that $\sum (2S_{\text{tot}}+1)=2^N$.}
    \label{fig:hsfoursitesu2}
  \end{center}
\end{figure}
The one-to-one correspondence between the Young tableaux and the
non-interacting many-spinon eigenstates of the Haldane--shastry model
is illustrated in Fig.~\ref{fig:hsfoursitesu2} for a chain with $N=4$
sites.  The rule is that in each Young tableau, we shift boxes to the
right such that each box is below or in the column to the right of the
box with the preceding number.  Each missing box in the resulting,
extended tableaux represents a spinon.  The extended tableaux provide
us with the total spin of each multiplet, which is given by the
representation specified by the original Young tableau, as well as the
number $L$ of spinons present and the individual spinon momentum
numbers $a_i$, which are just the numbers in the boxes above or below
the dots representing the spinons.  The single-spinon momenta are
obtained from those via
\begin{equation}
  \label{eq:hssinglespinonmom}
  p_i=\frac{\pi}{N}\,\left(a_i-\frac{1}{2}\right),
\end{equation}
which implies $\delta\le p_i\le\pi-\delta$ with
$\delta=\frac{\pi}{2N}\to 0$ for $N\to\infty$.

The total momentum and the total energies of the many-spinon
states are given by \eqref{eq:hsLspinonenergy} with
\begin{equation}
  \label{eq:hssutwopzeroezero}
  p_0=-\frac{\pi}{2}\:N,\quad 
  E_0=-\frac{\pi^2}{24}\left(N+\frac{5}{N}\right),
\end{equation}
and the single-spinon dispersion
\begin{equation}
  \label{eq:hssutwoepsilon}
  \epsilon(p)=\frac{1}{2}p\left(\pi-p\right)
    +\frac{\pi^2}{8N^2},
\end{equation}
where we use a convention according to which the ``vacuum'' state
$\ket{\dw\dw\ldots\dw}$ has momentum $p=0$ (and the empty state
$\ket{0}$ has $p=\pi (N-1)$).

\vspace{.5\baselineskip} This correspondence shows that spinons are
non-interacting, with momentum spacings appropriate for half-Fermions.
We may interpret the Haldane-Shastry model as a reparameterization of
a Hilbert space spanned by spin flips (\ref{eq:hsket}) into a basis
which consists of the Haldane-Shastry ground state plus all possible
many spinon states.  The reward for such a reparameterization is that
a highly non-trivial Hamiltonian in the original basis may be
approximately or exactly diagonal in the new basis, as this basis is
chosen in accordance with the quantum numbers of the elementary
excitations.

\section{The Moore--Read state and its parent Hamiltonian}
%\section{The Pfaffian state at $\nu=1/2$ and its parent Hamiltonian}
%\section{The Pfaffian state at even denominator fillings and its parent Hamiltonian}
\label{sec:3mod-pf}
\subsection{The Pfaffian state and its parent Hamiltonian}
\label{sec:3mod-pfintro}

The Pfaffian state at even denominator Landau level filling fractions
was introduced independently by Moore and Read~\cite{moore-91npb362}
as an example of a quantized Hall state which supports quasiparticle
excitations which obey non-Abelian statistics, and by Wen, Wilczek,
and ourselves~\cite{greiter-91prl3205,greiter-92npb567} as a candidate
for the observed plateau in Hall resistivity at Landau level filling
fraction $\nu=5/2$, \ie at $\nu=1/2$ in the second Landau
level~\cite{willet-87prl1776,pan-99prl3530,xia-04prl176809,pan-08prb075307,zhang-10prl166801},
a proposal which was subsequently
strengthened~\cite{morf98prl1505,moeller08prb075319,storni-10prl076803,thomale-10prl180502}
and which recently received experimental support through the direct
measurement of the quasiparticle
charge~\cite{dolev-08n829,radu-08s899}.

The wave function first proposed by Moore and Read~\cite{moore-91npb362} is
\begin{equation}
  \label{eq:pfpsi0}
  \psi_0(z_1,z_2,\ldots ,z_N) %[z_i]
  =\text{Pf}\left(\frac{1}{z_{i}-z_{j}}\right)
  \prod_{i<j}^{N}(z_i-z_j)^m
  \prod_{i=1}^N e^{-\frac{1}{4}|z_i|^2},          
\end{equation}
where the particle number $N$ is even, $m$ is even (odd) for fermions
(bosons), and the Pfaffian is is given by the fully antisymmetrized
sum over all possible pairings of the $N$ particle coordinates,
\begin{equation}
  \label{eq:pfpfaff}
  \text{Pf}\left(\frac{1}{z_i -z_j}\right)\equiv
  \mathcal{A}
  \left\{
    \frac{1}{z_1-z_2}\cdot\,\ldots\,\cdot\frac{1}{z_{N-1}-z_{N}}
  \right\}.
\end{equation}
The inverse Landau level filling fraction is given by 
\begin{equation}
  \label{eq:pfnu}
  \frac{1}{\nu}=\parder{N_\Phi}{N}=\parder{(m(N-1)-1)}{N}=m.
\end{equation}
The state describes a Laughlin state at $\nu=1/m$ supplemented by a
Pfaffian which implements p-wave pairing correlations.  Since the
Pfaffian is completely antisymmetric, it reverses the statistics from
bosons to fermions or vice versa, but does not change the Landau level
filling fraction.

The Pfaffian describes a BCS wave
function~\cite{bardeen-57pr162,Schrieffer64,deGennes66,Tinkham96} in
position space, obtained by projecting on a definite number of
particles~\cite{DysonQuotedinSchrieffer64,greiter05ap217}.  To see
this, first rewrite the (unnormalized) BCS wave function as
\begin{align}
  \nonumber \ket{\psi_\phi} 
  &=\prod_{\bs{k}}\left(1 +  e^{i\phi} \frac{v_{\bs{k}}}{u_{\bs{k}}} 
%  \prod_{\bs{k}}\left({u_{\bs{k}}} +  e^{i\phi} {v_{\bs{k}}} 
    c_{\bs{k}\up}^\dagger\,c_{-\bs{k}\dw}^\dagger\right)\vac
  \\[0.2\baselineskip] \nonumber
  &=\prod_{\bs{k}}\exp\Bigl(e^{i\phi}\frac{v_{\bs{k}}}{u_{\bs{k}}} 
    c_{\bs{k}\up}^\dagger\,c_{-\bs{k}\dw}^\dagger\Bigr)\vac 
  \\[0.2\baselineskip] \nonumber
  &=\exp\Bigl(e^{i\phi} \sum_{\bs{k}} \frac{v_{\bs{k}}}{u_{\bs{k}}} 
    c_{\bs{k}\up}^\dagger\,c_{-\bs{k}\dw}^\dagger\Bigr)\vac
  \\[0.2\baselineskip] %\nonumber
  &=\exp\bigl(e^{i\phi} b^\dagger\bigr)\vac ,
\end{align}
where the pair creation operator $b^\dagger$ is given by
\begin{align}
  \nonumber
  b^\dagger&\equiv \sum_{\bs{k}} \frac{v_{\bs{k}}}{u_{\bs{k}}} 
  c_{\bs{k}\up}^\dagger\,c_{-\bs{k}\dw}^\dagger
  \\[0.2\baselineskip] %\nonumber  
  &=\int\! \text{d}^{3\,}\!\bs{x}_1 \text{d}^{3\,}\!\bs{x}_{2}\,
  \varphi(\bs{x}_1-\bs{x}_2)\,
  \psi^\dagger_\up(\bs{x}_1)\psi^\dagger_\dw(\bs{x}_2)\vac .
\end{align}
The wave function for each of the individual pairs, which only
depends on the relative coordinate, is %(up to a normalization)
given by
\begin{equation}
  \label{eq:pfvarphi}
  \varphi(\bs{x})=\frac{1}{V}\sum_{\bs{k}} \frac{v_{\bs{k}}}{u_{\bs{k}}}
  e^{i\bs{k}\bs{x}}.  
\end{equation}
If we now project out a state with $N/2$
pairs~\cite{anderson66rmp298,DysonQuotedinSchrieffer64,greiter05ap217},
we obtain
\begin{align}
  \ket{\psi_N} 
  &= \frac{1}{2\pi}\int_0^{2\pi}\!\text{d}\phi\, e^{-iN\phi/2}\ket{\psi_\phi}
  \nonumber\\[0.2\baselineskip] 
  &=\frac{1}{2\pi}\int_0^{2\pi}\!\!\text{d}\phi\,
  e^{-iN\phi/2}\exp\bigl(e^{i\phi}b^\dagger\bigr)\vac
  \nonumber\\[0.2\baselineskip]  
  &=\frac{1}{\bigl(\frac{N}{2}\bigr)!} \bigl(b^\dagger\bigr)^{N/2}\vac ,
%  \frac{2\pi}{\bigl(\frac{N}{2}\bigr)!} \bigl(b^\dagger\bigr)^{N/2}\vac ,
%  \frac{\bigl(b^\dagger\bigr)^{N/2}}{\bigl(\frac{N}{2}\bigr)!} \vac ,
\end{align}
which is (up to a normalization) equivalent to 
\begin{align}
  \label{eq:pfbcsn}
  \ket{\psi_N} &=\int\!\! 
  \text{d}^{3\,}\!\bs{x}_1\!\ldots \text{d}^{3\,}\!\bs{x}_{N}\,
  \varphi(\bs{x}_1-\bs{x}_2)\cdot\ldots\cdot
  \varphi(\bs{x}_{N-1}\!-\bs{x}_{N})
  \nonumber\\[0.2\baselineskip] 
  &\hspace{30pt}  
  \cdot\,\psi^\dagger_\up(\bs{x}_1)\psi^\dagger_\dw(\bs{x}_2)\ldots
  \psi^\dagger_\up(\bs{x}_{N-1})\psi^\dagger_\dw(\bs{x}_{N})\vac \!.
\end{align}
This implies that the many-particle wavefunction is given by a
Pfaffian,
\begin{equation}
  \label{eq:pfbcspf}
  \psi(\bs{x}_1\ldots\bs{x}_{N})
  =\text{Pf}\left(\varphi(\bs{x}_i-\bs{x}_j)\right).
\end{equation}
This form nicely illustrates that all the pairs have condensed into
the same state, which is the essence of superfluidity.  For fermion
pairings with even relative angular momentum of the pairs, such as s- or
d-wave, the wave function $\varphi(\bs{x}_i-\bs{x}_j)$ of the pairs is
symmetric in real space, and antisymmetric in spin space (\ie a
singlet), while for pairings with odd angular momentum, such as p-wave,
$\varphi(\bs{x}_i-\bs{x}_j)$ is antisymmetric in real space and
symmetric in spin space (\ie a triplet).

In the quantized Hall state, the requirement of analyticity in the
complex coordinates constraints the possible form of the pair wave
function decisively.  Since the electrons are spin polarized, the only
possible choice is the p-wave pairing described by the Pfaffian with
$\varphi(z_i-z_j)=1/(z_i-z_j)$.  Note that this pair wave function
would not be normalizable if it were not multiplied by at least an
$m=1$ Laughlin state.

One of the most important mathematical properties of the Pfaffian is that
its square is equal to the determinant,
\begin{equation}
  \label{eq:pfpf2}
  \text{Pf}\left(\varphi(\bs{x}_i-\bs{x}_j)\right)^2=\det(M_{ij}),
\end{equation}
where 
\begin{equation}
  \label{eq:pfpf2Mij}
  M_{ij}=  
  \left\{\begin{array}{ll}
      0 &\text{for}\ i=j,\\
      \varphi(\bs{x}_i-\bs{x}_j) &\text{for}\ i\ne j .
    \end{array}\right.
\end{equation}
Another important identity, due to
Frobenius~\cite{frobenius-1882ram53}, is given by \eqref{eq:nacauchy}
in Section \ref{sec:napro} below.

The uniquely specifying property of the Pfaffian quantized Hall state
\eqref{eq:pfpsi0} is that the wave function vanishes as the
$(3m\hspace{-1pt}-\hspace{-1pt}1)$-th power as \emph{three} particles
approach each other.  This property simply reflects that there can be
at most only one pair among each triplet of particles.  This
observation has led Wen, Wilczek, and
ourselves~\cite{greiter-91prl3205,greiter-92npb577,greiter-92npb567}
to propose the parent Hamiltonian
\begin{equation}
  \label{eq:pfGWWham}
  V^{(m)}=\sum_{i,j<k}^N \left(\nabla_i^2\right)^{(m-1)}
  \left(\delta^{(2)}(z_i-z_j)\delta^{(2)}(z_i-z_k)\right),
\end{equation}
which, when supplemented with the kinetic Hamiltonian
\eqref{eq:qhSinglePartHam3} as well as all similar terms with smaller
powers of the Laplacian, singles out \eqref{eq:pfpsi0} as its unique
ground state.  For all practical purposes, however, it is best to
formulate our parent Hamiltonian in terms of three-body
pseudopotentials, as we will elaborate in Section
\ref{sec:3mod-pf-sph}.

\subsection{Quasiparticle excitations and the internal Hilbert space}

One of the key properties of superconductors is that the magnetic
vortices are quantized in units of one half of the Dirac flux quanta
$\Phi_0=2\pi\hbar c/e$, in accordance to the charge $-2e$ of the
Cooper pairs.  The paring correlations in the Pfaffian Hall state
have a similar effect on the vortices or quasiparticle excitations,
which carry one half of the flux and charge they would carry
without the pairing, \ie they carry charge $e^*=e/2m$.  The wave function
for two flux~\half  quasiholes at positions $\xi_1$ and $\xi_2$ is 
easily formulated.  We simply replace each factor in the Pfaffian
in \eqref{eq:pfpsi0} by
\begin{equation}
  \label{eq:pf2QH}
  \text{Pf}\left(\frac{1}{z_i-z_j}\right) 
%  \mapsto \text{Pf}\left(
  \to \text{Pf}\left(
    \frac{(z_i-\xi_1)(z_j-\xi_2)+(z_i\leftrightarrow z_j)}{z_i-z_j}
  \right),
\end{equation}
such that one member of each electron pair sees the additionally
inserted zero at $\xi_1$ and the other member sees it at $\xi_2$.  If
we set $\xi_1=\xi_2=\xi$, we will recover a regular quasihole in the
Laughlin fluid with charge $e^*=e/m$.

The internal Hilbert space spanned by the quasiparticle excitations
only emerges as we consider the wave function for four charge
$e^*=e/4$ quasiholes at positions $\xi_1,\ldots,\xi_4$, which is
obtained by replacing the Pfaffian in \eqref{eq:pfpsi0} by
\begin{align}
  \label{eq:pf4QH}
  \text{Pf}\left(\frac{1}{z_i-z_j}\right) 
%  \mapsto \text{Pf}\left(
  \to\text{Pf}\left(
    \frac{(z_i-\xi_1)(z_j-\xi_2)(z_i-\xi_3)(z_j-\xi_4)
      +(z_i\leftrightarrow z_j)}{z_i-z_j}
  \right).
  \nonumber\\[0.2\baselineskip] 
\end{align}
We see that $\xi_1$ and $\xi_3$ belong to one group in that they
constitute additional zeros seen by one member of each electron pair,
while $\xi_2$ and $\xi_4$ belong to another group as they constitute
zeros seen by the other members of each electron pair.  The wave
function is symmetric (or antisymmetric, depending on the number of
electron pairs) under interchange of both groups.  The state in the
internal Hilbert space spanned by the quasihole affiliations with the
two groups will change as we adiabatically interchange two quasiholes
belonging to different groups, say $\xi_3$ and $\xi_4$.  Naively, one
might think that the dimension of the internal Hilbert space is given
by the number of ways to partition the quasiholes at
$\xi_1,\ldots,\xi_{2n}$ into two different groups, \ie by $(2n-1)!!$
for $2n$ quasiholes.  Note that the number of quasiholes has to be
even on closed surfaces to satisfy the Dirac flux quantization
condition~\cite{dirac31prsla60}.  The true dimension of the internal
Hilbert space, however, is only
$2^{n-1}$~\cite{nayak-96npb529}.  %,fradkin-98npb704}.  
The reason for this is that the internal Hilbert space is spanned by
Majorana fermion states in the vortex cores~\cite{read-00prb10267}, as
we will elaborate in 
%Section \ref{sec:3mod-pf-nastat} below.
the following section.

The statistics is non-Abelian in the sense that the order according to
which we interchange quasiholes matters.  Let the matrix $M_{ij}$
describe the rotation of the internal Hilbert space state vector which
describes the adiabatic interchange two quasiholes at $\xi_i$ and
$\xi_j$:
\begin{equation*}
  \label{eq:pfMij}
  \ket{\psi}\to M_{ij}\ket{\psi}.
\end{equation*}
The statistics is non-Abelian if the matrices associated with
successive interchanges do not commute in general,
\begin{equation*}
  M_{ij}M_{jk}\ne M_{jk} M_{ij}.
\end{equation*}
Note that the internal state vector is protected in the sense that it
is insensitive to local perturbations---it can \emph{only} be
manipulated through braiding of the vortices.  For a sufficiently
large number of vortices, on the other hand, any unitary
transformation in this space can be approximated to arbitrary accuracy
through successive braiding operations~\cite{freedman-02cmp587}.
These properties together render non-Abelions preeminently suited for
applications as protected qubits in quantum
computation~\cite{dasSarma-05prl166802,nayak-08rmp1083,bishara-09prb155303,moore09p82,stern10n187}.

\subsection{Majorana fermions and non-Abelian statistics}
\label{sec:3mod-pf-nastat}

The key to understanding the non-Abelian statistics~\cite{stern10n187}
of the quasiparticle excitations of the Pfaffian state lies in the
Majorana fermion modes in the vortices of p-wave
superfluids~\cite{kopnin-91prb9667,read-00prb10267,ivanov01prl268,stern-04prb205338}.
The p-wave pairing symmetry implies that the order parameter for the
superfluid acquires a phase of $2\pi$ as we go around the Fermi
surface,
\begin{equation}
  \label{eq:pfpwaveOP}
%  \ev{\psi^\dagger\psi^\dagger}
%  =\Delta_0\,(k_\x+\text{i}k_\y).
  \langle {c_{\bs{k}}^\dagger\,c_{-\bs{k}}^\dagger} \rangle
  =\Delta_0(k)\cdot(k_\x+\text{i}k_\y),
\end{equation}
%where $\Delta_0$ can be choosen real.  
where $\Delta_0(k)$ can be chosen real.  
The Hamiltonian for a single vortex at the origin is given by
\begin{align}
  \label{eq:pfHvortex}
  H=\int\!\text{d}\bs{r}\left\{\psi^{\dagger}\!\left(\!-\frac{\bs{\nabla}^2}{2m}-
%      \mu
       \varepsilon_F\!
    \right)\!\psi
    +\psi^{\dagger}\left(
%      \exp(\text{i}\varphi)\Delta_0(r)*(\partial_x-\text{i}\partial_y)
      e^{\text{i}\varphi}\Delta_0(r)*(\partial_x-\text{i}\partial_y)
    \right)\psi^{\dagger}+\text{h.c.}\right\},
  \nonumber\\[0.2\baselineskip] 
\end{align}
where $A*B\equiv\frac{1}{2}\{A,B\}$ denotes the symmetrized product,
and $r$ and $\varphi$ are polar coordinates.  The order parameter
$\Delta_0(r)$ vanishes inside the vortex core.  We can obtain the
energy eigenstates localized inside the vortex by solving the
Bogoliubov--de Gennes equations~\cite{deGennes66} equations
\begin{equation}
  \label{eq:pfBogodeGennes}
  \comm{H}{\gamma_n^\dagger(\bs{x})}=E_n\gamma_n^\dagger(\bs{x}),
\end{equation}
where $n$ labels the modes and 
\begin{equation}
  \label{eq:pfBogoOp}
  \gamma_n^\dagger(\bs{x})
  =u_n(\bs{x})\psi^\dagger(\bs{x})+v_n(\bs{x})\psi(\bs{x})
\end{equation}
are the Bogoliubov quasiparticle operators.  The low energy spectrum
is given by~\cite{kopnin-91prb9667,read-00prb10267} 
\begin{equation}
  \label{eq:pfEn}
  E_n=n\omega_0,
\end{equation}
where $n$ is an integer and $\omega_0=\Delta^2/\varepsilon_F$ the
level spacing.  Note that while in an s-wave superfluid, the
Bogoliubov operators %$\gamma_\up$ 
\begin{equation}
  \label{eq:pfBogoOpswave}
  \gamma_{n\up}(\bs{x})
  =u_{n\up}(\bs{x})\psi^\dagger(\bs{x})+v_{n\dw}(\bs{x})\psi(\bs{x})
\end{equation}
combine $\up$-spin electron creation operators with $\dw$-spin
annihilation operators, in the p-wave superfluid, the
operators \eqref{eq:pfBogoOp} combine creation and annihilation
operators of the \emph{same} spinless (or spin-polarized) fermions.  Since 
the Bogoliubov--de Gennes equations are not able to distinguish
between particles and antiparticles, we obtain each physical
solution twice:  once with positive energy as a solution of the
Bogoliubov--de Gennes equation \eqref{eq:pfBogodeGennes} for the
creation operators, and once with negative energy as a solution of the
same equation for the annihilation operators,
\begin{equation}
  \label{eq:pfBogodeGennesann}
  \comm{H}{\gamma_n(\bs{x})}=-E_n\gamma_n(\bs{x}),
\end{equation}
which is obtained from \eqref{eq:pfBogodeGennes} by Hermitian
conjugation.  We resolve this technical artifact by discarding the
negative energy solutions as unphysical.  For the $n=0$ solution with
at $E_0=0$, it implies that we get one fermion solution when we
overcount by a factor of two.  The physical solution at $E=0$ is hence
given by one half of a fermion, or a Majorana fermion, as
\begin{equation}
  \label{eq:pfMajo1}
  \gamma_0^\dagger(\bs{x})=\gamma_0(\bs{x}).
\end{equation}

In general, one fermion $\psi,\psi^\dagger$ consists of two Majorana 
fermions,% $\gamma_1,\gamma_2$, 
%$\gamma_1=\gamma_1^\dagger$, $\gamma_2=\gamma_2^\dagger$,
\begin{equation}
  \label{eq:pfPsiGamma}
  \psi=\frac{1}{2}(\gamma_1+\text{i}\gamma_2),\qquad
  \psi^\dagger=\frac{1}{2}(\gamma_1-\text{i}\gamma_2),
\end{equation}
which in turn are given by the real and imaginary part of the fermion
operators,
\begin{equation}
  \label{eq:pfGammaPsi}
  \gamma_1%=\gamma_1^\dagger
  =\psi +\psi^\dagger,\qquad
  \gamma_2%=\gamma_2^\dagger
%  =i(\psi^\dagger -\psi).
  =-\i(\psi -\psi^\dagger).
\end{equation}
They obey the anticommutation relations
\begin{equation}
  \label{eq:pfMajoComm}
  \anticomm{\gamma_i}{\gamma_j}=2\delta_{ij},
\end{equation}
as one may easily verify with \eqref{eq:pfGammaPsi}.  Majorana
fermions are their own antiparticles, as $\gamma_i^\dagger=\gamma_i$.
If we write the basis for a single fermion as
$\{\vac,\psi^\dagger\vac\}$, we can write the fermion creation and
annihilation operators as 
\begin{equation}
  \label{eq:pf2compbasis}
  \psi^\dagger=\!\left(\!\!
    \begin{array}{cc}
      \,0\,&\,0\,\\1&0 
    \end{array}\!\!\right),\quad
  \psi=\!\left(\!\!
    \begin{array}{cc}
      \,0\,&\,1\,\\0&0 
    \end{array}\!\!\right).
\end{equation}
In this basis, the Majorana fermions are given by the first two Pauli
matrices,
\begin{equation}
  \label{eq:pf2compbasis2}
  \gamma_1=\!\left(\!\!
    \begin{array}{cc}
      \,0\,&\,1\,\\1&0 
    \end{array}\!\!\right)=\sigma_\x,\quad
  \gamma_2=\!\left(\!\!
    \begin{array}{cc}
      \,0\,&\,-\i\,\\\i&0 
    \end{array}\!\!\right)=\sigma_\y.
\end{equation}

\begin{figure}[t]
    \psfrag{gamma1}{{$\gamma_{i}$}}
    \psfrag{gamma2}{{$\gamma_{i+1}$}}
    \psfrag{branch}{{branch cut}}
    \psfrag{boundary}{{boundary}}
    \psfrag{asgamma}{{as $\gamma_{i+1}$ crosses the}} 
    \psfrag{cut}{{branch cut $\gamma_{i+1}\to-\gamma_{i+1}$}}
    \psfrag{ti}{{$T_i$}}
    \psfrag{time}{{time}}
  \begin{center}
    \includegraphics[scale =0.20]{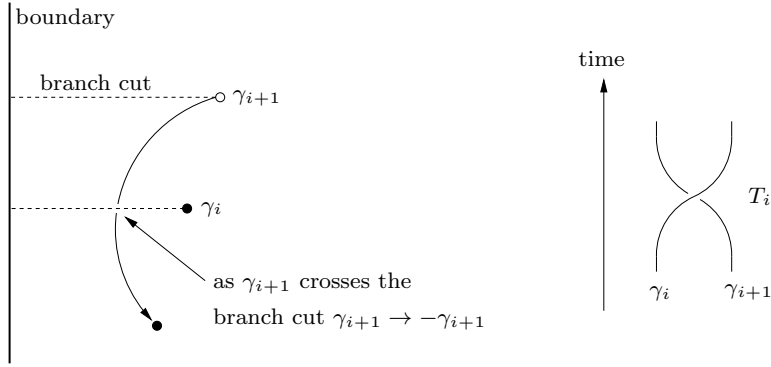}
  \end{center}
  \caption{The Majorana fermion $\gamma_{i+1}$ acquires a $-$ sign as
    it crosses the branch cut from another vortex.}
  % with Majorana fermion $\gamma_{i}$.}
  \label{fig:pfBranchCut}
\end{figure}

Returning to vortices in a p-wave superfluid, note that the order
parameter acquires by definition a phase of $2\pi$ as we go around a
vortex.  This implies that the electron creation and annihilation
operators acquire a phase $\pi$, or a minus sign, which implies via
\eqref{eq:pfGammaPsi} that the Majorana fermion states acquire
likewise %a phase $\pi$ or
a minus sign,
\begin{equation}
  \label{eq:pfMajoMinus}
  \gamma_i\to -\gamma_i,
\end{equation}
as we encircle a vortex.  By choice of gauge, we can implement the phase
change of $2\pi$ in the superconducting order parameter as a branch cut
connecting the vortices to the left boundary of the system, and assume 
a convention according to which the Majorana fermion in each vortex
crossing a branch cut acquires a minus sign, as illustrated in
Figure \ref{fig:pfBranchCut}.

\begin{figure}[t]
    \psfrag{gamma1}{{$\gamma_{i}$}}
    \psfrag{gamma2}{{$\gamma_{i+1}$}}
    \psfrag{gamma3}{{$\gamma_{i+2}$}}
    \psfrag{ti}{{$T_i$}}
    \psfrag{ti1}{{$T_{i+1}$}}
    \psfrag{eq}{{$=$}}
  \begin{center}
    \includegraphics[scale =0.20]{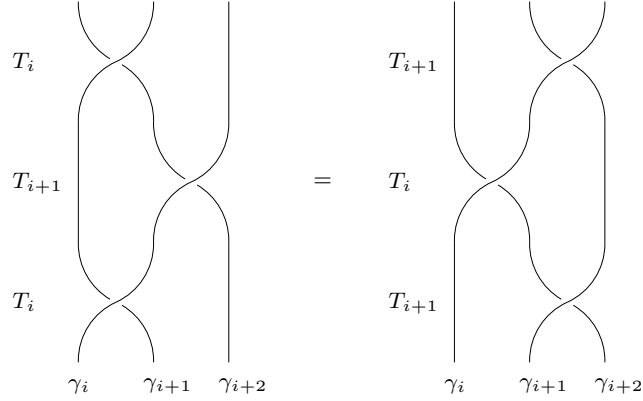}
  \end{center}
  \caption{Illustration of the defining algebra of the braid group $B_{2n}$: 
    $T_iT_{i+1}T_i = T_{i+1}T_iT_{i+1}$.}
  \label{fig:pfBraidingDef}
\end{figure}

To obtain the non-Abelian statistics, Ivanov~\cite{ivanov01prl268}
considered permutations of $2n$ vortices by braiding, which form the
braid group $B_{2n}$~\cite{Kauffman93}.  This group is generated by
counterclockwise interchanges $T_i$ of particles $i$ and $i+1$, which
are neighbors with regard to the positions of their branch cuts to the
boundary.  The algebra of the group is given by
% \begin{align}
%   \label{eq:pfBraidAlgebra}
%   T_iT_j \!&\!=\!&\! T_jT_i,\qquad\text{for}\ \,|i-j|>1\\
%   T_iT_jT_i \!&\!=\!&\! T_jT_iT_j,\qquad\text{for}\ \,|i-j|=1
% \end{align}
\begin{equation}
  \label{eq:pfBraidAlgebra}
  \begin{array}{r@{\ }c@{\ }ll}
    T_iT_j &= T_jT_i      \quad&\text{for}\ \,|i-j|>1,\\[8pt]
    T_iT_jT_i &= T_jT_iT_j\quad&\text{for}\ \,|i-j|=1,
  \end{array}
\end{equation}
as illustrated in Figure \ref{fig:pfBraidingDef}.  Note that the braid
group is different from the permutation group as  
\begin{equation*}
  T_i^{-1}\ne T_i.  
\end{equation*}
The convention for the minus signs acquired by the Majorana fermions
defined in Figure \ref{fig:pfBranchCut} implies the transformation rule
\begin{equation}
  \label{eq:pfTransRule}
  T_i(\gamma_j)=
  \left\{\begin{array}{@{}ll}
      \phantom{-}\gamma_{j+1} &\text{for}\ i=j,\\[2pt]
      -\gamma_{j-1} &\text{for}\ i=j-1,\\[2pt]
      \phantom{-}\gamma_{j} &\text{otherwise}.
    \end{array}\right.
\end{equation}
To describe the action of these transformations on the (internal)
state vectors, we hence need to find a representation $\tau(T_i)$ of
the braid group $B_{2n}$ such that
\begin{equation}
  \label{eq:pfRep}
  \tau(T_i)\gamma_{j}\tau(T_i)^{-1}= T_i(\gamma_j)
\end{equation}
with $T_i(\gamma_j)$ given by \eqref{eq:pfTransRule}.  The
solution is~\cite{ivanov01prl268} 
% \begin{equation}
%   \label{eq:pfRepSolution}
%   \tau(T_i)=\exp\left(\frac{\pi}{4}\gamma_{i+1}\gamma_{i}\right)
%   =\frac{1}{\sqrt{2}}(1+\gamma_{i+1}\gamma_{i}),
% \end{equation}
\begin{align}
  \label{eq:pfRepSolution}
  \tau(T_i)&=\exp\left(\frac{\pi}{4}\gamma_{i+1}\gamma_{i}\right)
  \nonumber\\[0.3\baselineskip]
  &=\cos\left(\frac{\pi}{4}\right) 
  + \gamma_{i+1}\gamma_{i}\,\sin\left(\frac{\pi}{4}\right),
  \nonumber\\[0.1\baselineskip]
  &=\frac{1}{\sqrt{2}}(1+\gamma_{i+1}\gamma_{i}),
\end{align}
as one can easily verify using $(\gamma_{i+1}\gamma_{i})^2=-1$.  The
inverse transformation is given by
\begin{equation}
  \label{eq:pfRepSolution-1}
  \tau(T_i)^{-1} %=\exp\left(-\frac{\pi}{4}\gamma_{i+1}\gamma_{i}\right)
  =\frac{1}{\sqrt{2}}(1-\gamma_{i+1}\gamma_{i}).
\end{equation}
A few steps of algebra yield
\begin{equation*}
  \tau(T_1)
  \left\{\begin{array}{@{}c@{}}\gamma_1\\\gamma_2 \end{array}\right\}
  \tau(T_1)^{-1}
  =\left\{\begin{array}{@{}r@{}}\gamma_2\\-\gamma_1 \end{array}\right\}.
\end{equation*}
This representation coincides with that of Nayak and
Wilczek~\cite{nayak-96npb529} for the statistics of the quasiholes in
the Pfaffian state.

The simplest examples of this representation are the cases of two and
four vortices~\cite{nayak-96npb529,fradkin-98npb704,ivanov01prl268},
which we will elaborate now.  In the case of two vortices, the two
Majorana fermions $\gamma_1$ and $\gamma_2$ can be combined into a
single fermion via \eqref{eq:pfPsiGamma}, and the ground state is
hence two-fold degenerate.  The braid group $B_2$ has only one
generator $T_1$ with representation
\begin{align}
  \label{eq:pfRepT1}
  \tau(T_1)&=\exp\left(\frac{\pi}{4}\gamma_{2}\gamma_{1}\right)
  \nonumber\\[0.2\baselineskip]
  &=\exp\left(-\text{i}\frac{\pi}{4}
    (\psi-\psi^\dagger)(\psi+\psi^\dagger)\right)
  \nonumber\\[0.2\baselineskip]
  &=\exp\left(\text{i}\frac{\pi}{4}
    (2\psi^\dagger\psi-1)\right)
  \nonumber\\[0.2\baselineskip]
  &=\exp\left(-\text{i}\frac{\pi}{4}\sigma_\z \right),
%   \nonumber\\[0.3\baselineskip]
%   &=\left(\!\!\begin{array}{cccc}
%        e^{-i\pi/4}&0\\ 
%        0& e^{i\pi/4} 
%      \end{array}\!\!\right),
\end{align}
where $\sigma_\z $
% \begin{equation}
%   \label{eq:pf2compbasis}
%   \sigma_\z =\!\left(\!\!
%     \begin{array}{cc}
%       1&0\\0&-1 
%     \end{array}\!\!\right)
% \end{equation}
is the third Pauli matrix \eqref{eq:qhsPauli} in the basis
$\big\{\vac,\psi^\dagger\vac\big\}$.  The braiding is hence diagonal in this
basis, and only gives an overall phase, which depends on whether the
fermion state is occupied or not.

The non-Abelian statistics manifests itself only once we consider four
vortices.  Following Ivanov~\cite{ivanov01prl268}, we combine the
four Majorana fermions into two fermions,
\begin{equation}
  \label{eq:pfPsiGamma1234}
  \psi_1=\frac{1}{2}(\gamma_1+\text{i}\gamma_2),\qquad
  \psi_2=\frac{1}{2}(\gamma_3+\text{i}\gamma_4),
%  \psi^\dagger=\frac{1}{2}(\gamma_1-\text{i}\gamma_2),
\end{equation}
and accordingly for the fermion creation operators $\psi_1^\dagger$,
$\psi_2^\dagger$.  The braid group $B_4$ has three generators $T_1$,
$T_2$, and $T_3$.  Their representations in a basis of fermion
occupation numbers
\begin{equation*}
  \big\{\vac,
  \psi_1^\dagger\vac,\psi_2^\dagger\vac,\psi_1^\dagger\psi_2^\dagger\vac\big\}
\end{equation*}
are given by two diagonal operators
\begin{align}
%  \label{eq:pfRepT13}
  \tau(T_1)&= %\hspace{-7pt}&=\hspace{-7pt}
  \exp\left(\frac{\pi}{4}\gamma_{2}\gamma_{1}\right)=
  \exp\left(-\text{i}\frac{\pi}{4}{\sigma_\z^{(1)}}\right)
  =\left(\!\!\begin{array}{cccc}
       e^{-\i\pi/4}&0&0&0\\ 
       0& e^{\i\pi/4}&0&0\\ 
       0&0&e^{-\i\pi/4}&0\\
       0&0&0& e^{\i\pi/4}
     \end{array}\!\!\right),
  \nonumber\\[0.5\baselineskip]
  \tau(T_3)&= %\hspace{-7pt}&=\hspace{-7pt}
  \exp\left(\frac{\pi}{4}\gamma_{4}\gamma_{3}\right)=
  \exp\left(-\text{i}\frac{\pi}{4}{\sigma_\z^{(2)}}\right)
  =\left(\!\!\begin{array}{cccc}
       e^{-\i\pi/4}&0&0&0\\ 
       0&e^{-\i\pi/4}&0&0\\ 
       0&0& e^{\i\pi/4}&0\\
       0&0&0& e^{\i\pi/4}
     \end{array}\!\!\right),
  \nonumber
\end{align}
and one off-diagonal operator,
\begin{align}
%  \label{eq:pfRepT2}
  \tau(T_2)&= %\hspace{-7pt}&=\hspace{-7pt}
  \exp\left(\frac{\pi}{4}\gamma_{3}\gamma_{2}\right)
  \nonumber\\ %[-0.5\baselineskip]
  &= %\hspace{-7pt}&=\hspace{-7pt}
  \frac{1}{\sqrt{2}}
  \left(1-\i(\psi_2+\psi_2^\dagger)(\psi_1-\psi_1^\dagger)\right)
%  \nonumber\\[0.3\baselineskip]
%  \hspace{-7pt}&=\hspace{-7pt}
  =
  \left(\!\!\begin{array}{cccc}
       1&0&0&\,-\i\,\\ 
       0&1&\,-\i\,&0\\ 
       0&\,-\i\,&1&0\\ 
       \,-\i\,&0&0&1
     \end{array}\!\!\right).
  \nonumber
\end{align}

Note that since the representations $\tau(T_i)$ given by
\eqref{eq:pfRepSolution} are even in the fermion operators, \ie change
the fermion numbers only by even integers, we may restrict them
%the representations 
to only even or odd sectors in the fermion numbers.
For the example of four vortices, these sectors are given by
$\{\vac,\psi_1^\dagger\psi_2^\dagger\vac\}$ and
$\{\psi_1^\dagger\vac,\psi_2^\dagger\vac\}$.  Each sector contains
$2^{n-1}$ states, which is the degeneracy found for a Pfaffian state
with an even number of electrons~\cite{nayak-96npb529}.  Physically,
this reflects that while the number of fermions is not a good quantum
number in a superfluid, the number of fermions modulo two, \ie whether
the number is even or odd, is a good quantum number.
  
Finally, note that the derivation of the non-Abelian statistics depends
only on (a) the vortices possessing Majorana fermion modes, and (b)
the Majorana fermions changing sign $\gamma_i\to -\gamma_i$ when the
order parameter phase changes by $2\pi$, as it does by definition when
we go around a vortex.

\subsection{The Pfaffian state and its parent Hamiltonian on the sphere}
\label{sec:3mod-pf-sph}

The Pfaffian state is readily formulated in the spherical
geometry~\cite{greiter-92npb567}.  The wave function for $N$ particles 
at Landau level filling $\nu=1/m$ on a sphere with $2s_0=m(N-1)-1$ 
magnetic flux quanta is given by
\begin{equation}
  \label{eq:qhspsiPf}
  \psi_0[u,v]=\text{Pf}\left(\frac{1}{u_iv_j-u_jv_i}\right)
  \prod_{i<j}^N(u_iv_j-u_jv_i)^m,
\end{equation}
where $m$ is even for fermions and odd for bosons.  Note that the
relation between flux and particle number implies that the states at
$\nu=1/2$ is not its own particle-hole
conjugate~\cite{levin-07prl236806,lee-07prl236807}.  The formulation
of quasihole excitations generalizes without incident from the planar
geometry.

As mentioned in Section \ref{sec:3mod-pfintro}, the uniquely
specifying property of the Pfaffian state \eqref{eq:qhspsiPf} is
that it vanishes as the $(3m\hspace{-1pt}-\hspace{-1pt}1)$-th power of
the distance as \emph{three} particles approach each other.  For the
spherical geometry, the corresponding parent
Hamiltonian %~\cite{greiter-91prl3205,greiter-92npb577,greiter-92npb567}
can be conveniently formulated using three-body
pseudopotentials~\cite{simon-07prb195306}.  In analogy to the
two-particle interaction Hamiltonian \eqref{eq:qhsVham2}, we write the
three-particle interaction Hamiltonian
\begin{equation}
  \label{eq:qhsVham3}
  H_{\s\text{int}}^{(3)}=\sum_{i<j<k}^N 
  \left\{\sum_{l}^{2s} V_l^{(3)} 
    \text{P}_{3s-l}(\bs{L}_i+\bs{L}_j+\bs{L}_k)\right\}.
\end{equation}
The three-body parent Hamiltonian proposed by Wen, Wilczek, and
ourselves \cite{greiter-91prl3205,%greiter-92npb577,
greiter-92npb567} then amounts to taking
\begin{equation}
  \label{eq:qhsPP3} 
  V_l^{(3)}=
  \left\{\begin{array}{ll}
    1\quad &\text{for}\ l<3m-1,\\[2pt]
    0 &\text{for}\ l\ge 3m-1 .
   \end{array}\right.
\end{equation}

The form \eqref{eq:qhsVham3} is not the most general one, as for $l\ge
6$ for bosons ($l\ge 9$ for fermions), the three particle state is no
longer uniquely described by the three body angular angular momentum
$l$, and one may assign different pseudopotential coefficients to the
different symmetric (antisymmetric), homogeneous, rotationally
invariant polynomials of degree $l$ describing the three body
states~\cite{simon-07prb195306}.  This, however, should not concern us
here as we are only interested in the case $m=1$ for bosons and $m=2$
for fermions.  Furthermore, as in the case of two-body
pseudopotentials, where $l$ had to be even for bosons and odd for
fermions, there exists a related restriction for the allowed values of
$l$ for three-body pseudopotentials.  Specifically, we have no state
with $l=1$ ($l=4$) for bosons (fermions).

For all practical purposes, we once again need to rewrite
\eqref{eq:qhsVham3} in terms of boson or fermion creation or
annihilation operators,
\begin{align}
  \label{eq:qhsVham3q}
  %\hspace{-25pt}
  H_{\s\text{int}}^{(3)}&=
  \sum_{m_1=-s}^s\, \sum_{m_2=-s}^s\, \sum_{m_3=-s}^s\, \sum_{m_4=-s}^s
  \, \sum_{m_5=-s}^s\, \sum_{m_6=-s}^s
%  \sum_{m_1}\sum_{m_2}\sum_{m_3}\sum_{m_4} 
  \ a_{m_1}^\dagger a_{m_2}^\dagger a_{m_3}^\dagger a_{m_4} a_{m_5} a_{m_6}\
  \nonumber\\*[0.3\baselineskip]
  &\quad 
  \cdot\,\delta_{m_1+m_2+m_3,m_4+m_5+m_6}
  \nonumber\\*[0.3\baselineskip]
  &\quad 
  \cdot\sum_{j=0}^{2s}\,\sum_{l=3s-(j+s)}^{3s-|j-s|} V_l^{(3)}\,
  \braket{s,m_1;s,m_2}{j,m_1+m_2}
  \nonumber\\*[-0.5\baselineskip]
  &\quad \hspace{76pt}\cdot
    \braket{j,m_1+m_2;s,m_3}{3s-l,m_1+m_2+m_3} 
  \nonumber\\*[0.3\baselineskip]
  &\quad \hspace{76pt}\cdot
  \braket{3s-l,m_4+m_5+m_6}{s,m_4;j,m_5+m_6} 
  \nonumber\\*[0.3\baselineskip]
  &\quad \hspace{76pt}\cdot
  \braket{j,m_5+m_6}{s,m_5;s,m_6},
\end{align}
where $a_m$ annihilates a boson or fermion in the properly normalized
single particle state
\begin{equation}
  \label{eq:qhsnormLLLbasis2}
  \psi_{m,0}^{s}(u,v) = 
  \sqrt{\frac{(2s+1)!}{4\pi\,(s+m)!\, (s-m)!}}\;
  u^{s+m}v^{s-m},
\end{equation}
and $\braket{s,m_1;s,m_2}{2s-l,m_1+m_2}$ \etc are Clebsch--Gordan
coefficients~\cite{baym69}.

\section{An $S=1$ spin liquid state described by a Pfaffian}
\label{sec:3mod-na}

\subsection{The ground state}
\label{sec:nags}

As for the Haldane--Shastry model, we consider a one-dimensional
lattice with periodic boundary conditions and an even number of sites
$N$ on a unit circle embedded in the complex plane.  The only
difference is that now the spin on each site is $S=1$:
\begin{center}
\begin{picture}(320,70)(-40,-35)
\put(0,0){\circle{100}}
%\put(20,0){\circle*{3}}
\put( 20.0,   .0){\circle*{3}}
\put( 17.3, 10.0){\circle*{3}}
\put( 10.0, 17.3){\circle*{3}}
\put(   .0, 20.0){\circle*{3}}
\put(-10.0, 17.3){\circle*{3}}
\put(-17.3, 10.0){\circle*{3}}
\put(-20.0,   .0){\circle*{3}}
\put(-17.3,-10.0){\circle*{3}}
\put(-10.0,-17.3){\circle*{3}}
\put(   .0,-20.0){\circle*{3}}
\put( 10.0,-17.3){\circle*{3}}
\put( 17.3,-10.0){\circle*{3}}
\qbezier[20]( 20.0,   .0)(  5.0,8.65)(-10.0, 17.3)
\put(50,12){\makebox(0,0)[l]
{$N$\ sites with spin 1 on unit circle: 
}}
%$\displaystyle \eta_\alpha=e^{i\frac{2\pi}{N}\alpha }$}}
\put(50,-12){\makebox(0,0)[l]
{$\displaystyle \eta_\alpha=e^{\text{i}\frac{2\pi}{N}\alpha }$
\ \ with\ $\alpha = 1,\ldots ,N$}}
\end{picture}
\end{center}

The ground state wave function we consider
here~\cite{greiter02jltp1029} is given by a
bosonic Pfaffian state in the complex lattice coordinates $z_i$
supplemented by a phase factor,
\begin{equation}
  \label{eq:napsi0}
  \psi^{S=1}_0(z_1,z_2,\ldots ,z_N) %[z_i]
  =\text{Pf}\left(\frac{1}{z_{i}-z_{j}}\right)
  \prod_{i<j}^{N}(z_i-z_j)\prod_{i=1}^{N}\,z_i.
\end{equation}
The Pfaffian is given by the fully antisymmetrized sum over all
possible pairings of the $N$ particle coordinates,
\begin{equation}
  \label{eq:napfaff}
  \text{Pf}\left(\frac{1}{z_i -z_j}\right)\equiv
  \mathcal{A}
  \left\{
    \frac{1}{z_1-z_2}\cdot\,\ldots\,\cdot\frac{1}{z_{N-1}-z_{N}}
  \right\}.
\end{equation}
The ``particles'' $z_i$ represent re-normalized spin flips
$\tilde{S}_{\alpha}^{+}$ acting on a vacuum with all spins in the
$S^{z}=-1$ state,
\begin{equation}
  \label{eq:naket}
  \ket{\psi^{S=1}_0}=\sum_{\{z_1,\dots,z_{N}\}} 
%  \psi_0[z_i]
  \psi^{S=1}_0(z_1,\dots,z_N)\
%  \tilde{S}_{z_1}^{+}\cdot\, \dots\, \cdot \tilde{S}_{z_{N}}^{+}\ 
  \tilde{S}_{z_1}^{+}\cdot\dots\cdot\tilde{S}_{z_{N}}^{+} 
%  \ket{1,-1}^{\otimes N},
%  \ket{1,-1}_N,
  \ket{-1}_N,
%  \ket{-1}^{\otimes N},
\end{equation}
where the sum extends over all possibilities of distributing the 
$N$ ``particles'' over the $N$ lattice sites allowing for double 
occupation, 
\begin{equation}
  \label{eq:naspinflip}
  \tilde{S}_{\alpha}^{+} \equiv\frac{{S}^{\rm{z}}_{\alpha}+1}{2} S_\alpha^{+},
\end{equation}
and
\begin{equation}
  \label{eq:navacuumket}
  \ket{-1}_N\equiv\otimes_{\alpha=1}^N \ket{1,-1}_{\alpha}.
\end{equation}
This state may be viewed as the one-dimensional analog of the
non-Abelian chiral spin liquid~\cite{greiter-09prl207203}.

Like the ground state of the Haldane--Shastry model, the $S=1$ state
\eqref{eq:napsi0} describes a critical spin liquid in one dimension,
with similarly algebraically decaying correlations.  It does not,
however, serve as a paradigm of the generic $S=1$ spin state, as the
generic state possesses a Haldane
gap~\cite{haldane83pl464,haldane83prl1153,affleck90proc,Fradkin91} in
the spin excitation spectrum due to linearly confining forces between
the
spinons~\cite{affleck-87prl799,affleck-88cmp477,greiter02jltp1029,greiter-07prb184441,greiter10np5}.

One of the objectives of this work is to identify a parent Hamiltonian
for which this state is the exact ground state, and hence accomplish
what Haldane and Shastry have accomplished for the spin %$S=\frac{1}{2}$
one-half Gutzwiller wave function.

\subsection{Symmetries}
\label{sec:na:sym}

%\vspace{\baselineskip} 
%\emph{Ground state momentum.}---%
\emph{Translational invariance.}---%
As for the Haldane--Shastry model, we obtain the ground state momentum
$p_0$ (in units of inverse lattice spacings $1/a$) by translating
the ground state by one lattice spacing around the unit circle,
\begin{equation}
  \label{eq:nat0}
  \boldsymbol{T} \ket{\psi^{S=1}_{0}} 
  = e^{\text{i}p_0} \ket{\psi^{S=1}_{0}}.
\end{equation}
With $\boldsymbol{T} z_i=\exp\left(\text{i}\frac{2\pi}{N}\right) z_i$ we find
\begin{equation}
  \label{eq:nagp00}
  p_0=\frac{2\pi}{N} \left(-\frac{N}{2}+\frac{N(N-1)}{2}+N\right) 
  = \pi N, 
\end{equation}
which implies $p_0=0$ as $N$ is even.

\vspace{\baselineskip} 
\emph{Invariance under SU(2) spin rotations.}---%
The proof of the singlet property is similar to the Haldane--Shastry 
model, but more instructive as it motivates the  
re-normalization of the spin-flip operators in \eqref{eq:naspinflip}.

% The proof of the singlet property is particularly instructive,
% as it also motivates the necessity for the re-normalization of the
% spin-flip operators \eqref{eq:naspinflip}.

Since ${S}^\z _\text{tot}\ket{\psi^{S=1}_0}=0$ by construction, it is
sufficient to show ${S}^-_\text{tot}\ket{\psi^{S=1}_0}=0$.  Note first that
when we substitute \eqref{eq:napsi0} with \eqref{eq:napfaff} into
\eqref{eq:naket}, we may replace the antisymmetrization $\mathcal{A}$
in \eqref{eq:napfaff} by an overall normalization factor, as it is
taken care by the commutativity of the bosonic operators
$\tilde{S}_{\alpha}$.  Let $\tilde\psi_0$ be $\psi^{S=1}_0$ without the
%operator $\mathcal{A}$ 
antisymmetrization
in \eqref{eq:napfaff},
\begin{equation}
  \label{eq:napsitilde}
  \tilde\psi_0[z_i]
%  \tilde\psi_0(z_1,\ldots ,z_N) 
  =(N-1)!!\,\left\{
    \frac{1}{z_1-z_2}\cdot\,\ldots\,\cdot\frac{1}{z_{N-1}-z_{N}}
  \right\}.
  \prod_{i<j}^{N}(z_i-z_j)\prod_{i=1}^{N}\,z_i.
\end{equation}
%
%We hence may replace \eqref{eq:nacsl} by
%\begin{equation}
%  \label{eq:psitilde}
%  \tilde\psi_0[z_i]
%  =\frac{1}{z_1\!-\!z_2}\,\frac{1}{z_3\!-\!z_4}\ldots\,
%  \prod_{i<j}^{N}\!(z_i-z_j)
%  \prod_{i=1}^{N}\!G(z_i)\,e^{-\frac{\pi}{2}|z_i|^2}.
%\end{equation}
Since $\tilde\psi_0(z_1,z_2,\ldots ,z_N)$ 
%$\tilde\psi_0[z_i]$ 
is still symmetric under interchange of pairs, we may assume that a
spin flip operator ${S}^-_\alpha$ acting on $\textket{\tilde\psi_0}$
will act on the pair $(z_1,z_2)$,
\begin{align}
  \label{eq:naSalphapsi}
    {S}^-_\alpha\bigket{\psi^{S=1}_0}&=\hspace{-13pt}
    \sum_{\{z_3,\dots,z_{N}\}}
    \biggl\{\sum_{z_2 (\ne \eta_\alpha)}
    \tilde\psi_0(\eta_\alpha,z_2,z_3,\dots)
    \,S_\alpha^-\, \tilde{S}_\alpha^+\tilde{S}_{z_2}^+\nonumber\\[2pt]
    &\quad \hspace{25pt}
    +\sum_{z_1 (\ne \eta_\alpha)}
    \tilde\psi_0(z_1,\eta_\alpha,z_3,\dots)
    \,S_\alpha^-\, \tilde{S}_{z_1}^+\tilde{S}_\alpha^+\nonumber\\
    &\quad \hspace{54pt}
    +\ \tilde\psi_0(\eta_\alpha,\eta_\alpha,z_3,\dots)
    \,S_\alpha^-\, (\tilde{S}_\alpha^+)^2\biggr\}
    \tilde{S}_{z_3}^+\dots\tilde{S}_{z_N}^+\ket{-1}_N\nonumber\\[5pt]
    &=\hspace{-13pt}
    \sum_{\{z_3,\dots,z_{N}\}}
    \biggl\{\sum_{z_2}
    2\tilde\psi_0(\eta_\alpha,z_2,z_3,\dots)\,
    \tilde{S}_{z_2}^+\biggr\}\tilde{S}_{z_3}^+\dots\tilde{S}_{z_N}^+\ket{-1}_N,
\end{align}
where we have used 
%$S_\alpha^-\,(\tilde{S}_\alpha^+)^n \ket{1,-1}_{\alpha}
%=n\, (\tilde{S}_\alpha^+)^{n-1}\ket{1,-1}_{\alpha}$.
\begin{equation}
  \label{eq:s-s+tilde}%\nonumber
  S_\alpha^-\,(\tilde{S}_\alpha^+)^n \ket{1,-1}_{\alpha}
  =n\, (\tilde{S}_\alpha^+)^{n-1}\ket{1,-1}_{\alpha},
\end{equation}
which follows directly form the definition \eqref{eq:naspinflip}.
% Note that we have implicitly assumed that each spin configuration in
% the sum over ${\{z_1,z_2,\ldots ,z_N\}}$ in \eqref{eq:hsket} appears
% only once (and not $N!$ times due to permutations of the $z_i$'s).

This implies 
\begin{align}
  \label{eq:naSpsi}
    {S}^-_\text{tot}\ket{\psi^{S=1}_0}
%    \hspace{-4pt}&=\hspace{-4pt}
    &= \sum_{\alpha=1}^N{S}^-_\alpha\bigket{\psi^{S=1}_0}\nonumber\\
    &= %\hspace{-10pt}
    2\sum_{\{z_2\dots,z_{N}\}}
    \underbrace{\sum_{\alpha=1}^N\tilde\psi_0(\eta_\alpha,z_2,\dots,z_{N})}_{=0}
    \,\tilde{S}_{z_2}^+\dots\tilde{S}_{z_N}^+\ket{-1}_N, %=0
\end{align}
since $\tilde\psi_{0}(\eta_\alpha,z_2,\ldots,z_N)$ contains
only powers $\eta_\alpha^1, \eta_\alpha^2,\ldots , \eta_\alpha^{N-1}$
in $\eta_\alpha$ and
\begin{equation*}
  \sum_{\alpha=1}^{N} \eta_\alpha^m =  N\delta_{m,0}\quad \text{mod}\ N.
    \label{eq:na-app-hs2}
\end{equation*}
%\newpage

\vspace{\baselineskip} 
\emph{Parity and time reversal invariance.}---%
To show that $\psi_0(z_1,\ldots ,z_N)$ is real, and hence  that
$\ket{\psi^{S=1}_{0}}$ is invariant
under parity, we calculate its complex conjugate,
\begin{align}
  \label{eq:napsistar}
  \Bigl(\psi^{S=1}_0[z]\Bigr)^*
%  \psi_0(z_1,z_2,\ldots ,z_N) %[z_i]
  &=\text{Pf}\left(\frac{1}{\frac{1}{z_{i}}-\frac{1}{z_{j}}}\right)
  \prod_{i<j}^{N}\left(\frac{1}{z_i}-\frac{1}{z_j}\right)
  \prod_{i=1}^{N}\,\frac{1}{z_i}
  \nonumber\\[0.2\baselineskip]
  &= (-1)^{\frac{N}{2}}\,\prod_{i=1}^{N}\,\frac{1}{z_i}\;
  (-1)^{\frac{N(N-1)}{2}}\,\prod_{i<j}^{N}\,\frac{1}{z_iz_j}\;
  \prod_{i=1}^{N}\,\frac{1}{z_i^2}\; \psi^{S=1}_0[z]
  \nonumber\\[0.2\baselineskip]
   &= \psi^{S=1}_0[z], 
\end{align}
as $N$ is even and $z_i^N=1$ for all $i$.  
Time reversal~\cite{gottfried66} transforms 
% $\text{i}\to -\text{i}$, $z_i\to\bar z_i$, $\bs{S}_\alpha\to -\bs{S}_\alpha$, and
% \begin{equation*}
%   \ket{s,m}\to \text{i}^{2m}\ket{s,-m}.
% \end{equation*}
\begin{equation*}
  \text{i}\to -\text{i},\quad 
  z_i\to\bar z_i,\quad
  \bs{S}_\alpha\to -\bs{S}_\alpha,\quad 
  \ket{s,m}\to \text{i}^{2m}\ket{s,-m},
\end{equation*}
which implies that the basis states in \eqref{eq:naket} transform according to
\begin{equation}
  \label{eq:naketT}
  \tilde{S}_{z_1}^{+}\cdot\dots\cdot\tilde{S}_{z_{N}}^{+}\ket{-1}_N
  \to\tilde{S}_{z_1}^{-}\cdot\dots\cdot\tilde{S}_{z_{N}}^{-}\ket{+1}_N,
\end{equation}
where 
% \begin{equation}
%   \label{eq:naspinflipT}
%   \tilde{S}_{\alpha}^{-} \equiv\frac{-{S}^{\rm{z}}_{\alpha}+1}{2} S_\alpha^{-},
% \end{equation}
% and
% \begin{equation}
%   \label{eq:navacuumketT}
%   \ket{+1}_N\equiv\otimes_{\alpha=1}^N \ket{1,+1}_{\alpha}.
% \end{equation}
\begin{equation}
  \label{eq:naspinflipT}
  \tilde{S}_{\alpha}^{-} \equiv\frac{-{S}^{\rm{z}}_{\alpha}+1}{2} S_\alpha^{-},
  \quad \ket{+1}_N\equiv\otimes_{\alpha=1}^N \ket{1,+1}_{\alpha}.
\end{equation}
Together with the singlet property, this implies that
$\ket{\psi^{S=1}_{0}}$ is invariant under time reversal.

All the symmetries properties discussed here will emerge almost trivially
when we generate the state $\ket{\psi^{S=1}_{0}}$ through projection form 
Gutzwiller (or Haldane--Shastry ground) states in Section \ref{sec:napro}.

\subsection{Schwinger bosons}
\label{sec:naschwinger}

Schwinger bosons~\cite{schwinger65proc,Auerbach94} constitute a way
to formulate spin-$S$ representations of an SU(2) algebra (which can
easily be generalized to SU($n$), see \eg~\cite{greiter-07prb184441}).
The spin operators
\begin{equation}
  \label{eq:naSschwb}
  \bs{S}=\frac{1}{2}\left(a^\dagger,b^\dagger\right)\,\bs{\sigma}\!\left(\!\!
    \begin{array}{c}
      \,a\, \\[2pt] b
    \end{array}\!\!\right)\!,
\end{equation}
where $\bs{\sigma}=(\sigma_\x,\sigma_\y,\sigma_\z )$ is the vector consisting of
the three Pauli matrices \eqref{eq:qhsPauli}, %(see \eqref{eq:qhsPauli}),
are given in terms of boson creation and annihilation
%where $a^\dagger,b^\dagger$ (and $a,b$) are boson creation (and annihilation)
operators which obey the usual commutation relations
% \begin{equation}
%   \begin{array}{c}
%     \comm{a}{a^\dagger}=\comm{b}{b^\dagger}=1, %\\
%     \quad
%     \comm{a^{\phantom{\dagger}}\!\!}{b}=\comm{a}{b^\dagger}
% %    =\comm{a^\dagger}{b}=\comm{a^\dagger}{b^\dagger}
%     =0\rule{0pt}{16pt}.
%   \end{array}
%   \label{eq:schwb}
% \end{equation}
\begin{equation}
  \begin{array}{c}
    \comm{a}{a^\dagger}=\comm{b}{b^\dagger}=1,\\
    \comm{a^{\phantom{\dagger}}\!\!}{b}=\comm{a}{b^\dagger}
    =\comm{a^\dagger}{b}=\comm{a^\dagger}{b^\dagger}
    =0\rule{0pt}{16pt}.
  \end{array}
  \label{eq:schwb}
\end{equation}
It is readily verified with 
\begin{equation}
  \label{eq:naPauliComm}
  \comm{\sigma_i}{\sigma_j}=2\text{i}\varepsilon^{ijk}\sigma_k  
%  \quad\hbox{where}\ i,j,k=x,y,\hbox{or}\ z, 
\end{equation}
and (\ref{eq:schwb}), that $S^\x$, $S^\y$, and
$S^\z $ satisfy the SU(2) algebra %\eqref{eq:su2algebra}.
\begin{equation}
  \label{eq:nasu2}
  \comm{S^i}{S^j}=\text{i}\varepsilon^{ijk}S^k. 
\end{equation}
Written out in components we have
%The components of $\bs{S}$ are
\begin{equation}
\begin{array}{r@{\hspace{5pt}}c@{\hspace{5pt}}c@{\hspace{5pt}}c@{\hspace{5pt}}l}
  S^\x + \text{i}S^\y &=& S^+ &=& a^\dagger b, \\\rule{0pt}{12pt} 
  S^\x - \text{i}S^\y &=& S^- &=& b^\dagger a, \\\rule{0pt}{12pt}
  && S^\z   &=& \frac{1}{2}(a^\dagger a - b^\dagger b). 
  \label{eq:schw}
\end{array}
\end{equation}
The spin quantum number $S$ is given by half the number of bosons,  
\begin{equation}
2S=a^\dagger a + b^\dagger b,
\label{eq:schwt}
\end{equation}
and the usual spin states (simultaneous eigenstates of $\boldsymbol{S}^2$
and $S^\z $) are given by
\begin{equation}
\ket{S,m} = \frac{(a^\dagger)^{S+m}}{\sqrt{(S+m)!}} \frac{(b^\dagger)^{S-m}}
{\sqrt{(S-m)!}} \vac.
\label{eq:schs}
\end{equation}
In particular, the spin-\half states are given by
\begin{equation}
\ket{\up}=c_{\up}^\dagger \vac =a^\dagger \vac ,\qquad 
\ket{\dw}=c_{\dw}^\dagger \vac =b^\dagger \vac ,
\label{eq:schwfun}
\end{equation}
%\begin{equation}
%\begin{array}{rcccl}
%\ket{\up} &= a^\dagger \vac &= c_{\up}^\dagger \vac \\ \rule{0pt}{14pt}
%\ket{\dw} &= b^\dagger \vac &= c_{\dw}^\dagger \vac ,
%\label{eq:schwh}
%\end{array}
%\end{equation}
\ie $a^\dagger$ and $b^\dagger$ act just like the fermion creation
operators $c^\dagger_\up$ and $c^\dagger_\dw$ in this case.  The
difference shows up only when two (or more) creation operators act on
the same site or orbital.  The fermion operators create an
antisymmetric or singlet configuration (in accordance with the Pauli
principle), 
\begin{equation}
\ket{0,0} = c_{\up}^\dagger c_{\dw}^\dagger \vac ,
\label{eq:schwfer}
\end{equation}
while the Schwinger bosons create a totally symmetric or
triplet (or higher spin if we create more than two bosons) configuration,
\begin{align}
  \label{eq:nas=1schw}
  \ket{1,1} &= \textstyle{\frac{1}{\sqrt{2}}} (a^\dagger)^{2}\vac 
  ,\nonumber\\[0.2\baselineskip]
  \ket{1,0} &= a^\dagger b^\dagger \vac , \\[0.2\baselineskip]
  \ket{1,-1} &= \textstyle{\frac{1}{\sqrt{2}}} (b^\dagger)^{2}\vac 
  .\nonumber
\end{align}
Representations of spin \half states in terms of Schwinger bosons
(rather than fermion creation operators or spin flips) are ideally
suited for the construction of higher spin states through projection
of $2S$ spin $\frac{1}{2}$'s onto the 
spin $S$ representations (\ie the symmetric representation) contained in
\begin{equation}
  \label{eq:nahalftimeshalftimes}
  \underbrace{\bs{\textstyle\frac{1}{2}}\otimes\bs{\textstyle\frac{1}{2}}
    \otimes\ldots\otimes\bs{\textstyle\frac{1}{2}}}
%  _{2S\ \text{spin} \frac{1}{2}\text{'s }} 
  _{2S} 
  = \bs{S} \oplus (2S-1)\,\bs{S\!-\!1} \oplus \ldots 
\end{equation}
Classic examples include the formulation of the
Affleck--Kennedy--Lieb--Tasaki (AKLT)
model~\cite{affleck-87prl799,affleck-88cmp477} in terms of Schwinger
bosons~\cite{arovas-88prl531,Auerbach94} as well as the $S=1$ chirality 
liquid~\cite{greiter02jltp1029}.

%\vspace{100pt}

%\subsection{Generation by projection from Haldane--Shastry or Gutzwiller states}
\subsection{Generation by projection from Gutzwiller states}
\label{sec:napro}

We will show now that the $S=1$ ground state \eqref{eq:napsi0} can
alternatively be generated by considering two (identical)
Haldane--Shastry or Gutzwiller states \eqref{eq:hspsi0} and 
projecting onto the triplet or $S=1$ configuration 
contained in
\begin{equation}
  \label{eq:nahalftimeshalf}
  \textstyle \bs{\frac{1}{2}}\otimes\bs{\frac{1}{2}} = \bs{0}\oplus \bs{1}
\end{equation}
at each site~\cite{greiter02jltp1029,greiter-09prl207203}.  To begin
with, we rewrite \eqref{eq:hsket}
% $\ket{\psi^{\s\text{HS}}_{0}}$
in terms of Schwinger bosons,
\begin{align}
  \label{eq:nahsket}
  \ket{\psi^{\s\text{HS}}_{0}}
  &=\sum_{\{z_1,z_2,\ldots ,z_M\}}
%  \psi^{\s\text{HS}}_{0}(z_1,\ldots ,z_M)\;
  \psi^{\s\text{HS}}_{0}[z]\,
  {S}^+_{z_1}\cdot\ldots\cdot {S}^+_{z_M} 
%  \big|\underbrace{\dw\dw\ldots\ldots\dw}_{\text{all\ } N \text{\ spins\ } \dw}
%  \big\rangle
  \ket{\dw\dw\ldots\dw} 
  \nonumber\\
  &=\sum_{\{z_1,\ldots ,z_M;w_1,\ldots,w_M\}}
%  \psi^{\s\text{HS}}_{0}(z_1,\ldots ,z_M)\;
%  \underbrace{\psi^{\s\text{HS}}_{0}[z]\;
  \psi^{\s\text{HS}}_{0}[z]\,
    {a}^+_{z_1}\ldots a^\dagger_{z_M}
    {b}^+_{w_1}\ldots b^\dagger_{w_M}%
%  }_{\displaystyle \equiv \Psi^{\s\text{HS}}_{0}[a^\dagger;b^\dagger] } 
  \vac\!\nonumber\\[5pt]
  &\equiv \Psi^{\s\text{HS}}_{0}[a^\dagger,b^\dagger] \vac\!,
\end{align}
where $M=\frac{N}{2}$ and the $w_k$'s are those lattice sites which are not
occupied by any of the $z_i$'s.  The $S=1$ state \eqref{eq:napsi0} is
then up to an overall normalization factor given by
\begin{equation}
  \label{eq:naop}
  \ket{\psi^{S=1}_0}=
  \Big(\Psi^{\s\text{HS}}_0\big[a^\dagger ,b^\dagger\big]\Big)^2\vac.
\end{equation}

%\vspace{\baselineskip} 
% \emph{Verification of \eqref{eq:naop}.}---%
% The proof is straightforward if we use the identity
To verify \eqref{eq:naop}, use the identity 
\begin{equation}
  \label{eq:na220=Pf} %\nonumber
  \mathcal{S}\Biggl\{
  \prod_{\substack{i,j=1\\[1pt] i<j}}^{M}\hspace{0pt}(z_i-z_j)^2\hspace{-2pt}
  \prod_{\substack{i,j=M+1\\[1pt] i<j}}^{2M}\hspace{-2pt}(z_i-z_j)^2\Biggr\}
  =\text{Pf}\left(\frac{1}{z_i-z_j}\right)
%  \prod_{\substack{i,j=1\\[1pt] i<j}}^{2M}\hspace{-8pt}(z_i-z_j)^2
  \prod_{i<j}^{2M}(z_i-z_j),
\end{equation}
where $\mathcal{S}$ indicates symmetrization over all the variables in
the curly brackets, and
\begin{equation}
  \label{eq:naStilde^n}
  \frac{1}{\sqrt{2}}(a^\dagger)^n (b^\dagger)^{(2-n)}\vac 
  =(\tilde{S}^+)^n\ket{1,-1},
\end{equation}
which is readily verified with \eqref{eq:schw}, \eqref{eq:nas=1schw}, 
and the definition \eqref{eq:naspinflip}.
To proof \eqref{eq:na220=Pf}, use the following identity due to
%Cauchy~\cite{MacDonald79},
Frobenius~\cite{frobenius-1882ram53},
\begin{equation}
  \label{eq:nacauchy}
  \det\left(\frac{1}{z_i-z_{M+j}}\right)
    =(-1)^{\textstyle\frac{M(M+1)}{2}}\,  
    \frac{\displaystyle 
%      \prod_{\substack{i,j=1\\[1pt] i<j}}^{M}(z_i-z_j)^2(z_{M+i}-z_{M+j})^2}{
      \prod_{\substack{i,j=1\\[1pt] i<j}}^{M}(z_i-z_j)
      \prod_{\substack{i,j=M+1\\[1pt] i<j}}^{2M}(z_i-z_j)
%      \prod_{i<j}^{M}(z_i-z_j)^2(z_{M+i}-z_{M+j})^2}{
    }{\displaystyle \prod_{i=1}^M\,\prod_{j=M+1}^{2M}(z_i-z_j)}.
\end{equation}
%\hfill $\Box$

%\vspace{\baselineskip} 
%\emph{Implication of \eqref{eq:naop}.}---%
The projective construction directly reveals several interesting features,
which were not nearly as obvious in the previous formulation:
\renewcommand{\theenumi}{\alph{enumi}}
\renewcommand{\labelenumi}{(\theenumi)}
% set the indent for the list:
\setlength{\svitemindent}{2\parindent}
\begin{enumerate}
\item Since the Haldane--Shastry ground state
  $\ket{\psi^{\s\text{HS}}_{0}}$ is translationally invariant with
  ground state momentum $p_0=0$ or $\pi$ (depending on whether
  $\frac{N}{2}$ is even or odd), the $S=1$ state $\ket{\psi^{S=1}_{0}}$
  is translationally invariant with $p_0=0$.\label{LA}
\item Since $\ket{\psi^{\s\text{HS}}_{0}}$ is a singlet, and the
  projection onto spin $S=1$ on each site commutes with spin
  rotations, $\ket{\psi^{S=1}_{0}}$ has to be a singlet as
  well.
\item Since %the Haldane--Shastry ground state wave function
  $\psi^{\s\text{HS}}_{0}(z_1,\ldots ,z_M)$ is real with the sign of
  each spin configuration given by 
   $\prod_{i=1}^M\,G(z_i)$,
%   \begin{equation*}
%     \prod_{i=1}^M\,G(z_i),
%   \end{equation*}
   the $S=1$ wave function $\psi^{S=1}_{0}(z_1,\ldots ,z_M)$ is
   likewise real with the sign given by $\prod_{i=1}^N\,G(z_i)$:
%  \begin{equation*}
%    \prod_{i=1}^N\,G(z_i).
%  \end{equation*}
%  We may hence write
  \begin{equation}
    \label{eq:napsi0real}
    \psi^{S=1}_0(z_1,\ldots ,z_N) %[z_i]
    =\Biggl|\,\text{Pf}\left(\frac{1}{z_{i}-z_{j}}\right)
    \prod_{i<j}^{N}(z_i-z_j)\,\Biggr|\,\prod_{i=1}^{N}\,G(z_i),
  \end{equation}
  with $G(\eta_\alpha)=\pm 1$ depending on whether $\alpha$ even or odd.
%    with
%   \begin{equation}
%     G(\eta_\alpha)=
%     % (\eta_\alpha)^\frac{N}{2}=
%     \left\{\begin{array}{ll} 
%         +1 &\quad \alpha\ \text{even}\\
%         -1 &\quad \alpha\ \text{odd}.
%       \end{array}\right.
%     \label{eq:napsi0g}
%   \end{equation}
\item Since $\ket{\psi^{\s\text{HS}}_{0}}$ is invariant under parity
  and and time reversal, $\ket{\psi^{S=1}_{0}}$ is invariant as well.
\end{enumerate}
\renewcommand{\theenumi}{\arabic{enumi}}
\renewcommand{\labelenumi}{\theenumi.}

%\newpage
\subsection{Topological degeneracies and non-Abelian statistics}
\label{sec:nana}

We have seen in Section \ref{sec:3mod-pf-nastat} that $2n$ spatially
well separated quasiparticle excitations or vortices carrying half of
a Dirac flux quanta each in the non-Abelian quantized Hall state
described by the Pfaffian will span an internal or topological Hilbert
space of dimensions $2^n$ ($2^{n-1}$ for either even or odd fermion
numbers), in accordance with the existence of one Majorana fermion
state at each vortex core.  The Majorana fermion
states can only be manipulated through braiding of the vortices, with
the interchanges being non-commutative or non-Abelian.

% For realistic interactions, the $2^L$ fold degeneracy associated with
% the internal Hilbert space becomes exact for infinitely separated
% vortices in an infinite system.

The question we wish to address in this section is whether there is
any manifestation of this topological space of dimension $2^n$, or the
$2n$ Majorana fermion states, in the spinon excitation Hilbert space
suggested by the $S=1$ ground state \eqref{eq:napsi0}.  In Section
\ref{sec:hsspinons}, we have seen that the fractional statistics of
the spinons in the Haldane--Shastry model, and presumably in any model
supporting one-dimensional anyons, is encoded in the momentum spacings
of the excitations.  This is not too surprising, as there are no other
suitable quantum numbers, like the relative angular momentum for
two-dimensional anyons, available.  We will propose now that the
topological degeneracies, or the occupation numbers of the $n$
fermions consisting of the $2n$ Majorana fermions, are once again
encoded in the momentum spacings between single spinon states.

In the Haldane--Shastry model, the spacings between neighboring
momenta were always half integer, in accordance with half-fermi
statistics, as the difference between consecutive spinon momentum
numbers  $a_i$ was always an odd integer,
\begin{equation}
  \label{eq:nahsspacings}
%  \text{Haldane--Shastry:}\qquad  
%  a_{i+1}-a_{i}=\text{odd integer}.
  a_{i+1}-a_{i}\,=\,\text{odd}.
\end{equation}
This follows directly from the construction of the extended Young
tableaux illustrated in Fig.\ \ref{fig:hsfoursitesu2}.  When two spinons
are in neighboring columns, the difference of the $a_i$ is one and
hence an odd integer; when we insert complete columns without spinons in
between, the number of boxes we insert is always even.

We will now show that for the $S=1$ chain with the Hilbert space
parameterized by the ground state $\ket{\psi^{S=1}_0}$ and spinon
excitations above it, the corresponding rule is
\begin{equation}
   \label{eq:naspacings}
   \begin{array}{rcl@{\hspace{20pt}}l}
     a_{i+1}-a_{i}&=&\text{even or odd,} &\text{for}\ i\ \text{odd},
     \\[2pt]
     a_{i+1}-a_{i}&=&\text{odd,}         &\text{for}\ i\ \text{even}.
   \end{array}
\end{equation}
%As $i$ runs from 1 to $2n$, 
As $i=1,2,\ldots,2n$, we have a total of $n$ spacings which can be
either even or odd, and another $n$ spacings which are always odd.
With the single spinon momenta given by
\begin{equation}
  \label{eq:nasinglespinonmom}
  p_i=\frac{\pi}{N}\,\left(a_i-\frac{1}{2}\right),
\end{equation}
this yields momentum spacings which can be either an integer or
an half-integer times $\frac{2\pi}{N}$ for $i$ odd.  This is a
topological distinction---for Abelian anyons, one choice corresponds
to bosons or fermions (which are for many purposes equivalent in one
dimension), and the other choice to half fermions.  For spinons which
are well separated in momentum space, the states spanning this in
total $2^n$ dimensional topological Hilbert space become degenerate as
we approach the thermodynamic limit.

\newcommand{\yt}[3]{\put(#1,#2){\framebox(0.94,0.94){#3}}}
\newcommand{\ye}[3]{\put(#1,#2){\makebox(0.94,0.94){#3}}}
\newcommand{\yd}[2]{\put(#1,#2){\makebox(1.38,0.92){\circle*{0.4}}}}
\begin{figure}[tb]
  \begin{center}
    \setlength{\unitlength}{\ytlength}
    \begin{picture}(0.3,0)(-1,9.4) \linethickness{0.5pt}
      \ye{-0.5}{14.5}{$=$}
      \yt{1}{15}{1}
      \yt{2}{15}{1}
      \yt{1}{14}{2}
      \yt{2}{14}{2}
      \ye{3.5}{14.5}{$\oplus$}
      \yt{5}{15}{1}
      \yd{5}{14}
      \yt{6}{15}{1}
      \yt{6}{14}{2}
      \yt{7}{15}{2}
      \yd{7}{14}
      \ye{1.5}{11.5}{$\oplus$}
      \yt{3}{12}{1}
      \yd{3}{11}
      \yt{4}{12}{1}
      \yd{4}{11}
      \yt{5}{12}{2}
      \yd{5}{11}
      \yt{6}{12}{2}
      \yd{6}{11}
     \end{picture}
    \begin{picture}(36,10.5)(1,9) \linethickness{0.5pt}
      \yt{1}{18}{1}
      \yt{2}{18}{1}
      \ye{3.5}{18}{$\otimes$}
      \yt{5}{18}{2}
      \yt{6}{18}{2}
      \ye{7.5}{18}{$\otimes$}
      \yt{9}{18}{3}
      \yt{10}{18}{3}
      \ye{12}{18}{$=$}
%      \ye{1.5}{16.5}{$\bs{1}$}
      \put(4,17){\makebox(0,0){$\underbrace{\phantom{\hspace{54pt}}}$}}
      \yt{14}{18}{1}
      \yt{15}{18}{1}
      \yt{14}{17}{2}
      \yt{15}{17}{2}
      \yt{16}{18}{3}
      \yt{17}{18}{3}
      \yd{16}{17}
      \yd{17}{17}
      \ye{15.5}{15.3}{$S=1$}
      \ye{18.5}{17.5}{$\oplus$}
      \yt{20}{18}{1}
      \yd{20}{17}
      \yt{21}{18}{1}
      \yt{21}{17}{2}
      \yt{22}{18}{2}
      \yt{22}{17}{3}
      \yt{23}{17}{3}
      \yd{23}{18}
      \ye{21.5}{15.3}{$S=0$}
      \ye{24.5}{17.5}{$\oplus$}
      \yt{26}{18}{1}
      \yd{26}{17}
      \yt{27}{18}{1}
      \yt{27}{17}{2}
      \yt{28}{18}{2}
      \yt{28}{17}{3}
      \yt{29}{18}{3}
      \yd{29}{17}
      \ye{27.5}{15.3}{$S=1$}
      \ye{30.5}{17.5}{$\oplus$}
      \yt{32}{18}{1}
      \yd{32}{17}
      \yt{33}{18}{1}
      \yt{33}{17}{2}
      \yt{34}{18}{2}
      \yd{34}{17}
      \yt{35}{18}{3}
      \yd{35}{17}
      \yt{36}{18}{3}
      \yd{36}{17}
      \ye{34}{15.3}{$S=2$}
      \ye{14.5}{11.5}{$\oplus$}
      \yt{16}{12}{1}
      \yd{16}{11}
      \yt{17}{12}{1}
      \yd{17}{11}
      \yt{18}{12}{2}
      \yt{18}{11}{3}
      \yt{19}{12}{2}
      \yt{19}{11}{3}
      \ye{17.5}{9.3}{$S=1$}
      \ye{20.5}{11.5}{$\oplus$}
      \yt{22}{12}{1}
      \yd{22}{11}
      \yt{23}{12}{1}
      \yd{23}{11}
      \yt{24}{12}{2}
      \yd{24}{11}
      \yt{25}{12}{2}
      \yt{25}{11}{3}
      \yt{26}{12}{3}
      \yd{26}{11}
      \ye{24}{9.3}{$S=2$}
      \ye{27.5}{11.5}{$\oplus$}
      \yt{29}{12}{1}
      \yd{29}{11}
      \yt{30}{12}{1}
      \yd{30}{11}
      \yt{31}{12}{2}
      \yd{31}{11}
      \yt{32}{12}{2}
      \yd{32}{11}
      \yt{33}{12}{3}
      \yd{33}{11}
      \yt{34}{12}{3}
      \yd{34}{11}
      \ye{31.5}{9.3}{$S=3$}
    \end{picture}
    \caption{Total spin representations of three $S=1$ spins
      in terms of extended Young tableaux.}
    \label{fig:nayoungdiagram}
  \end{center}
\end{figure}

\begin{figure}[tb]
  \begin{center}
    \setlength{\unitlength}{\ytlength}
% left row we start S=0
    \begin{picture}(19,4)(0,0) \linethickness{0.5pt}
      \ye{8.5}{4}{$S_{\text{tot}}$}
      \ye{12.5}{4}{$a_1,\dots,a_L$}
      \ye{16.5}{4}{$p_{\text{tot}}$}
      \yt{1}{2}{1}
      \yt{1}{1}{2}
      \yt{2}{2}{1}
      \yt{2}{1}{2}
      \yt{3}{2}{3}
      \yt{3}{1}{4}
      \yt{4}{2}{3}
      \yt{4}{1}{4}
      \ye{8.5}{1.5}{$0$}
      \put(10.5,2){\line(1,0){5}} 
      \multiput(11.5,1.85)(1,0){4}{\rule{0.5pt}{2pt}}
      \ye{16.5}{1.5}{$0$}
    \end{picture}
    \begin{picture}(19,5)(0,0) \linethickness{0.5pt}
      \ye{8.5}{4}{$S_{\text{tot}}$}
      \ye{12.5}{4}{$a_1,\dots,a_L$}
      \ye{16.5}{4}{$p_{\text{tot}}$}
      \yt{1}{2}{1}
      \yt{1}{1}{2}
      \yt{2}{2}{1}
      \yt{2}{1}{2}
      \yt{3}{2}{3}
      \yd{3}{1}
      \yt{4}{2}{3}
      \yd{4}{1}
      \yt{5}{2}{4}
      \yd{5}{1}
      \yt{6}{2}{4}
      \yd{6}{1}
      \ye{8.5}{1.5}{$2$}
      \put(10.5,2){\line(1,0){5}} 
      \multiput(11.5,1.85)(1,0){4}{\rule{0.5pt}{2pt}}
      \multiput(13.5,2)(1,0){2}{\circle*{0.5}}
      \multiput(13.5,2.6)(1,0){2}{\circle*{0.5}}
      \ye{13}{0.5}{3}
      \ye{14}{0.5}{4}
      \ye{16.5}{1.5}{$\pi$}
    \end{picture}
    \begin{picture}(19,3.5)(0,0) \linethickness{0.5pt}
      \yt{1}{2}{1}
      \yd{1}{1}
      \yt{2}{2}{1}
      \yt{2}{1}{2}
      \yt{3}{2}{2}
      \yt{3}{1}{3}
      \yt{4}{2}{3}
      \yt{4}{1}{4}
      \yd{5}{2}
      \yt{5}{1}{4}
      \ye{8.5}{1.5}{$0$}
      \put(10.5,2){\line(1,0){5}} 
      \multiput(11.5,1.85)(1,0){4}{\rule{0.5pt}{2pt}}
      \multiput(11.5,2)(3,0){2}{\circle*{0.5}}
%      \multiput(11.5,2.6)(3,0){2}{\circle*{0.5}}
      \ye{11}{0.5}{1}
      \ye{14}{0.5}{4}
      \ye{16.5}{1.5}{$\pi$}
    \end{picture}
    \begin{picture}(19,3.5)(0,0) \linethickness{0.5pt}
      \yt{1}{2}{1}
      \yd{1}{1}
      \yt{2}{2}{1}
      \yt{2}{1}{2}
      \yt{3}{2}{2}
      \yt{3}{1}{3}
      \yt{4}{2}{3}
      \yd{4}{1}
      \yt{5}{2}{4}
      \yd{5}{1}
      \yt{6}{2}{4}
      \yd{6}{1}
      \ye{8.5}{1.5}{$2$}
      \put(10.5,2){\line(1,0){5}} 
      \multiput(11.5,1.85)(1,0){4}{\rule{0.5pt}{2pt}}
      \multiput(11.5,2)(1,0){1}{\circle*{0.5}}
      \multiput(13.5,2)(1,0){2}{\circle*{0.5}}
      \multiput(14.5,2.6)(3,0){1}{\circle*{0.5}}
      \ye{11}{0.5}{1}
      \ye{13}{0.5}{3}
      \ye{14}{0.5}{4}
      \ye{16.5}{1.5}{$\frac{\pi}{2}$}
    \end{picture}
    \begin{picture}(19,3.5)(0,0) \linethickness{0.5pt}
      \yt{1}{2}{1}
      \yd{1}{1}
      \yt{2}{2}{1}
      \yd{2}{1}
      \yt{3}{2}{2}
      \yt{3}{1}{3}
      \yt{4}{2}{2}
      \yt{4}{1}{3}
      \yd{5}{2}
      \yt{5}{1}{4}
      \yd{6}{2}
      \yt{6}{1}{4}
      \ye{8.5}{1.5}{$0$}
      \put(10.5,2){\line(1,0){5}} 
      \multiput(11.5,1.85)(1,0){4}{\rule{0.5pt}{2pt}}
      \multiput(11.5,2)(3,0){2}{\circle*{0.5}}
      \multiput(11.5,2.6)(3,0){2}{\circle*{0.5}}
      \ye{11}{0.5}{1}
      \ye{14}{0.5}{4}
      \ye{16.5}{1.5}{$0$}
    \end{picture}
    \begin{picture}(19,3.5)(0,0) \linethickness{0.5pt}
      \yt{1}{2}{1}
      \yd{1}{1}
      \yt{2}{2}{1}
      \yd{2}{1}
      \yt{3}{2}{2}
      \yt{3}{1}{3}
      \yt{4}{2}{2}
      \yt{4}{1}{3}
      \yt{5}{2}{4}
      \yd{5}{1}
      \yt{6}{2}{4}
      \yd{6}{1}
      \ye{8.5}{1.5}{$2$}
      \put(10.5,2){\line(1,0){5}} 
      \multiput(11.5,1.85)(1,0){4}{\rule{0.5pt}{2pt}}
      \multiput(11.5,2)(3,0){2}{\circle*{0.5}}
      \multiput(11.5,2.6)(3,0){2}{\circle*{0.5}}
      \ye{11}{0.5}{1}
      \ye{14}{0.5}{4}
      \ye{16.5}{1.5}{$0$}
    \end{picture}
% left row we start S=1
    \begin{picture}(19,3.5)(0,0) \linethickness{0.5pt}
      \yt{1}{2}{1}
      \yt{1}{1}{2}
      \yt{2}{2}{1}
      \yt{2}{1}{2}
      \yt{3}{2}{3}
      \yd{3}{1}
      \yt{4}{2}{3}
      \yt{4}{1}{4}
      \yt{5}{2}{4}
      \yd{5}{1}
      \ye{8.5}{1.5}{$1$}
      \put(10.5,2){\line(1,0){5}} 
      \multiput(11.5,1.85)(1,0){4}{\rule{0.5pt}{2pt}}
      \multiput(13.5,2)(1,0){2}{\circle*{0.5}}
%      \multiput(13.5,2.6)(1,0){2}{\circle*{0.5}}
      \ye{13}{0.5}{3}
      \ye{14}{0.5}{4}
      \ye{16.5}{1.5}{$\frac{3\pi}{2}$}
    \end{picture}
    \begin{picture}(19,3.5)(0,0) \linethickness{0.5pt}
      \yt{1}{2}{1}
      \yd{1}{1}
      \yt{2}{2}{1}
      \yt{2}{1}{2}
      \yt{3}{2}{2}
      \yd{3}{1}
      \yt{4}{2}{3}
      \yd{4}{1}
      \yt{5}{2}{3}
      \yt{5}{1}{4}
      \yt{6}{2}{4}
      \yd{6}{1}
      \ye{8.5}{1.5}{$2$}
      \put(10.5,2){\line(1,0){5}} 
      \multiput(11.5,1.85)(1,0){4}{\rule{0.5pt}{2pt}}
      \multiput(11.5,2)(1,0){4}{\circle*{0.5}}
%      \multiput(11.5,2.6)(1,0){2}{\circle*{0.5}}
      \ye{11}{0.5}{1}
      \ye{12}{0.5}{2}
      \ye{13}{0.5}{3}
      \ye{14}{0.5}{4}
      \ye{16.5}{1.5}{$0$}
    \end{picture}
    \begin{picture}(19,3.5)(0,0) \linethickness{0.5pt}
      \yt{1}{2}{1}
      \yd{1}{1}
      \yt{2}{2}{1}
      \yt{2}{1}{2}
      \yt{3}{2}{2}
      \yt{3}{1}{3}
      \yt{4}{2}{3}
      \yt{4}{1}{4}
      \yt{5}{2}{4}
      \yd{5}{1}
      \ye{8.5}{1.5}{$1$}
      \put(10.5,2){\line(1,0){5}} 
      \multiput(11.5,1.85)(1,0){4}{\rule{0.5pt}{2pt}}
      \multiput(11.5,2)(3,0){2}{\circle*{0.5}}
%      \multiput(11.5,2.6)(3,0){2}{\circle*{0.5}}
      \ye{11}{0.5}{1}
      \ye{14}{0.5}{4}
      \ye{16.5}{1.5}{$\pi$}
    \end{picture}
    \begin{picture}(19,3.5)(0,0) \linethickness{0.5pt}
      \yt{1}{2}{1}
      \yd{1}{1}
      \yt{2}{2}{1}
      \yd{2}{1}
      \yt{3}{2}{2}
      \yd{3}{1}
      \yt{4}{2}{2}
      \yt{4}{1}{3}
      \yt{5}{2}{3}
      \yt{5}{1}{4}
      \yt{6}{2}{4}
      \yd{6}{1}
      \ye{8.5}{1.5}{$2$}
      \put(10.5,2){\line(1,0){5}}
      \multiput(11.5,1.85)(1,0){4}{\rule{0.5pt}{2pt}}
      \multiput(11.5,2)(1,0){2}{\circle*{0.5}}
      \multiput(11.5,2.6)(1,0){1}{\circle*{0.5}}
      \multiput(14.5,2)(1,0){1}{\circle*{0.5}}
      \put(11,0.5){\makebox(1,1){1}}
      \put(12,0.5){\makebox(1,1){2}}
      \put(14,0.5){\makebox(1,1){4}}
      \ye{16.5}{1.5}{$\frac{3\pi}{2}$}
    \end{picture}
    \begin{picture}(19,3.5)(0,0) \linethickness{0.5pt}
      \yt{1}{2}{1}
      \yd{1}{1}
      \yt{2}{2}{1}
      \yt{2}{1}{2}
      \yt{3}{2}{2}
      \yd{3}{1}
      \yt{4}{2}{3}
      \yt{4}{1}{4}
      \yt{5}{2}{3}
      \yt{5}{1}{4}
      \ye{8.5}{1.5}{$1$}
      \put(10.5,2){\line(1,0){5}} 
      \multiput(11.5,1.85)(1,0){4}{\rule{0.5pt}{2pt}}
      \multiput(11.5,2)(1,0){2}{\circle*{0.5}}
%      \multiput(11.5,2.6)(1,0){2}{\circle*{0.5}}
      \ye{11}{0.5}{1}
      \ye{12}{0.5}{2}
      \ye{13}{0.5}{3}
      \ye{14}{0.5}{4}
      \ye{16.5}{1.5}{$\frac{\pi}{2}$}
    \end{picture}
    \begin{picture}(19,3.5)(0,0) \linethickness{0.5pt}
      \yt{1}{2}{1}
      \yd{1}{1}
      \yt{2}{2}{1}
      \yd{2}{1}
      \yt{3}{2}{2}
      \yd{3}{1}
      \yt{4}{2}{2}
      \yd{4}{1}
      \yt{5}{2}{3}
      \yt{5}{1}{4}
      \yt{6}{2}{3}
      \yt{6}{1}{4}
      \ye{8.5}{1.5}{$2$}
      \put(10.5,2){\line(1,0){5}}
      \multiput(11.5,1.85)(1,0){4}{\rule{0.5pt}{2pt}}
      \multiput(11.5,2)(1,0){2}{\circle*{0.5}}
      \multiput(11.5,2.6)(1,0){2}{\circle*{0.5}}
      \put(11,0.5){\makebox(1,1){1}}
      \put(12,0.5){\makebox(1,1){2}}
%      \put(14,0.5){\makebox(1,1){4}}
      \ye{16.5}{1.5}{$\pi$}
    \end{picture}
% done with S=2 on the right
    \begin{picture}(19,3.5)(0,0) \linethickness{0.5pt}
      \yt{1}{2}{1}
      \yd{1}{1}
      \yt{2}{2}{1}
      \yt{2}{1}{2}
      \yt{3}{2}{2}
      \yt{3}{1}{3}
      \yd{4}{2}
      \yt{4}{1}{3}
      \yt{5}{2}{4}
      \yd{5}{1}
      \yt{6}{2}{4}
      \yd{6}{1}
      \ye{8.5}{1.5}{$1$}
      \put(10.5,2){\line(1,0){5}} 
      \multiput(11.5,1.85)(1,0){4}{\rule{0.5pt}{2pt}}
      \multiput(11.5,2)(1,0){1}{\circle*{0.5}}
      \multiput(13.5,2)(1,0){2}{\circle*{0.5}}
      \multiput(14.5,2.6)(3,0){1}{\circle*{0.5}}
      \ye{11}{0.5}{1}
      \ye{13}{0.5}{3}
      \ye{14}{0.5}{4}
      \ye{16.5}{1.5}{$\frac{\pi}{2}$}
    \end{picture}
    \begin{picture}(19,3.5)(0,0) \linethickness{0.5pt}
      \yt{1}{2}{1}
      \yd{1}{1}
      \yt{2}{2}{1}
      \yt{2}{1}{2}
      \yt{3}{2}{2}
      \yd{3}{1}
      \yt{4}{2}{3}
      \yd{4}{1}
      \yt{5}{2}{3}
      \yd{5}{1}
      \yt{6}{2}{4}
      \yd{6}{1}
      \yt{7}{2}{4}
      \yd{7}{1}
      \ye{9}{1.5}{$3$}
      \put(10.5,2){\line(1,0){5}} 
      \multiput(11.5,1.85)(1,0){4}{\rule{0.5pt}{2pt}}
      \multiput(11.5,2)(1,0){4}{\circle*{0.5}}
      \multiput(13.5,2.6)(1,0){2}{\circle*{0.5}}
      \ye{11}{0.5}{1}
      \ye{12}{0.5}{2}
      \ye{13}{0.5}{3}
      \ye{14}{0.5}{4}
      \ye{16.5}{1.5}{$\frac{3\pi}{2}$}
    \end{picture}
    \begin{picture}(19,3.5)(0,0) \linethickness{0.5pt}
      \yt{1}{2}{1}
      \yd{1}{1}
      \yt{2}{2}{1}
      \yd{2}{1}
      \yt{3}{2}{2}
      \yt{3}{1}{3}
      \yt{4}{2}{2}
      \yt{4}{1}{3}
      \yd{5}{2}
      \yt{5}{1}{4}
      \yt{6}{2}{4}
      \yd{6}{1}
      \ye{8.5}{1.5}{$1$}
      \put(10.5,2){\line(1,0){5}} 
      \multiput(11.5,1.85)(1,0){4}{\rule{0.5pt}{2pt}}
      \multiput(11.5,2)(3,0){2}{\circle*{0.5}}
      \multiput(11.5,2.6)(3,0){2}{\circle*{0.5}}
      \ye{11}{0.5}{1}
      \ye{14}{0.5}{4}
      \ye{16.5}{1.5}{$0$}
    \end{picture}
    \begin{picture}(19,3.5)(0,0) \linethickness{0.5pt}
      \yt{1}{2}{1}
      \yd{1}{1}
      \yt{2}{2}{1}
      \yd{2}{1}
      \yt{3}{2}{2}
      \yd{3}{1}
      \yt{4}{2}{2}
      \yt{4}{1}{3}
      \yt{5}{2}{3}
      \yd{5}{1}
      \yt{6}{2}{4}
      \yd{6}{1}
      \yt{7}{2}{4}
      \yd{7}{1}
      \ye{9}{1.5}{$3$}
      \put(10.5,2){\line(1,0){5}} 
      \multiput(11.5,1.85)(1,0){4}{\rule{0.5pt}{2pt}}
      \multiput(11.5,2)(1,0){4}{\circle*{0.5}}
      \multiput(11.5,2.6)(3,0){2}{\circle*{0.5}}
      \ye{11}{0.5}{1}
      \ye{12}{0.5}{2}
      \ye{13}{0.5}{3}
      \ye{14}{0.5}{4}
      \ye{16.5}{1.5}{$\pi$}
    \end{picture}
    \begin{picture}(19,3.5)(0,0) \linethickness{0.5pt}
      \yt{1}{2}{1}
      \yd{1}{1}
      \yt{2}{2}{1}
      \yd{2}{1}
      \yt{3}{2}{2}
      \yd{3}{1}
      \yt{4}{2}{2}
      \yt{4}{1}{3}
      \yt{5}{2}{3}
      \yt{5}{1}{4}
      \yd{6}{2}
      \yt{6}{1}{4}
      \ye{8.5}{1.5}{$1$}
      \put(10.5,2){\line(1,0){5}}
      \multiput(11.5,1.85)(1,0){4}{\rule{0.5pt}{2pt}}
      \multiput(11.5,2)(1,0){2}{\circle*{0.5}}
      \multiput(11.5,2.6)(1,0){1}{\circle*{0.5}}
      \multiput(14.5,2)(1,0){1}{\circle*{0.5}}
      \put(11,0.5){\makebox(1,1){1}}
      \put(12,0.5){\makebox(1,1){2}}
      \put(14,0.5){\makebox(1,1){4}}
      \ye{16.5}{1.5}{$\frac{3\pi}{2}$}
     \end{picture}
    \begin{picture}(19,3.5)(0,0) \linethickness{0.5pt}
      \yt{1}{2}{1}
      \yd{1}{1}
      \yt{2}{2}{1}
      \yd{2}{1}
      \yt{3}{2}{2}
      \yd{3}{1}
      \yt{4}{2}{2}
      \yd{4}{1}
      \yt{5}{2}{3}
      \yd{5}{1}
      \yt{6}{2}{3}
      \yt{6}{1}{4}
      \yt{7}{2}{4}
      \yd{7}{1}
      \ye{9}{1.5}{$3$}
      \put(10.5,2){\line(1,0){5}} 
      \multiput(11.5,1.85)(1,0){4}{\rule{0.5pt}{2pt}}
      \multiput(11.5,2)(1,0){4}{\circle*{0.5}}
      \multiput(11.5,2.6)(1,0){2}{\circle*{0.5}}
      \ye{11}{0.5}{1}
      \ye{12}{0.5}{2}
      \ye{13}{0.5}{3}
      \ye{14}{0.5}{4}
      \ye{16.5}{1.5}{$\frac{\pi}{2}$}
    \end{picture}
    \begin{picture}(22,3)(0,0.5) \linethickness{0.5pt}
      \yt{1}{2}{1}
      \yd{1}{1}
      \yt{2}{2}{1}
      \yd{2}{1}
      \yt{3}{2}{2}
      \yd{3}{1}
      \yt{4}{2}{2}
      \yd{4}{1}
      \yt{5}{2}{3}
      \yd{5}{1}
      \yt{6}{2}{3}
      \yd{6}{1}
      \yt{7}{2}{4}
      \yd{7}{1}
      \yt{8}{2}{4}
      \yd{8}{1}
      \ye{10.5}{1.5}{$4$}
      \put(12.5,2){\line(1,0){5}} 
      \multiput(13.5,1.85)(1,0){4}{\rule{0.5pt}{2pt}}
      \multiput(13.5,2)(1,0){4}{\circle*{0.5}}
      \multiput(13.5,2.6)(1,0){4}{\circle*{0.5}}
      \ye{13}{0.5}{1}
      \ye{14}{0.5}{2}
      \ye{15}{0.5}{3}
      \ye{16}{0.5}{4}
      \ye{18.5}{1.5}{$0$}
    \end{picture}
    \caption{Extended Young tableau decomposition 
% and the corresponding spinon states 
      for an $S=1$ spin chain with $N=4$ sites.  The dots represent
      the spinons.  The spinon momentum numbers $a_i$ are given by the
      numbers in the boxes of the same column.  Note that $\sum
      (2S_{\text{tot}}+1)=3^N$.}
    \label{fig:nafoursitesu2}
  \end{center}
\end{figure}
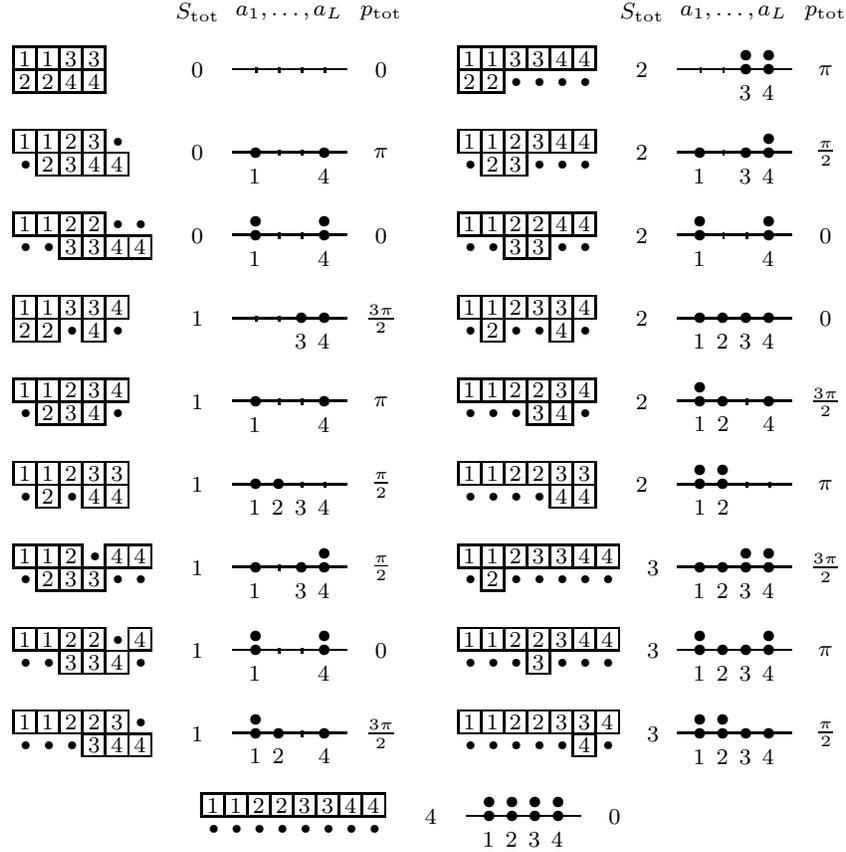

To derive \eqref{eq:naspacings}, we introduce a second formalism of
extended Young tableaux, %\cite{greiter-07prl237202},
this time for spin $S=1$.  The general rule we wish to propose for
obtaining the tableaux is illustrated in Fig.~\ref{fig:nayoungdiagram}
for three spins with $S=1$.  The construction is as follows.
For each of the $N$ spins, put a row of two adjacent boxes, which is
equivalent to the Young tableau for a single spin without any numbers
in the boxes.  Put these $N$ small tableaux on a line and number them
consecutively from left to right, with the same number in each pair of
boxes which represent a single spin.  To obtain the product of some
extended Young tableau representing spin ${S}_0$ on the left with a spin
$1$ tableau (\ie a row of two boxes with the same number in it) on the
right, we follow the rule
\begin{equation}
  \label{eq:nastimes1}
  \bs{S}_0\otimes \bs{1} = \left\{
      \begin{array}{ll} 
        \bs{1},                                          
        &\text{for}\ \, S_0=0, \\[2pt] 
        \bs{S}_0\bs{-1}\,\oplus\,\bs{S}_0\,\oplus\,\bs{S}_0\bs{+1},\quad 
        &\text{for}\ \, S_0=1,2,\ldots 
      \end{array}\right.
\end{equation}
\ie we obtain only one new tableau with both boxes from the right
added to the top row if the tableau on the left is a singlet, and
three new tableaux if it is has spin one or higher.  These three
tableaux are constructed by adding both boxes to the bottom row
(resulting in a representation $\bs{S}_0\bs{-1}$), by adding the first
box to the bottom row and the second box to the top row without
stacking them on top of each other (resulting in a representation
$\bs{S}_0$), and by adding both boxes to the top row (resulting in a
representation $\bs{S}_0\bs{+1}$).  In each extended tableau, the boxes
must be arranged such that the numbers are strictly increasing in each
column from top to bottom, and that they are not decreasing from left
to right in that the smallest number in each column cannot be smaller
than the largest number in the column to the left of it.  In analogy
to the Haldane--Shastry model, the empty spaces in between the boxes
are filled with dots representing spinons.  The spinon momentum number
$a_i$ associated with each spinon is given by the number in the box in
the same column.  A complete table of all the extended Young tableaux
for fours $S=1$ spins is shown in Fig.~\ref{fig:nafoursitesu2}.  The
assignment of physical single spinon momenta to the spinon momentum
numbers \eqref{eq:nasinglespinonmom} is identical to this assignment
for the
%$S=\frac{1}{2}$ Hilbert space of the exact eigenstates of the 
Haldane--Shastry model, as we can obtain the $3^N$ states of the $S=1$
Hilbert space by Schwinger boson projection (\ie by projecting on spin
$S=1$ on each site) from states contained in the $2^N\times 2^N$
dimensional Hilbert space of two $S=\frac{1}{2}$ models, a projection
which commutes with the total momentum.  The correctness of this
assignment has further been verified numerically up to $N=16$
sites~\cite{manuscriptinpreparationSG}.
 
With the tableau structure thus in place, all that is left to show is
that the momentum spacings are according to \eqref{eq:naspacings}.
Looking at any of the tableaux in Fig.~\ref{fig:nafoursitesu2}, we
note that from left to right, the spinons alternate
between being assigned to the first of the two boxes with a given
number and being assigned to the second of such two boxes.  This
follows simply form the fact that the number in between the columns
with the two neighboring spinons must be even.  The first spinon
momentum number $a_1$ is always odd, but all the other $a_i$'s can be
either even or odd.  The rule is therefore that if $i$ is odd, the
$i$-th spinon is assigned to the first of the two boxes with number
$a_i$, and the momentum spacing $a_{i+1}-a_i$ can be either even or
odd,
\begin{equation*}
%\begin{center}
  \setlength{\unitlength}{\ytlength}
  \begin{picture}(32,4)(0,0.5) \linethickness{0.5pt}
      \yt{1}{3}{\small 3}
      \yd{1}{2}
      \yt{2}{3}{\small 3}
      \yd{2}{2}
      \ye{1.5}{0.5}{\small even}
      \ye{4}{2.5}{or}
      \yt{6}{3}{\small 3}
      \yd{6}{2}
      \yt{7}{3}{\small 3}
      \yt{7}{2}{\small 4}
      \yt{8}{3}{\small 4}
      \yd{8}{2}
      \ye{7}{0.5}{\small odd}
      \ye{10}{2.5}{or}
      \yt{12}{3}{\small 3}
      \yd{12}{2}
      \yt{13}{3}{\small 3}
      \yt{13}{2}{\small 4}
      \yt{14}{3}{\small 4}
      \yt{14}{2}{\small 5}
      \yt{15}{3}{\small 5}
      \yd{15}{2}
      \ye{13.5}{0.5}{\small even}
      \ye{17}{2.5}{or}
      \ye{19}{2.2}{\ldots}
   \end{picture}
%\end{center}
\end{equation*}
If $i$ is even, however, the $i$-th spinon is assigned to the second of
the two boxes with number $a_i$, and the momentum spacing
$a_{i+1}-a_i$ has to be odd, as we can insert only an even number of
columns between the two spinons (recall that we cannot stack two boxes
with the same number in it on top of each other):
\begin{equation*}
%\begin{center}
  \setlength{\unitlength}{\ytlength}
  \begin{picture}(32,4)(0,0.5) \linethickness{0.5pt}
      \yt{1}{3}{\small 3}
      \yd{1}{2}
      \yt{2}{3}{\small 4}
      \yd{2}{2}
      \ye{1.5}{0.5}{\small odd}
      \ye{4}{2.5}{or}
      \yt{6}{3}{\small 3}
      \yd{6}{2}
      \yt{7}{3}{\small 4}
      \yt{7}{2}{\small 5}
      \yt{8}{3}{\small 4}
      \yt{8}{2}{\small 5}
      \yt{9}{3}{\small 6}
      \yd{9}{2}
      \ye{7.5}{0.5}{\small odd}
      \ye{11}{2.5}{or}
      \yt{13}{3}{\small 3}
      \yd{13}{2}
      \yt{14}{3}{\small 4}
      \yt{14}{2}{\small 5}
      \yt{15}{3}{\small 4}
      \yt{15}{2}{\small 5}
      \yt{16}{3}{\small 6}
      \yt{16}{2}{\small 7}
      \yt{17}{3}{\small 6}
      \yt{17}{2}{\small 7}
      \yt{18}{3}{\small 8}
      \yd{18}{2}
      \ye{15.5}{0.5}{\small odd}
      \ye{20}{2.5}{or}
      \ye{22}{2.2}{\ldots}
   \end{picture}
%\end{center}
\end{equation*}
The spacings between the single spinon momenta are hence as 
stated in \eqref{eq:naspacings}.

% OLD VERSION
% With the tableau structure thus in place, all that is left to show is
% that the momentum spacings are according to \eqref{eq:anspacings}.
% Looking at any of the tableaus in Fig.~\ref{fig:nafoursitesu2}, we see
% that the first spinon momentum number $a_1$ is always odd, while the
% second number $a_2$ can be even or odd.  If we cut a tableau to the
% right of the column of the second spinon, or in fact to the right of
% the $n$-th spinon where $n$ is even, we will also have an even number
% of boxes to the let of the cut, which implies that we will start with
% a new number to the right of the cut.  If the next spinon is in the
% column directly to the right of the cut, the numbers in the box to the
% left and to the right of the cut will differ by one, which is an odd
% number.  If the next spinon is after a second cut further to the
% right, the number of boxes in between both cuts will have to be even
% as we have no columns with the same numbers stacked on top of each other,
% and the difference of the momentum numbers for both spinons will likewise
% be even.  The same line of argument holds for every second momentum 
% spacing, and implies \eqref{eq:anspacings}.

%\vspace{100pt}

%\newpage
\subsection{Generalization to arbitrary spin S}
\label{sec:naarbs}

The projective generation introduced in Section \ref{sec:napro} can be
generalized to arbitrary spin $S=s$: %~\cite{greiter-09prl207203}:
\begin{equation}
  \label{eq:nacslopS}
   \ket{\psi^{S}_0}=
   \Big(\Psi^{\s\text{HS}}_0\big[a^\dagger ,b^\dagger\big]\Big)^{2s}\vac.
\end{equation} 
In order to write this state in a form similar to
\eqref{eq:napsi0}--\eqref{eq:navacuumket},
\begin{equation}
  \label{eq:naketS}
  \ket{\psi^S_0}\;=\sum_{\{z_1,\dots,z_{SN}\}} 
%  \psi_0[z_i]
  \psi^S_0(z_1,\dots,z_{SN})\
%  \tilde{S}_{z_1}^{+}\cdot\, \dots\, \cdot \tilde{S}_{z_{N}}^{+}\ 
  \tilde{S}_{z_1}^{+}\cdot\dots\cdot\tilde{S}_{z_{SN}}^{+} 
%  \ket{1,-1}^{\otimes N},
%  \ket{1,-1}_N,
  \ket{-s}_N,
%  \ket{-1}^{\otimes N},
\end{equation}
where 
\begin{equation}
  \label{eq:navacuumketS}
  \ket{-s}_N\equiv\otimes_{\alpha=1}^N \ket{s,-s}_{\alpha}
\end{equation}
is the ``vacuum'' state in which all the spins are maximally polarized
in the negative $\hat z$-direction, and we have to introduce %general
re-normalized spin flip operators $\tilde{S}^{+}$ which satisfy
\begin{equation}
  \label{eq:naStilde^nS}
  \frac{1}{\sqrt{(2s)!}}(a^\dagger)^n (b^\dagger)^{(2s-n)}\vac 
  =(\tilde{S}^+)^n\ket{s,-s}.
\end{equation}
%or 
%\begin{equation}
%  \label{eq:naStilde^nS}
%  (a^\dagger)^n (b^\dagger)^{(2s-n)}\vac =(\tilde{S}^+)^n(b^\dagger)^{2s}\vac.
%\end{equation}
If we assume a basis in which $S^\z $ is diagonal, we may write 
\begin{equation}
  \label{eq:naspinflipS}
  \tilde{S}^{+} 
  \,\equiv\,\frac{1}{b^\dagger b+1}\,a^\dagger b 
  \,=\, \frac{1}{s-{S}^\z +1}\, S^{+}.
\end{equation}
The wave function for the spin $S$ state \eqref{eq:nacslopS} is then
with $M=\frac{N}{2}$ given by 
\begin{equation}
  \label{eq:narr}
  \psi^S_0(z_1,\dots,z_{SN})
  =\prod_{m=1}^{2s}\left(
%    \prod_{i<j}^{M}(z_{i+(m-1)M}-z_{j+(m-1)M})^2 
    \prod_{\substack{i,j=(m-1)M+1\\[1pt] i<j}}^{mM}(z_i-z_j)^2 
  \right)\,
  \prod_{i=1}^{SN}z_i.
\end{equation}
% The wave function hence generalizes from a bosonic Pfaffian 
% state for $S=1$ to bosonic Read-Rezayi states~\cite{Read-99prb8084}
% for $S>1$.
Note that these states are similar to the Read-Rezayi
states~\cite{Read-99prb8084} in the quantized Hall effect.

In analogy to the $S=1$ state discussed in Section \ref{sec:napro},
the projective construction \eqref{eq:nacslopS} directly implies
several symmetries.  The state %\eqref{eq:nacslopS} 
$\ket{\psi^{S}_{0}}$ is translationally invariant with ground state
momentum $p_0=-\pi NS$, a spin singlet, and real:
\begin{equation}
  \label{eq:napsi0realS}
  \psi^S_0(z_1,\ldots ,z_{sN}) %[z_i]
  =\left|\psi^S_0(z_1,\ldots ,z_{sN})\right|\,\prod_{i=1}^{sN}\,G(z_i),
\end{equation}
with $G(z_i)$ given by \eqref{eq:hspsi0g}.

%%%%%%%%%%%%%%%%%%%%%%%%%%%%%%%%%%%%%%%%%%%%%%%%%%%%%%%%%%%%%%%%%%%%%%
%%%%%%%%%%%%%%%%%%%%%%%%%%%%%%%%%%%%%%%%%%%%%%%%%%%%%%%%%%%%%%%%%%%%%%
%%%%%%%%%%%%%%%%%%%%%%%%%%%%%%%%%%%%%%%%%%%%%%%%%%%%%%%%%%%%%%%%%%%%%%

\subsection{Momentum spacings and topological degeneracies for arbitrary spin S}
\label{sec:na:MSforSpinS}

In Section \ref{sec:nana}, we have shown that the non-Abelian
statistics of the Pfaffian state \eqref{eq:pfpsi0}, and in particular
the topological degeneracies associated with the Majorana fermion
states in the vortex cores discussed in Section
\ref{sec:3mod-pf-nastat}, manifests itself in topological choices for
the (kinematical) momentum spacings of the spinon excitations above
the $S=1$ ground state \eqref{eq:napsi0}.  Specifically, we found that
if we label the single spinon momenta in ascending order by
$p_i<p_{i+1}$, the spacings $p_{i+1}-p_i$ can be either even or odd
multiples of $\frac{\pi}{N}$ if $i$ is odd, while it has to be an odd
multiple if $i$ is even.

 \newcommand{\ytt}[3]{\put(#1,#2){\framebox(1.94,0.94){#3}}}
 \newcommand{\yttt}[3]{\put(#1,#2){\framebox(2.94,0.94){#3}}}

 In this Section, we formulate the corresponding restrictions for the
 general spin $S$ chain with ground state \eqref{eq:nacslopS}.  We
 will first state the rules and then motivate them.  Recall that
 spinons are represented by dots placed in the empty spaces of
 extended Young tableaux, and that the momentum number $a_i$ of spinon
 $i$ is given by the number in the box it shares a column with.  For
 general spin $S$, the tableau describing the representation on each
 site is given by
\begin{equation*}
%\begin{center}
  \setlength{\unitlength}{\ytlength}
  \begin{picture}(8,3.5)(0,0.8) \linethickness{0.5pt}
    \yt{1}{3}{}
    \yt{2}{3}{}
    \yt{3}{3}{}
    % \yt{4}{3}{}
    % \yt{5}{3}{}
    \ytt{4}{3}{}
    \yt{6}{3}{}
    \ye{7}{2.5}{,}
    \put(4,2){\makebox(0,0){$\underbrace{\phantom{\hspace{52pt}}}$}}
    \put(4,1){\makebox(0,0){\small $2S$ boxes}}
   \end{picture}
%\end{center}
\end{equation*}
\ie a horizontal array of $2S$ boxes indicating symmetrization, which
all contain the same number.  

If this number is $n$, the spinons we assign to any of these boxes
will have momentum number $a_i=n$.  Let us denote the number of the
box a given spinon $i$ with momentum number $a_i$ is assigned to, by
$b_i$, such that box number $b_i=1$ corresponds to the first, and box
number $b_i=2S$ to the last box with number $n$ in it:
\begin{equation*}
  \setlength{\unitlength}{\ytlength}
  \begin{picture}(8,5)(0,0.5) \linethickness{0.5pt}
    \yt{1}{3}{$n$}
    \yt{2}{3}{$n$}
    \yt{3}{3}{$n$}
    \ytt{4}{3}{}
    \yt{6}{3}{$n$}
    \ye{7}{2.5}{,}
    \yd{1}{2}
    \ye{1}{0.5}{\small $b_i=1$}
  \end{picture}
  \begin{picture}(10.5,5)(0,0.5) \linethickness{0.5pt}
    \yt{1}{3}{$n$}
    \yt{2}{3}{$n$}
    \yt{3}{3}{$n$}
    \ytt{4}{3}{}
    \yt{6}{3}{$n$}
    \ye{7}{2.5}{,}
    \ye{9}{2.5}{\ldots}
    \yd{2}{2}
    \ye{2}{0.5}{\small $b_i=2$}
   \end{picture}
  \begin{picture}(8,5)(0,0.5) \linethickness{0.5pt}
    \yt{1}{3}{$n$}
    \yt{2}{3}{$n$}
    \yt{3}{3}{$n$}
    \ytt{4}{3}{}
    \yt{6}{3}{$n$}
    \ye{7}{2.5}{.}
    \yd{6}{2}
    \ye{6}{0.5}{\small $b_i=2S$}
  \end{picture}
\end{equation*}
We will see below that if a representation of a spin $S$ chain
with $L$ spinons is written in terms of an extended Young tableau, the
first spinon with momentum number $a_1$ will always have box number
$b_1=1$, and the last spinon with $a_L$ will have $b_L=2S$.  The
restrictions corresponding to the non-abelian (SU(2) level $k=2S$)
statistics of the spinons are described by the flow diagram of the
numbers $b_i$ shown in Figure \ref{fig:biflow}.

\begin{figure}[tb]
\begin{center}
% see pst_ug.pdf  PSTricks Users Guide (by Timothy Van Zandt)
%
  \psset{unit=10pt,linewidth=.4pt,doublesep=1.25\pslinewidth,
    arrowsize=3.2pt,arrowlength=2,arrowinset=0,arcangleA=85,arcangleB=85}
  \begin{pspicture}(30,5.2)(-18.5,-3)
    \rput(-2,0){\rnode{a}{$1$}}
    \rput(2,0){\rnode{b}{$2$}}
    \ncline[doubleline=true,arrowsize=3.6pt,nodesep=6pt]{->}{a}{b}
    \ncarc[ncurv=1.57,nodesep=6pt]{<->}{a}{b}
    \rput(-4,0){$b_i=$}
    \rput(4,4){$a_{i+1}-a_i$\ \ odd}
    \rput(0,-2){$a_{i+1}-a_i$\ \ even}
    \rput(6,1){$S=1$}
    \psline[linewidth=.3pt]{-}(0,-1.3)(0,-.6)
    \rput(-16,4.5){(a)}
  \end{pspicture}
  \begin{pspicture}(30,10)(-18.5,-3)
    \rput(-12,0){\rnode{a}{$1$}}
    \rput(-8,0){\rnode{b}{$2$}}
    \rput(-4,0){\rnode{c}{$3$}}
    \rput(-1,0){\rnode{d}{}}
    \rput(4,0){\rnode{e}{}}
    \rput(8,0){\rnode{f}{$2S\!-\!1$}}
    \rput(12,0){\rnode{g}{$2S$}}
    \ncline[doubleline=true,arrowsize=3.6pt,nodesep=6pt]{->}{a}{b}
    \ncline[doubleline=true,nodesep=6pt]{->}{b}{c}
    \ncline[linestyle=dotted,arrowsize=3.6pt,nodesep=6pt]{-}{c}{d}
    \ncline[linestyle=dotted,arrowsize=3.6pt,nodesep=6pt]{-}{e}{f}
    \ncline[doubleline=true,nodesep=4pt]{->}{f}{g}
    \ncarc[nodesep=6pt]{<->}{a}{g}
    \ncarc[nodesep=6pt]{<->}{b}{f}
    \ncarc[nodesep=8pt,linestyle=dotted]{-}{c}{e}
    \rput(-14,0){$b_i=$}
    \rput(6,8){$a_{i+1}-a_i$\ \ odd}
    \rput(-6,-2){$a_{i+1}-a_i$\ \ even}
    \psline[linewidth=.3pt]{-}(-6,-1.3)(-6,-.6)
    \rput(-16,8.5){(b)}
  \end{pspicture}
  \end{center}
  \caption{Non-Abelian (SU(2) level $k=2S$) statistics in one
    dimension: flow diagram for the (auxiliary) box numbers $b_i$,
    which serve to describe the restrictions for the spinon momentum
    number spacings $a_{i+1}-a_i$ for the critical models of spin
    chains introduced in Sections \ref{sec:nags} and \ref{sec:naarbs}
    with (a) $S=1$, and (b) general spin $S$.  The unidirectional,
    horizontal arrows correspond to even integer momentum number
    spacings $a_{i+1}-a_i$, while the bidirectional, semicircle arrows
    correspond to odd integer spacings.  }
  \label{fig:biflow}
\end{figure}
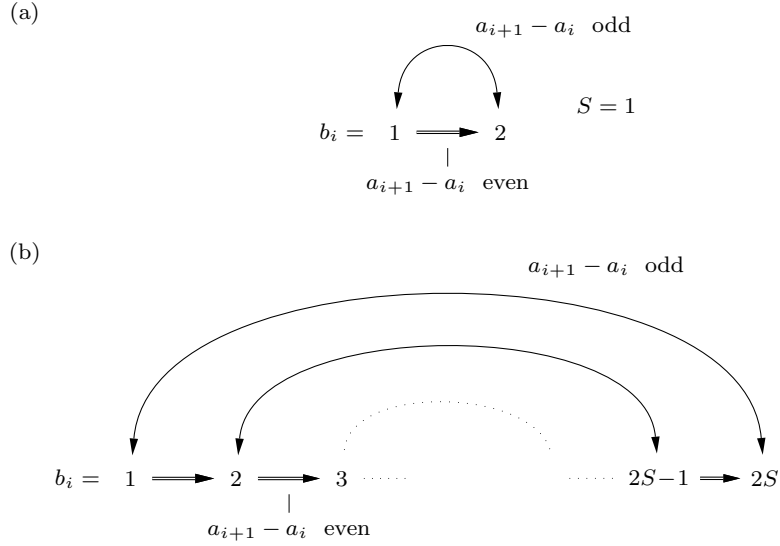

Let us elaborate this diagram first for the case $S=1$, which we
have already studied in Section \ref{sec:nana}.  In this case, 
\begin{equation}
  \label{eq:nastimes1}
   b_i=\left\{
      \begin{array}{ll} 
        1,\quad &\text{for}\ \, i\ \text{odd}, \\[2pt] 
        2,\quad &\text{for}\ \, i\ \text{even}.
      \end{array}\right.
\end{equation}
% In this case, $b_i=1$ for $i$ odd and $2$ or $i$ even, respectively.
For $i$ odd, we may move from $b_i=1$ to $b_{i+1}=2$ either via the
horizontal arrow or via the semicircle in Figure \ref{fig:biflow}a,
and $a_{i+1}-a_i$ may hence be either even or odd, respectively.  For
$i$ even or $b_{i}=2$, however, the semicircle is the only
available continuation, which implies that the spacing $a_{i+1}-a_i$
must be odd.
%The diagram shown in Figure \ref{fig:biflow}a implies that the
%semicircle is the only available continuation once we have
%reached $b_i=2$, and therefore that the spacing $a_{i+1}-a_i$
%must be odd if $i$ is even.  For even $i$, however, we can either
%follow the horizontal or the semicircular arrow, and obtain
%even or odd integer spacings, respectively.  

For general $S$, Figure \ref{fig:biflow}b implies that the spacings
can be even or odd until $b_i=2S$ is reached, which is then followed
by an odd integer spacing $a_{i+1}-a_i$, as the semicircular arrow is
the only possible continuation at this point.  Note that for $S\ge 1$,
the minimal number of spinons is two (these two spinons then have an
odd integer spacing $a_2-a_1$), and that we cannot have more than $2S$
spinons with the same momentum number $a_i=n$, as $a_{i+1}-a_i=0$ is
even.

We will now motivate this diagram.  To begin with, we generalize the
formalism of extended Young tableaux to arbitrary spin $S$.  The
construction is similar to the one for $S=1$ outlined in Section
\ref{sec:nana}.  For each of the $N$ spins, put a row of $2S$ adjacent
boxes.  Put these $N$ tableaux on a line and number them consecutively
from left to right, with the same number in each row of $2S$ boxes
representing a single spin.  To obtain the product of some extended Young
tableau representing spin ${S}_0$ on the left with a spin $S$ tableau
(\ie a row of $2S$ boxes with the same number in it) on the right, we
first recall
\begin{equation}
  \label{eq:naS0timesS}
  \bs{S}_0\otimes \bs{S} 
  = \,\vert\bs{S}_0\bs{-S}\vert\,
  \oplus\,\vert\bs{S}_0\bs{-S}\vert +1\,\oplus\,\ldots\,
%  \oplus\,\bs{S}_0\bs{+S-1}\,
  \oplus\,\,\bs{S}_0+\bs{S},
\end{equation}
which implies that we obtain either $2S_0+1$ or $2S+1$ new tableaux,
depending on which number is smaller.  In terms of extended Young tableaux,
\eqref{eq:nastimes1} translates into

\newcommand{\psyt}[3]{\rput(#1,#2){\psframe[dimen=middle](1,1)\rput(.5,.5){#3}}}
\newcommand{\psye}[3]{\rput(#1,#2){\rput(.5,.5){#3}}}
\newcommand{\psyd}[2]{\rput(#1,#2){\qdisk(.5,.5){0.16}}}
\newcommand{\psytt}[3]{\rput(#1,#2){\psframe[dimen=middle](2,1)\rput(.5,.5){#3}}}

\begin{equation}
  \label{eq:naS0timesSYT}
  \psset{unit=\ytlength,linewidth=.5pt}
  \setlength{\unitlength}{\ytlength}
    \begin{pspicture}(-0.5,5)(37,19) \linethickness{0.5pt}
     \rput(-0.5,  0){\psline[dimen=middle,fillcolor=lightgray,fillstyle=solid]
                    (-0.5,19)(2,19)(2,18)(1,18)(1,17)(-0.5,17)
                    \rput(0.75,15.75){$\bs{S}_0$}}
        \psye{2}{18}{$\otimes$}
     \rput(-0.5,  0){\psyt{4}{18}{$n$}\psyt{5}{18}{$n$}
                    \psytt{6}{18}{}   \psyt{8}{18}{$n$}
                    \rput(6.5,17){$\underbrace{\phantom{\hspace{43pt}}}$}
                    \rput(6.5,16){\small $2S$ boxes}
}%                    \rput(6.5,14.5){$\bs{S}$}}
   
       \rput(12,  0){\psye{-2.5}{17.5}{$=$}
                     \psline[fillcolor=lightgray,fillstyle=solid]
                     (-0.5,19)(2,19)(2,18)(1,18)(1,17)(-0.5,17)}
       \rput( 9,  0){\psyt{4}{17}{$n$}\psyt{5}{17}{$n$}\psyd{5}{18}
                    \psytt{6}{17}{}   \psyt{8}{17}{$n$}\psyd{8}{18}
                    \rput(5.75,15.25){\small for\ $S_0\ge S$}}
   
       \rput(21,  0){\psye{-2.25}{17.5}{$\oplus$}
                     \psline[fillcolor=lightgray,fillstyle=solid]
                     (-0.5,19)(2,19)(2,18)(1,18)(1,17)(-0.5,17)}
       \rput(18,  0){\psyt{4}{17}{$n$}\psyt{5}{17}{$n$}\psyd{5}{18}
                    \psytt{6}{17}{}   \psyt{8}{18}{$n$}
                    \rput(5.75,15.25){\small for\ $S_0\ge S-\frac{1}{2}$}}
   
       \rput(30,  0){\psye{-2.25}{17.5}{$\oplus$}
                     \psline[fillcolor=lightgray,fillstyle=solid]
                     (-0.5,19)(2,19)(2,18)(1,18)(1,17)(-0.5,17)}
       \rput(27,  0){\psyt{4}{17}{$n$}\psyt{7}{18}{$n$}
                    \psytt{5}{17}{}   \psyt{8}{18}{$n$}
                    \rput(5.75,15.25){\small for\ $S_0\ge S-1$}}

       \rput(12, -5){\psye{-2.5}{17.5}{$\oplus$}\psye{0.25}{17.25}{$\ldots$}}

       \rput(12, -9){\psye{-2.5}{17.5}{$\oplus$}
                     \psline[fillcolor=lightgray,fillstyle=solid]
                     (-0.5,19)(2,19)(2,18)(1,18)(1,17)(-0.5,17)}
       \rput( 9, -9){\psyt{4}{17}{$n$}\psyt{5}{17}{$n$}\psyd{5}{18}
                    \psytt{6}{18}{}   \psyt{8}{18}{$n$}
                    \rput(5.75,15.25){\small for\ $S_0\ge 1$}}

       \rput(21, -9){\psye{-2.25}{17.5}{$\oplus$}
                     \psline[fillcolor=lightgray,fillstyle=solid]
                     (-0.5,19)(2,19)(2,18)(1,18)(1,17)(-0.5,17)}
       \rput(18, -9){\psyt{4}{17}{$n$}\psyt{5}{18}{$n$}
                    \psytt{6}{18}{}   \psyt{8}{18}{$n$}
                    \rput(5.75,15.25){\small for\ $S_0\ge \frac{1}{2}$}}

       \rput(30, -9){\psye{-2.25}{17.5}{$\oplus$}
                     \psline[fillcolor=lightgray,fillstyle=solid]
                     (-0.5,19)(2,19)(2,18)(1,18)(1,17)(-0.5,17)}
       \rput(28, -9){\psyt{4}{18}{$n$}\psyt{5}{18}{$n$}\psyd{3}{17}
                    \psytt{6}{18}{}   \psyt{8}{18}{$n$}
                    \rput(5.75,15.25){\small always}}
   \end{pspicture}
\end{equation}
The first tableau on the right-hand side of \eqref{eq:naS0timesSYT}
exists only for $S_0\ge S$, the second only for $S_0\ge
S-\frac{1}{2}$, and so on.  Note that the shape of the right boundary
of the extended Young tableaux for $\bs{S}_0$ does not determine which
tableaux are contained in the expansion of $\bs{S}_0\otimes \bs{S}$,
as this depends only on the number $S_0-S$.  In the expansion
\eqref{eq:naS0timesSYT}, the $2S$ boxes representing a single spin $S$
always reside in adjacent columns.  In an extended tableau, the
numbers in the boxes are equal or increasing as we go from left to
right, and strictly increasing from top to bottom.  The empty spaces
we obtain as we build up the tableaux via this method represent the
spinons.  Note that we cannot take a given tableau and just add a pair
of spinons by inserting them somewhere, as the resulting tableau would
not occur in the expansion.
\begin{figure}[tb]
  \begin{center}
    \setlength{\unitlength}{\ytlength}
    \begin{picture}(32,10)(0,10) \linethickness{0.5pt}
%    \begin{picture}(34.5,10)(-1.5,10) \linethickness{0.5pt}
%      \ye{-1.5}{18}{(a)}
       \yt{1}{18}{1}
       \yt{2}{18}{1}
       \yt{3}{18}{1}
       \yt{4}{18}{1}
     \ye{2.5}{16.3}{$S=2$}
     \ye{5.5}{18}{$\otimes$}
       \yt{7}{18}{2}
       \yt{8}{18}{2}
       \yt{9}{18}{2}
      \yt{10}{18}{2}
     \ye{8.5}{16.3}{$S=2$}

      \ye{12}{17.5}{$=$}
      \yt{14}{18}{1}
      \yt{14}{17}{2}
      \yt{15}{18}{1}
      \yt{15}{17}{2}
      \yt{16}{18}{1}
      \yt{16}{17}{2}
      \yt{17}{18}{1}
      \yt{17}{17}{2}
    \ye{15.5}{15.3}{$S=0$}

    \ye{18.5}{17.5}{$\oplus$}
      \yt{20}{18}{1}
      \yd{20}{17}
      \yt{21}{18}{1}
      \yt{21}{17}{2}
      \yt{22}{18}{1}
      \yt{22}{17}{2}
      \yt{23}{18}{1}
      \yt{23}{17}{2}
      \yt{24}{18}{2}
      \yd{24}{17}
      \ye{22}{15.3}{$S=1$}

    \ye{25.5}{17.5}{$\oplus$}
      \yt{27}{18}{1}
      \yd{27}{17}
      \yt{28}{18}{1}
      \yd{28}{17}
      \yt{29}{18}{1}
      \yt{29}{17}{2}
      \yt{30}{18}{1}
      \yt{30}{17}{2}
      \yt{31}{18}{2}
      \yd{31}{17}
      \yt{32}{18}{2}
      \yd{32}{17}
    \ye{29.5}{15.3}{$S=2$}
 
      \put(-2,1){
      \ye{14}{11.5}{$\oplus$}
      \yt{16}{12}{1}
      \yd{16}{11}
      \yt{17}{12}{1}
      \yd{17}{11}
      \yt{18}{12}{1}
      \yd{18}{11}
      \yt{19}{12}{1}
      \yt{19}{11}{2}
      \yt{20}{12}{2}
      \yd{20}{11}
      \yt{21}{12}{2}
      \yd{21}{11}
      \yt{22}{12}{2}
      \yd{22}{11}

      \ye{19}{9.3}{$S=3$}
      \ye{23.5}{11.5}{$\oplus$}
      \yt{25}{12}{1}
      \yd{25}{11}
      \yt{26}{12}{1}
      \yd{26}{11}
      \yt{27}{12}{1}
      \yd{27}{11}
      \yt{28}{12}{1}
      \yd{28}{11}
      \yt{29}{12}{2}
      \yd{29}{11}
      \yt{30}{12}{2}
      \yd{30}{11}
      \yt{31}{12}{2}
      \yd{31}{11}
      \yt{32}{12}{2}
      \yd{32}{11}
      \ye{28.5}{9.3}{$S=4$}}
    \end{picture}
    \begin{picture}(32,5)(0,15) \linethickness{0.5pt}
%    \begin{picture}(34.5,10)(-1.5,10) \linethickness{0.5pt}
%      \ye{-1.5}{18}{(a)}
       \yt{1}{18}{1}
       \yt{2}{18}{1}
       \yt{3}{18}{1}
       \yt{4}{18}{1}
       \yt{1}{17}{2}
       \yt{2}{17}{2}
       \yt{3}{17}{2}
       \yt{4}{17}{2}
     \ye{2.5}{15.3}{$S=0$}
     \ye{5.5}{18}{$\otimes$}
       \yt{7}{18}{3}
       \yt{8}{18}{3}
       \yt{9}{18}{3}
      \yt{10}{18}{3}
     \ye{8.5}{16.3}{$S=2$}

      \ye{12}{17.5}{$=$}
      \yt{14}{18}{1}
      \yt{14}{17}{2}
      \yt{15}{18}{1}
      \yt{15}{17}{2}
      \yt{16}{18}{1}
      \yt{16}{17}{2}
      \yt{17}{18}{1}
      \yt{17}{17}{2}
      \yt{18}{18}{3}
      \yt{19}{18}{3}
      \yt{20}{18}{3}
      \yt{21}{18}{3}
      \yd{18}{17}
      \yd{19}{17}
      \yd{20}{17}
      \yd{21}{17}
    \ye{17.5}{15.3}{$S=2$}
    \end{picture}
    \begin{picture}(32,10)(1,10) \linethickness{0.5pt}
%    \begin{picture}(34.5,10)(-1.5,10) \linethickness{0.5pt}
%      \ye{-1.5}{18}{(b)}
      \put(1,0){\yt{0}{18}{1}
%                \yd{0}{17}
                \yt{1}{18}{1}
                \yt{1}{17}{2}
                \yt{2}{18}{1}
                \yt{2}{17}{2}
                \yt{3}{18}{1}
                \yt{3}{17}{2}
                \yt{4}{18}{2}
%                \yd{4}{17}
                \ye{2}{15.3}{$S=1$}}
      \ye{6.5}{18}{$\otimes$}
      \put(8,0){\yt{0}{18}{3}
                \yt{1}{18}{3}
                \yt{2}{18}{3}
                \yt{3}{18}{3}
              \ye{1.5}{16.3}{$S=2$}}
      \ye{13}{17.5}{$=$}
     \put(15,0){\yt{0}{18}{1}
                \yd{0}{17}
                \yt{1}{18}{1}
                \yt{1}{17}{2}
                \yt{2}{18}{1}
                \yt{2}{17}{2}
                \yt{3}{18}{1}
                \yt{3}{17}{2}
                \yt{4}{18}{2}
                \ye{4}{15.3}{$S=1$}}
     \put(19,0){\yt{0}{17}{3}
                \yt{1}{17}{3}
                \yd{1}{18}
                \yt{2}{18}{3}
                \yd{2}{17}
                \yt{3}{18}{3}   
                \yd{3}{17}}
     \ye{23.5}{17.5}{$\oplus$}
     \put(25,0){\yt{0}{18}{1}
                \yd{0}{17}
                \yt{1}{18}{1}
                \yt{1}{17}{2}
                \yt{2}{18}{1}
                \yt{2}{17}{2}
                \yt{3}{18}{1}
                \yt{3}{17}{2}
                \yt{4}{18}{2}
                \ye{3.5}{15.3}{$S=2$}}
     \put(29,0){\yt{0}{17}{3}
                \yt{1}{18}{3}
                \yd{1}{17}
                \yt{2}{18}{3}
                \yd{2}{17}
                \yt{3}{18}{3}   
                \yd{3}{17}}
     \ye{13}{12.5}{$\oplus$}
     \put(15,-5){\yt{0}{18}{1}
                \yd{0}{17}
                \yt{1}{18}{1}
                \yt{1}{17}{2}
                \yt{2}{18}{1}
                \yt{2}{17}{2}
                \yt{3}{18}{1}
                \yt{3}{17}{2}
                \yt{4}{18}{2}
                \yd{4}{17}
                \ye{4}{15.3}{$S=3$}}
     \put(20,-5){\yt{0}{18}{3}
                \yd{0}{17}
                \yt{1}{18}{3}
                \yd{1}{17}
                \yt{2}{18}{3}
                \yd{2}{17}
                \yt{3}{18}{3}   
                \yd{3}{17}}
    \end{picture}
    \begin{picture}(32,15)(2,5) \linethickness{0.5pt}
%    \begin{picture}(34.5,10)(-1.5,10) \linethickness{0.5pt}
%      \ye{-1.5}{18}{(b)}
      \put(1,0){\yt{0}{18}{1}
                \yt{1}{18}{1}
                \yt{2}{18}{1}
                \yt{2}{17}{2}
                \yt{3}{18}{1}
                \yt{3}{17}{2}
                \yt{4}{18}{2}
                \yt{5}{18}{2}
                \ye{2.5}{15.3}{$S=2$}}
      \ye{7.5}{18}{$\otimes$}
      \put(9,0){\yt{0}{18}{3}
                \yt{1}{18}{3}
                \yt{2}{18}{3}
                \yt{3}{18}{3}
              \ye{1.5}{16.3}{$S=2$}}
      \ye{14}{17.5}{$=$}
     \put(16,0){\yt{0}{18}{1}
                \yd{0}{17}
                \yt{1}{18}{1}
                \yd{1}{17}
                \yt{2}{18}{1}
                \yt{2}{17}{2}
                \yt{3}{18}{1}
                \yt{3}{17}{2}
                \yt{4}{18}{2}
                \yt{5}{18}{2}
                \ye{3.5}{15.3}{$S=0$}}
     \put(20,0){\yt{0}{17}{3}
                \yt{1}{17}{3}
                \yt{2}{17}{3}
                \yd{2}{18}
                \yt{3}{17}{3}   
                \yd{3}{18}}
     \ye{24.5}{17.5}{$\oplus$}
     \put(26,0){\yt{0}{18}{1}
                \yd{0}{17}
                \yt{1}{18}{1}
                \yd{1}{17}
                \yt{2}{18}{1}
                \yt{2}{17}{2}
                \yt{3}{18}{1}
                \yt{3}{17}{2}
                \yt{4}{18}{2}
                \yt{5}{18}{2}
                \ye{3.5}{15.3}{$S=1$}}
     \put(30,0){\yt{0}{17}{3}
                \yt{1}{17}{3}
                \yt{2}{17}{3}
                \yd{2}{18}
                \yt{3}{18}{3}   
                \yd{3}{17}}

     \ye{14}{12.5}{$\oplus$}
     \put(16,-5){\yt{0}{18}{1}
                \yd{0}{17}
                \yt{1}{18}{1}
                \yd{1}{17}
                \yt{2}{18}{1}
                \yt{2}{17}{2}
                \yt{3}{18}{1}
                \yt{3}{17}{2}
                \yt{4}{18}{2}
                \yt{5}{18}{2}
                \ye{3.5}{15.3}{$S=2$}}
     \put(20,-5){\yt{0}{17}{3}
                \yt{1}{17}{3}
                \yt{2}{18}{3}
                \yd{2}{17}
                \yt{3}{18}{3}   
                \yd{3}{17}}
     \ye{24.5}{12.5}{$\oplus$}
     \put(26,-5){\yt{0}{18}{1}
                \yd{0}{17}
                \yt{1}{18}{1}
                \yd{1}{17}
                \yt{2}{18}{1}
                \yt{2}{17}{2}
                \yt{3}{18}{1}
                \yt{3}{17}{2}
                \yt{4}{18}{2}
                \yd{4}{17}
                \yt{5}{18}{2}
                \ye{4}{15.3}{$S=3$}}
     \put(31,-5){\yt{0}{17}{3}
                \yd{1}{17}
                \yt{1}{18}{3}
                \yt{2}{18}{3}
                \yd{2}{17}
                \yt{3}{18}{3}   
                \yd{3}{17}}

     \ye{14}{7.5}{$\oplus$}
     \put(16,-10){\yt{0}{18}{1}
                \yd{0}{17}
                \yt{1}{18}{1}
                \yd{1}{17}
                \yt{2}{18}{1}
                \yt{2}{17}{2}
                \yt{3}{18}{1}
                \yt{3}{17}{2}
                \yt{4}{18}{2}
                \yd{4}{17}
                \yt{5}{18}{2}
                \yd{5}{17}
                \ye{4.5}{15.3}{$S=4$}}
     \put(22,-10){\yt{0}{18}{3}
                \yd{0}{17}
                \yt{1}{18}{3}
                \yd{1}{17}
                \yt{2}{18}{3}
                \yd{2}{17}
                \yt{3}{18}{3}   
                \yd{3}{17}}
       \end{picture}
    \caption{Examples of products of extended tableaux for an $S=2$
      spin chain.}
    \label{fig:na2S2}
  \end{center}
\end{figure}
In Figure \ref{fig:na2S2}, we illustrate the principle by writing
out a few terms in the expansion for an $S=2$ chain.

%\newpage o\newpage
We now turn to the question what this construction implies for the
momentum spacings of the spinons.  It is very easy to see from
Figure \ref{fig:na2S2} that $b_1=1$ and $a_1$ is odd, and that
$b_2=2S$ and $a_L$ is even (odd) for $N$ even (odd).

Let us assume we have a spinon $i$
with momentum number $a_i$ and box number $b_i$.  If we take $S=3$, $a_i=3$,
and $b_i=2$, this spinon would be represented by a dot which shares a
column with the second box with number $3$ in it,
\begin{equation*}
  \setlength{\unitlength}{\ytlength}
  \begin{picture}(8,5)(0,0.5) \linethickness{0.5pt}
    \yt{1}{3}{\small 3}
    \yt{2}{3}{\small 3}
    \yt{3}{3}{\small 3}
    \yt{4}{3}{\small 3}
    \yt{5}{3}{\small 3}
    \yt{6}{3}{\small 3}
%    \yt{7}{3}{\small 3}
%    \yt{8}{3}{\small 3}
    \ye{7}{2.5}{.}
    \yd{2}{2}
%    \ye{2.15}{0.8}{\small $i$}
    \ye{2.15}{0.8}{\small $b_i=2$}
%    \ye{2.15}{-0.2}{\small $a_i=3$}
  \end{picture}
\end{equation*}
For the box number $b_{i+1}$ of the next spinon, there are only two
possibilities:  
%First, 
\renewcommand{\theenumi}{\roman{enumi}}
\renewcommand{\labelenumi}{(\theenumi)}
\begin{enumerate}
\item $b_{i+1}=b_i+1$, which implies that $a_{i+1}-a_i$ is even.  The
  spinons either sit in neighboring columns with $a_{i+1}=a_i$, or
  contain an even number of spin $S$ representations (with $2S$ boxes
  each) in between them.  For our example, the corresponding tableaux
  are
  \begin{equation*}
    \setlength{\unitlength}{\ytlength}
    \begin{picture}(28,5.5)(0,-0.8) \linethickness{0.5pt}
      \put( 0,0){\yt{1}{3}{\small 3}
        \yt{2}{3}{\small 3}
        \yd{2}{2}
        \ye{2.15}{0.8}{\small $b_i$}
        \yt{3}{3}{\small 3}
        \yd{3}{2}
        \ye{3.75}{0.7}{\small $b_{i+1}$}
%        \ye{2.5}{-1}{\small $a_{i+1}=a_i$}
        \ye{3.3}{-1}{\small $a_{i+1}=a_i$}
        \yt{4}{3}{\small 3}
        \yt{5}{3}{\small 3}
        \yt{6}{3}{\small 3}}
      \yd{2}{2}
      \ye{8.5}{2.5}{and}

      \put(10,0){\yt{1}{3}{\small 3}
        \yt{2}{3}{\small 3}
        \yd{2}{2}
        \ye{2.15}{0.8}{\small $b_i$}
        \yt{3}{3}{\small 3}
        \yt{4}{3}{\small 3}
        \yt{5}{3}{\small 3}
        \yt{6}{3}{\small 3}}
      \put(12,0){\yt{1}{2}{\small 4}
        \yt{2}{2}{\small 4}
        \yt{3}{2}{\small 4}
        \yt{4}{2}{\small 4}
        \yt{5}{3}{\small 4}
        \yt{6}{3}{\small 4}}

      \put(16,0){\yt{1}{2}{\small 5}
        \yt{2}{2}{\small 5}
        \yt{3}{3}{\small 5}
        \yd{3}{2}
        \ye{3.75}{0.7}{\small $b_{i+1}$}
        \ye{-.5}{-1}{\small $a_{i+1}=a_i+2$}
        \yt{4}{3}{\small 5}
        \yt{5}{3}{\small 5}
        \yt{6}{3}{\small 5}}
      \ye{24.5}{2.5}{and}
      \ye{27}{2}{$\ldots$}
    \end{picture}
  \end{equation*}
  This possibility produces the unidirectional, horizontal arrows in
  Figure \ref{fig:biflow}.  If $b_i=2S$, this possibility does not
  exist, and there are either no further spinons or $a_{i+1}-a_i$ has
  to be odd.

\item $b_{i+1}=2S-b_i-1$, which implies that $a_{i+1}-a_i$ is odd.
  For our example, the tableaux are
  \begin{equation*}
    \setlength{\unitlength}{\ytlength}
    \begin{picture}(32,5.5)(0,-0.8) \linethickness{0.5pt}
      \put( 0,0){\yt{1}{3}{\small 3}
        \yt{2}{3}{\small 3}
        \yd{2}{2}
        \ye{2.15}{0.8}{\small $b_i$}
        \yt{3}{3}{\small 3}
        \yt{4}{3}{\small 3}
        \yt{5}{3}{\small 3}
        \yt{6}{3}{\small 3}}
      \put(2,0){\yt{1}{2}{\small 4}
        \yt{2}{2}{\small 4}
        \yt{3}{2}{\small 4}
        \yt{4}{2}{\small 4}
        \yt{5}{3}{\small 4}
        \yd{5}{2}
        \ye{5.75}{0.7}{\small $b_{i+1}$}
%        \ye{2.5}{-1}{\small $a_{i+1}=a_i+1$}
        \ye{3.2}{-1}{\small $a_{i+1}=a_i+1$}
        \yt{6}{3}{\small 4}}
      \ye{10.5}{2.5}{and}

      \put(12,0){\yt{1}{3}{\small 3}
        \yt{2}{3}{\small 3}
        \yd{2}{2}
        \ye{2.15}{0.8}{\small $b_i$}
        \yt{3}{3}{\small 3}
        \yt{4}{3}{\small 3}
        \yt{5}{3}{\small 3}
        \yt{6}{3}{\small 3}}
      \put(14,0){\yt{1}{2}{\small 4}
        \yt{2}{2}{\small 4}
        \yt{3}{2}{\small 4}
        \yt{4}{2}{\small 4}
        \yt{5}{3}{\small 4}
        \yt{6}{3}{\small 4}}

      \put(18,0){\yt{1}{2}{\small 5}
        \yt{2}{2}{\small 5}
        \yt{3}{3}{\small 5}
        \yt{4}{3}{\small 5}
        \yt{5}{3}{\small 5}
        \yt{6}{3}{\small 5}}
      \put(20,0){\yt{1}{2}{\small 6}
        \yt{2}{2}{\small 6}
        \yt{3}{2}{\small 6}
        \yt{4}{2}{\small 6}
        \yt{5}{3}{\small 6}
        \yd{5}{2}
        \ye{5.75}{0.7}{\small $b_{i+1}$}
        \ye{-0.5}{-1}{\small $a_{i+1}=a_i+3$}
        \yt{6}{3}{\small 6}}
      \ye{28.5}{2.5}{and}
      \ye{31}{2}{$\ldots$}
    \end{picture}
  \end{equation*}
  This possibility produces the bidirectional, semicircle arrows in
  Figure \ref{fig:biflow}.
\end{enumerate}
\renewcommand{\theenumi}{\arabic{enumi}}
\renewcommand{\labelenumi}{\theenumi.}
This concludes the motivation of the flow diagram in Figure
\ref{fig:biflow}b.  As in Sections \ref{sec:hsyt} and \ref{sec:nana},
the single spinon momenta are given by
\begin{equation}
  \label{eq:nasinglespinonmom}
  p_i=\frac{\pi}{N}\,\left(a_i-\frac{1}{2}\right).
\end{equation}
This yields momentum spacings $p_{i+1}-p_i$ which can be either an
integer or an half-integer times $\frac{2\pi}{N}$.

%\vspace{100pt}
%lp -dtkmcol -P 77-81 map.ps

%\vspace{100pt}
%lp -dtkmsek_colour -P 24-30 map.ps

%\include{ch-extra}
% lp -dtkmsek -P 40-46 map.ps 

\chapter{From a Laughlin state to the Haldane--Shastry model }
\label{sec:l2hs}

%\section{Comparism of the models}
\section{General considerations}
\label{sec:a:gen}
In this section, we wish to derive, or maybe better obtain, the
Haldane--Shastry model (see Section \ref{sec:hs})
% \eqref{eq:hsham}--\eqref{eq:hspsi0} 
from the bosonic $m=2$ Laughlin state 
% \eqref{eq:qhpsiLaugh} and its parent Hamiltonian \eqref{eq:qhTrugmanHboson}.  
and its parent Hamiltonian (see Section \ref{sec:qhlaughlin}).  At
first sight, this does not appear to be a sensible endeavor.  Let
us briefly recall both models.

\subsection{Comparison of the models}
%\subsection{Comparism of the models}
%\vspace{.5\baselineskip} 
%\emph{Comparison of the models.}---%
\label{sec:a:com}
The Haldane--Shastry model %(HSM)
describes a spin \half chain with periodic boundary conditions. %(PBCs).
The Hamiltonian is
\begin{equation}
  \label{eq:a:hsham}
  {H}^{\s\text{HS}} = \left(\frac{2\pi}{N}\right)^2
  \sum^N_{\alpha <\beta}\,
  \frac{{\boldsymbol{S}}_\alpha {\boldsymbol{S}}_\beta 
  }{\left|\eta_\alpha-\eta_\beta \right|^2}\,,
\end{equation}
where $\displaystyle \eta_\alpha=e^{\text{i}\frac{2\pi}{N}\alpha }$
with $\alpha = 1,\ldots ,N$ are sites on a unit circle embedded in the
complex plane.  Written as a wave function for the position of the
$M=\frac{N}{2}$ $\up$-spin coordinates $z_i$, the ground state is
given by
\begin{equation}
  \label{eq:a:hspsi0}
  \psi^{\s\text{HS}}_{0}(z_1,\ldots ,z_M) = 
  \prod_{i<i}^M\,(z_i-z_j)^2\,\prod_{i=1}^M\,z_i\,. 
\end{equation}
The bosonic $m=2$ Laughlin state for $M$ particles,
\begin{equation}
  \label{eq:a:psiLaughplane}
  \psi_0(z_1,\ldots,z_M)
  =\prod^M_{i<j}(z_i-z_j)^2\prod_{i=1}^M e^{-\frac{1}{4}|z_i|^2},
\end{equation}
is the exact ground state of the $\delta$-function potential
interaction Hamiltonian
\begin{equation}
  \label{eq:a:deltaplane}
  V=\sum_{i<j}^M \delta^{(2)}(z_i-z_j)
\end{equation}
in the lowest Landau level.  Obviously, both models share the factor
\begin{equation}
  \label{eq:a:jastrow2}
  \prod^M_{i<j}(z_i-z_j)^2
%  =\sum_{\{k_1,\ldots ,k_M\}} C_{k_1,\ldots ,k_M}\, z_1^{k_1}\ldots z_M^{k_M}
\end{equation}
in their ground state wave function, a connection which was exploited
recently by Thomale \etal~\cite{thomale-10prl116805} in their study of
the entanglement spectrum of spin chains, but this seems to be about
it.  The Gutz\-willer or Haldane-Shastry ground state is invariant
under P and T, under translations along the chain, and under global
SU(2) spin rotations (see Section \ref{sec:hsgs}).  The model further
possesses a Yangian symmetry and is integrable (see Section
\ref{sec:hssymm}).  The Laughlin ground state is up to a gauge
transformation invariant under rotations around the origin.  The
geometries of both models differ.
%The only connection between both Hamiltonians appears to be that
%Haldane~\cite{haldane88prl635,haldane83prl605} discovered both of
%them.

Let us proceed by clearing some obvious hurdles to our endeavor of
connecting the models.  To begin with, the circular droplet described
by the Laughlin wave function \eqref{eq:a:psiLaughplane} has a
boundary, while the Haldane--Shastry ground state describes a spin
liquid on a compact surface.  This problem, however, is easily
circumvented by formulating the quantum Hall model on the sphere
(see Section \ref{sec:laughsphere}).  Then the bosonic $m=2$ Laughlin
state for $M$ particles on a sphere with $2s=2M-2$ flux quanta is
given by
\begin{equation}
  \label{eq:a:psiLaugh}
  \psi_0[u,v]=\prod_{i<j}^M(u_iv_j-u_jv_i)^2.
\end{equation}
Within the lowest Landau level, it is the exact and unique zero-energy
ground state of the interaction Hamiltonian
\begin{align}
  \label{eq:a:Vham2s}
  V^{\s\text{qh}}
  &=\sum_{m_1=-s}^s\, \sum_{m_2=-s}^s\, \sum_{m_3=-s}^s\, \sum_{m_4=-s}^s
  \ a_{m_1}^\dagger a_{m_2}^\dagger 
  a_{m_3}^{\phantom{\dagger}} a_{m_4}^{\phantom{\dagger}}\,
  \delta_{m_1+m_2,m_3+m_4}
  \nonumber\\[0.5\baselineskip]
  &\quad\cdot
  \braket{s,m_1;s,m_2}{2s,m_1+m_2} \braket{2s,m_3+m_4}{sm_3,sm_4}\!,
\end{align}
where $a_m$ annihilates a boson in the properly normalized single
particle state
\begin{equation}
  \label{eq:a:qhsnormLLLbasis}
  \psi_{m,0}^{s}(u,v) 
%  =\braket{u,v}{m}
  =\bra{u,v} a_{m}^\dagger \vac
  =\sqrt{\frac{(2s+1)!}{4\pi\,(s+m)!\, (s-m)!}}\;
  u^{s+m}v^{s-m},
\end{equation}
and $\braket{s,m_1;s,m_2}{j,m_1+m_2}$ \etc are Clebsch--Gordan
coefficients~\cite{baym69}.  The Hamiltonian \eqref{eq:a:Vham2s}
assigns a finite energy cost whenever the relative angular momentum of
a pair of particles is zero.  The expansion coefficients of the
polynomial \eqref{eq:a:psiLaugh} are still identical to those of
\eqref{eq:a:jastrow2}.  
% The only connection between the Hamiltonians \eqref{eq:a:hsham} and
% \eqref{eq:a:Vham2s} appears to be that
% Haldane~\cite{haldane88prl635,haldane83prl605} discovered both of
% them.

\subsection{A hole at a pole}
\label{sec:a:holepole}
%\vspace{.5\baselineskip} 
%\emph{A hole at a pole.}---%
The Haldane--Shastry ground state wave function
\eqref{eq:a:hspsi0}, however, contains an additional factor $\prod_{i}z_i$.
This is related to another problem.  The dimension of the single
particle Hilbert space for the bosons on the sphere is $2s+1=2M-1$,
while the dimension of the single particle Hilbert space for the spin
flips on the unit circle is equal to the number of sites, $N=2M$.  The
Hilbert space dimensions of both models hence do not match.  We can
adapt the quantum Hall state by insertion of a
quasihole at the south pole $(\alpha,\beta)=(0,1)$ of the sphere.  This
leads to the wave function
\begin{equation}
  \label{eq:a:psiqH}
  \psi^{\s\text{qH}}_0[u,v]=\prod_{i<j}^M(u_iv_j-u_jv_i)^2\,\prod_{i=1}^M u_i,
\end{equation}
on a sphere with $2s+1=2M$ single particle states.  It is the exact
and unique ground state of
\begin{equation}
  \label{eq:a:HqH}
  H^{\s\text{qH}}=V^{\s\text{qH}}+U^{\s\text{qH}}
\end{equation}
with 
\begin{equation}
  \label{eq:a:UqH}
  U^{\s\text{qH}}=U_0\, a_{-s}^\dagger a_{-s}
\end{equation}
for $U_0>0$ if we restrict our Hilbert space again to the lowest
Landau level.  In \eqref{eq:a:HqH}, we have added a local repulsive
potential $U_0$ for the single particle state with $m=-s$, \ie the
state at the south pole, to the interaction Hamiltonian
\eqref{eq:a:Vham2s}.  Note that both $V^{\s\text{qH}}$ and
$U^{\s\text{qH}}$ annihilate the ground state \eqref{eq:a:psiqH}
individually.  The single particle Hilbert space dimensions match now,
$2s+1=2M=N$, and the expansion coefficients $C_{q_1,\ldots ,q_M}$ for
the polynomials
\begin{equation}
  \label{eq:a:psiHSq}
  \psi^{\s\text{HS}}_{0}[z] %(z_1,\ldots ,z_M) = 
   =\sum_{\{q_1,\ldots ,q_M\}} C_{q_1,\ldots ,q_M}\, z_1^{q_1}\ldots z_M^{q_M}
\end{equation}
and
\begin{equation}
  \label{eq:a:psiLaughq}
  \psi^{\s\text{qH}}_0[u,v] %=\prod_{i<j}^M(u_iv_j-u_jv_i)^2\,\prod_{i=1}^M u_i,
  =\sum_{\{q_1,\ldots ,q_M\}} C_{q_1,\ldots ,q_M}\, 
  u_1^{q_1}v_1^{2s-q_1}\ldots u_M^{q_M}v_M^{2s-q_M}
\end{equation}
% \begin{eqnarray}
%   \label{eq:a:psiLaughq}
%   \psi^{\s\text{qH}}_0[u,v] %=\prod_{i<j}^M(u_iv_j-u_jv_i)^2\,\prod_{i=1}^M u_i,
%   \hspace{-6pt}&=&\hspace{-6pt}\sum_{\{q_1,\ldots ,q_M\}} C_{q_1,\ldots ,q_M}\, 
%   u_1^{q_1}v_1^{2s-q_1}\ldots u_M^{q_M}v_M^{2s-q_M}
%   \nonumber\\[0.5\baselineskip]
%   \hspace{-6pt}&=&\hspace{-6pt}\sum_{\{m_1,\ldots ,m_M\}} C_{m_1+s,\ldots ,m_M+s}\,\,
%   u_1^{s+m_1}v_1^{s-m_1}\ldots u_M^{s+m_M}v_M^{s-m_M}\qquad
% \end{eqnarray}
are identical.  Note that in both states, the amplitudes are non-zero
only for $1\le q_i\le N$, \ie the $q=0$ state is never occupied.  For
the Haldane--Shastry ground state, this means that we never flip a
spin $S^+_q$ with momentum $q=0$, which is a necessary requirement for
the singlet property (see Section \ref{sec:hsgs}).  For the quantized
Hall state, it means no particle occupies the $m=-s$ state at the
south pole of the sphere.  Note further that the Hamiltonians for the
sphere \eqref{eq:a:HqH} and for the spin chain \eqref{eq:a:hsham} are
formulated in different spaces.  The Hamiltonian \eqref{eq:a:HqH} with
\eqref{eq:a:Vham2s} on the sphere scatters bosons in a basis of
(angular) momentum eigenstates $m$, while the Haldane--Shastry
Hamiltonian \eqref{eq:a:hsham} scatters bosonic spin-flips in a
position space basis of sites $\eta_\alpha$.

% The equivalence of the coefficients in the expansions
% \eqref{eq:a:psiHSq} and \eqref{eq:a:psiLaughq} was recently exploited
% by Thomale \etal~\cite{thomale-10prl116805}, who proposed that the
% entanglement spectrum can be used to define order in gapless systems
% such as the $S=\frac{1}{2}$ spin chain.

\section{Hilbert space renormalization}
\label{sec:a:hilbert}
%\vspace{.5\baselineskip} 
%\emph{Hilbert space renormalization.}---%
There is yet another significant difference between both models.  We
noted above that the coefficients in the polynomial expansions of the
ground states \eqref{eq:a:psiHSq} and \eqref{eq:a:psiLaughq} are
identical.  The expansions of both states in terms of single particle
states, however, are different, due to different normalizations of the
polynomials.  In the Haldane--Shastry model, the wave function acts on
a Hilbert space constructed out of spin flips at positions $z_i$,
\begin{equation}
  \label{eq:a:hsket}
  \ket{\psi^{\s\text{HS}}_{0}}\,=
  \sum_{\{z_1,\ldots ,z_M\}}\psi^{\s\text{HS}}_{0} 
  (z_1,\ldots ,z_M)\,{S}^+_{z_1}\cdot\ldots\cdot {S}^+_{z_M} 
  \big|\underbrace{\dw\dw\ldots\ldots\dw}_{\text{all\ } N \text{\ spins\ } \dw}
  \big\rangle.
\end{equation}
% Multiplied by $\frac{1}{\sqrt{N}}$, 
The polynomial $\frac{1}{\sqrt{N}}z^q$ describes the normalized single
particle state
\begin{equation}
  \label{eq:a:z^qstate}
  \frac{1}{\sqrt{N}}\sum_{\{z\}} z^q {S}^+_z \ket{\dw\dw\ldots\dw}
  =\frac{1}{\sqrt{N}}\sum_\alpha 
  \eta_\alpha^q {S}^+_{\alpha} \ket{\dw\dw\ldots\dw}
  \equiv {\hat S}^+_{q}\ket{\dw\dw\ldots\dw},
\end{equation}
and we can rewrite the state vector \eqref{eq:a:hsket} in terms of
\eqref{eq:a:psiHSq} as
\begin{equation}
  \label{eq:a:hsCket}
  \ket{\psi^{\s\text{HS}}_{0}}\,
  =\sum_{\{q_1,\ldots ,q_M\}} C_{q_1,\ldots ,q_M}\,
  {\hat S}^+_{q_1}\cdot\ldots\cdot {\hat S}^+_{q_M} 
  \ket{\dw\dw\ldots\ldots\dw}.
\end{equation}
The polynomials $u^{s+m}v^{s-m}$, by contrast, describe the
unnormalized single particle states
\begin{equation}
  \label{eq:uvstate}
  g_m \ket{\psi_{m,0}^{s}}
%\equiv g_m \ket{m}
  = g_m a_m^\dagger \vac, 
\end{equation}
where 
\begin{equation}
  \label{eq:a:gm}
  g_m=\sqrt{\frac{4\pi\,(s+m)!\, (s-m)!}{(2s+1)!}}
\end{equation}
is the normalization factor from \eqref{eq:a:qhsnormLLLbasis} and
$a_m^\dagger$ is the associated, properly normalized creation
operator.  The state vector for the quantum Hall state is hence given by
\begin{equation}
  \label{eq:a:psiCLaughq}
  \ket{\psi^{\s\text{qH}}_0}
%   =\sum_{\{q_1,\ldots ,q_M\}} C_{q_1,\ldots ,q_M}\,
%   g_{q_1-s}\ldots g_{q_M-s} \, 
%   {a}^\dagger_{q_1-s}\ldots {a}^\dagger_{q_M-s} \vac
  =\sum_{\{m_1,\ldots ,m_M\}} C_{m_1+s,\ldots ,m_M+s}\,\,
  g_{m_1}\ldots g_{m_M} \,
  {a}^\dagger_{m_1}\ldots {a}^\dagger_{m_M} \vac
\end{equation}
This means that not only the Hamiltonians, but also the coefficients
in the ground state vectors, are different.  In particular, we we
diagonalize the Haldane-Shastry Hamiltonian \eqref{eq:a:hsham} for a
finite chain, we obtain a ground state vector which is quite different
from the ground state vector of \eqref{eq:a:HqH} with
\eqref{eq:a:Vham2s}.  If two models have different symmetries,
different Hamiltonian and different ground states, it is not clear
what the connection should be.
%The symmetries mentioned above are not the only differences of the models.

If we think about the problem from a scholarly perspective, the
conclusion would probably be to abandon our undertaking.  The
scholarly approach, however, is not always the most fruitful one.
Haldane~\cite{haldane83prl605} invented the parent Hamiltonian
%\eqref{eq:qhsVham2s} with 
\eqref{eq:qhsPP} because he was looking for an economical way to write
out the coefficients of a Laughlin state for a significant number of
particles without expanding the polynomial, which he could then
compare to the ground state for Coulomb interactions.  Along these
lines, note that since the single particle normalizations
%\eqref{eq:a:gm} 
$g_m$ are known, it is easy to obtain the coefficients in
\eqref{eq:a:hsCket} from the coefficients in \eqref{eq:a:psiCLaughq}
and vice versa.  So regardless of how different the two states are
from a scholarly point of view, there may be practical benefit in
exploring the common features.

%\newpage
In fact, even though the quantum Hall Hamiltonian \eqref{eq:a:HqH}
with \eqref{eq:a:Vham2s} cannot be used directly to obtain the
Haldane--Shastry ground state \eqref{eq:a:hsCket}, we can construct a
parent Hamiltonian for \eqref{eq:a:hsCket} from \eqref{eq:a:HqH}.  To do so,
consider first the following theorem.

%\newtheorem{theorem}{Theorem}
% \vspace{0.7\baselineskip}\noindent {\bf Theorem. }{\it 
\begin{theorem} 
  \addtolength{\leftskip}{\parindent}
  \addtolength{\rightskip}{\parindent}
  Let $\ket{\psi_0}$ be the exact and
  non-degenerate zero-energy\linebreak ground state of $H$,
  \begin{align}\nonumber
    H\ket{\psi_0}=0,\quad \bra{\psi}H\ket{\psi} &\ge 0\quad\forall\, \ket{\psi},
    \\\nonumber
    &=0\quad\text{if and only if}\ \ket{\psi}=\ket{\psi_0},
  \end{align}
  and let $G$ be an invertible matrix, $G^{-1}G=1$.  Then
  $G^{-1}\ket{\psi_0}$ is the exact and non-degenerate zero-energy
  ground state of $G^\dagger HG$. %}\\ %[-.3\baselineskip]
\end{theorem}

%\begin{proof}
\addtolength{\leftskip}{\parindent}
\addtolength{\rightskip}{\parindent}
  \vspace{.5\baselineskip}\noindent {\sc Proof.}  Trivially, 
  $G^\dagger HG\,G^{-1}\ket{\psi_0}=0$.  With 
  $\ket{\psi'}\equiv G\ket{\psi}$, we have
  \begin{equation*}
    \hspace{8pt}\begin{array}{rrlr}
      \bra{\psi}G^\dagger HG\ket{\psi}
      =\bra{\psi'}H\ket{\psi'}
      &\multicolumn{2}{l}
      {\ge 0\quad\forall\,\ket{\psi'}\ \text{and hence}\ \forall\,\ket{\psi},}
      \\[4pt]
      &=0\quad\text{if and only if}&\ket{\psi'}=\ket{\psi_0},
      \\[4pt]
      &\text{i.e.,}&\ket{\psi}=G^{-1}\ket{\psi_0}. 
%      \hspace{30pt} &\Box
      \quad &\Box
%     \qed
    \end{array}
  \end{equation*}
%\end{proof}
\setlength{\leftskip}{0pt}
\setlength{\rightskip}{0pt}
Note that this transformation is not just a rotation of the basis.  It
completely changes the Hamiltonian, but has the benefit instructing us
how to obtain the zero energy ground state of the new Hamiltonian from
the original one.

While this theorem points in the right direction, we are not aware of
any way of arriving at a convenient parent Hamiltonian by employing it
directly.  On the positive side, if we choose
\begin{equation}
  \label{eq:a:g}
  G = \prod_{m=-s}^s \, (g_m)^{a_m^\dagger a_m^{\phantom{\dagger}}}, \ \
  G^{-1} = \prod_{m=-s}^s 
  \left(\frac{1}{g_m}\right)^{a_m^\dagger a_m^{\phantom{\dagger}}},
\end{equation}
we obtain
\begin{equation}
  \label{eq:a:psiLaughGC}
  G^{-1}\ket{\psi^{\s\text{qH}}_0}
  =\sum_{\{m_1,\ldots ,m_M\}} C_{m_1+s,\ldots ,m_M+s}\,\,
  {a}^\dagger_{m_1}\ldots {a}^\dagger_{m_M} \vac,
\end{equation}
which is identical to the Haldane--Shastry ground state
\eqref{eq:a:hsCket} if we were to substitute\footnote{It is not clear
  whether such a substitution is sensible, since the operators
  $a_m^\dagger$ and ${\hat S}^+_{s+m}$ obey different commutation
  relations.  For this reason, we do not implement it, but merely
  mention the possibility.  We will see below %in Section \ref{sec:a:four}
  that a similar transition from the Fourier transforms of
  $a_m^\dagger$ to local spin flips ${S}^+_\alpha$ can be implemented
  sensibly.} $a_m^\dagger\to {\hat S}^+_{s+m}$.  On the negative
side, the Hamiltonian $G^\dagger H^{\s\text{qH}} G$ is unnecessarily
complicated.  To obtain a convenient parent Hamiltonian for
\eqref{eq:a:psiLaughGC}, we avail ourselves of another theorem.

% The problem with it is that $G^\dagger H^{\s\text{qH}} G$ constitutes
% an unnecessarily complicated Hamiltonian.  To obtain a convenient
% parent Hamiltonian for \eqref{eq:a:psiLaughGC}, we avail ourselves of
% the following theorem.

% \vspace{0.7\baselineskip}\noindent {\bf Theorem. }{\it 
\begin{theorem}\label{th:second} 
  \addtolength{\leftskip}{\parindent}
  \addtolength{\rightskip}{\parindent}
  Let $\ket{\psi_0}$
%   \begin{equation}
%     \ket{\psi_0}=\sum_{\{m_1,\ldots ,m_M\}} \phi_{m_1,\ldots ,m_M}\,\,
%     {a}^\dagger_{m_1}\ldots {a}^\dagger_{m_M} \vac
%   \end{equation}
  be a zero-energy eigenstate of the interaction Hamiltonian
  \begin{equation*}
    H=\sum_{\{m_1,m_2,m_3,m_4\}} 
    a_{m_1}^\dagger a_{m_2}^\dagger \, V_{m_1,m_2,m_3,m_4}\,\,
    a_{m_3}^{\phantom{\dagger}} a_{m_4}^{\phantom{\dagger}},
  \end{equation*}
%  $H\ket{\psi_0}=0$, 
  and let $G$ be an invertible matrix, $G^{-1}G=1$.  Then
  $G^{-1}\ket{\psi_0}$ is a zero-energy eigenstate of
  \begin{equation*}
    H'=\sum_{\{m_1,m_2,m_3,m_4\}} 
    G^\dagger a_{m_1}^\dagger a_{m_2}^\dagger {G^\dagger}^{-1}
    \, V_{m_1,m_2,m_3,m_4} \,\,
    G^{-1} a_{m_3}^{\phantom{\dagger}} a_{m_4}^{\phantom{\dagger}} G.
  \end{equation*} %}\\[-.5\baselineskip]
% $H G^{-1}\ket{\psi_0}=0$.
\end{theorem}

\addtolength{\leftskip}{\parindent}
\addtolength{\rightskip}{\parindent}
  \vspace{.5\baselineskip}\noindent {\sc Proof.} 
  The property $H\ket{\psi_0}=0$ implies
  \begin{equation*}
    \sum_{\{m_3,m_4\}} V_{m_1,m_2,m_3,m_4}\,\,
    a_{m_3} a_{m_4}\, \ket{\psi_0}=0
    \quad\forall\, m_1,m_2,
  \end{equation*}
  and hence
  \begin{equation*}
    \sum_{\{m_3,m_4\}} V_{m_1,m_2,m_3,m_4}\,
     G^{-1}\, a_{m_3} a_{m_4}\, G\, G^{-1}\ket{\psi_0}=0
    \quad\forall\, m_1,m_2,
  \end{equation*}
  which in turn implies $H'G^{-1}\ket{\psi_0}=0$. 
%  \hfill $\Box$\\[-.5\baselineskip]
  \quad $\Box$\\[-.5\baselineskip]

 \vspace{0.7\baselineskip}\noindent {\sc Remark. }{\it 
   The Theorem holds for %not only for two, but also for $n$-body interactions.
   $n$-body interactions as well.
 }\\ %[-.5\baselineskip]

\setlength{\leftskip}{0pt}
\setlength{\rightskip}{0pt}
\noindent 
The choice \eqref{eq:a:g} implies $G^\dagger=G$ and
\begin{align}\hspace{20pt} 
  G^{-1}\, a_m\, G       &=g_m a_m,&             
  G\, a_m\, G^{-1}       &=\frac{1}{g_m} a_m,\hspace{20pt} \\
  G^{-1}\, a_m^\dagger\, G&=\frac{1}{g_m} a_m^\dagger,&   
  G\, a_m^\dagger\, G^{-1}&= g_m a_m^\dagger .
\end{align}
Theorem \ref{th:second} implies that the ``renormalized'' quantum Hall
state \eqref{eq:a:psiLaughGC} is a zero-energy eigenstate of
\begin{align}
  \label{eq:a:V1}
  V&=\sum_{m_1=-s}^s\, \sum_{m_2=-s}^s\, \sum_{m_3=-s}^s\, \sum_{m_4=-s}^s
  \ a_{m_1}^\dagger a_{m_2}^\dagger 
  a_{m_3}^{\phantom{\dagger}} a_{m_4}^{\phantom{\dagger}}\,
  \delta_{m_1+m_2,m_3+m_4}
  \nonumber\\[0.5\baselineskip]
  &\quad\cdot\, g_{m_1}g_{m_2}
  \!\braket{s,m_1;s,m_2}{2s,m_1+m_2} \braket{2s,m_3+m_4}{sm_3,sm_4}\!
  g_{m_3}g_{m_4}.
  \nonumber\\[0.2\baselineskip]
%   \\ \label{eq:a:U1} %[0.5\baselineskip]
%   U\hspace{-4pt}&=&\hspace{-4pt}
%   U_0\, a_{-s}^\dagger a_{-s}^{\phantom{\dagger}},
%   \hspace{8pt}
\end{align}
Since \eqref{eq:a:psiLaughGC} is likewise annihilated by
\eqref{eq:a:UqH}, it is also a zero energy state of
\begin{equation}
  \label{eq:a:H1}
  H%^{\s\text{rqH}} 
  = V%^{\s\text{rqH}}
  +U^{\s\text{qH}}.
\end{equation}
We will see in Section \ref{sec:a:ren} that \eqref{eq:a:psiLaughGC} is
a ground state of \eqref{eq:a:H1}, but we have not been able to deduce
this from the considerations presented so far.  For our purposes,
however, it is sufficient to know that \eqref{eq:a:H1} annihilates the
state \eqref{eq:a:psiLaughGC}.

With \eqref{eq:a:gm} and the explicit formula
% \begin{eqnarray}
%   \label{eq:a:cg1}
%   \braket{s,m_1;s,m_2}{2s,m} 
%   \hspace{-4pt}&=&\hspace{-4pt} 
%   \frac{\sqrt{(2s-m)!\,(2s+m)!}}{\sqrt{(s-m_1)!\,(s+m_1)!\,(s-m_2)!\,(s+m_2)!}}
%   \nonumber\\[0.5\baselineskip] &&\hspace{-6pt}
%   \cdot\,\frac{\sqrt{s}\cdot (2s-1)!}{\sqrt{(4s-1)!}}\cdot\delta_{m,m_1+m_2},
% \end{eqnarray}
\begin{align}
  \label{eq:a:cg1}
  \braket{s,m_1;s,m_2}{2s,m_1+m_2} 
  &=  \frac{\sqrt{(2s-m_1+m_2)!\,(2s+m_1+m_2)!}}
       {\sqrt{(s-m_1)!\,(s+m_1)!\,(s-m_2)!\,(s+m_2)!}}
  \nonumber\\[0.3\baselineskip] 
  &\quad\cdot\,\frac{\sqrt{s}\cdot (2s-1)!}{\sqrt{(4s-1)!}}
\end{align}
for the Clebsch--Gordan coefficients~\cite{baym69}, we obtain
% \begin{equation}
%   \label{eq:a:gmcg}
%   g_{m_1}g_{m_2}\!\braket{s,m_1;s,m_2}{2s,m}
%   =\frac{2\pi\cdot \delta_{m,m_1+m_2}}{(2s+1)\sqrt{s\cdot (4s-1)!}}
%   \sqrt{(2s-m)!\,(2s+m)!} 
% \end{equation}
%
% \begin{eqnarray}
%   \label{eq:a:gmcg}
%   g_{m_1}g_{m_2}\!\braket{s,m_1;s,m_2}{2s,m}
%   \hspace{-4pt}&=&\hspace{-4pt} 
%   \sqrt{(2s-m)!\,(2s+m)!}
%   \nonumber\\[0.2\baselineskip] &&\hspace{-6pt}
%   \cdot\,\frac{2\pi}{(2s+1)\sqrt{s\,(4s-1)!}}
%   \cdot\delta_{m,m_1+m_2}.
% \end{eqnarray}
\begin{align}
  \label{eq:a:gmcg}
  g_{m_1}g_{m_2}\!\braket{s,m_1;s,m_2}{2s,m_1+m_2}
  &=\sqrt{(2s-m_1+m_2)!\,(2s+m_1+m_2)!}
  \nonumber\\[0.3\baselineskip] 
  &\quad\cdot\,\frac{2\pi}{(2s+1)\sqrt{s\,(4s-1)!}}.
%  \cdot\delta_{m,m_1+m_2}.
\end{align}
The second factor in \eqref{eq:a:gmcg} does not depend on any $m_i$ and
can hence be absorbed by rescaling $V$ accordingly.  This yields
\begin{equation}
  \label{eq:a:V2}
  V%^{\s\text{rqH}}
  =\sum_{m_1=-s}^s\,\sum_{m_2=-s}^s\,\sum_{m_3=-s}^s\,\sum_{m_4=-s}^s\,
  a_{m_1}^\dagger a_{m_2}^\dagger 
  a_{m_3}^{\phantom{\dagger}} a_{m_4}^{\phantom{\dagger}}\,
  V_{m_1,m_2,m_3,m_4}
%  + U_0\, a_{-s}^\dagger a_{-s}^{\phantom{\dagger}}
\end{equation}
with 
\begin{gather}%{equation}
  \label{eq:a:Vmmmm}
    V_{m_1,m_2,m_3,m_4}=V_{m_1+m_2}\cdot\delta_{m_1+m_2,m_3+m_4},
    \\[0.3\baselineskip]
  \label{eq:a:Vm}
    V_m={(2s-m)!\,(2s+m)!}.
\end{gather}%{equation}
The essential simplification we have encountered so far is that the
scattering matrix elements $V_{m_1,m_2,m_3,m_4}$ in \eqref{eq:a:V2}
depend only on the conserved total value of $L^\z$, $m_1+m_2=m_3+m_4$,
and not on the (angular) momentum transfer.

Even though the Hamiltonian \eqref{eq:a:H1} with \eqref{eq:a:V2} and
\eqref{eq:a:UqH} annihilates the Haldane--Shastry ground state
\eqref{eq:a:psiLaughGC}, we a still very far from having derived the
Haldane--Shastry Hamiltonian \eqref{eq:a:hsham}.
First, \eqref{eq:a:V2} scatters single particle states in momentum
space, since $m=q-s$ is effectively a momentum quantum number.
Second, \eqref{eq:a:H1} is not likely to share the symmetries of
\eqref{eq:a:hsham}.
Third, we do not even know whether \eqref{eq:a:psiLaughGC} is the
(non-degenerate) ground state of \eqref{eq:a:H1}.

\section{Fourier transformation}
\label{sec:a:four}
%\subsection{Fourier transformation of operators}
\subsection{Particle creation and annihilation operators}
%\vspace{.5\baselineskip} 
%\emph{Fourier transformation.}---%
We proceed by transforming the interaction Hamiltonian \eqref{eq:a:V2}
into Four\-ier space.  To this end, we define the transformations
\begin{equation}
  \label{eq:a:aFT}
  a_m
  =\frac{1}{\sqrt{N}}\sum_{\alpha=1}^N (\bar\eta_\alpha)^{s+m} a_\alpha,\quad
  a_m^\dagger
  =\frac{1}{\sqrt{N}}\sum_{\alpha=1}^N (\eta_\alpha)^{s+m} a_\alpha^\dagger,
\end{equation}
where $N=2s+1$, %and $\eta_\alpha=e^{\text{i}\frac{2\pi}{N}\alpha}$.
$\eta_\alpha=e^{\text{i}\frac{2\pi}{N}\alpha}$, and
$\bar\eta_\alpha=e^{-\text{i}\frac{2\pi}{N}\alpha}$.  We may interpret
$\alpha$ as site indices of a periodic chain with $N$ sites, and
$\eta_\alpha$ as the positions of these sites when the periodic chain
is embedded as a unit circle in the complex plane.

The Fourier transformation yields
\begin{equation}
  \label{eq:a:V3}
  V = \frac{1}{N^2}\sum_{\{\alpha_1,\alpha_2,\alpha_3,\alpha_4\}}
  a_{\alpha_4}^\dagger a_{\alpha_3}^\dagger 
  a_{\alpha_2}^{\phantom{\dagger}} a_{\alpha_1}^{\phantom{\dagger}}\,
  V_{\alpha_1,\alpha_2,\alpha_3,\alpha_4}
\end{equation}
with
\begin{align}
  \label{eq:a:Valpha0}
    V_{\alpha_1,\alpha_2,\alpha_3,\alpha_4}
    &=\sum_{m_1=-s}^s\,\sum_{m_2=-s}^s\,\sum_{m_3=-s}^s\,\sum_{m_4=-s}^s
    V_{m_1+m_2}\,\delta_{m_1+m_2,m_3+m_4}\,
    \nonumber\\[0.3\baselineskip]
    &\hspace{30pt}\cdot\, (\eta_{\alpha_4})^{s+m_4}(\eta_{\alpha_3})^{s+m_3}
        (\bar\eta_{\alpha_2})^{s+m_2}(\bar\eta_{\alpha_1})^{s+m_1}
\end{align}
for the interaction Hamiltonian \eqref{eq:a:V2} and 
\begin{align}
  \label{eq:a:psialpha1}
  \ket{\psi_0}%^{\s\text{rqH}}}
  &=G^{-1}\ket{\psi^{\s\text{qH}}_0}
  \nonumber\\[0.2\baselineskip]
  &=\sum_{\{\alpha_1,\ldots ,\alpha_M\}} 
  \frac{1}{\sqrt{N}^M}
  \sum_{\{m_1,\ldots ,m_M\}} 
%  C_{s+m_1,\ldots ,s+m_M}\,
  C_{m_1+s,\ldots ,m_M+s}\,
  (\eta_{\alpha_1})^{s+m_1}\ldots (\eta_{\alpha_M})^{s+m_M}
  \nonumber\\[0.2\baselineskip]
  &\hspace{30pt}\cdot\,
  {a}^\dagger_{\alpha_1}\ldots {a}^\dagger_{\alpha_M} \vac
  \nonumber\\[0.5\baselineskip]
  &=\sum_{\{\alpha_1,\ldots ,\alpha_M\}} 
  \psi^{\s\text{HS}}_0(\eta_{\alpha_1},\ldots,\eta_{\alpha_M})\,
  {a}^\dagger_{\alpha_1}\ldots {a}^\dagger_{\alpha_M} \vac
\end{align}
for the ground state it annihilates.  In \eqref{eq:a:psialpha1}, we
have used the definition of the coefficients $C_{m_1+s,\ldots ,m_M+s}$
from \eqref{eq:a:hsket}--\eqref{eq:a:hsCket}.  Since
$\psi^{\s\text{HS}}_0(\eta_{\alpha_1},\ldots,\eta_{\alpha_M})$
vanishes identically whenever two coordinates $\eta_{\alpha}$
coincide, we are allowed to discard configurations with multiply
occupied sites. % from the Hilbert space.
This yields a reduced Hilbert space in which the boson creation and
annihilation operators $a^\dagger$ and $a$ obey the same commutation
relations as the spin flip operators ${S}^+$ and ${S}^-$.  We may
hence substitute one for the each other.

If we substitute $a_{\alpha_i}^\dagger\to {S}^+_{z_i}$,
$a_{\alpha_i}\to {S}^-_{z_i}$, in \eqref{eq:a:V3} and
\eqref{eq:a:psialpha1}, we find that the Haldane--Shastry ground state
\eqref{eq:a:hsket} with \eqref{eq:a:hspsi0} is annihilated by the
interaction Hamiltonian
\begin{equation}
  \label{eq:a:Vspin1}
  V = \frac{1}{N^2}\sum_{\{\alpha_1,\alpha_2,\alpha_3,\alpha_4\}}
  {S}^+_{\alpha_4} {S}^+_{\alpha_3} 
  {S}^-_{\alpha_2} {S}^-_{\alpha_1}\,
  V_{\alpha_1,\alpha_2,\alpha_3,\alpha_4}\,
\end{equation}
with the matrix elements \eqref{eq:a:Valpha0}.  For the on-site
potential term \eqref{eq:a:UqH}, Fourier transformation and subsequent
substitution yields
\begin{equation}
  \label{eq:a:UqHsub}
  U^{\s\text{qH}} = \frac{1}{N} U_0\, {S}_{\text{tot}}^+ {S}_{\text{tot}}^-,
\end{equation}
where ${\bs{S}}_{\text{tot}}$ is defined in \eqref{eq:hsspinsymmetry}.
%\begin{equation}
%  \label{eq:a:Stot}
%  {\bs{S}}_{\text{tot}} = \sum_{\alpha=1}^N {\bs{S}}_\alpha.
%\end{equation}
This term annihilates any singlet state, and in particular the
Haldane--Shastry ground state \eqref{eq:a:hsket} with
\eqref{eq:a:hspsi0}.  It will not be helpful in constructing a parent
Hamiltonian, but it might be useful to keep in mind that this term was
required to single out the ground state wave function on the quantum
Hall sphere.

These observations, and in particular \eqref{eq:a:Vspin1} with
\eqref{eq:a:Valpha0} and \eqref{eq:a:Vm}, are the results of the
considerations presented so far, and the starting point for the
analysis below.

%\subsection{Fourier transformation of the renormalized matrix elements}
\subsection{Renormalized matrix elements}
%\section{Fourier transformation of the renormalized matrix elements}
\label{sec:a:ren}

In this section, we wish to obtain an explicit expression for the
scattering matrix elements %$V_{\alpha_1,\alpha_2,\alpha_3,\alpha_4}$
\eqref{eq:a:Valpha0}
% \begin{equation}
%   \label{eq:a:Valpha0a}
%   \begin{split}
%     V_{\alpha_1,\alpha_2,\alpha_3,\alpha_4}
%     &=\sum_{m_1=-s}^s\,\sum_{m_2=-s}^s\,\sum_{m_3=-s}^s\,\sum_{m_4=-s}^s
%     V_{m_1+m_2}\,\delta_{m_1+m_2,m_3+m_4}\,
%     \\[0.5\baselineskip]
%     &\hspace{30pt}\cdot\, (\eta_{\alpha_4})^{s+m_4}(\eta_{\alpha_3})^{s+m_3}
%         (\bar\eta_{\alpha_2})^{s+m_2}(\bar\eta_{\alpha_1})^{s+m_1}
%         \\[0.3\baselineskip] %\hspace{40pt}
%   \end{split}
% \end{equation}
of \eqref{eq:a:Vspin1} for general $V_m$ by direct evaluation.  For
convenience, we assume $\alpha_1\ne\alpha_2$ and
$\alpha_3\ne\alpha_4$, as enforced by the spin flips in
\eqref{eq:a:Vspin1}.

This transformation may look trivial at first, but it is not.  When we
perform a conventional Fourier transform from real space into momentum
space or vice versa, both spaces are periodic.  In particular, if we
scatter a momentum across the boundary at one end of the Brillouin
zone, it will just reappear at the other boundary.  The distinguishing
feature of the $L^\z$ angular momentum quantum number $m$ is that it is
not subject to periodic, but to hard wall boundary conditions if we
attempt to scatter $m$ to values smaller than $-s$ or larger than $s$.
%beyond the interval from $-s$ to $s$. 
This does not preclude a Fourier transformation, but it does lead to phase
space restrictions we have to take into account.

The $\delta$-function in 
%\eqref{eq:a:Valpha0a} 
\eqref{eq:a:Valpha0}
allows us to eliminate the two summations over $m_3$ and $m_4$ in
favor of a single summation,
\begin{align}
  \label{eq:a:Valpha1}
    V_{\alpha_1,\alpha_2,\alpha_3,\alpha_4}
    &=\sum_{m_1=-s}^s\,\sum_{m_2=-s}^s V_{m_1+m_2}\,{\sum_q}'
    \nonumber\\[0.3\baselineskip]
    &\hspace{30pt}\cdot\, (\eta_{\alpha_4})^{s+m_2-q}(\eta_{\alpha_3})^{s+m_1+q} 
        (\bar\eta_{\alpha_2})^{s+m_2}(\bar\eta_{\alpha_1})^{s+m_1},
\end{align}
where %we have set $m_3=m_1+q$, $m_4=m_2-q$, and
\begin{equation*}
  \label{eq:a:sum'1}
  {\sum_q}'\equiv \sum_{q=\max\{-s-m_1,-s+m_2\}}^{\min\{s-m_1,s+m_2\}}.
\end{equation*}
With $m=m_2+m_1$, $p=m_2-m_1$, we write
\begin{equation*}
  \label{eq:a:sum'2}
  {\sum_q}'=\sum_{q=-s+\frac{1}{2}\max\{p-m,p+m\}}^{s+\frac{1}{2}\min\{p-m,p+m\}}
  =\sum_{q=-s+\frac{p}{2}+\frac{|m|}{2}}^{s+\frac{p}{2}-\frac{|m|}{2}}.
\end{equation*}
% This implies
% \begin{equation*}
%   \label{eq:a:sum'3}
%   {\sum_q}' (\eta_{34})^{q}
%   =\sum_{q=-s+\frac{1}{2}}^{s+\frac{1}{2}} (\eta_{34})^{q}
%   -\sum_{q=-s+\frac{1}{2}}^{-s+\frac{p}{2}+\frac{|m|}{2}-1} (\eta_{34})^{q}
%   -\sum_{q=s+\frac{p}{2}-\frac{|m|}{2}+1}^{s+\frac{1}{2}} (\eta_{34})^{q}.
% \end{equation*}
% where $\eta_{34}\equiv \eta_{\alpha_3-\alpha_4}=\eta_{\alpha_3}\bar\eta_{\alpha_4}$.
% Note that since $N=2s+1$ is even, $s$, $m_1=\frac{m-p}{2}$, and 
% $m_2=\frac{m+p}{2}$ are half-integer.
% %
% % \begin{equation*}
% %   \label{eq:a:sum'3}
% %   {\sum_q}' (\eta_{\alpha_3-\alpha_4})^{q}
% %   =\sum_{q=-s}^s (\eta_{\alpha_3-\alpha_4})^{q}
% %   -\sum_{q=-s}^{-s+\frac{p+|m|}{2}-1} (\eta_{\alpha_3-\alpha_4})^{q}
% %   -\sum_{q=s+\frac{p-|m|}{2}+1}^{s} (\eta_{\alpha_3-\alpha_4})^{q}.
% % \end{equation*}
% % where $\eta_{\alpha_3-\alpha_4}=\eta_{\alpha_3}\bar\eta_{\alpha_4}$.
% %
% The first term gives $N\delta_{\alpha_3,\alpha_4}$, and vanishes for
% $\alpha_3\ne\alpha_4$.  Using the periodicity in Fourier
% space,
% \begin{equation}
%   \label{eq:a:-s->s+1}
%   (\eta_{\alpha})^{-s+n}=(\eta_{\alpha})^{s+n+1},
% \end{equation}
% we can combine the remaining two terms into
% \begin{equation*}
%   \label{eq:a:sum'4}
%   {\sum_q}' (\eta_{34})^{q}
% %  =\delta_{\alpha_3,\alpha_4}
%   = -\sum_{q=s+\frac{p}{2}-\frac{|m|}{2}+1}^{s+\frac{p}{2}+\frac{|m|}{2}} 
%   (\eta_{34})^{q}.
% \end{equation*}
With
\begin{equation}
  \label{eq:a:x^qSum}
  \sum_{q=a}^b x^q = \frac{x^{b+1}-x^a}{x-1}
  = \frac{x^{b+\frac{1}{2}}-x^{a-\frac{1}{2}}}{x^{\frac{1}{2}}-x^{-\frac{1}{2}}},
  \qquad b\ge a,
\end{equation}
we obtain
\begin{equation*}
  \label{eq:a:sum'5}
  \begin{split}
%    \sum_{q=-s+\frac{p}{2}+\frac{|m|}{2}}^{s+\frac{p}{2}-\frac{|m|}{2}}    
    {\sum_q}' 
    (\eta_{34})^{q} 
    &=\frac{(\eta_{34})^{s+\frac{p}{2}-\frac{|m|}{2}+\frac{1}{2}}
           -(\eta_{34})^{-s+\frac{p}{2}+\frac{|m|}{2}-\frac{1}{2}}}
           {(\eta_{34})^{\frac{1}{2}}-(\eta_{34})^{-\frac{1}{2}}},
  \end{split}
\end{equation*}
where $\eta_{34}\equiv
\eta_{\alpha_3-\alpha_4}=\eta_{\alpha_3}\bar\eta_{\alpha_4}$.  Note
that $\eta_{34}\ne 1$ as $\alpha_3\ne\alpha_4$.  Using the periodicity
in Fourier space,
\begin{equation}
  \label{eq:a:-s->s+1}
  (\eta_{\alpha})^{-s}=(\eta_{\alpha})^{s+1},
\end{equation}
we can rewrite the second term in the numerator, and obtain
\begin{equation*}
  \label{eq:a:sum'6}
  \begin{split}
%    \sum_{q=-s+\frac{p}{2}+\frac{|m|}{2}}^{s+\frac{p}{2}-\frac{|m|}{2}}    
    {\sum_q}' 
    (\eta_{34})^{q} 
    &= -(\eta_{34})^{s+\frac{p}{2}+\frac{1}{2}} J(|m|,\alpha_3-\alpha_4),
  \end{split}
\end{equation*}
where
\begin{equation}
  \label{eq:a:Jm}
    J(|m|,\alpha)\equiv 
     \frac{(\eta_{\alpha})^{\frac{|m|}{2}}-(\eta_{\alpha})^{-\frac{|m|}{2}}}
          {(\eta_{\alpha})^{\frac{1}{2}}-(\eta_{\alpha})^{-\frac{1}{2}}}.
\end{equation}
Substitution into \eqref{eq:a:Valpha1} yields
\begin{align}
  \label{eq:a:Valpha2}
    V_{\alpha_1,\alpha_2,\alpha_3,\alpha_4}
    &=\sum_{m_1=-s}^s\,\sum_{m_2=-s}^s V_{m}\cdot
    (\eta_{42})^{s+m_2}(\eta_{31})^{s+m_1}
    \nonumber\\*[0.3\baselineskip]
    &\hspace{50pt}\cdot\,
    (-1)\cdot(\eta_{34})^{s+\frac{p}{2}+\frac{1}{2}}\, J(|m|,\alpha_3-\alpha_4).
\end{align}
% where $m=m_1+m_2$, as defined above.  

With $m_1=\frac{m-p}{2}$, $m_2=\frac{m+p}{2}$, we can rewrite the sums as
%We can rewrite the sums as
\begin{equation*}
  \sum_{m_1=-s}^s\,\,\sum_{m_2=-s}^s=
  \sum_{m=-2s}^{2s}\,\,\sum_{\substack{p=-2s+|m|\\[2pt] 
%                                      \text{$p$ even or odd,}\\[2pt]
%                                      \text{depending on $m$}}}^{2s-|m|}
                                      \text{even or odd}}}^{2s-|m|},
\end{equation*}
where the last sum extends only over even (odd) values of $p$ for $m$
odd (even).  (Since $N=2s+1$ is even, $2s$ is odd.)  This
yields
\begin{align}
  \label{eq:a:Valpha3}
    V_{\alpha_1,\alpha_2,\alpha_3,\alpha_4}
    &=\sum_{m=-2s}^{2s} V_{m}\cdot (-1)\cdot J(|m|,\alpha_3-\alpha_4)
    \nonumber\\[0.3\baselineskip]
    &\hspace{20pt}\cdot\,
    \sum_{\substack{p=-2s+|m|\\[2pt]\text{even or odd}}}^{2s-|m|}
    (\eta_{42})^{s+\frac{m}{2}+\frac{p}{2}}\,
    (\eta_{31})^{s+\frac{m}{2}-\frac{p}{2}}\,
    (\eta_{34})^{s+\frac{p}{2}+\frac{1}{2}}.
\end{align}
We proceed by evaluating the sum over the terms which depend on $p$,
\begin{equation*}
  \begin{split}
    \sum_{\substack{p=-2s+|m|\\[2pt]\text{even or odd}}}^{2s-|m|}
    (\eta_{42})^{\frac{p}{2}}\, (\eta_{31})^{-\frac{p}{2}}\,
    (\eta_{34})^{\frac{p}{2}}
%    &=\sum_{\substack{p=-2s+|m|\\[2pt]\text{even or odd}}}^{2s-|m|}
%    (\eta_{12})^{\frac{p}{2}}
%    \\[0.3\baselineskip]
    &=\sum_{k=-s+\frac{|m|}{2}}^{s-\frac{|m|}{2}}(\eta_{12})^k
    \\[0.3\baselineskip]
    &=\frac{(\eta_{12})^{ s-\frac{|m|}{2}+\frac{1}{2}}
           -(\eta_{12})^{-s+\frac{|m|}{2}-\frac{1}{2}}}
           {(\eta_{12})^{\frac{1}{2}}
           -(\eta_{12})^{-\frac{1}{2}}}
    \\[0.5\baselineskip]
    &= -(\eta_{12})^{s+\frac{1}{2}} J(|m|,\alpha_1-\alpha_2),
  \end{split}
\end{equation*}
where we have used $(\eta_{12})^{-s}=(\eta_{12})^{s+1}$ and $\eta_{12}\ne 1$.
Substitution into \eqref{eq:a:Valpha3} yields
\begin{align}
  \label{eq:a:Valpha4}
    V_{\alpha_1,\alpha_2,\alpha_3,\alpha_4}
    &=\sum_{m=-2s}^{2s} V_{m}\cdot 
    J(|m|,\alpha_3-\alpha_4)\,J(|m|,\alpha_1-\alpha_2)
    \nonumber\\[0.4\baselineskip]
    &\hspace{60pt}\cdot\,
    (\eta_{42}\eta_{31})^{s+\frac{m}{2}}\,
    (\eta_{34}\eta_{12})^{s+\frac{1}{2}}.
\end{align}
Writing out the factors in the second line yields
\begin{equation*}
  (\et{4})^{\frac{m}{2}-\frac{1}{2}}
%  \underbrace{(\eta_3)^{2s+\frac{m}{2}+\frac{1}{2}}}_{(\eta_3)^{\frac{m}{2}-\frac{1}{2}}}
  (\et{3})^{2s+\frac{m}{2}+\frac{1}{2}}
  (\et{2})^{-2s-\frac{m}{2}-\frac{1}{2}}
  (\et{1})^{-\frac{m}{2}+\frac{1}{2}}
%  =(\eta_{42}\eta_{31})^{\frac{m}{2}-\frac{1}{2}}.
  =\left(\frac{\et{4}\et{3}}{\et{2}\et{1}}\right)^{\frac{m}{2}-\frac{1}{2}}.
\end{equation*}
With the definition \eqref{eq:a:Jm} we obtain
\begin{align}
  \label{eq:a:Valpha5}
    V_{\alpha_1,\alpha_2,\alpha_3,\alpha_4}
    &=\sum_{m=-2s}^{2s} V_{m}\cdot 
     \frac{(\eta_{34})^{\frac{|m|}{2}}
          -(\eta_{34})^{-\frac{|m|}{2}}}
          {(\eta_{34})^{\frac{1}{2}}
          -(\eta_{34})^{-\frac{1}{2}}}\cdot
     \frac{(\eta_{12})^{\frac{|m|}{2}}
          -(\eta_{12})^{-\frac{|m|}{2}}}
          {(\eta_{12})^{\frac{1}{2}}
          -(\eta_{12})^{-\frac{1}{2}}}\,
    \nonumber\\[0.4\baselineskip]
    &\hspace{60pt}\cdot\,
%    (\eta_{42}\eta_{31})^{\frac{m}{2}-\frac{1}{2}}
    \left(\frac{\et{4}\et{3}}{\et{2}\et{1}}\right)^{\frac{m}{2}-\frac{1}{2}}.
\end{align}
Note that we may omit the absolute value signs from $m$, as both
fractions in \eqref{eq:a:Valpha5} change their sign with $m$.
This yields
\begin{equation}
  \label{eq:a:Valpha6}
  \begin{split}
    V_{\alpha_1,\alpha_2,\alpha_3,\alpha_4}
    &=\sum_{m=-2s}^{2s} V_{m}\cdot 
     \frac{\et{4}^{m}-\et{3}^{m}}{\et{4}^{ }-\et{3}^{ }}\cdot
     \frac{\etb{2}^{m}-\etb{1}^{m}}
          {\etb{2}^{ }-\etb{1}^{ }}.
    \\[0.2\baselineskip] %\hspace{40pt}
  \end{split}
\end{equation}

\subsection{An alternative derivation}
\label{sec:a:renalt}
Inspired by the result \eqref{eq:a:Valpha6}, we realize that there is
an alternative derivation, which will lend itself to generalization to
the case $S=1$.  To begin with, note that the matrix elements
\eqref{eq:a:Valpha0} may be written
\begin{equation}
  \label{eq:a:Valpha1a}
  \begin{split}
    V_{\alpha_1,\alpha_2,\alpha_3,\alpha_4}
    &=\sum_{m=-2s}^{2s} V_{m}\cdot 
    \bar A_{m;\alpha_4,\alpha_3}\, %V_{m}\, 
    A_{m;\alpha_2,\alpha_1},
    \\[0.2\baselineskip] %\hspace{40pt}
  \end{split}
\end{equation}
where we have defined the sums
\begin{eqnarray}
  \label{eq:a:amaa}
  A_{m;\alpha_1,\alpha_2}
  \hspace{-4pt}&=&\hspace{-4pt}
  \sum_{m_1=-s}^s\,\sum_{m_2=-s}^s\,(\etb{2})^{s+m_2}(\etb{1})^{s+m_1}\,
  \delta_{m,m_1+m_2},
  \nonumber\\[0.2\baselineskip]
  \bar A_{m;\alpha_3,\alpha_4}
  \hspace{-4pt}&=&\hspace{-4pt}
  \sum_{m_3=-s}^s\,\sum_{m_4=-s}^s\,(\et{4})^{s+m_4}(\et{3})^{s+m_3}\,
  \delta_{m,m_3+m_4}.
\end{eqnarray}
As these sums are complex conjugates to each other, it is sufficient to
evaluate $\bar A_{m;\alpha_1,\alpha_2}$.  With $m_2=m-m_1$ and the restriction
$-s\le m_2\le s$, we find $-s+m\le m_1\le s+m$.  This yields
\begin{equation}
  \label{eq:a:amaa1}
  \begin{split}
    A_{m;\alpha_1,\alpha_2}
    &=(\etb{2})^{2s+m}\,
    \sum_{m_1=\max\{-s,-s+m\}}^{\min\{s,s+m\}} (\etba_{12})^{s+m_1},
  \end{split}
\end{equation}
where $\etba_{12}=\etba_{\alpha_1-\alpha_2}$.  With \eqref{eq:a:x^qSum},
the sum gives for $0\le m\le 2s$ 
\begin{equation*}
  \sum_{m_1=-s+m}^{s}(\etba_{12})^{s+m_1}
  =\frac{(\etba_{12})^{2s+1}-(\etba_{12})^{m}}{\etba_{12}-1},
\end{equation*}
and for $-2s\le m<0$
\begin{equation*}
  \sum_{m_1=-s}^{s+m}(\etba_{12})^{s+m_1}
  =\frac{(\etba_{12})^{2s+1+m}-1}{\etba_{12}-1}.
\end{equation*}
% \begin{equation*}
%   \begin{split}
%   \sum_{m_1=-s+m}^{s}(\etba_{12})^{s+m_1}
%   &=\frac{(\etba_{12})^{2s+1}-(\etba_{12})^{m}}{\etba_{12}-1}\qquad\text{for}\ m>0
%      \\[0.2\baselineskip] 
%   \sum_{m_1=-s}^{s+m}(\etba_{12})^{s+m_1}
%   &=\frac{(\etba_{12})^{2s+1+m}-1}{\etba_{12}-1}\qquad\text{for}\ m<0
%   \end{split}
% \end{equation*}
With $(\etb{})^{2s+1}=1$, we obtain
\begin{align}
  \label{eq:a:amaa2}
    A_{m;\alpha_1,\alpha_2}
    &=-\sign(m)\cdot(\etb{2})^{m-1}\,
    \frac{(\etb{1}\et{2})^{m}-1}{\etb{1}\et{2}-1}
    \nonumber\\[0.2\baselineskip] 
    &=-\sign(m)\cdot\frac{\etb{1}^{m}-\etb{2}^{m}}{\etb{1}-\etb{2}},
\end{align}
where 
%$\sign(m)=\pm m$ for $m\gtrless 0$, $\sign(0)=0$.
%$\sign(m)$ is the sign of $m$.  
we have defined 
\begin{equation}
  \label{eq:a:sign}
  \sign(m)\equiv\left\{
    \begin{aligned} 
       1&,\quad m>0, \\[-.2\baselineskip] 
       0&,\quad m=0, \\[-.2\baselineskip] 
      -1&,\quad m<0.
    \end{aligned}\right.
\end{equation}
Since the signs cancels in the sum \eqref{eq:a:Valpha1a}, we obtain
\eqref{eq:a:Valpha6}.

\section{The \defining condition for the Gutzwiller state}
%\section{The characteristic condition for the Gutzwiller state}
%\section{The \defining condition for the Haldane--Shastry ground state}
%\section{The characteristic condition for the Haldane--Shastry ground state}
\label{sec:a:con}
%\subsection{The condition}
\subsection{Annihilation operators}
\label{sec:a:annhiop}
So far, we have shown that the Gutzwiller or Haldane--Shastry ground
state $\ket{\psi^{\s\text{HS}}_{0}}$ given by \eqref{eq:a:hsket} with
\eqref{eq:a:hspsi0} above is annihilated by the interaction
Hamiltonian
\begin{equation}
  \label{eq:a:Vspin2}
  V = \frac{1}{N^2}\sum_{\{\alpha_1,\alpha_2,\alpha_3,\alpha_4\}}
  {S}_{\alpha_4}^+ {S}_{\alpha_3^+} 
  {S}_{\alpha_2}^- {S}_{\alpha_1}^-\,
  V_{\alpha_1,\alpha_2,\alpha_3,\alpha_4}\,
\end{equation}
with the matrix elements \eqref{eq:a:Valpha6} and 
\begin{equation*}
  \label{eq:a:Vm1}
  V_m={(2s-m)!\,(2s+m)!}.
\end{equation*}
If we now define an operator
\begin{equation}
  \label{eq:a:Am}
  A_m\equiv \frac{1}{N}\sum_{\alpha_1\ne\alpha_2}^N
     \frac{\etb{2}^{m}-\etb{1}^{m}}{\etb{2}-\etb{1}}
  {S}^-_{\alpha_2} {S}^-_{\alpha_1}
  = \frac{2}{N}\sum_{\alpha_1\ne\alpha_2}^N
     \frac{\etb{1}^{m}}{\etb{1}-\etb{2}}
  {S}^-_{\alpha_2} {S}^-_{\alpha_1},
\end{equation}
we may rewrite \eqref{eq:a:Vspin2} as 
\begin{equation}
  \label{eq:a:Vspin3}
  V = \sum_{m=-2s}^{2s} V_{m} A_m^\dagger A_m.
\end{equation}
The fact that $V$ annihilates the Gutzwiller state
$\ket{\psi^{\s\text{HS}}_{0}}$ implies
\begin{align}
  \label{eq:a:<V>}
    \bra{\psi^{\s\text{HS}}_{0}} V \ket{\psi^{\s\text{HS}}_{0}} 
    &=\sum_{m=-2s}^{2s} V_{m} \bra{\psi^{\s\text{HS}}_{0}}
    A_m^\dagger A_m \ket{\psi^{\s\text{HS}}_{0}}
    \nonumber\\[0.4\baselineskip] 
    &=\sum_{m=-2s}^{2s} V_{m} 
    \bigl\Vert A_m \ket{\psi^{\s\text{HS}}_{0}} \bigr\Vert^2=0.
%    \\[0.2\baselineskip]
%    &=0
\end{align}
Since all the values $V_m$ for $-2s\le m\le 2s$ are positive, and the
norms of the vectors by definition non-negative, \eqref{eq:a:<V>}
implies that the vectors $A_m \ket{\psi^{\s\text{HS}}_{0}}$ must vanish
for all values of $m\in [-2s,2s]$.  Since $A_m$ is further periodic 
under $m\to m+N$ and $N\le 4s+1$, we have
\begin{equation}
  \label{eq:a:Ampsi0}
  A_m \ket{\psi^{\s\text{HS}}_{0}} = 0 \quad\forall\, m.
\end{equation}
This a much stronger condition than we could have hoped to obtain.
As an aside, the form \eqref{eq:a:Vspin3} implies that the spectrum of
$V$ is positive semi-definite, \ie all the eigenvalues are non-negative,
and hence that $\ket{\psi^{\s\text{HS}}_{0}}$ is a ground state.  Of course,
we do not know whether it is the only ground state.

Since the Gutzwiller or Haldane--Shastry state
$\ket{\psi^{\s\text{HS}}_{0}}$ is real or invariant under parity, \ie
under $\et{}\to\etb{}$, as shown in Section \ref{sec:hsgs}, it is also
annihilated by the complex conjugates $\bar A_m$ of $A_m$ for all $m$.

The state $\ket{\psi^{\s\text{HS}}_{0}}$ is further annihilated
by the operators
\begin{align}
  \label{eq:a:Omegadef}
  \OaHS &\equiv \frac{1}{2}\sum_{m=0}^{N} \eab^{m} \bar A_m
  \nonumber\\[0.2\baselineskip]
  &=\sum_{\substack{\beta=1\\[2pt]\beta\ne\alpha}}^N 
  \frac{1}{\ea-\eb} \Sa^- \Sb^-,\qquad 
  \OaHS \ket{\psi^{\s\text{HS}}_{0}} = 0 \quad\forall\, \alpha,
\end{align}
which are obtained from the complex conjugate of \eqref{eq:a:Am} by
Fourier transformation, as well as their complex conjugates:
\begin{equation}
  \label{eq:a:barOmegadef}
    \OabHS =\sum_{\substack{\beta=1\\[2pt]\beta\ne\alpha}}^N 
    \frac{1}{\eab-\ebb} \Sa^- \Sb^-,\qquad 
    \OabHS \ket{\psi^{\s\text{HS}}_{0}} = 0 \quad\forall\, \alpha.
\end{equation}
Note that we would not need to exclude configurations with
$\beta=\alpha$, as the spin operators exclude these automatically.

In Section \ref{sec:a:ham}, we will use the operators $\Om{}$ to
construct a parent Hamiltonian, which is translationally invariant,
invariant under P and T, and invariant under SU(2) spin rotations, for
the Gutzwiller state $\ket{\psi^{\s\text{HS}}_{0}}$.  Not
surprisingly, this Hamiltonian will turn out to be the Haldane-Shastry
Hamiltonian \eqref{eq:a:hsham} plus a constant to account for the
ground state energy \eqref{eq:hse0}.

This implies that the Haldane-Shastry Hamiltonian is completely
specified by the condition \eqref{eq:a:Omegadef} plus the symmetries
mentioned in the previous paragraph.  Therefore, we will refer to
\eqref{eq:a:Omegadef} as the \emph{\defining condition} of the
Gutzwiller or Haldane--Shastry ground state.  The universality of this
condition is such that both the parent Hamiltonian of the bosonic
Laughlin state and the Haldane-Shastry Hamiltonian secretly use
\eqref{eq:a:Omegadef} or \eqref{eq:a:barOmegadef} to single out the
Jastrow polynomial \eqref{eq:a:jastrow2} as their ground state.

%\vspace{.5\baselineskip} 
\subsection{Direct verification}
Before proceeding, however, we wish to verify the \defining condition
\eqref{eq:a:Omegadef} directly for the Haldane--Shastry ground state
\eqref{eq:a:hspsi0}.  This only takes a few lines, and is reassuring
after the acrobatics we performed to derive it.  We have
\begin{align}
  \label{eq:a:Omegapsi}
    \hspace{0pt}\OaHS\ket{\psi^{\s\text{HS}}_{0}}
    &=\sum_{\substack{\beta=1\\[2pt]\beta\ne\alpha}}^N
    \frac{1}{\ea-\eb} \Sa^- \Sb^-
    \sum_{\{z_1,\ldots z_M\}} \psi^{\s\text{HS}}_{0}(z_1,z_2,\ldots z_M)\, 
    {S}^+_{z_1}\ldots {S}^+_{z_M}\ket{\dw\ldots\dw} 
    \nonumber\\[0.3\baselineskip]
    &=\sum_{\{z_1,\ldots z_M\}}
    \underbrace{\sum_{\substack{\beta=1\\[2pt]\beta\ne\alpha}}^N
%    \frac{1}{\ea-\eb}\,\psi^{\s\text{HS}}_{0}(\ea,\eb,z_3,\ldots z_M)\,
    \frac{\psi^{\s\text{HS}}_{0}(\ea,\eb,z_3,\ldots z_M)}{\ea-\eb}}_{=0}\,
    {S}^+_{z_3}\ldots {S}^+_{z_M}\ket{\dw\ldots\dw},
    \nonumber\\[-0.8\baselineskip]
\end{align}
since 
\begin{equation*}
  \frac{\psi^{\s\text{HS}}_{0}(\ea,\eb,z_3,\ldots z_M)}{\ea-\eb}
  =(\ea-\eb)\ea\eb\prod_{i=3}^M(\ea-z_i)^2(\eb-z_i)^2z_i
  \prod_{3\le i< j}^M(z_i-z_j)^2 
\end{equation*}
vanishes for $\beta=\alpha$ and contains only powers $\eb^1,\eb^2,\ldots , 
\eb^{N-2}$.  Note that the calculation for $\OabHS$ is almost identical,
since
\begin{equation*}
  \frac{\psi^{\s\text{HS}}_{0}(\ea,\eb,z_3,\ldots z_M)}{\eab-\ebb}
  =-\ea\eb
  \frac{\psi^{\s\text{HS}}_{0}(\ea,\eb,z_3,\ldots z_M)}{\ea-\eb}
%  =-(\ea-\eb)\ea^2\eb^2\prod_{i=3}^M(\ea-z_i)^2(\eb-z_i)^2z_i
%  \prod_{3\le i< j}^M(z_i-z_j)^2 
\end{equation*}
vanishes also for $\beta=\alpha$ and 
contains only powers $\eb^2,\eb^3,\ldots , \eb^{N-1}$.

%\newpage
\subsection{The role of the hole}

In Section \ref{sec:a:holepole}, we introduced a quasihole at the south
pole of the quantum Hall sphere, such that the quantum Hall and the
Haldane--Shastry ground state wave functions would resemble each more
closely and the Hilbert space dimensions of both models would match.
We introduced an additional term \eqref{eq:a:UqH} for the quantum Hall
Hamiltonian, which morphed into the total spin term
\eqref{eq:a:UqHsub} under Fourier transformation, and has played no
role since.  

The attentive reader will have noticed that 
% In fact, 
the creation of the quasihole has played no role in our analysis
up to \eqref{eq:a:Ampsi0} whatsoever.  In other words, if we had not
created it, instead of \eqref{eq:a:Ampsi0} we would have found 
that the state
\begin{align}
  \label{eq:a:hsketN=2M-1}
  \ket{\psi^{N=2M-1}_{0}}\,=
  \sum_{\{z_1,\ldots ,z_M\}}\psi^{N=2M-1}_{0} 
  (z_1,\ldots ,z_M)\,{S}^+_{z_1}\cdot\ldots\cdot {S}^+_{z_M} 
  \big|\underbrace{\dw\dw\ldots\ldots\dw}_{\text{all\ } N \text{\ spins\ } \dw}
  \big\rangle
  \nonumber\\* %[0.2\baselineskip]
\end{align}
with
\begin{equation}
  \label{eq:a:hspsiN=2M-1}
  \psi^{N=2M-1}_{0}(z_1,\ldots ,z_M) = 
  \prod_{i<i}^M\,(z_i-z_j)^2 %\,\prod_{i=1}^M\,z_i\,. 
\end{equation}
on a unit circle with $N=2M-1$ sites is annihilated by $A_m$ as
defined in \eqref{eq:a:Am},
\begin{equation}
%  \label{eq:a:AmpsiN=2M-1}
  \bar A_m \ket{\psi^{N=2M-1}_{0}} = 0 \quad\forall\, m.
%  \bar A_m \ket{\psi^{\s\text{nohole}}_{0}} = 0 \quad\forall\, m
\end{equation}
% In the definition $\displaystyle
% \eta_\alpha=e^{\text{i}\frac{2\pi}{N}\alpha }$, $N$ is of course
% throughout the analysis by one smaller than with the hole.  
The state \eqref{eq:a:hsketN=2M-1} is likewise annihilated by $\OabHS
\forall\,\alpha$, which can easily be verified directly along the
lines of \eqref{eq:a:Omegapsi}, as
%\begin{equation*}
\begin{multline*}
  \frac{\psi^{N=2M-1}_{0}(\ea,\eb,z_3,\ldots z_M)}{\eab-\ebb}\\[-.5\baselineskip]
  =-\ea\eb (\ea-\eb)\prod_{i=3}^M(\ea-z_i)^2(\eb-z_i)^2 \prod_{3\le
    i< j}^M(z_i-z_j)^2
\end{multline*}
%\end{equation*}
vanishes for $\beta=\alpha$ and contains only powers
$\eb^1,\eb^2,\ldots , \eb^{N-1}$ for $N=2M-1$.

The state \eqref{eq:a:hsketN=2M-1} with \eqref{eq:a:hspsiN=2M-1},
however, is neither real (and hence not invariant under P and T) nor a
spin singlet.  It is not annihilated by $\OaHS$ for any $\alpha$.  It is
not a sensible spin liquid, and we have no symmetries to construct a
Hamiltonian.  We conclude that while the quasihole is not essential to
the mapping of the model itself, it is essential to obtaining a
sensible spin model via this mapping.

\section{Rotations and spherical tensor operators }
\label{sec:a:tensorops}

As mentioned in Section \ref{sec:a:annhiop} above, we intend to use the
\defining condition \eqref{eq:a:Omegadef} to formulate a parent
Hamiltonian for the Gutzwiller ground state \eqref{eq:a:hspsi0}.  We
wish the Hamiltonian to be invariant under translations, parity and time
reversal transformations, and SU(2) spin rotations.  This last
invariance states that the Hamiltonian must transform as a scalar
under spin rotations, while $\OaHS$ transforms as a tensor of 2nd
order.  When we construct the Hamiltonian, we will project out certain
tensor components (like the scalar or vector component) from operators
which do not have simple transformation properties (\ie which consist
of tensor components of different orders).  For example, the operator
$\Sa^+\Sb^-+\Sa^-\Sb^+$ consists of both a scalar component and a 2nd
order tensor component.  
In Section \ref{sec:pf2s1}, we will have to analyze the tensor content
of more complicated operators, like ${\Sa^\z}^2(\Sb^+\Sc^-+\Sb^-\Sc^+)$.

% When we construct a parent Hamiltonian for the $S=1$ spin liquid
% \ref{eq:napsi0} in Section \ref{sec:b:ham}, we will have to project
% out the scalar component of the operators
% ${\Sa^\z}^2(\Sb^+\Sc^-+\Sb^-\Sc^+)$.

In this Section, we review the rotation properties of tensor
operators~\cite{gottfried66,baym69} including the use of Clebsch--Gordan
coefficients for projections onto certain tensor components.

\subsection{Representations of rotations}
%\subsection{Rotations}

%Since rotations are generated by the angular momentum operator $\bs{J}$, 
The angular momentum operator $\bs{J}$ is the generator of SU(2)
rotations.  Specifically, the operator
\begin{equation}
  \label{eq:a:R}
  R_{\bs{\omega}}=e^{-\text{i}\bs{J}\bs{\omega}},
\end{equation}
rotates a state vector by an angle
$|\bs{\omega}|$ around the axis $\bs{\omega}$.  Let $\ket{j,m}$ be an
eigenstate of $\bs{J}^2$ and $J^\z$ with eigenvalues $j(j+1)$ and $m$,
respectively.  Since \eqref{eq:a:R} commutes with the total angular
momentum,
%\begin{equation*}
%  \comm{R_{\bs{\omega}}}{\bs{J}^2}=0,
%\end{equation*}
%and the set of states $\ket{j,m}$ form a complete set,
the action of $R_{\bs{\omega}}$ on this state can only change $m$, \ie
\begin{equation}
  \label{eq:b:Rjmket}
  R_{\bs{\omega}}\ket{j,m}=\sum_{m'=-j}^j \ket{j,m'} d^{(j)}_{m'm}(\bs{\omega}).
\end{equation}
Since the states $\ket{j,m}$ form a complete basis set which does not
contain any subgroup of states which only transform under themselves,
the matrices
\begin{equation}
  \label{eq:a:dm'm}
  d^{(j)}_{m'm}(\bs{\omega})=\bra{j,m'}e^{-\text{i}\bs{J}\bs{\omega}}\ket{j,m}
\end{equation}
describe an irreducible, $2j+1$ dimensional representation of the
group SU(2)\footnote{For half integer $j$, these matrices constitute
  double valued representation of the rotation group O(3), and a
  single valued representation of the larger group SU(2).  For integer
  $j$, they are single valued representations of both groups.}.

\subsection{Tensor operators}

We can further use the operators \eqref{eq:a:R} %$R_{\bs{\omega}}$
to rotate operators,
\begin{equation}
  \label{eq:a:RAR}
  A\to R_{\bs{\omega}} A R_{\bs{\omega}}^{-1},
\end{equation}
such that the expectation value of an operator $A$ in a state
$\ket{\psi}$ is equal to the expectation value of the rotated operator
$R_{\bs{\omega}} A R_{\bs{\omega}}^{-1}$ in the rotated state
$R_{\bs{\omega}}\ket{\psi}$.  Certain operators transform as scalars
under rotations, which means that they commute with $\bs{J}$ and
remain unchanged under \eqref{eq:a:RAR}.  Other operators, like the
position vector $\bs{r}$ or the angular momentum operator $\bs{J}$,
transform as vectors.  In general, an irreducible tensor operator
$T^{(j)}$ of order $j$ has $2j+1$ components ${T^{(j)}}^{m}$,
$m=-j,\ldots,j$, which transform among themselves under rotations
according to
\begin{equation}
  \label{eq:b:Rjmtensor}
  R_{\bs{\omega}}\, {T^{(j)}}^m R_{\bs{\omega}}^{-1}
  =\sum_{m'=-j}^j {T^{(j)}}^{m'} d^{(j)}_{m'm}(\bs{\omega}),
\end{equation}
where the coefficients $d^{(j)}_{m'm}(\bs{\omega})$ are given by
\eqref{eq:a:dm'm}.  Clearly, a scalar is an irreducible tensor of
order $j=0$, and a vector is an irreducible tensor of order $j=1$.

If we write out \eqref{eq:b:Rjmtensor} for infinitesimal rotations %$\epsilon$
\begin{equation}
  \label{eq:a:Repsilon}
  R_{\bs{\epsilon}}
  =e^{-\text{i}\bs{J}\bs{\epsilon}}
  \approx 1-\text{i}\bs{J}\bs{\epsilon},
\end{equation}
and compare coefficients to first order in $\bs{\epsilon}$, we obtain
\begin{equation}
  \label{eq:b:commJjmtensor}
  \comm{\bs{J}}{{T^{(j)}}^m}
  =\sum_{m'=-j}^j {T^{(j)}}^{m'}\! \bra{j,m'}\bs{J}\ket{j,m}.
\end{equation}
With \eqref{eq:app-am:Jpmketlm}, this implies
%and hence with ...:
%\begin{eqnarray}
%  \label{eq:a:commJzT}
%  \comm{J^\z}{{T^{(j)}}^m}\hspace{-6pt}&=&\hspace{-6pt}m\; {T^{(j)}}^m\\[6pt]
%  \label{eq:a:commJ+-T}
%  \comm{J^\pm}{{T^{(j)}}^m}\hspace{-6pt}&=&\hspace{-6pt}
%  \sqrt{j(j+1)-m(m\pm 1)}\; {T^{(j)}}^{m\pm 1}
%\end{eqnarray}
\begin{gather}
  \label{eq:a:commJzT}
  \comm{J^\z}{{T^{(j)}}^m} = m\, {T^{(j)}}^m, \\[10pt]
  \label{eq:a:commJ+-T}
  \comm{J^\pm}{{T^{(j)}}^m} =
  \sqrt{j(j+1)-m(m\pm 1)}\,\, {T^{(j)}}^{m\pm 1},
\end{gather}
where $J^\pm\equiv J^\x\pm iJ^\y$.  Equations \eqref{eq:a:commJzT}
and \eqref{eq:a:commJ+-T} are fully equivalent to \eqref{eq:b:Rjmtensor},
but much more convenient to use in practise.  

Since a vector operator $\bs{V}$ obeys the commutation relations
\begin{equation*}
  \comm{J^i}{V^j}=\text{i}\epsilon^{ijk}V^k,
\end{equation*}
\eqref{eq:a:commJzT} and \eqref{eq:a:commJ+-T} imply that the tensor
components are (up to an overall normalization factor) given by
\begin{equation}
  \label{eq:a:vectorrep}
    V^{m=1}=-\frac{V^\x+\text{i}V^\y}{\sqrt{2}},\quad
    V^{m=0}=V^\z,\quad
    V^{m=-1}=\frac{V^\x-\text{i}V^\y}{\sqrt{2}}.
\end{equation}
% \begin{equation*}
%   \begin{split}
%     V^{m=1}&=-\frac{V^\x+\text{i}V^\y}{\sqrt{2}},\\[0.2\baselineskip]
%     V^{m=0}&=V^\z,\\[0.2\baselineskip]
%     V^{m=-1}&=\frac{V^\x-\text{i}V^\y}{\sqrt{2}}.
%   \end{split}
% \end{equation*}

Note that the $J^\z$ eigenvalue of
\begin{equation*}
  {T^{(j)}}^m \ket{j',m'}
\end{equation*}
is $m+m'$, as one can easily verify by either considering a rotation
\eqref{eq:b:Rjmtensor} around the $z$-axis, or directly with
\eqref{eq:a:commJzT},  
\begin{equation}
  \label{eq:a:comJzTket}
  \begin{split}
    J^\z\, {T^{(j)}}^m\! \ket{j',m'}
    &= \comm{J^\z}{{T^{(j)}}^{m}}\ket{j',m'} + {T^{(j)}}^m J^\z \ket{j',m'}
    \\[0.4\baselineskip]
    &= (m+m')\;{T^{(j)}}^m \ket{j',m'}.
  \end{split}
\end{equation}
The tensor operator ${T^{(j)}}^m$ hence increases the eigenvalue of
$J^\z$ by $m$.

%\subsection{Combinations of tensor operators}
\subsection{Products of tensor operators}

Similarly, the $J^\z$ quantum number $m$ of a product of two tensors 
\begin{equation}
  \label{eq:a:TT}
  {{T^{(j_1)}}^{m_1}{T^{(j_2)}}^{m_2}}
\end{equation}
is simply the sum of the $J^\z$ quantum numbers of the individual
tensors, $m=m_1+m_2$.  We can again verify this by considering a rotation 
\eqref{eq:b:Rjmtensor} around the $z$-axis, or directly with
\eqref{eq:a:commJzT}. %:
% \begin{equation}
%   \label{eq:a:comJzTT}
% %  \comm{J^\z}{{T^{(j_1)}}^{m_1}{T^{(j_2)}}^{m_2}} 
% %  = (m_1+m_2)\,{{T^{(j_1)}}^{m_1}{T^{(j_2)}}^{m_2}}.
%   \begin{split}
%     \comm{J^\z}{{T^{(j_1)}}^{m_1}{T^{(j_2)}}^{m_1}} 
%     &= \comm{J^\z}{{T^{(j_1)}}^{m_1}}{T^{(j_2)}}^{m_2}
%      + {T^{(j_1)}}^{m_1}\! \comm{J^\z}{{T^{(j_2)}}^{m_2}}
%     \\[0.4\baselineskip]
%     &= (m_1+m_2)\;{{T^{(j_1)}}^{m_1}{T^{(j_2)}}^{m_2}}.
%   \end{split}
% \end{equation}
The product \eqref{eq:a:TT}, however, is not an irreducible tensor,
but in general rather a sum of irreducible tensors of orders 
$|j_1-j_2|,\ldots,j_1+j_2$.

We can combine two tensors using Clebsch--Gordan coefficients,
however, to obtain a tensor of well-defined order $j$.  Specifically,
we can write
\begin{equation}
  \label{eq:a:TjmCG}
  {T^{(j)}}^m = \sum_{m_1=-j_1}^{j_1}\sum_{m_2=-j_2}^{j_2}
  {{T^{(j_1)}}^{m_1}{T^{(j_2)}}^{m_2}} \braket{j_1,m_1;j_2,m_2}{j,m},
\end{equation}
where $\braket{j_1,m_1;j_2,m_2}{j,m}$ are Clebsch--Gordan
coefficients.  To verify that the left-hand side of \eqref{eq:a:TjmCG}
is an irreducible tensor of order $j$, consider its transformation
properties under a rotation \eqref{eq:b:Rjmtensor} with coefficient
matrices \eqref{eq:a:dm'm}:
\begin{equation}
  \label{eq:a:RTR=TT}
  \begin{split}
    \hspace{20pt}&\hspace{-20pt}
    R_{\bs{\omega}}\, {T^{(j)}}^m R_{\bs{\omega}}^{-1}
    \\[0.2\baselineskip]
    &=\sum_{m_1,m_2}
    R_{\bs{\omega}}\,{T^{(j_1)}}^{m_1} R_{\bs{\omega}}^{-1}
    R_{\bs{\omega}}\,{T^{(j_2)}}^{m_2} R_{\bs{\omega}}^{-1}
    \braket{j_1,m_1;j_2,m_2}{j,m}
    \\[0.2\baselineskip]
    &=\sum_{m'_1,m'_2} {{T^{(j_1)}}^{m'_1}{T^{(j_2)}}^{m'_2}}
    \\[0.2\baselineskip]
    &\hspace{25pt}\cdot\sum_{m_1,m_2} 
    \bra{j_1,m'_1;j_2,m'_2} e^{-\text{i}\bs{J}\bs{\omega}}\ket{j_1,m_1;j_2,m_2}
    \braket{j_1,m_1;j_2,m_2}{j,m}
    \\[0.2\baselineskip]
    &=\sum_{m'_1,m'_2} {{T^{(j_1)}}^{m'_1}{T^{(j_2)}}^{m'_2}}
%    \\[0.2\baselineskip]
%    &\hspace{20pt}\cdot
    \sum_{j',m'} 
    \braket{j_1,m'_1;j_2,m'_2}{j',m'} 
    \bra{j',m'} e^{-\text{i}\bs{J}\bs{\omega}}\ket{j,m}
%    \\[0.2\baselineskip]
%    &=\sum_{m'}\,{T^{(j)}}^{m'}\! 
%    \bra{j,m'} e^{-\text{i}\bs{J}\bs{\omega}}\ket{j,m}
    \\[0.2\baselineskip]
    &=\sum_{m'}\,{T^{(j)}}^{m'}\! d^{(j)}_{m'm}(\bs{\omega}).
    \\[-\baselineskip]
\end{split}
\end{equation}
Here we have used the completeness relations 
\begin{gather}
  \label{eq:a:CGm1m2}
  \sum_{m_1=-j_1}^{j_1}\sum_{m_2=-j_2}^{j_2}  
%  \sum_{m_1,m_2} 
  \ket{j_1,m_1;j_2,m_2}\bra{j_1,m_1;j_2,m_2} = 1,
  \\[0.2\baselineskip]
  \label{eq:a:CGjm}
%  \sum_{j'=|j_1-j_2|}^{j_1+j_2}\sum_{m'=-j'}^{j'} \ket{j',m'}\bra{j',m'} = 1,
  \sum_{j=|j_1-j_2|}^{j_1+j_2}\sum_{m=-j}^{j} \ket{j,m}\bra{j,m} = 1,
\end{gather}
of the Clebsch--Gordan algebra, which are understood to be valid 
in a Hilbert space with fixed $j_1$ and $j_2$.

We can use the relations \eqref{eq:a:CGm1m2} and \eqref{eq:a:CGjm}
further to invert \eqref{eq:a:TjmCG}.  This yields
\begin{equation}
  \label{eq:a:Tm1m2CG}
  {{T^{(j_1)}}^{m_1}{T^{(j_2)}}^{m_2}} = \sum_{j=|j_1-j_2|}^{j_1+j_2}\sum_{m=-j}^{j}
  {T^{(j)}}^m \braket{j,m}{j_1,m_1;j_2,m_2}.
\end{equation}

Let us denote the projection of a tensor $A$ onto its $j$-th order
component tensor  by $\{A\}_j$.  Then \eqref{eq:a:Tm1m2CG} implies
\begin{equation}
  \label{eq:a:Tm1m2CGj}
  \begin{split}
    \{ {T^{(j_1)}}^{m_1}{T^{(j_2)}}^{m_2}\}_j &= {T^{(j)}}^{m_1+m_2}
    \braket{j,m_1+m_2}{j_1,m_1;j_2,m_2},
  \end{split}
\end{equation}
where ${T^{(j)}}^{m_1+m_2}$ is given by \eqref{eq:a:TjmCG}, \ie
\begin{equation}
  \label{eq:a:TjmCGj}
  {T^{(j)}}^m = \sum_{m_1=\max\{-j_1,-j_2+m\}}^{\min\{j_1,j_2+m\}}
  {{T^{(j_1)}}^{m_1}{T^{(j_2)}}^{m-m_1}} \braket{j_1,m_1;j_2,m-m_1}{j,m}.
\end{equation}
For $m=m_1=m_2=0$, we obtain
\begin{equation}
  \label{eq:a:Tm1m2CGj0}
  \begin{split}
    \hspace{20pt}&\hspace{-20pt}
    \{ {T^{(j_1)}}^{0}{T^{(j_2)}}^{0}\}_j =
    \\[0.2\baselineskip]
    &\braket{j,0}{j_1,0;j_2,0}\sum_{m=-\min\{j_1,j_2\}}^{\min\{j_1,j_2\}}
    {{T^{(j_1)}}^{m}{T^{(j_2)}}^{-m'}}\! \braket{j_1,m;j_2,-m}{0,0}.
  \end{split}
\end{equation}
%a formula we will use repeatedly below.
We will use this formula repeatedly below.

The tensors we can form out of up to three spin operators, and the tensor
decomposition of expressions like %$S_1^\z S_2^\z$, 
$S_1^+S_2^-$ or $S_1^\z S_2^+S_3^-$, are given in Appendix~\ref{sec:app-t}.

%\newpage
\section{Construction of a parent Hamiltonian for the Gutzwiller state}
\label{sec:a:ham}

We now turn to the construction of a parent Hamiltonian for the
Gutzwiller state \eqref{eq:a:hsket} with \eqref{eq:a:hspsi0} using the
annihilation operator \eqref{eq:a:Omegadef}, \ie
\begin{equation}
  \label{eq:a:Omegapro}
    \OaHS \ket{\psi^{\s\text{HS}}_{0}} = 0 \quad\forall\, \alpha,
    \quad\text{where}\quad
    \OaHS =\sum_{\substack{\beta=1\\\beta\ne\alpha}}^N 
    \frac{1}{\ea-\eb} \Sa^- \Sb^-.
\end{equation}
The Hamiltonian has to be Hermitian, and we wish it to be invariant
under translations, time reversal (T), parity (P), and SU(2) spin
rotations.

%\newpage
\subsection{Translational, time reversal, and parity symmetry}
\label{sec:a:ham:transTP}

The operator ${\OaHS}^\dagger\OaHS$ is Hermitian and positive semi-definite,
meaning that all the eigenvalues are non-negative.  A translationally
invariant operator is given by
\begin{align}
  \label{eq:a:h0}
  H_0=\sum_{\a=1}^N {\OaHS}^\dagger\OaHS
%  \nonumber\\[0.2\baselineskip] 
  &=\sum_{\substack{\a,\b,\c\\ \a\ne \b,\c}}
    \frac{1}{\eab-\ebb}\frac{1}{\ea-\ec} 
    \Sa^+\Sa^-\Sb^+\Sc^-
  \nonumber\\[0.2\baselineskip] 
  &=\sum_{\substack{\a,\b,\c\\ \a\ne \b,\c}}
  \omega_{\a\b\c}\left(\Sa^\z + \frac{1}{2}\right)\Sb^+\Sc^-,
\end{align}
where we have defined 
\begin{equation}
  \label{eq:a:omegaabc}
  \omega_{\a\b\c}\equiv \frac{1}{\eab-\ebb}\frac{1}{\ea-\ec}. 
\end{equation}

The transformation properties of the individual entities 
in \eqref{eq:a:h0} under time reversal (T) are~\cite{gottfried66}
\begin{align}
  \text{T:}\quad
%  \i \,\to\, \Pi\i\Pi = -\i,\quad
%  \i\to -\i,\quad
  \ea  \,\to\, \Theta\ea \Theta = \eab,\quad 
  \bsS \,\to\, \Theta\bsS\Theta = -\bsS, 
\end{align}
and hence
\begin{align}
  \omega_{\a\b\c}\to\omega_{\a\c\b},\quad
  S^+\to -S^-,\quad S^-\to -S^+,\quad S^\z\to -S^\z.
%   \Theta\,\omega_{\a\b\c}\,\Theta = \omega_{\a\c\b},\quad
%   \Theta S^+ \Theta= -S^-,\quad 
%   \Theta S^- \Theta= -S^=,\quad 
%   \Theta S^\z \Theta= -S^\z.
\end{align}
The operator \eqref{eq:a:h0} transforms into
\begin{align}
  \label{eq:a:Th0T}
  \Theta H_0\Theta & =\sum_{\substack{\a,\b,\c\\ \a\ne \b,\c}}
  \omega_{\a\b\c}\left( -\Sa^\z + \frac{1}{2}\right)\Sc^-\Sb^+.
\end{align}
We proceed with the T invariant operator
\begin{align}
  \label{eq:a:h0T}
  H_0^{\text{T}}=\frac{1}{2}\left(H_0+\Theta H_0 \Theta\right)
  = H_0^{\text{T}=}+H_0^{\text{T}\ne },
\end{align}
where
\begin{align}
  \label{eq:a:h0T=}
  H_0^{\text{T}=}&=\sum_{\substack{\a,\b\\ \a\ne \b}}\omega_{\a\b\b}
  \left(\frac{1}{2}\Sa^\z \bigcomm{\Sb^+}{\Sb^-}
    +\frac{1}{4}\biganticomm{\Sb^+}{\Sb^-}\right)
  \nonumber\\[0.2\baselineskip] 
  &=\sum_{\substack{\a,\b\\ \a\ne \b}}\omega_{\a\b\b}
  \left(\Sa^\z\Sb^\z + \frac{1}{4}\right),
  \\[0.4\baselineskip] 
  \label{eq:a:h0Tnot=}
  H_0^{\text{T}\ne} &=\frac{1}{2}\sum_{\substack{\a,\b,\c\\ \a\ne\b\ne\c\ne\a}}
  \omega_{\a\b\c}\Sb^+\Sc^-.
\end{align}

The transformation properties of the individual operators under
%one-dimensional 
parity (P) are~\cite{gottfried66}
\begin{align}
  \text{P:}\quad
%  \i \,\to\, \Pi\i\Pi = \i,\quad
  \ea  \,\to\, \Pi\ea\Pi = \eab,\quad 
  \bsS \,\to\, \Theta\bsS\Theta = \bsS,  
\end{align}
and hence $\omega_{\a\b\c}\to\omega_{\a\c\b}$.
% \begin{align}
%   \omega_{\a\b\c}\to\omega_{\a\c\b}.
% \end{align}
We proceed with the P and T invariant operator
\begin{align}
  \label{eq:a:h0PT}
  H_0^{\text{PT}}=\frac{1}{2}\left(H_0^{\text{T}}
    +\Pi\hspace{1pt} H_0^{\text{T}}\hspace{1pt} \Pi\right)
  = H_0^{\text{PT}=}+H_0^{\text{PT}\ne },
\end{align}
where
\begin{align}
  \label{eq:a:h0PT1}
  H_0^{\text{PT}=}&=H_0^{\text{T}=}, 
  \hspace{-10pt}&\hspace{-10pt}
%  \\[0.4\baselineskip] 
%  \label{eq:a:h0Tn}
  H_0^{\text{PT}\ne}&=
  \frac{1}{4}\sum_{\substack{\a,\b,\c\\ \a\ne\b\ne\c\ne\a}}\omega_{\a\b\c}
  \big(\Sb^+\Sc^- + \Sb^-\Sc^+\big).
\end{align}
Since the operator $\Sb^+\Sc^- + \Sb^-\Sc^+$ is symmetric under
interchange of $\b$ and $\c$, we can use \eqref{eq:app-hs15} from
Appendix~\ref{sec:app-hssums} to obtain
\begin{align}
  \label{eq:a:h0PT2}
  H_0^{\text{PT}\ne}
  =\frac{1}{2}\sum_{\a\ne\b}
  \omega_{\a\b\b}\big(\Sa^+\Sb^- + \Sa^-\Sb^+\big)
  -\frac{1}{8}\sum_{\a\ne\b}
  \big(\Sa^+\Sb^- + \Sa^-\Sb^+\big).
\end{align}
Adding \eqref{eq:a:h0T=} and \eqref{eq:a:h0PT2} together, we obtain
with \eqref{eq:app-hs12}
\begin{align}
  \label{eq:a:h0PT3}
  H_0^{\text{PT}}
%    &=\sum_{\a\ne\b}
%    \omega_{\a\b\b}\left(\bSa\bSb+\frac{1}{4}\right)
%    -\frac{1}{8}\sum_{\a\ne\b}
%    \big(\Sa^+\Sb^- + \Sa^-\Sb^+\big).
%    \nonumber\\[0.2\baselineskip] 
  &=\sum_{\a\ne\b}
  \frac{\bSa\bSb}{\vert\ea-\eb\vert^2} + \frac{N(N^2-1)}{48}
  -\frac{1}{8}\sum_{\a\ne\b}
  \big(\Sa^+\Sb^- + \Sa^-\Sb^+\big).
\end{align}

%\subsection{Projecting out the scalar component}
\subsection{Spin rotation symmetry}

The Haldane--Shastry ground state
$\ket{\psi^{\s\text{HS}}_{0}}$ 
%given by \eqref{eq:a:hsket} with \eqref{eq:a:hspsi0} 
is annihilated by \eqref{eq:a:h0PT3}, and is also
a spin singlet.  Since the different tensor components of
\eqref{eq:a:h0PT3} yield states 
%with different transformation properties  
which transform according to different representations 
under SU(2) spin rotations when we act with them on
$\ket{\psi^{\s\text{HS}}_{0}}$, each tensor component must annihilate
$\ket{\psi^{\s\text{HS}}_{0}}$ individually.

With the exception of the last term, \eqref{eq:a:h0PT3}
transforms like a scalar under spin rotations.  With
\eqref{eq:app-t-S+-S-+}, we find that the scalar component of the last
term of \eqref{eq:a:h0PT3} is given by
\begin{align}
   -\frac{1}{6}\sum_{\a\ne\b} \bSa\bSb
   =-\frac{1}{6}\bsS_{\text{tot}}^2+\frac{1}{6}\sum_{\a}\bSa^2
   =-\frac{1}{6}\bsS_{\text{tot}}^2+\frac{N}{8}.
\end{align}
The scalar component of \eqref{eq:a:h0PT3} is therefore given by
\begin{align}
  \label{eq:a:h0PT0}
%  H_0^{\text{PT\{0}\}}\equiv
%  H_0^{\text{PT0}}\equiv
  \left\{H_0^{\text{PT}}\right\}_0
  &=\sum_{\a\ne\b}\frac{\bSa\bSb}{\vert\ea-\eb\vert^2} + \frac{N(N^2+5)}{48}
  -\frac{\bsS_{\text{tot}}^2}{6}.
\end{align}
We have hence derived that $\ket{\psi^{\s\text{HS}}_{0}}$ is an
eigenstate of
\begin{align}
  \label{eq:a:h0PT0_1}
  H^{\s\text{HS}}
  =\frac{2\pi^2}{N^2}\sum_{\a\ne\b}\frac{\bSa\bSb}{\vert\ea-\eb\vert^2}
\end{align}
with energy eigenvalue
\begin{align}
  \label{eq:a:e0PT0_1}
  E_0^{\s\text{HS}}=-\frac{2\pi^2}{N^2}\frac{N(N^2+5)}{48}.
\end{align}
In other words, we have derived the Haldane--Shastry model.
% 
% This implies that $\ket{\psi^{\s\text{HS}}_{0}}$ is an eigenstate of
% % \begin{align}
% %   \label{eq:a:h0PT0_1}
% %   \frac{2\pi^2}{N^2}
% %   \left[\left\{H_0^{\text{PT}}\right\}_0+\frac{\bsS_{\text{tot}}^2}{6}\right]
% %    +E_0
% %   =\frac{2\pi^2}{N^2}
% %    \sum_{\a\ne\b}\frac{\bSa\bSb}{\vert\ea-\eb\vert^2}
% %   =H^{\s\text{HS}}
% % \end{align}
% \begin{align}
%   \label{eq:a:h0PT0_1}
%   H^{\s\text{HS}}
%   &=\left(\frac{2\pi}{N}\right)^2
%   \sum_{\a<\b}\frac{\bSa\bSb}{\vert\ea-\eb\vert^2}
% %  =\frac{2\pi^2}{N^2}\sum_{\a\ne\b}\frac{\bSa\bSb}{\vert\ea-\eb\vert^2}
%   \\[0.2\baselineskip]
%   &=E_0 + \frac{2\pi^2}{N^2}
%   \left(\left\{H_0^{\text{PT}}\right\}_0+\frac{\bsS_{\text{tot}}^2}{6}\right)
% \end{align}
% with energy 
% \begin{equation}
%   \label{eq:a:e0}
%   E_0=-\frac{2\pi^2}{N^2}\,\frac{N(N^2+5)}{48}
% %  =-\frac{\pi^2}{24}\left(N+\frac{5}{N}\right).
% \end{equation}
% We have hence derived the Haldane--Shastry model.

This derivation by (conceptually) straightforward projection onto
the scalar component is instructive as we will employ this method for
the $S=1$ spin chain in Section \ref{sec:b:ham}.  It has the
disadvantage, however, that the information regarding the
semi-positive definiteness has been lost.  There are two ways to
restore this information.  The first is via an alternative derivation
of the model without projection from \eqref{eq:a:h0PT3}, we will
explain now.  The second way is to derive first a vector annihilation
operator for $\ket{\psi^{\s\text{HS}}_{0}}$, and then construct the
Hamiltonian from there, as explained in Section \ref{sec:a:vec}.

\subsection{An alternative derivation}

The operators $H_0$, $H_0^{\text{T}}$, and $H_0^{\text{PT}}$
constructed in Section \ref{sec:a:ham:transTP} are all sums of terms
of the form $A^\dagger A$, and are hence all positive semi-definite,
\ie have only non-negative eigenvalues.  Since
$\ket{\psi^{\s\text{HS}}_{0}}$ is an eigenstate with eigenvalue zero,
it is also a ground state of these operators when we view them as 
Hamiltonians.

We now wish to employ \eqref{eq:a:h0PT3} to derive that
$\ket{\psi^{\s\text{HS}}_{0}}$ is not only an eigenstate of
\eqref{eq:a:h0PT0_1} with energy \eqref{eq:a:e0PT0_1}, but also a
ground state.  For this purpose, we rewrite \eqref{eq:a:h0PT3} as
\begin{align}
  \label{eq:a:h0PT4}
  H_0^{\text{PT}}&+\frac{1}{8}
  \big(S_{\text{tot}}^+S_{\text{tot}}^-+S_{\text{tot}}^-S_{\text{tot}}^+\big)
  \nonumber\\*[0.2\baselineskip] 
  &=\sum_{\a\ne\b}
  \frac{\bSa\bSb}{\vert\ea-\eb\vert^2} + \frac{N(N^2-1)}{48}
  +\frac{1}{8}\sum_{\a}
  \big(\Sa^+\Sa^- + \Sa^-\Sa^+\big)
  \nonumber\\[0.2\baselineskip] 
  &=\sum_{\a\ne\b}
  \frac{\bSa\bSb}{\vert\ea-\eb\vert^2} + \frac{N(N^2+5)}{48},
\end{align}
where we have used $\Sa^+\Sa^- + \Sa^-\Sa^+=1$ for spin \half.
%$S^+S^- + S^-S^+=1$ for spin \half.
Since the left-hand side of \eqref{eq:a:h0PT4} is a sum of positive
semi-definite operators which annihilate
$\ket{\psi^{\s\text{HS}}_{0}}$, $\ket{\psi^{\s\text{HS}}_{0}}$ has to
be a zero energy ground state of the right-hand side as well, \ie a
ground state of \eqref{eq:a:h0PT0_1} with energy \eqref{eq:a:e0PT0_1}.

%\newpage
%\section{The rapidity operator and other annihilation operators}
\section{The rapidity operator and more}
%\section{A vector annihilation operator}
\label{sec:a:vec}

%\subsection{T even operators}
\subsection{Annihilation operators which transform even under T}
\label{sec:a:vecTeven}

We can use the \defining condition \eqref{eq:a:Omegapro} further to
construct a vector annihilation operator.  First note that since
\begin{equation*}
  \OaHS \ket{\psi^{\s\text{HS}}_{0}} = 0 \quad\forall\, \alpha,
\end{equation*}
$\ket{\psi^{\s\text{HS}}_{0}}$ is also annihilated by the Hermitian
operator 
\begin{align}
  \label{eq:a:ha}
%   D_{0,\a}={\OaHS}^\dagger\OaHS
%   &=\sum_{\substack{\b,\c\\ \b,\c\ne\a }}
%     \frac{1}{\eab-\ebb}\frac{1}{\ea-\ec} 
%     \Sa^+\Sa^-\Sb^+\Sc^-
%   \nonumber\\[0.2\baselineskip] 
  H_{\a}={\OaHS}^\dagger\OaHS
  &=\sum_{\substack{\b,\c\\ \b,\c\ne\a }}
    \omega_{\b\c}\left(\Sa^\z + \frac{1}{2}\right)\Sb^+\Sc^-,
\end{align}
which is just the operator \eqref{eq:a:h0} without the sum over $\a$.
Constructing an operator which is even under T,
\begin{equation}
  \label{eq:a:haT}
  H_\a^{\text{T}}=\frac{1}{2}\left(H_\a+\Theta H_\a \Theta\right)
  = H_\a^{\text{T}=}+H_\a^{\text{T}\ne },
\end{equation}
with 
\begin{align}
  H_\a^{\text{T}=} &=\sum_{\substack{\b\\ \b\ne\a}}\omega_{\a\b\b}
  \left(\Sa^\z\Sb^\z + \frac{1}{4}\right),
  \hspace{-6pt}&\hspace{-6pt}
%  H_\a^{\text{T}\ne} &=\frac{1}{2}\sum_{\substack{\b,\c\\ \a\ne\b\ne\c\ne\a}}
  H_\a^{\text{T}\ne} &=\frac{1}{2}\sum_{\substack{\b\ne\c\\ \b,\c\ne\a}}
  \omega_{\a\b\c}\Sb^+\Sc^-,
\end{align}
% \begin{align}
%   \label{eq:a:haT=}
%   H_\a^{\text{T}=} &=\sum_{\substack{\b\\ \b\ne\a}}\omega_{\a\b\b}
%   \left[\Sa^\z\Sb^\z + \frac{1}{4}\right],
%   \\[0.4\baselineskip]
%   \label{eq:a:haTnot=}
%   H_\a^{\text{T}\ne} 
%   &=\frac{1}{2}\sum_{\substack{\b,\c\\ \b,\c\ne\a\\ \b\ne\c}}
%   \omega_{\a\b\c}\Sb^+\Sc^-,
% \end{align}
and odd under P, we obtain 
% along the lines of the construction in Section \ref{sec:a:ham:transTP}
\begin{align}
  \label{eq:a:habarPT}
  H_\a^{\rm {\bar P}T}=\frac{1}{2}
  \left(H_\a^{\text{T}} - \Pi\hspace{1pt} H_\a^{\text{T}}\hspace{1pt} \Pi\right)
  = H_\a^{{\rm {\bar P}T}=} + H_\a^{{\rm {\bar P}T}\ne},
\end{align}
where
\begin{align}
  \label{eq:a:habarPT1}
  H_\a^{{\rm {\bar P}T}=}&=0, 
  \hspace{-10pt}&\hspace{-10pt}
%  \\[0.4\baselineskip] 
%  \label{eq:a:h0Tn}
  H_\a^{{\rm {\bar P}T}\ne}&=
  \frac{1}{4}\sum_{\substack{\b\ne\c\\ \b,\c\ne\a}}\omega_{\a\b\c}
  \big(\Sb^+\Sc^- - \Sb^-\Sc^+\big).
\end{align}
With 
\begin{align}
  \label{eq:a:omega_abc-omega_acb}
  \omega_{\a\b\c}-\omega_{\a\c\b}
  &=\frac{1}{\eab-\ebb}\frac{1}{\ea-\ec}-\frac{1}{\ea-\eb}\frac{1}{\eab-\ecb}
  \nonumber\\[0.2\baselineskip] 
  &=(-\ea\eb-\ea\ec)\,\frac{1}{\ea-\eb}\frac{1}{\ea-\ec}
  \nonumber\\[0.2\baselineskip] 
  &=\ea\big((\ea-\eb)-(\ea-\ec)\big)\,\frac{1}{\ea-\eb}\frac{1}{\ea-\ec}
  \nonumber\\[0.2\baselineskip] 
  &=\frac{\ea}{\ea-\ec}-\frac{1}{2}-\left(\frac{\ea}{\ea-\eb}-\frac{1}{2}\right)
  \nonumber\\[0.2\baselineskip] 
  &=-\frac{1}{2}\left(\frac{\ea+\eb}{\ea-\eb}-\frac{\ea+\ec}{\ea-\ec}\right)
\end{align}
and $\Sb^+\Sc^- - \Sb^-\Sc^+=-2\i (\bSb\times\bSc)^\z$ (\cf
\eqref{eq:app-t-SSS-1abc}), we obtain
\begin{align}
  \label{eq:a:habarPT2}
  H_\a^{{\rm {\bar P}T}}
%   &=\frac{1}{4}\sum_{\substack{\b\ne\c\\ \b,\c\ne\a}}\omega_{\a\b\c}
%   \big(\Sb^+\Sc^- - \Sb^-\Sc^+\big)
%   \nonumber\\[0.2\baselineskip] 
%
%   &=-\frac{\i}{4}\sum_{\substack{\b\ne\c\\ \b,\c\ne\a}}
%   (\omega_{\a\b\c}-\omega_{\a\c\b})(\bSb\times\bSc)^\z
%   \nonumber\\[0.2\baselineskip] 
  &=\frac{\i}{4}\sum_{\substack{\b\ne\c\\ \b,\c\ne\a}}
  \frac{\ea+\eb}{\ea-\eb} (\bSb\times\bSc)^\z
  \nonumber\\[0.2\baselineskip] 
  &=\frac{\i}{4}\sum_{\substack{\b\\ \b\ne\a}}
  \frac{\ea+\eb}{\ea-\eb} 
  \big(\bSb\times(\bsS_{\text{tot}}-\bSa-\bSb)\big)^\z
  \nonumber\\[0.2\baselineskip] 
  &=\frac{\i}{4}\sum_{\substack{\b\\ \b\ne\a}}
  \frac{\ea+\eb}{\ea-\eb} 
  \big((\bSa\times\bSb)-\i\bSb\big)^\z
  +\frac{\i}{4}\sum_{\substack{\b\\ \b\ne\a}}
  \frac{\ea+\eb}{\ea-\eb} 
  \big(\bSb\times\bsS_{\text{tot}}\big)^\z,
\end{align}
where we have used $\bSb\times\bSb=i\bSb$.  Since
$\ket{\psi^{\s\text{HS}}_{0}}$ is a spin singlet, it is trivially
annihilated by the second term in the last line of
\eqref{eq:a:habarPT2}, and hence also annihilated by the first term,
which is the $z$ component of a vector.  The singlet property of
the ground state implies that $\ket{\psi^{\s\text{HS}}_{0}}$ is
annihilated by all the components of this vector, \ie
\begin{align}
  \label{eq:a:doperator}
  \bs{D}_\a&=\frac{\i}{2}\sum_{\substack{\b=1\\ \b\ne\a}}^N
  \frac{\ea+\eb}{\ea-\eb} 
  \Big[(\bSa\times\bSb)-\i\bSb\Big],
  \quad \bs{D}_\a\ket{\psi^{\s\text{HS}}_{0}} = 0 \quad \forall\, \alpha.
\end{align}
This is exactly the auxiliary operator \eqref{eq:hsdoperator} we
introduced in \ref{sec:hsfactorization}, where we have further shown
that
% \begin{equation}
%   \label{eq:a:factorization}
%   H^{\s\text{HS}}=\frac{2\pi^2}{N} \Biggl[ 
%   \frac{2}{9}\sum_{\alpha=1}^N \bs{D}_\alpha^\dagger %\cdot 
%   \bs{D}_\alpha
% %  - \frac{ N ( N^2 + 5 )}{ 48} 
%   + \frac{ N +1}{ 12} \bs{S}_{\text{tot}}^2 \Biggr] +E_0.
% \end{equation}
\begin{equation*}
  \frac{2}{9}\sum_{\alpha=1}^N\bs{D}_{\alpha}^{\dagger}\bs{D}_{\alpha}
  +\frac{N+1}{12}\bs{S}_{\text{tot}}^2
  =\sum^N_{\alpha\neq\beta}
  \frac{\bs{S}_{\alpha}\bs{S}_{\beta}}{\vert\eta_{\alpha}-\eta_{\beta}\vert^2}
  +\frac{N(N^2+5)}{48}.
\end{equation*}
This proofs once more that $\ket{\psi^{\s\text{HS}}_{0}}$ is
a ground state of \eqref{eq:a:h0PT0_1} with energy
\eqref{eq:a:e0PT0_1}.  

Equation \eqref{eq:a:doperator} implies that the Haldane-Shastry
ground state is further annihilated by
\begin{align}
  \label{eq:a:lambdaoperator}
  \bs{\Lambda} = \sum_{\alpha=1}^N\bs{D}_{\alpha}
  =\frac{\i}{2}\sum_{\substack{\a\ne\b}}^N
  \frac{\ea+\eb}{\ea-\eb} (\bSa\times\bSb),
%  \quad \bs{\Lambda} \ket{\psi^{\s\text{HS}}_{0}} = 0,
\end{align}
where we have used \eqref{eq:app-hs13}.  This is the rapidity operator
\eqref{eq:hsrapidityoperator} from Section \ref{sec:hssymm}, which
together with the total spin operator generates the Yangian symmetry
algebra of the Haldane--Shastry model.

For completeness, we further wish to mention the scalar operator we can
construct from \eqref{eq:a:haT}, which transforms even under P,
and which yields the Hamiltonian \eqref{eq:a:h0PT0} when we sum over
$\a$.  This operator is given by 
\begin{align}
  \label{eq:a:haPT}
  H_\a^{\rm PT}=\frac{1}{2}
  \left(H_\a^{\text{T}} + \Pi\hspace{1pt} H_\a^{\text{T}}\hspace{1pt} \Pi\right)
  = H_\a^{{\rm PT}=} + H_\a^{{\rm PT}\ne},
\end{align}
where
\begin{align}
  \label{eq:a:haPT1}
  H_\a^{{\rm {P}T}=}&=H_\a^{{\rm T}=},
  \hspace{-6pt}&\hspace{-6pt}
%  H_\a^{\text{PT}=} &=\sum_{\substack{\b\\ \b\ne\a}}\omega_{\a\b\b}
%  \left(\Sa^\z\Sb^\z + \frac{1}{4}\right), &
  H_\a^{{\rm PT}\ne}&=
  \frac{1}{4}\sum_{\substack{\b\ne\c\\ \b,\c\ne\a}}\omega_{\a\b\c}
  \big(\Sb^+\Sc^- + \Sb^-\Sc^+\big).
\end{align}
The scalar component of this operator is with \eqref{eq:app-t-SzSz}
and \eqref{eq:app-t-S+-S-+} given by
% \begin{align}
%   \label{eq:a:habarPT2}
%   \big\{H_\a^{{\rm PT}=}\big\}_0&=
%   \frac{1}{3}\sum_{\substack{\b\\ \b\ne\a}}\omega_{\a\b\b}\bSa\bSb
%   + \frac{N^2-1}{48}, 
%   \\[0.2\baselineskip] 
%   \big\{H_\a^{{\rm PT}\ne}\big\}_0&=
%   \frac{1}{3}\sum_{\substack{\b\ne\c\\ \b,\c\ne\a}}\omega_{\a\b\c}\bSb\bSc.
% \end{align}
\begin{align}
  \label{eq:a:haPT0}
  \big\{H_\a^{{\rm PT}}\big\}_0
  &=\frac{1}{3}\sum_{\substack{\b\\ \b\ne\a}}
  \frac{\bSa\bSb}{\vert\ea-\eb\vert^2}
  + \frac{1}{3}\sum_{\substack{\b\ne\c\\ \b,\c\ne\a}}%\omega_{\a\b\c}
  \frac{\bSb\bSc}{(\eab-\ebb)(\ea-\ec)}
  + \frac{N^2-1}{48},
%  \nonumber\\[0.2\baselineskip]
%  \big\{H_\a^{{\rm PT}}\big\}_0
%  &\ket{\psi^{\s\text{HS}}_{0}} = 0 \ \ \forall\, \alpha.
\end{align}
and annihilates the  Gutzwiller state,
\begin{align*}
  \big\{H_\a^{{\rm PT}}\big\}_0
  \ket{\psi^{\s\text{HS}}_{0}} = 0 \quad \forall\, \alpha.
\end{align*}
%We do not consider this operator to be of importance.
We do not believe that this operator is useful.

\subsection{Annihilation operators which transform odd under T}

Finally, we consider annihilation operators we can construct from
\eqref{eq:a:ha}, and which transform odd under T,
\begin{equation}
  \label{eq:a:habarT}
  H_\a^{\rm\bar T}=\frac{1}{2}\left(H_\a-\Theta H_\a \Theta\right)
  = H_\a^{\rm\bar T=} + H_\a^{\rm\bar T\ne}
\end{equation}
with 
\begin{align}
  \label{eq:a:habarT=}
  H_\a^{\rm\bar T=} &=\sum_{\substack{\b\\\b\ne\a}}\omega_{\a\b\b}
  \left(\frac{1}{2}\Sa^\z \biganticomm{\Sb^+}{\Sb^-}
    +\frac{1}{4}\bigcomm{\Sb^+}{\Sb^-}\right)
  \nonumber\\[0.2\baselineskip] 
  &=\frac{1}{2}\sum_{\substack{\b\\\b\ne\a}}\omega_{\a\b\b}
  \left(\Sa^\z+\Sb^\z\right)
  \nonumber\\[0.2\baselineskip] 
  &= \frac{N^2-1}{24}\, \Sa^\z
  + \frac{1}{2}\sum_{\substack{\b\\\b\ne\a}}\omega_{\a\b\b}\Sb^\z,
  \\[0.4\baselineskip] 
  \label{eq:a:habarTnot=}
  H_\a^{\rm\bar T\ne} &=\sum_{\substack{\b\ne\c\\ \b,\c\ne\a}}\omega_{\a\b\c},
  \Sa^\z\Sb^+\Sc^-
\end{align}
where we have used \eqref{eq:app-hs12}.
$\ket{\psi^{\s\text{HS}}_{0}}$ is hence annihilated by all the tensor
components of \eqref{eq:a:habarT}, which are readily obtained with
\eqref{eq:app-t-Sz+-}, \eqref{eq:app-t-SSS-0}, and
\eqref{eq:app-t-SSS-1abc}.  Let us consider first the scalar operator
\begin{align}
  \label{eq:a:habarTnot=0v1}
  \big\{H_\a^{\rm\bar T}\big\}_0
  &=-\frac{\i}{3}\sum_{\substack{\b\ne\c\\ \b,\c\ne\a}}
  \frac{\bSa(\bSb\times\bSc)}{(\eab-\ebb)(\ea-\ec)},
\end{align}
which is odd under P.
With \eqref{eq:a:omega_abc-omega_acb}, we obtain
\begin{align}
  \label{eq:a:habarTnot=0v2}
  \big\{H_\a^{\rm\bar T}\big\}_0
  &=\frac{\i}{6}\sum_{\substack{\b\ne\c\\ \b,\c\ne\a}}
  \frac{\ea+\eb}{\ea-\eb}\bSa\big(\bSb\times\bSc\big)
  \nonumber\\[0.2\baselineskip] 
  &=\frac{\i}{6}\sum_{\substack{\b\\ \b\ne\a}}
  \frac{\ea+\eb}{\ea-\eb}
  \bSa\big(\bSb\times(\bsS_{\text{tot}}-\bSa-\bSb)\big)
  \nonumber\\[0.2\baselineskip] 
  &=\frac{\i}{6}\sum_{\substack{\b\\ \b\ne\a}}
  \frac{\ea+\eb}{\ea-\eb}
  \bSa\big(\bSb\times\bsS_{\text{tot}}\big),
\end{align}
where we have used 
\begin{align}
  \label{eq:a:SaSbSaSb}
  \bSa\big(\bSb\times(-\bSa-\bSb)\big)
  &=\bSb\big(\bSa\times\bSa\big)-\bSa\big(\bSb\times\bSb\big)=0.
\end{align}
The operator \eqref{eq:a:habarTnot=0v2} annihilates every spin
singlet, and is therefore useless in the present context.
% not of any use to the model.

The vector component of \eqref{eq:a:habarT}, however, constitutes
a viable annihilation operator for the Haldane--Shastry ground state,
%of $\ket{\psi^{\s\text{HS}}_{0}}$,
\begin{align}
  \label{eq:a:aoperator}
%  \bs{A}_\a &\equiv 5\big\{H_\a^{\rm\bar T}\big\}_{\bs{1}}
  \bs{A}_\a &\equiv 5\Big(\big\{H_\a^{\rm\bar T=}\big\}_{\bs{1}}
  + \{H_\a^{\rm\bar T\ne}\big\}_{\bs{1}}\big)
  \nonumber\\[0.2\baselineskip] 
%  &=\frac{5}{2}\sum_{\substack{\b\\\b\ne\a}}\frac{\bSb}{\vert\ea-\eb\vert^2}
  &=\frac{5}{2}\sum_{\substack{\b\\\b\ne\a}}\frac{\bSa+\bSb}{\vert\ea-\eb\vert^2}
  +\sum_{\substack{\b\ne\c\\ \b,\c\ne\a}}
  \frac{4\bSa(\bSb\bSc)-\bSb(\bSa\bSc)-\bSc(\bSa\bSb)}{(\eab-\ebb)(\ea-\ec)},
%  \nonumber\\[0.2\baselineskip] 
%  &\quad+\frac{5\,(N^2-1)}{24}\,\bSa
  \nonumber\\[0.5\baselineskip] 
  \bs{A}_\a %\big\{H_\a^{\rm\bar T}\big\}_{\bs{1}}
  &\ket{\psi^{\s\text{HS}}_{0}} = 0 \quad \forall\, \alpha.
\end{align}
This operator is even under P.  Summing over $\a$, we find that
the first term annihilates every singlet, since
\begin{align*}
  \frac{1}{2}\sum_{\substack{\a,\b\\\a\ne\b}}\frac{\bSa+\bSb}{\vert\ea-\eb\vert^2}
  &=\sum_\a\bSa\sum_{\substack{\b\\\b\ne\a}}\omega_{\a\b\b}
  =\frac{N^2-1}{12} \bsS_{\text{tot}}.
\end{align*}
% \begin{align}
%   \label{eq:a:haPbarT=1v2}
%   \sum_\a \big\{H_\a^{\rm\bar T=}\big\}_{\bs{1}}
% %  =\frac{1}{2}\sum_{\substack{\a,\b\\\a\ne\b}}
% %   \frac{\bSa+\bSb}{\vert\ea-\eb\vert^2}
%   =\sum_\a \bSa \sum_{\substack{\b\\\b\ne\a}}\omega_{\a\b\b}
%   =\frac{N^2-1}{12} \bsS_{\text{tot}}.
% \end{align}
This implies that $\ket{\psi^{\s\text{HS}}_{0}}$ is further annihilated by
the vector operator
\begin{align}
  \label{eq:a:upsilonoperator}
%  \big\{H_0^{\rm\bar T}\big\}_{\bs{1}}
  \bs{\Upsilon}
  &=5\sum_{\a}\big\{H_\a^{\rm\bar T\ne}\big\}_{\bs{1}}
%  \nonumber\\[0.2\baselineskip] 
%  &
  =\sum_{\substack{\a,\b,\c\\ \a\ne\b\ne\c\ne\a}}%\hspace{-10pt}
  \frac{4\bSa(\bSb\bSc)-\bSb(\bSa\bSc)-\bSc(\bSa\bSb)}{(\eab-\ebb)(\ea-\ec)}.
  \nonumber\\[-0.4\baselineskip]
\end{align}
This is a three spin operator, and has to our knowledge
not been considered before.

\section{Concluding remarks}

\renewcommand{\strut}{\rule[-5pt]{0pt}{16pt}}
\begin{table}[t]
  \centering
  \caption{Annihilation operators for the 
    Haldane--Shastry ground state.  With the exception of the \defining
    operator $\OaHS$, which is the $m=2$ component of a 2nd order tensor,
    we have only included scalar and vector annihilation operators.}
  \label{tab:a:annihilationops}
  \centering
  \begin{tabular}{p{22mm}p{22mm}p{13mm}p{13mm}p{26mm}p{18mm}}\hline
  \multicolumn{6}{c}{{\bf Annihilation operators for} 
    $\ket{\psi^{\s\text{HS}}_{0}}$}\strut \\[2pt]\hline
   Operator &Equation 
   &\multicolumn{4}{l}{Symmetry transformation properties}\strut 
   \\[2pt]\cline{3-6}
   &&T &P &order of tensor &transl.~inv.\strut\\[2pt]\hline %\svhline
   $\bsS_{\text{tot}}$&\eqref{eq:hsspinsymmetry}&$-$&$+$&vector&yes\rule[0pt]{0pt}{12pt}\\[2pt]
   $\OaHS$&\eqref{eq:a:Omegadef}&no&no&2nd&no\\[4pt]
   $\big\{H_\a^{{\rm PT}}\big\}_0$&\eqref{eq:a:haPT0}&$+$&$+$&scalar&no\\[4pt]
   $H^{\s\text{HS}}-E_0^{\s\text{HS}}$&\eqref{eq:a:h0PT0_1}&$+$&$+$&scalar&yes\\[2pt]
   $\bs{D}_\a$&\eqref{eq:a:doperator}&$+$&$-$&vector&no\\[2pt]
   $\bs{\Lambda}$&\eqref{eq:a:lambdaoperator}&$+$&$-$&vector&yes\\[2pt]
   $\bs{A}_\a$&\eqref{eq:a:aoperator}&$-$&$+$&vector&no\\[2pt]
   $\bs{\Upsilon}$&\eqref{eq:a:upsilonoperator}&$-$&$+$&vector&yes\\[3pt]
  \hline
  \end{tabular}
\end{table}

The various annihilation operators for the Haldane--Shastry model
are summarized in Table \ref{tab:a:annihilationops}.

The Haldane--Shastry model, including the operators presented in
Section \ref{sec:a:vecTeven}, have been known for a long time.  In the
work of Haldane and Shastry, however, the model was discovered, while
we derived it here.  Unlike the discovery, the derivation we presented
here lends itself to a generalization to higher spins, which is what
we will pursue in the following \chap . %\ref{sec:pf2s1}.

It is worth noting that the derivation of the model presented
in Section \ref{sec:a:ham:transTP}, which only assumes the \defining
condition \eqref{eq:a:Omegadef}, is significantly simpler than the
previously established verification of the model reviewed in Section
\ref{sec:hsexsol} with Appendix~\ref{sec:app-hssums}.  The
disadvantage of the present derivation, however, is that it is not
clear how to extract information regarding excitations via the
formalism employed.

\vspace{100pt}
\newpage
% lp -dtkmsek -P 26-30 map.ps 

% lp -dtkmsek -P 38-38 map.ps 

\chapter{From a bosonic Pfaffian state to an $S=1$ spin chain}
%\chapter{From a bosonic Moore--Read state to an $\bs{S=1}$ spin chains}
%\section{From a bosonic Pfaffian state to an exact model of a critical
%\section{From a Moore--Read state to an exact model of a critical
%$\bs{S=1}$ spin chain}
\label{sec:pf2s1}
\section{General considerations}
In this section, we wish to use the bosonic Pfaffian state at Landau
level filling fraction $\nu=1$ and its parent Hamiltonian (see Section
\ref{sec:3mod-pf}), to construct a parent Hamiltonian for the critical
$S=1$ spin liquid state introduced in Section \ref{sec:3mod-na}.  The
Hamiltonian we construct should be invariant under all the trivial
symmetries of the spin liquid ground state described in Section
\ref{sec:na:sym}, \ie under space translations, P and T, and SU(2)
spin rotations.
This task would probably be beyond our means if we had not established
a suitable technique in Section \ref{sec:l2hs}, when we derived the
Haldane--Shastry Hamiltonian from a bosonic Laughlin state and its
parent Hamiltonian.  The purpose of this derivation was really to
establish the technique which we will fruitfully use in the present 
analysis.
%
% ??? profitably use in the present context.
%
% As for the technique, we will borrow heavily from Section
% \ref{sec:l2hs}, in which we derived the Haldane--Shastry Hamiltonian
% from a bosonic Laughlin state and its parent Hamiltonian.
% 
% The technique we will employ will be similar to the technique we
% employed in Section \ref{sec:l2hs}, when we derived the
% Haldane--Shastry Hamiltonian from a bosonic Laughlin state and its
% parent Hamiltonian.  

To begin with, we briefly recall the quantum Hall model and the spin
liquid ground state.

\subsection{A model and a ground state}

The wave function for the bosonic $m=1$ Pfaffian Hall state~\cite{moore-91npb362,greiter-91prl3205,greiter-92npb567} 
%reviewed in Section \ref{sec:3mod-pf} 
%The wave function for the bosonic $m=1$ Moore--Read state is
\begin{equation}
  \label{eq:b:psi0MR}
  \psi_0(z_1,z_2,\ldots ,z_N)
  =\text{Pf}\left(\frac{1}{z_{i}-z_{j}}\right)
  \prod_{i<j}^{N}(z_i-z_j)
  \prod_{i=1}^N e^{-\frac{1}{4}|z_i|^2},          
\end{equation}
where the particle number $N$ is even, and the Pfaffian is is given by
the fully antisymmetrized sum over all possible pairings of the $N$
particle coordinates,
\begin{equation}
  \label{eq:b:pfaff}
  \text{Pf}\left(\frac{1}{z_i -z_j}\right)\equiv
  \mathcal{A}
  \left\{
    \frac{1}{z_1-z_2}\cdot\,\ldots\,\cdot\frac{1}{z_{N-1}-z_{N}}
  \right\}.
\end{equation}
It is the exact ground state of the three-body Hamiltonian~\cite{greiter-91prl3205,greiter-92npb567}
\begin{equation}
  \label{eq:b:GWWham}
  V=\sum_{i,j<k}^N \delta^{(2)}(z_i-z_j)\delta^{(2)}(z_i-z_k).
\end{equation}

In Section \ref{sec:3mod-na}, we introduced an $S=1$ spin liquid state
described by a Pfaffian.  We considered a one-dimensional lattice with periodic
boundary conditions and an even number of sites $N$ on a unit circle
embedded in the complex plane, $\displaystyle
\eta_\alpha=e^{\text{i}\frac{2\pi}{N}\alpha }$ with $\alpha = 1,\ldots
,N$.  The wave function is given by a bosonic Pfaffian state in the
complex lattice coordinates $z_i$ supplemented by a phase factor,
\begin{equation}
  \label{eq:b:psi0}
  \psi^{S=1}_0(z_1,z_2,\ldots ,z_N) %[z_i]
  =\text{Pf}\left(\frac{1}{z_{i}-z_{j}}\right)
  \prod_{i<j}^{N}(z_i-z_j)\prod_{i=1}^{N}\,z_i.
\end{equation}
The ``particles'' $z_i$ represent re-normalized spin flips
%$\tilde{S}_{\alpha}^{+}$ 
\begin{equation}
  \label{eq:b:tildeS+}
  \tilde{S}_{\alpha}^{+} \equiv\frac{{S}^{\rm{z}}_{\alpha}+1}{2} S_\alpha^{+},
\end{equation}
which act on a vacuum with all spins in the
$S^{\rm{z}}=-1$ state,
\begin{equation}
  \label{eq:b:ket}
  \ket{\psi^{S=1}_0}=\sum_{\{z_1,\dots,z_{N}\}} 
%  \psi_0[z_i]
  \psi^{S=1}_0(z_1,\dots,z_N)\
%  \tilde{S}_{z_1}^{+}\cdot\, \dots\, \cdot \tilde{S}_{z_{N}}^{+}\ 
  \tilde{S}_{z_1}^{+}\cdot\dots\cdot\tilde{S}_{z_{N}}^{+} 
%  \ket{1,-1}^{\otimes N},
%  \ket{1,-1}_N,
  \ket{-1}_N,
%  \ket{-1}^{\otimes N},
\end{equation}
where the sum runs over all possibilities of distributing the $N$
``particles'' over the $N$ lattice sites allowing for double
occupation, and
\begin{equation}
  \label{eq:b:vacuumket}
  \ket{-1}_N\equiv\otimes_{\alpha=1}^N \ket{1,-1}_{\alpha}.
\end{equation}

As for the Laughlin state in Section \ref{sec:a:com}, the circular
droplet described by the %Moore--Read
quantum Hall wave function \eqref{eq:b:psi0MR} has a boundary, while
the $S=1$ ground state \eqref{eq:b:ket} with \eqref{eq:b:psi0}
describes a spin liquid on a compact surface.  To circumvent this
problem, we formulate the quantum Hall model on the sphere (see
Section \ref{sec:laughsphere}).  Then the bosonic $m=1$ Pfaffian state
for $N$ particles on a sphere with $2s=N-2$ flux quanta is given by
\begin{equation}
  \label{eq:b:psiMRsp}
  \psi_0[u,v]=\text{Pf}\left(\frac{1}{u_iv_j-u_jv_i}\right)
  \prod_{i<j}^N(u_iv_j-u_jv_i).
\end{equation}
Within the lowest Landau level, it is the exact and unique zero-energy
ground state of the interaction Hamiltonian
\begin{equation}
  \label{eq:b:V3bodysphere}
  \begin{split}
%    V_{\s\text{int}}^{(3)}
    V^{\s\text{qh}}
    =&\sum_{m_1=-s}^s\, \sum_{m_2=-s}^s\, \sum_{m_3=-s}^s\, \sum_{m_4=-s}^s
    \, \sum_{m_5=-s}^s\, \sum_{m_6=-s}^s
    \\[0.3\baselineskip]
    &\cdot a_{m_1}^\dagger a_{m_2}^\dagger a_{m_3}^\dagger a_{m_4}a_{m_5}a_{m_6}\,
    \delta_{m_1+m_2+m_3,m_4+m_5+m_6}\,
    \\[0.3\baselineskip]
    &\cdot\braket{s,m_1;s,m_2}{2s,m_1+m_2}
    \braket{2s,m_1+m_2;s,m_3}{3s,m_1+m_2+m_3}  
    \\[0.3\baselineskip]
    &\cdot\braket{3s,m_4+m_5+m_6}{s,m_4;2s,m_5+m_6} 
    \braket{2s,m_5+m_6}{s,m_5;s,m_6},\quad
  \end{split}
\end{equation}
where $a_m$ annihilates a boson in the properly normalized single
particle state
\begin{equation}
  \label{eq:b:qhsnormLLLbasis}
  \psi_{m,0}^{s}(u,v) 
  =\bra{u,v} a_{m}^\dagger \vac
%  =\sqrt{\frac{(2s+1)!}{4\pi\,(s+m)!\, (s-m)!}}\; u^{s+m}v^{s-m},
  =\frac{1}{g_m}\; u^{s+m}v^{s-m},
\end{equation}
with
\begin{equation}
  \label{eq:b:gm}
  g_m=\sqrt{\frac{4\pi\,(s+m)!\, (s-m)!}{(2s+1)!}},
\end{equation}
and $\braket{s,m_1;s,m_2}{2s-l,m_1+m_2}$ \etc are Clebsch--Gordan
coefficients~\cite{baym69}.

The differences between the Pfaffian Hall state \eqref{eq:b:psiMRsp}
and the spin liquid state \eqref{eq:b:psi0} are almost in exact
correspondence to the differences between the Laughlin state
\eqref{eq:a:psiLaugh} and the Haldane--Shastry ground state
\eqref{eq:a:hspsi0}.  We will employ the same techniques to
adapt the quantum Hall model to the spin chain.

%\pagebreak
\subsection{Creation of a quasihole}

The wave function of the spin liquid state \eqref{eq:b:psi0} differs
from the quantum Hall state in that it contains an additional factor
$\prod_{i}z_i$.  We can adapt the quantum Hall state by insertion of 
a quasihole at the south pole of the sphere.  This yields 
\begin{equation}
  \label{eq:b:psiqH}
  \psi^{\s\text{qH}}_0[u,v]=\text{Pf}\left(\frac{1}{u_iv_j-u_jv_i}\right)
  \prod_{i<j}^N(u_iv_j-u_jv_i)\,\prod_{i=1}^N u_i,
\end{equation}
on a sphere with $2s=N-1$.  It is the exact and unique ground state
of
\begin{equation}
  \label{eq:b:HqH}
  H^{\s\text{qH}}=V^{\s\text{qH}}+U^{\s\text{qH}}
\end{equation}
with 
\begin{equation}
  \label{eq:b:UqH}
  U^{\s\text{qH}}=U_0\, a_{-s}^\dagger a_{-s}
\end{equation}
for $U_0>0$ if we restrict our Hilbert space again to the lowest
Landau level.  Note that both $V^{\s\text{qh}}$ and $U^{\s\text{qh}}$
annihilate the ground state \eqref{eq:b:psiqH} individually.  The
single particle Hilbert space dimension of the bosons on the sphere is
now equal to the dimension dimension of the single particle Hilbert
space for the spin flips on the unit circle, $2s+1=N$.  The expansion
coefficients $C_{q_1,\ldots ,q_N}$ for the polynomials
\begin{equation}
  \label{eq:b:psiS1q}
  \psi^{S=1}_0[z] %(z_1,\ldots ,z_M) = 
   =\sum_{\{q_1,\ldots ,q_N\}} C_{q_1,\ldots ,q_N}\, z_1^{q_1}\ldots z_N^{q_N}
\end{equation}
and
\begin{equation}
  \label{eq:b:psiqHq}
  \psi^{\s\text{qH}}_0[u,v] %=\prod_{i<j}^M(u_iv_j-u_jv_i)^2\,\prod_{i=1}^M u_i,
  =\sum_{\{q_1,\ldots ,q_N\}} C_{q_1,\ldots ,q_N}\, 
  u_1^{q_1}v_1^{2s-q_1}\ldots u_N^{q_N}v_N^{2s-q_N}
\end{equation}
are identical.  

%\pagebreak
\section{Hilbert space renormalization}
\label{sec:b:hilbert}

While the coefficients in the polynomial expansions of the ground
states \eqref{eq:b:psiS1q} and \eqref{eq:b:psiqHq} are identical, the
expansions of both states in terms of single particle states are not.
The state vector for the quantum Hall state is given by
\begin{equation}
  \label{eq:b:psiMRC}
  \ket{\psi^{\s\text{qH}}_0}
  =\sum_{\{m_1,\ldots ,m_N\}} C_{m_1+s,\ldots ,m_N+s}\,\,
  g_{m_1}\ldots g_{m_M} \,
  {a}^\dagger_{m_1}\ldots {a}^\dagger_{m_N} \vac
\end{equation}
where $g_m$
% \begin{equation}
%   \label{eq:b:gm}
%   g_m=\sqrt{\frac{4\pi\,(s+m)!\, (s-m)!}{(2s+1)!}}
% \end{equation}
are the normalizations \eqref{eq:b:gm} of the polynomials $u^{s+m}v^{s-m}$ 
in \eqref{eq:b:qhsnormLLLbasis}.  In the spin chain, the polynomials $z^q$
require no such normalization factors, as discussed in Section 
\eqref{sec:a:hilbert}.

To adjust the quantum Hall state, we renormalize the Hilbert space
using Theorem \ref{th:second} of Section \ref{sec:a:hilbert} with
the same operators $G$ given in \eqref{eq:a:g}.  This yields that
\begin{equation}
  \label{eq:b:psiMRGC}
  G^{-1}\ket{\psi^{\s\text{qH}}_0}
  =\sum_{\{m_1,\ldots ,m_N\}} C_{m_1+s,\ldots ,m_N+s}\,\,
%  g_{m_1}\ldots g_{m_M} \,
  {a}^\dagger_{m_1}\ldots {a}^\dagger_{m_N} \vac
\end{equation}
is an exact zero-energy eigenstate of 
\begin{equation}
  \label{eq:b:V1}
  \begin{split}
%    V_{\s\text{int}}^{(3)}
    V %^{\s\text{qh}}
    =&\sum_{m_1=-s}^s\, \sum_{m_2=-s}^s\, \sum_{m_3=-s}^s\, \sum_{m_4=-s}^s
    \, \sum_{m_5=-s}^s\, \sum_{m_6=-s}^s 
    \\[0.3\baselineskip]
    &\cdot a_{m_1}^\dagger a_{m_2}^\dagger a_{m_3}^\dagger a_{m_4}a_{m_5}a_{m_6}\,
    \delta_{m_1+m_2+m_3,m_4+m_5+m_6}\, g_{m_1}g_{m_2}g_{m_3}
    \\[0.3\baselineskip]
    &\cdot\braket{s,m_1;s,m_2}{2s,m_1+m_2}
    \braket{2s,m_1+m_2;s,m_3}{3s,m_1+m_2+m_3}  
    \\[0.3\baselineskip]
    &\cdot\braket{3s,m_4+m_5+m_6}{s,m_4;2s,m_5+m_6} 
    \braket{2s,m_5+m_6}{s,m_5;s,m_6},\quad
    \\[0.3\baselineskip]
    &\cdot g_{m_4}g_{m_5}g_{m_6}\\[-0.5\baselineskip]
  \end{split}
\end{equation}
Since \eqref{eq:b:psiMRGC} is likewise annihilated by
\eqref{eq:b:UqH}, it is also a zero energy state of
\begin{equation}
  \label{eq:b:H1}
  H%^{\s\text{rqH}} 
  = V%^{\s\text{rqH}}
  +U^{\s\text{qH}}.
\end{equation}
With \eqref{eq:a:gm}, \eqref{eq:a:gmcg} and the explicit formula
\begin{align}
  \label{eq:b:cg2}
    \hspace{20pt}&\hspace{-20pt}\braket{2s,m_1+m_2;s,m_3}{3s,m_1+m_2+m_3}
    \nonumber\\[0.3\baselineskip]
    &=\frac{\sqrt{(3s-m_1-m_2-m_3)!\,(3s+m_1+m_2+m_3)!}}
           {\sqrt{(2s-m_1+m_2)!\,(2s+m_1+m_2)!\,(s-m_3)!\,(s+m_3)!}}
    \nonumber\\[0.3\baselineskip]
    &\quad\cdot\,2\,\sqrt{\frac{s\,(2s-1)!\,(4s-1)!}{3\cdot(6s-1)!}}
\end{align}
for the second set of Clebsch--Gordan coefficients~\cite{baym69}, we obtain
\begin{align}
  \label{eq:b:gmcg2}
    &g_{m_1}g_{m_2}g_{m_3}\!\braket{s,m_1;s,m_2}{2s,m_1+m_2}
    \braket{2s,m_1+m_2;s,m_3}{3s,m_1+m_2+m_3}
%     \\[0.3\baselineskip]
%     &\hspace{5pt}=\sqrt{(3s-m_1-m_2-m_3)!\,(3s+m_1+m_2+m_3)!}
%     \\[0.3\baselineskip]
%     &\hspace{10pt}\cdot\,2\sqrt{\frac{s\,(2s-1)!\,(4s-1)!}{3}}\cdot\,
%     \frac{2\pi}{(2s+1)\sqrt{s\,(4s-1)!}}
%     \sqrt{\frac{4\pi}{(2s+1)!}}
    \nonumber\\[0.3\baselineskip]
    &\hspace{0pt}=\sqrt{(3s-m_1-m_2-m_3)!\,(3s+m_1+m_2+m_3)!}
%    \cdot\,\frac{2}{\sqrt{3s}}\,\sqrt{\frac{2\pi}{2s+1}}^3
    \,\frac{2}{\sqrt{3s\,(6s-1)!}}\left(\frac{2\pi}{2s+1}\right)^{\!\!\frac{3}{2}}.
    \nonumber\\[0.2\baselineskip]
\end{align}
The last two factors in \eqref{eq:b:gmcg2} do not depend on any $m_i$ and
can hence be absorbed by rescaling $V$ accordingly.  This yields
% \begin{equation}
%   \label{eq:b:V3}
%   \begin{split}
%     V%^{\s\text{qh}}
%     =&\sum_{m_1=-s}^s\, \sum_{m_2=-s}^s\, \sum_{m_3=-s}^s\, \sum_{m_4=-s}^s
%     \, \sum_{m_5=-s}^s\, \sum_{m_6=-s}^s
% %    a_{m_1}^\dagger a_{m_2}^\dagger a_{m_3}^\dagger a_{m_4}a_{m_5}a_{m_6}
%     \\[0.3\baselineskip]
%     &\cdot a_{m_1}^\dagger a_{m_2}^\dagger a_{m_3}^\dagger a_{m_4}a_{m_5}a_{m_6}\,
%     V_{m_1,m_2,m_3,m_4,m_5,m_6}
%   \end{split}
% \end{equation}
\begin{align}
  \label{eq:b:V2}
    V%^{\s\text{qh}}
    =&\sum_{m_1=-s}^s\, \sum_{m_2=-s}^s\, \sum_{m_3=-s}^s\, \sum_{m_4=-s}^s
    \, \sum_{m_5=-s}^s\, \sum_{m_6=-s}^s
    a_{m_1}^\dagger a_{m_2}^\dagger a_{m_3}^\dagger a_{m_4}a_{m_5}a_{m_6}
    \nonumber\\[0.3\baselineskip]
    &\hspace{180pt}\cdot 
    % a_{m_1}^\dagger a_{m_2}^\dagger a_{m_3}^\dagger a_{m_4}a_{m_5}a_{m_6}\,
    V_{m_1,m_2,m_3,m_4,m_5,m_6}
\end{align}
with 
\begin{gather}%{equation}
  \label{eq:b:V6m}
    V_{m_1,m_2,m_3,m_4,m_5,m_6}=V_{m_1+m_2+m_3}\cdot\delta_{m_1+m_2+m_3,m_4+m_5+m_6},
    \\[0.3\baselineskip]
  \label{eq:b:Vm}
    V_m={(3s-m)!\,(3s+m)!}.
\end{gather}%{equation}
Note that the scattering matrix elements $V_{m_1,m_2,m_3,m_4,m_5,m_6}$
in \eqref{eq:b:V2} depend once again only on the conserved total value
of $L^\z$, $m_1+m_2+m_3=m_4+m_5+m_6$, and not on any of the (angular)
momentum transfers.  This constitutes an enormous simplification.

\section{Fourier transformation}
\label{sec:b:four}

\subsection{Particle creation and annihilation operators}

We proceed by transforming the Hamiltonian \eqref{eq:b:V2} into Fourier 
space, using the transformations
\begin{equation}
  \label{eq:b:aFT}
  a_m
  =\frac{1}{\sqrt{N}}\sum_{\alpha=1}^N (\bar\eta_\alpha)^{s+m} a_\alpha,\quad
  a_m^\dagger
  =\frac{1}{\sqrt{N}}\sum_{\alpha=1}^N (\eta_\alpha)^{s+m} a_\alpha^\dagger,
\end{equation}
where $N=2s+1$, and $\eta_\alpha=e^{\text{i}\frac{2\pi}{N}\alpha}$.
We again interpret $\alpha$ as site indices of a periodic chain with
$N$ sites, and $\eta_\alpha$ as the positions of these sites when the
periodic chain is embedded as a unit circle in the complex plane.

The Fourier transformation yields
\begin{equation}
  \label{eq:b:V3}
  V = \frac{1}{N^3}\sum_{\{\alpha_1,\alpha_2,\alpha_3,\alpha_4,\alpha_5,\alpha_6\}}
  a_{\alpha_6}^\dagger a_{\alpha_5}^\dagger a_{\alpha_4}^\dagger 
  a_{\alpha_3}^{\phantom{\dagger}}
  a_{\alpha_2}^{\phantom{\dagger}} a_{\alpha_1}^{\phantom{\dagger}}\,
  V_{\alpha_1,\alpha_2,\alpha_3,\alpha_4,\alpha_5,\alpha_6}
\end{equation}
with
\begin{equation}
  \label{eq:b:Valpha0}
  \begin{split}
    V_{\alpha_1,\alpha_2,\alpha_3,\alpha_4,\alpha_5,\alpha_6}
    &=\sum_{m_1=-s}^s\,\sum_{m_2=-s}^s\,\sum_{m_3=-s}^s\,\sum_{m_4=-s}^s
    \, \sum_{m_5=-s}^s\, \sum_{m_6=-s}^s
    \\[0.4\baselineskip]
    &V_{m_1+m_2+m_3}\,\delta_{m_1+m_2+m_3,m_4+m_5+m_6}\,
    \\[0.4\baselineskip]
    &(\et{6})^{s+m_6}(\et{5})^{s+m_5}(\et{4})^{s+m_4}
    (\etb{3})^{s+m_3}(\etb{2})^{s+m_2}(\etb{1})^{s+m_1}
%    &\cdot\, (\et{6})^{s+m_6}(\et{5})^{s+m_5}(\et{4})^{s+m_4}
%    (\etb{3})^{s+m_3}(\etb{2})^{s+m_2}(\etb{1})^{s+m_1}
%    (\eta_{\alpha_4})^{s+m_4}(\eta_{\alpha_3})^{s+m_3}
%    (\bar\eta_{\alpha_2})^{s+m_2}(\bar\eta_{\alpha_1})^{s+m_1}
    \\[0.3\baselineskip] %\hspace{40pt}
  \end{split}
\end{equation}
and $V_m$ given by \eqref{eq:b:Vm} for the interaction Hamiltonian,
and 
\begin{equation}
  \label{eq:b:psialpha1}
  \begin{split}
    \ket{\psi_0}%^{\s\text{rqH}}}
  &=G^{-1}\ket{\psi^{\s\text{qH}}_0}
  \\[0.2\baselineskip]
  &=\sum_{\{\alpha_1,\ldots ,\alpha_N\}} 
  \frac{1}{\sqrt{N}^N}
  \sum_{\{m_1,\ldots ,m_N\}} 
%  C_{s+m_1,\ldots ,s+m_N}\,
  C_{m_1+s,\ldots ,m_N+s}\,
  (\eta_{\alpha_1})^{s+m_1}\ldots (\eta_{\alpha_N})^{s+m_N}
  \\[0\baselineskip]
  &\hspace{230pt}\cdot\,
  {a}^\dagger_{\alpha_1}\ldots {a}^\dagger_{\alpha_N} \vac
  \\[0.6\baselineskip]
  &=\sum_{\{\alpha_1,\ldots ,\alpha_N\}} 
  \psi^{S=1}_0(\eta_{\alpha_1},\ldots,\eta_{\alpha_N})\,
  {a}^\dagger_{\alpha_1}\ldots {a}^\dagger_{\alpha_N} \vac,
  \end{split}
\end{equation}
where $\psi^{S=1}_0(\eta_{\alpha_1},\ldots,\eta_{\alpha_N})$ is given
by \eqref{eq:b:psi0}, for the ground state it annihilates.  In
\eqref{eq:b:psialpha1}, we have used the definition of the
coefficients $C_{m_1+s,\ldots ,m_N+s}$ from \eqref{eq:b:psiS1q}.

\subsection{Substitution of spin flip operators for boson operators}

The formulation of the model in terms of position space operators
allows us to substitute spin flip operators for the creation and
annihilation operators, and thus to turn our boson model into a spin
model.  For the $S=1$ model, this step is not as trivial as for the
$S=\frac{1}{2}$ model treated in Section \ref{sec:l2hs}, as the
usual spin flip operators do not obey the same commutation relations
as bosonic ladder operators in the subspace where each site can be
doubly occupied at most.  The relation % \eqref{eq:s-s+tilde},
\begin{equation}
  \label{eq:b:s-s+tilde}%\nonumber
  S_\alpha^-\,(\tilde{S}_\alpha^+)^n \ket{1,-1}_{\alpha}
  =n\, (\tilde{S}_\alpha^+)^{n-1}\ket{1,-1}_{\alpha},
  \quad\text{for}\ n=0,1,2,
\end{equation}
which follows directly form the definition \eqref{eq:b:tildeS+},
instructs us how to proceed.  Since
\begin{equation*}
  a\,(a^\dagger)^n\vac =n\,(a^\dagger)^{n-1}\vac ,
\end{equation*}
%\eqref{eq:b:psialpha1} is still annihilated by \eqref{eq:b:V3} if we
we may substitute $a_{\alpha_i}^\dagger\to {S}^+_{\alpha_i}$,
$a_{\alpha_i}\to {S}^-_{\alpha_i}$, in the Hamiltonian and
$a_{\alpha_i}^\dagger\to {\tilde S}^+_{\alpha_i}$, $\vac\to
\ket{-1}_N$ in the ground state.  In other words, the non-Abelian
$S=1$ spin liquid state \eqref{eq:b:ket} with \eqref{eq:b:psi0}
introduced in Section \ref{sec:3mod-na}, is annihilated by
\begin{equation}
  \label{eq:b:Vspin1}
  V = \frac{1}{N^3}\sum_{\{\alpha_1,\alpha_2,\alpha_3,\alpha_4,\alpha_5,\alpha_6\}}
  {S}^+_{\alpha_6} {S}^+_{\alpha_5} {S}^+_{\alpha_4}\, 
  {S}^-_{\alpha_3} {S}^-_{\alpha_2} {S}^-_{\alpha_1}\,
  V_{\alpha_1,\alpha_2,\alpha_3,\alpha_4,\alpha_5,\alpha_6}
\end{equation}
with the matrix elements \eqref{eq:b:Valpha0}.  For the on-site
potential term \eqref{eq:b:UqH}, Fourier transformation and subsequent
substitution yields again
\begin{equation}
  \label{eq:b:UqHsub}
  U^{\s\text{qH}} = \frac{1}{N} U_0\, {S}_{\text{tot}}^+ {S}_{\text{tot}}^-.
\end{equation}
This term annihilates any singlet state, and will not be helpful in
constructing a parent Hamiltonian.  We will keep in mind, however,
that the original term was required to single out the ground state wave
function \eqref{eq:b:psiqH} on the quantum Hall sphere.

Note that this substitution does not just amount to a renaming of
operators, as it did for the spin \half chain discussed in Section
\ref{sec:l2hs}.  In the present case, it effectively renormalizes the
single particle Hilbert spaces once more, and hence leads to a
different model.  To see this, compare the normalizations of
``unoccupied'', ``singly occupied'', and ``doubly occupied'' sites in
the $S=1$ spin chain,
\begin{gather*}
%  \bra{1,-1} {\tilde S}^- {\tilde S}^+\ket{1,-1}=\frac{1}{2},\\
%  \bra{1,-1} ({\tilde S}^-)^2 ({\tilde S}^+)^2\ket{1,-1}=1,
  \bra{1,-1} {\tilde S}_\alpha^- {\tilde S}_\alpha^+\ket{1,-1}=\frac{1}{2},\\
  \bra{1,-1} ({\tilde S}_\alpha^-)^2 ({\tilde S}_\alpha^+)^2\ket{1,-1}=1,
\end{gather*}
to those of bosons,
\begin{equation}
  \label{eq:b:aadagnorm}
%  \bac a^n {(a^\dagger)}^n\vac = n!,
  \bac a^n {a^\dagger}^n\vac = n!.
\end{equation}
The difference does not just amount to a different overall
normalizations of the states.  If we were, for example, to renormalize
the already renormalized spin operators ${\tilde S}_\alpha\to
\sqrt{2}{\tilde S}_\alpha$, we would obtain
\begin{gather*}
%  \bra{1,-1}  {{\tilde S}}^-     {{\tilde S}}^+\ket{1,-1}=1,\\
%  \bra{1,-1} ({{\tilde S}}^-)^2 ({{\tilde S}}^+)^2\ket{1,-1}=4.
  \bra{1,-1}  {{\tilde S}}_\alpha^-     {{\tilde S}}_\alpha^+\ket{1,-1}=1,\\
  \bra{1,-1} ({{\tilde S}}_\alpha^-)^2 ({{\tilde S}}_\alpha^+)^2\ket{1,-1}=4.
\end{gather*}
This would match \eqref{eq:b:aadagnorm} for $n=1$, but not for $n=2$.
The amplitudes of the individual spin configurations in the
spin state vector are hence different from those of the corresponding
amplitudes in the boson state vector.

\subsection{Many body annihilation operators}
\label{sec:b:MBannihilation}

Since the scatting elements \eqref{eq:b:Valpha0} %in \eqref{eq:b:Vspin1}
depend only on the total angular momentum quantum number $m$, we can 
rewrite \eqref{eq:b:Vspin1} as
\begin{equation}
  \label{eq:b:Vspin3}
  V = \sum_{m=-3s}^{3s} V_{m} B_m^\dagger B_m,
\end{equation}
where $V_m$ is given by \eqref{eq:b:Vm}, and
\begin{equation}
  \label{eq:b:Bm}
  B_m=B_m^{\ne} +\,B_m^=
\end{equation}
with
\begin{align}
  \label{eq:b:Bmnot=}
%    B_m&=B_m^{\ne} +3\,B_m^=,
%    \\[0.5\baselineskip]
    B_m^{\ne} 
    &=\frac{1}{\sqrt{N^3}}
    \sum_{\substack{\alpha_1,\alpha_2,\alpha_3=1\\ 
        \alpha_1\ne\alpha_2\ne\alpha_3\ne\alpha_1}}^N 
    B_{m;\alpha_1,\alpha_2,\alpha_3}^{\ne}
    {S}^-_{\alpha_3} {S}^-_{\alpha_2} {S}^-_{\alpha_1},
    \\[0.3\baselineskip]\label{eq:b:Bm=}
    B_m^= 
    &=\frac{3}{\sqrt{N^3}}
    \sum_{\substack{\alpha_1,\alpha_2=1\\ \alpha_1\ne\alpha_2}}^N 
    B_{m;\alpha_1,\alpha_2}^=
    \left({S}^-_{\alpha_2}\right)^2 {S}^-_{\alpha_1}.
\end{align}
The coefficients in \eqref{eq:b:Bmnot=} and \eqref{eq:b:Bm=} are given by
% \begin{equation}
%   \label{eq:b:Bmaaa}
%   \begin{split}
%     B_{m;\alpha_1,\alpha_2,\alpha_3}^{\ne}
%     &=\sum_{m_1=-s}^s\,\sum_{m_2=-s}^s\,\sum_{m_3=-s}^s\,
%     (\etb{3})^{s+m_3}(\etb{2})^{s+m_2}(\etb{1})^{s+m_1}\,
%     \\[0\baselineskip]
%     &\hspace{102pt} \cdot\delta_{m,m_1+m_2+m_3},
%     \\[0.4\baselineskip]
%     B_{m;\alpha_1,\alpha_2}^=
%     &=\sum_{m_1=-s}^s\,\sum_{m_2=-s}^s\,\sum_{m_3=-s}^s\,
%     (\etb{2})^{s+m_3}(\etb{2})^{s+m_2}(\etb{1})^{s+m_1}\,
% %    (\etb{2})^{2s+m_2+m_3}(\etb{1})^{s+m_1}\,
%     \\[0\baselineskip]
%     &\hspace{102pt} \cdot\delta_{m,m_1+m_2+m_3}.
%   \end{split}
% \end{equation}
\begin{align}
  \label{eq:b:Bmnot=aaa}
    B_{m;\alpha_1,\alpha_2,\alpha_3}^{\ne}
    &=
    \sum_{m_1=-s}^s\,\sum_{m_2=-s}^s\,\sum_{m_3=-s}^s\,
    (\etb{3})^{s+m_3}(\etb{2})^{s+m_2}(\etb{1})^{s+m_1}\,
    \nonumber\\[0\baselineskip]
    &\hspace{102pt} \cdot\,\delta_{m,m_1+m_2+m_3},
    \\[0.4\baselineskip]\label{eq:b:Bm=aaa}
    B_{m;\alpha_1,\alpha_2}^=
    &=
    \sum_{m_1=-s}^s\,\sum_{m_2=-s}^s\,\sum_{m_3=-s}^s\,
    (\etb{2})^{s+m_3}(\etb{2})^{s+m_2}(\etb{1})^{s+m_1}\,
%    (\etb{2})^{2s+m_2+m_3}(\etb{1})^{s+m_1}\,
    \nonumber\\[0\baselineskip]
    &\hspace{102pt} \cdot\,\delta_{m,m_1+m_2+m_3}.
\end{align}
The factor 3 in the definition \eqref{eq:b:Bm=} of $B_m^=$ stems from
the three possibilities of two coordinates being equal.

%\subsection{Evaluation of the coefficients $B_{m;\alpha_1,\alpha_2,\alpha_3}^{\ne}$}

\subsection{Evaluation of $B_{m;\alpha_1,\alpha_2,\alpha_3}^{\ne}$}
\label{sec:b:Bnot=}

In this section, we evaluate
\begin{align}
  \label{eq:b:Bmnot=1}
    B_{m;\alpha_1,\alpha_2,\alpha_3}^{\ne}
    &=\sum_{m_1=-s}^s\,\sum_{m_2=-s}^s\,\sum_{m_3=-s}^s\,
    (\etb{3})^{s+m_3}(\etb{2})^{s+m_2}(\etb{1})^{s+m_1}\,
    \nonumber\\[0\baselineskip]
    &\hspace{102pt} \cdot\delta_{m,m_1+m_2+m_3}
\end{align}
subject to the condition that none the coordinates $\alpha_1$, $\alpha_2$, 
and $\alpha_3$ coincide.

To begin with, we carry out the sum over $m_3$, % in \eqref{eq:b:Bm1},
and obtain
\begin{align}
  \label{eq:b:Bmnot=2}
    B_{m;\alpha_1,\alpha_2,\alpha_3}^{\ne}
    &={\sum_{m_1}}'\,(\etb{1})^{s+m_1}\,
%    {\sum_{m_2}}'\,(\etb{3})^{s+m-m_1-m_2}(\etb{2})^{s+m_2}
    \underbrace{{\sum_{m_2}}'\,(\etb{3})^{s+m-m_1-m_2}(\etb{2})^{s+m_2}}_{
      \textstyle \equiv I_{m_1} \rule{0pt}{.6\baselineskip}},
\end{align}
where the primed sums are restricted such that all the exponents of the
$\etb{}$'s are between $0$ and $2s$.  With $-s\le {m-m_1-m_2}\le s$,
we have 
\begin{align}
  \label{eq:b:Bmnot=m2sum1}
%    {\sum_{m_2}}'\,&(\etb{3})^{s+m-m_1-m_2}(\etb{2})^{s+m_2}
%    \\[0.2\baselineskip]
    I_{m_1}
    &=(\etb{3})^{2s+m-m_1} 
    \sum_{m_2=\max\{-s,-s+m-m_1\}}^{\min\{s,s+m-m_1\}}(\etba_{23})^{s+m_2}
    \nonumber\\[0.4\baselineskip]
    &=\left\{
    \begin{alignedat}{2} 
       &A_{m-m_1;\alpha_2,\alpha_3} &\quad &\text{for}\ -2s\le m-m_1\le 2s, 
       \\[.2\baselineskip] 
       &\,0                           &\quad &\text{otherwise},
    \end{alignedat}\right.
    \nonumber\\[0.6\baselineskip]
    &=\left\{
    \begin{alignedat}{2} 
       &\frac{\etb{2}^{m-m_1}-\etb{3}^{m-m_1}}{\etb{2}-\etb{3}}&
       \quad &\text{for}\ m\le m_1\le 2s+m, 
       \\[.2\baselineskip] 
%       0&,\quad m_1=m, \\[-.2\baselineskip] 
       -&\frac{\etb{2}^{m-m_1}-\etb{3}^{m-m_1}}{\etb{2}-\etb{3}}&
       \quad &\text{for}\, -2s+m\le m_1\le m-1, 
       \\[.2\baselineskip] 
       &\,0&\quad &\text{otherwise},
    \end{alignedat}\right.
\end{align}
where we have defined $\etba_{23}\equiv
\etba_{\alpha_2-\alpha_3}=\etb{2}\et{3}$ and used the result
\eqref{eq:a:amaa2} for the sum \eqref{eq:a:amaa1} from Section
\ref{sec:a:renalt}.

For the evaluation of the sum over $m_1$, we consider three different
regimes for $m$.

%\vspace{.5\baselineskip}%\noindent 
\renewcommand{\labelenumi}{\alph{enumi})}
\begin{enumerate}
\item $-s<m\le s\,$.  In this regime, $m-m_1$ changes sign as we sum over
$m_1$.  We obtain
\begin{align}
  \label{eq:b:Bmnot=3}
    B_{m;\alpha_1,\alpha_2,\alpha_3}^{\ne}
    &=
    \frac{\etb{2}^{s+m}}{\etb{2}-\etb{3}}
    \left(-\sum_{m_1=-s}^{m-1} \etba_{12}^{s+m_1}
          +\sum_{m_1=m}^{s} \etba_{12}^{s+m_1}\right)%\hspace{65pt}
    \nonumber\\*[0.3\baselineskip]
    &\quad +\ \text{same term with}\,\ \etb{2}\leftrightarrow \etb{3}
    \nonumber\\[0.4\baselineskip]
    &=
    \frac{\etb{2}^{s+m}}{\etb{2}-\etb{3}}
    \left(-\frac{\etba_{12}^{s+m}-1}{\etba_{12}-1}
          +\frac{1-\etba_{12}^{s+m}}{\etba_{12}-1}\right)
    \nonumber\\*[0.3\baselineskip]
    &\quad +\ \text{same term with}\,\ \etb{2}\leftrightarrow \etb{3}
    \nonumber\\[0.4\baselineskip]
%    \phantom{B_{m;\alpha_1,\alpha_2,\alpha_3}^{\ne}}
    &=
    -\frac{2\etb{2}}{\etb{2}-\etb{3}}\cdot
    \frac{\etb{1}^{s+m}-\etb{2}^{s+m}}{\etb{1}-\etb{2}}
    +\ \text{same term with}\,\ \etb{2}\leftrightarrow \etb{3}
%    \nonumber\\*[0.3\baselineskip]
%    &&\quad +\ \text{same term with}\,\ \etb{2}\leftrightarrow \etb{3}
    \nonumber\\[0.6\baselineskip]
    &=
    \frac{2\etb{2}^{s+m+1}}{(\etb{1}-\etb{2})(\etb{2}-\etb{3})}
     +\frac{2\etb{3}^{s+m+1}}{(\etb{2}-\etb{3})(\etb{3}-\etb{1})}
    \nonumber\\*[0.4\baselineskip]
    &\quad -\frac{2\etb{1}^{s+m}}{\etb{2}-\etb{3}}
    \left(\frac{\etb{2}}{\etb{1}-\etb{2}}-\frac{\etb{3}}{\etb{1}-\etb{3}}\right)
    \nonumber\\[0.6\baselineskip]
    &=
    \frac{2\etb{2}^{s+m+1}}{(\etb{1}-\etb{2})(\etb{2}-\etb{3})}
     +\frac{2\etb{3}^{s+m+1}}{(\etb{2}-\etb{3})(\etb{3}-\etb{1})}
    \nonumber\\*[0.4\baselineskip]
    &\quad +\frac{2\etb{1}^{s+m+1}}{(\etb{3}-\etb{1})(\etb{1}-\etb{2})}
    \nonumber\\[0.4\baselineskip]
    &\equiv
    2Q^{\ne}_{m;\alpha_1,\alpha_2,\alpha_3},
\end{align}
where $Q^{\ne}_{m;\alpha_1,\alpha_2,\alpha_3}$ is strictly periodic under
$m\to m+N$ with $N=2s+1$.
%\begin{equation}
%  \label{eq:b:Qperiodic}
%  Q^{\ne}_{m+N;\alpha_1,\alpha_2,\alpha_3}=Q^{\ne}_{m;\alpha_1,\alpha_2,\alpha_3}
%\end{equation}

\item  $-3s\le m\le-s$.  Since $-s\le m_1\le s$, this implies that we are always
in the first regime in \eqref{eq:b:Bmnot=m2sum1}, $m\le m_1\le 2s+m$.
This yields
\begin{align}
  \label{eq:b:Bmnot=5}
    B_{m;\alpha_1,\alpha_2,\alpha_3}^{\ne}
    &=\frac{\etb{2}^{s+m}}{\etb{2}-\etb{3}}
    \sum_{m_1=-s}^{2s+m} \etba_{12}^{s+m_1}
    +\ \text{same term with}\,\ \etb{2}\leftrightarrow \etb{3}
    \nonumber\\[0.4\baselineskip]
    &=\frac{\etb{2}^{s+m}}{\etb{2}-\etb{3}}
    \frac{\etba_{12}^{s+m}-1}{\etba_{12}-1}
    +\ \text{same term with}\,\ \etb{2}\leftrightarrow \etb{3}
    \nonumber\\[0.4\baselineskip]
    &=-Q^{\ne}_{m;\alpha_1,\alpha_2,\alpha_3}.
\end{align}

\item $s<m\le 3s$.  Since $-s\le m_1\le s$, this implies that we are always
in the second regime in \eqref{eq:b:Bmnot=m2sum1}, $-2s+m\le m_1\le m-1$.
This yields
\begin{align}
  \label{eq:b:Bmnot=7}
    B_{m;\alpha_1,\alpha_2,\alpha_3}^{\ne}
    &=-\frac{\etb{2}^{s+m}}{\etb{2}-\etb{3}}
    \sum_{m_1=-2s+m}^{s} \etba_{12}^{s+m_1}
    +\, \text{same term with}\,\ \etb{2}\leftrightarrow \etb{3}
    \nonumber\\[0.4\baselineskip]
    &=-\frac{\etb{2}^{s+m}}{\etb{2}-\etb{3}}
    \sum_{m_1=-2s+m-1}^{s} \etba_{12}^{s+m_1}\nonumber\\[0.2\baselineskip]
    &\hspace{100pt}+\ \text{same term with}\,\ \etb{2}\leftrightarrow \etb{3}
%    &\quad +\ \text{same term with}\,\ \etb{2}\leftrightarrow \etb{3}
    \nonumber\\[0.4\baselineskip]
    &=-\frac{\etb{2}^{s+m}}{\etb{2}-\etb{3}}
    \frac{1-\etba_{12}^{s+m}}{\etba_{12}-1}
    +\ \text{same term with}\,\ \etb{2}\leftrightarrow \etb{3}
    \nonumber\\[0.4\baselineskip]
    &=-Q^{\ne}_{m;\alpha_1,\alpha_2,\alpha_3}.
\end{align}

\end{enumerate}

Note that since $Q^{\ne}_{\pm s;\alpha_1,\alpha_2,\alpha_3}=0$, it does not
matter with which regime we associate the cases $m=\pm s$.
As a curiosity, note further that $Q^{\ne}_{s+2;\alpha_1,\alpha_2,\alpha_3}=1$
(\cf \eqref{eq:app-hsidentity} of \ref{sec:app-hssums}).

\subsection{Evaluation of $B_{m;\alpha_1,\alpha_2}^=$}
\label{sec:b:B=}

We now evaluate
\begin{align}
  \label{eq:b:Bm=1}
    B_{m;\alpha_1,\alpha_2}^{=}
    &=\sum_{m_1=-s}^s\,\sum_{m_2=-s}^s\,\sum_{m_3=-s}^s\,
    (\etb{2})^{s+m_3}(\etb{2})^{s+m_2}(\etb{1})^{s+m_1}\,
%    (\etb{2})^{2s+m_2+m_3}(\etb{1})^{s+m_1}\,
    \nonumber\\[0\baselineskip]
    &\hspace{102pt} \cdot\delta_{m,m_1+m_2+m_3}
\end{align}
subject to the condition $\alpha_1\ne\alpha_2$.

To begin with, we carry out the sum over $m_3$, % in \eqref{eq:b:Bm1},
and obtain 
\begin{align}
  \label{eq:b:Bm=2}
    B_{m;\alpha_1,\alpha_2}^{=}
    &={\sum_{m_1}}'\,(\etb{1})^{s+m_1}\,
    {\sum_{m_2}}'\,(\etb{2})^{2s+m-m_1},
%    \sum_{m_2=\max\{-s,-s+m-m_1\}}^{\min\{s,s+m-m_1\}}.
    \nonumber\\[0.4\baselineskip]
%    &={\sum_{m_1}}'\,(\etb{1})^{s+m_1}\,(\etb{2})^{2s+m-m_1}\,
%    {\sum_{m_2}}'\,,
    &=(\etb{2})^{s+m-1}\,{\sum_{m_1}}'\,(\etba_{12})^{s+m_1}\,
    {\sum_{m_2}}'\,1,
\end{align}
where the primed sums are restricted such that all the exponents of
the original $\etb{}$'s in \eqref{eq:b:Bm=1} are between $0$ and $2s$.
With $-s\le {m-m_1-m_2}\le s$, we have
\begin{align}
  \label{eq:b:Bm=m2sum1}
    {\sum_{m_2}}'\,1 
    &=\sum_{m_2=\max\{-s,-s+m-m_1\}}^{\min\{s,s+m-m_1\}}1 
    \nonumber\\[0.6\baselineskip]
%     &=\left\{
%     \begin{alignedat}{2} 
%        &N-|m-m_1| &\quad &\text{for}\ -2s\le m-m_1\le 2s, 
%        \\ %[.2\baselineskip] 
%        &\,0       &\quad &\text{otherwise}.
%     \end{alignedat}\right.
%     \\[0.6\baselineskip]
    &=\left\{
    \begin{alignedat}{2} 
       &N+m-m_1 &\quad &\text{for}\ m\le m_1\le 2s+m, 
       \\ %[.2\baselineskip] 
       &N-m+m_1&\quad &\text{for}\, -2s+m\le m_1\le m-1, 
       \\ %[.2\baselineskip] 
       &\,0&\quad &\text{otherwise},
    \end{alignedat}\right.
%    \\[0.4\baselineskip]
\end{align}

For the evaluation of the sum over $m_1$, we again consider three different
regimes for $m$.

%\vspace{.5\baselineskip}%\noindent 
\renewcommand{\labelenumi}{\alph{enumi})}
\begin{enumerate}
\item $-s<m\le s\,$.  In this regime, $m-m_1$ changes sign as we sum over
$m_1$.  We obtain
%\newpage
\begin{align}
  \label{eq:b:Bm=3}
%    \frac{B_{m;\alpha_1,\alpha_2}^{=}}{(\etb{2})^{s+m-1}}
    \frac{B_{m;\alpha_1,\alpha_2}^{=}}{\etb{2}^{s+m-1}}
    &=\sum_{m_1=-s}^{m-1} (N-m+m_1)\, \etba_{12}^{s+m_1}
     +\sum_{m_1=m}^{s}(N+m-m_1)\, \etba_{12}^{s+m_1}
    \nonumber\\[0.3\baselineskip]
    &=N\underbrace{\sum_{m_1=-s}^{s}\etba_{12}^{s+m_1}}_{=0}
    -\, m\left(\frac{\etba_{12}^{s+m}-1}{\etba_{12}-1}
            -\frac{1-\etba_{12}^{s+m}}{\etba_{12}-1}\right)
    \nonumber\\[0\baselineskip]
    &\quad +\sum_{m_1=-s}^{m-1} m_1 \etba_{12}^{s+m_1}
     -\sum_{m_1=m}^{s}m_1 \etba_{12}^{s+m_1}.
\end{align}
With the formula
\begin{equation}
  \label{eq:b:qx^qSum}
  \sum_{q=a}^b\, q\, x^q = 
  \frac{(b+1)\,x^{b+1}-a\,x^a}{x-1}-\frac{x^{b+2}-x^{a+1}}{(x-1)^2},
  \qquad b\ge a,
\end{equation}
we obtain for the last two sums in \eqref{eq:b:Bm=3},
\begin{align*}
  \sum_{m_1=-s}^{m-1} m_1 \etba_{12}^{s+m_1}
  &=
  \frac{m\,\etba_{12}^{s+m}+s}{\etba_{12}-1}
  -\frac{\etba_{12}^{s+m+1}-\etba_{12}}{(\etba_{12}-1)^2}
  \\[0.4\baselineskip]
  -\sum_{m_1=m}^{s}m_1 \etba_{12}^{s+m_1}
  &=
  -\frac{(s+1)-m\,\etba_{12}^{s+m}}{\etba_{12}-1}
  +\frac{\etba_{12}-\etba_{12}^{s+m+1}}{(\etba_{12}-1)^2}.
\end{align*}
Summing up all the terms we find
\begin{align}
  \label{eq:b:Bm=4}
    B_{m;\alpha_1,\alpha_2}^{=}
    &=\etb{2}^{s+m-1}\left(\frac{2m-1}{\etba_{12}-1}
      -2\,\frac{\etba_{12}^{s+m+1}-\etba_{12}}{(\etba_{12}-1)^2}\right)
    \nonumber\\[0.4\baselineskip]
    &=\etb{2}^{s+m-1}\left(\frac{2m+1}{\etba_{12}-1}
      -2\,\frac{\etba_{12}^{s+m+1}-1}{(\etba_{12}-1)^2}\right)
    \nonumber\\[0.4\baselineskip]
    &=(2m+1)\,\frac{\etb{2}^{s+m}}{\etb{1}-\etb{2}}
    -2\frac{\etb{1}^{s+m+1}-\etb{2}^{s+m+1}}{(\etb{1}-\etb{2})^2}
    \nonumber\\[0.4\baselineskip]
    &=(2m+1)\,P_{m;\alpha_1,\alpha_2}+2Q^=_{m;\alpha_1,\alpha_2},
\end{align}
where we have defined 
% \begin{align}
%   \label{eq:b:Pdef}
%   P_{m;\alpha_1,\alpha_2}
%   \hspace{-5pt}&\equiv&\hspace{-5pt}
%   \frac{\etb{2}^{s+m}}{\etb{1}-\etb{2}},
%   \\[0.4\baselineskip]\label{eq:b:Rdef}
%   Q^=_{m;\alpha_1,\alpha_2}
%   \hspace{-5pt}&\equiv&\hspace{-5pt}
%   -\etb{1}\,\frac{\etb{1}^{s+m}-\etb{2}^{s+m}}{(\etb{1}-\etb{2})^2}
% \end{align}
\begin{equation}
  \label{eq:b:PRdef}
  P_{m;\alpha_1,\alpha_2}\equiv \frac{\etb{2}^{s+m}}{\etb{1}-\etb{2}},
  \quad
  Q^=_{m;\alpha_1,\alpha_2}\equiv 
  -\frac{\etb{1}^{s+m+1}-\etb{2}^{s+m+1}}{(\etb{1}-\etb{2})^2}.
\end{equation}

\vspace{0.5\baselineskip}
\item $-3s\le m\le-s$.  Since $-s\le m_1\le s$, this implies that we
  are always in the first regime in \eqref{eq:b:Bm=m2sum1}, $m\le
  m_1\le 2s+m$.  This yields
\begin{align}
  \label{eq:b:Bm=5}
    \frac{B_{m;\alpha_1,\alpha_2}^{=}}{\etb{2}^{s+m-1}}
%    B_{m;\alpha_1,\alpha_2,\alpha_3}^{\ne}
    &=%\etb{2}^{s+m-1}
    \sum_{m_1=-s}^{2s+m} (N+m-m_1)\,\etba_{12}^{s+m_1}
    \nonumber\\[0.4\baselineskip]
    &=(N+m)\,\frac{\etba_{12}^{s+m}-1}{\etba_{12}-1}
%    \\[0.4\baselineskip]
%    &\quad 
    -\frac{(N+m)\,\etba_{12}^{s+m}+s}{\etba_{12}-1}
    +\frac{\etba_{12}^{s+m+1}-\etba_{12}}{(\etba_{12}-1)^2}
%     \\[0.4\baselineskip]
%     &=-\frac{N+m+s}{\etba_{12}-1}
%     +\frac{\etba_{12}^{s+m+1}-\etba_{12}}{(\etba_{12}-1)^2},
    \nonumber\\[0.4\baselineskip]
    &=-\frac{N+m+s+1}{\etba_{12}-1}
    +\frac{\etba_{12}^{s+m+1}-1}{(\etba_{12}-1)^2},
\end{align}
and 
\begin{align}
  \label{eq:b:Bm=6}
    B_{m;\alpha_1,\alpha_2}^{=}
    &=-(N+m+s+1)\,\frac{\etb{2}^{s+m}}{\etb{1}-\etb{2}}
    +\frac{\etb{1}^{s+m+1}-\etb{2}^{s+m+1}}{(\etb{1}-\etb{2})^2}
    \nonumber\\[0.4\baselineskip]
    &=-(N+m+s+1)\,P_{m;\alpha_1,\alpha_2}-Q^=_{m;\alpha_1,\alpha_2}.
\end{align}

\item $s<m\le 3s$.  Since $-s\le m_1\le s$, this implies that we are always
in the second regime in \eqref{eq:b:Bm=m2sum1}, $-2s+m\le m_1\le m-1$.
This yields
\begin{align}
  \label{eq:b:Bm=7}
    \frac{B_{m;\alpha_1,\alpha_2}^{=}}{\etb{2}^{s+m-1}}
%    B_{m;\alpha_1,\alpha_2,\alpha_3}^{\ne}
    &=%\etb{2}^{s+m-1}
    \sum_{m_1=-2s+m}^{s} (N-m+m_1)\,\etba_{12}^{s+m_1}
    \nonumber\\[0.4\baselineskip]
    &=\sum_{m_1=-2s+m-1}^{s} (N-m+m_1)\,\etba_{12}^{s+m_1}
    \nonumber\\[0.4\baselineskip]
    &=(N-m)\,\frac{1-\etba_{12}^{s+m}}{\etba_{12}-1}
    \nonumber\\[0.4\baselineskip]
    &\quad 
    +\frac{(s+1)+(N-m)\,\etba_{12}^{s+m}}{\etba_{12}-1}
    -\frac{\etba_{12}-\etba_{12}^{s+m+1}}{(\etba_{12}-1)^2}
%     \\[0.4\baselineskip]
%     &=\frac{N-m+s+1}{\etba_{12}-1}
%     +\frac{\etba_{12}^{s+m+1}-\etba_{12}}{(\etba_{12}-1)^2},
    \nonumber\\[0.4\baselineskip]
    &=\frac{N-m+s}{\etba_{12}-1}
    +\frac{\etba_{12}^{s+m+1}-1}{(\etba_{12}-1)^2},
\end{align}
and 
\begin{align}
  \label{eq:b:Bm=8}
    B_{m;\alpha_1,\alpha_2}^{=}
    &=(N-m+s)\,\frac{\etb{2}^{s+m}}{\etb{1}-\etb{2}}
    +\frac{\etb{1}^{s+m+1}-\etb{2}^{s+m+1}}{(\etb{1}-\etb{2})^2}
    \\[0.4\baselineskip]
    &=(N-m+s)\,P_{m;\alpha_1,\alpha_2}-Q^=_{m;\alpha_1,\alpha_2}.
\end{align}
\end{enumerate}
Note that since $Q^=_{s;\alpha_1,\alpha_2}=0$ and
\begin{equation*}
  Q^=_{-s;\alpha_1,\alpha_2}
  =-\frac{\etb{1}-\etb{2}}{(\etb{1}-\etb{2})^2}
  =-\frac{1}{(\etb{1}-\etb{2})}
  =-P_{-s;\alpha_1,\alpha_2},
\end{equation*}
it does not matter with which regimes we associate the cases $m=\pm
s$.  The expressions \eqref{eq:b:Bm=4} and \eqref{eq:b:Bm=6} are equal
for $m=-s$, and \eqref{eq:b:Bm=4} and \eqref{eq:b:Bm=8} are equal for
$m=s$.

\section{The \defining condition for the $S=1$ Pfaffian chain}

\subsection{Derivation}

In Section \ref{sec:b:four}, we have shown that the non-Abelian
$S=1$ spin liquid state \eqref{eq:b:ket} with \eqref{eq:b:psi0}
introduced in Section \ref{sec:3mod-na}, is annihilated by
\begin{equation}
  \label{eq:b:VspinCon}
  V = \sum_{m=-3s}^{3s} V_{m} B_m^\dagger B_m,
\end{equation}
where %$V_m$ is given by \eqref{eq:b:Vm}, 
\begin{equation}
  \label{eq:b:VmCon}
    V_m={(3s-m)!\,(3s+m)!}
\end{equation}
and
\begin{equation}
  \label{eq:b:BmCon}
  B_m=B_m^{\ne} +B_m^=
\end{equation}
with
\begin{align}
  \label{eq:b:Bmnot=Con}
    B_m^{\ne} 
    &=\frac{1}{\sqrt{N^3}}
    \sum_{\substack{\alpha_1,\alpha_2,\alpha_3=1\\ 
        \alpha_1\ne\alpha_2\ne\alpha_3\ne\alpha_1}}^N 
    B_{m;\alpha_1,\alpha_2,\alpha_3}^{\ne}
    {S}^-_{\alpha_3} {S}^-_{\alpha_2} {S}^-_{\alpha_1},
    \\[0.3\baselineskip]\label{eq:b:Bm=Con}
    B_m^= 
    &=\frac{3}{\sqrt{N^3}}
    \sum_{\substack{\alpha_1,\alpha_2=1\\ \alpha_1\ne\alpha_2}}^N 
    B_{m;\alpha_1,\alpha_2}^=
    \left({S}^-_{\alpha_2}\right)^2 {S}^-_{\alpha_1}.
\end{align}
We calculated the coefficients in \eqref{eq:b:Bmnot=Con} and
\eqref{eq:b:Bm=Con} in Sections \ref{sec:b:Bnot=} and \ref{sec:b:B=},
respectively, and found
\begin{equation}
  \label{eq:b:Bmnot=result}
    B_{m;\alpha_1,\alpha_2,\alpha_3}^{\ne}
    =\left\{
    \begin{alignedat}{3} 
       -&Q^{\ne}_{m;\alpha_1,\alpha_2,\alpha_3}&
       \quad &\text{for}\, &s<&m\le 3s,
       \\[.2\baselineskip] 
       2&Q^{\ne}_{m;\alpha_1,\alpha_2,\alpha_3}&
       \quad &\text{for}\, &-s<&m\le s, 
       \\[.2\baselineskip] 
       -&Q^{\ne}_{m;\alpha_1,\alpha_2,\alpha_3}&
       \quad &\text{for}\, &-3s<&m\le -s, 
    \end{alignedat}\right.
\end{equation}
and
\begin{equation}
  \label{eq:b:Bm=result}
  \begin{split}
    B_{m;\alpha_1,\alpha_2}^{=}
    &=\left\{
    \begin{alignedat}{4} 
      (N-m+s)\,&P_{m;\alpha_1,\alpha_2}&-&&Q^=_{m;\alpha_1,\alpha_2}
      \quad &\text{for}\, &s<&m\le 3s,
      \\[.2\baselineskip] 
      (2m+1)\,&P_{m;\alpha_1,\alpha_2}\,&+&\,2&Q^=_{m;\alpha_1,\alpha_2}
      \quad &\text{for}\, &-s<&m\le s, 
      \\[.2\baselineskip] 
      -(N+m+s+1)\,&P_{m;\alpha_1,\alpha_2}&-&&Q^=_{m;\alpha_1,\alpha_2}
      \quad &\text{for}\, &-3s<&m\le -s. 
    \end{alignedat}\right.
%    \\[.2\baselineskip] 
  \end{split}
\end{equation}
$Q^{\ne}_{m;\alpha_1,\alpha_2,\alpha_3}$ is defined in \eqref{eq:b:Bmnot=3},
and $P_{m;\alpha_1,\alpha_2}$ and $ Q^=_{m;\alpha_1,\alpha_2}$ are defined
in \eqref{eq:b:PRdef}.  All three are periodic functions of $m$, \ie
\begin{equation}
  \label{eq:b:QPRperiodic}
  \begin{split}
    Q^{\ne}_{m+N;\alpha_1,\alpha_2,\alpha_3}&=Q^{\ne}_{m;\alpha_1,\alpha_2,\alpha_3},
    \\[.2\baselineskip] 
    P_{m+N;\alpha_1,\alpha_2}&=P_{m;\alpha_1,\alpha_2},
    \\[.2\baselineskip] 
    Q^=_{m+N;\alpha_1,\alpha_2}&=Q^=_{m;\alpha_1,\alpha_2}.
  \end{split}
\end{equation}

The property that $\ket{\psi^{S=1}_0}$ is annihilated by $V$ implies
with \eqref{eq:b:VspinCon} that
\begin{align}
  \label{eq:b:<V>}
    \bra{\psi^{S=1}_{0}} V \ket{\psi^{S=1}_{0}} 
    &=\sum_{m=-3s}^{3s} V_{m} \bra{\psi^{S=1}_{0}}
    B_m^\dagger B_m \ket{\psi^{S=1}_{0}}
    \nonumber\\[0.2\baselineskip]
    &=\sum_{m=-3s}^{3s} V_{m} 
    \bigl\Vert B_m \ket{\psi^{S=1}_{0}} \bigr\Vert^2=0.
\end{align}
Since all the values $V_m$ for $-3s\le m\le 3s$ are positive, and the
norms of the vectors by definition non-negative, \eqref{eq:b:<V>}
implies that the vectors $B_m \ket{\psi^{S=1}_{0}}$ must vanish
for all allowed values of $m$.  In other words,
\begin{equation}
  \label{eq:b:Bmpsi0}
  B_m \ket{\psi^{S=1}_{0}} = 0 \quad\forall\, m\in [-3s,3s].
\end{equation}
This implies that $\ket{\psi^{S=1}_{0}}$ is further annihilated by 
any linear combination of the $B_m$'s, and in particular also
those in which the terms involving $Q^{\ne}_{m+N;\alpha_1,\alpha_2,\alpha_3}$
and $Q^=_{m+N;\alpha_1,\alpha_2}$ cancel.  These include for $-s<m\le s$
\begin{align}
  \label{eq:b:Bmdiff1}
    B_m+2B_{m-N}
%    &=\underbrace{\!\bigl[(2m+1)-2(s+m+1)\bigr]\rule[-4pt]{0pt}{0pt}\!}_{-N} 
    &=\bigl[(2m+1)-2(s+m+1)\bigr]
    \sum_{\alpha_1\ne\alpha_2}^N P_{m;\alpha_1,\alpha_2}
    \left({S}^-_{\alpha_2}\right)^2 {S}^-_{\alpha_1}
    \nonumber\\[0.2\baselineskip]
    &=-N %P_{m},
    \sum_{\alpha_1\ne\alpha_2}^N P_{m;\alpha_1,\alpha_2}
    \left({S}^-_{\alpha_2}\right)^2 {S}^-_{\alpha_1},
\end{align}
and for $m=s+1$
\begin{equation}
  \label{eq:b:Bmdiff2}
  \begin{split}
    B_{s+1}-B_{-s}
    &=2N
%    &=\underbrace{\!\bigl[N-(-N)\bigr]\rule[-4pt]{0pt}{0pt}\!}_
%    {2N} 
%    &\left[N-(-N)\right]
    \sum_{\alpha_1\ne\alpha_2}^N 
    P_{s+1;\alpha_1,\alpha_2} \left({S}^-_{\alpha_2}\right)^2 {S}^-_{\alpha_1}.
  \end{split}
\end{equation}
Given the periodicity of $P_{m;\alpha_1,\alpha_2}$ in $m$,
\eqref{eq:b:Bmdiff1} and \eqref{eq:b:Bmdiff2} imply that
\begin{equation}
  \label{eq:b:Pmpsi0}
    P_{m} \ket{\psi^{S=1}_{0}} = 0\quad\forall\, m,
\end{equation}
where we have defined
\begin{align}
  \label{eq:b:Pmdef}
    P_{m}&\equiv\sum_{\alpha_1\ne\alpha_2}^N 
    P_{m;\alpha_1,\alpha_2}\left({S}^-_{\alpha_2}\right)^2 {S}^-_{\alpha_1}
    \nonumber\\[0.2\baselineskip]
    &=\sum_{\alpha_1\ne\alpha_2}^N\frac{\etb{2}^{s+m}}{\etb{1}-\etb{2}}
    \left({S}^-_{\alpha_2}\right)^2 {S}^-_{\alpha_1}.
\end{align}
Since the spin liquid state $\ket{\psi^{S=1}_{0}}$
is invariant under parity, \ie under $\et{}\to\etb{}$ (see Section
\ref{sec:na:sym}), it is also annihilated by the complex conjugates 
$\bar P_m$ of $P_m$ for all $m$.

The non-Abelian $S=1$ spin liquid state \eqref{eq:b:ket} with
\eqref{eq:b:psi0} is further annihilated by the operators
% \begin{equation}
%   \label{eq:b:Thetadef}
%   \begin{split}
%     \Oa &\equiv -\frac{1}{N}\sum_{m=0}^{N} \eab^{s+m} \bar P_m
%     \\[0.2\baselineskip]
%      &=\sum_{\substack{\beta=1\\[2pt]\beta\ne\alpha}}^N 
%     \frac{1}{\ea-\eb} (\Sa^-)^2 \Sb^-,\qquad 
%     \Oa \ket{\psi^{S=1}_{0}} = 0 \quad\forall\, \alpha,
%  \end{split}
% \end{equation}
\begin{align}
  \label{eq:b:Omegadef}
    \OaSone 
    &\equiv
    -\frac{1}{N}\sum_{m=0}^{N} \eab^{s+m} \bar P_m
    \nonumber\\[0.2\baselineskip]
    &=
     \sum_{\substack{\beta=1\\[2pt]\beta\ne\alpha}}^N 
    \frac{1}{\ea-\eb} (\Sa^-)^2 \Sb^-,\qquad 
    \OaSone \ket{\psi^{S=1}_{0}} = 0 \quad\forall\, \alpha,
\end{align}
which are obtained from the complex conjugate of \eqref{eq:b:Pmdef} by
Fourier transformation, as well as their complex conjugates,
\begin{equation}
  \label{eq:b:barThetadef}
    \OabSone =\sum_{\substack{\beta=1\\[2pt]\beta\ne\alpha}}^N 
    \frac{1}{\eab-\ebb} (\Sa^-)^2 \Sb^-,\qquad 
    \OabSone \ket{\psi^{S=1}_{0}} = 0 \quad\forall\, \alpha.
\end{equation}
Note that we would not need to exclude configurations with
$\beta=\alpha$, as the spin operators take care of this automatically.

In Section \ref{sec:b:ham}, we will use the operators $\OaSone$ to
construct a parent Hamiltonian, which is translationally invariant,
invariant under P and T, and invariant under SU(2) spin rotations, for
the non-Abelian $S=1$ spin liquid state $\ket{\psi^{S=1}_{0}}$.  The
analysis will imply that $\ket{\psi^{S=1}_{0}}$ is completely
specified by the condition \eqref{eq:b:Omegadef} plus the the
mentioned symmetries.  Therefore, we will refer to
\eqref{eq:b:Omegadef} as the \emph{\defining condition} of non-Abelian
$S=1$ spin chain we introduce in Section \ref{sec:3mod-na}.

\subsection{A second condition}
\label{sec:c:2ndcond}

It is worth noting that the condition \eqref{eq:b:Pmpsi0} with 
\eqref{eq:b:Pmdef} implies that the remaining terms in $B_m$ annihilate
$\ket{\psi^{S=1}_{0}}$ as well.  In particular, we have
\begin{equation}
  \label{eq:b:Qmpsi0}
    Q_{m} \ket{\psi^{S=1}_{0}} = \bar Q_{m} \ket{\psi^{S=1}_{0}} = 0
    \quad\forall\, m,
\end{equation}
where we have defined
\begin{align} 
  \label{eq:b:Qmdef}
    Q_{m}\,&\equiv\, \frac{1}{3}\hspace{-5pt}
    \sum_{\substack{\alpha_1,\alpha_2,\alpha_3=1\\ 
        \alpha_1\ne\alpha_2\ne\alpha_3\ne\alpha_1}}^N %\hspace{-5pt}
    Q^{\ne}_{m;\alpha_1,\alpha_2,\alpha_3}
    {S}^-_{\alpha_3} {S}^-_{\alpha_2} {S}^-_{\alpha_1}
    +\sum_{\alpha_1\ne\alpha_2}^N 
    Q^=_{m;\alpha_1,\alpha_2}\left({S}^-_{\alpha_2}\right)^2 {S}^-_{\alpha_1}
    \nonumber\\[0.3\baselineskip]
    &=\sum_{\substack{\alpha_1,\alpha_2,\alpha_3=1\\ 
        \alpha_1\ne\alpha_2\ne\alpha_3\ne\alpha_1}}^N %\hspace{-5pt}
    \frac{\etb{1}^{s+m+1}}{(\etb{3}-\etb{1})(\etb{1}-\etb{2})}
    {S}^-_{\alpha_3} {S}^-_{\alpha_2} {S}^-_{\alpha_1}
    \nonumber\\[-0.1\baselineskip]
    &\hspace{16.5pt}\hspace{80pt} -\sum_{\alpha_1\ne\alpha_2}^N
    \frac{\etb{1}^{s+m+1}-\etb{2}^{s+m+1}}{(\etb{1}-\etb{2})^2}
    \left({S}^-_{\alpha_2}\right)^2 {S}^-_{\alpha_1}
    \nonumber\\[0.3\baselineskip]
    &=\sum_{\substack{\alpha_1,\alpha_2,\alpha_3=1\\ 
         \alpha_1\ne\alpha_2\\\alpha_1\ne\alpha_3}}^N %\hspace{-5pt}
%        \alpha_1\ne\alpha_2,\alpha_3}}^N %\hspace{-5pt}
    \frac{\etb{1}^{s+m+1}}{(\etb{3}-\etb{1})(\etb{1}-\etb{2})}
    {S}^-_{\alpha_3} {S}^-_{\alpha_2} {S}^-_{\alpha_1}
    \nonumber\\[-0.5\baselineskip]
    &\hspace{16.5pt}\hspace{80pt} 
    +\sum_{\alpha_1\ne\alpha_2}^N
    \frac{\etb{2}^{s+m+1}}{(\etb{1}-\etb{2})^2}
    \left({S}^-_{\alpha_2}\right)^2 {S}^-_{\alpha_1}.
%     \\[0.3\baselineskip]
%     &=\sum_{\substack{\alpha_1,\alpha_2,\alpha_3=1\\ 
%          \alpha_1\ne\alpha_2\\ \alpha_3\ne\alpha_2}}^N %\hspace{-5pt}
% %        \alpha_2\ne\alpha_1,\alpha_3}}^N %\hspace{-5pt}
%     \frac{\etb{2}^{s+m+1}}{(\etb{1}-\etb{2})(\etb{2}-\etb{3})}
%     {S}^-_{\alpha_3} {S}^-_{\alpha_2} {S}^-_{\alpha_1}
%     \\[-0.1\baselineskip]
%     &\hspace{16.5pt}\hspace{80pt} 
%     +\sum_{\alpha_1\ne\alpha_2}^N
%     \frac{\etb{2}^{s+m+1}}{(\etb{1}-\etb{2})^2}
%     \left({S}^-_{\alpha_2}\right)^2 {S}^-_{\alpha_1}.
\end{align}
The non-Abelian $S=1$ spin liquid state $\ket{\psi^{S=1}_{0}}$ is
further annihilated by the operators
\begin{equation}
  \label{eq:b:Thetadef}
  \begin{split}
    \Xi_{\a} &\equiv -\frac{1}{N}\sum_{m=0}^{N} \eab^{s+m+1} \bar Q_m
    \\[0.2\baselineskip]
%     &=\sum_{\substack{\beta,\gamma=1\\\beta\ne\alpha\\\gamma\ne\alpha}}^N 
     &=\sum_{\substack{\beta,\gamma=1\\\beta,\gamma\ne\alpha}}^N 
    \frac{\Sa^- \Sb^- \Sc^-}{(\ea-\eb)(\ea-\ec)}
%    \\[-0.1\baselineskip]
%    &\hspace{16.5pt}\hspace{80pt} 
    -\sum_{\substack{\beta=1\\[2pt]\beta\ne\alpha}}^N
    \frac{(\Sa^-)^2 \Sb^-}{(\ea-\eb)^2} ,\quad 
%    \\[0.4\baselineskip]
%    \Ta &\ket{\psi^{S=1}_{0}} = 0 \quad\forall\, \alpha,
    \Xi_{\a} \ket{\psi^{S=1}_{0}} = 0 \quad\forall\, \alpha,
 \end{split}
\end{equation}
which are obtained from the complex conjugate of \eqref{eq:b:Qmdef} by
Fourier transformation, as well as their complex conjugates $\bar\Xi_{\a}$.
These operators, however, do not appear promising for the construction
of a simple parent Hamiltonian for the state.

\subsection{Direct verification}

In this section, we wish to verify the \defining condition
\eqref{eq:b:Omegadef} directly for the $S=1$ ground state
\eqref{eq:b:psi0}.  The method will be similar to the proof of the
singlet property in Section \ref{sec:na:sym}.  To begin with, we again
notice that when we substitute \eqref{eq:b:psi0} with
\eqref{eq:b:pfaff} into \eqref{eq:b:ket}, we may replace the
antisymmetrization $\mathcal{A}$ in \eqref{eq:b:pfaff} by an overall
normalization factor 9which we ignore), as it is taken care by the
commutativity of the bosonic operators $\tilde{S}_{\alpha}$.  Let
$\tilde\psi_0$ be $\psi^{S=1}_0$ without the antisymmetrization in
\eqref{eq:b:pfaff},
\begin{equation}
  \label{eq:b:psitilde}
%  \tilde\psi_0[z_i]
  \tilde\psi_0(z_1,\ldots ,z_N) 
%  =(N-1)!!\,
  =\left\{
    \frac{1}{z_1-z_2}\cdot\,\ldots\,\cdot\frac{1}{z_{N-1}-z_{N}}
  \right\}.
  \prod_{i<j}^{N}(z_i-z_j)\prod_{i=1}^{N}\,z_i.
\end{equation}
%where we have not kept track of overall normalization factors.
% for the wave functions.
Since $\tilde\psi_0(z_1,z_2,\ldots ,z_N)$ 
%$\tilde\psi_0[z_i]$ 
is still symmetric under interchange of pairs, we may assume that the
spin flip operators $(\Sa^-)^2$ and $\Sb^-$ of \eqref{eq:b:Omegadef}
will act on the pairs $(z_1,z_2)$ and $(z_3,z_4)$, respectively:
\begin{align}
  \label{eq:Sa^2Sbpsi}
(\Sa^-)^2 \Sb^-\bigket{\psi^{S=1}_0}&=
%    \frac{N}{2}\left(\frac{N}{2}-1\right)
    \sum_{\{z_5,\dots,z_{N}\}}(\Sa^-)^2(\Sa^+)^2\nonumber\\[2pt] 
    &\quad
    \biggl\{\sum_{z_4 (\ne \eb)}
    \tilde\psi_0(\ea,\ea,\eb,z_4,z_5,\ldots,z_N)
    \,\Sb^-\, \tilde{S}_\b^+ \tilde{S}_{z_4}^+\nonumber\\[4pt]
    &\hspace{-4.5pt}\quad
    +\sum_{z_3 (\ne \eb)}
    \tilde\psi_0(\ea,\ea,z_3,\eb,z_5,\ldots,z_N)
    \,\Sb^-\, \tilde{S}_{z_3}^+ \tilde{S}_\b^+\nonumber\\[0pt]
    &\hspace{29.5pt}
    +\ \ \tilde\psi_0(\ea,\ea,\eb,\eb,z_5,\ldots,z_N)
    \,\Sb^-\, (\tilde{S}_\b^+)^2\biggr\}\nonumber\\[4pt]
    &\hspace{152pt}
    \cdot\,\tilde{S}_{z_5}^+\dots\tilde{S}_{z_N}^+\ket{-1}_N\nonumber\\[6pt]
    &\hspace{-67pt}
    =4\sum_{\{z_5,\dots,z_{N}\}}
    \biggl\{\sum_{z_4}
    \tilde\psi_0(\ea,\ea,\eb,z_4,z_5,\ldots,z_N)\,
    \tilde{S}_{z_4}^+\biggr\}\,
    \tilde{S}_{z_5}^+\dots\tilde{S}_{z_N}^+\ket{-1}_N,\nonumber\\[-6pt]
\end{align}
where we have used 
\begin{equation}
  \label{eq:b:s-s+tilde2}%\nonumber
  S_\alpha^-\,(\tilde{S}_\alpha^+)^n \ket{1,-1}_{\alpha}
  =n\, (\tilde{S}_\alpha^+)^{n-1}\ket{1,-1}_{\alpha},
\end{equation}
which follows directly form the definition \eqref{eq:b:tildeS+}.
% Note that we have implicitly assumed that each spin configuration in
% the sum over ${\{z_1,z_2,\ldots ,z_N\}}$ in \eqref{eq:hsket} appears
% only once (and not $N!$ times due to permutations of the $z_i$'s).
We hence obtain
\begin{equation}
  \label{eq:b:Omegapsi}
  \begin{split}
    \OaSone\bigket{\psi^{S=1}_0}
    &=\sum_{\{z_4,\dots,z_{N}\}}
    \underbrace{\sum_{\substack{\beta=1\\[2pt]\beta\ne\alpha}}^N
    \frac{\tilde\psi_0(\ea,\ea,\eb,z_4,\ldots,z_N)}{\ea-\eb}}_{=0}\,
    \tilde{S}_{z_4}^+\dots\tilde{S}_{z_N}^+\ket{-1}_N,
  \end{split}
\end{equation}
since 
\begin{equation*}
  \begin{split}
    \frac{\tilde\psi_0(\ea,\ea,\eb,z_4,\ldots,z_N)}{\ea-\eb}& %\hspace{-80pt}&
    =\;(\ea-\eb)\ea^2\eb
    \\[0.2\baselineskip]
    &\quad \cdot
    \prod_{i=4}^N(\ea-z_i)^2 z_i \prod_{i=5}^N(\eb-z_i) 
    \prod_{4\le i< j}^N(z_i-z_j)  
    \\[0.2\baselineskip]
    &\quad \cdot
%    \frac{1}{\eb-z_4}\cdot
    \frac{1}{z_5-z_6}\cdot\,\ldots\,\cdot\frac{1}{z_{N-1}-z_{N}}
  \end{split}
\end{equation*}
vanishes for $\beta=\alpha$ and contains only powers $\eb^1,\eb^2,\ldots , 
\eb^{N-2}$.  Note that the calculation for $\OabSone$ is almost identical,
since
\begin{equation*}
  \frac{\tilde\psi_0(\ea,\ea,\eb,z_4,\ldots,z_N)}{\eab-\ebb}
  =-\ea\eb
  \frac{\tilde\psi_0(\ea,\ea,\eb,z_4,\ldots,z_N)}{\ea-\eb}
%  =-(\ea-\eb)\ea^2\eb^2\prod_{i=3}^M(\ea-z_i)^2(\eb-z_i)^2z_i
%  \prod_{3\le i< j}^M(z_i-z_j)^2 
\end{equation*}
vanishes also for $\beta=\alpha$ and 
contains only powers $\eb^2,\eb^3,\ldots , \eb^{N-1}$.

\section{Construction of a parent Hamiltonian}
\label{sec:b:ham}

We will now construct a parent Hamiltonian for the non-Abelian $S=1$
spin liquid state \eqref{eq:b:ket} with \eqref{eq:b:psi0} using the
annihilation operator \eqref{eq:b:Omegadef}, \ie
\begin{equation}
  \label{eq:b:Omegapro}
    \OaSone \ket{\psi^{S=1}_{0}} = 0 \quad\forall\, \alpha,
    \quad\text{where}\quad
    \OaSone = \sum_{\substack{\beta=1\\\beta\ne\alpha}}^N 
    \frac{1}{\ea-\eb} \big(\Sa^-\big)^2\Sb^-.
\end{equation}
The Hamiltonian has to be Hermitian, and we wish it to be invariant
under translations, time reversal (T), parity (P), and SU(2) spin
rotations.

%\newpage
\subsection{Translational, time reversal, and parity symmetry}
\label{sec:b:ham:transTP}

The operator ${\OaSone}^\dagger\OaSone$ is Hermitian and positive semi-definite,
meaning that all the eigenvalues are non-negative.  A translationally
invariant operator is given by
\begin{align}
  \label{eq:b:h0}
  H_0=\frac{1}{2}\sum_{\a=1}^N {\OaSone}^\dagger\OaSone
%  \nonumber\\[0.2\baselineskip] 
  &=\frac{1}{2}\sum_{\substack{\a,\b,\c\\ \a\ne \b,\c}}
    \frac{1}{\eab-\ebb}\frac{1}{\ea-\ec} 
    \big(\Sa^+\big)^2\big(\Sa^-\big)^2\Sb^+\Sc^-
  \nonumber\\[0.2\baselineskip] 
  &=\sum_{\substack{\a,\b,\c\\ \a\ne \b,\c}}\omega_{\a\b\c}\, 
  \Sa^\z\left(\Sa^\z+1\right)\Sb^+\Sc^-,
\end{align}
where $\omega_{\a\b\c}$ is defined in \eqref{eq:a:omegaabc},
% \begin{equation*}
%   \label{eq:b:omegaabc}
%   \omega_{\a\b\c}\equiv \frac{1}{\eab-\ebb}\frac{1}{\ea-\ec}. 
% \end{equation*}
and we have used that 
\begin{equation*}
  \big(\Sa^+\big)^2\big(\Sa^-\big)^2=2\Sa^\z\left(\Sa^\z+1\right)
%  \quad\text{for}\quad S=1.
\end{equation*}
for $S=1$, which is readily verified with \eqref{eq:app-am:Jpmketlm}.
With the transformation properties under time reversal,
\begin{align*}
  \text{T:}\quad
%  \i \,\to\, \Pi\i\Pi = -\i,\quad
%  \i\to -\i,\quad
  \ea  \,\to\, \Theta\ea \Theta = \eab,\quad 
  \bsS \,\to\, \Theta\bsS\Theta = -\bsS, 
\end{align*}
and hence
\begin{align*}
  \omega_{\a\b\c}\to\omega_{\a\c\b},\quad
  S^+\to -S^-,\quad S^-\to -S^+,\quad S^\z\to -S^\z,
\end{align*}
the operator \eqref{eq:b:h0} transforms into
\begin{align}
  \label{eq:b:Th0T}
  \Theta H_0\Theta & =\sum_{\substack{\a,\b,\c\\ \a\ne \b,\c}}\omega_{\a\b\c}\, 
  \Sa^\z\left(\Sa^\z-1\right)\Sc^-\Sb^+.
\end{align}
%\newpage
We proceed with the T invariant operator
\begin{align}
  \label{eq:b:h0T}
  H_0^{\text{T}}=\frac{1}{2}\left(H_0+\Theta H_0 \Theta\right)
  = H_0^{\text{T}=}+H_0^{\text{T}\ne },
\end{align}
where
\begin{align}
  \label{eq:b:h0T=}
  H_0^{\text{T}=}&=\frac{1}{2}\sum_{\substack{\a,\b\\ \a\ne \b}}\omega_{\a\b\b}
  \left({\Sa^\z}^2 \biganticomm{\Sb^+}{\Sb^-}
    +\Sa^\z\bigcomm{\Sb^+}{\Sb^-}\right)
  \nonumber\\[0.2\baselineskip] 
  &=\frac{1}{2}\sum_{\substack{\a\ne \b}}\omega_{\a\b\b}
   \left({\Sa^\z}^2 \biganticomm{\Sb^+}{\Sb^-} + 2\Sa^\z\Sb^\z\right),
  \\[0.4\baselineskip] 
  \label{eq:b:h0Tnot=}
  H_0^{\text{T}\ne} &=\sum_{\substack{\a,\b,\c\\ \a\ne\b\ne\c\ne\a}}
  \omega_{\a\b\c}\,{\Sa^\z}^2 \Sb^+\Sc^-.
\end{align}
With the transformation properties under parity,
\begin{align}
  \text{P:}\quad
%  \i \,\to\, \Pi\i\Pi = \i,\quad
  \ea  \,\to\, \Pi\ea\Pi = \eab,\quad 
  \bsS \,\to\, \Theta\bsS\Theta = \bsS,  
\end{align}
and hence $\omega_{\a\b\c}\to\omega_{\a\c\b}$, we obtain the P and T
invariant operator
\begin{align}
  \label{eq:b:h0PT}
  H_0^{\text{PT}}=\frac{1}{2}\left(H_0^{\text{T}}
    +\Pi\hspace{1pt} H_0^{\text{T}}\hspace{1pt} \Pi\right)
  = H_0^{\text{PT}=}+H_0^{\text{PT}\ne },
\end{align}
where
\begin{align}
  \label{eq:b:h0PT1}
  H_0^{\text{PT}=}&=H_0^{\text{T}=},
  \hspace{-3pt}&\hspace{-3pt}
%  \\[0.4\baselineskip] 
%  \label{eq:b:h0Tn}
  H_0^{\text{PT}\ne}&=\frac{1}{2}
  \hspace{-3pt}\sum_{\substack{\a,\b,\c\\ \a\ne\b\ne\c\ne\a}}\hspace{-3pt}
  \omega_{\a\b\c} {\Sa^\z}^2\big(\Sb^+\Sc^- + \Sb^-\Sc^+\big).
\end{align}

%\subsection{Projecting out the scalar component}
\subsection{Spin rotation symmetry}

Since the non-Abelian spin liquid state $\ket{\psi^{S=1}_0}$ is a spin
singlet, the property that it is annihilated by \eqref{eq:b:h0PT}
with \eqref{eq:b:h0PT1} and \eqref{eq:b:h0T=} implies that it is 
annihilated by each tensor component of \eqref{eq:b:h0PT} individually.

With the tensor decompositions \eqref{eq:app-t-S+-S-+},
\eqref{eq:app-t-SzSz}, and $\bsS^2=2$ for $S=1$, we can rewrite the
two contributions as
\begin{align}
  \label{eq:b:h0PT=}
  H_0^{\text{PT}=}&=\sum_{\substack{\a\ne \b}}\omega_{\a\b\b}
  \bigg[\left(\frac{2}{3} + \frac{1}{\sqrt{6}}\,T_{\a\a}^0\right)
    \left(\frac{4}{3} - \frac{1}{\sqrt{6}}\,T_{\b\b}^0\right)
    + \frac{1}{3}\bSa\bSb + \frac{1}{\sqrt{6}}\,T_{\a\b}^0\bigg],
%  \nonumber\\[0.2\baselineskip] 
%  &\hspace{100pt} + \frac{1}{3}\bSa\bSb + \frac{1}{\sqrt{6}}\,T_{\a\b}^0\bigg]
  \\[0.2\baselineskip] 
  \label{eq:b:h0PTnot=}
  H_0^{\text{PT}\ne} 
  &=\hspace{-3pt}\sum_{\substack{\a,\b,\c\\ \a\ne\b\ne\c\ne\a}}\hspace{-3pt}
  \omega_{\a\b\c} \left(\frac{2}{3} + \frac{1}{\sqrt{6}}\,T_{\a\a}^0\right)
    \left(\frac{2}{3}\bSb\bSc - \frac{1}{\sqrt{6}}\,T_{\b\c}^0\right).
\end{align}
Projecting out the scalar components under SU(2) spin rotations yields
\begin{align}
  \label{eq:b:h0PT=0}
%  \frac{1}{2}
  \big\{H_0^{\text{PT}=}\big\}_0
  &=\sum_{\substack{\a\ne \b}}\omega_{\a\b\b}
  \left(\frac{8}{9} - \frac{1}{6}\left\{T_{\a\a}^0 T_{\b\b}^0\right\}_0   
    + \frac{1}{3}\bSa\bSb \right),
  \\[0.2\baselineskip] 
  \label{eq:b:h0PTnot=0}
  \big\{H_0^{\text{PT}\ne}\big\}_0
  &=\sum_{\substack{\a,\b,\c\\ \a\ne\b\ne\c\ne\a}}\omega_{\a\b\c} 
  \left(\frac{4}{9}\bSb\bSc 
    - \frac{1}{6}\left\{T_{\a\a}^0 T_{\b\c}^0\right\}_0\right).
\end{align}
The next step is to calculate the scalar component of the tensor
products in \eqref{eq:b:h0PT=0}.

\subsection{Evaluation of $\big\{T_{\a\a}^0 T_{\b\c}^0\big\}_0$}
\label{sec:b::5TaaTbc_0}

We evaluate the scalar component of the tensor product of $T_{\a\a}^0$
and $T_{\b\c}^0$ with $\a\ne\b,\c$ using \eqref{eq:a:Tm1m2CGj0},
\begin{align}
  \label{eq:b:TaaTbc_j}
  \big\{T_{\a\a}^0T_{\b\c}^0 \big\}_j &= 
  \braket{j,0}{2,0;2,0}\sum_{m=-2}^{2}
    T_{\a\a}^{m} T_{\b\c}^{-m} \braket{2,m;2,-m}{j,0}.
\end{align}
With \eqref{eq:app-t-SS-2} and the Clebsch--Gordan coefficients
\begin{equation}
  \label{eq:b:CG220}
  \braket{2,m;2,-m}{0,0}=\frac{(-1)^m}{\sqrt{5}} ,
\end{equation}
we obtain
\begin{align}
  \label{eq:b:TaaTbc_0}
  5\,\big\{T_{\a\a}^0T_{\b\c}^0 \big\}_0 
%  &= 5 \braket{j,0}{2,0;2,0}\sum_{m=-2}^{2}
%  T_{\a\a}^{m} T_{\b\c}^{-m} \braket{2,m;2,-m}{j,0}.
  \hspace{0pt}&\hspace{0pt}= \sum_{m=-2}^{2} (-1)^m T_{\a\a}^{m} T_{\b\c}^{-m}
  \nonumber\\[0.2\baselineskip] 
  &= \Sa^-\Sa^-\Sb^+\Sc^+
  \nonumber\\[0.2\baselineskip] 
  &\quad+ \bigl(\Sa^\z\Sa^- + \Sa^-\Sa^\z\bigr)\bigl(\Sb^\z\Sc^+ + \Sb^+\Sc^\z\bigr)
  \nonumber\\[0.2\baselineskip] 
  &\quad+ \frac{1}{6} \bigl(4 \Sa^\z\Sa^\z - \Sa^+\Sa^- - \Sa^-\Sa^+\bigr)
                 \bigl(4 \Sb^\z\Sc^\z - \Sb^+\Sc^- - \Sb^-\Sc^+\bigr)
  \nonumber\\[0.2\baselineskip] 
  &\quad+ \bigl(\Sa^\z\Sa^+ + \Sa^+\Sa^\z\bigr)\bigl(\Sb^\z\Sc^- + \Sb^-\Sc^\z\bigr)
  \nonumber\\[0.2\baselineskip] 
  &\quad+ \Sa^+\Sa^+\Sb^-\Sc^-.
\end{align}
We wish to write this in a more convenient form, which directly displays
that it transforms as a scalar under spin rotations.  Since
\begin{equation*}
  \bs{1}\otimes\bs{1}\otimes\bs{1}\otimes\bs{1}
  = 3\cdot \bs{0}\oplus 6\cdot\bs{1}\oplus 6\cdot\bs{2}\oplus 3\cdot\bs{3},
  \oplus\bs{4},
\end{equation*}
we can only form three scalars from four spin operators.  For
$\a\ne\b\ne\c\ne\a$, three such scalars are
\begin{equation*}
  \bSa^2(\bSb\bSc),\quad (\bSa\bSb)(\bSa\bSc),
  \quad\text{and}\quad (\bSa\bSc)(\bSa\bSb).
\end{equation*}
%\begin{equation*}
%  \bSa^2(\bSb\bSc),\ (\bSa\bSb)(\bSa\bSc),
%  \ \text{and}\ (\bSa\bSc)(\bSa\bSb).
%\end{equation*}
For $\a\ne\b=\c$, the latter two are identical, but we have the
additional scalar $\bSa\bSb$.  For $\a\ne\b,\c$ in general, we
write
\begin{align}
  \label{eq:b:TaaTbc_0=}
  5\,\big\{T_{\a\a}^0T_{\b\c}^0 \big\}_0\hspace{0pt}
  &= a\, \bSa^2(\bSb\bSc)
  + b\, \bigl[(\bSa\bSb)(\bSa\bSc)+ (\bSa\bSc)(\bSa\bSb)\bigr]
  \nonumber\\[0.2\baselineskip] 
  &\quad + c\, \delta_{\b\c}\,\bSa\bSb,
\end{align}
where we have used the invariance of the tensor product under
interchange of $\b$ and $\c$.  The coefficients $a$ and $b$ may depend
on whether $\b=\c$ or not.

%\newpage
Since the $\Sa^-\Sa^-$ term in \eqref{eq:b:TaaTbc_0} has to come form
the second term in \eqref{eq:b:TaaTbc_0=}, we can immediately infer 
$b=2$.  To obtain $a$ and $c$, we first write out the second term in
\eqref{eq:b:TaaTbc_0=} for $\a\ne\b,\c$,
\begin{align}
  \label{eq:b:TaaTbc_0bterm1}
  \hspace{40pt}&\hspace{-40pt}
  2\,\bigl[(\bSa\bSb)(\bSa\bSc)+ (\bSa\bSc)(\bSa\bSb)\bigr]
  \nonumber\\[0.2\baselineskip] 
  &=\frac{1}{2}\bigl(2 \Sa^\z\Sb^\z + \Sa^+\Sb^- + \Sa^-\Sb^+\bigr)
               \bigl(2 \Sa^\z\Sc^\z + \Sa^+\Sc^- + \Sa^-\Sc^+\bigr)
  \nonumber\\*[0\baselineskip]&\hspace{60pt} 
               +\,\text{same with}\ \b\leftrightarrow\c
  \nonumber\\[0.2\baselineskip] 
  &=\Sa^+\Sa^+\Sb^-\Sc^- + \Sa^-\Sa^-\Sb^+\Sc^+
  \nonumber\\[0.2\baselineskip] 
  &\quad+\Sa^\z\Sb^\z \bigl(\Sa^+\Sc^- + \Sa^-\Sc^+\bigr)
   +\Sa^\z\Sc^\z \bigl(\Sa^+\Sb^- + \Sa^-\Sb^+\bigr)
  \nonumber\\[0.2\baselineskip] 
  &\quad+\bigl(\Sa^+\Sb^- + \Sa^-\Sb^+\bigr) \Sa^\z\Sc^\z
   +\bigl(\Sa^+\Sc^- + \Sa^-\Sc^+\bigr) \Sa^\z\Sb^\z
  \nonumber\\[0.2\baselineskip] 
  &\quad+\frac{1}{2} \Sa^+\Sa^- \bigl(\Sb^-\Sc^+ + \Sc^-\Sb^+\bigr)
   +\frac{1}{2} \Sa^-\Sa^+ \bigl(\Sb^+\Sc^- + \Sc^+\Sb^-\bigr)
  \nonumber\\[0.2\baselineskip] 
  &\quad+4\Sa^\z\Sa^\z\Sb^\z\Sc^\z,
\end{align}
and order the terms such that the $\Sb$ operators are to the left of
the $\Sc$ operators,
\begin{align}
  \label{eq:b:TaaTbc_0bterm2}
  \hspace{40pt}&\hspace{-40pt}
  2\,\bigl[(\bSa\bSb)(\bSa\bSc)+ (\bSa\bSc)(\bSa\bSb)\bigr]
  \nonumber\\[0.2\baselineskip] 
  &= \Sa^+\Sa^+\Sb^-\Sc^- + \Sa^-\Sa^-\Sb^+\Sc^+
  \nonumber\\[0.2\baselineskip] 
  &\quad+ \Sa^\z\Sa^+\Sb^\z\Sc^- + \Sa^\z\Sa^-\Sb^\z\Sc^+
   + \Sa^\z\Sa^+\Sb^-\Sc^\z + \Sa^\z\Sa^-\Sb^+\Sc^\z
  \nonumber\\[0.2\baselineskip] 
  &\quad -\Sa^\z\Sa^+\bigcomm{\Sb^-}{\Sc^\z}- \Sa^\z\Sa^-\bigcomm{\Sb^+}{\Sc^\z}
  \nonumber\\[0.2\baselineskip] 
  &\quad+\Sa^+\Sa^\z\Sb^-\Sc^\z + \Sa^-\Sa^\z\Sb^+\Sc^\z 
   +\Sa^+\Sa^\z\Sb^\z\Sc^- + \Sa^-\Sa^\z\Sb^\z\Sc^+ 
  \nonumber\\[0.2\baselineskip] 
  &\quad -\Sa^+\Sa^\z\bigcomm{\Sb^\z}{\Sc^-}
         -\Sa^-\Sa^\z\bigcomm{\Sb^\z}{\Sc^+}
  \nonumber\\[0.2\baselineskip] 
%   &\quad+\frac{1}{2} \Sa^+\Sa^- \bigl(\Sb^-\Sc^+ + \Sb^+\Sc^- -
%   \bigcomm{\Sb^+}{\Sc^-}\bigr)
%   \nonumber\\[0.2\baselineskip] 
%   &\quad+\frac{1}{2} \Sa^-\Sa^+ \bigl(\Sb^+\Sc^- + \Sb^-\Sc^+ -
%   \bigcomm{\Sb^-}{\Sc^+}\bigr)
%   \nonumber\\[0.2\baselineskip] 
  &\quad+\frac{1}{2} \Sa^+\Sa^- \bigl(\Sb^-\Sc^+ + \Sb^+\Sc^-\bigr)
   +\frac{1}{2} \Sa^-\Sa^+ \bigl(\Sb^+\Sc^- + \Sb^-\Sc^+\bigr)
  \nonumber\\[0.2\baselineskip] 
  &\quad -\frac{1}{2}\bigcomm{\Sa^+}{\Sa^-}\bigcomm{\Sb^+}{\Sc^-}
  \nonumber\\[0.2\baselineskip] 
  &\quad+4\Sa^\z\Sa^\z\Sb^\z\Sc^\z.
\end{align}
With
\begin{align*}
  &\bigcomm{\Sa^\z}{\Sa^+}\bigcomm{\Sb^-}{\Sb^\z}
  +\bigcomm{\Sa^\z}{\Sa^-}\bigcomm{\Sb^+}{\Sb^\z}
  +\frac{1}{2}\bigcomm{\Sa^+}{\Sa^-}\bigcomm{\Sb^+}{\Sb^-}
  \nonumber\\[0.2\baselineskip] 
  &\hspace{50pt}=\Sa^+\Sb^- + \Sa^-\Sb^+ + 2\Sa^\z\Sb^\z 
  = 2\bSa\bSb,
\end{align*}
we finally obtain
\begin{align}
  \label{eq:b:TaaTbc_0bcterm3}
  \hspace{40pt}&\hspace{-40pt}
  2\,\bigl[(\bSa\bSb)(\bSa\bSc)+ (\bSa\bSc)(\bSa\bSb)\bigr]
  +2\delta_{\b\c}\,\bSa\bSb\,
  \nonumber\\[0.2\baselineskip] 
  &= \Sa^+\Sa^+\Sb^-\Sc^- + \Sa^-\Sa^-\Sb^+\Sc^+
  \nonumber\\[0.2\baselineskip] 
  &\quad+ \Sa^\z\Sa^+\Sb^\z\Sc^- + \Sa^\z\Sa^-\Sb^\z\Sc^+
   + \Sa^\z\Sa^+\Sb^-\Sc^\z + \Sa^\z\Sa^-\Sb^+\Sc^\z
  \nonumber\\[0.2\baselineskip] 
  &\quad+\Sa^+\Sa^\z\Sb^-\Sc^\z + \Sa^-\Sa^\z\Sb^+\Sc^\z 
   +\Sa^+\Sa^\z\Sb^\z\Sc^- + \Sa^-\Sa^\z\Sb^\z\Sc^+ 
  \nonumber\\[0.2\baselineskip] 
%  &\quad+\frac{1}{2} \Sa^+\Sa^- \bigl(\Sb^-\Sc^+ + \Sb^+\Sc^-\bigr)
%   +\frac{1}{2} \Sa^-\Sa^+ \bigl(\Sb^+\Sc^- + \Sb^-\Sc^+\bigr)
  &\quad+\frac{1}{2} \bigl(\Sa^+\Sa^- + \Sa^-\Sa^+\bigr)
                \bigl(\Sb^+\Sc^- + \Sb^-\Sc^+\bigr)
  \nonumber\\[0.2\baselineskip] 
  &\quad+4\,\Sa^\z\Sa^\z\Sb^\z\Sc^\z.
\end{align}
Subtracting this from \eqref{eq:b:TaaTbc_0}, we obtain
\begin{align}
  \label{eq:b:TaaTbc_0-bcterm4}
  5\,\big\{T_{\a\a}^0T_{\b\c}^0 \big\}_0 
  \hspace{0pt}&\hspace{0pt}
  - 2\,\bigl[(\bSa\bSb)(\bSa\bSc)+ (\bSa\bSc)(\bSa\bSb)\bigr]
  - 2\delta_{\b\c}\,\bSa\bSb\,
  \nonumber\\[0.2\baselineskip] 
  &= \frac{1}{6} \bigl(4 \Sa^\z\Sa^\z - \Sa^+\Sa^- - \Sa^-\Sa^+\bigr)
                 \bigl(4 \Sb^\z\Sc^\z - \Sb^+\Sc^- - \Sb^-\Sc^+\bigr)
  \nonumber\\[0.2\baselineskip] 
  &\quad-\frac{1}{2} \bigl(\Sa^+\Sa^- + \Sa^-\Sa^+\bigr)
                \bigl(\Sb^+\Sc^- + \Sb^-\Sc^+\bigr)
                -4\,\Sa^\z\Sa^\z\Sb^\z\Sc^\z.
  \nonumber\\[0.2\baselineskip] 
%   &=-\frac{4}{3}\Sa^\z\Sa^\z\Sb^\z\Sc^\z
%   \nonumber\\[0.2\baselineskip] 
%   &\quad-\frac{2}{3}\bigl(\Sa^+\Sa^- + \Sa^-\Sa^+\bigr)\Sb^\z\Sc^\z
%    -\frac{2}{3}\Sa^\z\Sa^\z\bigl(\Sb^+\Sc^- + \Sb^-\Sc^+\bigr)
%   \nonumber\\[0.2\baselineskip] 
%   &\quad-\frac{1}{3}\bigl(\Sa^+\Sa^- + \Sa^-\Sa^+\bigr)
%                \bigl(\Sb^+\Sc^- + \Sb^-\Sc^+\bigr)
%   \nonumber\\[0.2\baselineskip] 
  &= -\frac{4}{3} \bSa^2(\bSb\bSc),
\end{align}
or 
\begin{align}
  \label{eq:b:5TaaTbc_0}
  5\,\big\{T_{\a\a}^0T_{\b\c}^0 \big\}_0 
  \hspace{0pt}&\hspace{0pt}
  = -\frac{4}{3} \bSa^2(\bSb\bSc)
  + 2\,\bigl[(\bSa\bSb)(\bSa\bSc)+ (\bSa\bSc)(\bSa\bSb)\bigr]
  \nonumber\\[0.2\baselineskip] 
  &\quad +2\delta_{\b\c}\,\bSa\bSb\,
\end{align}

As an aside, since the Clebsch--Gordan coefficient
\begin{equation}
  \label{eq:b:CG221}
  \braket{2,0;2,0}{1,0}=0,
\end{equation}
\eqref{eq:b:TaaTbc_j} implies that the tensor product of
$T_{\a\a}^0$ and $T_{\b\c}^0$ has no vector component, \ie 
\begin{align}
  \label{eq:b:TaaTbc_1}
  \big\{T_{\a\a}^0T_{\b\c}^0 \big\}_1=0.
\end{align}
This implies that we cannot obtain a vector annihilation operator
which is even under P and T from the operator $H_0$ defined in
\eqref{eq:b:h0}.

%\subsection{Collecting the terms}
%\subsection{The Hamiltonian}
\subsection{Writing out the Hamiltonian}
\label{sec:b:ham:writeout}

Substitution of \eqref{eq:b:5TaaTbc_0} into \eqref{eq:b:h0PT=0} and
\eqref{eq:b:h0PTnot=0} yields
\begin{align}
  \label{eq:b:h0PT=0v1}
  \big\{H_0^{\text{PT}=}\big\}_0
%   &=\sum_{\substack{\a\ne \b}}\omega_{\a\b\b}
%   \biggl[\frac{8}{9} 
%   - \frac{1}{30}\left(\!\!-\frac{16}{3} + 4\,(\bSa\bSb)^2 + 2\,\bSa\bSb\right)
%   + \frac{1}{3}\bSa\bSb \biggr]
%   \nonumber\\[0.2\baselineskip] 
  &=\frac{1}{15}\sum_{\substack{\a\ne \b}}\omega_{\a\b\b}
  \Bigl[16 - 2\,\big(\bSa\bSb\big)^2 + 4\,\bSa\bSb \Bigr],   
  \\[0.2\baselineskip] 
  \label{eq:b:h0PTnot=0v1}
  \big\{H_0^{\text{PT}\ne}\big\}_0
%   &=\sum_{\substack{\a,\b,\c\\ \a\ne\b\ne\c\ne\a}}\omega_{\a\b\c} 
%   \bigg[\frac{4}{9}\bSb\bSc - \frac{1}{30} \bigg(\!\!-\frac{8}{3} \bSb\bSc
%   + 2\,(\bSa\bSb)(\bSa\bSc)
%   \nonumber\\[-0.5\baselineskip] 
%   &\hspace{186.5pt} + 2\,(\bSa\bSc)(\bSa\bSb)\bigg)\bigg]
%   \nonumber\\[0.2\baselineskip] 
  &=\frac{1}{15}
  \hspace{-5pt}\sum_{\substack{\a,\b,\c\\ \a\ne\b\ne\c\ne\a}}\hspace{-5pt}
  \omega_{\a\b\c} 
  \Bigl[8\,\bSb\bSc - (\bSa\bSb)(\bSa\bSc) - (\bSa\bSc)(\bSa\bSb)\Bigr].
\end{align}
With \eqref{eq:app-hs15}, we rewrite the first term in
\eqref{eq:b:h0PTnot=0v1} as
\begin{align*}
  8\hspace{-5pt}\sum_{\substack{\a,\b,\c\\ \a\ne\b\ne\c\ne\a}}\hspace{-5pt}
  \omega_{\a\b\c}\bSb\bSc 
  &=16\sum_{\substack{\a\ne \b}}\omega_{\a\b\b} \bSa\bSb
   -4\sum_{\substack{\a\ne \b}} \bSa\bSb
  \\ %[0.2\baselineskip] 
  &=16\sum_{\substack{\a\ne \b}}\omega_{\a\b\b} \bSa\bSb
   -4\bsS_{\text{tot}}^2 + 8N.
\end{align*}
Collecting all the terms we obtain
\begin{align}
  \label{eq:b:h0PT0v1}
  15\big\{H_0^{\text{PT}}\big\}_0
%   &=15\big\{H_0^{\text{PT}=}\big\}_0
%    +15\big\{H_0^{\text{PT}\ne}\big\}_0
%   \nonumber\\[0.4\baselineskip] 
  &=\sum_{\substack{\a\ne \b}}\frac{1}{\vert\ea-\eb\vert^2}
  \Bigl[20\,\bSa\bSb - 2\,\big(\bSa\bSb\big)^2\Bigr]
  \nonumber\\ %[0.2\baselineskip] 
  &\quad-\hspace{-5pt}\sum_{\substack{\a,\b,\c\\ \a\ne\b\ne\c\ne\a}}\hspace{-5pt}
  \frac{1}{\eab-\ebb}\frac{1}{\ea-\ec}
  \Bigl[(\bSa\bSb)(\bSa\bSc) + (\bSa\bSc)(\bSa\bSb)\Bigr]
  \nonumber\\ %[0.2\baselineskip] 
  &\quad- 4\bsS_{\text{tot}}^2 
   + \frac{4 N (N^2-1)}{3} + 8N,
\end{align}
and finally
\begin{align}
  \label{eq:b:h0PT0v2}
%  H^{S=1}&=
  \frac{3}{4}\big\{H_0^{\text{PT}}\big\}_0
%  \nonumber\\*[0.4\baselineskip] 
  &=\sum_{\substack{\a\ne \b}}\frac{1}{\vert\ea-\eb\vert^2}
  \Bigl[\bSa\bSb - \frac{1}{10}\big(\bSa\bSb\big)^2\Bigr]
  \nonumber\\* %[0.2\baselineskip] 
  &-\frac{1}{20}
  \hspace{-5pt}\sum_{\substack{\a,\b,\c\\ \a\ne\b\ne\c\ne\a}}\hspace{-5pt}
  \frac{1}{(\eab-\ebb)(\ea-\ec)}
  \Bigl[(\bSa\bSb)(\bSa\bSc) + (\bSa\bSc)(\bSa\bSb)\Bigr]
  \nonumber\\* %[0.2\baselineskip] 
  &- \frac{1}{5}\bsS_{\text{tot}}^2 
   + \frac{N (N^2+5)}{15}\,.
\end{align}
Note that the second term in the first line of \eqref{eq:b:h0PT0v2}
is equal to what we would get if we were to take $\b=\c$ on the 
term in the second line.

In conclusion, we have derived that the non-Abelian $S=1$ Pfaffian
spin liquid state $\ket{\psi^{S=1}_0}$ introduced in Section
\ref{sec:3mod-na} is an exact eigenstate of 
\begin{align}
  \label{eq:b:h}
%   H^{S=1}&=\sum_{\substack{\a\ne \b}}\frac{\bSa\bSb}{\vert\ea-\eb\vert^2} 
%   -\frac{1}{20}\sum_{\substack{\a,\b,\c\\ \a\ne\b,\c}}
%   \frac{(\bSa\bSb)(\bSa\bSc) + (\bSa\bSc)(\bSa\bSb)}{(\eab-\ebb)(\ea-\ec)}
  H^{S=1}&=\frac{2\pi^2}{N^2}
  \Bigg[
  \sum_{\substack{\a\ne \b}}\frac{\bSa\bSb}{\vert\ea-\eb\vert^2} 
  -\frac{1}{20}\sum_{\substack{\a,\b,\c\\ \a\ne\b,\c}}
  \frac{(\bSa\bSb)(\bSa\bSc) + (\bSa\bSc)(\bSa\bSb)}{(\eab-\ebb)(\ea-\ec)}
  \Bigg]
\end{align}
with energy eigenvalue 
\begin{align}
  \label{eq:b:e0}
  E_0^{S=1}=-\frac{2\pi^2}{N^2}\frac{N (N^2+5)}{15}
  =-\frac{2\pi^2}{15}\left(N+\frac{5}{N}\right).
\end{align}
The information regarding the positive semi-definiteness of
$H_0^{\text{PT}}$, which was still intact on the level of
\eqref{eq:b:h0PT=} and \eqref{eq:b:h0PTnot=}, has unfortunately been
lost as we carried out the projection onto the scalar components
\eqref{eq:b:h0PT=0} and \eqref{eq:b:h0PTnot=0}.  We will recover this
information in Section \ref{sec:c:factorization}.  Exact diagonalization
studies~\cite{manuscriptinpreparationTRSG11} carried out numerically
for up to $N=18$ sites further show that $\ket{\psi^{S=1}_0}$ is the
unique ground states of \eqref{eq:b:h}, and that the model is gapless.

% \newpage
\section{Vector annihilation operators}
%\section{The rapidity operator and more}
%\section{A vector annihilation operator}
\label{sec:b:vec}

%\subsection{T even operators}
\subsection{Annihilation operators which transform even under T}
\label{sec:b:vecTeven}

We can use the \defining condition \eqref{eq:b:Omegapro} further to
construct a vector annihilation operator.  First note that since
\begin{equation*}
    \OaSone \ket{\psi^{S=1}_{0}} = 0 \quad\forall\, \alpha,
\end{equation*}
$\ket{\psi^{S=1}_{0}}$ is also annihilated by the Hermitian operator
\begin{align}
  \label{eq:b:ha}
  H_{\a}=\frac{1}{2} {\OaSone}^\dagger\OaSone
  &=\sum_{\substack{\b,\c\\ \a\ne \b,\c}}\omega_{\a\b\c}\, 
  \Sa^\z\left(\Sa^\z+1\right)\Sb^+\Sc^-,
\end{align}
which is just the operator \eqref{eq:b:h0} without the sum over $\a$.
Constructing an operator which is even under T,
\begin{equation}
  \label{eq:b:haT}
  H_\a^{\text{T}}=\frac{1}{2}\left(H_\a+\Theta H_\a \Theta\right)
  = H_\a^{\text{T}=}+H_\a^{\text{T}\ne },
\end{equation}
with 
\begin{align}
  \label{eq:b:haT=}
  H_\a^{\text{T}=} 
%   &=\sum_{\substack{\b\\ \b\ne\a}}\omega_{\a\b\b}
%   \left({\Sa^\z}^2 \biganticomm{\Sb^+}{\Sb^-}
%     +\Sa^\z\bigcomm{\Sb^+}{\Sb^-}\right)
  &=\frac{1}{2}\sum_{\substack{\b\\ \b\ne\a}}\omega_{\a\b\b}
  \left({\Sa^\z}^2 \biganticomm{\Sb^+}{\Sb^-} + 2\Sa^\z\Sb^\z\right),
  \\[0.4\baselineskip] 
  \label{eq:b:haTnot=}
  H_\a^{\text{T}\ne} 
  &=\sum_{\substack{\b\ne\c\\ \b,\c\ne\a}}\omega_{\a\b\c}
  \,{\Sa^\z}^2\Sb^+\Sc^-.
\end{align}
and odd under P, we obtain 
% along the lines of the construction in Section \ref{sec:a:ham:transTP}
\begin{align}
  \label{eq:b:habarPT}
  H_\a^{\rm {\bar P}T}=\frac{1}{2}
  \left(H_\a^{\text{T}} - \Pi\hspace{1pt} H_\a^{\text{T}}\hspace{1pt} \Pi\right)
  = H_\a^{{\rm {\bar P}T}=} + H_\a^{{\rm {\bar P}T}\ne},
\end{align}
% \newpage
where
\begin{align}
  \label{eq:b:habarPT1}
  H_\a^{{\rm {\bar P}T}=}&=0, 
  \hspace{-10pt}&\hspace{-10pt}
%  \\[0.4\baselineskip] 
%  \label{eq:b:h0Tn}
  H_\a^{{\rm {\bar P}T}\ne}&=
  \frac{1}{2}\sum_{\substack{\b\ne\c\\ \b,\c\ne\a}}\omega_{\a\b\c}
  \,{\Sa^\z}^2\big(\Sb^+\Sc^- - \Sb^-\Sc^+\big).
\end{align}
With \eqref{eq:a:omega_abc-omega_acb} %,
% \begin{align*}
%   \omega_{\a\b\c}-\omega_{\a\c\b}
%   &=-\frac{1}{2}\left(\frac{\ea+\eb}{\ea-\eb}-\frac{\ea+\ec}{\ea-\ec}\right),
% \end{align*}
% \vspace{50pt}
and $\Sb^+\Sc^- - \Sb^-\Sc^+=-2\i (\bSb\times\bSc)^\z$ (\cf
\eqref{eq:app-t-SSS-1abc}), we obtain
\begin{align}
  \label{eq:b:habarPT2}
  H_\a^{{\rm {\bar P}T}}
%   &=\frac{1}{2}\sum_{\substack{\b\ne\c\\ \b,\c\ne\a}}\omega_{\a\b\c}
%   \,{\Sa^\z}^2\big(\Sb^+\Sc^- - \Sb^-\Sc^+\big)
%   \nonumber\\[0.2\baselineskip] 
%   &=-\frac{\i}{2}\sum_{\substack{\b\ne\c\\ \b,\c\ne\a}}
%   (\omega_{\a\b\c}-\omega_{\a\c\b})\,{\Sa^\z}^2(\bSb\times\bSc)^\z
%   \nonumber\\[0.2\baselineskip] 
  &=\frac{\i}{2}\sum_{\substack{\b\ne\c\\ \b,\c\ne\a}}
  \frac{\ea+\eb}{\ea-\eb}\,{\Sa^\z}^2 (\bSb\times\bSc)^\z.
\end{align}
With \eqref{eq:app-t-SzSzSz} and \eqref{eq:app-t-SSS-1abc} we find that
the scalar component of the product of the $z$-components of three vectors
vanishes identically, while the vector component is given by
\begin{align*}
  5\big\{ {\Sa^\z}^2 (\bSb\times\bSc)^\z \big\}_1 & = \Sa^\z \big(\bSa
  (\bSb\times\bSc)\big) + \big(\bSa (\bSb\times\bSc)\big) \Sa^\z + 2
  (\bSb\times\bSc)^\z,
\end{align*}
where we have used $\a\ne\b,\c$ and $\bSa^2=2$.  The Pfaffian spin liquid
state $\ket{\psi^{S=1}_{0}}$ is hence annihilated by the vector operator
\begin{align}
  \label{eq:b:habarPT_1}
  5\big\{ H_\a^{{\rm {\bar P}T}} \big\}_{\bs{1}}
  = %\frac{\i}{2}
  \i\sum_{\substack{\b\ne\c\\ \b,\c\ne\a}}\frac{\ea+\eb}{\ea-\eb}
  \bigg[
  (\bSb\times\bSc)&+\frac{1}{2} \bSa \big(\bSa(\bSb\times\bSc)\big)
  \nonumber\\[-0.6\baselineskip] 
                  &+\frac{1}{2} \big(\bSa(\bSb\times\bSc)\big)\bSa
  \bigg].
\end{align}
With
\begin{align*}
  \sum_{\substack{\c\\ \c\ne\a,\b}}\bSc = \bsS_{\text{tot}}-\bSa-\bSb,
%  \qquad\text{and}\qquad \bSb\times\bSb=i\bSb,
\end{align*}
% $\bsS_{\text{tot}}\ket{\psi^{S=1}_{0}}=0$,\ \ 
$\bSb\times\bSb=i\bSb$, and
% \begin{align*}
%   \bSa\big(\bSb\times(-\bSa-\bSb)\big)
%   &=\bSb\big(\bSa\times\bSa\big)-\bSa\big(\bSb\times\bSb\big)=0,
% \end{align*}
\eqref{eq:a:SaSbSaSb}, we find from \eqref{eq:b:habarPT_1} that
$\ket{\psi^{S=1}_{0}}$ is also annihilated by
\begin{align}
  \label{eq:b:habarPT_1v1}
%  \frac{\i}{2}
  \i\sum_{\substack{\b\\ \b\ne\a}}\frac{\ea+\eb}{\ea-\eb}
  \bigg[(\bSa\times\bSb) - i\bSb 
  + \frac{1}{2}\big(\bSa(\bSb\times\bsS_{\text{tot}})\big)\bSa\bigg].
\end{align}
We can rewrite the product of the four spin operators in the last term as
% \begin{align*}
%   \big(\bSa(\bSb\times\bsS_{\text{tot}})\big)\Sa^d
%   =\sum_{\c=1}^N\varepsilon^{abc}\Sa^a\Sb^b\comm{\Sc^c}{\Sa^d}
% %  +\varepsilon^{abc}\Sa^a\Sb^b\Sa^d S_{\text{tot}}^c 
%   +\varepsilon^{abc}\Sa^a\Sb^b\Sa^d\sum_{\c=1}^N\Sc^c,
% \end{align*}
\begin{align}
  \label{eq:b:SaSbtimesStotSa1}
  \big(\bSa(\bSb\times\bsS_{\text{tot}})\big)\Sa^d
  =\sum_{\c=1}^N\varepsilon^{abc}\Sa^a\Sb^b\comm{\Sc^c}{\Sa^d}
  +\text{something}\cdot S_{\text{tot}}^c,
\end{align}
where the second term annihilates every singlet. The first term yields
\begin{align}
  \label{eq:b:SaSbtimesStotSa2}
%  \sum_{\c=1}^N\varepsilon^{abc}\Sa^a\Sb^b\comm{\Sc^c}{\Sa^d}
  \varepsilon^{abc}\Sa^a\Sb^b\comm{\Sa^c}{\Sa^d}
  &=\i\varepsilon^{abc}\varepsilon^{cde}\Sa^a\Sb^b\Sa^e
  \nonumber\\[0.2\baselineskip] 
  &=\i\big(\delta^{ad}\delta^{be}-\delta^{ae}\delta^{bd}\big)\Sa^a\Sb^b\Sa^e
  \nonumber\\[0.2\baselineskip] 
  &=\i\Sa^d\big(\bSa\bSb\big) - \i\Sb^d \bSa^2,
%   \\[0.2\baselineskip] 
%   &=\i\Sa^d\big(\bSb\bSa\big) - 2\i\Sb^d.
\end{align}
where we have used $\a\ne\b$.  The Pfaffian spin liquid state
\eqref{eq:b:ket} with \eqref{eq:b:psi0} is therefore also annihilated
by
% \begin{align}
%   \label{eq:b:habarPT_1v2}
% %  \frac{\i}{2}
%   \sum_{\substack{\b\\ \b\ne\a}}\frac{\ea+\eb}{\ea-\eb}
%   \bigg[i(\bSa\times\bSb) + 2\bSb -
%   \frac{1}{2}\bSa\big(\bSa\bSb\big)\bigg].
% \end{align}
\begin{align}
  \label{eq:b:doperator}
  &\bs{D}_\a^{S=1}
  =\frac{1}{2}\sum_{\substack{\b\\ \b\ne\a}}\frac{\ea+\eb}{\ea-\eb}
  \bigg[\i(\bSa\times\bSb) + 2\bSb - \frac{1}{2}\bSa\big(\bSa\bSb\big)\bigg],
  \nonumber\\[0.2\baselineskip] 
  &\bs{D}_\a^{S=1}\ket{\psi^{S=1}_{0}} = 0 \quad\forall\, \alpha.
\end{align}
This is the analog of the auxiliary operator 
\eqref{eq:hsdoperator} or \eqref{eq:a:doperator} 
of the Haldane--Shastry model.

Equation \eqref{eq:b:doperator} implies that $\ket{\psi^{S=1}_{0}}$
% the $S=1$ spin liquid ground state 
is further annihilated by
\begin{align}
  \label{eq:b:lambdaoperator}
  \bs{\Lambda}^{S=1} &= \sum_{\alpha=1}^N\bs{D}_{\alpha}^{S=1}
%  \nonumber\\[0.2\baselineskip] &
  =\frac{1}{2}\sum_{\substack{\a\ne\b}}\frac{\ea+\eb}{\ea-\eb}
  \bigg[\i(\bSa\times\bSb) - \frac{1}{2}\bSa\big(\bSa\bSb\big)\bigg],
%  \quad \bs{\Lambda}^{S=1} \ket{\psi^{\s\text{HS}}_{0}} = 0,
\end{align}
where we have used \eqref{eq:app-hs13}.  This is the analog of the
rapidity operator \eqref{eq:hsrapidityoperator} or
\eqref{eq:a:lambdaoperator} of the Haldane-Shastry model.  In contrast
to the Haldane--Shastry model, however, the operator
\eqref{eq:b:lambdaoperator} does not commute with the Hamiltonian
\eqref{eq:b:h}.  

%\newpage
\subsection{Annihilation operators which transform odd under T}

Finally, we consider annihilation operators we can construct from
\eqref{eq:b:ha}, and which transform odd under T,
\begin{equation}
  \label{eq:b:habarT}
  H_\a^{\rm\bar T}=\frac{1}{2}\left(H_\a-\Theta H_\a \Theta\right)
  = H_\a^{\rm\bar T=} + H_\a^{\rm\bar T\ne}
\end{equation}
with 
\begin{align}
  \label{eq:b:habarT=}
  H_\a^{\rm\bar T=}&=\frac{1}{2}\sum_{\substack{\b\\\b\ne\a}}\omega_{\a\b\b}
  \big({\Sa^\z}^2 \bigcomm{\Sb^+}{\Sb^-}
    +\Sa^\z\biganticomm{\Sb^+}{\Sb^-}\big)
  \nonumber\\[0.2\baselineskip] 
  &=\sum_{\substack{\b\\\b\ne\a}}\omega_{\a\b\b}
  \left({\Sa^\z}^2\Sb^\z + \Sa^\z\big(\bSb^2-{\Sb^\z}^2\big)\right),
  \\[0.4\baselineskip] 
  \label{eq:b:habarTnot=}
  H_\a^{\rm\bar T\ne} &=\sum_{\substack{\b\ne\c\\ \b,\c\ne\a}}\omega_{\a\b\c}
  \,\Sa^\z\Sb^+\Sc^-.
\end{align}
Let us first look at the component which transforms odd under P,
\begin{align}
  \label{eq:b:habarPbarT}
  H_\a^{\rm \bar P\bar T}=\frac{1}{2}
  \big(H_\a^{\rm\bar T} - \Pi\hspace{1pt} H_\a^{\rm\bar T}\hspace{1pt} \Pi\big)
  = H_\a^{{\rm \bar P\bar T}=} + H_\a^{{\rm \bar P\bar T}\ne},
\end{align}
where
\begin{align}
  \label{eq:b:habarPbarT1}
  H_\a^{{\rm {\bar P}\bar T}=}&=0, 
  \hspace{-10pt}&\hspace{-10pt}
%  \\[0.4\baselineskip] 
%  \label{eq:b:h0Tn}
  H_\a^{{\rm {\bar P}\bar T}\ne}&=
  \frac{1}{2}\sum_{\substack{\b\ne\c\\ \b,\c\ne\a}}\omega_{\a\b\c}
  \,{\Sa^\z}\big(\Sb^+\Sc^- - \Sb^-\Sc^+\big).
\end{align}
This operator has no vector component.  With \eqref{eq:app-t-Sz+-w-}, 
we obtain the scalar component
\begin{align}
  \label{eq:b:habarTnot=0v1}
  \big\{H_\a^{\rm\bar T}\big\}_0
  &=-\frac{\i}{3}\sum_{\substack{\b\ne\c\\ \b,\c\ne\a}}
  \frac{\bSa(\bSb\times\bSc)}{(\eab-\ebb)(\ea-\ec)},
\end{align}
It is identical to \eqref{eq:a:habarTnot=0v1} in the Haldane--Shastry
model, and annihilates every spin singlet.  We will not consider it
further.

We will now turn the component which transforms even under P,
\begin{align}
  \label{eq:b:haPbarT}
  H_\a^{\rm P\bar T}=\frac{1}{2}
  \big(H_\a^{\rm\bar T} + \Pi\hspace{1pt} H_\a^{\rm\bar T}\hspace{1pt} \Pi\big)
  = H_\a^{{\rm P\bar T}=} + H_\a^{{\rm P\bar T}\ne},
\end{align}
where
\begin{align}
  \label{eq:b:haPbarT1}
  H_\a^{{\rm {P}\bar T}=}&=H_\a^{{\rm \bar T}=}, 
  \hspace{-6pt}&\hspace{-6pt}
%  \\[0.4\baselineskip] 
%  \label{eq:b:haPbarTne}
  H_\a^{{\rm {P}\bar T}\ne}&=
  \frac{1}{2}\sum_{\substack{\b\ne\c\\ \b,\c\ne\a}}\omega_{\a\b\c}
  \,{\Sa^\z}\big(\Sb^+\Sc^- + \Sb^-\Sc^+\big),
\end{align}
which has no scalar, but a vector component.  With
\eqref{eq:app-t-SzSzSz} and \eqref{eq:app-t-SSS-1abc}, we write
  \begin{align}
  \label{eq:b:haPbarT=1v1}
  \big\{H_\a^{\rm\bar T=}\big\}_{1}
  &=\frac{1}{5}\sum_{\substack{\b\\\b\ne\a}}\omega_{\a\b\b}
  \Big[\Sa^\z(\bSa\bSb) + \bSa(\Sa^\z)\bSb + \bSa^2\Sb^\z
  \nonumber\\[-0.4\baselineskip] 
  &\hspace{71.5pt} +4\Sa^\z\bSb^2 - \bSa(\Sb^\z)\bSb - (\bSa\bSb)\Sb^\z\Big].
\end{align}
Writing out the second term, we obtain 
\begin{align}
  \label{eq:b:commSaSb,Sa}
  \bSa(\Sa^\z)\bSb
  &=\frac{1}{2}\big( \Sa^-\Sa^\z\Sb^+ + \Sa^+\Sa^\z\Sb^-\big) + \Sa^\z\Sa^\z\Sb^\z,
  \nonumber\\[0.2\baselineskip] 
  &=\Sa^\z(\bSa\bSb)+
  \frac{1}{2}\Big(\bigcomm{\Sa^-}{\Sa^\z}\Sb^+ +\bigcomm{\Sa^+}{\Sa^\z}\Sb^-\Big)
%   \nonumber\\[0.2\baselineskip] 
%   &=\Sa^\z(\bSa\bSb)+
%   \frac{1}{2}\Big(\Sa^-\Sb^+ - \Sa^+\Sb^-\Big)
  \nonumber\\[0.3\baselineskip] 
  &=\Sa^\z(\bSa\bSb)+ \i (\bSa\times\bSb)^\z.
\end{align}
% Similarly, we obtain
% \begin{align*}
%   (\bSa\bSb)\Sb^\z=\bSa(\Sb^\z)\bSb -\i (\bSa\times\bSb)^\z. 
% \end{align*}
% Collecting the terms, we obtain 
% \begin{align}
%   \label{eq:b:haPbarT=1v2}
%   \big\{H_\a^{\rm\bar T=}\big\}_{1}
%   &=\frac{2}{5}\sum_{\substack{\b\\\b\ne\a}}\omega_{\a\b\b}
%   \Big[4\Sa^\z + \Sb^\z +(\Sa^\z-\Sb^\z)(\bSa\bSb) + \i (\bSa\times\bSb)^\z\Big].
% \end{align}
% where we have used $\bSa^2=\bSb^2=2$.
% 
Similarly, the fifth term gives
\begin{align}
  \label{eq:b:commSaSb,Sb}
  \bSa(\Sb^\z)\bSb
%   &=\frac{1}{2}\big( \Sa^-\Sb^\z\Sb^+ + \Sa^+\Sb^\z\Sb^-\big) + \Sa^\z\Sb^\z\Sb^\z,
%   \nonumber\\[0.2\baselineskip] 
%   &=(\bSa\bSb)\Sb^\z+
%   \frac{1}{2}\Big(\Sa^-\bigcomm{\Sb^\z}{\Sb^+} + \Sa^+\bigcomm{\Sb^\z}{\Sb^-}\Big)
% %   \nonumber\\[0.2\baselineskip] 
% %   &=(\bSa\bSb)\Sa^\z+
% %   \frac{1}{2}\Big(\Sa^-\Sb^+ - \Sa^+\Sb^-\Big).
%   \nonumber\\[0.3\baselineskip] 
  &=(\bSa\bSb)\Sb^\z+ \i (\bSa\times\bSb)^\z.
\end{align}
Collecting the terms, we obtain 
\begin{align}
  \label{eq:b:haPbarT=1v2}
  \big\{H_\a^{\rm\bar T=}\big\}_{1}
  &=\frac{2}{5}\sum_{\substack{\b\\\b\ne\a}}\omega_{\a\b\b}
  \Big[4\Sa^\z + \Sb^\z + \Sa^\z(\bSa\bSb) - (\bSa\bSb)\Sb^\z\Big],
\end{align}
where we have used $\bSa^2=\bSb^2=2$.
With \eqref{eq:app-t-Sz+-w+}, we further obtain
\begin{align}
  \label{eq:b:haPbarTnot=1}
  \big\{H_\a^{\rm\bar T\ne}\big\}_{1}
  =\frac{1}{5}\sum_{\substack{\b\ne\c\\ \b,\c\ne\a}}\omega_{\a\b\c}
%  \,{\Sa^\z}\big(\Sb^+\Sc^- + \Sb^-\Sc^+\big).
  \Big[4\Sa^\z(\bSb\bSc) - \Sb^\z(\bSa\bSc) - \Sc^\z(\bSa\bSb)\Big].
\end{align}
Combining \eqref{eq:b:haPbarT=1v2} and \eqref{eq:b:haPbarTnot=1}, we finally
obtain the vector annihilation operator
\begin{align}
  \label{eq:b:aoperator}
%  \bs{A}_\a &\equiv 5\big\{H_\a^{\rm\bar T}\big\}_{\bs{1}}
  \bs{A}_\a^{S=1} &\equiv 5\Big(\big\{H_\a^{\rm\bar T=}\big\}_{\bs{1}} +
  \{H_\a^{\rm\bar T\ne}\big\}_{\bs{1}}\big)
  \nonumber\\*[0.2\baselineskip] 
  &=2\sum_{\substack{\b\\\b\ne\a}}
%  \frac{\Sa^\z\big(4 + \bSa\bSb\big) + \big(1 -\bSa\bSb\big)\Sb^\z }
  \frac{4\bSa + \bSb + \bSa(\bSa\bSb) - (\bSa\bSb)\bSb}
  {\vert\ea-\eb\vert^2}
  \nonumber\\*[0.2\baselineskip] 
  &\quad %\hspace{50pt}
  +\sum_{\substack{\b\ne\c\\ \b,\c\ne\a}}
  \frac{4\bSa(\bSb\bSc)-\bSb(\bSa\bSc)-\bSc(\bSa\bSb)}{(\eab-\ebb)(\ea-\ec)},
  \nonumber\\[0.2\baselineskip] 
  \bs{A}_\a^{S=1} &\ket{\psi^{S=1}_{0}} = 0 \quad \forall\, \alpha.
\end{align}
This operator is rather complicated, but does simplify as we sum over $\a$.
From \eqref{eq:b:habarT=}, we obtain
\begin{align*}
%  \label{eq:b:h0PbarT=1}
  \sum_\a \big\{H_\a^{\rm\bar T=}\big\}_{\bs{1}}
  &=2\sum_\a\bSa\sum_{\substack{\b\\\b\ne\a}}\omega_{\a\b\b}
  =\frac{N^2-1}{6} \bsS_{\text{tot}}.
\end{align*}
This implies that $\ket{\psi^{S=1}_{0}}$ is also annihilated by
\begin{align}
  \label{eq:b:upsilonoperator}
%  \big\{H_0^{\rm\bar T}\big\}_{\bs{1}}
  \bs{\Upsilon}^{S=1}
  &=5\sum_{\a}\big\{H_\a^{\rm\bar T\ne}\big\}_{\bs{1}}
%  &=\sum_{\a}\bs{A}_\a
%  \nonumber\\[0.2\baselineskip] 
%  &
  =\sum_{\substack{\a,\b,\c\\ \a\ne\b\ne\c\ne\a}}%\hspace{5pt}
  \frac{4\bSa(\bSb\bSc)-\bSb(\bSa\bSc)-\bSc(\bSa\bSb)}{(\eab-\ebb)(\ea-\ec)}.
  \nonumber\\[-0.4\baselineskip]
\end{align}

\section{Concluding remarks}

\renewcommand{\strut}{\rule[-5pt]{0pt}{16pt}}
\begin{table}[t]
  \centering
  \caption{Annihilation operators for the 
    $S=1$ spin liquid ground state.  With the exception of the \defining
    operator $\OaSone$ and $\Xi_{\a}$, which are the $m=3$ components of 3rd order tensors,
    we have only included scalar and vector annihilation operators.}
  \label{tab:b:annihilationops}
  \centering
  \begin{tabular}{p{24mm}p{20mm}p{12mm}p{12mm}p{26mm}p{18mm}}\hline
%  \begin{tabular}{|c|c|c|c|c|c|}\hline
  \multicolumn{6}{c}{{\bf Annihilation operators for} 
    $\ket{\psi^{S=1}_{0}}$}\strut \\[2pt]\hline
   Operator &Equation 
   &\multicolumn{4}{l}{Symmetry transformation properties}\strut 
   \\[3pt]\cline{3-6}
   &&T &P &order of tensor &transl.~inv.\strut\\[3pt]\hline %\svhline
   $\bsS_{\text{tot}}$&\eqref{eq:hsspinsymmetry}&$-$&$+$&vector&yes\rule[0pt]{0pt}{13pt}\\[3pt]
   $\OaSone$&\eqref{eq:b:Omegadef}&no&no&3rd&no\\[3pt]
   $\Xi_{\a}$&\eqref{eq:b:Thetadef}&no&no&3rd&no\\[3pt]
   $H^{S=1}-E_0^{S=1}$&\eqref{eq:b:h}&$+$&$+$&scalar&yes\\[3pt]
   $\bs{D}_\a^{S=1}$&\eqref{eq:b:doperator}&$+$&$-$&vector&no\\[3pt]
   $\bs{\Lambda}^{S=1}$&\eqref{eq:b:lambdaoperator}&$+$&$-$&vector&yes\\[3pt]
   $\bs{A}_\a^{S=1}$&\eqref{eq:b:aoperator}&$-$&$+$&vector&no\\[3pt]
   $\bs{\Upsilon}^{S=1}$&\eqref{eq:b:upsilonoperator}&$-$&$+$&vector&yes\\[3pt]
  \hline
  \end{tabular}
\end{table}

The various annihilation operators for the $S=1$ model derived in this
section are summarized in Table \ref{tab:b:annihilationops}.  

The main result, of course, is the Hamiltonian $H^{S=1}$ given by
\eqref{eq:b:h}.  It is a three-spin operator.  The three-body
interaction terms fall off as $1/(r_{12}r_{13})$, which makes the
model long-ranged.  Since the wave function \eqref{eq:napsi0}
introduced in Section \ref{sec:3mod-na} is critical, \ie has
algebraically decaying correlations, it is not surprising that we need
a Hamiltonian with long-ranged interaction to single it out as unique
and exact ground states.  Hamiltonians with only short-ranged
interactions, like the Heisenberg model, tend to single out states
with exponentially decaying correlations, and a Haldane gap in the
excitation
spectrum~\cite{haldane83pl464,haldane83prl1153,affleck90proc,Fradkin91}.

The most intriguing feature of the $S=1$ Pfaffian spin liquid state we
have elevated into an exactly soluble model here is that the spinon
excitations obey a novel form of quantum statistics, which is
presumably the closest analog to non-Abelian statistics one can define
in one dimensions.  As explained in Section \ref{sec:nana}, there is
an internal, topological Hilbert space of dimension $2^n$ associated
with a state with $2n$ spinons.  In the thermodynamic limit, all the
states in this internal Hilbert space become degenerate.  We assume
that the information regarding the internal state is encoded in
fractional shifts in the momentum spacings between the individual the
spinons (see Section \ref{sec:nana}).  These shifts are topological
quantum numbers, and are hence insensitive against local, external
perturbations.  This makes this model, and presumably a range of
models of critical $S=1$ spin chains, suited for applications as
protected qubits in quantum computing.

Preliminary numerical work~\cite{manuscriptinpreparationTRSG11}
indicates that the rapidity operator $\bs{\Lambda}$ given in
\eqref{eq:b:lambdaoperator} does not commute with $H^{S=1}$. The model
hence does not appear to share the integrability structure of the spin \half
Haldane--Shastry model.  We conjecture that the reason for this is
related to the rich internal structure of the Hilbert space, which
makes the universality class of the states we introduce here both much
less accessible and much more interesting than the Abelian
$S=\frac{1}{2}$ Heisenberg model.

In the following \chap , we will employ the theoretical method
developed here to generalize the model to arbitrary spin, \ie to
identify a parent Hamiltonian for the state \eqref{eq:nacslopS}
introduced in Section \ref{sec:naarbs}.

\vspace{50pt}\newpage
% lp -dtkmsek -P 55-59 map.ps 

\chapter{Generalization to arbitrary spin $\bs S$}
\label{sec:gen2s}

%\section{A critical spin $S$ liquid state}
%\section{Formulation of a critical spin liquid with spin $S$}
\section{A critical spin liquid state with spin $S$}

\subsection{Generation through projection of Gutzwiller states}
\label{sec:c:genbyGutz}

In this section, we wish to generalize the model introduced and
derived in the previous section for spin $S=1$ to arbitrary spin $S=s$.
The generalization of the $S=1$ ground state \eqref{eq:naket} with
\eqref{eq:napsi0} was introduced in Section \ref{sec:naarbs}.
In essence, we combine $2s$ identical copies of the Gutzwiller or
Haldane--Shastry ground state with spin $\frac{1}{2}$, 
and project the spin on each site onto spin $s$,
\newcommand{\struta}{\rule[-6pt]{0pt}{0pt}}
\newcommand{\strutb}{\rule[6pt]{0pt}{0pt}}
\begin{align*}
  \underbrace{\textstyle \bs{\frac{1}{2}}\otimes\bs{\frac{1}{2}}\otimes
    \ldots\otimes\bs{\frac{1}{2}}\struta}_{2s\strutb}
  =\bs{s}\oplus (2s-1)\cdot \bs{s\!-\!1}\oplus\ldots
\end{align*}
The projection onto the completely symmetric representation can be
carried out conveniently using Schwinger bosons (see Section
\ref{sec:naschwinger}).  In particular, if we write the
Haldane-Shastry ground state as 
 \begin{align}
  \label{eq:c:hsket}
  \ket{\psi^{\s\text{HS}}_{0}}
  &=\sum_{\{z_1,z_2,\ldots ,z_M\}}
  \psi^{\s\text{HS}}_{0}(z_1,\ldots ,z_M)\;
%  \psi^{\s\text{HS}}_{0}[z]\,
  {S}^+_{z_1}\cdot\ldots\cdot {S}^+_{z_M} 
%  \big|\underbrace{\dw\dw\ldots\ldots\dw}_{\text{all\ } N \text{\ spins\ } \dw}
%  \big\rangle
  \ket{\dw\dw\ldots\dw} 
  \nonumber\\[0.4\baselineskip] 
  &=\sum_{\{z_1,\ldots ,z_M;w_1,\ldots,w_M\}}
  \psi^{\s\text{HS}}_{0}(z_1,\ldots ,z_M)\;
%  \psi^{\s\text{HS}}_{0}[z]\,
    {a}^+_{z_1}\ldots a^\dagger_{z_M}
    {b}^+_{w_1}\ldots b^\dagger_{w_M}%
  \vac\!
  \nonumber\\[0.4\baselineskip] 
  &\equiv \Psi^{\s\text{HS}}_{0}[a^\dagger,b^\dagger] \vac\!,
\end{align}
where $\psi^{\s\text{HS}}_{0}(z_1,\ldots ,z_M)$ is given by \eqref{eq:hspsi0},
$M=\frac{N}{2}$ and the $w_k$'s are those lattice sites which are not occupied
by any of the $z_i$'s, we can write the spin $S$ state obtained by the
mentioned projection as
(\cf \ref{eq:nacslopS})
\begin{equation}
  \label{eq:c:psi0schwinger}
  \ket{\psi^{S}_0}
  =\Big(\Psi^{\s\text{HS}}_0\big[a^\dagger ,b^\dagger\big]\Big)^{2s}\vac.
\end{equation} 
In Section \ref{sec:naarbs}, we mentioned that the state can alternatively 
be written as 
\begin{equation}
  \label{eq:c:psi0ket}
  \ket{\psi^S_0}\;=\sum_{\{z_1,\dots,z_{SN}\}} 
%  \psi_0[z_i]
  \psi^S_0(z_1,\dots,z_{SN})\
%  \tilde{S}_{z_1}^{+}\cdot\, \dots\, \cdot \tilde{S}_{z_{N}}^{+}\ 
  \tilde{S}_{z_1}^{+}\cdot\dots\cdot\tilde{S}_{z_{SN}}^{+} 
%  \ket{1,-1}^{\otimes N},
%  \ket{1,-1}_N,
  \ket{-s}_N,
%  \ket{-1}^{\otimes N},
\end{equation}
where $N$ is the number of lattice sites,
\begin{equation}
  \label{eq:c:vacket}
  \ket{-s}_N\equiv\otimes_{\alpha=1}^N \ket{s,-s}_{\alpha}
\end{equation}
is the ``vacuum'' state in which all the spins are maximally polarized
in the negative $\hat z$-direction, and $\tilde{S}^{+}$ are
re-normalized spin flip operators $\tilde{S}^{+}$ which satisfy
\begin{equation}
  \label{eq:c:Stilde^n}
  \frac{1}{\sqrt{(2s)!}}(a^\dagger)^n (b^\dagger)^{(2s-n)}\vac 
  =(\tilde{S}^+)^n\ket{s,-s}.
\end{equation}
In a basis in which $S^\z$ is diagonal, we may write 
\begin{equation}
  \label{eq:c:Stilde}
  \tilde{S}^{+} 
  \,\equiv\,\frac{1}{b^\dagger b+1}\,a^\dagger b 
  \,=\, \frac{1}{S-{S}^\z+1}\, S^{+}.
\end{equation}
Note that \eqref{eq:c:Stilde^n} implies
\begin{align}
  \label{eq:c:S-Stilde^n}
  S^-(\tilde{S}^+)^n\ket{s,-s}
  &=b^\dagger a \frac{1}{\sqrt{(2s)!}}(a^\dagger)^n (b^\dagger)^{(2s-n)}\vac 
  \nonumber\\[0.2\baselineskip] 
  &=n(\tilde{S}^+)^{n-1}\ket{s,-s}.
\end{align}
The wave function for the spin $S$ state \eqref{eq:c:psi0schwinger} are then
given by 
\begin{equation}
  \label{eq:psirr}
  \psi^S_0(z_1,\dots,z_{SN})
  =\prod_{m=1}^{2s}\left(
%    \prod_{i<j}^{M}(z_{i+(m-1)M}-z_{j+(m-1)M})^2 
    \prod_{\substack{i,j=(m-1)M+1\\[1pt] i<j}}^{mM}(z_i-z_j)^2 
  \right)\,
  \prod_{i=1}^{sN}z_i.
\end{equation}
Note the similarity to Read--Rezayi states~\cite{Read-99prb8084} in
the quantized Hall effect.

For the purposes in Sections \eqref{sec:c:directsinglet} and
\eqref{sec:c:directdefining}, it is convenient to write
the state in the form
\begin{align}
  \label{eq:c:psi0prodket}
  \ket{\psi^{S}_0}
  &=%\Bigg(
  \left[
    \sum_{\{z_1,\ldots ,z_M\}}
    \psi^{\s\text{HS}}_{0}(z_1,\ldots ,z_M)\;
    \tilde{S}_{z_1}^{+}\cdot\ldots\cdot\tilde{S}_{z_{M}}^{+} %\Bigg)
  \right]^{2s}\vac.
%   \nonumber\\[0.4\baselineskip] 
%   &=\prod_{n=1}^{2S}%\Bigg[
%     \left[
%     \sum_{\{z_{1,n},\ldots ,z_{M,n}\}}
%     \psi^{\s\text{HS}}_{0}(z_{1,n},\ldots ,z_{M,n})\;
%     \tilde{S}_{z_{1,n}}^{+}\cdot\dots\cdot\tilde{S}_{z_{M,n}}^{+} %\Bigg)
%     %\Bigg]%
%     \right]
%     \vac,
\end{align}
%where the coordinates of the re-normalized spin flips are labeled as
%$z_{i,n}$, with $i=1,\ldots,M$ and $n=1,\ldots, 2S$.

%\vspace{50pt}\newpage
%\newpage
\subsection{Direct verification of the singlet property}
\label{sec:c:directsinglet}

The singlet property of $\ket{\psi^S_0}$ is manifest from the method
we employed to construct it by combining $2s$ copies of states which
are singlets, and in particular through \eqref{eq:c:psi0schwinger}.
It is nonetheless instructive to proof it directly from
\eqref{eq:c:psi0prodket}, as the proof of the \defining condition for
the state in Section \ref{sec:c:directdefining} will proceed along
similar lines.

Since the $S_{\text{tot}}^\z$ component of \eqref{eq:c:psi0prodket} is
trivially equal to zero, it is sufficient to show that
$\ket{\psi^s_0}$ is annihilated by $S_{\text{tot}}^-$. 
% $S_{\text{tot}}^-=\sum_\a S_\a^-$.
As we act with $S_\a^-$ on \eqref{eq:c:psi0prodket}, we have to
distinguish between configurations with $n=0,1,2,\ldots,2s$
re-normalized spin flips $\tilde{S}_\a^+$ at site $\a$.  Since the
state is symmetric under interchange of the $2s$ copies of
$\psi^{\s\text{HS}}_{0}$, we may assume that the $n$ spin flips are
present in the first $n$ copies, and account for the restriction
through ordering by a combinatorial factor.  This yields
\renewcommand{\struta}{\rule[-18pt]{0pt}{0pt}}
\begin{align}
  \label{eq:c:Stotpsi0}
  \hspace{5pt}&\hspace{-5pt}\sum_\a S_\a^-\ket{\psi^{S}_0}
  \nonumber\\*[-0.2\baselineskip] 
  &=\sum_{\a=1}^N S_\a^-\sum_{n=0}^{2s}\binom{2s}{n}
  \left[\sum_{\{z_2,\ldots ,z_M\}}
    \psi^{\s\text{HS}}_{0}(\ea,z_2,\ldots ,z_M)\,
    \tilde{S}_{\a}^{+}\tilde{S}_{z_2}^{+}\cdot\ldots\cdot\tilde{S}_{z_{M}}^{+}
  \right]^n
  \nonumber\\*[0.2\baselineskip] 
  &\hspace{83pt}\cdot\left[\sum_{\substack{\{z_1,\ldots ,z_M\}\ne\ea}}
    \psi^{\s\text{HS}}_{0}(z_1,\ldots ,z_M)\,
    \tilde{S}_{z_1}^{+}\cdot\ldots\cdot\tilde{S}_{z_{M}}^{+}
  \right]^{2s-n}\vac
  \nonumber\\[0.6\baselineskip] 
  &=2s\sum_{\a=1}^N \left[
    \sum_{\{z_2,\ldots ,z_M\}}\psi^{\s\text{HS}}_{0}(\ea,z_2,\ldots ,z_M)
    \,\tilde{S}_{z_2}^{+}\cdot\ldots\cdot\tilde{S}_{z_{M}}^{+}\right]
%  \sum_{n=1}^{2s}\binom{2s\!-\!1}{m\!-\!1}
  \nonumber\\* %[0.0\baselineskip] 
  &\quad\cdot
  \sum_{n=1}^{2s}\binom{2s-1}{n-1}
  \left[\sum_{\{z_2,\ldots ,z_M\}}
    \psi^{\s\text{HS}}_{0}(\ea,z_2,\ldots ,z_M)\,
    \tilde{S}_{\a}^{+}\tilde{S}_{z_2}^{+}\cdot\ldots\cdot\tilde{S}_{z_{M}}^{+}
  \right]^{n-1}
  \nonumber\\*[0.2\baselineskip] 
  &\hspace{72.13pt}\cdot\left[\sum_{\substack{\{z_1,\ldots ,z_M\}\ne\ea}}
    \psi^{\s\text{HS}}_{0}(z_1,\ldots ,z_M)\,
    \tilde{S}_{z_1}^{+}\cdot\ldots\cdot\tilde{S}_{z_{M}}^{+}
  \right]^{2s-n}\vac
  \nonumber\\[0.6\baselineskip] 
  &=2s\left[\sum_{\{z_2,\ldots ,z_M\}}\right.
  \,\underbrace{\sum_{\a=1}^N \psi^{\s\text{HS}}_{0}(\ea,z_2,\ldots ,z_M)\struta}_{=0}\,
  \,\tilde{S}_{z_2}^{+}\cdot\ldots\cdot\tilde{S}_{z_{M}}^{+}
  \left]\phantom{\sum_{\{z_2\}}}\right.
  \nonumber\\*%[0.2\baselineskip] 
  &\hspace{72.13pt}\cdot\left[\sum_{\substack{\{z_1,\ldots ,z_M\}}}
    \psi^{\s\text{HS}}_{0}(z_1,\ldots ,z_M)\,
    \tilde{S}_{z_1}^{+}\cdot\ldots\cdot\tilde{S}_{z_{M}}^{+}
  \right]^{2s-1}\!\vac,
  \nonumber\\[-0.2\baselineskip] 
\end{align}
%where the restriction $\{z_1,\ldots ,z_M\}\ne\ea$ means that all of
%the $z_i$'s must be different from $\ea$, and 
where we have used \eqref{eq:c:S-Stilde^n} and that
$\psi^{\s\text{HS}}_{0}(\ea,z_2,\ldots ,z_M)$ contains only powers
$\eb^1,\eb^2,\ldots , \eb^{N-1}$.

% \begin{align}
%   \label{eq:c:S-Stilde^n}
%   S^-(\tilde{S}^+)^n\ket{S,-S}
%   &=b^\dagger a \frac{1}{\sqrt{(2s)!}}(a^\dagger)^n (b^\dagger)^{(2s-n)}\vac 
%   \nonumber\\[0.2\baselineskip] 
%   &=n(\tilde{S}^+)^{n-1}\ket{S,-S}.
% \end{align}
% and 
% \begin{align*}
%   \sum_\a \psi^{\s\text{HS}}_{0}(\ea,z_2,\ldots ,z_M)=0
% \end{align*}
% (\cf \eqref{eq:hsStot-psi0}  and \eqref{eq:hssumeta^m}).

\section{The \defining condition for the spin $S$ chain}
\label{sec:c:defining}

\subsection{Statement}

The \defining condition for the spin $S$ state is by direct generalization
of \eqref{eq:a:Omegadef} and \eqref{eq:b:Omegadef} given by
\begin{align}
  \label{eq:c:Omegadef}
    \OaS 
    &=\sum_{\substack{\beta=1\\[2pt]\beta\ne\alpha}}^N 
    \frac{1}{\ea-\eb} (\Sa^-)^{2s} \Sb^-,\qquad 
    \OaS \ket{\psi^S_{0}} = 0 \quad\forall\, \alpha.
\end{align}
Since the state is real, it is also annihilated by the complex conjugate
of $\OaS$,
\begin{align}
  \label{eq:c:barOmegadef}
    \OabS 
    &=\sum_{\substack{\beta=1\\[2pt]\beta\ne\alpha}}^N 
    \frac{1}{\eab-\ebb} (\Sa^-)^{2s} \Sb^-,\qquad 
    \OabS \ket{\psi^S_{0}} = 0 \quad\forall\, \alpha.
\end{align}

\subsection{Direct verification}
\label{sec:c:directdefining}

Unlike for the cases of spin \half and spin one, we have not derived
the \defining condition \eqref{eq:c:Omegadef} from the parent
Hamiltonian of a quantized Hall state.  The direct and explicit
verification presented here does therefore not just serve to check the
validity of the previous analysis, but is an essential part of the
entire argument we present.

Let us consider the action of $(\Sa^-)^{2s} \Sb^-$ on 
$\ket{\psi^S_{0}}$ written in the form 
\eqref{eq:c:psi0prodket}.  Since $\psi^{\s\text{HS}}_{0}(z_1,\ldots ,z_M)$
vanishes whenever two arguments $z_i$ coincide, one of the $z_i$'s
in each of the $2s$ copies 
%of $\psi^{\s\text{HS}}_{0}(z_1,\ldots ,z_M)$ 
in \eqref{eq:c:psi0prodket} must equal $\ea$; since
$\psi^{\s\text{HS}}_{0}(z_1,\ldots ,z_M)$ is symmetric under
interchange of the $z_i$'s and we count each distinct configuration in
the sums over ${\{z_1,\ldots ,z_M\}}$ only once, we may take
$z_1=\ea$.  Regarding the action of $\Sb^-$ on
\eqref{eq:c:psi0prodket}, we have to distinguish between
configurations with $n=0,1,2,\ldots,2s$ re-normalized spin flips
$\tilde{S}_\b^+$ at site $\b$.  Since the state is symmetric under
interchange of the $2s$ copies% of $\psi^{\s\text{HS}}_{0}$
, we may assume that the $n$ spin flips are present in the first $n$
copies, and account for the restriction through ordering by a
combinatorial factor.  This yields
\begin{align}
  \label{eq:c:SaSbpsi0}
  \hspace{5pt}&\hspace{-5pt}(\Sa^-)^{2s} \Sb^-\ket{\psi^{S}_0}
  \nonumber\\*[-0.2\baselineskip] 
  &=(\Sa^-)^{2s} \Sb^-\sum_{n=0}^{2s}\binom{2s}{n}\hspace{-3pt}
  \left[\sum_{\{z_3,\ldots ,z_M\}}
    \psi^{\s\text{HS}}_{0}(\ea,\eb,z_3,\ldots)\,
    \tilde{S}_{\a}^{+}\tilde{S}_{\b}^{+}
    \tilde{S}_{z_3}^{+}\cdot\ldots\cdot\tilde{S}_{z_{M}}^{+}
  \right]^n
  \nonumber\\*[0.2\baselineskip] 
  &\hspace{94pt}\cdot\left[\sum_{\substack{\{z_2,\ldots ,z_M\}\ne\eb}}
    \psi^{\s\text{HS}}_{0}(\ea,z_2,\ldots )\,
    \tilde{S}_{\a}^{+}\tilde{S}_{z_2}^{+}\cdot\ldots\cdot\tilde{S}_{z_{M}}^{+}
  \right]^{2s-n}\!\vac
  \nonumber\\[0.6\baselineskip] 
%  &=(2s)!\cdot 2s\cdot
  &=(2s)!\, 2s\!
  \left[\sum_{\{z_2,\ldots ,z_M\}}\psi^{\s\text{HS}}_{0}(\ea,\eb,z_3,\ldots ,z_M)
  \,\tilde{S}_{z_3}^{+}\cdot\ldots\cdot\tilde{S}_{z_{M}}^{+}\right]
  \nonumber\\* %[0.0\baselineskip] 
  &\qquad\cdot
  \sum_{n=1}^{2s}\binom{2s-1}{n-1}
  \left[\sum_{\{z_3,\ldots ,z_M\}}
    \psi^{\s\text{HS}}_{0}(\ea,\eb,z_3,\ldots ,z_M)\,
    \tilde{S}_{\b}^{+}\tilde{S}_{z_3}^{+}\cdot\ldots\cdot\tilde{S}_{z_{M}}^{+}
  \right]^{n-1}
  \nonumber\\*[0.2\baselineskip] 
  &\quad\hspace{72.13pt}\cdot\left[\sum_{\substack{\{z_2,\ldots ,z_M\}\ne\eb}}
    \psi^{\s\text{HS}}_{0}(\ea,z_2,\ldots ,z_M)\,
    \tilde{S}_{z_2}^{+}\cdot\ldots\cdot\tilde{S}_{z_{M}}^{+}
  \right]^{2s-n}\vac
  \nonumber\\[0.6\baselineskip] 
%  &=(2s)!\cdot 2s\cdot
  &=(2s)!\, 2s\!
  \left[\sum_{\{z_3,\ldots ,z_M\}}\psi^{\s\text{HS}}_{0}(\ea,\eb,z_3,\ldots ,z_M)
  \,\tilde{S}_{z_3}^{+}\cdot\ldots\cdot\tilde{S}_{z_{M}}^{+}\right]
  \nonumber\\*[0.2\baselineskip] 
  &\phantom{=(2s)!\, 2s\!}\hspace{-2.7pt}
    \cdot\left[\sum_{\substack{\{z_2,\ldots ,z_M\}}}
    \psi^{\s\text{HS}}_{0}(\ea,z_2,\ldots ,z_M)\,
    \tilde{S}_{z_2}^{+}\cdot\ldots\cdot\tilde{S}_{z_{M}}^{+}
  \right]^{2s-1}\vac ,
\end{align}
where we have used \eqref{eq:c:S-Stilde^n}.  This implies
\renewcommand{\struta}{\rule[-18pt]{0pt}{0pt}}
\begin{align}
  \hspace{5pt}&\hspace{-5pt}\OaS \ket{\psi^{S}_0}
  =\frac{1}{(2s)!\, 2s}\,\sum_{\substack{\beta=1\\[2pt]\beta\ne\alpha}}^N 
  (\Sa^-)^{2s} \Sb^- \ket{\psi^{S}_0}
  \nonumber\\[0.2\baselineskip] 
  &=(2s)!\, 2s\!\left[\sum_{\{z_3,\ldots ,z_M\}}\right.
    \underbrace{\sum_{\substack{\beta=1}}^N
    \frac{\psi^{\s\text{HS}}_{0}(\ea,\eb,z_3,\ldots ,z_M)}{\ea-\eb}\struta}_{=0}
    \,\tilde{S}_{z_3}^{+}\cdot\ldots\cdot\tilde{S}_{z_{M}}^{+}
  \left]\phantom{\sum_{\{z_2\}}}\right.
  \nonumber\\*[0.2\baselineskip] 
  &\phantom{=(2s)!\, 2s\!}\hspace{-2.7pt}
    \cdot\left[\sum_{\substack{\{z_2,\ldots ,z_M\}}}
    \psi^{\s\text{HS}}_{0}(\ea,z_2,\ldots ,z_M)\,
    \tilde{S}_{z_2}^{+}\cdot\ldots\cdot\tilde{S}_{z_{M}}^{+}
  \right]^{2s-1}\vac,
\end{align}
%
% \renewcommand{\struta}{\rule[-18pt]{0pt}{0pt}}
% \begin{align}
%   \hspace{35pt}&\hspace{-35pt}\frac{1}{(2s)!\, 2s}\, \OaS \ket{\psi^{S}_0}
%   =\frac{1}{(2s)!\, 2s}\,\sum_{\substack{\beta=1\\[2pt]\beta\ne\alpha}}^N 
%   \frac{1}{\ea-\eb} (\Sa^-)^{2s} \Sb^- \ket{\psi^{S}_0}
%   \nonumber\\[0.2\baselineskip] 
%   &=\left[\sum_{\{z_3,\ldots ,z_M\}}\right.
%     \underbrace{\sum_{\substack{\beta=1}}^N
%     \frac{\psi^{\s\text{HS}}_{0}(\ea,\eb,z_3,\ldots ,z_M)}{\ea-\eb}\struta}_{=0}
%     \,\tilde{S}_{z_3}^{+}\cdot\ldots\cdot\tilde{S}_{z_{M}}^{+}
%   \left]\phantom{\sum_{\{z_2\}}}\right.
%   \nonumber\\*[0.2\baselineskip] 
%   &\hspace{6pt}
%     \cdot\left[\sum_{\substack{\{z_2,\ldots ,z_M\}}}
%     \psi^{\s\text{HS}}_{0}(\ea,z_2,\ldots ,z_M)\,
%     \tilde{S}_{z_2}^{+}\cdot\ldots\cdot\tilde{S}_{z_{M}}^{+}
%   \right]^{2s-1}\vac,
% \end{align}
where we have used that 
\begin{equation*}
  \frac{\psi^{\s\text{HS}}_{0}(\ea,\eb,z_3,\ldots z_M)}{\ea-\eb}
  =(\ea-\eb)\ea\eb\prod_{i=3}^M(\ea-z_i)^2(\eb-z_i)^2z_i
  \prod_{3\le i< j}^M(z_i-z_j)^2 
\end{equation*}
vanishes for $\beta=\alpha$ and contains only powers
$\eb^1,\eb^2,\ldots , \eb^{N-2}$.  Note that the calculation for
$\OabS$ is almost identical, since
\begin{equation*}
  \frac{\psi^{\s\text{HS}}_{0}(\ea,\eb,z_3,\ldots z_M)}{\eab-\ebb}
  =-\ea\eb
  \frac{\psi^{\s\text{HS}}_{0}(\ea,\eb,z_3,\ldots z_M)}{\ea-\eb}
%  =-(\ea-\eb)\ea^2\eb^2\prod_{i=3}^M(\ea-z_i)^2(\eb-z_i)^2z_i
%  \prod_{3\le i< j}^M(z_i-z_j)^2 
\end{equation*}
vanishes also for $\beta=\alpha$ and 
contains only powers $\eb^2,\eb^3,\ldots , \eb^{N-1}$.

\section{Construction of a parent Hamiltonian}
\label{sec:c:ham}

\subsection{Translational symmetry}

\def\prefac{\frac{1}{2a_0}}

A Hermitian and translationally invariant operator which annihilates
$\ket{\psi^{S}_0}$ is given by
\begin{align}
  \label{eq:c:h0}
  H_0&=\prefac\sum_{\a=1}^N {\OaS}^\dagger\OaS
%  \nonumber\\[0.2\baselineskip] 
%  &
  =\prefac\sum_{\substack{\a,\b,\c\\ \a\ne \b,\c}}
    \frac{1}{\eab-\ebb}\frac{1}{\ea-\ec} (\Sa^+)^{2s}(\Sa^-)^{2s}\Sb^+\Sc^-,
\end{align}
where $a_0$ is a parameter we will conveniently choose below.
We wish the Hamiltonian to be further invariant under P, T , and spin 
rotations.  From \eqref{eq:app-t-S+S-}, the tensor content of
$\Sb^+\Sc^-$ is
\begin{align}
  \label{eq:c:Sb+Sc-}
  \Sb^+\Sc^-&=\frac{2}{3}\bSb\bSc - \text{i}(\bSb\times\bSc)^\z
    -\frac{1}{\sqrt{6}}\,T_{\b\c}^0,
\end{align}
where 
\begin{align}
  \label{eq:c:Tbc}
  T_{\b\c}^0
  &=\frac{1}{\sqrt{6}}\big(4 \Sb^\z\Sc^\z - \Sb^+\Sc^- - \Sb^-\Sc^+\big)
  \nonumber\\[0.2\baselineskip] 
  &=\frac{1}{\sqrt{6}}\big(6 \Sb^\z\Sc^\z - 2\bSb\bSc\big).
\end{align}
This implies that we only have to know the scalar, vector and 2nd
order tensor components of $(\Sa^+)^{2s}(\Sa^-)^{2s}$
in order to obtain the scalar component of $H_0$.

%\subsection{Tensor decomposition of $\big(\Sa^+\big)^{2s}\big(\Sa^-\big)^{2s}$}
%\vspace{50pt}
\subsection{Tensor decomposition of $(S^+)^{2s}(S^-)^{2s}$}
\label{sec:c:S+^2sS-^2s}

\vspace{0.2\baselineskip}
Since $(S^+)^{2s}(S^-)^{2s}$ contains only a
single spin operator $\bsS$ with Casimir $\bsS^2=s(s+1)$, its scalar component
$U$ must be a constant, its vector proportional to %$V^0=S^\z$ 
\begin{align}
  \label{eq:c:V^0}
  V^0=S^\z
\end{align}
(\cf \eqref{eq:app-t-S}), its 2nd order tensor component proportional to
\begin{align}
  \label{eq:c:T^0} 
  T^0=\frac{1}{\sqrt{6}}\big(4 S^\z S^\z - S^+S^- - S^-S^+\big)
  =\frac{2}{\sqrt{6}}\,\Big[3{S^\z}^2 - s(s+1)\Big]
\end{align}
(\cf \eqref{eq:app-t-SS-2}), and its 3rd order tensor component
proportional to
\begin{align}
  \label{eq:c:W^0}
  W^0&=-\frac{1}{\sqrt{5}}\big(
  S^-S^+S^\z + S^+S^\z S^- + S^\z S^-S^+
  \nonumber\\ %[0.2\baselineskip]
  &\hspace{30.6pt}     
  + S^+S^-S^\z  + S^-S^\z S^+ + S^\z S^+S^- \big)
  + \frac{4}{\sqrt{5}} S^\z S^\z S^\z ,
  \nonumber\\[0.2\baselineskip]
  &=\frac{2}{\sqrt{5}}\Big[5{S^\z}^2 - 3s(s+1) + 1\Big] S^\z
\end{align}
(\cf \eqref{eq:app-t-SSS-3m=0}).  Our task in this section is to 
calculate the constants of proportionality in the expansion
% \begin{align}
%   \label{eq:c:S+^2sS-^2s}
%   \hspace{5pt}&\hspace{-5pt}(S^+)^{2s}(S^-)^{2s}
%   \nonumber\\[0.2\baselineskip]
%   &=a_0\hspace{1pt}\Big\{ 1 + a S^\z + b\hspace{1pt}\Big[3{S^\z}^2 - 2s(s+1)\Big]
%   + c\hspace{1pt}\Big[5{S^\z}^2 - 3s(s+1) + 1\Big] S^\z \Big\} 
%   \nonumber\\[0.2\baselineskip]
%   &\quad + \ \text{tensors of order $>3$}. 
% \end{align}
\begin{align}
  \label{eq:c:S+^2sS-^2s}
  (S^+)^{2s}(S^-)^{2s}
  &=a_0\,\Big\{ 1 + a\, V^0 + b\,T^0  + c\,W^0  \Big\} 
%  &=a_0\hspace{1pt}\Big\{ 1 + a\hspace{1pt} V^0 + b\hspace{1pt}T^0
%  + b\hspace{1pt}W^0  \Big\} 
  \nonumber\\[0.2\baselineskip]
  &\quad 
  + \ \text{tensors of order $>3$}. 
\end{align}

To begin with, note that $(S^+)^{2s}$ and $(S^-)^{2s}$ are up to
a sign equal to the tensor components with $m=\pm 2s$ of one and the same
tensor of order $2s$,
% \begin{gather}
%   {T^{(2s)}}^{m=2s} = (-1)^{2s}\,(S^+)^{2s},
%   \nonumber\\[0.2\baselineskip]
%   {T^{(2s)}}^{m=-2s} = (S^-)^{2s}.
% \end{gather}
% \begin{align}
%   &{T^{(2s)}}^{m=2s} = (-1)^{2s}\,(S^+)^{2s},
%   \nonumber\\[0.2\baselineskip]
%   &{T^{(2s)}}^{m=-2s} = (S^-)^{2s}.
% \end{align}
\begin{align}
  \label{eq:c:T^(2s)^2s}
  \begin{split}
    &{T^{(2s)}}^{2s} = (-1)^{2s}\,(S^+)^{2s},
%    &{T^{(2s)}}^{m=2s} = (-1)^{2s}\,(S^+)^{2s},
    \\[0.2\baselineskip]
    &{T^{(2s)}}^{-2s} = (S^-)^{2s}.
%    &{T^{(2s)}}^{m=-2s} = (S^-)^{2s}.
  \end{split}
\end{align}
%Using 
%With 
Recalling \eqref{eq:a:Tm1m2CG}, we write
\begin{align}
  \label{eq:c:T2s-2sCG}
%  {{T^{(2s)}}^{(2s)}}^{m=2s}{{T^{(2s)}}^{(2s)}}^{m=-2s}
  {T^{(2s)}}^{2s}\,{T^{(2s)}}^{-2s}
  &=\sum_{j=1}^{4s} {T^{(j)}}^{0} \braket{j,0}{2s,2s;2s,-2s},
\end{align}
where ${T^{(j)}}^{0}$ is with \eqref{eq:a:TjmCG} given by
\begin{align} 
  \label{eq:c:TjmCG}
  {T^{(j)}}^{0}
  &=\sum_{m=-2s}^{2s} {T^{(2s)}}^{m}\, {T^{(2s)}}^{-m} \braket{2s,m;2s,-m}{j,0}.
\end{align}
With \eqref{eq:a:commJ+-T}, we can calculate the components ${T^{(2s)}}^{m}$
from ${T^{(2s)}}^{\pm 2s}$,
\begin{align}
  \label{eq:c:commJ+-T}
  {T^{(2s)}}^{m\pm 1}
  =\frac{1}{\sqrt{2s(2s+1)-m(m\pm 1)}}\,\Bigcomm{S^\pm}{{T^{(2s)}}^m}.
\end{align}

Specifically, ${T^{(2s)}}^{2s-n}$ is given in terms of ${T^{(2s)}}^{2s}$
by
\renewcommand{\struta}{\rule[-10pt]{0pt}{0pt}}
\renewcommand{\strutb}{\rule[8pt]{0pt}{0pt}}
\begin{align}
  \label{eq:c:T^{2s-n}}
  {T^{(2s)}}^{2s-n}
  &=\left(\,\prod_{i=1}^n \frac{1}{\sqrt{2s(2s+1)-(2s-i+1)(2s-i)}}\,\right)
  \\[0.5\baselineskip]
  &\quad\cdot\hspace{3pt}\underbrace{\hspace{-3pt}
  \Bigcomm{S^-}{\Bigcomm{S^-}{\ldots\Bigcomm{S^-}{{T^{(2s)}}^{2s}}\ldots}}
  \hspace{-3pt}\struta}_{\textstyle n\ \text{operators}\ S^-\strutb}
  \hspace{3pt}.
\end{align}
% With 
% \begin{align}
%   \prod_{i=1}^n \big(2s(2s+1)-(2s-i+1)(2s-i)\big)
%   &=\frac{(4s)!\cdot n!}{(4s-n)!},
% \end{align}
To evaluate the first term, we use
\begin{align}
  \label{eq:c:(S^+)^n(S^-)^nformula}
%  \prod_{i=1}^n \Big(l(l+1)-(l-i+1)(l-i)\Big)
%  &=\frac{(2l)!\cdot n!}{(2l-n)!},
  \prod_{i=1}^n \Big(s(s+1)-(s-i+1)(s-i)\Big)
  &=\prod_{i=1}^n (2s-i+1) i 
   =\frac{(2s)!\cdot n!}{(2s-n)!},
\end{align}
which holds for $1\le n\le 2s$, $2s$ and $n$ integer.
% , and is readily verified by evaluation of
% $\bra{s,s}(S^+)^n(S^-)^n\ket{s,s}$ once with
% \eqref{eq:app-am:Jpmketlm}, and once with Schwinger bosons
% \eqref{eq:schw}.
This yields
\begin{align}
  \label{eq:c:T^{2s-n}v2}
  {T^{(2s)}}^{2s-n}
  &=\sqrt{\frac{(4s-n)!}{(4s)!\cdot n!}}\;
%  \\[0.5\baselineskip]
%  &\quad\cdot
  \sum_{k=0}^n\binom{n}{k} (-1)^{n-k}\, (S^-)^k\, {T^{(2s)}}^{2s}\, (S^-)^{n-k}.
\end{align}
Similarly, we find
\begin{align}
  \label{eq:c:T^{-2s+n}v2}
  {T^{(2s)}}^{-2s+n}
  &=\sqrt{\frac{(4s-n)!}{(4s)!\cdot n!}}\;
%  \\[0.5\baselineskip]
%  &\quad\cdot
  \sum_{k=0}^n\binom{n}{k} (-1)^{k}\, (S^+)^{n-k}\, {T^{(2s)}}^{-2s}\, (S^+)^{k}.
\end{align}
Note that \eqref{eq:c:T^{2s-n}v2} and \eqref{eq:c:T^{-2s+n}v2} hold
for $0\le n\le 4s$.  With \eqref{eq:c:T^(2s)^2s} and the shorthand
$\ket{m}\equiv\ket{s,m}$, we can write
\begin{align*}
  \hspace{15pt}&\hspace{-15pt}(S^-)^k\, {T^{(2s)}}^{2s}\, (S^-)^{n-k}
  \nonumber\\[0.2\baselineskip]
  &=(-1)^{2s}\,\ket{s-k}\bra{-s+n-k}
  \nonumber\\[0.2\baselineskip]
  &\quad\cdot\bra{s-k}(S^-)^k\ket{s}\bra{s}(S^+)^{2s}\ket{-s}
   \bra{-s}(S^-)^{k-n}\ket{-s+n-k},
\end{align*}
and similarly
\begin{align*}
  \hspace{15pt}&\hspace{-15pt}(S^+)^{n-k}\, {T^{(2s)}}^{-2s}\,(S^+)^k 
  \nonumber\\[0.2\baselineskip]
  &=(-1)^{2s}\,\ket{-s+n-k}\bra{s-k}
  \nonumber\\[0.2\baselineskip]
  &\quad\cdot\bra{-s+n-k}(S^+)^{n-k}\ket{-s}\bra{-s}(S^-)^{2s}\ket{s}
   \bra{s}(S^+)^k\ket{s-k}.
\end{align*}
This implies that in the product
${T^{(2s)}}^{2s-n}{T^{(2s)}}^{-2s+n}$, only terms with matching values
of $k$ in the sums \eqref{eq:c:T^{2s-n}v2} and
\eqref{eq:c:T^{-2s+n}v2} contribute.  With
\eqref{eq:c:(S^+)^n(S^-)^nformula} we obtain
\begin{align}
  \bra{s}(S^+)^k(S^-)^k\ket{s}&=\left\{
    \begin{alignedat}{2}
      &\frac{(2s)!\cdot k!}{(2s-k)!}&\quad&\text{for}\ 0\le k\le 2s,
      \\[.2\baselineskip] 
      &\, 0&&\text{otherwise},
    \end{alignedat}\right.
  \nonumber\\[0.6\baselineskip]
  \bra{-s}(S^-)^{k-n}(S^+)^{n-k}\ket{-s}&=\left\{
    \begin{alignedat}{2}
      &\frac{(2s)!\cdot (n-k)!}{(2s-n+k)!}&\quad&\text{for}\ 0\le n-k\le 2s,
      \\[.2\baselineskip] 
      &\, 0&&\text{otherwise},
    \end{alignedat}\right.
  \nonumber\\[0.6\baselineskip]
  \bra{s}(S^+)^{2s}(S^-)^{2s}\ket{s}&={(2s)!}^2.
\end{align}
This yields
\begin{align}
  \label{eq:c:T^{2s-n}T^{-2s+n}}
  \hspace{10pt}&\hspace{-10pt} {T^{(2s)}}^{2s-n}\,{T^{(2s)}}^{-2s+n}
  \nonumber\\[0.4\baselineskip]
  &={(2s)!}^2\,(-1)^{2s+n}\, 
  \nonumber\\*[0.2\baselineskip]
  &\quad\cdot
  \frac{(4s-n)!}{(4s)!\cdot n!}\,
  \sum_{k=\max(n-2s,0)}^{\min(n,2s)}\! \binom{n}{k}\binom{n}{k}
  \frac{(2s)!\cdot k!}{(2s-k)!}\,\frac{(2s)!\cdot (n-k)!}{(2s-n+k)!}
  \ket{s-k}\bra{s-k}
  \nonumber\\[0.4\baselineskip]
  &={(2s)!}^2\,(-1)^{2s+n}\, {\binom{4s}{n}}^{\! -1}
%  \nonumber\\[0.2\baselineskip]
%  &\quad\cdot
  \sum_{k=\max(n-2s,0)}^{\min(n,2s)} 
  \binom{2s}{k}\binom{2s}{n-k} \ket{s-k}\bra{s-k}.
\end{align}
%
%In order to evaluate ${T^{(j)}}^{m=0}$, we substitute into
%\eqref{eq:c:T^{2s-n}T^{-2s+n}} into \eqref{eq:c:TjmCG}.  This yields
Substitution into \eqref{eq:c:TjmCG} yields
\renewcommand{\struta}{\rule[-7pt]{0pt}{0pt}}
\renewcommand{\strutb}{\rule[11pt]{0pt}{0pt}}
\begin{align} 
  \label{eq:c:Tjm1}
  {T^{(j)}}^{0}
  &=\sum_{n=0}^{4s} {T^{(2s)}}^{2s-n}\,{T^{(2s)}}^{-2s+n}\,
  \underbrace{\!\braket{2s,2s-n;2s,-2s+n}{j,0}\!\struta}_{
    \textstyle\equiv C^{2s-n}_j\strutb}.
\end{align}
%\newpage
With
\begin{align*}
  \sum_{n=0}^{4s}\ \sum_{k=\max(n-2s,0)}^{\min(n,2s)}
  =\sum_{k=0}^{2s}\,\sum_{n=k}^{2s+k},
\end{align*}
we obtain
\begin{align} 
  \label{eq:c:Tjm2}
  {T^{(j)}}^{0}
  &={(2s)!}^2\,(-1)^{2s}\sum_{k=0}^{2s}\left\{
    \sum_{n=k}^{2s+k}C^{2s-n}_j(-1)^n 
    \frac{\displaystyle\binom{2s}{k}\binom{2s}{n-k}}{\displaystyle\binom{4s}{n}}
  \right\}\ket{s-k}\bra{s-k}
  \nonumber\\[0.6\baselineskip]
  &={(2s)!}^2\,(-1)^{2s}\sum_{k=0}^{2s}\left\{
    \sum_{p=0}^{2s}C^{2s-k-p}_j(-1)^{k+p} 
    \frac{\displaystyle\binom{2s}{k}\binom{2s}{p}}{\displaystyle\binom{4s}{k+p}}
  \right\}\ket{s-k}\bra{s-k}.
%   \nonumber\\[0.6\baselineskip]
%   &={(2s)!}^2\sum_{m=-s}^{s}\left\{
%     \sum_{l=-s}^{s}C^{m+l}_j(-1)^{m+l} 
%     \frac{\displaystyle\binom{2s}{s-m}\binom{2s}{s-l}}{
%       \displaystyle\binom{4s}{2s-m-l}}
%   \right\}\ket{m}\bra{m}.
\end{align}
The individual tensors in the decomposition 
\begin{align}
  \label{eq:c:S+^2sS-^2sjsum}
  (S^+)^{2s}(S^-)^{2s}
  &=\sum_{j=1}^{4s}\Big\{(S^+)^{2s}(S^-)^{2s}\Big\}_j
\end{align}
are hence with \eqref{eq:c:T^(2s)^2s}, \eqref{eq:c:T2s-2sCG}, and the
definition of $C^{2s-n}_j$ in \eqref{eq:c:Tjm1} given by
\begin{align}
  \label{eq:c:S+^2sS-^2s_j}
  \Big\{(S^+)^{2s}(S^-)^{2s}\Big\}_j
  &=(-1)^{2s}\,C^{2s}_j\,{T^{(j)}}^{0}
  \nonumber\\[0.0\baselineskip]
%  &=\frac{{(2s)!}^2}{2s+1}\,\sum_{m=-s}^{s}\Gamma^{2s,m}_j\, \ket{m}\bra{m}
  &=\frac{{(2s)!}^2}{2s+1}\,\sum_{k=0}^{2s}P^k_j\, \ket{s-k}\bra{s-k},
\end{align}
where we have defined
\begin{align}
  \label{eq:c:P^k_jdef}
  P^k_j&=(2s+1)\,C^{2s}_j
  \sum_{p=0}^{2s}C^{2s-k-p}_j(-1)^{k+p} 
  \frac{\displaystyle\binom{2s}{k}\binom{2s}{p}}{\displaystyle\binom{4s}{k+p}}.
\end{align}
% 
% \begin{align}
% %  \nonumber\\[0.0\baselineskip]
%   \Gamma^{2s,m}_j
%   &=(2s+1)\,C^{2s}_j
%     \sum_{l=-s}^{s}C^{m+l}_j(-1)^{2s-m-l} 
%     \frac{\displaystyle\binom{2s}{s-m}\binom{2s}{s-l}}{
%       \displaystyle\binom{4s}{2s-m-l}}
% \end{align}
We are not aware of any method to evaluate this sum analytically.  We
have used \emph{Mathematica} to evaluate it for $k=0$ and $j=0,1,2,3$
as a function of $s$, and then obtain the coefficients in the
expansion \eqref{eq:c:S+^2sS-^2s} from these terms.

With the Clebsch--Gordan coefficients
% ORIGINAL FROM MATHEMATICA
% \begin{align}
%   C^{2s,2s,m}_0&=\frac{1}{\sqrt{4 s+1}} (-1)^{2 s-m}
%   \nonumber\\[0.2\baselineskip]
%   C^{2s,2s,m}_1&=\sqrt{\frac{3}{2}} m 
%   \frac{1}{\sqrt{s (1 + 2 s) (1 + 4 s)}} (-1)^{2 s-m}
%   \nonumber\\[0.2\baselineskip]
%   C^{2s,2s,m}_2&=\sqrt{\frac{5}{2}} 
%   \sqrt{\frac{1}{s (1 + 2 s) (-1 + 4 s) (1 + 4 s) (3 + 4 s)}}
%   \left(-(-1)^{2 s-m}\right)
%   \left(-3 m^2+4 s^2+2 s\right)
%   \nonumber\\[0.2\baselineskip]
%   C^{2s,m}_3&=-\frac{1}{2} \sqrt{7} m 
%   \sqrt{\frac{1}{s (1 + s)(-1 + 2 s)(1 + 2 s)(-1 + 4 s)(1 + 4 s)(3 + 4 s)}}
%   \nonumber\\[0.2\baselineskip]
%   &\quad (-1)^{2 s-m} \left(-5 m^2+12 s^2+6 s-1\right)
% \end{align}
\begin{align*}
  C^{m}_0&=\frac{(-1)^{2 s-m}}{\sqrt{4 s+1}}, 
  \nonumber\\[0.2\baselineskip]
  C^{m}_1&=\frac{\sqrt{3}\,(-1)^{2 s-m}\cdot m}{\sqrt{2s (2 s + 1) (4 s + 1)}},
  \nonumber\\[0.2\baselineskip]
  C^{m}_2&=
  \frac{\sqrt{5}\,(-1)^{2 s-m}\cdot\left(3 m^2- 2s (2 s + 1)\right)}{
    \sqrt{2s (2 s + 1) (4 s - 1) (4 s + 1) (4 s + 3)}},
  \nonumber\\[0.2\baselineskip]
  C^{m}_3&=
  \frac{\sqrt{7}\,(-1)^{2 s-m}\cdot m\left(5 m^2+1 - 6s(2s + 1)\right)}{
    2\sqrt{s (s + 1) (2 s - 1) (2 s + 1) (4 s - 1) (4 s + 1) (4 s + 3)}},
\end{align*}
we find
\begin{align}
  \label{eq:c:P^0_j}
  P^0_0&=1,
  \nonumber\\*[0.2\baselineskip]
  P^0_1&=\frac{3s}{s+1},
  \nonumber\\*[0.2\baselineskip]
  P^0_2&=\frac{5s(2s-1)}{(s+1)(2s+3)},
  \nonumber\\*[0.2\baselineskip]
  P^0_3&=\frac{7s(2s-1)(s-1)}{(s+1)(2s+3)(s+2)}.
\end{align}
Comparing \eqref{eq:c:S+^2sS-^2s_j} with \eqref{eq:c:P^0_j} to the
coefficients of $\ket{s}\bra{s}$ we obtain from \eqref{eq:c:V^0},
\eqref{eq:c:T^0}, and \eqref{eq:c:W^0},
\begin{align}
  V^0&= s\ket{s}\bra{s}+\ldots,
  \nonumber\\*[0.2\baselineskip]
  T^0&= \frac{2}{\sqrt{6}}\,s(2s-1)\ket{s}\bra{s}+\ldots,
  \nonumber\\*[0.2\baselineskip]
  W^0&= \frac{2}{\sqrt{5}}\,s(2s-1)(s-1)\ket{s}\bra{s}+\ldots,
\end{align}
% \begin{align}
%   V^0&= s\ket{s}\bra{s}+(s-1)\ket{s-1}\bra{s-1}\ldots,
%   \nonumber\\*[0.2\baselineskip]
%   T^0&= \frac{2}{\sqrt{6}}\Big[s(2s-1)\ket{s}\bra{s}
%   +(2s-1)(s-3)\ket{s-1}\bra{s-1}+\ldots\Big],
%   \nonumber\\*[0.2\baselineskip]
%   W^0&= \frac{2}{\sqrt{5}}\Big[s(2s-1)(s-1)\ket{s}\bra{s}
%   +s(2s-1)(s-6)s\ket{s-1}\bra{s-1}+\ldots\Big],
% \end{align}
we obtain
\begin{align}
  \label{eq:c:a_0abd} 
  a_0&=\frac{{(2s)!}^2}{2s+1},
  \nonumber\\*[0.2\baselineskip]
  a&=\frac{3}{s+1},
  \nonumber\\*[0.2\baselineskip]
  b&=\frac{\sqrt{6}}{2}\frac{5}{(s+1)(2s+3)},
  \nonumber\\*[0.2\baselineskip]
  c&=\frac{\sqrt{5}}{2}\,\frac{7}{(s+1)(2s+3)(s+2)}
\end{align}
for the coefficients in the expansion \eqref{eq:c:S+^2sS-^2s}.

% \newpage
% \subsection{Time reversal, parity, and spin rotation symmetry}
\subsection{Time reversal and parity symmetry}

The for the scalar and vector component relevant part of the operator
$H_0$ introduced \eqref{eq:c:h0} is with \eqref{eq:a:omegaabc} and 
\eqref{eq:c:S+^2sS-^2s}
%, \eqref{eq:c:a_0abd}, and \eqref{eq:c:Sb+Sc-} 
given by
\begin{align}
  \label{eq:c:h0'}
  H_0'
  &=\frac{1}{2}\sum_{\substack{\a,\b,\c\\ \a\ne \b,\c}}\omega_{\a\b\c}\, 
  \Big\{ 1 + a\, V^0_{\a} + b\,T^0_{\a\a}  + c\,W^0_{\a\a\a}  \Big\} \Sb^+\Sc^-.
%   \nonumber\\*[-0.5\baselineskip]&\quad\hspace{60pt} 
%   \Big\{ \frac{2}{3}\bSb\bSc - \i(\bSb\times\bSc)^\z
%     -\frac{1}{\sqrt{6}}\,T_{\b\c}^0 \Big\} 
%   \nonumber\\*[0.5\baselineskip]
%   &=H_0'^= + H_0'^\ne
\end{align}
From now on, we omit the prime.  
With the transformation properties under time reversal,
\begin{align*}
  \text{T:}\quad
%  \i \,\to\, \Pi\i\Pi = -\i,\quad
%  \i\to -\i,\quad
  \ea  \,\to\, \Theta\ea \Theta = \eab,\quad 
  \bsS \,\to\, \Theta\bsS\Theta = -\bsS, 
\end{align*}
and hence
\begin{gather*}
  \omega_{\a\b\c}\to\omega_{\a\c\b},\quad
  S^+\to -S^-,\quad S^-\to -S^+,\quad S^\z\to -S^\z,
  \nonumber\\[0.2\baselineskip] 
  V^0\to -V^0,\quad T^0\to T^0,\quad W^0\to -W^0,
\end{gather*}
the operator \eqref{eq:c:h0'} transforms into
\begin{align}
  \label{eq:c:Th0T}
  \Theta H_0\Theta & =\frac{1}{2}\sum_{\substack{\a,\b,\c\\ \a\ne \b,\c}}\omega_{\a\b\c}\, 
  \Big\{ 1 - a\, V^0_{\a} + b\,T^0_{\a\a}  - c\,W^0_{\a\a\a}  \Big\} \Sc^-\Sb^+,
\end{align}
We proceed with the T invariant operator
\begin{align}
  \label{eq:c:h0T}
  H_0^{\text{T}}=\frac{1}{2}\left(H_0+\Theta H_0 \Theta\right)
  = H_0^{\text{T}=}+H_0^{\text{T}\ne },
\end{align}
where
\begin{align}
  \label{eq:c:h0T=}
  H_0^{\text{T}=}&=\frac{1}{2}\sum_{\substack{\a,\b\\ \a\ne \b}}\,\omega_{\a\b\b}
  \left[\big(1 + b\,T^0_{\a\a}\big)\frac{1}{2}\biganticomm{\Sb^+}{\Sb^-}
    +\big(a\,V^0_{\a} + c\,W^0_{\a\a\a}\big)
      \frac{1}{2}\bigcomm{\Sb^+}{\Sb^-}\right]
  \nonumber\\[0.2\baselineskip] 
  &=\frac{1}{2}\sum_{\substack{\a\ne \b}}\omega_{\a\b\b}
  \Bigg[ \big(1 + b\,T^0_{\a\a}\big)
         \left(\frac{2s(s+1)}{3}-\frac{1}{\sqrt{6}}\,T_{\b\b}^0\right)
  \nonumber\\ %[-0.2\baselineskip] 
  &\hspace{140pt}     + \big(a\,\Sa^\z + c\,W^0_{\a\a\a}\big) \Sb^\z \Bigg]
  \\[0.4\baselineskip] 
  \label{eq:c:h0Tnot=}
  H_0^{\text{T}\ne} 
  &=\frac{1}{2}\sum_{\substack{\a,\b,\c\\ \a\ne\b\ne\c\ne\a}}\omega_{\a\b\c}
  \,\big(1 + b\,T^0_{\a\a}\big)\, \Sb^+\Sc^-.
  \nonumber\\[0.2\baselineskip] 
  &=\frac{1}{2}\sum_{\substack{\a,\b,\c\\ \a\ne\b\ne\c\ne\a}}\omega_{\a\b\c}
  \,\big(1 + b\,T^0_{\a\a}\big) 
  \left(\frac{2}{3}\bSb\bSc - \text{i}(\bSb\times\bSc)^\z
    -\frac{1}{\sqrt{6}}\,T_{\b\c}^0\right).
\end{align}
With the transformation properties under parity,
\begin{align}
  \text{P:}\quad
%  \i \,\to\, \Pi\i\Pi = \i,\quad
  \ea  \,\to\, \Pi\ea\Pi = \eab,\quad 
  \bsS \,\to\, \Theta\bsS\Theta = \bsS,  
\end{align}
and hence $\omega_{\a\b\c}\to\omega_{\a\c\b}$, we obtain the P and T
invariant operator
\begin{align}
  \label{eq:c:h0PT}
  H_0^{\text{PT}}=\frac{1}{2}\left(H_0^{\text{T}}
    +\Pi\hspace{1pt} H_0^{\text{T}}\hspace{1pt} \Pi\right)
  = H_0^{\text{PT}=}+H_0^{\text{PT}\ne },
\end{align}
where
\begin{align}
  \label{eq:c:h0PT=}
  H_0^{\text{PT}=}&=H_0^{\text{T}=},
  \\[0.4\baselineskip] 
  \label{eq:c:h0PTnot=}
  H_0^{\text{PT}\ne}
  &=\frac{1}{2}\sum_{\substack{\a,\b,\c\\ \a\ne\b\ne\c\ne\a}}\omega_{\a\b\c}
  \,\big(1 + b\,T^0_{\a\a}\big)\, 
  \left(\frac{2}{3}\bSb\bSc -\frac{1}{\sqrt{6}}\,T_{\b\c}^0\right).
\end{align}

\subsection{Spin rotation symmetry}

Since the critical spin liquid state $\ket{\psi^S_0}$ introduced in 
Sections \ref{sec:naarbs} and \ref{sec:c:genbyGutz} is a spin
singlet, the property that it is annihilated by \eqref{eq:c:h0PT}
with \eqref{eq:c:h0PT=} and \eqref{eq:c:h0PTnot=} implies that it is 
annihilated by each tensor component of \eqref{eq:c:h0PT} individually.

Since we wish to construct a Hamiltonian which is invariant under 
SU(2) spin rotations, we proceed by projecting out the scalar component.
This yields
\begin{align}
  \label{eq:c:h0PT=0}
  \big\{H_0^{\text{PT}=}\big\}_0
  &=\frac{1}{2}\sum_{\substack{\a\ne \b}}\omega_{\a\b\b}
  \left[\frac{2s(s+1)}{3} 
    - \frac{b}{\sqrt{6}}\left\{T_{\a\a}^0 T_{\b\b}^0\right\}_0   
    + \frac{a}{3}\bSa\bSb \right],
  \\[0.4\baselineskip] 
  \label{eq:c:h0PTnot=0}
  \big\{H_0^{\text{PT}\ne}\big\}_0
  &=\frac{1}{2}\sum_{\substack{\a,\b,\c\\ \a\ne\b\ne\c\ne\a}}\omega_{\a\b\c} 
  \left[\frac{2}{3}\bSb\bSc 
    - \frac{b}{\sqrt{6}}\left\{T_{\a\a}^0 T_{\b\c}^0\right\}_0\right].
\end{align}
With \eqref{eq:b:5TaaTbc_0}, or specifically
\begin{align*}
%  \label{eq:c:5TaaTbc_0=}
  5\,\big\{T_{\a\a}^0T_{\b\b}^0 \big\}_0 
  &= -\frac{4}{3} s^2(s+1)^2 + 4(\bSa\bSb)^2 + 2\bSa\bSb,\quad \a\ne\b,
\end{align*}
and \eqref{eq:c:a_0abd}, we obtain
\begin{align*}
  -\frac{b}{\sqrt{6}}\left\{T_{\a\a}^0 T_{\b\b}^0\right\}_0
%   &=\frac{1}{(s+1)(2s+3)}
%   \left(\frac{2}{3} s^2(s+1)^2-\bSa\bSb-2(\bSa\bSb)^2\right)
%   \nonumber\\[0.2\baselineskip] 
  &=\frac{2s^2(s+1)}{3\,(2s+3)}
  -\frac{\bSa\bSb+2(\bSa\bSb)^2}{(s+1)(2s+3)}, 
  % \left(\bSa\bSb+2(\bSa\bSb)^2\right),
  \\[0.2\baselineskip] 
  \frac{a}{3}\bSa\bSb &= \frac{\bSa\bSb}{(s+1)},
\end{align*}
and hence
\begin{align}
  \label{eq:c:h0PT=0v1}
  \big\{H_0^{\text{PT}=}\big\}_0 
  &=\sum_{\substack{\a\ne\b}}\omega_{\a\b\b} 
  \frac{1}{2s+3}\,\bigg[s(s+1)^2+\bSa\bSb-\frac{(\bSa\bSb)^2}{(s+1)}\bigg].
\end{align}
% To evaluate \eqref{eq:c:h0PTnot=0}
Similarly, we use
\begin{align*}
%  \label{eq:b:5TaaTbc_0not=}
  5\,\big\{T_{\a\a}^0T_{\b\c}^0 \big\}_0 
%  \hspace{0pt}&\hspace{0pt}
  = -\frac{4s(s+1)}{3} \bSb\bSc
  + 2\,\bigl[(\bSa\bSb)(\bSa\bSc)+& (\bSa\bSc)(\bSa\bSb)\bigr],
  \nonumber\\[0\baselineskip] 
  &\a\ne\b\ne\c\ne\a,
\end{align*}
to obtain
\begin{align*}
  - \frac{b}{\sqrt{6}}\left\{T_{\a\a}^0 T_{\b\c}^0\right\}_0
  &=\frac{2s\,\bSb\bSc}{3\,(2s+3)}
  + \frac{(\bSa\bSb)(\bSa\bSc)+(\bSa\bSc)(\bSa\bSb)}{(s+1)(2s+3)}
\end{align*}
and hence
\begin{align}
  \label{eq:c:h0PTnot=0v1}
  \hspace{6pt}&\hspace{-6pt}\big\{H_0^{\text{PT}\ne}\big\}_0 
  \nonumber\\[0.2\baselineskip] 
  &=\hspace{-3pt}\sum_{\substack{\a,\b,\c\\ \a\ne\b\ne\c\ne\a}}\hspace{-3pt}
  \omega_{\a\b\c}\frac{1}{2s+3}\, 
  \bigg[(s+1)\bSb\bSc
   -\frac{(\bSa\bSb)(\bSa\bSc)+(\bSa\bSc)(\bSa\bSb)}{2(s+1)}\bigg].
\end{align}
With \eqref{eq:app-hs15}, we rewrite the first sum in
\eqref{eq:c:h0PTnot=0v1} as
\begin{align*}
  \hspace{-5pt}\sum_{\substack{\a,\b,\c\\ \a\ne\b\ne\c\ne\a}}\hspace{-5pt}
  \omega_{\a\b\c}\bSb\bSc 
  &=2\sum_{\substack{\a\ne \b}}\omega_{\a\b\b} \bSa\bSb
  -\frac{1}{2}\bsS_{\text{tot}}^2 + \frac{s(s+1)}{2}\,N.
\end{align*}
Collecting all the terms we obtain
\begin{align}
  \label{eq:c:h0PT0}
  \big\{H_0^{\text{PT}}\big\}_0
  &=\sum_{\substack{\a\ne\b}}\omega_{\a\b\b} 
  \bigg[\bSa\bSb - \frac{(\bSa\bSb)^2}{(s+1)(2s+3)}\bigg]
  \nonumber\\[0.2\baselineskip] 
  &\quad 
  +\hspace{-3pt}\sum_{\substack{\a,\b,\c\\ \a\ne\b\ne\c\ne\a}}\hspace{-3pt}
  \omega_{\a\b\c}\, 
  \frac{(\bSa\bSb)(\bSa\bSc)+(\bSa\bSc)(\bSa\bSb)}{2(s+1)(2s+3)}
  \nonumber\\[0.2\baselineskip] 
  &\quad - \frac{s+1}{2(2s+3)}\bsS_{\text{tot}}^2 
  + \frac{s(s+1)^2}{2s+3}\,\frac{N (N^2+5)}{12}.
\end{align}
Note that the second term in the first line of \eqref{eq:c:h0PT0}
is equal to what we would get if we were to take $\b=\c$ on the 
term in the second line.

The spin $S$ spin liquid state $\ket{\psi^{S}_0}$ introduced in 
Sections \eqref{sec:naarbs} and \eqref{sec:c:genbyGutz} is hence
an exact eigenstate of 
\begin{align}
  \label{eq:c:h}
  H^{S} =\frac{2\pi^2}{N^2}
  \Bigg[
  &\sum_{\substack{\a\ne\b}} \frac{\bSa\bSb}{\vert\ea-\eb\vert^2}
  \nonumber\\[0.2\baselineskip] 
  &  -\frac{1}{2(s+1)(2s+3)}\sum_{\substack{\a,\b,\c\\ \a\ne\b,\c}}
  \frac{(\bSa\bSb)(\bSa\bSc) + (\bSa\bSc)(\bSa\bSb)}{(\eab-\ebb)(\ea-\ec)}
  \Bigg]
  \nonumber\\[0.4\baselineskip] 
  =\frac{2\pi^2}{N^2}
  \Bigg[
  &\sum_{\substack{\a\ne\b}} \frac{\bSa\bSb}{\vert\ea-\eb\vert^2}
%  \nonumber\\[0.2\baselineskip] 
%  &  
  -\sum_{\substack{\a,\b,\c\\ \a\ne\b,\c}}
  \Re\bigg\{\frac{1}{(\eab-\ebb)(\ea-\ec)}\bigg\}
%  \Re\bigg(\frac{1}{(\eab-\ebb)(\ea-\ec)}\bigg)
  \frac{(\bSa\bSb)(\bSa\bSc)}{(s+1)(2s+3)}
  \Bigg],
\end{align}
where $\Re$ denotes the real part.  The energy eigenvalue is given by
\begin{align}
  \label{eq:c:E_0}
  E_0^{S} &=-\frac{2\pi^2}{N^2}\frac{s(s+1)^2}{2s+3}\,\frac{N (N^2+5)}{12}
%  \nonumber\\[0.2\baselineskip] &
  = -\frac{\pi^2}{6}\frac{s(s+1)^2}{2s+3}\left(N+\frac{5}{N}\right).
\end{align}
This is the main result of this work.  We will show in Section
\ref{sec:c:factorization} that $\ket{\psi^{S}_0}$ is also a ground
state of \eqref{eq:c:h0PT0}, \ie that all the eigenvalues of
$H^{S}-E_0^{S}$ are non-negative.  Exact diagonalization
studies~\cite{manuscriptinpreparationTRSG11} carried out numerically
for up to $N=16$ sites for the $S=1$ model and for up to $N=10$ sites
for the $S=\frac{3}{2}$ model further show that
$\bigket{\psi^{S=1}_0}$ and $\bigket{\psi^{S=\frac{3}{2}}_0}$ are the
unique ground states of \eqref{eq:c:h}, and that the models are
gapless.  We assume this property to hold for general spin $S$.

\section{Vector annihilation operators}
\label{sec:c:vec}

\subsection{Annihilation operators which transform even under T}
\label{sec:c:vecTeven}

We can use the \defining condition \eqref{eq:c:Omegadef} further to
construct a vector annihilation operator.  First note that since
\begin{equation*}
    \OaS \ket{\psi^{S}_{0}} = 0 \quad\forall\, \alpha,
\end{equation*}
$\ket{\psi^{S}_{0}}$ is also annihilated by the Hermitian operator
\begin{align}
  \label{eq:c:ha}
  H_{\a}=\frac{1}{2a_0} {\OaS}^\dagger\OaS
  &=\frac{1}{2a_0}\sum_{\substack{\b,\c\\ \a\ne \b,\c}}\omega_{\a\b\c}\, 
  (\Sa^+)^{2s}(\Sa^-)^{2s}\Sb^+\Sc^-,
\end{align}
and therefore also by the scalar and the vector components of 
\begin{align}
  \label{eq:c:ha'}
  H_a'
  &=\frac{1}{2}\sum_{\substack{\a,\b,\c\\ \a\ne \b,\c}}\omega_{\a\b\c}\, 
  \Big\{ 1 + a\, V^0_{\a} + b\,T^0_{\a\a}  + c\,W^0_{\a\a\a}  \Big\} \Sb^+\Sc^-,
\end{align}
which is just the operator \eqref{eq:c:h0'} without the sum over $\a$.
From now on, we omit the prime.  
Constructing an operator which is even under T,
\begin{equation}
  \label{eq:c:haT}
  H_\a^{\text{T}}=\frac{1}{2}\left(H_\a+\Theta H_\a \Theta\right)
  = H_\a^{\text{T}=}+H_\a^{\text{T}\ne },
\end{equation}
where
\begin{align}
  \label{eq:c:h0T=v1}
  H_\a^{\text{T}=}
%   &=\frac{1}{2}\sum_{\substack{\b\\ \b\ne\a}}\omega_{\a\b\b}
%   \left[\big(1 + b\,T^0_{\a\a}\big)\frac{1}{2}\biganticomm{\Sb^+}{\Sb^-}
%     +\big(a\,V^0_{\a} + c\,W^0_{\a\a\a}\big)
%       \frac{1}{2}\bigcomm{\Sb^+}{\Sb^-}\right]
%   \nonumber\\[0.2\baselineskip] 
  &=\frac{1}{2}\sum_{\substack{\b\\ \b\ne\a}}\omega_{\a\b\b}
  \Bigg[ \big(1 + b\,T^0_{\a\a}\big)
         \left(\frac{2s(s+1)}{3}-\frac{1}{\sqrt{6}}\,T_{\b\c}^0\right)
  \nonumber\\[-0.4\baselineskip] 
  &\hspace{140pt}     + \big(a\,\Sa^\z + c\,W^0_{\a\a\a}\big) \Sb^\z \Bigg],
  \\[0.4\baselineskip] 
  \label{eq:c:h0Tnot=v1}
  H_\a^{\text{T}\ne} 
%   &=\frac{1}{2}\sum_{\substack{\b\ne\c\\ \b,\c\ne\a}}\omega_{\a\b\c}
%   \,\big(1 + b\,T^0_{\a\a}\big)\, \Sb^+\Sc^-.
%   \nonumber\\[0.2\baselineskip] 
  &=\frac{1}{2}\sum_{\substack{\b\ne\c\\ \b,\c\ne\a}}\omega_{\a\b\c}
  \,\big(1 + b\,T^0_{\a\a}\big) 
  \left(\frac{2}{3}\bSb\bSc - \text{i}(\bSb\times\bSc)^\z
    -\frac{1}{\sqrt{6}}\,T_{\b\c}^0\right),
\end{align}
and odd under P, we obtain 
\begin{align}
  \label{eq:c:habarPT}
  H_\a^{\rm {\bar P}T}=\frac{1}{2}
  \left(H_\a^{\text{T}} - \Pi\hspace{1pt} H_\a^{\text{T}}\hspace{1pt} \Pi\right)
  = H_\a^{{\rm {\bar P}T}=} + H_\a^{{\rm {\bar P}T}\ne},
\end{align}
% \newpage
where
\begin{align}
  \label{eq:c:habarPT1}
  H_\a^{{\rm {\bar P}T}=}&=0\hspace{1pt}, 
  \hspace{-6pt}&\hspace{-6pt}
%  \\[0.4\baselineskip] 
%  \label{eq:b:h0Tn}
  H_\a^{{\rm {\bar P}T}\ne}&=
  -\frac{\i}{2}\sum_{\substack{\b\ne\c\\ \b,\c\ne\a}}\omega_{\a\b\c}
  \,\big(1 + b\,T^0_{\a\a}\big) (\bSb\times\bSc)^\z.
\end{align}
With \eqref{eq:a:omega_abc-omega_acb}, we obtain
\begin{align}
  \label{eq:c:habarPT2}
  H_\a^{{\rm {\bar P}T}}
  &=\frac{\i}{4}\sum_{\substack{\b\ne\c\\ \b,\c\ne\a}}\frac{\ea+\eb}{\ea-\eb}
  \,\big(1 + b\,T^0_{\a\a}\big) (\bSb\times\bSc)^\z.
\end{align}
While the scalar component of \eqref{eq:c:habarPT2} vanishes, the vector
component does not.  With \eqref{eq:c:T^0} and \eqref{eq:c:a_0abd}, we write
\begin{align}
  \label{eq:c:1+bTaa}
  1+b\,T^0_{\a\a}
%  &=1+\frac{\sqrt{6}}{2}\frac{5}{(s+1)(2s+3)}
%    \frac{2}{\sqrt{6}}\,\Big[3{S^\z}^2 - s(s+1)\Big]
%  \nonumber\\[0.2\baselineskip] 
  &=1+\frac{5}{(s+1)(2s+3)}\Big[3{S^\z}^2 - s(s+1)\Big]
  \nonumber\\[0.2\baselineskip] 
%  &=1+\frac{15{S^\z}^2 - 5 s(s+1)}{(s+1)(2s+3)}
%  \nonumber\\[0.2\baselineskip] 
%  &=\frac{15{S^\z}^2}{(s+1)(2s+3)}+\frac{2s+3}{2s+3} -\frac{5s}{2s+3}
%  \nonumber\\[0.2\baselineskip] 
  &=\frac{15{S^\z}^2}{(s+1)(2s+3)}-\frac{3(s-1)}{2s+3}.
\end{align}
With \eqref{eq:app-t-SzSzSz} and \eqref{eq:app-t-SSS-1abc} we find for the
vector component of the product of the $z$-components
\begin{align*}
  \hspace{25pt}&\hspace{-25pt}5\big\{ {\Sa^\z}^2 (\bSb\times\bSc)^\z \big\}_1 
  \nonumber\\[0.2\baselineskip] 
  & = \Sa^\z \big(\bSa(\bSb\times\bSc)\big) 
    + \big(\bSa (\bSb\times\bSc)\big) \Sa^\z 
    + s(s+1)(\bSb\times\bSc)^\z.
\end{align*}
Substitution into \eqref{eq:c:habarPT2} yields
\renewcommand{\struta}{\rule[-14pt]{0pt}{0pt}}
\begin{align}
  \label{eq:c:habarPT_1}
%  \frac{2s+3}{3}\,
  \big\{H_\a^{{\rm {\bar P}T}}\big\}_{\bs{1}}
  =\frac{\i}{4}\,\frac{3}{2s+3}
  \sum_{\substack{\b\ne\c\\ \b,\c\ne\a}}\frac{\ea+\eb}{\ea-\eb}
  \bigg[
  (\bSb\times\bSc)^\z &+\frac{1}{s+1} \bSa \big(\bSa(\bSb\times\bSc)\big)
  \nonumber\\[-0.6\baselineskip] 
                 &+\frac{1}{s+1} \big(\bSa(\bSb\times\bSc)\big)\bSa
  \struta\bigg].
%  \nonumber\\[0.2\baselineskip]
\end{align}
With
\begin{align*}
  \sum_{\substack{\c\\ \c\ne\a,\b}}\bSc = \bsS_{\text{tot}}-\bSa-\bSb,
%  \qquad\text{and}\qquad \bSb\times\bSb=i\bSb,
\end{align*}
% $\bsS_{\text{tot}}\ket{\psi^{S=1}_{0}}=0$,\ \ 
$\bSb\times\bSb=i\bSb$, and
% \begin{align*}
%   \bSa\big(\bSb\times(-\bSa-\bSb)\big)
%   &=\bSb\big(\bSa\times\bSa\big)-\bSa\big(\bSb\times\bSb\big)=0,
% \end{align*}
\eqref{eq:a:SaSbSaSb}, we find from \eqref{eq:c:habarPT_1} that
$\ket{\psi^{S}_{0}}$ is also annihilated by
\begin{align}
  \label{eq:c:habarPT_1v1}
  \frac{\i}{2}
  \sum_{\substack{\b\\ \b\ne\a}}\frac{\ea+\eb}{\ea-\eb}
  \bigg[(\bSa\times\bSb) - i\bSb 
  + \frac{1}{s+1}\big(\bSa(\bSb\times\bsS_{\text{tot}})\big)\bSa\bigg].
\end{align}
With \eqref{eq:b:SaSbtimesStotSa1} and \eqref{eq:b:SaSbtimesStotSa2},
we rewrite the product of the four spin operators in the last term as
\begin{align}
  \big(\bSa(\bSb\times\bsS_{\text{tot}})\big)\bSa
  &=\i\bSa\big(\bSa\bSb\big) - \i\bSa^2\bSb
  \nonumber\\[0.2\baselineskip] 
  &+\text{term which annihilates every spin singlet},
  \nonumber\\ %[-0.2\baselineskip] 
\end{align}
which holds for $\a\ne\b$.  The spin liquid state
\eqref{eq:c:psi0schwinger} is therefore also annihilated by
% \begin{align}
%   \label{eq:c:habarPT_1v2}
% %  \frac{\i}{2}
%   \sum_{\substack{\b\\ \b\ne\a}}\frac{\ea+\eb}{\ea-\eb}
%   \bigg[i(\bSa\times\bSb) + 2\bSb -
%   \frac{1}{2}\bSa\big(\bSa\bSb\big)\bigg].
% \end{align}
\begin{align}
  \label{eq:c:doperator}
  &\bs{D}_\a^{S}
%   =\frac{\i}{2}\sum_{\substack{\b\\ \b\ne\a}}\frac{\ea+\eb}{\ea-\eb}
%   \bigg[(\bSa\times\bSb) -\i(s+1)\bSb 
%   + \frac{\i}{s+1}\bSa\big(\bSa\bSb\big)\bigg],
  =\frac{1}{2}\sum_{\substack{\b\\ \b\ne\a}}\frac{\ea+\eb}{\ea-\eb}
  \bigg[\i(\bSa\times\bSb) + (s+1)\,\bSb 
  - \frac{1}{s+1}\bSa\big(\bSa\bSb\big)\bigg],
  \nonumber\\[0.2\baselineskip] 
  &\bs{D}_\a^{S}\ket{\psi^{S}_{0}} = 0 \quad\forall\, \alpha.
\end{align}
This is the generalization of the auxiliary operator
%\eqref{eq:hsdoperator} or 
\eqref{eq:a:doperator} of the Haldane--Shastry model.

Equation \eqref{eq:c:doperator} implies that the spin liquid state
$\ket{\psi^{S}_{0}}$ is further annihilated by
\begin{align}
  \label{eq:c:lambdaoperator}
  \bs{\Lambda}^{S} &= \sum_{\alpha=1}^N\bs{D}_{\alpha}^{S}
%  \nonumber\\[0.2\baselineskip] &
%  =\frac{\i}{2}\sum_{\substack{\a\ne\b}}\frac{\ea+\eb}{\ea-\eb}
%  \bigg[(\bSa\times\bSb) + \frac{\i}{s+1}\bSa\big(\bSa\bSb\big)\bigg],
  =\frac{1}{2}\sum_{\substack{\a\ne\b}}\frac{\ea+\eb}{\ea-\eb}
%  \bigg[\i(\bSa\times\bSb) - \frac{\bSa\big(\bSa\bSb\big)}{s+1}\bigg],
  \bigg[\i(\bSa\times\bSb) - \frac{1}{s+1}\bSa\big(\bSa\bSb\big)\bigg],
%  \quad \bs{\Lambda}^{S} \ket{\psi^{S}_{0}} = 0,
\end{align}
where we have used \eqref{eq:app-hs13}.  This is the analog of the
rapidity operator \eqref{eq:hsrapidityoperator} or
\eqref{eq:a:lambdaoperator} of the Haldane-Shastry model.  In contrast
to the Haldane--Shastry model, however, the operator
\eqref{eq:c:lambdaoperator} does not commute with the Hamiltonian
\eqref{eq:c:h}.  The model is hence not likely to share the
integrability structure of the Haldane--Shastry model.  It is
possible, however, that the model is integrabel in the thermodynamic
limit $N\to\infty$.

%\newpage
\subsection{Annihilation operators which transform odd under T}

Finally, we consider annihilation operators we can construct from
\eqref{eq:c:ha'}, and which transform odd under T,
\begin{equation}
  \label{eq:c:habarT}
  H_\a^{\rm\bar T}=\frac{1}{2}\left(H_\a-\Theta H_\a \Theta\right)
  = H_\a^{\rm\bar T=} + H_\a^{\rm\bar T\ne},
\end{equation}
where
\begin{align}
  \label{eq:c:habarT=}
  H_\a^{\rm\bar T=}&=\frac{1}{2}\sum_{\substack{\b\\\b\ne\a}}\omega_{\a\b\b}
  \left[\big(a\,V^0_{\a} + c\,W^0_{\a\a\a}\big)
    \frac{1}{2}\biganticomm{\Sb^+}{\Sb^-}
    +\big(1 + b\,T^0_{\a\a}\big)
      \frac{1}{2}\bigcomm{\Sb^+}{\Sb^-}\right]
  \nonumber\\[0.2\baselineskip] 
  &=\frac{1}{2}\sum_{\substack{\b\\\b\ne\a}}\omega_{\a\b\b}
  \Bigg[ \big(a\,\Sa^\z + c\,W^0_{\a\a\a}\big)
         \left(\frac{2s(s+1)}{3}-\frac{1}{\sqrt{6}}\,T_{\b\b}^0\right)
  \nonumber\\[-0.2\baselineskip] 
  &\hspace{180pt}     + \big(1 + b\,T^0_{\a\a}\big) \Sb^\z \Bigg],
 \\[0.4\baselineskip] 
  \label{eq:c:habarTnot=}
  H_\a^{\rm\bar T\ne} 
  &=\frac{1}{2}\sum_{\substack{\b\ne\c\\ \b,\c\ne\a}}\omega_{\a\b\c}
  \,\big(a\,\Sa^\z + c\,W^0_{\a\a\a}\big)
  \left(\frac{2}{3}\bSb\bSc - \text{i}(\bSb\times\bSc)^\z
    -\frac{1}{\sqrt{6}}\,T_{\b\c}^0\right).
\end{align}

Let us first look at the component which transforms odd under P,
\begin{align}
  \label{eq:c:habarPbarT}
  H_\a^{\rm \bar P\bar T}=\frac{1}{2}
  \big(H_\a^{\rm\bar T} - \Pi\hspace{1pt} H_\a^{\rm\bar T}\hspace{1pt} \Pi\big)
  = H_\a^{{\rm \bar P\bar T}=} + H_\a^{{\rm \bar P\bar T}\ne},
\end{align}
where
\begin{align}
  \label{eq:c:habarPbarT=}
  H_\a^{{\rm {\bar P}\bar T}=}&=0, 
&
%  \hspace{-10pt}&\hspace{-10pt}
%  \\[0.4\baselineskip] 
%   \label{eq:c:habarPbarTnot=}
  H_\a^{{\rm {\bar P}\bar T}\ne}&=
  -\frac{\i}{2}\sum_{\substack{\b\ne\c\\ \b,\c\ne\a}}\omega_{\a\b\c}
  \,\big(a\,\Sa^\z + c\,W^0_{\a\a\a}\big)\,(\bSb\times\bSc)^\z.
\end{align}
This operator has no vector component.  With \eqref{eq:app-t-SzSz} and
\eqref{eq:c:a_0abd}, we obtain the scalar component
\begin{align}
  \label{eq:c:habarTnot=0v1}
  \big\{H_\a^{\rm\bar T}\big\}_0
  &=-\frac{\i}{2(s+1)}\sum_{\substack{\b\ne\c\\ \b,\c\ne\a}}
  \frac{\bSa(\bSb\times\bSc)}{(\eab-\ebb)(\ea-\ec)}.
\end{align}
This is is identical to \eqref{eq:a:habarTnot=0v1} in the
Haldane--Shastry model, and annihilates every spin singlet by the line
of reasoning pursued in \eqref{eq:a:habarTnot=0v2}.  We will not
consider it further.

We will now turn the component which transforms even under P,
\begin{align}
  \label{eq:c:haPbarT}
  H_\a^{\rm P\bar T}=\frac{1}{2}
  \big(H_\a^{\rm\bar T} + \Pi\hspace{1pt} H_\a^{\rm\bar T}\hspace{1pt} \Pi\big)
  = H_\a^{{\rm P\bar T}=} + H_\a^{{\rm P\bar T}\ne},
\end{align}
where
\begin{align}
  \label{eq:c:haPbarT=}
  H_\a^{\rm P\bar T=} &=H_\a^{{\rm \bar T}=}, 
%%%
%   THE FOLOWING LINES SHOULD EVENTUALLY BE COMMENTED OUT!
%%%
  \nonumber\\[0.4\baselineskip] 
  &=\frac{1}{2}\sum_{\substack{\b\\\b\ne\a}}\omega_{\a\b\b} \Bigg[
  \big(a\,\Sa^\z + c\,W^0_{\a\a\a}\big)
  \left(\frac{2s(s+1)}{3}-\frac{1}{\sqrt{6}}\,T_{\b\b}^0\right)
  \nonumber\\[-0.2\baselineskip] 
  &\hspace{180pt}     + \big(1 + b\,T^0_{\a\a}\big) \Sb^\z \Bigg].
%%%
  \\[0.4\baselineskip] 
  \label{eq:c:haPbarTnot=}
  H_\a^{\rm P\bar T\ne} 
  &=\frac{1}{2}\sum_{\substack{\b\ne\c\\ \b,\c\ne\a}}\omega_{\a\b\c}
  \,\big(a\,\Sa^\z + c\,W^0_{\a\a\a}\big)
  \left(\frac{2}{3}\bSb\bSc -\frac{1}{\sqrt{6}}\,T_{\b\c}^0\right).
\end{align}
which has no scalar, but a vector component.  The vector components of
\eqref{eq:c:haPbarT=} and \eqref{eq:c:haPbarTnot=} are given 
% with \eqref{eq:c:T^0}, \eqref{eq:c:1+bTaa}, and \eqref{eq:c:a_0abd} 
by
\begin{align}
  \label{eq:c:haPbarT=_1}
  \big\{H_\a^{\rm P\bar T=}\big\}_{1} 
  &=\frac{1}{2}\sum_{\substack{\b\\\b\ne\a}}\omega_{\a\b\b} \bigg[
%  a\,\big\{\Sa^\z \big(s(s+1)-{\Sb^\z}^2 \big)\big\}_{1}
  \frac{a}{2}\,\big\{\Sa^\z \big(\Sb^+\Sb^-+\Sb^-\Sb^+\big)\big\}_{1}
  - \frac{c}{\sqrt{6}}\,\big\{W_{\a\a\a}^0\,T_{\b\b}^0 \big\}_{1}
  \nonumber\\[-0.2\baselineskip] 
  &\hspace{180pt}+ \big\{\big(1 + b\,T^0_{\a\a}\big) \Sb^\z\big\}_{1}\bigg],
  \nonumber\\[0.2\baselineskip] 
%   \nonumber\\[0.2\baselineskip] 
%   &=\frac{1}{2}\sum_{\substack{\b\\\b\ne\a}}\omega_{\a\b\b} \Bigg[
%   3s\, \Sa^\z -\frac{3}{s+1}\,\big\{\Sa^\z{\Sb^\z}^2\big\}_{1}
%   \nonumber\\[-0.4\baselineskip] 
%   &\hspace{80pt}-\frac{7}{2(s+1)(2s+3)(s+2)}\sqrt{\frac{5}{6}}\,
%   \,\big\{W_{\a\a\a}^0\,T_{\b\b}^0 \big\}_{1}
%   \nonumber\\[0.2\baselineskip] 
%   &\hspace{80pt}+\frac{15}{(s+1)(2s+3)}\big\{ {\Sa^\z}^2\Sb^\z\big\}_{1}
%   -\frac{3(s-1)}{2s+3}\Sb^\z \Bigg]
%
  \\[0.4\baselineskip] 
  \label{eq:c:haPbarTnot=_1}
  \big\{H_\a^{\rm P\bar T\ne}\big\}_{1} 
  &=\frac{1}{2}\sum_{\substack{\b\ne\c\\ \b,\c\ne\a}}\omega_{\a\b\c}\bigg[
  \frac{a}{2}\,\big\{\Sa^\z \big(\Sb^+\Sc^-+\Sb^-\Sc^+ \big)\big\}_{1}
  - \frac{c}{\sqrt{6}}\,\big\{W_{\a\a\a}^0\,T_{\b\c}^0 \big\}_{1}\bigg],
\end{align}
where we have rewitten the first term in the way we originally
obtained it.  For $S=\frac{1}{2}$ or $S=1$, these expressions simplify
significantly as $W^0_{\a\a\a}=0$, which follows directly from
$W^{3}_{\a\a\a}=-(\Sa^+)^{3}=0$ for $S<\frac{3}{2}$.  For general $S=s$,
however, we have to evaluate $\big\{W_{\a\a\a}^0\,T_{\b\c}^0 \big\}_{1}$.

%\vspace{\baselineskip} 
%\emph{Evaluation of $\big\{W_{\a\a\a}^0\,T_{\b\c}^0 \big\}_{1}$.}---%
\subsection{Evaluation of $\big\{W_{\a\a\a}^0\,T_{\b\c}^0 \big\}_{\bs{1}}$}

We evaluate the vector component of the tensor product of
$W_{\a\a\a}^0$ and $T_{\b\c}^0$ with $\a\ne\b,\c$ using
\eqref{eq:a:Tm1m2CGj0},
\begin{align}
  \label{eq:b:WaaaTbc_1}
  \big\{W_{\a\a\a}^0T_{\b\c}^0 \big\}_1 &= 
  \braket{1,0}{3,0;2,0}\sum_{m=-2}^{2} 
    W_{\a\a\a}^{m} T_{\b\c}^{-m}\, \braket{3,m;2,-m}{1,0} .
\end{align}
From either \eqref{eq:app-t-SSS-3} or directly from
\eqref{eq:a:commJ+-T}, we obtain
\begin{align}
  \label{eq:c:Waaa}
  W_{\a\a\a}^3&=-\Sa^+\Sa^+\Sa^+,
  \nonumber\\[0.3\baselineskip]
  W_{\a\a\a}^2&= \frac{1}{\sqrt{6}}\comm{\Sa^-}{W_{\a\a\a}^3}
  =\sqrt{6}\, \Sa^+\Sa^+ \big(\Sa^\z+1\big),
  \nonumber\\[0.3\baselineskip]
  W_{\a\a\a}^1&= \frac{1}{\sqrt{10}}\comm{\Sa^-}{W_{\a\a\a}^2}
  =-\sqrt{\frac{3}{5}}\, \Sa^+ \Big[5{\Sa^\z}(\Sa^\z+1)-s(s+1)+2\Big],
  \nonumber\\[0.3\baselineskip]
  W_{\a\a\a}^0&= \frac{1}{\sqrt{12}}\comm{\Sa^-}{W_{\a\a\a}^1}
%  \nonumber\\[0.2\baselineskip]
  =\frac{2}{\sqrt{5}}\, \Big[5{\Sa^\z}^2-3s(s+1)+1\Big]\Sa^\z
  \\[0.3\baselineskip]\nonumber
  W_{\a\a\a}^{-1}&= \frac{1}{\sqrt{12}}\comm{\Sa^-}{W_{\a\a\a}^0}
  =\sqrt{\frac{3}{5}}\, \Sa^- \Big[5{\Sa^\z}(\Sa^\z-1)-s(s+1)+2\Big],
  \\[0.3\baselineskip]\nonumber
  W_{\a\a\a}^{-2}&= \frac{1}{\sqrt{10}}\comm{\Sa^-}{W_{\a\a\a}^{-1}}
  =\sqrt{6}\, \Sa^-\Sa^- \big(\Sa^\z-1\big),
  \\[0.3\baselineskip]\nonumber
  W_{\a\a\a}^{-3}&= \frac{1}{\sqrt{6}}\comm{\Sa^-}{W_{\a\a\a}^{-2}}
  =\Sa^-\Sa^-\Sa^-.
\end{align}
With \eqref{eq:app-t-SS-2} and the Clebsch--Gordan coefficients
\begin{align}
  \braket{3,m;2,-m}{1,0}&=\frac{1}{\sqrt{7\cdot5}}\left\{
    \begin{alignedat}{3}
      &\sqrt{5}&\quad&\text{for}&\quad m&=\pm 2,
      \\ %[.2\baselineskip] 
      &-2\sqrt{2}&&\text{for}& m&=\pm 1,
      \\ %[.2\baselineskip] 
      &3&&\text{for}& m&=0,
    \end{alignedat}\right.
\end{align}
%\newpage\noindent
we obtain
\begin{align}
  \label{eq:b:WaaaTbc_1v1}
  \hspace{15pt}&\hspace{-15pt}
%  Q_{\a\b\c}&\equiv
  \frac{7\cdot 5}{3}\sqrt{\frac{5}{6}}\,\big\{W_{\a\a\a}^0T_{\b\c}^0 \big\}_1 
  \nonumber\\[0.4\baselineskip] &
  =5\, \Sa^+\Sa^\z\Sa^+\Sb^-\Sc^-
  \nonumber\\[0.2\baselineskip]
  &\quad
  +\Big[5\big(\Sa^+\Sa^\z\Sa^\z+\Sa^\z\Sa^\z\Sa^+\big)-\big(2s(s+1)+1\big)\Sa^+\Big]
  \big(\Sb^\z\Sc^-+\Sb^-\Sc^\z\big)
  \nonumber\\[0.2\baselineskip]
  &\quad+\Big[5{\Sa^\z}^2-3s(s+1)+1\Big]\,\Sa^\z\, 
    \Big[4\Sb^\z\Sc^\z-\Sb^+\Sc^--\Sb^-\Sc^+\Big]
  \nonumber\\[0.2\baselineskip]
  &\quad
  +\Big[5\big(\Sa^-\Sa^\z\Sa^\z+\Sa^\z\Sa^\z\Sa^-\big)-\big(2s(s+1)+1\big)\Sa^-\Big]
  \big(\Sb^\z\Sc^++\Sb^+\Sc^\z\big)
  \nonumber\\[0.2\baselineskip]
  &\quad
  +5\, \Sa^-\Sa^\z\Sa^-\Sb^+\Sc^+.
\end{align}
We wish to write this in a more convenient form, which directly displays
that it transforms as a vector under spin rotations.

Let us consider first the case $\b\ne\c$, and try an Ansatz of the
form\footnote{Note that there is no relation between these
  coefficients and those introduced in \eqref{eq:c:S+^2sS-^2s}}
\begin{align}
  \label{eq:b:WaaaTbc_1ansatz}
%  Q_{\a\b\c}^\ne 
%  \frac{7\cdot 5}{3}\sqrt{\frac{6}{5}}\,\big\{W_{\a\a\a}^0T_{\b\c}^0 \big\}_1 
%  &=
  &2a\,\bigl[(\bSa\bSb)\Sa^\z(\bSa\bSc)+ (\bSa\bSc)\Sa^\z(\bSa\bSb)\bigr]
  \nonumber\\*[0.2\baselineskip]
  &\quad + 2b\,\bigl[\Sb^\z(\bSa\bSc) + (\bSa\bSb)\Sc^\z\bigr] 
  + 2c\,\Sa^\z(\bSb\bSc).
%  \nonumber\\[0.2\baselineskip]
\end{align}
Comparing the coefficients of the (five-spin) terms containing
$\Sb^-\Sc^-$ and $\Sb^-\Sc^+$ yields $a=5$.  Comparing the
coefficients of the three-spin terms containing $\Sb^\z\Sc^-$,
$\Sb^-\Sc^\z$, $\Sb^\z\Sc^+$, and $\Sb^+\Sc^\z$ yields
$b=-2s(s+1)-1$.  
% $b=-\big(2s(s+1)+1\big)$.  
If compare the coefficients of both the three-spin and the
five-spin terms containing $\Sb^+\Sc^-$ and $\Sb^-\Sc^+$ terms,
\begin{align*}
  -\big(5{\Sa^\z}^2-3s(s+1)+1\big)\Sa^\z
  &=\frac{a}{2}\big(\Sa^+\Sa^\z\Sa^-+\Sa^-\Sa^\z\Sa^+\big) + c\Sa^\z 
  \nonumber\\[0.2\baselineskip]
  &=\frac{5}{2}\big(\Sa^+\Sa^-(\Sa^\z-1)+\Sa^-\Sa^+(\Sa^\z+1)\big) + c\Sa^\z 
  \nonumber\\[0.2\baselineskip]
%  &=\big(2s(s+1)-2{\Sa^\z}^2\big)\Sa^\z - \comm{\Sa^+}{\Sa^-}
  &=\frac{5}{2}\big(\big(2s(s+1)-2{\Sa^\z}^2\big)\Sa^\z - 2\Sa^\z\big) + c\Sa^\z 
  \nonumber\\[0.2\baselineskip]
  &=-5{\Sa^\z}^3 + \big(5s(s+1)-5\big)\Sa^\z + c\Sa^\z, 
\end{align*}
% with 
% \begin{align*}
%   \frac{a}{4}\big(\Sa^+\Sa^\z\Sa^-+\Sa^-\Sa^\z\Sa^+\big)
%   &=\frac{5}{2}\big(\Sa^+\Sa^-(\Sa^\z-1)+\Sa^-\Sa^+(\Sa^\z+1)\big)
%   \nonumber\\[0.2\baselineskip]
% %  &=\big(2s(s+1)-2{\Sa^\z}^2\big)\Sa^\z - \comm{\Sa^+}{\Sa^-}
%   &=\frac{5}{2}\big(\big(2s(s+1)-2{\Sa^\z}^2\big)\Sa^\z - 2\Sa^\z\big)
%   \nonumber\\[0.2\baselineskip]
%   &=-5{\Sa^\z}^3 + \big(5s(s+1)-5\big)\Sa^\z 
% \end{align*}
we obtain $c=-2s(s+1)+4$.  With these choices, the coefficients of
both the three-spin and the five-spin terms containing $\Sb^\z\Sc^\z$
agree as well,
\begin{align*}
  \big(20{\Sa^\z}^2-12s(s+1)+4\big)\Sa^\z&=4a\,\Sa^\z\Sa^\z\Sa^\z + (4b+2c)\Sa^\z. 
\end{align*}
Finally, the coefficients of the five-spin terms containing
$\Sb^\z\Sc^-$, $\Sb^-\Sc^\z$, $\Sb^\z\Sc^+$, and $\Sb^+\Sc^\z$, in
\eqref{eq:b:WaaaTbc_1ansatz},
\begin{align*}
  a\Big[&\Sa^\z\Sb^\z \Sa^\z\bigl(\Sa^+\Sc^- + \Sa^-\Sc^+\bigr)
  +\Sa^\z\Sc^\z \Sa^\z\bigl(\Sa^+\Sb^- + \Sa^-\Sb^+\bigr)
  \nonumber\\[0.2\baselineskip] 
  &+\bigl(\Sa^+\Sb^- + \Sa^-\Sb^+\bigr) \Sa^\z\Sa^\z\Sc^\z
  +\bigl(\Sa^+\Sc^- + \Sa^-\Sc^+\bigr) \Sa^\z\Sa^\z\Sb^\z\Big], 
\end{align*}
agree with those in \eqref{eq:b:WaaaTbc_1v1}.

For the equivalence to hold for the case $\b=\c$ as well, we need to
order the spin operators in all terms in \eqref{eq:b:WaaaTbc_1ansatz}
such that the $\Sb$'s are to the left of the $\Sc$'s, as this is the
order of the spin operators in \eqref{eq:b:WaaaTbc_1v1}.  We hence
have to replace the second term in the first bracket in 
\eqref{eq:b:WaaaTbc_1ansatz} by
\begin{align*}
  \bSa\big(\Sa^\z(\bSa\bSb)\big)\bSc \equiv \Sa^i\Sa^\z\Sa^j\Sb^j\Sc^i,
\end{align*}
or equivalently add a term
\begin{align}
  \hspace{35pt}&\hspace{-35pt}
  \bSa\big(\Sa^\z(\bSa\bSb)\big)\bSc-(\bSa\bSc)\Sa^\z(\bSa\bSb)
  \nonumber\\[0.2\baselineskip] 
%   &  =\Sa^i\Sa^\z\Sa^j\Sb^j\Sc^i-\Sa^i\Sa^\z\Sa^j\Sc^i\Sb^j
%   \nonumber\\[0.2\baselineskip] 
  &=\Sa^i\Sa^\z\Sa^j\bigcomm{\Sb^j}{\Sc^i}
  \nonumber\\[0.2\baselineskip] 
  &= -\delta_{\b\c}\,\i\varepsilon^{ijk}\Sa^i\Sa^\z\Sa^j\Sb^k
  \nonumber\\[0.2\baselineskip] 
  &= -\delta_{\b\c}\big(\i\varepsilon^{ijk}\Sa^\z\Sa^i\Sa^j\Sb^k
  + \i\varepsilon^{ijk}\bigcomm{\Sa^i}{\Sa^\z}\Sa^j\Sb^k\big)
  \nonumber\\[0.2\baselineskip] 
  &= -\delta_{\b\c}\big(\Sa^\z\,\i(\bSa\times\bSa)\bSb
  - \varepsilon^{ijk}\varepsilon^{izl}\Sa^l\Sa^j\Sb^k\big)
  \nonumber\\[0.2\baselineskip] 
  &= -\delta_{\b\c}\big(-\Sa^\z(\bSa\bSb)
  - \big(\delta^{jz}\delta^{kl}-\delta^{jl}\delta^{kz}\big) \Sa^l\Sa^j\Sb^k\big)
  \nonumber\\[0.2\baselineskip] 
  &= \delta_{\b\c}\big(\Sa^\z(\bSa\bSb) + (\bSa\bSb)\Sa^\z - s(s+1)\Sb^\z\big).
%  \nonumber\\[0.2\baselineskip] 
\end{align}

%\newpage
Taking all the terms together, we finally obtain
\begin{align}
  \label{eq:b:WaaaTbc_1v2}
%  \hspace{15pt}&\hspace{-15pt}
  \frac{7\cdot 5}{2\cdot 3}\sqrt{\frac{5}{6}}\,
  \big\{W_{\a\a\a}^0T_{\b\c}^0 \big\}_1 
%  \nonumber\\[0.4\baselineskip] &
  &=5\,\bigl[(\bSa\bSb)\Sa^\z(\bSa\bSc)+ (\bSa\bSc)\Sa^\z(\bSa\bSb)\bigr]
  \nonumber\\[0.2\baselineskip]
  &\quad -\big(2s(s+1)+1\big)\,\bigl[\Sb^\z(\bSa\bSc) + (\bSa\bSb)\Sc^\z\bigr] 
  \nonumber\\[0.2\baselineskip]
  &\quad -\big(2s(s+1)-4\big)\,\Sa^\z(\bSb\bSc)
  \nonumber\\[0.2\baselineskip]
  &\quad +5\,\delta_{\b\c}
  \big[\Sa^\z(\bSa\bSb) + (\bSa\bSb)\Sa^\z - s(s+1)\Sb^\z\big].
  \nonumber\\[0.2\baselineskip]
\end{align}
%\newpage

\subsection{Annihilation operators which transform odd under T (continued)}

Substitution of \eqref{eq:b:WaaaTbc_1v2} into \eqref{eq:c:haPbarT=_1}
%and \eqref{eq:c:haPbarTnot=_1} 
yields with \eqref{eq:c:a_0abd}, \eqref{eq:app-t-Sz+-w+},
\eqref{eq:c:1+bTaa} and \eqref{eq:app-t-SzSzSz}
\begin{align}
  \label{eq:c:haPbarT=_1v1}
  \hspace{10pt}&\hspace{-10pt}
  \big\{H_\a^{\rm P\bar T=}\big\}_{1} 
%   =\frac{1}{2}\sum_{\substack{\b\\\b\ne\a}}\omega_{\a\b\b} \Big[
%   \frac{a}{2}\,\big\{\Sa^\z \big(\Sb^+\Sb^-+\Sb^-\Sb^+\big)\big\}_{1}
%   - \frac{c}{\sqrt{6}}\,\big\{W_{\a\a\a}^0\,T_{\b\b}^0 \big\}_{1}
%   \nonumber\\[-0.2\baselineskip] 
%   &\hspace{210pt}+ \big\{\big(1 + b\,T^0_{\a\a}\big) \Sb^\z\big\}_{1}\Big],
%   \nonumber\\[0.2\baselineskip] 
  =\frac{1}{2}\sum_{\substack{\b\\\b\ne\a}}\omega_{\a\b\b} \Bigg[
  \frac{3}{2(s+1)}\,\big\{\Sa^\z \big(\Sb^+\Sb^-+\Sb^-\Sb^+\big)\big\}_{1}
  \nonumber\\*[-0.4\baselineskip] 
  &\hspace{100pt}-\frac{7}{2(s+1)(2s+3)(s+2)}\sqrt{\frac{5}{6}}\,
  \,\big\{W_{\a\a\a}^0\,T_{\b\b}^0 \big\}_{1}
  \nonumber\\*[0.2\baselineskip] 
  &\hspace{100pt}+\frac{15}{(s+1)(2s+3)}\big\{ {\Sa^\z}^2\Sb^\z\big\}_{1}
  -\frac{3(s-1)}{2s+3}\Sb^\z \Bigg]
  \nonumber\\[0.5\baselineskip] 
  &=\frac{3}{2(s+1)}\sum_{\substack{\b\\\b\ne\a}}\omega_{\a\b\b} \Bigg[
  %\bigg[
  \frac{4}{5}s(s+1)\Sa^\z-\frac{1}{5}\Sb^\z(\bSa\bSb)-\frac{1}{5}(\bSa\bSb)\Sb^\z
  %\bigg]
  \nonumber\\[0.2\baselineskip] 
  &\hspace{46pt}-\frac{1}{5(2s+3)(s+2)}\,\Big[
  10(\bSa\bSb)\Sa^\z(\bSa\bSb)
  \nonumber\\*[0.2\baselineskip] 
  &\hspace{135pt}
  -\big(2s(s+1)+1\big)\,\bigl[\Sb^\z(\bSa\bSb) + (\bSa\bSb)\Sb^\z \bigr]
  \nonumber\\*[0.2\baselineskip] 
  &\hspace{135pt}-\big(2s(s+1)-4\big)\,s(s+1)\,\Sa^\z
  \nonumber\\*[0.2\baselineskip] 
  &\hspace{135pt}+5\big[\Sa^\z(\bSa\bSb) + (\bSa\bSb)\Sa^\z - s(s+1)\Sb^\z\big]
  \Big]
  \nonumber\\[0.2\baselineskip] 
  &\hspace{46pt}+\frac{1}{2s+3}\,\Big[
  \Sa^\z(\bSa\bSb) + (\bSa\bSb)\Sa^\z + s(s+1)\Sb^\z
  \Big]  -\frac{s^2-1}{2s+3}\Sb^\z \Bigg]
  \nonumber\\[0.5\baselineskip] 
  &=\frac{3}{2(2s+3)(s+2)}\sum_{\substack{\b\\\b\ne\a}}\omega_{\a\b\b} \bigg[
  - \frac{2}{s+1}(\bSa\bSb)\Sa^\z(\bSa\bSb)  
%  +2s(s+1)(s+2)\,\Sa^\z + 2(s+1)\,\Sb^\z
  \nonumber\\*[-0.4\baselineskip] 
  &\hspace{131pt}+ \big[\Sa^\z(\bSa\bSb) + (\bSa\bSb)\Sa^\z\big]
  \nonumber\\*[0.2\baselineskip] 
  &\hspace{131pt}- \big[\Sb^\z(\bSa\bSb) + (\bSa\bSb)\Sb^\z\big]
  \nonumber\\*%[0.2\baselineskip] 
  &\hspace{131pt}
%  - \frac{2}{s+1}(\bSa\bSb)\Sa^\z(\bSa\bSb)\bigg]
%  - \frac{2}{s+1}(\bSa\bSb)\Sa^\z(\bSa\bSb)  
  +2s(s+1)(s+2)\,\Sa^\z + 2(s+1)\,\Sb^\z\bigg]
  \nonumber\\[0.5\baselineskip] 
  &=\frac{3}{(2s+3)(s+2)}\sum_{\substack{\b\\\b\ne\a}}\omega_{\a\b\b} \bigg[
  -\frac{1}{s+1}(\bSa\bSb)\Sa^\z(\bSa\bSb)
%  \frac{(\bSa\bSb)\Sa^\z(\bSa\bSb)}{s+1}
  \nonumber\\*[-0.4\baselineskip] 
  &\hspace{128pt}+ \Sa^\z(\bSa\bSb) - (\bSa\bSb)\Sb^\z
  \nonumber\\*%[0.2\baselineskip] 
  &\hspace{128pt}+ s(s+1)(s+2)\,\Sa^\z + (s+1)\,\Sb^\z\bigg].
\end{align}
%  \\[0.4\baselineskip] 
%\newpage 
Similarly, substitution of \eqref{eq:b:WaaaTbc_1v2} into
\eqref{eq:c:haPbarTnot=_1} yields with \eqref{eq:c:a_0abd} and
\eqref{eq:app-t-Sz+-w+}
\begin{align}
  \label{eq:c:haPbarTnot=_1v1}
  \hspace{10pt}&\hspace{-10pt}
  \big\{H_\a^{\rm P\bar T\ne}\big\}_{1} 
%   =\frac{1}{2}\sum_{\substack{\b\ne\c\\ \b,\c\ne\a}}\omega_{\a\b\c}\bigg[
%   \frac{a}{2}\,\big\{\Sa^\z \big(\Sb^+\Sc^-+\Sb^-\Sc^+ \big)\big\}_{1}
%   - \frac{c}{\sqrt{6}}\,\big\{W_{\a\a\a}^0\,T_{\b\c}^0 \big\}_{1}\bigg]
%    \nonumber\\[0.2\baselineskip] 
  =\frac{1}{2}\sum_{\substack{\b\ne\c\\ \b,\c\ne\a}}\omega_{\a\b\c}\bigg[
  \frac{3}{2(s+1)}\,\big\{\Sa^\z \big(\Sb^+\Sc^-+\Sb^-\Sc^+\big)\big\}_{1}
  \nonumber\\*[-0.4\baselineskip] 
  &\hspace{105pt}-\frac{7}{2(s+1)(2s+3)(s+2)}\sqrt{\frac{5}{6}}\,
  \,\big\{W_{\a\a\a}^0\,T_{\b\c}^0 \big\}_{1}\bigg]
  \nonumber\\[0.5\baselineskip] 
  &=\frac{3}{2(s+1)}\sum_{\substack{\b\ne\c\\ \b,\c\ne\a}}\omega_{\a\b\c}\Bigg[
  %\bigg[
  \frac{4}{5}\Sa^\z(\bSb\bSc)
  -\frac{1}{5}\Sb^\z(\bSa\bSc)-\frac{1}{5}(\bSa\bSb)\Sc^\z
  %\bigg]
  \nonumber\\[0.2\baselineskip] 
  &\hspace{46pt} - \frac{1}{5(2s+3)(s+2)}\,\Big[
  5\,\bigl[(\bSa\bSb)\Sa^\z(\bSa\bSc)+ (\bSa\bSc)\Sa^\z(\bSa\bSb)\bigr]
  \nonumber\\*[0.2\baselineskip] 
  &\hspace{136pt}
  - \big(2s(s+1)+1\big)\,\bigl[\Sb^\z(\bSa\bSc) + (\bSa\bSb)\Sc^\z \bigr]
  \nonumber\\*[0.2\baselineskip] 
  &\hspace{137pt} - \big(2s(s+1)-4\big)\,\Sa^\z(\bSb\bSc)
  \Big]\Bigg]
%  \nonumber\\[0.2\baselineskip] 
%
  \nonumber\\[0.5\baselineskip] 
  &=\frac{3}{2(2s+3)(s+2)}
%  \nonumber\\[0.2\baselineskip] &\quad
  \sum_{\substack{\b\ne\c\\ \b,\c\ne\a}}\omega_{\a\b\c}\bigg[
  -\frac{(\bSa\bSb)\Sa^\z(\bSa\bSc) + (\bSa\bSc)\Sa^\z(\bSa\bSb)}{s+1}\,
%  \frac{1}{s+1}\,
%  \bigl[(\bSa\bSb)\Sa^\z(\bSa\bSc) + (\bSa\bSc)\Sa^\z(\bSa\bSb)\bigr]
  \nonumber\\*[-0.4\baselineskip] 
  &\hspace{137pt} +  2(s+2)\,\Sa^\z(\bSb\bSc) 
  \nonumber\\*[0.2\baselineskip] 
  &\hspace{137pt} - \Sb^\z(\bSa\bSc) - (\bSa\bSb)\Sc^\z\bigg]. 
%  \nonumber\\[0.2\baselineskip] 
\end{align}
Combining \eqref{eq:c:haPbarT=_1v1} and \eqref{eq:c:haPbarTnot=_1v1}, we finally
obtain the vector annihilation operator
\begin{align}
  \label{eq:c:aoperator}
  \bs{A}_\a^{S} &\equiv \frac{2(2s+3)(s+2)}{3}
  \Big(\big\{H_\a^{\rm P\bar T=}\big\}_{\bs{1}} +
  \{H_\a^{\rm P\bar T\ne}\big\}_{\bs{1}}\big)
  \nonumber\\[0.6\baselineskip] 
  &=\sum_{\substack{\b\\\b\ne\a}} %\omega_{\a\b\b} \Big[
%  \bSa(\bSa\bSb) + (\bSa\bSb)\bSa + 2(s+1)\,\bSb\Big]
  \frac{\bSa(\bSa\bSb) + (\bSa\bSb)\bSa + 2(s+1)\,\bSb}{\vert\ea-\eb\vert^2}
  \nonumber\\[0.2\baselineskip] 
  &+\sum_{\substack{\b,\c\\ \b,\c\ne\a}}%\omega_{\a\b\c}\bigg[
  \frac{1}{(\eab-\ebb)(\ea-\ec)}\bigg[
  -\frac{(\bSa\bSb)\bSa(\bSa\bSc) + (\bSa\bSc)\bSa(\bSa\bSb)}{s+1}\,
%  -\frac{1}{s+1}\,
%  \bigl[(\bSa\bSb)\bSa(\bSa\bSc) + (\bSa\bSc)\bSa(\bSa\bSb)\bigr]
  \nonumber\\*[-0.4\baselineskip] 
  &\hspace{102pt} +  2(s+2)\,\bSa(\bSb\bSc) 
  - \bSb(\bSa\bSc) - (\bSa\bSb)\bSc\bigg], 
  \nonumber\\[0.2\baselineskip] 
  \bs{A}_\a^{S} &\ket{\psi^{S}_{0}} = 0 \quad \forall\, \alpha.
\end{align}
This operator is even more complicated than the corresponding operator
\eqref{eq:b:aoperator} for $S=1$, and only simplifies moderately if we
summ over $\a$. From \eqref{eq:c:haPbarT=_1v1}, we obtain
\begin{align*}
%  \label{eq:c:h0PbarT=_1}
  \hspace{20pt}&\hspace{-20pt}
  \frac{2(2s+3)(s+2)}{3}\sum_\a \big\{H_\a^{\rm P\bar T=}\big\}_{\bs{1}}
  \nonumber\\*[0.2\baselineskip]& 
  =-\frac{2}{s+1}\sum_{\substack{\a\ne\b}}\omega_{\a\b\b}\,
  (\bSa\bSb)\Sa^\z(\bSa\bSb)
  \,+\, 2(s+1)^3\sum_\a\bSa\sum_{\substack{\b\\\b\ne\a}}\omega_{\a\b\b}
  \nonumber\\*[0.2\baselineskip]& 
  =-\frac{2}{s+1}\sum_{\substack{\a\ne\b}}\omega_{\a\b\b}\,
  (\bSa\bSb)\Sa^\z(\bSa\bSb)
  \,+\, (s+1)^3\,\frac{N^2-1}{6} \bsS_{\text{tot}}.
\end{align*}
This implies that $\ket{\psi^{S}_{0}}$ is also annihilated by
\begin{align}
  \label{eq:c:upsilonoperator}
  \bs{\Upsilon}^{S}
  &=-\frac{1}{s+1}\sum_{\substack{\a,\b,\c\\ \b,\c\ne\a}}%\omega_{\a\b\c}
%  \frac{(\bSa\bSb)\bSa(\bSa\bSc) + (\bSa\bSc)\bSa(\bSa\bSb)}{s+1}\,
  \frac{(\bSa\bSb)\bSa(\bSa\bSc) + (\bSa\bSc)\bSa(\bSa\bSb)}
  {(\eab-\ebb)(\ea-\ec)}
  \nonumber\\*[0.4\baselineskip] 
  &\quad +\sum_{\substack{\a,\b,\c\\ \a\ne\b\ne\c\ne\a}}%\omega_{\a\b\c}\Big[
  \frac{2(s+2)\,\bSa(\bSb\bSc) - \bSb(\bSa\bSc) - (\bSa\bSb)\bSc}
  {(\eab-\ebb)(\ea-\ec)} .
\end{align}
Whether this operator is of any practical use for further study of the
model, however, remains an open question.  The derivation of it
concludes our study of non-trivial scalar and vector operators we can
obtain from the \defining condition \eqref{eq:c:Omegadef} for the
critical spin liquid state \eqref{eq:c:psi0ket}.  These operators are
summarized in Table \ref{tab:c:annihilationops}.

\renewcommand{\strut}{\rule[-5pt]{0pt}{16pt}}
\begin{table}[t]
  \centering
  \caption{Annihilation operators for the 
    general spin liquid ground state.  With the exception of the \defining
    operator $\OaS$, which is the $m=2s+1$ component of a tensor of order
    $2s+1$, we have only included scalar and vector annihilation operators.}
  \label{tab:c:annihilationops}
  \centering
  \begin{tabular}{p{22mm}p{22mm}p{13mm}p{13mm}p{26mm}p{18mm}}\hline
  \multicolumn{6}{c}{{\bf Annihilation operators for} 
    $\ket{\psi^{S}_{0}}$}\strut \\[2pt]\hline
   Operator &Equation 
   &\multicolumn{4}{l}{Symmetry transformation properties}\strut 
   \\[2pt]\cline{3-6}
   &&T &P &order of tensor &transl.~inv.\strut\\[3pt]\hline %\svhline
   $\bsS_{\text{tot}}$&\eqref{eq:hsspinsymmetry}&$-$&$+$&vector&yes\rule[0pt]{0pt}{12pt}\\[3pt]
   $\OaS$&\eqref{eq:c:Omegadef}&no&no&$2s+1$&no\\[3pt]
   $H^{S}-E_0^{S}$&\eqref{eq:c:h}&$+$&$+$&scalar&yes\\[3pt]
%   &\eqref{eq:a:e0PT0_1}&&&& \\[2pt]
   $\bs{D}_\a^{S}$&\eqref{eq:c:doperator}&$+$&$-$&vector&no\\[3pt]
   $\bs{\Lambda}^{S}$&\eqref{eq:c:lambdaoperator}&$+$&$-$&vector&yes\\[3pt]
   $\bs{A}_\a^{S}$&\eqref{eq:c:aoperator}&$-$&$+$&vector&no\\[3pt]
   $\bs{\Upsilon}^{S}$&\eqref{eq:c:upsilonoperator}&$-$&$+$&vector&yes\\[3pt]
  \hline
  \end{tabular}
\end{table}

\section{Scalar operators constructed from vectors}
\label{sec:c:scalarfromvec}

We see from Table \ref{tab:c:annihilationops} that there are two
simple ways of constructing translationally, parity, and time reversal
invariant scalar operators which annihilate $\ket{\psi^{S}_{0}}$ from
vector operators.  These operators are
\begin{align}
  \sum_{\a} {\bs{D}_\a^{S}}^\dagger\bs{D}_\a^{S}
  \qquad\text{and}\qquad
  \sum_{\a} \bSa \bs{A}_\a^{S}.
\end{align}
These could potentially lead to alternative parent Hamiltonians for
$\ket{\psi^{S}_{0}}$.  If we just recover \eqref{eq:c:h}, the evaluation
of the first operator will show that $H^{S}-E_0$ is positive semi-definite,
or in other words, that $\ket{\psi^{S}_{0}}$ is a ground state of $H^{S}$.

%\newpage
\subsection{Factorization of the Hamiltonian}
\label{sec:c:factorization}

In this section, we will evaluate 
\begin{align*}
  \sum_{\a} {\bs{D}_\a^{S}}^\dagger\bs{D}_\a^{S},
\end{align*}
with  $\bs{D}_\a^{S}$ given by \eqref{eq:c:doperator}, or explicitly
% As in Section \ref{sec:hsfactorization}, we define
% \begin{align*}
%   \theta_{\a\b} \equiv \frac{\ea+\eb}{\ea-\eb},
% \end{align*}
% and write 
\begin{align*}
%  \label{eq:c:ddoperator}
  {\bs{D}_\a^{S}}^\dagger
  &=\frac{1}{2}\sum_{\substack{\b\\ \b\ne\a}}\frac{\ea+\eb}{\ea-\eb}
  \bigg[\i(\bSa\times\bSb) - (s+1)\,\bSb 
  + \frac{1}{s+1}\big(\bSa\bSb\big)\bSa\bigg],
  \nonumber\\[0.2\baselineskip] 
  \bs{D}_\a^{S}
  &=\frac{1}{2}\sum_{\substack{\c\\ \c\ne\a}}\frac{\ea+\ec}{\ea-\ec}
  \bigg[\i(\bSa\times\bSc) + (s+1)\,\bSc 
  - \frac{1}{s+1}\bSa\big(\bSa\bSc\big)\bigg].
%  \nonumber\\[0.2\baselineskip] 
\end{align*}
With $\a\ne\b,\c$ and 
\begin{align}
  \hspace{20pt}&\hspace{-20pt}
  \i(\bSa\times\bSb)\i(\bSa\times\bSc)
  =\varepsilon^{ijk}\varepsilon^{ilm}\Sb^j\Sa^k\Sa^l\Sc^m
  \nonumber\\[0.2\baselineskip] 
  &=\delta^{jl}\delta^{km}
  \big(\Sb^j\Sa^l\Sa^k\Sc^m - \Sb^j\comm{\Sa^l}{\Sa^k}\Sc^m\big)
  -\delta^{jm}\delta^{kl}\Sb^j\Sa^k\Sa^l\Sc^m
%   \nonumber\\[0.2\baselineskip] 
%   &=(\bSa\bSb)(\bSa\bSc)-\i\varepsilon^{lkn}\Sb^l\Sa^n\Sc^m
%   -s(s+1)\bSb\bSc
  \nonumber\\[0.2\baselineskip] 
  &=(\bSa\bSb)(\bSa\bSc)-\i\bSa(\bSb\times\bSc)-s(s+1)\bSb\bSc,
\end{align}
%multiplication of the two square brackets yields 
we obtain for the product of the two square brackets
\begin{align}
  \label{eq:c:prodsquarebrackets}
  &(\bSa\bSb)(\bSa\bSc)-\i\bSa(\bSb\times\bSc)-s(s+1)\bSb\bSc
  \nonumber\\[0.2\baselineskip] 
  &  + 2(s+1)\i\bSa(\bSb\times\bSc)
%  \nonumber\\[0.2\baselineskip] &
  - \frac{2}{s+1}\big(\bSa\bSb\big)\big(\bSa\bSc\big)
  \nonumber\\[0.2\baselineskip] &
  - (s+1)^2\,\bSb\bSc
%  \nonumber\\[0.2\baselineskip] &
  + 2\big(\bSa\bSb\big)\big(\bSa\bSc\big)
%  \nonumber\\[0.2\baselineskip] &
  -\frac{s}{s+1}\big(\bSa\bSb\big)\big(\bSa\bSc\big)
  \nonumber\\[0.4\baselineskip] &
  =(2s+1)\bigg[\frac{1}{s+1}(\bSa\bSb)(\bSa\bSc) + \i\bSa(\bSb\times\bSc)
  - (s+1)\bSb\bSc\bigg].
\end{align}
The product of the prefactors is given by
\begin{align}
  \label{eq:c:prodprefactors}
  \hspace{10pt}&\hspace{-10pt}
  \frac{\ea+\eb}{\ea-\eb}\cdot\frac{\ea+\ec}{\ea-\ec}
  \nonumber\\[0.3\baselineskip] 
  &=\frac{1}{2}\bigg[ 1+\frac{2\eb}{\ea-\eb}\bigg]
               \bigg[-1+\frac{2\ea}{\ea-\ec}\bigg]
   +\frac{1}{2}\bigg[-1+\frac{2\ea}{\ea-\eb}\bigg]
               \bigg[ 1+\frac{2\ec}{\ea-\ec}\bigg]
  \nonumber\\[0.3\baselineskip] 
%   &=\frac{1}{2}\bigg( 1+\frac{2\eb}{\ea-\eb}\bigg)
%                \bigg(-1+\frac{2\ea}{\ea-\ec}\bigg)
%    +\frac{1}{2}\bigg(-1+\frac{2\ea}{\ea-\eb}\bigg)
%                \bigg( 1+\frac{2\ec}{\ea-\ec}\bigg)
%   \nonumber\\[0.2\baselineskip] 
  &=-1+1+1+2\frac{\ea\eb}{(\ea-\eb)(\ea-\ec)}+2\frac{\ea\ec}{(\ea-\eb)(\ea-\ec)}
  \nonumber\\[0.3\baselineskip] 
  &=1-2(\omega_{\a\b\c}+\omega_{\a\c\b}).
\end{align}
We now define
\begin{align}
  \label{eq:c:bdoperator}
  \bs{B}_\a^{S}
  &\equiv\frac{\i}{2}\sum_{\substack{\c\\ \c\ne\a}}
  \bigg[\i(\bSa\times\bSc) + (s+1)\,\bSc 
  - \frac{1}{s+1}\bSa\big(\bSa\bSc\big)\bigg]
%%  \nonumber\\[0.2\baselineskip] 
%%  &=\frac{1}{2}\bigg[\i\big(\bSa\times(\bStot-\bSa)\big) 
%%  + (s+1)\,(\bStot-\bSa) 
%%  - \frac{\bSa\big(\bSa(\bStot-\bSa)\big)}{s+1}\bigg]
%   \nonumber\\[0.2\baselineskip] 
%   &=\frac{\i}{2}\bigg[\i\big(\bSa\times\bStot\big)+\bSa 
%   + (s+1)\,\bStot-(s+1)\,\bSa 
%   - \frac{\bSa\big(\bSa\bStot\big)}{s+1}+s\bSa\bigg]
  \nonumber\\[0.2\baselineskip] 
  &=\frac{\i}{2}\bigg[\i\big(\bSa\times\bStot\big)
  + (s+1)\,\bStot - \frac{\bSa\big(\bSa\bStot\big)}{s+1}\bigg],
\end{align}
and its Hermitian conjugate,
\begin{align*}
%  \label{eq:c:ddoperator}
  {\bs{B}_\a^{S}}^\dagger
  &=\frac{\i}{2}\sum_{\substack{\b\\ \b\ne\a}}
  \bigg[\i(\bSa\times\bSb) - (s+1)\,\bSb 
  + \frac{1}{s+1}\big(\bSa\bSb\big)\bSa\bigg].
\end{align*}
Obviously, $\bs{B}_\a^{S}$ annihilates every spin singlet, and
$\ket{\psi^{S}_0}$ in particular.  
With \eqref{eq:c:bdoperator}, we may write
\begin{align}
  \label{eq:c:sumDD+BB}
  \hspace{10pt}&\hspace{-10pt}
  \frac{1}{2s+1}\,\sum_{\a} \Big({\bs{D}_\a^{S}}^\dagger\bs{D}_\a^{S}
  + {\bs{B}_\a^{S}}^\dagger\bs{B}_\a^{S}\Big)
  \nonumber\\[0.4\baselineskip] 
%  &=-\frac{1}{2}\sum_{\substack{\a,\b,\c\\ \b,\c\ne\a}}
%  (\omega_{\a\b\c}+\omega_{\a\c\b})
  &=-\sum_{\substack{\a,\b,\c\\ \b,\c\ne\a}}
  \frac{\omega_{\a\b\c}+\omega_{\a\c\b}}{2}
%  \nonumber\\ %[0.2\baselineskip] 
%  &\hspace{38pt} 
%  \bigg[\frac{1}{s+1}(\bSa\bSb)(\bSa\bSc) 
  \bigg[\frac{(\bSa\bSb)(\bSa\bSc)}{s+1} 
  + \i\bSa(\bSb\times\bSc) - (s+1)\bSb\bSc\bigg]
  \nonumber\\[0.4\baselineskip] 
  &= -\sum_{\substack{\a\ne\b}}\omega_{\a\b\b}
  \bigg[\frac{(\bSa\bSb)(\bSa\bSc)}{s+1}- \bSa\bSb - s(s+1)^2\bigg] 
  \nonumber\\[0.2\baselineskip] 
  &\quad- \sum_{\substack{\a,\b,\c\\ \a\ne\b\ne\c\ne\a}}\omega_{\a\b\c}
  \bigg[\frac{(\bSa\bSb)(\bSa\bSc)+(\bSa\bSc)(\bSa\bSb)}{2(s+1)}
  - (s+1)\bSb\bSc\bigg] 
  \nonumber\\[0.4\baselineskip] 
  &= (2s+3)\sum_{\substack{\a\ne\b}}\omega_{\a\b\b}\bSa\bSb
%  +s(s+1)^2 \frac{N^2-1}{12} - \frac{s+1}{2}\sum_{\substack{\a\ne\b}}\bSa\bSb
%  \nonumber\\[0.2\baselineskip] &\quad
  - \sum_{\substack{\a,\b,\c\\ \b,\c\ne\a}}\omega_{\a\b\c}
  \frac{(\bSa\bSb)(\bSa\bSc)+(\bSa\bSc)(\bSa\bSb)}{2(s+1)}
  \nonumber\\[0.2\baselineskip] 
%   &\quad+s(s+1)^2\frac{N(N^2-1)}{12}
% %  -\frac{s+1}{2}\sum_{\substack{\a\ne\b}}\bSa\bSb
%   -\frac{s+1}{2}\big(\bStot^2-s(s+1)N\big),
  &\quad+s(s+1)^2\frac{N(N^2+5)}{12}-\frac{s+1}{2}\bStot^2,
\end{align}
where we have used \eqref{eq:app-hs12} and \eqref{eq:app-hs15}.
With the Hamiltonian \eqref{eq:c:h} and the ground state
energy \eqref{eq:c:E_0} derived in Section \ref{sec:c:ham}, we may write
\begin{align}
  \label{eq:c:DdagD+BdagB}
  \hspace{5pt}&\hspace{-5pt}
  \frac{1}{(2s+1)(2s+3)}\,\sum_{\a} \Big({\bs{D}_\a^{S}}^\dagger\bs{D}_\a^{S}
  + {\bs{B}_\a^{S}}^\dagger\bs{B}_\a^{S}\Big)
  + \frac{s+1}{2(2s+3)}\bStot^2
  \nonumber\\[0.6\baselineskip] 
  &=\sum_{\substack{\a\ne\b}} \frac{\bSa\bSb}{\vert\ea-\eb\vert^2}
%  \nonumber\\[0.2\baselineskip] 
%  &\quad  
  -\frac{1}{2(s+1)(2s+3)}\sum_{\substack{\a,\b,\c\\ \a\ne\b,\c}}\!
  \frac{(\bSa\bSb)(\bSa\bSc) + (\bSa\bSc)(\bSa\bSb)}{(\eab-\ebb)(\ea-\ec)}
  \nonumber\\[0.2\baselineskip] 
  &\quad+\frac{s(s+1)^2}{2s+3}\,\frac{N (N^2+5)}{12}
  \nonumber\\[0.6\baselineskip] &
  =\frac{N^2}{2\pi^2}\, \Big[H^{S}-E_0^{S}\Big].
\end{align}
Since all the operators on the left hand side of
\eqref{eq:c:DdagD+BdagB} are positive semi-definite, \ie have only
non-negative eigenvalues, the operator $H^{S}-E_0^{S}$ on the right
has to be positive semi-definite as well.  Furthermore, since all the
operators on the left annihilate $\ket{\psi^{S}_0}$
%
% , which follows from \eqref{eq:c:doperator} and the fact that
% $\bs{B}_\a^{S}$ and $\bStot$ annihilate every spin singlet,
%
we have shown that $\ket{\psi^{S}_0}$ is a zero energy ground state of
$H^{S}-E_0^{S}$.  Exact diagonalization
studies~\cite{manuscriptinpreparationTRSG11} carried out numerically
for up to $N=18$ sites for the $S=1$ model and for up to $N=12$ sites
for the $S=\frac{1}{2}$ model further show that $\ket{\psi^{S=1}_0}$
and $\ket{\psi^{S=\frac{3}{2}}_0}$ are the unique ground states of
\eqref{eq:c:h}.  We assume this property to hold for general spin $S$,
but are not aware of any method to prove this analytically.

Note that the derivation using the operators
$\bs{D}_\a^{S}$ is actually the simplest derivation of \eqref{eq:c:h}
we are aware of.  As compared to our original derivation in Section
\ref{sec:c:ham}, it has the advantage that, except for the tensor
decomposition of $(S^+)^{2s}(S^-)^{2s}$ spelled out in Section
\ref{sec:c:S+^2sS-^2s}, we only needed the formula
\eqref{eq:app-t-SzSzSz} for the vector content of $S_1^\z S_2^\z S_3^\z$,
but not the significantly more complicated formula
\eqref{eq:b:5TaaTbc_0} for the scalar component of
$T_{\a\a}^0T_{\b\c}^0$ derived in Section \ref{sec:b::5TaaTbc_0}.
That we have arrived at the same model twice using different methods
gives us some confidence in the uniqueness of the final Hamiltonian
\eqref{eq:c:h}.

\subsection{A variation of the model}
\label{sec:c:variation}

The analysis in the previous section suggests that another, closely
related Hamiltonian is positive semi-definite as well.  Writing the
product of prefactors \eqref{eq:c:prodprefactors} as
\begin{align}
  \label{eq:c:prodprefactors_v1}
  \frac{\ea+\eb}{\ea-\eb}\cdot\frac{\ea+\ec}{\ea-\ec}
  &=-2\bigg(\omega_{\a\b\c}+\omega_{\a\c\b}-\frac{1}{2}\bigg),
\end{align}
we can derive a model directly from
\begin{align*}
  \sum_{\a} {\bs{D}_\a^{S}}^\dagger\bs{D}_\a^{S},
\end{align*}
without any need to introduce the operators $\bs{B}_\a^{S}$ and
${\bs{B}_\a^{S}}^\dagger$.  This yields
\begin{align}
  \label{eq:c:sumDD}
  \hspace{10pt}&\hspace{-10pt}
  \frac{1}{2s+1}\,\sum_{\a} {\bs{D}_\a^{S}}^\dagger\bs{D}_\a^{S}
  \nonumber\\[0.4\baselineskip] 
%   &=-\sum_{\substack{\a,\b,\c\\ \b,\c\ne\a}}
%   \bigg[\frac{\omega_{\a\b\c}+\omega_{\a\c\b}}{2}-\frac{1}{4}\bigg]
%   \bigg[\frac{(\bSa\bSb)(\bSa\bSc)}{s+1} 
%   + \i\bSa(\bSb\times\bSc) - (s+1)\bSb\bSc\bigg]
%   \nonumber\\[0.4\baselineskip] 
  &= -\sum_{\substack{\a\ne\b}}\bigg(\omega_{\a\b\b}-\frac{1}{4}\bigg)
  \bigg[\frac{(\bSa\bSb)(\bSa\bSc)}{s+1}- \bSa\bSb - s(s+1)^2\bigg] 
  \nonumber\\[0.2\baselineskip] 
  &\quad-\hspace{-5pt}\sum_{\substack{\a,\b,\c\\ \a\ne\b\ne\c\ne\a}}\hspace{-5pt}
  \bigg(\omega_{\a\b\c}-\frac{1}{4}\bigg)
  \bigg[\frac{(\bSa\bSb)(\bSa\bSc)+(\bSa\bSc)(\bSa\bSb)}{2(s+1)}
  - (s+1)\bSb\bSc\bigg] 
  \nonumber\\[0.4\baselineskip] 
  &= (2s+3)\sum_{\substack{\a\ne\b}}
  \bigg(\omega_{\a\b\b}-\frac{1}{4}\bigg)\bSa\bSb
  \nonumber\\[0.2\baselineskip] 
  &\quad %\hspace{24pt}
  - \sum_{\substack{\a,\b,\c\\ \b,\c\ne\a}}
  \bigg(\omega_{\a\b\c}-\frac{1}{4}\bigg)
  \frac{(\bSa\bSb)(\bSa\bSc)+(\bSa\bSc)(\bSa\bSb)}{2(s+1)}
  \nonumber\\[0.2\baselineskip] 
  &\quad+s(s+1)^2\sum_{\substack{\a\ne\b}}\omega_{\a\b\b}
  -\frac{1}{4}(s+1)\sum_{\substack{\a,\b,\c\\ \b,\c\ne\a}}\bSb\bSc
  \nonumber\\[0.4\baselineskip] 
  &= (2s+3)\sum_{\substack{\a\ne\b}}
  \bigg(\omega_{\a\b\b}-\frac{1}{4}\bigg)\bSa\bSb
  \nonumber\\[0.2\baselineskip] 
  &\quad %\hspace{24pt}
  - \sum_{\substack{\a,\b,\c\\ \b,\c\ne\a}}
  \bigg(\omega_{\a\b\c}-\frac{1}{4}\bigg)
  \frac{(\bSa\bSb)(\bSa\bSc)+(\bSa\bSc)(\bSa\bSb)}{2(s+1)}
  \nonumber\\*[0.2\baselineskip] 
  &\quad+s(s+1)^2\frac{N(N^2-4)}{12}-\frac{(s+1)(N-2)}{4}\bStot^2,
\end{align}
where we have used \eqref{eq:app-hs12} and \eqref{eq:app-hs15}. % again.
If we now define the alternative model 
\begin{align}
  \label{eq:c:htilde}
  \tilde H^{S} &\equiv\frac{2\pi^2}{N^2}
  \Bigg[
  \sum_{\substack{\a\ne\b}} 
  \bigg(\frac{1}{\vert\ea-\eb\vert^2}-\frac{1}{4}\bigg)\bSa\bSb
  \nonumber\\[0.2\baselineskip] 
  &\quad -\sum_{\substack{\a,\b,\c\\ \a\ne\b,\c}}
  \bigg(\frac{1}{(\eab-\ebb)(\ea-\ec)}-\frac{1}{4}\bigg)
  \frac{(\bSa\bSb)(\bSa\bSc) + (\bSa\bSc)(\bSa\bSb)}{2(s+1)(2s+3)}
  \Bigg]
\end{align}
with energy eigenvalue
\begin{align}
  \label{eq:c:E_0tilde}
  \tilde E_0^{S} 
  &= -\frac{2\pi^2}{N^2}\frac{s(s+1)^2}{2s+3}\,\frac{N (N^2-4)}{12}
%  \nonumber\\[0.2\baselineskip] &
  = -\frac{\pi^2}{6}\frac{s(s+1)^2}{2s+3}\left(N-\frac{4}{N}\right),
\end{align}
we may rewrite \eqref{eq:c:sumDD} as
\begin{align}
  \label{eq:c:DdagD}
  \hspace{5pt}&\hspace{-5pt}
  \frac{2\pi^2}{N^2}\,\Bigg[
  \frac{1}{(2s+1)(2s+3)}\,\sum_{\a}{\bs{D}_\a^{S}}^\dagger\bs{D}_\a^{S}
  + \frac{(s+1)(N-2)}{4(2s+3)}\bStot^2\Bigg]
%  \nonumber\\[0.4\baselineskip] &
%  = \frac{N^2}{2\pi^2}\,\Big[\tilde H^{S}-\tilde E_0^{S}\Big]
  = \tilde H^{S}-\tilde E_0^{S}.
\end{align}
This implies that $\ket{\psi^{S}_0}$ is also a ground state of 
$\tilde H^{S}$ with energy $\tilde E_0^{S}$, as defined in 
\eqref{eq:c:htilde} and \eqref{eq:c:E_0tilde}, respectively.

Since the maximal distance of $\ea$ and $\eb$ on the unit circle is 2,
%given by 
%$\abs{\ea-(-\ea)}=2$, 
%$\abs{\ea-\eta_{\a+N/2})}=2$, 
the shift in the coefficients in \eqref{eq:c:htilde} (as compared to
\eqref{eq:c:h}) effects that these coefficients go to zero as the sites
$\ea$ and $\eb,\ec$ are maximally separated on the unit circle.  The
alternative model \eqref{eq:c:htilde} is hence more local than the
original model \eqref{eq:c:h}.
It is possible that the alternative model \eqref{eq:c:htilde} possesses
symmetries (or even an integrability structure) the original model does not 
share.

\subsection{The third derivation}

Finally, another translationally, parity, and time reversal invariant 
scalar operators which annihilates $\ket{\psi^{S}_0}$ is given by
\begin{align}
  \sum_{\a} \bSa \bs{A}_\a^{S},
\end{align}
where $\bs{A}_\a^{S}$ is given by \eqref{eq:c:aoperator},
\begin{align*}
%  \label{eq:c:aoperator}
  \bs{A}_\a^{S} 
  &=\sum_{\substack{\b\\\b\ne\a}} \omega_{\a\b\b} \Big[
  \bSa(\bSa\bSb) + (\bSa\bSb)\bSa + 2(s+1)\,\bSb\Big]
%  \frac{\bSa(\bSa\bSb) + (\bSa\bSb)\bSa + 2(s+1)\,\bSb}{\vert\ea-\eb\vert^2}
  \nonumber\\[0.2\baselineskip] 
  &\quad +\sum_{\substack{\b,\c\\ \b,\c\ne\a}}\omega_{\a\b\c}\bigg[
%  \frac{1}{(\eab-\ebb)(\ea-\ec)}\bigg[
  -\frac{(\bSa\bSb)\bSa(\bSa\bSc) + (\bSa\bSc)\bSa(\bSa\bSb)}{s+1}\,
%  -\frac{1}{s+1}\,
%  \bigl[(\bSa\bSb)\bSa(\bSa\bSc) + (\bSa\bSc)\bSa(\bSa\bSb)\bigr]
  \nonumber\\*[-0.4\baselineskip] 
  &\hspace{74pt} +  2(s+2)\,\bSa(\bSb\bSc) 
  - \bSb(\bSa\bSc) - (\bSa\bSb)\bSc\bigg]. 
\end{align*}
With 
\begin{align*}
  \bSa(\bSa\bSb)\bSa
  &=\bSa\big[\bSa(\bSa\bSb)+i\bSa\times\bSb\big]
  \nonumber\\[0.2\baselineskip] 
  &=\big(s(s+1)-1\big)\bSa\bSb,
  \nonumber\\[0.5\baselineskip]  
  \bSa(\bSa\bSb)\bSc
  &=\bSa\big[\bSc(\bSa\bSb)-i\delta_{\b\c}\,\bSa\times\bSb\big]
  \nonumber\\[0.3\baselineskip] 
  &=(\bSa\bSc)(\bSa\bSb)+\delta_{\b\c}\,\bSa\bSb,
\end{align*}
which follows from \eqref{eq:b:commSaSb,Sa}, \eqref{eq:b:commSaSb,Sb}
and holds for $\a\ne\b,\c$, we obtain
%\newpage
\begin{align}
%  \hspace{5pt}&\hspace{-5pt}
  \sum_{\a} \bSa \bs{A}_\a^{S}
  &=\sum_{\substack{\a\ne\b}}\omega_{\a\b\b} 
  \Big[ 2s(s+1)-1 + 2(s+1) - 1 \Big]\,\bSa\bSb
  \nonumber\\[0.2\baselineskip] 
  &+\sum_{\substack{\a,\b,\c\\ \a\ne\b,\c}}\omega_{\a\b\c}\bigg[
  -\big(s(s+1)-1\big)\frac{(\bSa\bSb)(\bSa\bSc) + (\bSa\bSc)(\bSa\bSb)}{s+1}\,
  \nonumber\\* %[-0.4\baselineskip] 
  &\hspace{14pt} +  2s(s+1)(s+2)\,\bSb\bSc 
  - (\bSa\bSb)(\bSa\bSc) - (\bSa\bSc)(\bSa\bSb)\bigg]. 
  \nonumber\\[0.2\baselineskip] 
\end{align}
With \eqref{eq:app-hs12} and \eqref{eq:app-hs15}, we find
\begin{align*}
  \sum_{\substack{\a,\b,\c\\ \a\ne\b,\c}}\omega_{\a\b\c}\,\bSb\bSc
  &=s(s+1)\,\frac{N (N^2+5)}{12}
%  &=s(s+1)\,\frac{N (N^2-1)}{12}
  +2\sum_{\substack{\a\ne\b}}\omega_{\a\b\b}\bSa\bSb-\frac{1}{2}\bStot^2,
%  +\frac{1}{2}s(s+1)N
\end{align*}
and therewith
\begin{align}
%  \hspace{5pt}&\hspace{-5pt}
  \sum_{\a} \bSa \bs{A}_\a^{S}
  &=2s(s+2)(2s+3)\sum_{\substack{\a\ne\b}}\omega_{\a\b\b}\,\bSa\bSb
  \nonumber\\[0.2\baselineskip] 
  &-\frac{s(s+2)}{s+1}
  \sum_{\substack{\a,\b,\c\\ \a\ne\b,\c}}\omega_{\a\b\c}
  \Big[(\bSa\bSb)(\bSa\bSc) + (\bSa\bSc)(\bSa\bSb)\Big]
  \nonumber\\* %[-0.4\baselineskip] 
  &\hspace{14pt} +  2s(s+2)s(s+1)^2\,\frac{N (N^2+5)}{12}
  -s(s+1)(s+2)\,\bStot^2.
  \nonumber\\[0.2\baselineskip]. 
\end{align}
We may rewrite this
\begin{align}
  \hspace{5pt}&\hspace{-5pt}
  \frac{1}{2s(s+2)(2s+3)}\sum_{\a} \bSa \bs{A}_\a^{S}
  +\frac{s+1}{2(2s+3)}\,\bStot^2
  \nonumber\\[0.2\baselineskip] 
  &=\sum_{\substack{\a\ne\b}} \frac{\bSa\bSb}{\vert\ea-\eb\vert^2}
%  \nonumber\\[0.2\baselineskip] 
%  &\quad  
  -\frac{1}{2(s+1)(2s+3)}\sum_{\substack{\a,\b,\c\\ \a\ne\b,\c}}\!
  \frac{(\bSa\bSb)(\bSa\bSc) + (\bSa\bSc)(\bSa\bSb)}{(\eab-\ebb)(\ea-\ec)}
  \nonumber\\[0.2\baselineskip] 
  &\quad+\frac{s(s+1)^2}{2s+3}\,\frac{N (N^2+5)}{12}.
%  \nonumber\\[0.6\baselineskip] &
%  =\frac{N^2}{2\pi^2}\, \big[H^{S}-E_0^{S}\Big].
\end{align}
In other words, we obtain the model Hamiltonian \eqref{eq:c:h} for a third
time.  The present derivation is the most complicated one, and does
not yield any new insights, except that it further strengthens the case
that there is a certain uniqueness to our Hamiltonian.

%\section{The special case $S=\frac{1}{2}$ or the Haldane--Shastry
\section{The case $S=\frac{1}{2}$ 
%or the Haldane--Shastry model 
once more}
\label{sec:c:hs}

Finally, we wish to demonstrate that the general spin $S$ model
introduced and derived in this section includes the Haldane--Shastry
model as the special case $S=\frac{1}{2}$.

For $S=\frac{1}{2}$, the higher order interaction terms in the
Hamiltonian \eqref{eq:c:h} simplify, as
\begin{align}
  \label{eq:c:halfSaSb^2}
  (\bSa\bSb)^2=-\frac{1}{2}\bSa\bSb + \frac{3}{16}, \quad \a\ne\b,
\end{align}
and
\begin{align}
  \label{eq:c:halfSaSbSaSc}
  (\bSa\bSb)(\bSa\bSc) + (\bSa\bSc)(\bSa\bSb)=\frac{1}{2}\bSb\bSc,
  \quad \a\ne\b\ne\c\ne\a .
\end{align}
% We can verify \eqref{eq:c:halfSaSb^2} directly with
% \begin{align}
%   \label{eq:c:halfSaSb=P}
%   \bSa\bSb=\frac{1}{2}\bigg(P_{\a\b}-\frac{1}{2}\bigg),
% \end{align}
% where $P_{\a\b}$ permutes the spins on the sites $\a$ and $\b$,
% and verify \eqref{eq:c:halfSaSbSaSc} with \eqref{eq:c:halfSaSb=P} and
% \begin{align}
%   \label{eq:c:halfPsum}
%   1-P_{\a\b}-P_{\a\c}-P_{\b\c}+P_{\a\b}P_{\b\c}+P_{\b\c}P_{\a\b}=0.
% \end{align}
% Both \eqref{eq:c:halfSaSb=P} and \eqref{eq:c:halfPsum} hold only for
% $S=\frac{1}{2}$.
We can verify \eqref{eq:c:halfSaSb^2} and \eqref{eq:c:halfSaSbSaSc} with
\begin{align}
  (\bSa\bSb)(\bSa\bSc)&=\Sb^i\Sa^i\Sa^j\Sc^j
  \nonumber\\[0.2\baselineskip] 
  &=\Sb^i\bigg(
  \frac{1}{4}\delta^{ij} + \frac{\i}{2}\varepsilon^{ijk}\Sa^k
  \bigg)\Sc^j
  \nonumber\\[0.2\baselineskip] 
  &=\frac{1}{4}\bSb\bSc + \frac{\i}{2}\bSa(\bSb\times\bSc),
\end{align}
which holds only for $S=\frac{1}{2}$ and $\a\ne\b,\c$.
Alternatively, since ${\Sa^+}^2=0$ for $S=\frac{1}{2}$, $T_{\a\a}^m=0$
for all $m$, and \eqref{eq:b:5TaaTbc_0} reduces to
\begin{align}
%  -\frac{4}{3} \bSa^2(\bSb\bSc)
  -\bSb\bSc 
  + 2\,\bigl[(\bSa\bSb)(\bSa\bSc)+ (\bSa\bSc)(\bSa\bSb)\bigr]
  +2\delta_{\b\c}\,\bSa\bSb=0.
%  \nonumber\\[0.2\baselineskip] 
%  &\quad +2\delta_{\b\c}\,\bSa\bSb\,
\end{align}
For $\b=\c$ and $\b\ne\c$, this yields \eqref{eq:c:halfSaSb^2} and
\eqref{eq:c:halfSaSbSaSc}, respectively.

Substitution of \eqref{eq:c:halfSaSb^2}, \eqref{eq:c:halfSaSbSaSc},
and $s=\frac{1}{2}$ into the general Hamiltonian \eqref{eq:c:h}
yields
\begin{align}
  \label{eq:c:hhalf}
  H^{S=\frac{1}{2}} &=\frac{2\pi^2}{N^2}\Bigg[
  \sum_{\substack{\a\ne\b}} \frac{1}{\vert\ea-\eb\vert^2}
  \bigg(\bSa\bSb+\frac{1}{12}\bSa\bSb-\frac{1}{32}\bigg)
  \nonumber\\[0.2\baselineskip] 
  &\hspace{80pt}-\frac{1}{24}\sum_{\substack{\a,\b,\c\\ \a\ne\b\ne\c\ne\a}}
  \frac{\bSb\bSc}{(\eab-\ebb)(\ea-\ec)}\Bigg]
  \nonumber\\[0.2\baselineskip] 
  &=\frac{2\pi^2}{N^2}\Bigg[
  \sum_{\substack{\a\ne\b}} \frac{\bSa\bSb}{\vert\ea-\eb\vert^2}
  - \frac{1}{32}\frac{N (N^2-1)}{12} + \frac{1}{48}\bStot^2 
  - \frac{N}{64}
  \Bigg]
  \nonumber\\[0.2\baselineskip] 
  &={H}^{\s\text{HS}}- \frac{1}{32}\frac{N (N^2+5)}{12}+ \frac{1}{48}\bStot^2.
\end{align}
The energy of the Haldane--Shastry ground state $\ket{\psi^{\s\text{HS}}_{0}}$
is hence with \eqref{eq:c:E_0} given by 
\begin{align*}
  E_0^{\s\text{HS}} 
  &= E_0^{S=\frac{1}{2}} + \frac{2\pi^2}{N^2}\frac{1}{32}\frac{N (N^2+5)}{12}
%  \nonumber\\[0.2\baselineskip] &
  =-\frac{2\pi^2}{N^2}\,\frac{N (N^2+5)}{48},
\end{align*}
which agrees with \eqref{eq:hse0}.  Note that as the derivation in
\eqref{eq:c:hhalf} stands, we have lost the information that
${H}^{\s\text{HS}}-E_0^{\s\text{HS}}$ is positive semi-definite, due
to the $\bStot^2$ term.  This information, however, can be recovered
if we take the last term on the left-hand side of
\eqref{eq:c:DdagD+BdagB}
% , which amounts to $\frac{3}{32}\bStot^2$ for $S=\frac{1}{2}$,
into account.  The spin $S$ model we have derived here hence includes
the Haldane--Shastry model as the special case $S=\frac{1}{2}$.

\vspace{50pt}\newpage

% lp -dtkmsek -P 25-25 map.ps 
% lp -dtkmcol -P 44 map.ps
% lp -dtkmsek -P 144-148 map-full.ps
% lp -dtkmcol -P 142-147 map-full.ps

\numberwithin{equation}{chapter}
\chapter{Conclusions and unresolved issues}
\label{sec:concl}

%\vspace{.5\baselineskip} 
\emph{The model.}---%
In this \paper , we have presented an exact model of a critical spin
chain with spin $S$.  The Hamiltonian is given by 
\begin{align}
  \label{eq:e:c:h}
  H^{S} =\frac{2\pi^2}{N^2}
  \Bigg[
  &\sum_{\substack{\a\ne\b}}^N \frac{\bSa\bSb}{\vert\ea-\eb\vert^2}
  \nonumber\\[0.2\baselineskip] 
  &  -\frac{1}{2(S+1)(2S+3)}\sum_{\substack{\a,\b,\c\\ \a\ne\b,\c}}^N
  \frac{(\bSa\bSb)(\bSa\bSc) + (\bSa\bSc)(\bSa\bSb)}{(\eab-\ebb)(\ea-\ec)}
  \Bigg],
\end{align}
where $\eta_\alpha=e^{\text{i}\frac{2\pi}{N}\alpha }$,
$\a=1,\ldots,N$, are the coordinates of $N$ sites on a unit circle
embedded in the complex plane.  If we write the ground state of the
Haldane--Shastry model~\cite{haldane88prl635,shastry88prl639}, which
is equivalent to the Gutzwiller state obtained by projection of filled
bands~\cite{gutzwiller63prl159,metzner-87prl121,gebhardt-87prl1472},
in terms of Schwinger bosons,
\begin{align}
  \label{eq:e:c:hsket}
  \ket{\psi^{\s\text{HS}}_{0}}
  &=\sum_{\{z_1,\ldots ,z_M;w_1,\ldots,w_M\}}
  \psi^{\s\text{HS}}_{0}(z_1,\ldots ,z_M)\;
    {a}^+_{z_1}\ldots a^\dagger_{z_M}
    {b}^+_{w_1}\ldots b^\dagger_{w_M}%
  \vac\!
  \nonumber\\[0.4\baselineskip] 
  &\equiv \Psi^{\s\text{HS}}_{0}[a^\dagger,b^\dagger] \vac\!,
\end{align}
where $M=\frac{N}{2}$ and the $w_k$'s are those coordinates on
the unit circle which are not occupied by any of the $z_i$'s, then the
exact ground state of our model Hamiltonian \eqref{eq:e:c:h} is given
by
\begin{equation}
  \label{eq:e:c:psi0schwinger}
  \ket{\psi^{S=1}_0}=
  \Big(\Psi^{\s\text{HS}}_0\big[a^\dagger ,b^\dagger\big]\Big)^S\vac.
\end{equation}
The ground state energy is
\begin{align}
  \label{eq:e:c:E_0}
  E_0^{S} &=-\frac{2\pi^2}{N^2}\frac{S(S+1)^2}{2S+3}\,\frac{N (N^2+5)}{12}.
\end{align}
For $S=\frac{1}{2}$, the model \eqref{eq:e:c:h} reduces to the
Haldane--Shastry model.  Since the model describes a critical spin
chain with spin $S$, the low energy effective field theory is given by
the SU(2) level $k=2S$ Wess-Zumino-Witten
model~\cite{wess-71plb95,witten84cmp455}.

The Hamiltonian was constructed from the condition 
\begin{align}
  \label{eq:e:c:Omegadef}
  \OaS 
  &=\sum_{\substack{\beta=1\\[2pt]\beta\ne\alpha}}^N 
  \frac{1}{\ea-\eb} (\Sa^-)^{2S} \Sb^-,\qquad 
  \OaS \ket{\psi^S_{0}} = 0 \quad\forall\, \alpha,
\end{align}
which we obtained for $S=\frac{1}{2}$ and for $S=1$ from the two- and
three-body parent Hamiltonians of bosonic Laughlin and Moore--Read
states in quantum Hall systems, respectively, and then generalized to
arbitrary spin.  

\vspace{.5\baselineskip} 
\emph{Uniqueness and the quest for integrability.}---%
Starting with the defining condition \eqref{eq:e:c:Omegadef}, we
constructed a total of three translationally, parity, and time
reversal invariant scalar annihilation operator for the state
\eqref{eq:e:c:psi0schwinger}---one directly, and two by taking the
scalar products of vector operators.  All three operators yielded the
parent Hamiltonian \eqref{eq:e:c:h}.  This attests are certain
uniqueness to the model.

Nonetheless, it is clear that the model is not completely unique.
First, the ground state \eqref{eq:e:c:psi0schwinger} is trivially
annihilated by all terms which annihilate every spin singlet.
For example, we could add the term
\begin{align}
  \sum_\a \big(\bSa\bStot\big)^2
\end{align}
with an arbitrary coefficient to \eqref{eq:e:c:h}.  Then
\eqref{eq:e:c:psi0schwinger} would remain the ground state as long as
the operator $H^{S}-E_0^{S}$ were to remain positive semi-definite.
(This ambivalence was exploited in Section \eqref{sec:c:variation},
when we derived the alternative Hamiltonian \eqref{eq:c:htilde}.)
Another three-spin term which annihilates every spin singlet is given
by \eqref{eq:c:habarTnot=0v1}, even though this term is not suitable
as it violates both parity and time reversal symmetry.  If we allow
for four-spin interactions, there is a plethora of parity and time
reversal invariant scalar operators we could add.

Second, we could construct another parent Hamiltonian from the
annihilation operator
\begin{align}
  \label{eq:e:b:Thetadef}
    \Xi_{\a} 
     &=\sum_{\substack{\beta,\gamma=1\\\beta,\gamma\ne\alpha}}^N 
    \frac{\Sa^- \Sb^- \Sc^-}{(\ea-\eb)(\ea-\ec)}
    -\sum_{\substack{\beta=1\\[2pt]\beta\ne\alpha}}^N
    \frac{(\Sa^-)^2 \Sb^-}{(\ea-\eb)^2} ,\quad 
    \Xi_{\a} \ket{\psi^{S=1}_{0}} = 0 \ \forall\, \alpha,
\end{align}
which we derived in Section \ref{sec:c:2ndcond}.  This Hamiltonian
will presumably contain five-spin interactions.

The issue of uniqueness of the model is relevant to the question of
whether the model, or a closely related model, is integrable.
Preliminary numerical work~\cite{manuscriptinpreparationTRSG11}
indicates that the model \eqref{eq:e:c:h} is not integrable for finite
system sizes, while the data are consistent with integrability in the
thermodynamic limit.

\vspace{.5\baselineskip} 
\emph{Momentum spacings and topological degeneracies.}---%
The other highly important, unresolved issue regarding the model
concerns the momentum spacings of the spinon excitations.  In Section
\ref{sec:nana}, we proposed that the spacings for the $S=1$ model
would alternate between being odd multiples of $\frac{\pi}{N}$ and
being either odd or even multiples of $\frac{\pi}{N}$.  (Recall that
odd multiples of $\frac{\pi}{N}$ correspond to half-fermions in one
dimension, while even multiples represent either fermions or bosons.)
Whenever we have a choice between even and odd, this choice represents
a topological quantum number, which is insensitive to local
perturbations.  These topological quantum numbers span an internal or
topological Hilbert space of dimension $2^L$ when $2L$ spinons are
present.  All the states in this space are degenerate in the
thermodynamic limit.  This topological Hilbert space is the
one-dimensional analog of the topological Hilbert space spanned by the
Majorana fermion
states~\cite{read-00prb10267,nayak-96npb529,ivanov01prl268,stern-04prb205338}
in the vortex cores of the Moore--Read
state~\cite{moore-91npb362,greiter-91prl3205,greiter-92npb567} or the
non-abelian chiral spin liquid~\cite{greiter-09prl207203}.  In Section
\ref{sec:na:MSforSpinS}, we generalized these conditions for the momentum
spacings to the models with arbitrary spin $S>1$.

The first unresolved issue with regard to our proposal is whether it
is correct.  In view of the established momentum
spacings~\cite{greiter-07prl237202} for the spinons in the
Haldane--Shastry model, the construction of the state suggests that it
is.  Since the model \eqref{eq:e:c:h} is presumably not integrable,
however, the momenta of the individual spinons will not be good
quantum numbers when more than one spinon is present. (This is always
the case, as the minimal number of spinons for the models with $S\ge
1$ is two.)  Nonetheless, the topological shifts can still be good
quantum numbers.  In this regard, the situation is similar to the
Moore--Read state, where, when long-ranged interaction are present,
the state vectors in the internal Hilbert space are degenerate in the
thermodynamic limit only.

Assuming that our assignment of the momentum spacings is correct, the
next question to ask is whether the picture applies only to the exact
model we have constructed in this \paper , or to a whole range of
critical spin chain models with $S=1$.  If it applies to a range of
models, as we believe, the topological space spanned by the spinons
may be useful in applications as protected cubits.  The internal state
vector can probably be manipulated though measurements involving
several spinons simultaneously, but it is far from clear how to do so
efficiently.

To study the spinon excitations systematically, it would be highly
desirable to apply the method reviewed in Section \eqref{sec:hsexsol}
for the Haldane--Shastry model to the general model \eqref{eq:e:c:h}.
% Unfortunately, our efforts in doing so have so far not been successful.  
Unfortunately, this does not appear straightforward.  The problem
arises when we write out the $\Sb^+\Sa^-\,\Sc^+\Sa^-$ term along the
lines of \eqref{eq:hss+s-psi}--\eqref{eq:hs3}.  When we evaluated the
$\Sa^+\Sb^-$ term in the Haldane--Shastry 
model, we used the Taylor series expansion \eqref{eq:hs2} to shift
the variable $\eb$ in the function
\begin{align*}
  \frac{\psi (z_{1},\ldots,z_{j-1},\eb,z_{j+1},\ldots,z_{M})}{\eb}
\end{align*}
in \eqref{eq:hs1} to $z_j$.  When we evaluate the action of the
$\Sb^+\Sa^-\,\Sc^+\Sa^-$ term in the $S=1$ model, we need to shift ``two
variables'' $\ea$ in the function 
\begin{align*}
  \frac{\psi(z_{1},\ldots,z_{j-1},\ea,z_{j+1},\ldots,z_{k-1},\ea,z_{k+1},\ldots, 
    z_{M})}{\ea^2}
\end{align*}
via Taylor expansions, one to $z_j$ and one to $z_k$.  This yields
for $z_j\ne z_k$
\begin{align}
  \label{eq:e:S=1Taylor}
  \hspace{25pt}&\hspace{-25pt}
  \frac{\psi(z_{1},\ldots,\ea,\ldots,\ea,\ldots,z_{M})}{\ea^2}
  \nonumber\\[0.2\baselineskip] 
  &=\sum_{l = 0}^{N-1}\frac{(\ea - z_j)^l}{l!}
  \sum_{m = 0}^{N-1}\frac{(\ea - z_k)^m}{m!}
  \frac{\partial^l}{\partial z_j^l}\frac{\partial^m}{\partial z_k^m}
  \frac{\psi(z_{1},\ldots,z_{M})}{z_j z_k}.
\end{align}
The sum over $\a$ we need to evaluate is hence 
\begin{align}
    \sum_{\substack{\a=1\\[1pt]\ea \neq z_j,z_k}}^N
    \frac{\ea^2(\ea - z_j)^l(\ea - z_k)^m}{(\ea - z_j)(\eab - \bar z_k)},
    \quad 0\le l,m\le N-1. 
\end{align}
In the Haldane--Shastry model, the corresponding sum \eqref{eq:hs3}
is non-zero only for $l=0,1,\ \text{and}\ 2$.  In the present case,
however, further terms arise for $l+m+1=N$.  These yield terms
with very high derivatives when substituted in \eqref{eq:e:S=1Taylor}.
It is not clear whether an analysis along these lines is feasible.

\vspace{.5\baselineskip} 
\emph{Static spin correlations.}---%
Another open issue is the static spin correlation functions of the
ground state \eqref{eq:e:c:psi0schwinger}.  We conjecture that it can be
evaluated via a generalization of the method employed by Metzner and
Vollhardt~\cite{metzner-87prl121} for the Gutzwiller wave function.

\vspace{.5\baselineskip} 
\emph{Generalization to symmetric representations of SU($n$).}---%
The generalization of the model to symmetric representations of
SU($n$), like the representations $\bs{6}$ or $\bs{10}$ of SU(3),
appears to follow without incident.  If we write the SU(3) Gutz\-willer or
Haldane--Shastry ground
state~\cite{kawakami92prb1005,kawakami92prb3191} in terms of SU(3)
Schwinger bosons $b^\dagger,r^\dagger,g^\dagger$ (for blue, red, and
green; see \eg\cite{greiter-07prb184441}),
\begin{align}
  \label{eq:e:su3hsket}
  \bigket{\psi^{\s\text{HS}}_{0}}
%   &=\sum_{\{z_1,\ldots ,z_M;w_1,\ldots,w_M\}}
%   \psi^{\s\text{HS}}_{0}(z_1,\ldots ,z_M)\;
%     {a}^+_{z_1}\ldots a^\dagger_{z_M}
%     {b}^+_{w_1}\ldots b^\dagger_{w_M}%
%   \vac\!
%   \nonumber\\[0.4\baselineskip] 
  &\equiv \Psi^{\s\text{HS}}_{0}[b^\dagger,r^\dagger,g^\dagger] \vac\!,
\end{align}
the generalizations to the SU(3) representation $\bs{6}$ and $\bs{10}$
are given by 
\begin{equation}
  \label{eq:e:su3ket}
  \bigket{\psi^{k}_0} = \Big(\Psi^{\s\text{HS}}_0
  \big[b^\dagger ,r^\dagger,g^\dagger\big]\Big)^k\vac,
\end{equation}
with $k=2$ and $k=3$, respectively.  The generalization of the
defining condition \eqref{eq:e:c:Omegadef} is 
\begin{align}
  \label{eq:e:c:su3Omegadef}
  \Omega_{\alpha}^{k}
  &=\sum_{\substack{\beta=1\\[2pt]\beta\ne\alpha}}^N 
  \frac{1}{\ea-\eb} (I_\a^-)^{k} I_\b^-,\qquad 
  \Omega_{\alpha}^{k} \bigket{\psi^{k}_0} = 0 \quad\forall\, \alpha,
\end{align}
where $I^-\equiv r^\dagger b$ is one of the three ``lowering''
operators for the SU(3) spins.  We assume that the construction of a
parent Hamiltonian along the lines of 
\Chaps~\ref{sec:pf2s1} and~\ref{sec:gen2s} will proceed without
incident.  The momentum spacings in the SU($n$) models are likely to
follow patterns which have no analog in quantum Hall systems, and have
hence not been studied before.

\vspace{.5\baselineskip} 
\emph{Generalization to include mobile holes.}---%
It appears likely that the model can be generalized to include mobile
holes as well, a task which has been accomplished for the
$S=\frac{1}{2}$ model by Kuramoto and
Yokoyama~\cite{kuramoto-91prl1338}.

\vspace{.5\baselineskip} 
\emph{Conclusion.}---%
We have introduced an exact model of critical spin chains with
arbitrary spin $S$.  For $S=\frac{1}{2}$, the model reduces to one
previously discovered by Haldane~\cite{haldane88prl635} and
Shastry~\cite{shastry88prl639}.  The spinon excitations obey
non-abelian statistics for $S\ge 1$, with the internal Hilbert space
spanned by topological spacings of the single spinon momenta.  There
is a long list of unresolved issues, including the quest for
integrability and the viability of potential applications as protected
cubits in quantum computation.

%\vspace{50pt}\newpage 

\numberwithin{equation}{section}

%lp -dtkmsek -P 156-159 map-full.ps
%lp -dtkmcol -P 154-157 map-full.ps
\def\myitemsep{.6\baselineskip} 
\numberwithin{equation}{chapter}
\renewcommand{\theequation}{\Alph{chapter}.\arabic{equation}}

%\allowdisplaybreaks[3]
\appendix
%\addcontentsline{toc}{chapter}{Appendices}

\chapter{Spherical coordinates}
\label{sec:app-sphere}

The formalism for Landau level quantization on the sphere
developed in Section \ref{sec:qhs} requires vector analysis in
spherical coordinates.  In this appendix, we will briefly review the
conventions.
Vectors and vector fields are given by
%\begin{eqnarray*}
\begin{gather}
  \label{eq:app-spr}
  \bs{r}=r \bs{e}_\r, \\
  \bs{v}(\bs{r})= v_\r\bs{e}_\r + v_\theta\bs{e}_\theta + v_\varphi\bs{e}_\varphi,
\end{gather}
%\end{eqnarray*}
with
\begin{equation}
  \label{eq:app-spe}
  \bs{e}_\r = \left(
    \begin{array}{c}
      \cos\varphi\,\sin\theta \\
      \sin\varphi\,\sin\theta \\
      \cos\theta
    \end{array}\right),
  \quad
  \bs{e}_\theta = \left(
    \begin{array}{c}
      \cos\varphi\,\cos\theta \\
      \sin\varphi\,\cos\theta \\
      -\sin\theta
    \end{array}\right),
  \quad
  \bs{e}_\varphi = \left(
    \begin{array}{c}
      -\sin\varphi\\
      \cos\varphi \\
      0
    \end{array}\right).
\end{equation}
where $\varphi\in[0,2\pi[$ and $\theta\in[0,\pi]$.  This implies
\begin{equation}
  \label{eq:app-sprhs}
  \bs{e}_\r\times\bs{e}_\theta=\bs{e}_\varphi,\quad
  \bs{e}_\theta\times\bs{e}_\varphi=\bs{e}_\r,\quad
  \bs{e}_\varphi\times\bs{e}_\r=\bs{e}_\theta,
\end{equation}
and
\begin{align}
  \parder{\bs{e}_\r}{\theta}&=\bs{e}_\theta,&
  \parder{\bs{e}_\theta}{\theta}&=-\bs{e}_\r,&
  \parder{\bs{e}_\varphi}{\theta}&=0,\nonumber\\[0.5\baselineskip]
  \parder{\bs{e}_\r}{\varphi}&=\sin\theta\,\bs{e}_\varphi,&
  \parder{\bs{e}_\theta}{\varphi}&=\cos\theta\,\bs{e}_\varphi,&
  \parder{\bs{e}_\varphi}{\varphi}&=-\sin\theta\,\bs{e}_\r-\cos\theta\,\bs{e}_\theta.
  \label{eq:app-spderivatives}
\end{align}
%\begin{align}
%  \parder{\bs{e}_\r}{\theta}&=\bs{e}_\theta,&
%  \parder{\bs{e}_\r}{\varphi}&=\sin\varphi\,\bs{e}_\varphi,\\[0.5\baselineskip]
%  \parder{\bs{e}_\theta}{\theta}&=-\bs{e}_\r,&
%  \parder{\bs{e}_\theta}{\varphi}&=\cos\theta\,\bs{e}_\varphi,\\[0.5\baselineskip]
%  \parder{\bs{e}_\varphi}{\theta}&=0,&
%  \parder{\bs{e}_\varphi}{\varphi}&-\sin\theta\,\bs{e}_\r-\cos\theta\,\bs{e}_\theta.
%\end{align}
With
\begin{equation}
   \label{eq:app-spnabla}
   \nabla\;=\;\bs{e}_\r\parder{}{r}
   +\bs{e}_\theta\frac{1}{r}\parder{}{\theta}
   +\bs{e}_\varphi\frac{1}{r\sin\theta}\parder{}{\varphi}
\end{equation}
we obtain
\begin{eqnarray}
  \label{eq:app-spdiv}
     \nabla \bs{v}\!&\!=\!&\!\frac{1}{r^2}\parder{(r^2v_\r)}{r}
%     \nabla \bs{v}\!&\!=\!&\!\parder{v_\r}{r}+\frac{2v_\r}{r}
     +\frac{1}{r\sin\theta}\parder{(\sin\theta v_\theta)}{\theta}
     +\frac{1}{r\sin\theta}\parder{v_\varphi}{\varphi},
     \\[0.5\baselineskip]\label{eq:app-sprot}
     \nabla\times\bs{v}\!&\!=\!&\!
     \bs{e}_\r\frac{1}{r\sin\theta}\left(\parder{(\sin\theta v_\varphi)}{\theta}
       -\parder{v_\theta}{\varphi}\right)\nonumber\\*[0.5\baselineskip]
     \!&\!+\!&\!
     \bs{e}_\theta\left(\frac{1}{r\sin\theta}\parder{v_\r}{\varphi}
       -\frac{1}{r}\parder{(rv_\varphi)}{r}\right)\nonumber\\*[0.5\baselineskip]
     \!&\!+\!&\!
     \bs{e}_\varphi\left(\frac{1}{r}\parder{(rv_\theta)}{r}
       -\frac{1}{r}\parder{v_\r}{\theta}\right),
     \\[0.5\baselineskip]\label{eq:app-splaplace}
     \nabla^2\!&\!=\!&\!
     \frac{1}{r^2}\parder{}{r}\left(r^2\parder{}{r}\right)
%     \parder{^2}{r^2}+\frac{2}{r}\parder{}{r}
     +\frac{1}{r^2\sin\theta}\parder{}{\theta}
     \left(\sin\theta\parder{}{\theta}\right)
     +\frac{1}{r^2\sin^2\theta}\parder{^2}{\varphi^2}.\quad
%     \nonumber\\*[0.5\baselineskip]
%     &&\!\phantom{\parder{^2}{r^2}}
\end{eqnarray}

\chapter{Fourier sums for one-dimensional lattices}
\label{sec:app-hssums}

In this appendix we collect and proof some useful formulas for the
explicit calculations of the Haldane--Shastry model.  In particular,
we provide the Fourier sums required for the evaluation of the
coefficients $A_l$ in \eqref{eq:hs3} using two different methods,
first by contour integration loosely following Laughlin
\etal~\cite{laughlin-00proc}, and second by Feynmanesque algebra.

%\section{Evaluation by contour integration}

\vspace{.5\baselineskip}
For $\eta_\alpha=e^{i\frac{2\pi}{N}\alpha }$ with $\alpha = 1,\ldots ,N$
the following hold:
\begin{enumerate}
\setlength{\itemsep}{\myitemsep}

\item
  \begin{equation}
    \eta_{\alpha}^N=1.
    \label{eq:app-hs1}
  \end{equation}

\item
%   \begin{equation}
%     \sum_{\alpha=1}^N \eta_{\alpha}^m=\left\{
%       \begin{aligned} N&,\quad m=kN ,\; k\in\mathbb{Z}, \\ 
%         0&, \quad\text{otherwise}.
%       \end{aligned}\right.
%     \label{eq:app-hs2}
%   \end{equation}
  \begin{equation}
    \sum_{\alpha=1}^{N} \eta_\alpha^m = 
    % \left\{\begin{array}{cl} 
    %     N &\ \ \text{for}\ m=0\ \text{mod}\ N\\
    %     0 &\ \ \text{otherwise.}
    %   \end{array}\right.
    N\delta_{m,0}\quad \text{mod}\ N.
    \label{eq:app-hs2}
  \end{equation}

\item
  \begin{equation}
    \prod_{\alpha=1}^N(\eta-\eta_{\alpha})=\eta^N-1.
    \label{eq:app-hs3}
  \end{equation}
  {\em Proof:} The $\eta_\alpha$ are by
  definition roots of $1$. \hfill $\Box$

\item
  \begin{equation}
    \sum_{\alpha=1}^N\frac{1}{\eta-\eta_{\alpha}}=\frac{N\eta^{N-1}}{\eta^N-1}.
    \label{eq:app-hs4}
  \end{equation}
  {\em Proof:} Take $\frac{\partial}{\partial\eta}$ of
  \eqref{eq:app-hs3} and divide both sides by $\eta^N-1$.\hfill $\Box$

\item
  \begin{equation}
    \sum_{\alpha=1}^N\frac{\eta_{\alpha}}{\eta-\eta_{\alpha}}
    =\frac{N}{\eta^N-1}\;.
    \label{eq:app-hs5}
  \end{equation}
  {\em Proof:} Substitute $\eta_{\alpha}\to\frac{1}{\eta_{\alpha}}$, %and
  $\eta\to\frac{1}{\eta}$
  in \eqref{eq:app-hs4} and divide by $(-\eta)$.\hfill $\Box$

\item
  \begin{equation}
    \sum_{\substack{\alpha,\beta,\gamma=1\\[1pt] 
        \alpha\neq\beta\neq\gamma\neq\alpha}}^{N}
    \frac{\eta_{\gamma}^2}
    {(\eta_{\alpha}-\eta_{\gamma})(\eta_{\beta}-\eta_{\gamma})}
    =\frac{N (N-1)(N-2)}{3}.
%    =\frac{1}{3} {N (N-1)(N-2)}.
    \label{eq:app-hs6}
  \end{equation}
  {\em Proof:}  Use the algebraic identity
  \begin{equation}
    \label{eq:app-hsidentity}
    \frac{a^2}{(a-b)(a-c)}+\frac{b^2}{(b-a)(b-c)}+\frac{c^2}{(c-a)(c-b)} =1.
  \end{equation} %\\[-2\baselineskip]
  \vspace{-0.5\baselineskip}  \hfill{$\Box$}

\item
  \begin{equation}
    \sum_{\substack{\alpha,\beta=1\\[1pt] \alpha\neq\beta}}^{N-1}
    \frac{1}{(\eta_{\alpha}-1)(\eta_{\beta}-1)}
    =\frac{(N-1)(N-2)}{3}.
    \label{eq:app-hs7}
  \end{equation}
  {\em Proof:}  Substitute $\eta_{\alpha}\to{\eta_{\alpha}}{\eta_{\gamma}}$,
  $\eta_{\beta}\to{\eta_{\beta}}{\eta_{\gamma}}$ in \eqref{eq:app-hs6}

% FIRST IMPORTANT SUM

\item
  \begin{equation}
%    S_1(m)\equiv
    \sum_{\alpha=1}^{N-1}\frac{\eta_\alpha^m}{\eta_\alpha -1}=\frac{N+1}{2}-m,
    \quad 1\le m \le N
    \label{eq:app-hsfouriersum1}
  \end{equation}

  \begin{figure}[t]
    \begin{center}
      \includegraphics[scale =0.18]{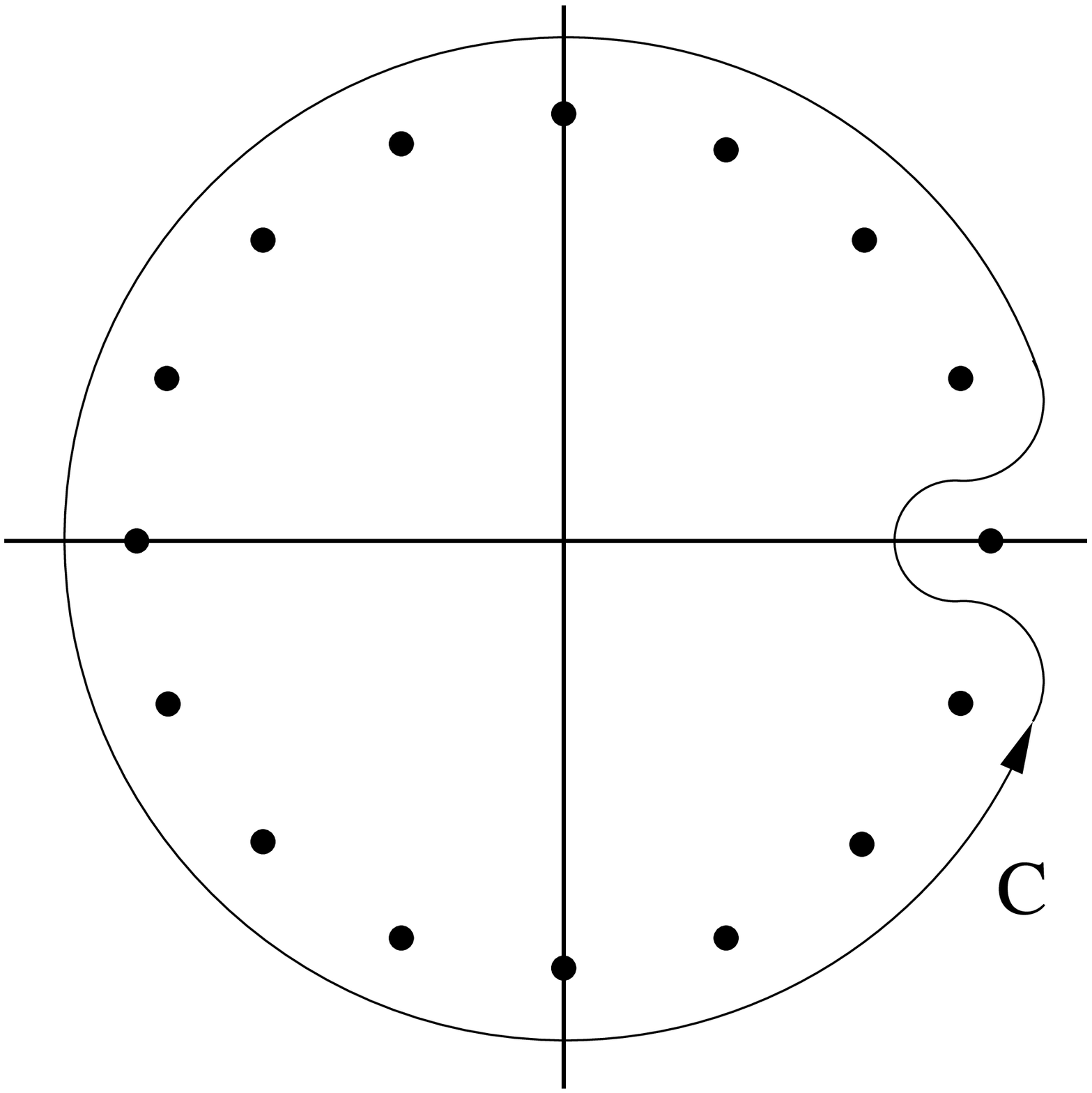}\hspace{1cm}
      \includegraphics[scale =0.18]{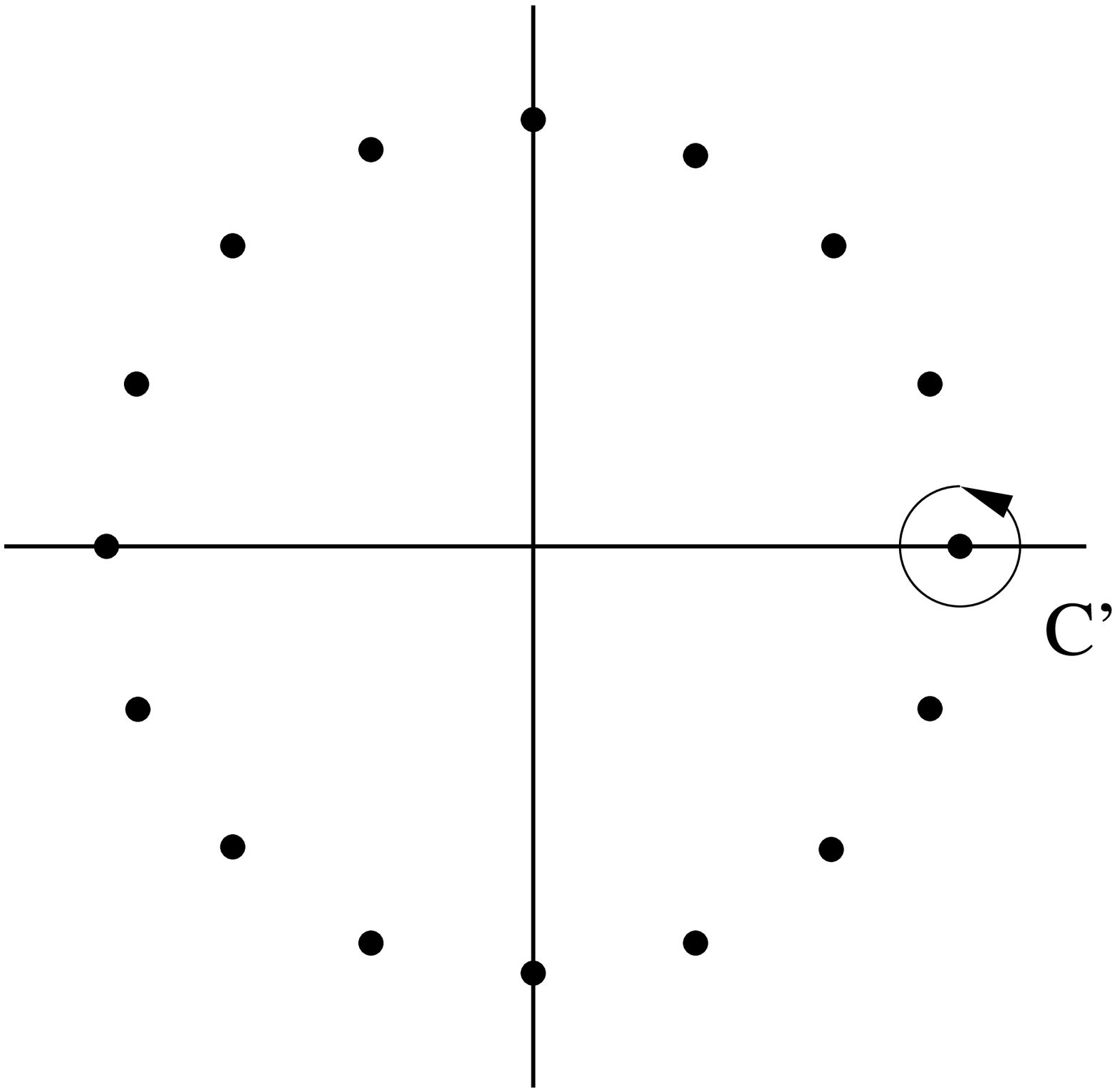}
    \end{center}
    \caption{Contours for integrations}\label{fig:app-hscontours}
  \end{figure}

  {\em Proof by contour integration:} 
  Use Cauchy's theorem~\cite{Lang85} for the function
  \begin{displaymath}
    f(z)=\frac{z^{m-1}}{z-1},\quad N \ge 2,
  \end{displaymath}
  with the contours shown in Fig.~\ref{fig:app-hscontours} yields
  \begin{eqnarray*}
    \sum_{\alpha=1}^{N-1}\frac{\eta_\alpha^m}{\eta_\alpha -1}
    &=&\frac{1}{2\pi \text{i}}\sum_{\alpha=1}^{N-1}
    \oint_C \frac{z^{m-1}}{z-1}\,\frac{\eta_\alpha}{z-\eta_\alpha} \text{d}z \\[2pt]
%    &\stackrel{\rm (i)}{=}&
    &=&\frac{N}{2\pi \text{i}}
    \oint_C 
    \frac{z^{m-1}}{(z-1)(z^N-1)}
    \text{d}z \\[2pt]
%    &\stackrel{\rm (ii)}{=}&
    &=&-\frac{N}{2\pi \text{i}}
    \oint_{C'}\underbrace{\frac{z^{m-1}}{(z-1)(z^N-1)}}_{=f(z)} \text{d}z 
  \end{eqnarray*}
  where we have first used \eqref{eq:app-hs5} and then deformed the
  contour C such that the radius of circle goes to infinity, used that
  the circle at infinity does not contribute to the integral as the
  integrand falls off as at least $1/z^2$ for $m\le N$, and finally
  reversed the direction of integration to replace C by C'.

  Since $f(z)$ has a pole of second order at $z=1$, the residue is given by
  \begin{equation*}
   c_{-1}
   =\lim_{z\to 1}\,\frac{\text{d}}{\text{d}z}\underbrace{(z-1)^2 f(z)}_{=1/g(z)}
   =-\lim_{z\to 1}\frac{g'(z)}{g^2(z)}.
  \end{equation*}
  With
  \begin{eqnarray}
    g(z)\!&=&\!\frac{1}{z^{m-1}}\frac{z^N-1}{z-1}=\sum_{k=1}^{N}z^{k-m}
    \ \xrightarrow{z\to 1}\ N, \label{eq:app-gz}\\
    g'(z)\!&=&\!\sum_{k=1}^{N}(k-m)\, z^{k-m-1}
    \ \xrightarrow{z\to 1}\ \frac{N(N+1)}{2}-mN, \label{eq:app-gz'}
  \end{eqnarray}
  we obtain
%   \begin{equation*}
%     c_{-1}=\lim_{z\to 1}\,\frac{d}{dz}\left((z-1)^2 f(z)\right)
%     \equiv\lim_{z\to 1}\,\frac{d}{dz}\,\frac{1}{g(z)}
%     =-\lim_{z\to 1}\frac{1}{g^2(z)}\frac{dg(z)}{dz}.
%   \end{equation*}
%   \begin{equation*}
%     g(z)=\frac{1}{z^{m-1}}\frac{z^N-1}{z-1}=\sum_{k=1}^{N}z^{k-m},\quad
%     \frac{dg(z)}{dz}=\sum_{k=1}^{N}(k-m) z^{k-m-1},\quad
%   \end{equation*}
%   we obtain
%   \begin{equation*}
%     \lim_{z\to 1}g(z)=N,\quad
%     \lim_{z\to 1}\,\frac{dg(z)}{dz}=\frac{N(N+1)}{2}-mN,
%   \end{equation*}
%   and hence
  \begin{equation*}
    \sum_{\alpha=1}^{N-1}\frac{\eta_\alpha^m}{\eta_\alpha -1}
    =-Nc_{-1}=\frac{(N+1)}{2}-m.
  \end{equation*}\hfill $\Box$ 

  \vspace{\baselineskip} 
  {\em Proof by algebra:} With the definition
  \begin{equation*}
    S_m\equiv\sum_{\alpha=1}^{N-1}\frac{\eta_\alpha^m}{\eta_\alpha -1},
    \label{eq:app-hsalgebra0}
  \end{equation*}
  we find
  \begin{equation*}
    S_{m+1}-S_m =\sum_{\alpha=1}^{N-1}\eta_\alpha^m
    =\begin{cases}
      \phantom{N}-1,& 1\le m\le N-1, \\
      N-1,& m=0.
    \end{cases}
    \label{eq:app-hsalgebra1}
  \end{equation*}
  and
  \begin{equation*}
    S_0=\sum_{\alpha=1}^{N-1}\frac{1}{\eta_\alpha -1}
    =-\sum_{\alpha=1}^{N-1}\frac{\eta_\alpha}{\eta_\alpha -1}=-S_1,
    \label{eq:app-hsalgebra2}
  \end{equation*}
  where we substituted $\eta_\alpha\to\frac{1}{\eta_\alpha}$.  
%
%  This directly implies $S_1=-S_0=\frac{N-1}{2}$ and
%  $S_m=\frac{(N+1)}{2}-m$ for $1\le m\le N$. \hfill $\Box$ 
%
%  \eqref{eq:app-hsfouriersum1}.\hfill $\Box$ 
%
  This directly implies 
  \begin{equation*}
    S_1=-S_0=\frac{N-1}{2}\quad\text{and}\quad 
    S_m=\frac{(N+1)}{2}-m,\quad 1\le m\le N.
  \end{equation*}

\item
  \begin{equation}
     \sum_{\alpha=1}^{N-1}\frac{1}{\eta_\alpha -1}=-\frac{N-1}{2}.
    \label{eq:app-hs9}
  \end{equation}
  {\em Proof:} Use \eqref{eq:app-hsfouriersum1} with $m=N$.\hfill $\Box$

% SECOND IMPORTANT SUM
%\pagebreak

\item
  \begin{equation}
    \sum_{\alpha=1}^{N-1}\frac{\eta_\alpha^m}{(\eta_\alpha -1)^2}=
    -\frac{N^2-1}{12}+\frac{(m-1)(N-m+1)}{2},
    \quad 1\le m \le N.
    \label{eq:app-hsfouriersum2}
  \end{equation}
  {\em Proof by contour integration:} 
  In analogy to the proof of \eqref{eq:app-hsfouriersum1} we write
  \begin{eqnarray*}
    \sum_{\alpha=1}^{N-1}\frac{\eta_\alpha^{m}}{(\eta_\alpha -1)^2}
    &=&\frac{1}{2\pi \text{i}}\sum_{\alpha=1}^{N-1}
    \oint_C \frac{z^{m-1}}{(z-1)^2}\,\frac{\eta_\alpha}{z-\eta_\alpha}
    \text{d}z\\[2pt]
    &=&\frac{N}{2\pi \text{i}}
    \oint_C     \frac{z^{m-1}}{(z-1)^2(z^N-1)}
    \text{d}z \\[2pt]
    &=&-\frac{N}{2\pi \text{i}}
    \oint_{C'}\underbrace{\frac{z^{m-1}}{(z-1)^2(z^N-1)}}_{=h(z)} \text{d}z 
  \end{eqnarray*}
  where we have again used \eqref{eq:app-hs5} and replaced the
  contour C by C'.
  As $h(z)$ has a now pole of third order at $z=1$, the residue is given by
  \begin{equation*}
    c_{-1}=\frac{1}{2}
    \lim_{z\to 1}\,
    \frac{\text{d}^2}{\text{d}z^2}\underbrace{(z-1)^3 h(z)}_{=1/g(z)}
%    =\lim_{z\to 1}\left(-\frac{g''}{2g^2}+\frac{(g')^2}{g^3}\right).
    =\lim_{z\to 1}\!\left(-\frac{g''(z)}{2g^2(z)} 
      + \frac{\left(g'(z)\right)^2}{g^3(z)}\right).
  \end{equation*}
  With $g(1)$ and $g'(1)$ as given by \eqref{eq:app-gz} and
  \eqref{eq:app-gz'} and
  \begin{eqnarray*}
    g''(z)\!\!\!&=&\!\!\!\sum_{k=1}^{N}(k-m)(k-m-1) z^{k-m-2}\\
    &\xrightarrow{z\to 1}& 
    \frac{N(N+1)(2N+1)}{6}-(2m+1)\frac{N(N+1)}{2}+m(m+1), 
  \end{eqnarray*}
  we find after some algebra that $-Nc_{-1}$ equals the expression on
  the right of \eqref{eq:app-hsfouriersum2}.\hfill $\Box$

  \vspace{\baselineskip} 
  {\em Proof by algebra:} With the definition
  \begin{equation*}
    R_m\equiv\sum_{\alpha=1}^{N-1}\frac{\eta_\alpha^m}{(\eta_\alpha -1)^2},
    \label{eq:app-hsalgebra3}
  \end{equation*}
  we find
  \begin{equation*}
    R_{m+1}-R_m =\sum_{\alpha=1}^{N-1}\frac{\eta_\alpha^m}{(\eta_\alpha -1)}
    =S_m
    \label{eq:app-hsalgebra4}
  \end{equation*}
  and
  \begin{eqnarray*}
    R_0 &=&\sum_{\alpha=1}^{N-1}\frac{1}{(\eta_\alpha -1)^2}\\
   &=&\sum_{\alpha=1}^{N-1}\sum_{\beta=1}^{N-1}
    \frac{1}{(\eta_\alpha -1)(\eta_\beta -1)}
   -\sum_{\substack{\alpha,\beta=1\\[1pt] (\alpha\neq\beta)}}^{N-1}
   \frac{1}{(\eta_{\alpha}-1)(\eta_{\beta}-1)}\\
%    &=&\frac{(N-1)^2}{4}-\frac{(N-1)(N-2)}{3}\\
%    &=&\underbrace{\sum_{\alpha=1}^{N-1}\frac{1}{(\eta_\alpha -1)}}_{-(N-1)/2}
%    \underbrace{\sum_{\beta=1}^{N-1}\frac{1}{(\eta_\beta -1)}}_{-(N-1)/2}
%    -\underbrace{\sum_{\substack{\alpha,\beta=1\\[1pt] (\alpha\neq\beta)}}^{N-1}
%    \frac{1}{(\eta_{\alpha}-1)(\eta_{\beta}-1)}}_{(N-1)(N-2)/3}\\
    &=&-\frac{(N-1)(N-5)}{12},
    \label{eq:app-hsalgebra5}
  \end{eqnarray*}
  where we have used \eqref{eq:app-hs9} and \eqref{eq:app-hs7}.  This implies
%  \begin{equation*}
%    R_{1}=R_0+S_0=-\frac{N^2-1}{12},
%    \label{eq:app-hsalgebra6}
%  \end{equation*}
%  and hence
  \begin{eqnarray*}
%     R_{m+1}&=&R_{1}+\sum_{n=1}^{m}S_n\\
     R_{m+1}&=&R_0+S_0+\sum_{n=1}^{m}S_n\\
%     &=&-\frac{N^2-1}{12}+\frac{(N+1)m}{2}-\sum_{n=1}^{m} n\\
     &=&-\frac{N^2-1}{12}+\sum_{n=1}^{m}\left(\frac{(N+1)}{2}- n \right)\\
     &=&-\frac{N^2-1}{12}+\frac{m(N-m)}{2}.
  \end{eqnarray*}
  for $1\le m \le N$.\hfill $\Box$ 

\item
  \begin{equation}
     \sum_{\alpha=1}^{N-1}\frac{1}{(\eta_\alpha -1)^2}=-\frac{(N-1)(N-5)}{12}.
    \label{eq:app-hs11}
  \end{equation}
  {\em Proof:} Use \eqref{eq:app-hsfouriersum2} with $m=N$.\hfill $\Box$

\item
  \begin{equation}
     \sum_{\alpha=1}^{N-1}\frac{\eta_\alpha^m}{\vert\eta_\alpha -1\vert^2}
     =\frac{N^2-1}{12}-\frac{m(N-m)}{2},
    \quad 0\le m \le N.
    \label{eq:app-hs12}
  \end{equation}
  {\em Proof:} Use \eqref{eq:app-hsfouriersum2} with $m\to m+1$.\hfill $\Box$

\item
  \begin{equation}
    \sum_{\substack{\alpha=1\\[1pt]\alpha\neq\beta}}^N
    \frac{\eta_\alpha +\eta_\beta}{\eta_\alpha - \eta_\beta}=0.
    \label{eq:app-hs13}
  \end{equation}
  {\em Proof:}  Substitute $\eta_\alpha\to\frac{1}{\eta_\alpha}$,
  $\eta_\beta\to\frac{1}{\eta_\beta}$ in one of the terms or
  use \eqref{eq:app-hsfouriersum1} and \eqref{eq:app-hs9}.\hfill $\Box$ 

\item
  \begin{equation}
    \label{eq:app-hs14}
    \sum_{\substack{\alpha=1\\[1pt]\alpha\neq\beta,\gamma\\[1pt]\beta\neq\gamma}}^N
 %    \sum_{\substack{\alpha=1\\[1pt]\alpha\neq\beta\neq\gamma\neq\alpha}}^N
   \frac{1}{\bar\eta_\alpha - \bar\eta_\beta}
    \frac{1}{\eta_\alpha - \eta_\gamma}
    = -\frac{\eta_\beta}{\eta_\beta - \eta_\gamma}
    +\frac{2}{\vert\eta_\beta -\eta_\gamma\vert^2}.
  \end{equation}
  {\em Proof:} With
  \begin{equation}
    \label{eq:app-hspartialfraction}
    \frac{1}{(\eta_{\alpha}-\eta_{\beta})(\eta_{\alpha}-\eta_{\gamma})}
    =\frac{1}{\eta_{\beta}-\eta_{\gamma}}
    \left(\frac{1}{\eta_{\alpha}-\eta_{\beta}}-
      \frac{1}{\eta_{\alpha}-\eta_{\gamma}}\right)  
  \end{equation}
  and
  \begin{equation}
    \sum_{\substack{\alpha=1\\[1pt]\alpha\neq\beta,\gamma\\[1pt]\beta\neq\gamma}}^N
    \frac{\eta_{\alpha}}{\eta_{\alpha}-\eta_{\beta}}
    =\frac{N-1}{2}-\frac{\eta_{\gamma}}{\eta_{\gamma}-\eta_{\beta}},
  \end{equation}
  which follows directly from \eqref{eq:app-hsfouriersum1}, we write
  \begin{eqnarray*}
    \sum_{\substack{\alpha=1\\[1pt]\alpha\neq\beta,\gamma\\[1pt]\beta\neq\gamma}}^N
 %    \sum_{\substack{\alpha=1\\[1pt]\alpha\neq\beta\neq\gamma\neq\alpha}}^N
   \frac{1}{\bar\eta_\alpha - \bar\eta_\beta}
    \frac{1}{\eta_\alpha - \eta_\gamma}
    &=& -\sum_{\substack{\alpha=1\\[1pt]
        \alpha\neq\beta,\gamma\\[1pt]\beta\neq\gamma}}^N
    \frac{\eta_\beta}{\eta_{\beta}-\eta_{\gamma}}
    \left(\frac{\eta_{\alpha}}{\eta_{\alpha}-\eta_{\beta}}-
      \frac{\eta_{\alpha}}{\eta_{\alpha}-\eta_{\gamma}}\right)\\
    &=& -\frac{\eta_{\beta}}{\eta_{\beta}-\eta_{\gamma}}
    \left(-\frac{\eta_{\gamma}}{\eta_{\gamma}-\eta_{\beta}}
      +\frac{\eta_{\beta}}{\eta_{\beta}-\eta_{\gamma}}\right)\\%[-1.1\baselineskip]
    &=& -\frac{\eta_{\beta}}{\eta_{\beta}-\eta_{\gamma}}
    \left(1+\frac{2\eta_{\gamma}}{\eta_{\beta}-\eta_{\gamma}}\right)
    \\[2pt]
    &=& -\frac{\eta_{\beta}}{\eta_{\beta}-\eta_{\gamma}}
    +\frac{2}{\vert\eta_\beta -\eta_\gamma\vert^2}.
  \end{eqnarray*}
  \hfill $\Box$ 

% \item For symmetric operators ${O}_{\beta\gamma}={O}_{\gamma\beta}$ it holds:
%   \begin{equation}
%     \label{eq:app-hs15}
% %    \sum_{\substack{\alpha=1\\[1pt]
% %        \alpha\neq\beta,\gamma\\[1pt]\beta\neq\gamma}}^N
%      \sum_{\substack{\alpha,\beta,\gamma=1\\[1pt]
%          \alpha\neq\beta\neq\gamma\neq\alpha}}^N
% \frac{{O}_{\beta\gamma}}
%    {(\bar\eta_\alpha - \bar\eta_\beta)(\eta_\alpha - \eta_\gamma)}
%    = \sum_{\beta\neq\gamma}
%    \frac{2{O}_{\beta\gamma}}{\vert\eta_\beta -\eta_\gamma\vert^2}
%    -\frac{1}{2}\sum_{\beta\neq\gamma}{O}_{\beta\gamma}.
% %     = \sum_{\beta\neq\gamma}
% %     \left(\frac{2}{\vert\eta_\beta -\eta_\gamma\vert^2}-\frac{1}{2}\right)
% %     {O}_{\beta,\gamma}
%   \end{equation}
% %  for symmetric operators ${O}_{\beta\gamma}={O}_{\gamma\beta}$.

\item For symmetric operators ${A}_{\beta\gamma}={A}_{\gamma\beta}$ it holds:
  \begin{equation}
    \label{eq:app-hs15}
    \sum_{\substack{\alpha,\beta,\gamma=1\\[1pt]
        \alpha\neq\beta\neq\gamma\neq\alpha}}^N
    \frac{{A}_{\beta\gamma}}
    {(\bar\eta_\alpha - \bar\eta_\beta)(\eta_\alpha - \eta_\gamma)}
    = \sum_{\beta\neq\gamma}
    \frac{2{A}_{\beta\gamma}}{\vert\eta_\beta -\eta_\gamma\vert^2}
    -\frac{1}{2}\sum_{\beta\neq\gamma}{A}_{\beta\gamma}.
  \end{equation}

  \vspace{.5\baselineskip}
  {\em Proof:} Use \eqref{eq:app-hs14}.\hfill $\Box$ 

\end{enumerate}

\chapter{Angular momentum algebra}
\label{sec:app-am}

In this appendix, we review a few very well known relations for
angular momentum operators~\cite{gottfried66,baym69}.  %We set $\hbar=1$.
The components of the angular momentum operator $\bs{J}$ obey
the SU(2) Lie algebra
\begin{equation}
  \label{eq:app-am:SU2}
%  \comm{J^i}{J^j}=\text{i}\epsilon^{ijk}J^k 
%  \quad \text{for}\quad i,j,k = x,y,z. 
  \comm{J^a}{J^b}=\text{i}\varepsilon^{abc}J^c 
  \quad \text{for}\quad a,b,c = \x,\y,\z. 
\end{equation}
Since $\comm{\bs{J}^2}{J^\z}=0$,
%\begin{equation*}
%  \comm{\bs{J}^2}{J^\z}=0,
%\end{equation*}
we can choose a basis of simultaneous eigenstates of $\bs{J}^2$ and
${J^\z}$,
\begin{equation}
  \begin{split}
    \bs{J}^2\ket{j,m}&=^2 j(j+1)\ket{j,m},    
    \\[0.2\baselineskip]
    J^\z\ket{j,m}&= m\ket{j,m},    
  \end{split}
\end{equation}
where $m=-j,\ldots,j$.  With $J^\pm\equiv J^\x\pm \text{i}J^\y$, we have
\begin{equation}
  \label{eq:app-am:commJzJ-}
  \comm{J^\z}{J^\pm}=\pm J^\pm.
\end{equation}
We further have
\begin{equation}
  \label{eq:app-am:J+J-}
  \begin{split}
    J^+ J^- & 
    =(J^\x)^2+(J^\y)^2-\text{i}\comm{J^\x}{J^\y}
    =\bs{J}^2-(J^\z)^2+ J^\z,
    \\[0.2\baselineskip]
    J^- J^+ &= \bs{J}^2-(J^\z)^2- J^\z,
  \end{split}
\end{equation}
and therefore
\begin{equation}
  \label{eq:app-am:commJ+J-}
  \comm{J^+}{J^-}=2 J^\z.
\end{equation}
Equations \eqref{eq:app-am:commJzJ-} and \eqref{eq:app-am:J+J-}
further imply
%\footnote{In principle, we can choose the phase between $J^-\ket{j,m}$ 
% and $\ket{j,m-1}$ for each $m\in [-l+1,l]$ arbitrarily.}
\begin{equation}
  \label{eq:app-am:Jpmketlm}
  \begin{split}
    J^\pm\ket{j,m}&=\sqrt{j(j+1)-m(m\pm 1)}\,\ket{j,m-1},
%    \\[0.2\baselineskip]
  \end{split}
\end{equation}
where we have chosen the phases between $J^-\ket{j,m}$ and $\ket{j,m-1}$
real.

\numberwithin{equation}{chapter}
\renewcommand{\theequation}{\Alph{chapter}.\arabic{section}.\arabic{equation}}

\chapter{Tensor decompositions of spin operators}
\label{sec:app-t}

In this appendix, we will write out the tensor
components~\cite{gottfried66,baym69} of all the tensors of different
order we can form from one, two, or three spins operators.

\section{One spin operator}
\label{sec:app-t-S}

A single spin $\bs{S}$ transforms as a vector under rotations, which
we normalize such that the $m=0$ component equals $S^\z$ (see
\eqref{eq:a:vectorrep} in Section \ref{sec:a:tensorops}).  The
components of $V^{m}$ are
\begin{equation}
  \label{eq:app-t-S}
  \begin{split}
    V^{1}&=-\frac{1}{\sqrt{2}}S^+,\\[0.2\baselineskip]
    V^{0}&=\frac{1}{\sqrt{2}} \comm{S^-}{V^{1}} = S^\z,\\[0.2\baselineskip]
    V^{-1}&=\frac{1}{\sqrt{2}} \comm{S^-}{V^{0}} = \frac{1}{\sqrt{2}}S^-.
  \end{split}
\end{equation}

\section{Two spin operators}
\label{sec:app-t-SS}

Since each spin operator transforms as a vector, and the representation
content of four vectors is given by
\begin{equation*}
  \bs{1}\otimes\bs{1}=\bs{0}\oplus\bs{1}\oplus\bs{2},
\end{equation*}
we can form one scalar, one vector, and one tensor of second order from
two spin operators $\bs{S}_1$ and $\bs{S}_2$.  The scalar is %trivially
given by
\begin{equation}
  \label{eq:app-t-SS-0}
   U_{12}=
  \bs{S}_1\bs{S}_2=\frac{1}{2}\left(S_1^+S_2^-+S_1^-S_2^+\right)+S_1^\z S_2^\z
\end{equation}
and the vector by $-\text{i} \left(\bsS_1\times\bsS_2\right)$.  Written out
in components, we obtain
\begin{equation}
  \label{eq:app-t-SS-1}
  \begin{split}
    V^{1}_{12}=\frac{\text{i}}{\sqrt{2}}(\bsS_1\times\bsS_2)^+
    &=\frac{1}{\sqrt{2}}\left(S_1^+S_2^\z-S_1^\z S_2^+\right),
    \\[0.2\baselineskip]
    V^{0}_{12}%=\frac{1}{\sqrt{2}} \comm{S_1^-+S_2^-}{V^{1}} 
    = -\text{i}(\bsS_1\times\bsS_2)^\z
    &=\frac{1}{2}\left(S_1^+S_2^--S_1^-S_2^+\right),
    \\[0.2\baselineskip]
    V^{-1}_{12}=-\frac{\text{i}}{\sqrt{2}}(\bsS_1\times\bsS_2)^-
    &=\frac{1}{\sqrt{2}}\left(S_1^-S_2^\z-S_1^\z S_2^-\right).
  \end{split}
\end{equation}

With regard to the 2nd order tensor, note that $S_1^+S_2^+$ is the
only operator we can construct with two spin operators which raises
the $S_{\s\text{tot}}^\z$ quantum number by two.  It must hence be
proportional to the $m=2$ component of the 2nd order tensor.  As there
is no particularly propitious way to normalize this tensor, we simply
set the $m=2$ component equal to $S_1^+S_2^+$, and then obtain the
other components using \eqref{eq:a:commJ+-T}.  
This yields\footnote{We denote general tensors of order $j$ with
  ${T^{(j)}}$ and 2nd order tensors with $T$.}
\begin{align}
  \label{eq:app-t-SS-2}
  T_{12}^2&=S_1^+S_2^+,
  \nonumber\\[0.2\baselineskip] 
  T_{12}^1&=
  \frac{1}{2} \comm{S_1^-+S_2^-}{T_{12}^2} = 
  - S_1^\z S_2^+ - S_1^+S_2^\z,
  \nonumber\\[0.2\baselineskip] 
  T_{12}^0&=
  \frac{1}{\sqrt{6}} \comm{S_1^-+S_2^-}{T_{12}^1} = 
  \frac{1}{\sqrt{6}} \left(4 S_1^\z S_2^\z - S_1^+S_2^- - S_1^-S_2^+\right),
  \\[0.2\baselineskip]\nonumber
  T_{12}^{-1}&=
  \frac{1}{\sqrt{6}} \comm{S_1^-+S_2^-}{T_{12}^0} = 
  S_1^\z S_2^- + S_1^-S_2^\z,
  \\*[0.2\baselineskip]\nonumber
  T_{12}^{-2}&=
  \frac{1}{2} \comm{S_1^-+S_2^-}{T_{12}^{-1}} = 
  S_1^-S_2^-.
\end{align}
Equations \eqref{eq:app-t-SS-0} and \eqref{eq:app-t-SS-2} imply
\begin{align}
  \label{eq:app-t-S+-S-+}
  \frac{1}{2}\left( S_1^+S_2^- + S_1^-S_2^+ \right) &=
  \frac{2}{3}\bs{S}_1\bs{S}_2 - \frac{1}{\sqrt{6}}\,T_{12}^0,
  \\[0.2\baselineskip]
  \label{eq:app-t-SzSz}
  S_1^\z S_2^\z &=
  \frac{1}{3}\bs{S}_1\bs{S}_2 + \frac{1}{\sqrt{6}}\,T_{12}^0.
\end{align}
Combining \eqref{eq:app-t-S+-S-+} with \eqref{eq:app-t-SS-1} yields
\begin{equation}
  \begin{split}
    \label{eq:app-t-S+S-}
    S_1^+S_2^-&=\frac{2}{3}\bs{S}_1\bs{S}_2 - \text{i}(\bsS_1\times\bsS_2)^\z
    -\frac{1}{\sqrt{6}}\,T_{12}^0,
    \\[0.2\baselineskip]
    S_1^-S_2^+&=\frac{2}{3}\bs{S}_1\bs{S}_2 + \text{i}(\bsS_1\times\bsS_2)^\z
    -\frac{1}{\sqrt{6}}\,T_{12}^0.
  \end{split}
\end{equation}
For $\bs{S}_1=\bs{S}_2$, \eqref{eq:app-t-S+S-} reduces 
with $\bs{S}_1\times\bs{S}_1=\text{i}\bs{S}_1$ to
\begin{equation}
  \begin{split}
    \label{eq:app-t-S1+S1-}
    S_1^+S_1^-&=\frac{2}{3}\bs{S}_1^2 + S_1^\z - \frac{1}{\sqrt{6}}\,T_{11}^0,
    \\[0.2\baselineskip]
    S_1^-S_1^+&=\frac{2}{3}\bs{S}_1^2 - S_1^\z - \frac{1}{\sqrt{6}}\,T_{11}^0.
  \end{split}
\end{equation}
%where $\bs{S}_1^2=s(s+1)$.
% For $\bs{S}_1=\bs{S}_2=\bs{S}$, \eqref{eq:app-t-S+S-} reduces 
% with $\bs{S}\times\bs{S}=\text{i}\bs{S}$ to
% \begin{equation}
%   \begin{split}
%     \label{eq:app-t-S1+S1-}
%     S^+S^-&=\frac{2}{3}\bs{S}^2 + S^\z- T^0,
%     \\[0.2\baselineskip]
%     S^-S^+&=\frac{2}{3}\bs{S}^2 - S^\z - T^0,
%   \end{split}
% \end{equation}
% where $\bs{S}^2=s(s+1)$.

\section{Three spin operators}
\label{sec:app-t-SSS}

Since 
\begin{equation*}
  \bs{1}\otimes\bs{1}\otimes\bs{1}
  =\bs{0}\oplus 3\cdot\bs{1}\oplus 2\cdot\bs{2}\oplus \bs{3},
\end{equation*}
we can form one scalar, three vectors, two tensors of second order, and
one tensor of third order, from three spin operators $\bs{S}_1$,
$\bs{S}_2$, and $\bs{S}_3$.

%\noindent 
The scalar is given by 
%\begin{align}
  \begin{equation}
    \label{eq:app-t-SSS-0}%
    \begin{split}
      U_{123}&=-\text{i}\bsS_1(\bsS_2\times\bsS_3)
      \\[0.2\baselineskip]
      &=\frac{1}{2} S_1^\z \left(S_2^+S_3^--S_2^-S_3^+\right) +\
      \text{2 cyclic permutations}
      \\[0.2\baselineskip]
      &=\frac{1}{2}\big( S_1^\z S_2^+S_3^- + S_1^+S_2^-S_3^\z +
      S_1^-S_2^\z S_3^+
      \\[0.4\baselineskip]
      &\hspace{13pt} -S_1^\z S_2^-S_3^+ - S_1^-S_2^+S_3^\z -
      S_1^+S_2^\z S_3^- \big).
      % \\[0.4\baselineskip]
      % &=\frac{\text{i}}{2}\big( S_1^\z S_2^+S_3^- + S_1^+S_2^-S_3^\z  +
      % S_1^-S_2^\z S_3^+ -S_1^\z S_2^-S_3^+ - S_1^-S_2^+S_3^\z  -
      % S_1^+S_2^\z S_3^- \big).
    \end{split}
  \end{equation}
%\end{align}

%If we assume that $\bsS_1$, $\bsS_2$, and $\bsS_3$ are all different,
The three vectors are given by
\begin{equation}
    \label{eq:app-t-SSS-1}
   \bsS_1(\bsS_2\bsS_3),\quad \bsS_1(\bsS_2)\bsS_3,
   \quad \text{and}\quad (\bsS_1\bsS_2)\bsS_3, 
\end{equation}
where the scalar product in the second expression is understood to
contract $\bsS_1$ and $\bsS_3$.  The components for each $m$ are
according to the conventions specified in \eqref{eq:app-t-S}.  For
later purposes, we write for the $m=0$ components,
\begin{align}
  \label{eq:app-t-SSS-1abc}
  {V}_{a,123}^{0} =S_1^\z(\bsS_2\bsS_3)
  &=\frac{1}{2}\big( S_1^\z S_2^+S_3^- + S_1^\z S_2^-S_3^+\big) + S_1^\z S_2^\z S_3^\z,
  \nonumber\\[0.4\baselineskip]
%  \label{eq:app-t-SSS-1b}
  {V}_{b,123}^{0} =\bsS_1(S_2^\z)\bsS_3
  &=\frac{1}{2}\big( S_1^-S_2^\z S_3^+ + S_1^+S_2^\z S_3^-\big) + S_1^\z S_2^\z S_3^\z ,
  \\[0.4\baselineskip]\nonumber
%  \label{eq:app-t-SSS-1c}
  {V}_{c,123}^{0} =(\bsS_1\bsS_2)S_3^\z
  &=\frac{1}{2}\big( S_1^+S_2^-S_3^\z + S_1^-S_2^+S_3^\z\big) + S_1^\z S_2^\z S_3^\z.
\end{align}

To obtain a tensor operator of second order, or more precisely the
\mbox{$m=2$} component of it, all we need to do is to form the product of the
\mbox{$m=1$} components of two vector operators constructed out of the three
spins, like $\bsS_1$ and $-\text{i}(\bsS_2\times\bsS_3)$ or
$-\text{i}(\bsS_1\times\bsS_2)$ and $\bsS_3$.  In this way, we
construct the tensor operators of second order
\begin{equation}
  \begin{split}
    T_{a,123}^2 &= -\text{i}S_1^+ (\bsS_2\times\bsS_3)^+,
    \\[0.4\baselineskip]
    T_{b,123}^2 &= -\text{i}(\bsS_1\times\bsS_2)^+ S_3^+. 
  \end{split}
\end{equation}
The other components are obtained as in \eqref{eq:app-t-SS-2}.  As we
are primarily interested in the $m=0$ component, we may use 
\eqref{eq:app-t-SS-2} directly to write
\begin{equation}
  \begin{split}
    \label{eq:app-t-SSS-2a}
    T_{a,123}^0
    &=-\frac{\text{i}}{\sqrt{6}} \big[
    4 S_1^\z(\bsS_2\times\bsS_3)^\z
    - S_1^+(\bsS_2\times\bsS_3)^- 
    - S_1^-(\bsS_2\times\bsS_3)^+\big]
    \\[0.4\baselineskip]
    &=\frac{1}{\sqrt{6}} \big[
      2 S_1^\z\left(S_2^+S_3^- - S_2^-S_3^+\right)
      - S_1^+S_2^-S_3^\z + S_1^+S_2^\z S_3^-
    \\[0.2\baselineskip] 
    &\hspace{127pt}
      + S_1^-S_2^+S_3^\z - S_1^-S_2^\z S_3^+\big],
\end{split}
\end{equation}
and similarly for $T_{b,123}^0$, which can be obtained from $T_{b,123}^0$
by a cyclical permutation of the superscripts $+,-,z$.  Note that there is
no third tensor of this kind, as the sum of the three tensors obtained
from \eqref{eq:app-t-SSS-2a} by cyclic permutations of the superscripts
equals zero.

We obtain the tensor of third order with the method we used to obtain
the second order tensor \eqref{eq:app-t-SS-2} formed by two spins:
\begin{align}
  \label{eq:app-t-SSS-3}
  W_{123}^3&=-S_1^+S_2^+S_3^+,
  \nonumber\\[0.3\baselineskip]
  W_{123}^2&= \frac{1}{\sqrt{6}}\comm{S_1^-+S_2^-+S_3^-}{W_{123}^3}
  \nonumber\\[0.2\baselineskip]
  &= -\frac{1}{\sqrt{6}} 
%  \underbrace{\comm{S_1^-}{S_1^+}}_{\textstyle =-2S_1^\z} S_2^+S_3^+ 
  \comm{S_1^-}{S_1^+} S_2^+S_3^+ +\ \text{2 cycl.\ permutations}
  \nonumber\\[0.2\baselineskip]
  &= \sqrt{\frac{2}{3}}\, S_1^\z S_2^+S_3^+ +\ \text{2 cycl.\ permutations},
  \nonumber\\[0.3\baselineskip]
  W_{123}^1&= \frac{1}{\sqrt{10}} \comm{S_1^-+S_2^-+S_3^-}{W_{123}^2}
  \nonumber\\[0.2\baselineskip]
  &= \frac{1}{\sqrt{15}}\big( \comm{S_1^-}{S_1^\z} S_2^+S_3^+
  +  S_1^\z\comm{S_2^-+S_3^-}{S_2^+S_3^+}\big) +\ \text{2 cycl.\ perms.}
%  +  S_1^\z\comm{S_2^-+S_3^-}{S_2^+S_3^+}\big) +\ \text{2 cycl.\ permutations}
  \nonumber\\[0.2\baselineskip]
  &= \frac{1}{\sqrt{15}}\big(S_1^-S_2^+S_3^+
  - 4 S_1^\z S_2^\z S_3^+\big) +\ \text{2 cycl.\ permutations},
  \nonumber\\[0.3\baselineskip]
  W_{123}^0&= \frac{1}{\sqrt{12}} \comm{S_1^-+S_2^-+S_3^-}{W_{123}^1}
  \nonumber\\[0.2\baselineskip]
  &= \frac{1}{6\sqrt{5}}\Big( 
  S_1^-\comm{S_2^-+S_3^-}{S_2^+S_3^+}
  - 4  \comm{S_1^-+S_2^-}{S_1^\z S_2^\z} S_3^+
  \nonumber\\*[0.2\baselineskip]
  &\hspace{40pt} - 4 S_1^\z S_2^\z \comm{S_3^-}{S_3^+}\Big)
  +\ \text{2 cycl.\ permutations}
  \nonumber\\[0.3\baselineskip]
  &= -\frac{1}{\sqrt{5}}\big(S_1^-S_2^+S_3^\z +\ \text{5 permutations}\big)
  + \frac{4}{\sqrt{5}} S_1^\z S_2^\z S_3^\z,
  \\[0.3\baselineskip]\nonumber
  W_{123}^{-1}&= \frac{1}{\sqrt{12}} \comm{S_1^-+S_2^-+S_3^-}{W_{123}^0}
  \\[0.2\baselineskip]\nonumber
%  &= \frac{1}{2\sqrt{15}}\big( 
%  - S_1^-\comm{S_2^-}{S_2^+}S_3^\z - S_1^-S_2^+ \comm{S_3^-}{S_3^\z}
%  + \ \text{5 permutations}
%  \\[0.2\baselineskip]\nonumber
%  &\hspace{40pt} + 4 \comm{S_1^-}{S_1^\z} S_2^\z S_3^\z 
%  +\ \text{2 cycl.\ permutations}\big)
%  \\[0.2\baselineskip]\nonumber
  &= -\frac{1}{\sqrt{15}}\big(S_1^-S_2^-S_3^+
  - 4 S_1^-S_2^\z S_3^\z \big) +\ \text{2 cycl.\ permutations},
  \\[0.3\baselineskip]\nonumber
  W_{123}^{-2}&= \frac{1}{\sqrt{10}}\comm{S_1^-+S_2^-+S_3^-}{W_{123}^{-1}}
  \\[0.2\baselineskip]\nonumber
%  &= -\frac{1}{\sqrt{10\cdot 15}}\big( 
%  S_1^-S_2^-\comm{S_3^-}{S_3^+} - 4 S_1^-\comm{S_2^-+S_3^-}{S_2^\z S_3^\z}\big)
%  \nonumber\\[0.2\baselineskip]
  &= \sqrt{\frac{2}{3}}\, S_1^-S_2^-S_3^\z +\ \text{2 cycl.\ permutations},
  \\[0.3\baselineskip]\nonumber
  W_{123}^{-3}&= \frac{1}{\sqrt{6}}\comm{S_1^-+S_2^-+S_3^-}{W_{123}^{-2}}
  \\[0.2\baselineskip]\nonumber
  &=S_1^-S_2^-S_3^-.
\end{align}
The permutations here always refer to permutations of the superscripts
$+,-,z$, as otherwise we would have to assume again that none of the
three spin operators are identical.  In particular, writing out the
$m=0$ yields
%\begin{equation}
%  \label{eq:app-t-SSS-3m=0}
%  \begin{split}
%    W_{123}^0&=\frac{4}{\sqrt{5}} S_1^\z S_2^\z S_3^\z
%    \\[0.2\baselineskip]
%    &\hspace{-10pt} -\frac{1}{\sqrt{5}}\big(
%    S_1^-S_2^+S_3^\z + S_1^+S_2^\z S_3^- + S_1^\z S_2^-S_3^+
%    + S_1^+S_2^-S_3^\z + S_1^-S_2^\z S_3^+ + S_1^\z S_2^+S_3^-
%    \big)
%  \end{split}
%\end{equation}
\begin{equation}
  \label{eq:app-t-SSS-3m=0}
  \begin{split}
    W_{123}^0&=-\frac{1}{\sqrt{5}}\Big(
    S_1^-S_2^+S_3^\z + S_1^+S_2^\z S_3^- + S_1^\z S_2^-S_3^+
    \\[0.2\baselineskip]
    &\hspace{30.6pt}     
    + S_1^+S_2^-S_3^\z + S_1^-S_2^\z S_3^+ + S_1^\z S_2^+S_3^- \Big)
    + \frac{4}{\sqrt{5}} S_1^\z S_2^\z S_3^\z.
  \end{split}
\end{equation}

Combining \eqref{eq:app-t-SSS-1abc}  and \eqref{eq:app-t-SSS-3m=0},
we obtain
\begin{equation}
  \label{eq:app-t-SzSzSz}
  S_1^\z S_2^\z S_3^\z
  = \frac{1}{5}\big( V_{a,123}^0 + V_{b,123}^0 + V_{c,123}^0 \big) 
  + \frac{1}{2\sqrt{5}}\,W_{123}^0,
\end{equation}
and hence
\begin{align}
  \label{eq:app-t-Sz+-w+}
  \frac{1}{2} S_1^\z \left( S_2^+S_3^- + S_2^-S_3^+ \right)
  &= V_{a,123}^0 - S_1^\z S_2^\z S_3^\z
  \nonumber\\[0.2\baselineskip]
  &= \frac{4}{5}\, V_{a,123}^0 
  - \frac{1}{5}\, V_{b,123}^0 -\frac{1}{5}\, V_{c,123}^0 - 
  \frac{1}{2\sqrt{5}}\,W_{123}^0.
%  - \frac{1}{5}\big(V_{231}^0 + V_{312}^0 \big) - W_{123}^0
%  &= \frac{4V_{123}^0-V_{231}^0- V_{312}^0}{5}  - W_{123}^0
\end{align}
From \eqref{eq:app-t-SSS-0} and \eqref{eq:app-t-SSS-2a} we obtain
\begin{align}
  \label{eq:app-t-Sz+-w-}
  \frac{\text{1}}{2} S_1^\z \left( S_2^+S_3^- - S_2^-S_3^+ \right)
  &= \frac{1}{3}\, U_{123} + \frac{1}{\sqrt{6}}\,T_{a,123}^0. 
%  \phantom{\;\, - \frac{1}{5}\, V_{b,123}^0 -\frac{1}{5}\, V_{c,123}^0.}
\end{align}
Combining  \eqref{eq:app-t-Sz+-w+} and \eqref{eq:app-t-Sz+-w-} we finally
obtain
\begin{align}
  \label{eq:app-t-Sz+-}
  S_1^\z S_2^+S_3^- &= +\frac{1}{3}\, U_{123} +\frac{1}{5}
  \left(4\;\!V_{a,123}^0-V_{b,123}^0- V_{c,123}^0\right)
  +\frac{1}{\sqrt{6}}\, T_{a,123}^0 -
  \frac{1}{2\sqrt{5}}\,W_{123}^0
  \nonumber\\*[0.2\baselineskip]
  &= +\frac{1}{3}\bsS_1(\bsS_2\times\bsS_3) +\frac{1}{5}
  \big[4\;\!  S_1^\z(\bsS_2\bsS_3) - \bsS_1(S_2^\z)\bsS_3 -
  (\bsS_1\bsS_2)S_3^\z \big]
  \nonumber\\*[0.2\baselineskip]
  &\quad +\frac{1}{\sqrt{6}}\, T_{a,123}^0 -
  \frac{1}{2\sqrt{5}}\,W_{123}^0,
  \\[0.4\baselineskip]
  \label{eq:app-t-Sz-+}
  S_1^\z S_2^-S_3^+
  &= -\frac{1}{3}\, U_{123} 
     +\frac{1}{5} \left(4\;\!V_{a,123}^0-V_{b,123}^0- V_{c,123}^0\right)
     -\frac{1}{\sqrt{6}}\, T_{a,123}^0 - \frac{1}{2\sqrt{5}}\,W_{123}^0
  \nonumber\\*[0.2\baselineskip]
  &= -\frac{1}{3}\bsS_1(\bsS_2\times\bsS_3)
     +\frac{1}{5} \big[4\;\! 
     S_1^\z(\bsS_2\bsS_3) - \bsS_1(S_2^\z)\bsS_3 - (\bsS_1\bsS_2)S_3^\z \big]   
  \nonumber\\*[0.2\baselineskip]
  &\quad -\frac{1}{\sqrt{6}}\, T_{a,123}^0 - \frac{1}{2\sqrt{5}}\,W_{123}^0.
\end{align}

%lp -dtkmsek -P 51-54 map.ps

\backmatter%%%%%%%%%%%%%%%%%%%%%%%%%%%%%%%%%%%%%%%%%%%%%%%%%%%%%%%%%%
%\include{glossary}
%\include{solutions}
%\printindex

%%%%%%%%%%%%%%%%%%%%%%%%%%%%%%%%%%%%%%%%%%%%%%%%%%%%%%%%%%%%%%%%%%%%%

%\newpage
%\bibliographystyle{spmpsci-habil}
%\bibliographystyle{spmpsci}
%\bibliographystyle{spphys}
%\bibliographystyle{../../../bib/prsty}
%\bibliographystyle{../../../bib/phdsty}
%\bibliography{../../../bib/book,../../../bib/paper,../../../bib/proc,../../../bib/unpub,../../../bib/htc}

%\include{references.tex}

\end{document}